\documentclass[phd,bottom,nosig]{usbthesis}
\usepackage{graphicx}
\usepackage{color}
\usepackage{bm}
\usepackage{amsmath}
\usepackage{amssymb}
\usepackage{setspace}
\usepackage{array}
\usepackage{tabularx}
\usepackage{multirow}
\usepackage{afterpage}
\usepackage{hhline}
\usepackage{url}
\usepackage{slashed}
\usepackage{booktabs}
\usepackage[colorlinks=true,colorlinks=true, linkcolor=blue]{hyperref} 
\usepackage{enumitem}
\usepackage[caption=false]{subfig}
\usepackage{float}
\usepackage{hyperref} 
\usepackage{footmisc} 
\usepackage{xspace}
\usepackage{dsfont}
\usepackage[numbers,sort&compress]{natbib}
\usepackage{pdfpages}
\usepackage{indentfirst}

\newcommand{\ket}[1]{\left| #1 \right\rangle}
\newcommand{\bra}[1]{\left\langle #1 \right|}
\newcommand{\braket}[2]{\left\langle #1 \middle| #2 \right\rangle}

\def\simge{%
    \mathrel{\rlap{\raise 0.511ex
    \hbox{$>$}}{\lower 0.511ex \hbox{$\sim$}}}}
\def\simle{%
    \mathrel{\rlap{\raise 0.511ex
    \hbox{$<$}}{\lower 0.511ex \hbox{$\sim$}}}}

\DeclareMathAccent{\ring}{\mathalpha}{operators}{"17}

\providecommand{\renewoperator}[3]{\renewcommand*{#1}{\mathop{#2}#3}}
\renewoperator{\Re}{\mathrm{Re}}{\nolimits}
\renewoperator{\Im}{\mathrm{Im}}{\nolimits}

\makeatletter
\providecommand*{\diff}{\@ifnextchar^{\DIfF}{\DIfF^{}}}
\def\DIfF^#1{\mathop{\mathrm{\mathstrut d}}\nolimits^{#1}\gobblespace}
\def\gobblespace{\futurelet\diffarg\opspace}
\def\opspace{%
    \let\DiffSpace\!%
    \ifx\diffarg(%
        \let\DiffSpace\relax
    \else
        \ifx\diffarg[%
            \let\DiffSpace\relax
        \else
            \ifx\diffarg\{%
                \let\DiffSpace\relax
            \fi\fi\fi\DiffSpace}

\author{Wei-Yang Liu}%
\title{A Quantitative Framework of Nonperturbative QCD from Topological Vacuum with Application to Parton Structures}%

\month{\textbf{May}}
\year{\textbf{2026}}%
\program{Physics}%
\director{\textbf{Ismail Zahed}}{Professor, Department of Physics and Astronomy}%
\chairman{\textbf{Edward Shuryak}}{Professor, Department of Physics and Astronomy}%
\fstmember{\textbf{Abhay Deshpande}}{Professor, Department of Physics and Astronomy}%
\sndmember{\textbf{Alexander Abanov}}{Professor, Department of Physics and Astronomy}%
\outmember{\textbf{Yoshitaka Hatta}}{Scientist}{Brookhaven National Laboratory}%
\dean{\textbf{Celia Marshik}}%
\begin{document}

\singlespacing %
\pagenumbering{roman} %
\maketitle %
\makeapproval %

\begin{abstract}
This dissertation develops a quantitative framework that provides a physical picture for
first-principles QCD in continuum limit based on the topological structure of the infrared
QCD and constrained by lattice QCD and experimental data. By integrating out the
ultraviolet degrees of freedom, the infrared gluon configurations are modeled as a liquid
ensemble of instantons and anti-instantons, which induce nonlocal effective interactions
among light quarks. This framework captures the origin of trace and axial anomalies through
the infrared distributions of QCD, dynamically breaks chiral symmetry, and describes the
confinement at intermediate distance scales. More specifically, this framework can be
formulated in two ways: in one, we construct a statistical ensemble with weights defined by
the instanton action and quark determinant, while in the other we formulate an effective
field theory by rewriting the quark determinant as effective quark interactions.
By reformulating the framework on the light front, we explicitly construct the light-front
wave functions and derive distribution amplitudes (DAs), parton distribution functions
(PDFs) for light hadrons. We further embed the framework into TMD factorization and
establish a vacuum origin for rapidity evolution by computing the soft functions and
transverse momentum dependent distributions (TMDs). The resulting partonic structure
matches its Euclidean counterpart after analytic continuation, establishing a direct
connection between vacuum structure and light-front dynamics.
We extend this approach to various form factors in light hadrons, including scalar,
pseudoscalar, and energy-momentum tensor (EMT), as well as higher-twist color force and
multigluon correlations, with applications to hadron mass and spin decomposition, near-threshold quarkonium production, and strong CP problem, highlighting the importance of the vacuum origin in
hadron structures. Overall, this work demonstrates that the topological QCD vacuum
provides a quantitatively crucial description of hadronic structure from low to moderate
resolution, bridging nonperturbative dynamics with partonic phenomenology.
\end{abstract}

\begin{dedication}
\it To my parents and those who never stopped believing in me.
\end{dedication}

\tableofcontents %
\listoffigures %
\listoftables %
\begin{acknowledgements}
    As my Ph.D. journey comes to an end, I would like to acknowledge all those who have supported me along the way, for this work would not have been possible without them.

My advisor, Ismail Zahed, has always been a beacon of support, guiding me through the times when I felt lost in the swamp of research. He has nurtured my curiosity about the physical world and consistently encouraged me to pursue new ideas with confidence and rigor. His mentorship has been instrumental in shaping my development as an independent theoretical physicist, while also keeping me from straying too far off course in research. I have truly valued working with such an inspiring advisor, whose dedication to hadron and nuclear physics reflects a shared pursuit of understanding the fundamental building block of our visible matter in the universe.

I would like to thank the rest of my defense committee, Edward Shuryak, Abhay Deshpande, Alexander Abanov, and Yoshitaka Hatta, for their time and guidance on my defense.

I would also like to thank my longtime collaborator and mentor, Yong Zhao, for his constant support throughout my Ph.D. journey. He has been not only an outstanding research partner but also a wonderful friend. Through our collaboration, I have learned a great deal about perturbative QCD techniques and their interplay with lattice QCD calculations. He has always looked out for me and has made every effort to seek out opportunities for young researchers.


Xiangdong Ji has provided many valuable opportunities that have shaped my development in physics. I truly appreciate to work with such a passionate and insightful physicist, and his guidance has offered profound perspectives on my research journey in nuclear physics and quantum field theory.

I would also like to thank Jinchen He and Yushan Su for inspiring me to pursue new and interesting research directions during my Ph.D studies. They have also been wonderful friends, and I have greatly enjoyed our many discussions across a wide range of topics in physics, especially quantum field theory and computational physics. I wish them the very best in their research careers and look forward to our paths crossing again.

My dearest and first American friends from the CFNS summer school at Stony Brook, Nicholas Baldonado, Brandon Manly, and Richard Whitehill, made my Ph.D journey more memorable. The time shared with them was truly enjoyable, and our many discussions offered valuable insights into TMDs and small-$x$ physics from phenomenological, computational, and theoretical perspectives. These experiences provided a broader view of QCD and were especially meaningful for my academic career. I have missed them dearly while being apart across the country. I wish their research journeys continue to flourish, and I look forward to meeting again somewhere along the way.

I would also like to thank my fellow postdocs and colleagues in the Nuclear Theory Group at Stony Brook University, in particular David Franklakh, Sebastian Grieninger, Fangcheng He, Florian Hechenberger, Kazuki Ikeda, Shuzhe Shi, and Rajeev Singh, for their kindness, support, and meaningful discussions throughout my Ph.D. I really enjoyed the few trips with some of them over the past few years. 
Florian also shared many experiences about how to navigate academic life as a postdoc in different countries. I am also grateful to David for his advice and his experience as a senior student. Their company made my Ph.D journey especially enjoyable. Without them, I believe my journey would have been much more difficult. 

Ming-Wei Chou, my dearest Taiwanese friend from the Department of Chemistry at Stony Brook University, has been a constant support outside my department. Through stories from daily lab life, shared PhD struggles, and reflective philosophical conversations, her companionship has been an oasis amid the long and often solitary journey of a Ph.D.

My former husband, Tso-Chun Chen, has offered constant support throughout my Ph.D. journey. I am deeply grateful for his patience, kindness, and care for my well-being, as well as the steady emotional support he provided during a particularly challenging period of my academic life, always giving his best. Although our paths eventually parted, the unwavering support he offered during this journey remains deeply meaningful to me and is something I will always carry with me.

Last but certainly not least, I would like to thank my dearest companion and partner, Austin Bryan. Without him, I would not have been able to complete my journey toward pursuing my dream of working in QCD. He has been there for me during my lowest moments and has shared in the joy of my highest ones. For his patience, love, and unwavering support, I will be forever grateful. I am deeply thankful to have him as such a wonderful partner.
\end{acknowledgements}
\pagestyle{thesis}
\newpage
\pagenumbering{arabic}
\chapter{Introduction}

\begin{figure}[t]
    \centering
    \includegraphics[width=0.7\textwidth]{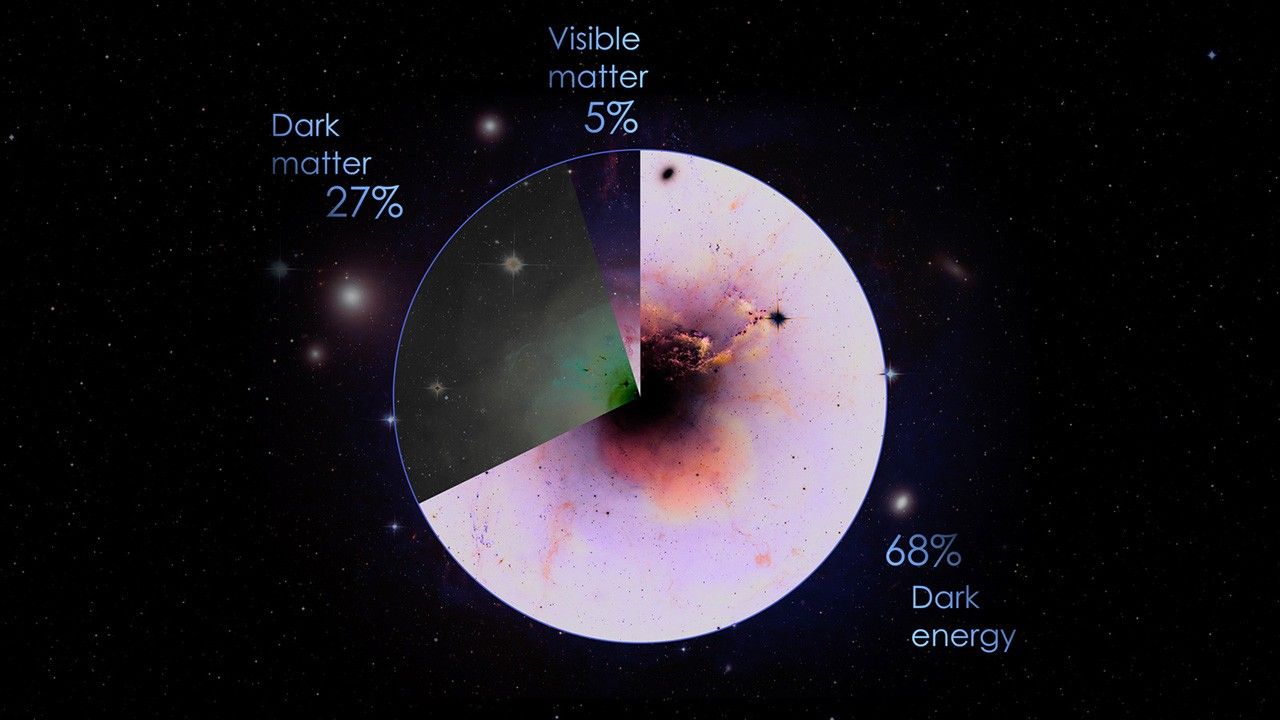}
    \caption{The universe is made up of three components: visible matter (5\%), dark matter (27\%), and dark energy (68\%). Image credit: \href{https://science.nasa.gov/dark-matter/}{NASA's Goddard Space Flight Center}.}
    \label{fig:universe_composition}
\end{figure}

In the universe, the overwhelming majority of its composition, namely dark matter and dark energy, remains largely unknown, leaving the approximately 5\% of visible matter as the only sector currently accessible through direct experiments, as illustrated in Fig.~\ref{fig:universe_composition}. At the heart of this visible world are hadrons, whose structure is governed by the strong interaction. Quantum chromodynamics (QCD), as the fundamental theory describing how the strong interaction binds quarks and gluons into hadrons, therefore occupies a central role in our understanding of matter. Unraveling its intricate and highly nontrivial dynamics is essential not only for explaining the emergence of hadronic structure, but also for establishing a robust foundation upon which potential signals of physics beyond the Standard Model can be identified.

Although hadron physics is fundamentally rooted in QCD, its low-energy degrees of freedom differ from those in the high-energy regime. Despite more than five decades of development, the nonperturbative structure of QCD remains only partially understood, and establishing a quantitative connection between hadronic observables and the underlying quark and gluon dynamics continues to be a central challenge. From a perspective of a quantum field theory (QFT), the QCD vacuum represents the ground state of the theory, while hadrons correspond to its lowest excitations. Consequently, the dependence of hadronic properties on quantum numbers provides valuable insight into the structure of the vacuum itself. The gauge fields that constitute the QCD vacuum can be generally categorized into perturbative gluon fluctuations, which resemble quantum excitations around the trivial vacuum, and nonperturbative fields with amplitudes $\sim\mathcal{O}(1/g)$, associated with extended topological configurations of the gauge fields. These include instantons and related solitonic objects, which are distinct from small-amplitude gluonic plane waves and reflect the inherently nontrivial topology of the vacuum.

Since the 1980s, the mainstream for exploring the QCD vacuum has been the large-scale numerical simulations performed on supercomputers. In parallel with the development of lattice QCD as a first-principles approach, a variety of semiclassical models have been proposed, describing the vacuum as an ensemble of solitonic gauge-field configurations. The earliest discovery of such objects in non-Abelian gauge theories were magnetic monopoles \cite{tHooft:1974kcl,Polyakov:1974ek}, followed by the discovery of instantons \cite{Belavin:1975fg}. More recently, instanton-dyons \cite{Kraan:1998sn}, also known as instanton monopoles, have emerged as a unifying framework that is particularly relevant in the vicinity of QCD phase transitions.

\begin{figure}
 \centering
\subfloat[\label{fig:vac_Biddle}]{\includegraphics[width=0.6\linewidth]{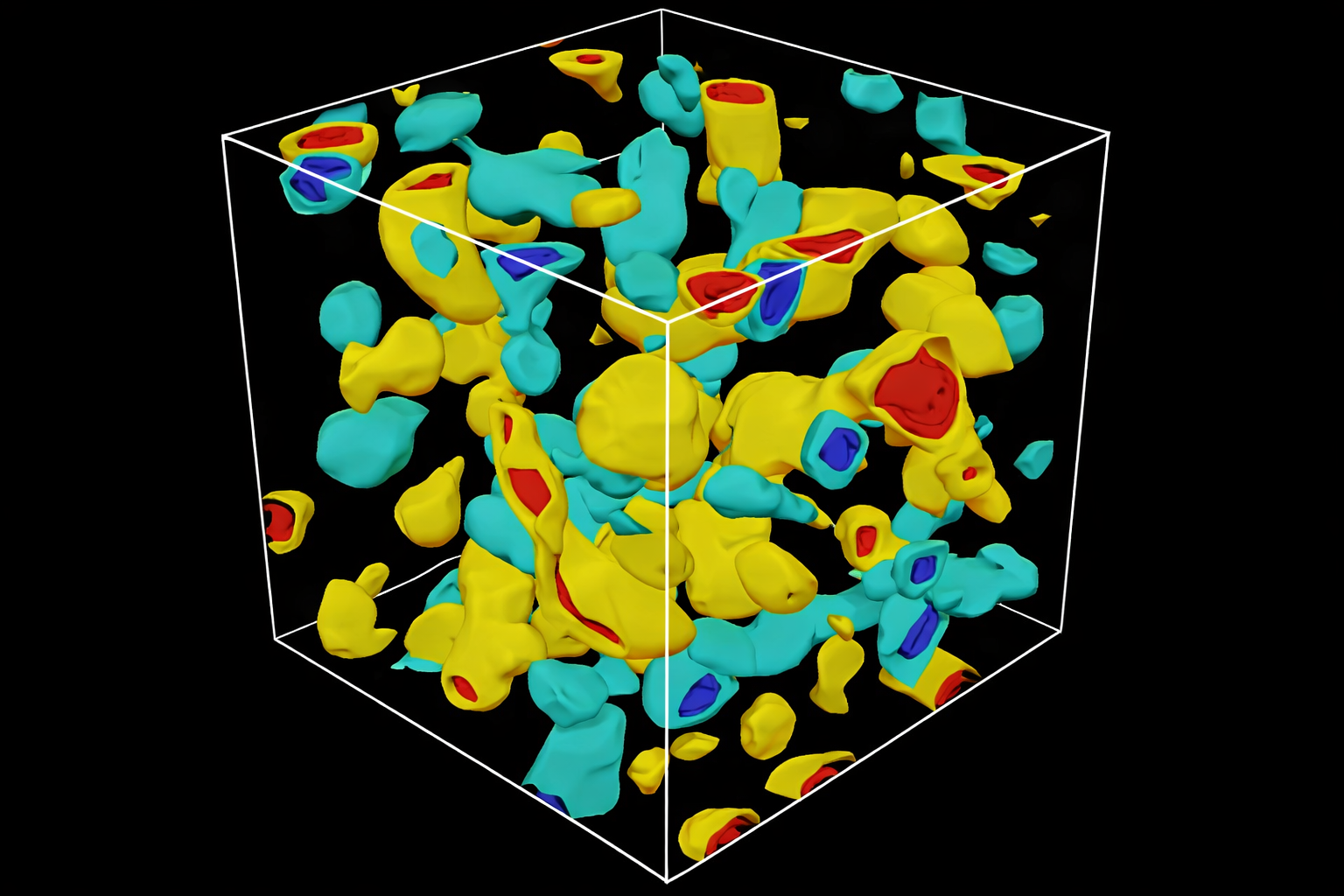}}
\hfill
\subfloat[\label{fig:vac_q}]{\includegraphics[width=0.4\linewidth]{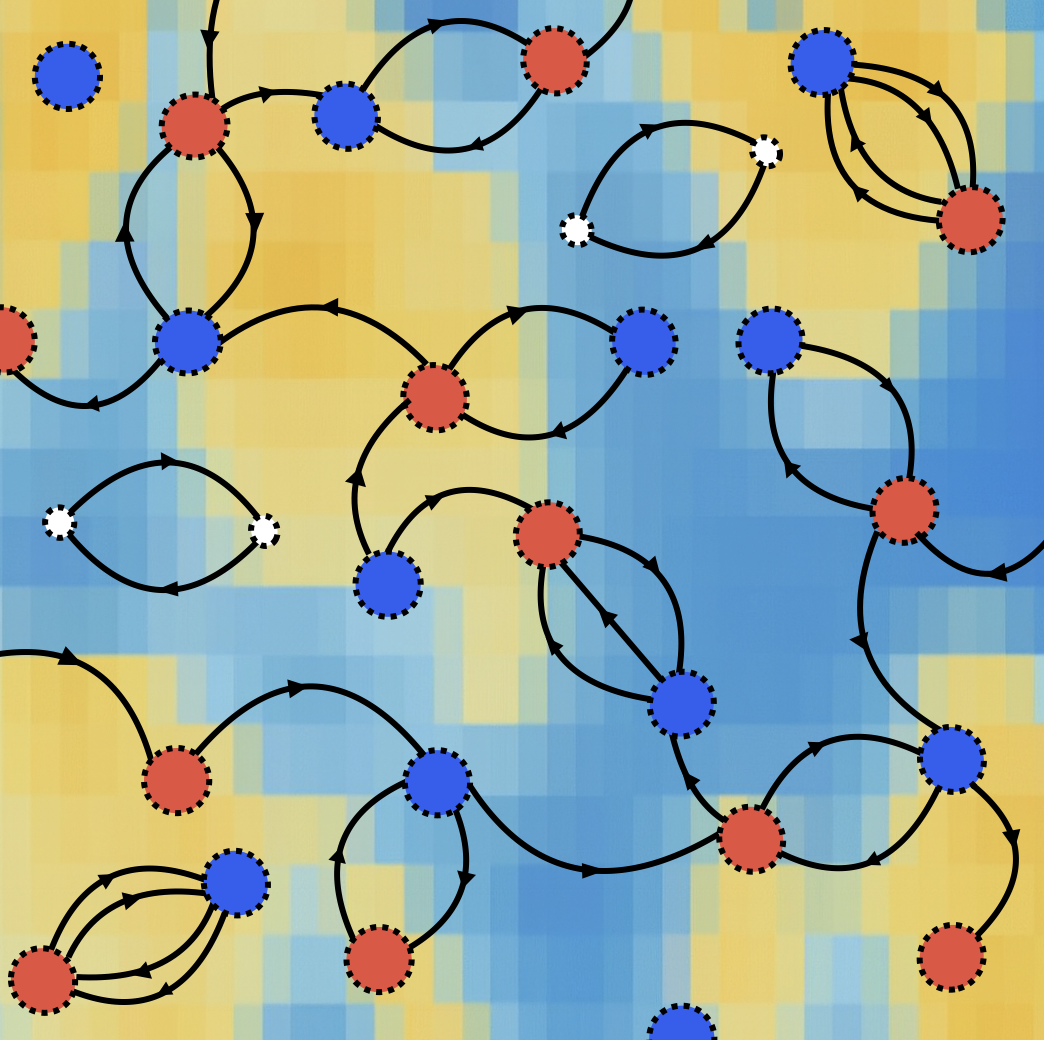}}
    \caption{(a) Lattice visualization of a YM vacuum (Image credit: \href{https://www.bu.edu/tech/support/research/whats-happening/highlights/lattice-qcd09/}{LSD collaboration: C. Rebbi, A. Avakian, R. Babich, R. Brower, M. Clark, J. Osborn, D. Schaich, R. Gasser}). (b) the cartoon illustration for an unquenched vacuum.}
    \label{fig:vacuum}
\end{figure}

The QCD vacuum consists of semi-classical and topological instantons and anti-instantons (pseudoparticles) \cite{Diakonov:1983hh}, that are described by an ensemble of instantons and anti-instantons (quenched) with additional fermionic determinantal interactions (unquenched)~\cite{Callan:1976je, Callan:1977gz}. The key features of this landscape are~\cite{Shuryak:1981ff}
\begin{equation}
n_{I+A}\equiv \frac 1{R^4}\approx 1~{\rm fm}^{-4},  \qquad\qquad\rho\approx \frac 13 \,\mathrm{fm}   \label{eqn_ILM}
\end{equation}
for the instanton plus anti-instanton density and their average size, respectively. The hadronic scale $R=1\,{\rm fm}$ emerges as the mean quantum tunneling rate of the pseudoparticles which gives the dimensionless packing fraction $\kappa=n_{I+A}\pi^2\rho^4\approx0.1$. 
A large body of analytical and numerical work, from both phenomenological and theoretical studies, has demonstrated that such topological vacuum structures play a crucial role in governing nonperturbative QCD. These include mechanisms underlying axial and trace anomalies, spontaneous chiral symmetry breaking, and QCD phase transitions at finite temperature and density, as well as the emergence of hadronic structure, including many properties of the pions and the anomalously heavy $\eta'$ meson \cite{Diakonov:1985eg,Vainshtein:1981wh,Diakonov:1995ea,Miesch:2023hvl,Shuryak:2022wtk,Shuryak:2022thi,tHooft:1986ooh,Shuryak:2026pqt,Shuryak:2018fjr} and a comprehensive review \cite{Schafer:1996wv} (and references therein).

From the perspective of lattice QCD, large-scale numerical simulations provide a cornerstone for exploring the nonperturbative regime of the theory, offering a systematically improvable framework for first-principles calculations. However, while lattice QCD enables increasingly precise determinations of physical observables, it offers limited direct insight into the underlying physical mechanisms governing the vacuum structure. This limitation can introduce ambiguities in the interpretation and extraction of physical quantities. These considerations motivate the development of complementary nonperturbative frameworks that provide a more transparent physical picture of the QCD vacuum.

Among such approaches, the Instanton Liquid Model (ILM) offers a particularly compelling description of the underlying gauge-field configurations at low resolution, capturing essential aspects of the vacuum structure in terms of an interacting ensemble of topological objects. This perspective naturally raises the question of how such vacuum structures emerge within the lattice formulation itself. Indeed, a large body of substantial lattice evidence highlights the importance of topological configurations \cite{Michael:1994uu,Michael:1995br,Leinweber:1999cw,Bakas:2010by,Biddle:2018bst,Hasenfratz:2019hpg,Athenodorou:2018jwu,Biddle:2020eec,Zimmermann:2024mar}, with some studies even providing indications consistent with the instanton liquid picture \cite{Ringwald:1999ze,Faccioli:2003qz} (and references therein).


\section{Overview of Dissertation}

The main contributions of this dissertation are organized as follows:
\begin{itemize}
    \item Chapter~\ref{ch:vac} reviews the QCD vacuum structure under gredient flow renormalization group, its topological origin, and their role in nonperturbative dynamics.
    
    \item Chapter~\ref{ch:ILM} reviews the theoretical foundations of the ILM, including both YM theory and QCD (YM with quarks). It introduces the standard formulation and summarizes prior developments as well as extensions explored in our recent work in Refs.~\cite{Liu:2024rdm,Liu:2025ldh}.

   \item Chapter~\ref{ch:low_QCD} presents the chiral aspects of low-energy QCD emerging from the instanton vacuum and the resulting spectroscopy and chiral dynamics of mesons and baryons within the ILM framework. The connection to light-front quantization is also discussed. Parts of this chapter are based on Refs.~\cite{Liu:2023yuj,Liu:2023fpj} and are further extended with preliminary results presented here for future reference.
   
   \item Chapter~\ref{ch:FF} presents a systematic study of hadronic form factors (QCD local operators) within the ILM framework, covering both quark and gluonic operators across different kinematic regimes of $Q^2$. It examines the breakdown of perturbative factorization at intermediate $Q^2$ and demonstrate the emergence of nonperturbative relation among quark and gluon form factors driven by the instanton vacuum. Particular emphasis is placed on scalar, pseudoscalar, axial-vector, and energy–momentum tensor (EMT) form factors, as well as their connections to the trace and axial anomalies and their implications for nucleon mass and spin decomposition. Parts of this chapter are based on Refs.~\cite{Liu:2024rdm,Liu:2024jno,Liu:2024vkj}. The new re-analysis for \cite{Liu:2024rdm,Liu:2024jno,Liu:2024vkj} upon several minor corrections are also presented.

\item Chapter~\ref{ch:wilson} extends the ILM framework to nonlocal QCD operators, focusing on Wilson lines in both quenched and unquenched QCD vacuum. Particular emphasis is placed on the role of topological configurations and their impact on the nonperturbative structure of Wilson loops. Parts of this chapter build upon Ref.~\cite{Liu:2024sqj}.

\item With parameters constrained by experiments and lattice QCD, Chapters~\ref{ch:near_th}, \ref{ch:twist3}, and \ref{ch:CP} apply the preceding ILM framework to several phenomenological topics: heavy meson photoproduction near threshold based on \cite{Liu:2024yqa}, the color Lorentz force and twist-3 dynamics based on \cite{Liu:2025ypg}, and strong $CP$ via the nucleon electric dipole moment based on \cite{Liu:2025kuc}. Several important nontrivial relations between gluonic and quark operators, which are not apparent in perturbative QCD, are established, providing new perspectives for future lattice QCD studies and experiments. The new re-analysis for \cite{Liu:2024yqa} upon several minor corrections are also presented.

\item Chapter~\ref{ch:had-to-par} establishes the connection between the ILM and parton physics by formulating light-front observables at finite resolution within a gradient-flow picture, and examines their relation to PDFs, DAs, TMDs, and GPDs. It further discusses the matching procedure to perturbative QCD and the relation to Large Momentum Effective Theory (LaMET), demonstrating how low-resolution observables can be systematically related to high-energy parton distributions. Parts of this chapter are based on Refs.~\cite{Liu:2023yuj,Liu:2023fpj}. The new re-analysis for~\cite{Liu:2023yuj,Liu:2023fpj} upon several minor corrections are also presented. While the conceptual framework is established, a rigorous demonstration and quantitative implementation remain in progress.

\item Chapter~\ref{ch:tmd} develops TMD factorization within the ILM framework, emphasizing TMD PDFs and soft functions and their connection to high-energy semi-inclusive processes and current phenomenological studies. Parts of this chapter are based on Refs.~\cite{Liu:2024sqj,Liu:2025mbl,Liu:2025fuf}

\item Chapter~\ref{ch:conclusion} presents the conclusions.

\item In the appendices, we summarize the conventions and notations in the ILM (Appendices~\ref{App:conv}, \ref{App:singular}, \ref{App:ZM}), technical details (Appendices~\ref{App:NZM}, \ref{App:average}, \ref{App:pair}), supplementary derivations and mathematical identities used throughout this dissertation (Appendices~\ref{app:mole}, \ref{App:pol}, \ref{app:Gordan}, \ref{app:vac_pol}, \ref{App:Fierz}, \ref{app:Baryon}), as well as explicit forms of light-front wave functions (Appendix~\ref{app:LCWF}).

\end{itemize}

\chapter{QCD vacuum}
\label{ch:vac}
\section{QCD vacuum in gradient flow}
\label{sec:vac_GF}
\begin{figure}
    \centering
\subfloat[\label{GF1}]{\includegraphics[width=0.16\linewidth]{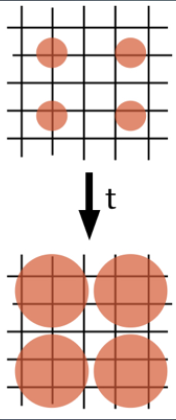}}
\hfill
\subfloat[\label{GF2}]{\includegraphics[width=0.75\linewidth]{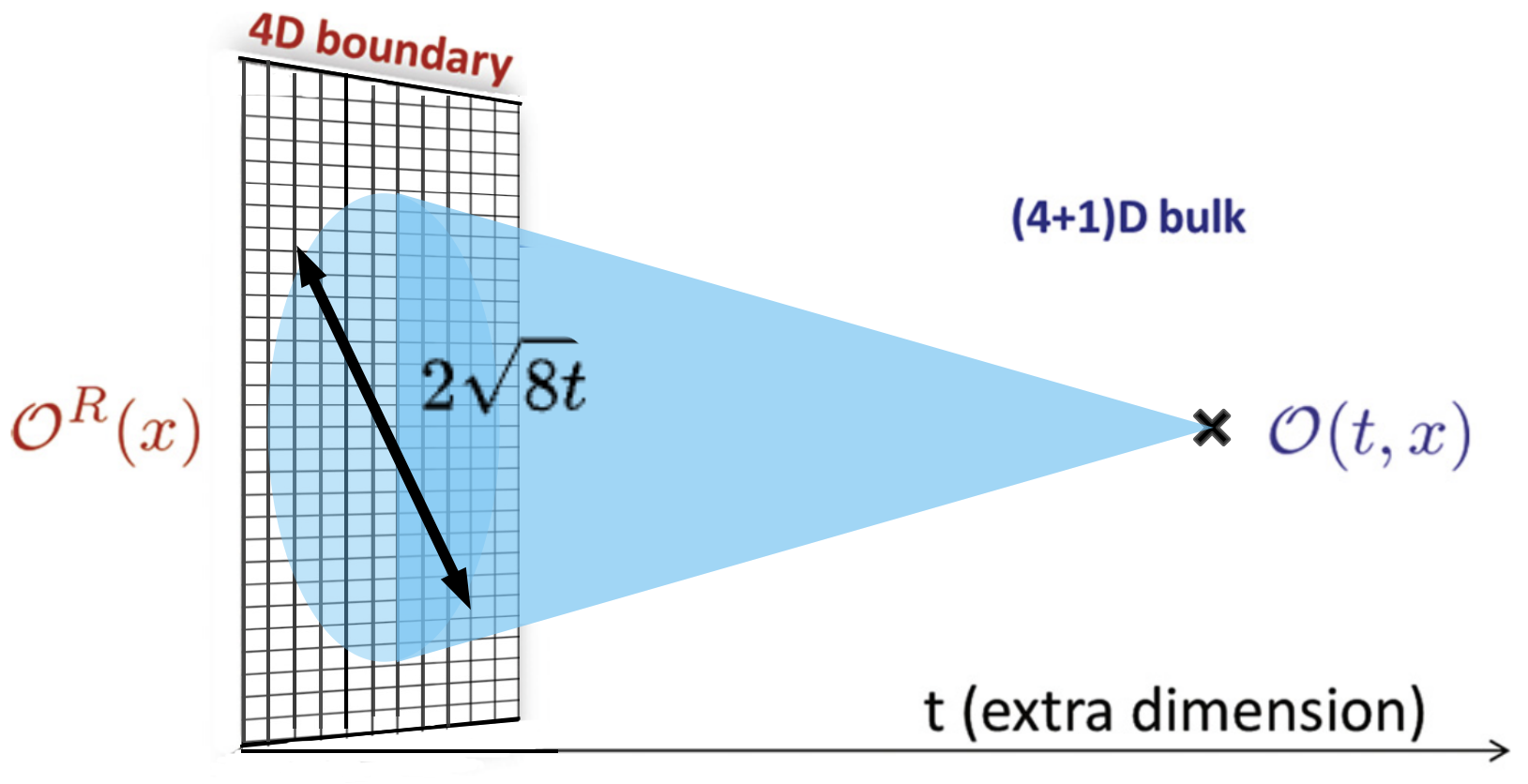}}
    \caption{A schematic illustration of the Yang–Mills gradient flow toward the direction of the fictitious time }
    \label{fig:GF}
\end{figure}

The structure of the QCD vacuum is inherently dependent on the resolution scale. An intuitive and gauge-invariant way to implement this scale dependence is through Lüscher gradient flow~\cite{Luscher:2009eq}, which is a renormalization group (RG) evolution governed by a diffusive equation of motion with gauge fields evolving in a fictitious flow time $t$

\begin{equation}
   \partial_t A_\mu=\frac{\delta S_{QCD}}{\delta A_\mu}[A_\nu]=D_\nu F_{\nu\mu}
\end{equation}

In this RG scheme, the topological content of the gauge field is preserved during the flow as these configurations correspond to local minima of the action. As the flow time $t$ increases, ultraviolet (UV) gluonic fluctuations above a given scale are mostly removed. Thus, a clean topological landscape can be revealed at sufficiently large $t$.

As illustrated in Fig.~\ref{GF1}, the gradient flow RG smooths gauge fields over a characteristic radius $\sim \sqrt{8t}$, exponentially suppressing UV modes carrying momenta $p \gg 1/\sqrt{8t}$. As a result, operators evaluated at finite flow time are automatically UV finite, providing a controlled and continuous way to remove short-distance fluctuations while preserving long-distance topological structures. The gradient flow scheme is conceptually equivalent to the Wilsonian RG flow, offering a nonperturbative realization of RG evolution for first-principles calculations. More specifically, the RG equation can be derived by

\begin{equation}
\langle \mathcal{O}(e^{2\xi} t, e^{\xi} x) \rangle_{g}
=
Z(\xi)\,\langle \mathcal{O}(t, x) \rangle_{g(\xi)},
\end{equation}
which mirrors the structure of Wilsonian RG transformations 

As illustrated in Fig.~\ref{GF2}, the operator constructed from flowed gauge fields can be mapped to a smooth renormalized operator by a perturbative expansion at small flow time~\cite{Luscher:2011bx}, allowing a direct connection between flowed operators $\mathcal{O}(t,x)$ and renormalized local operators $\mathcal{O}^R(x)$. The small flow-time expansion (SFTE) can be understood as an operator product expansion (OPE) along the additional flow-time direction $t$, where the flow time acts as a resolution scale,

\begin{equation}
\mathcal{O}(t,x) \xrightarrow[t \to 0]{} \sum_k c_k(t)\,\mathcal{O}_k^{R}(x) + \text{(powers in } t\text{)},
\end{equation}
where the coefficients $c_k(t)$ can be computed perturbatively due to asymptotic freedom. For more details, see \cite{Shuryak:2026pqt,Luscher:2013vga,Makino:2018rys,Narayanan:2006rf} (and references therein).

\subsection{QCD vacuum in different flow-time limits}

\begin{figure}
 \centering
\includegraphics[width=.85\linewidth]{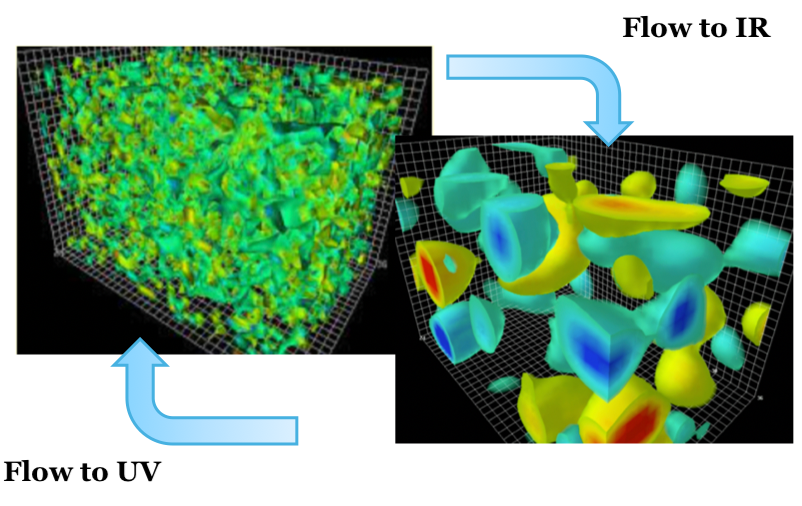}
\caption{Visualization of the QCD vacuum in different resolutions}
    \label{fig:vacuum_flow}
\end{figure}

As illustrated in Fig.~\ref{fig:vacuum_flow}, gauge configurations are initially dominated by UV gluonic modes at a resolution scale of approximately $\frac{1}{10}~\mathrm{fm}$ (top).  In lattice QCD, the short  wavelengths of those noisy modes are characterized by a cut-off scale set by the lattice spacing $\sim a$. After gradient-flow cooling, a coarser resolution of about $\frac{1}{3}~\mathrm{fm}$ is reached (bottom). The vacuum becomes significantly smoother, revealing the emergence of pseudoparticle structures~\cite{Moran:2008xq}. As the flow time increases, the perturbative gluons are gradually removed  while topological objects such as instantons persist due to their stability as (approximate) local minima of the action~\cite{Moran:2008xq,Athenodorou:2018jwu}. 
After a moderate amount of cooling, the gauge field typically resembles a dense ensemble of strongly correlated instanton–anti-instanton pairs. With further flow, these pairs are gradually annihilated, leaving behind a more dilute ensemble of individual instantons and anti-instantons that remain robust even under deep cooling. This picture provides the basis for the ILM, for which comprehensive reviews can be found in \cite{Schafer:1996wv,Shuryak:2018fjr,Musakhanov:2023dsn,hutter2001instantonsqcdtheoryapplication,Liu:2025ldh}.

\begin{figure}
    \centering
\subfloat[\label{rho}]{\includegraphics[width=0.5\linewidth]{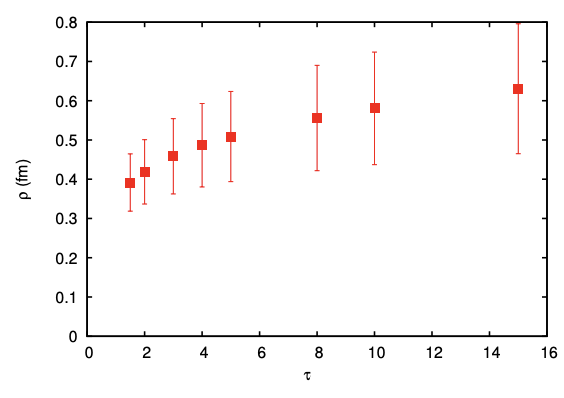}}
\hfill
\subfloat[\label{cool2}]{\includegraphics[width=0.5\linewidth]{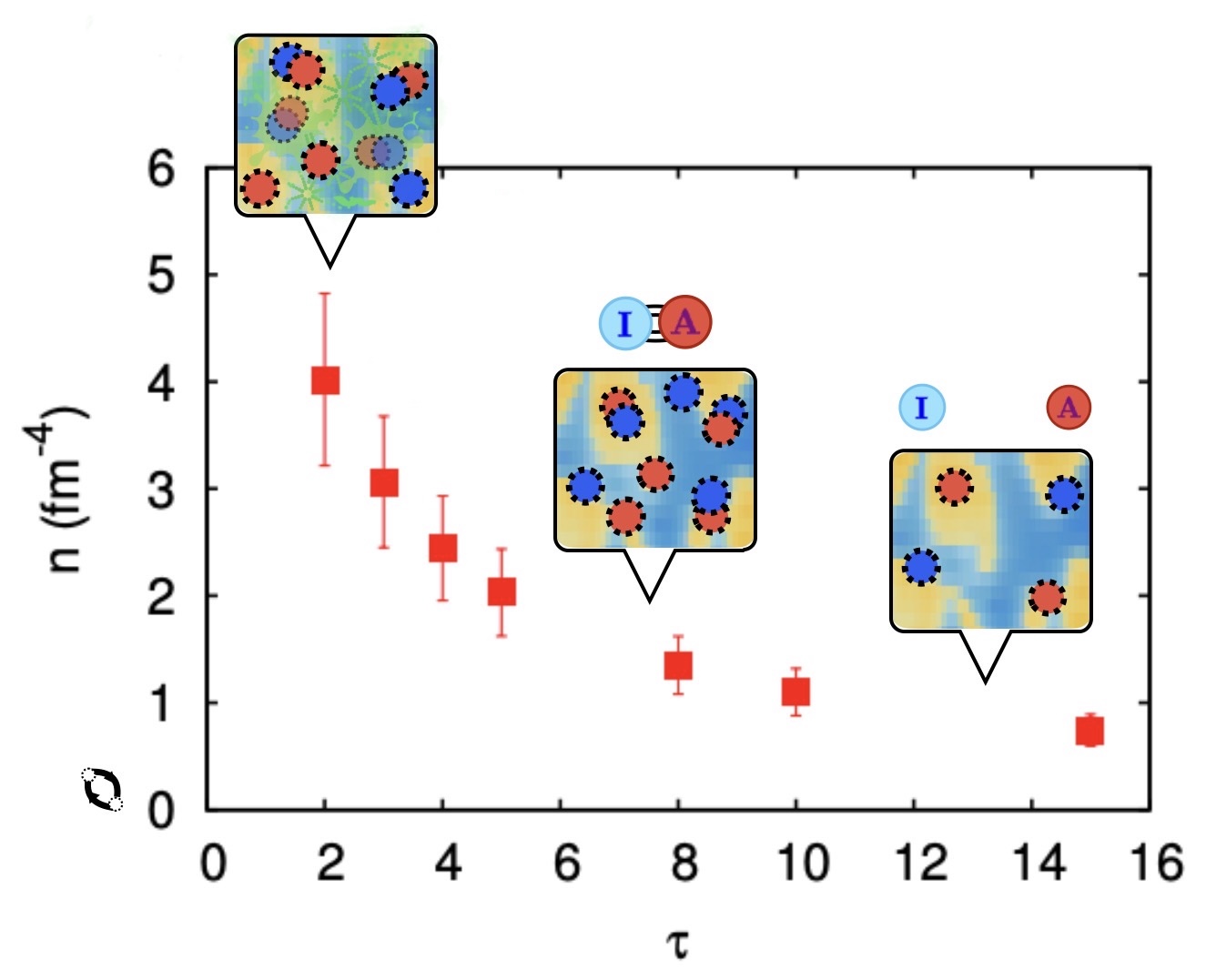}}
    \caption{Dependence of the mean instanton size $\rho$ (left panel) and density $n=n_{I+A}$ (right panel) on the dimensionless gradient flow time $\tau=t/a^2$ with lattice spacing $a=0.046$ fm~\cite{Athenodorou:2018jwu}, determined by $\sqrt{8t_0}=0.1$ fm, or $1/\sqrt{8t_0}\sim2$ GeV. The quantum vacuum corresponds to the extrapolation $\tau\rightarrow0$}
    \label{fig:cool2}
\end{figure}

The dependence of the mean instanton size and density on the gradient-flow time $t$ determined from the lattice analysis in \cite{Athenodorou:2018jwu} is presented in Fig.~\ref{fig:cool2}. The key values upon extrapolation to zero flow time ($t \to 0$) are
\begin{equation}
\rho \;\to\; \frac{1}{3}\,\mathrm{fm}, 
\qquad 
n_{I+A} \;\to\; 10\,\mathrm{fm}^{-4},
\end{equation}
indicating that the density of nonperturbative gauge fields is significantly larger than the standard estimate of the ILM. While the ILM is recovered in the deep-cooling regime (large $t$), where $n_{I+A} \sim 1\,\mathrm{fm}^{-4}$.  More specifically, by relating cooling time $t$ to the renormalization scale by $\mu\sim 1/\sqrt{8t}$, we found out, deep in the cooling time ($\tau=9$) or low resolution $\mu=0.51~\mathrm{GeV}\sim 1/\rho$, the tunnelings are sparse, well described by the ILM  with a dilute packing fraction $ \kappa\equiv \pi^2\rho^4 n_{I+A}\approx 0.1$, as illustrated in Fig.~\ref{cool2}.
In this regime, most instantons are annihilated with anti-instantons, leaving a dilute ensemble of isolated instantons and anti-instantons stable under action minimization~\cite{Shuryak:1981ff}. This corresponds to the regime where the spontaneous breaking of chiral symmetry is commonly observed. At shorter cooling times ($\tau=0.6$) or high resolution $\mu=2~\mathrm{GeV}$, the larger density $n_{I+A}\sim 7.46~\mathrm{fm}^{-4}$ is reached as more instanton-anti-instanton pairs are present.

Between these two limits of flow time, the full quantum vacuum ($t\rightarrow0$) and the dilute instanton liquid ensemble ($t\rightarrow \rho^2/8\sim0.01125$ fm$^2$) there exists an intermediate regime characterized by strong gauge-field fluctuations that are neither self-dual nor directly associated with near-zero Dirac eigenmodes. These configurations were not included in the original ILM, as they do not contribute directly to spontaneous chiral symmetry breaking. Nevertheless, they represent genuine nonperturbative dynamics and can play an important role in observables sensitive to gauge fields, such as Wilson lines (see Ch.~\ref{ch:wilson}).

A particularly important subset of such configurations consists of closely spaced instanton–anti-instanton pairs, often referred to as “molecules.” These structures have been observed in lattice simulations and are especially relevant in regimes where chiral symmetry is restored. For example, at temperatures above the critical temperature $T_c$ (in the chiral limit), instanton–anti-instanton molecules become the dominant surviving topological configurations. Their role has been extensively studied in the context of QCD phase transitions, as well as in dense matter, where they contribute to quark pairing and color superconductivity \cite{Rapp:1997zu,Rapp:1999qa}. A unified framework incorporating both isolated (“atomic”) instantons and molecular configurations was first developed in \cite{Ilgenfritz:1988dh}, and more recent studies have explored their impact on nonperturbative contributions to hadronic observables, including mesonic form factors and matching kernels \cite{Liu:2023fpj,Liu:2024rdm,Shuryak:2021fsu}.

The interaction between instantons and anti-instantons is closely related to topological transitions in gauge theories, such as sphaleron processes. The streamline configurations connecting such pairs have been studied extensively, beginning in quantum mechanics \cite{Shuryak:1987tr} and later generalized to non-Abelian gauge theories \cite{Balitsky:1986qn}. These configurations have been analyzed both analytically and numerically \cite{Yung:1987zp,Verbaarschot:1991sq}, with notable results showing that the Yung ansatz provides a remarkably accurate description even at relatively small separations. Many of these analyses exploit conformal transformations that map the instanton–anti-instanton pair to a common center, thereby simplifying the study of their interaction dynamics. More details about instanton molecules will be further discussed in Secs.~\ref{sec:landscape} and \ref{sec:mole}.

The observed dramatic dropping of the instanton density in the gradient flow smearing can be primarily attributed to pair annihilation, leading to the equal decreasing rates of both $n_I$ and $n_A$. If we assume this is a first order process based on the collision picture, the flow time evolution of instanton and anti-instanton density will be given by

\begin{equation}
\frac{dn_I}{d\tau} = \frac{dn_A}{d\tau} = -\lambda(\tau) n_I(\tau) n_A(\tau)
\end{equation}
Here the rate constant $\lambda$ may vary with the flow time $\tau$ via the instanton size and inter-pseudoparticle distance. For simplicity, we assume that it
is well described by a constant. By assuming the initial condition $n_I=n_A=n_{I+A}/2$, we have \cite{Athenodorou:2018jwu}

\begin{equation}
n_{I,A}(\tau) = \frac{n_{I,A}(0)}{1 + \lambda n_{I,A}(0) \tau}
\end{equation}
where the numerical fitting of Fig.~\ref{cool2} indicates $\lambda=0.1678$ fm$^4$.

\section{Topology of QCD vacuum}

The classical vacuum of a gauge theory corresponds to configurations with vanishing field strength,
\begin{equation}
F_{\mu\nu}^a = 0,
\end{equation}
and hence zero energy. In an Abelian theory, such vacuum configuration corresponds to constant gauge potentials. 
However, in a non-Abelian gauge theory, such condition does not eliminate the nontrivial pure gauge field. For instance, in the temporal gauge $A^0 = 0$, 
\begin{equation}
A_i(x) = i\, U^\dagger(x) \partial_i U(x),
\end{equation}
where $U(x) \in SU(N_c)$ is a gauge transformation. 

The set of all possible vacuum configurations is therefore classified by the set of gauge transformations $U(x)$. 
These configurations are not all equivalent: they can be grouped into distinct topological sectors according to their boundary behavior. The physical boundary condition
\begin{equation}
U(\mathbf{x}) \to \mathbf{1} \quad \text{as} \quad |\mathbf{x}| \to \infty,
\end{equation}
compactifies $\mathbb{R}^3$ to $S^3$, so that vacuum configurations are classified by mappings
\begin{equation}
U: S^3 \to SU(N_c).
\end{equation}

These mappings fall into homotopy classes characterized by an integer-valued winding number $n_w \in \mathbb{Z}$,
\begin{equation}
\pi_3\big(SU(N_c)\big) = \mathbb{Z}.
\end{equation}
where the winding number is given by
\begin{equation}
n_w = \frac{1}{24\pi^2} \int d^3\vec{x} \, \epsilon^{ijk} \,
\mathrm{Tr} \left[
(U^\dagger \partial_i U)
(U^\dagger \partial_j U)
(U^\dagger \partial_k U)
\right].
\end{equation}

These winding number $n_w$ can be further identified with the Chern--Simons number in terms of the gauge fields in temperal gauge,
\begin{equation}
\label{eq:winding}
n_w = N_{\mathrm{CS}}= \frac{1}{16\pi^2} \int d^3x \, \epsilon^{ijk}
\left(
A_i^a \, \partial_j A_k^a 
+ \frac{1}{3} f^{abc} A_i^a A_j^b A_k^c
\right).
\end{equation}
The winding number is invariant under continuous (small) deformations of the gauge field as long as the deformation does not change the topology. Thus, smooth variations of the gauge fields cannot change the value of $N_{\mathrm{CS}}$. The vacuum structure of a non-Abelian gauge theory consists of an infinite set of distinct sectors labeled by the integer $n_w \in \mathbb{Z}$. These vacua are topologically inequivalent and cannot be continuously connected by gauge transformation within the space of pure-gauge (zero field strength) configurations. Transitions between sectors therefore requires finite-action configurations (nonzero field strength), corresponding to quantum tunneling processes. 

To better illustrate this, we first generalize Eq.~\eqref{eq:winding} to a Chern--Simons current in covariant form. 

\begin{equation}
K^{\mu} = \frac{g^2}{32\pi^2}\epsilon^{\mu\nu\rho\lambda}
\left(
A^a_{\nu} F^a_{\rho\lambda}
- \frac{g}{3} f^{abc} A^a_{\nu} A^b_{\rho} A^c_{\lambda}
\right).
\end{equation}
where the associated Chern--Simons number is defined through
\begin{equation}
N_{\rm CS} =  \int d^3 x \, K^0,
\end{equation}
Although $K^\mu$ and $N_{\rm CS}$ are in general not gauge invariant for finite-action fields, the total derivative of $K^\mu$ is gauge invariant and defines the topological charge density
\begin{equation}
\partial_\mu K^\mu =\,\frac{g^2}{32\pi^2}F^a_{\mu\nu} \tilde{F}^{a\mu\nu},
\end{equation}
where the dual field strength is defined as
$
\tilde{F}^{a\mu\nu} = \frac{1}{2}\epsilon^{\mu\nu\alpha\beta} F^a_{\alpha\beta}$. Thus the topological charge is related to the transition in Chern--Simons number,
\begin{equation}
Q_{\rm top} 
= \frac{g^2}{32\pi^2} \int d^4 x \, F^a_{\mu\nu} \tilde{F}^{a\mu\nu}
= N_{\rm CS}(+\infty) - N_{\rm CS}(-\infty).
\end{equation}

In Euclidean Yang--Mills (YM) theory, one can rewrite the action as
\begin{equation}
S = \frac{1}{4g^2} \int d^4 x \, F^a_{\mu\nu} F^{a\mu\nu}=\frac{1}{4g^2} \int d^4 x
\left[
\pm F^a_{\mu\nu} \tilde F^{a\mu\nu}
+ \frac12 \left(F^a_{\mu\nu} \mp \tilde F^a_{\mu\nu}\right)^2
\right].
\end{equation}
This action is minimized for (anti-)self-dual fields,
\begin{equation}
F^a_{\mu\nu} = \pm \tilde F^a_{\mu\nu},
\end{equation}
corresponding to (anti-)instantons, 
and their actions are determined entirely by non-trivial topological transition between vacua,
\begin{equation}
S = \frac{8\pi^2 |Q_{\rm top}|}{g^2}.
\end{equation}

Therefore, instantons provide the semiclassical realization of such transitions, interpolating between vacua with different Chern–Simons numbers and capturing the nontrivial topological structure of the QCD vacuum. These structures play a central role in nonperturbative chiral dynamics, giving rise to observable phenomena such as the axial anomaly, which will be discussed in Sec.~\ref{sec:InstYM}.

Although the QCD vacuum sectors are classified by the Chern–Simons number $n=N_{\rm CS}$, this quantity is not gauge invariant and therefore does not by itself define a physical state. Physical observables must instead be formulated in a gauge-invariant manner. In non-Abelian gauge theories, the vacuum structure is periodic, and instantons connect different topological sectors. As a result, the true ground state of QCD cannot be identified with a single topological vacuum $\ket{n}$, but must instead be a coherent superposition of all such vacua characterized by the integer-valued topological charge $Q=Q_{\rm top}$. It is analogous to the Bloch wavefunction of an electron in a periodic potential, leading to the $\theta$-vacuum,
\begin{equation}
\ket{\theta} = \sum_n e^{i n \theta} \ket{n}.
\end{equation}
In this framework, the vacuum transition amplitude can be expressed schematically as
\begin{align}
\label{eq:vac_Q}
\braket{0}{0}
&= \sum_{Q} e^{-i\theta Q}\int \mathcal{D}A \mathcal{D}\psi \mathcal{D}\bar\psi\,
e^{i S[\psi,\bar{\psi},A]}
\delta\left(
Q - \frac{g^2}{32\pi^2} \int d^4 x , F^a_{\mu\nu} \tilde{F}^{a\mu\nu}
\right)
,
\end{align}
where topological charge $Q$ labels the gauge-invariant topological sector.

\subsection{Topological landscape of gauge fields}
\label{sec:landscape}
The energy landscape with different $\theta$ vacua can be contructed straightforwardly within instanton and anti-instanton vacuum picture. We first consider the transition amplitude between two vacua with fixed topological charge $Q = n' - n$,
\begin{equation}
\bra{n'}e^{-Ht}\ket{n}=\int \mathcal{D}A \mathcal{D}\psi \mathcal{D}\bar\psi\,
e^{i S[\psi,\bar{\psi},A]}
\delta\left(
Q - \frac{g^2}{32\pi^2} \int d^4 x , F^a_{\mu\nu} \tilde{F}^{a\mu\nu}
\right)
\end{equation}
This amplitude receives contributions from configurations containing $N_+$ instantons and $N_-$ anti-instantons where $Q=N_+-N_-$, so the amplitude can be written as

\begin{equation}
\begin{aligned}
\bra{n'}e^{-Ht}\ket{n}
=& \sum_{N_+,N_-} 
\frac{1}{N_+! \, N_-!}
\left( K t e^{-S_c} \right)^{N_+ + N_-}
\delta_{N_+ - N_-,\, n' - n},
\end{aligned}
\end{equation}
where $K$ is a prefactor from fluctuations and $S_c$ is the instanton action.

Using the integral representation of the Kronecker delta,
\begin{equation}
\begin{aligned}
\delta_{N_+ - N_-,\, Q}
=&\frac{1}{2\pi} \int_0^{2\pi} d\theta \,
e^{-i\theta (N_+ - N_- -Q)},
\end{aligned}
\end{equation}
we obtain
\begin{align}
\bra{n+Q}{e^{-Ht}}\ket{n}
&= \frac{1}{2\pi} \int_0^{2\pi} d\theta \, e^{i\theta Q}
\sum_{N_+,N_-}
\frac{1}{N_+! \, N_-!}
\left( K t e^{-S} e^{-i\theta} \right)^{N_+}
\left( K t e^{-S} e^{i\theta} \right)^{N_-} \\
&= \frac{1}{2\pi} \int_0^{2\pi} d\theta \, e^{i\theta Q}
\exp\!\left( 2 K t e^{-S_c} \cos\theta \right).
\end{align}

Compared with Eq.~\eqref{eq:vac_Q} in asymptotic large time, we have
\begin{equation}
E(\theta) = -2 K e^{-S_c} \cos\theta.
\end{equation}
This result shows the energy landscape of true eigenstates $\theta$-vacua. The lowest energy state corresponds to $\theta = 0$ and its negative energy eigenvalue reflects tunneling events lower the ground-state energy.

\begin{figure}
    \centering
    \includegraphics[width=\linewidth]{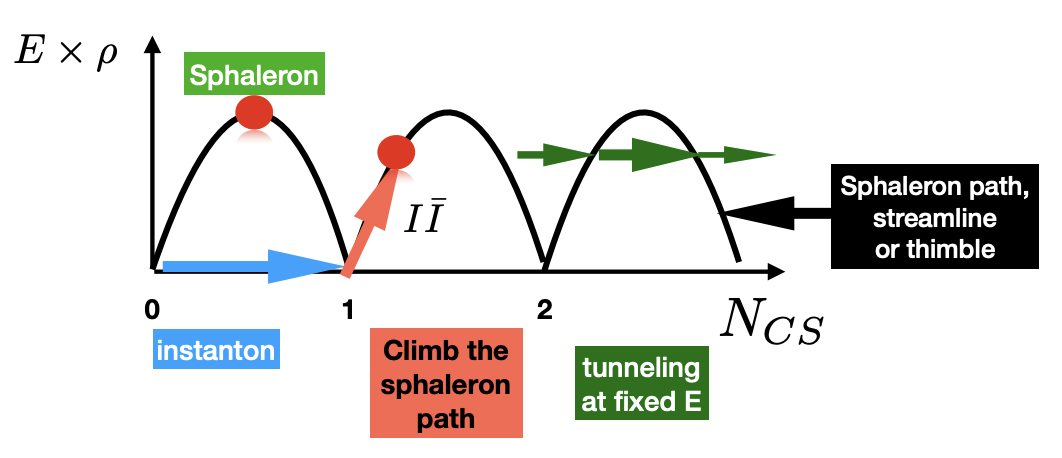}
    \caption{Topological landscape of gauge fields}
    \label{fig:landscape}
\end{figure}

Recently, a richer topological landscape is given by the set of minimal-energy gauge-field configurations parameterized with two key variables, as shown in Fig.~\ref{fig:landscape} from Refs.~\cite{Shuryak:2026pqt,Shuryak:2021fsu}. The first variable is the topological Chern--Simons number $N_{\rm CS}$ and the second is the root-mean-square size of the configuration,
\begin{equation}
\rho 
= \frac{\int d^3\vec{x}\, |\vec{x}|^2 G_{\mu\nu}^2 }{\int d^3\vec{x}\,  G_{\mu\nu}^2}.
\end{equation}

For fixed $\rho$, the energy landscape can be parameterized by a variable $k$~\cite{Ostrovsky:2002cg},
\begin{equation}
E_{\min}(k,\rho) = (1-k^2)^2 \frac{3\pi^2}{g^2 \rho},
\qquad
N_{\mathrm{CS}}(k) = \frac{1}{4}\,\mathrm{sign}(k)\,(1-|k|)^2 (2+|k|).
\end{equation}
The energy vanishes at $k=\pm1$, corresponding to vacuum configurations, while $k=0$ defines the sphaleron configuration~\cite{Klinkhamer:1984di}, an unstable field configuration that sits at the top of the energy barrier between topologically distinct vacua.

Connecting configurations at arbitrary $k$ forms the \emph{sphaleron path}, or \emph{streamline}. They correspond to static, three-dimensional magnetic fields and can be interpreted as turning points in the tunneling process, where the semiclassical momentum vanishes. These configurations are local minima in all directions except along the $\rho$ and $N_{\mathrm{CS}}$ directions, which remain fixed.

Instantons and anti-instantons correspond to tunneling trajectories connecting neighboring minima of the landscape. 
They carry topological charge $Q=\pm1$ and induce transitions 
\begin{equation}
\Delta N_{\mathrm{CS}} = \pm 1.
\end{equation}

However, instanton paths are not the only relevant topological fluctuations. Two additional classes of trajectories can be identified in Fig.~\ref{fig:landscape}:
\begin{enumerate}[label=(\roman*)]
    \item Paths that follow the gradient of the energy landscape, often referred to as the sphaleron path or streamline.
    \item Trajectories that penetrate the barrier at fixed energy, corresponding to tunneling processes occurring at nonzero energy.
\end{enumerate}

The first class is described by a constrained YM equation with a nonvanishing source term, which drives the configuration along the gradient. 
Mathematically, these paths correspond to Lefschetz thimbles connecting extrema of the action. For more details, see Sec.~\ref{sec:mole}.

The second class satisfies the YM equations without external sources and occurs at fixed energy. 
Such trajectories pass through turning points twice, with a Euclidean-time segment in between, representing tunneling between vacua at nonzero energy. These paths should be supplemented by Minkowski-time evolution before and after the turning points. It is a zigzag path stitched by real-time evolution and tunneling, indicating a transition from real to imaginary time and back again. For more details, see reviews in \cite{Shuryak:2021iqu}.

\subsection{Quark sector and axial charge}


In principle, distinct topological sectors of the gauge field configurations can be physically distinquished via light quarks. The axial charges they are carrying serves as a natural probe, as it is directly sensitive to the topology of the gauge configuration.

In QCD, the divergence of the axial $U(1)$ current is related to the Adler--Bell--Jackiw (ABJ) anomaly together with explicit breaking of quark masses,
\begin{equation}
\label{eq:ABJ_clean}
\partial_{\mu} J_5^{\mu} 
= \frac{N_f g^2}{16\pi^2} \, F^a_{\mu\nu}\tilde{F}^{a\mu\nu}
+ 2m \, \bar{\psi} i\gamma^5 \psi,
\end{equation}
where $J_5^\mu = \bar{\psi}\gamma^\mu\gamma^5\psi$ and $N_f$ is the number of light flavors. 
This anomalous relation is one-loop exact. 

In the chiral limit ($m \to 0$), the anomaly implies that the change in axial charge is directly tied to topology,
\begin{equation}
\Delta Q_5 
= \int d^4x \, \partial_\mu J_5^\mu 
= 2N_f \, Q_{\rm top}.
\end{equation}
Although the combination
\begin{equation}
\partial_\mu \left(J_5^\mu - 2N_f K^\mu \right) = 0
\end{equation}
is mathematically conserved, it does not correspond to a physical symmetry due to the non-gauge-invariance of $K^\mu$. 
Consequently, the presence of massless poles in correlation functions of $K^\mu$ does not imply a physical Goldstone boson, resolving the $U(1)_A$ problem.

A deeper understanding can also be provided by the Atiyah--Singer index theorem~\cite{Atiyah:1963index}, which relates the topological charge to the zero modes of the Dirac operator,
\begin{equation}
Q_{\rm top} = n_L - n_R,
\end{equation}
where $n_L$ and $n_R$ are the numbers of left- and right-handed fermionic zero modes. 
Thus,
\begin{equation}
\Delta Q_5 = 2N_f (n_L - n_R),
\end{equation}
demonstrating that nontrivial topology induces chirality violation in the fermionic sector. 
In particular, a single instanton ($Q_{\rm top}=+1$) produces one left-handed zero mode per flavor, while an anti-instanton ($Q_{\rm top}=-1$) produces a right-handed one.

\section{Other topological configurations}
\label{sec:topo_config}
\begin{figure}
    \centering
\subfloat[\label{fig:collimate}]{\includegraphics[height=4.2cm,width=.38\linewidth]{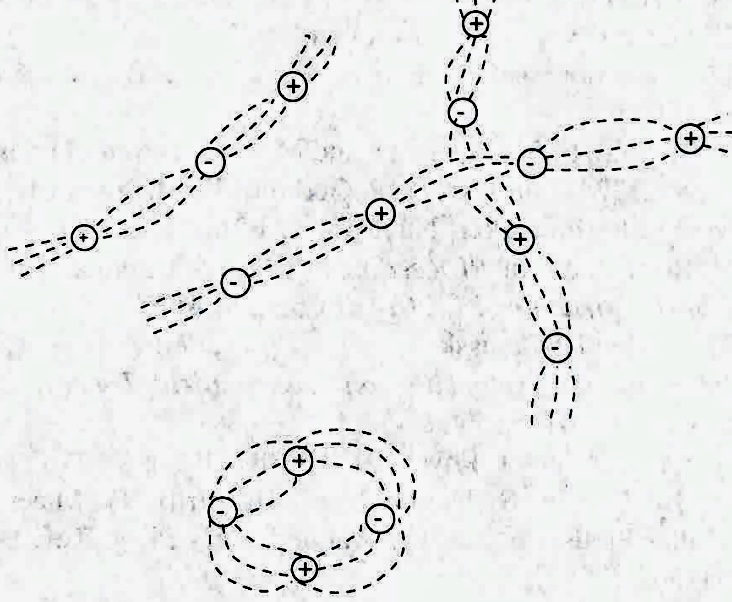}} 
\hfill
\subfloat[]{\includegraphics[height=4.2cm,width=.57\linewidth]{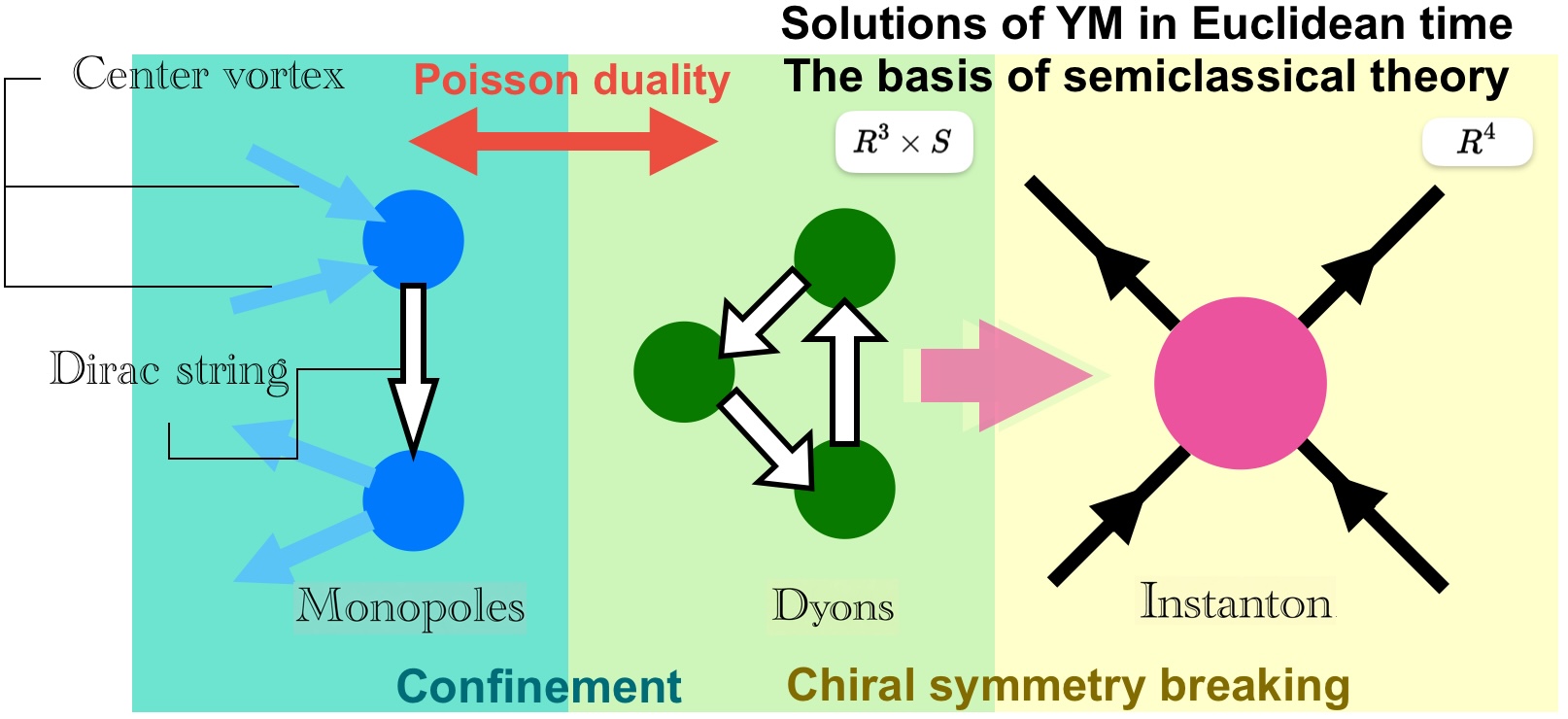}}   
    \caption{(a) Cartoon picture for the coexistence of center vortices and monopoles \cite{Greensite:2016pfc}. (c) The relation among center vortices, monopoles, dyons, and instantons.}
    \label{fig:topo_field}
\end{figure}

On top of the pseudoparticles we discussed in previous section, a variety of topological gauge field configurations in the vacuum also live in the infrared (IR) vacuum and are deeply related to one another, as depicted in Fig.~\ref{fig:topo_field}. In this section, we will focus on some other topological configurations that are relevant to confinement, namely monopoles, dyons, and center vortices. 



\subsection{Center P-vortices and monopoles}

\begin{figure}
    \centering
    \includegraphics[width=1\linewidth]{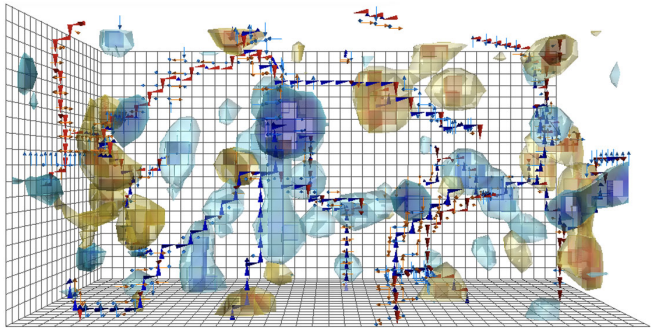}
    \caption{Instanton (yellow) and anti-instanton (blue) configurations in the deep-cooled YM vacuum, threaded by center P-vortices using center projection on lattice.}
    \label{fig:p-v}
\end{figure}

The center P-vortex is a $Z_{N_c}$ flux tube configuration, characterized by a number of branching points (monopoles). The Wilson loop pierced by them picks up a non-trivial value from the group center $Z_{N_c}$. The analytical string-like structure of center vortices can be found in \cite{Diakonov:1999gg,Diakonov:2002bx,Maas:2005qt}. For instance, 

\begin{equation}
    \epsilon_{\mu\nu}n_\mu A_\nu^a(r) = \delta^{a3} \frac{\mu(r)}{r}
\end{equation}
where $n$ is the unit radial vector in the $xy$ plane with the radial coordinate $r=n\cdot x$ and $\mu(r)$ is the profile of the vortex with $\mu(0)=0$.  

In a 4D gauge theory, monopoles are represented by world lines and center vortices are world sheets. When projected to a 3D subspace, monopoles are viewed as points and center vortices are lines. As shown in Fig.~\ref{fig:p-v}, instanton (yellow) and anti-instanton (blue) configurations in the deep-cooled YM vacuum are threaded by center P-vortices on lattice \cite{Biddle:2019gke,Biddle:2020eec}. The topological configurations in the vacuum form the primordial gluon epoxy (hard glue) that underpins the origin of light hadron masses and the string-like center P-vortices play a key role in confinement. In $SU(2)$, the intersection of two center vortices where their fluxes merge, correspond to a monopole.  Each $SU(2)$ vortex carries a quantized flux of $\pi$. When two such vortices merge, their fluxes add to $2\pi$ (topologically trivial), which can be viewed as a thin flux tube terminating on a magnetic monopole. This string is referred to as a Dirac string.  Although it is not a physical observable due to the gauge dependence, it ensures the monopole consistently carries magnetic flux.   

In the case of $SU(3)$, as shown in Fig.~\ref{fig:p-v} with lattice simulations through gradient-flow smearing, the structure of the QCD vacuum is revealed as an intricate network of string-like center vortices, which behave as anchors of topological pseudoparticles. In Euclidean spacetime, these vortices form extended two-dimensional world sheets that fluctuate and percolate. Their essential feature is the presence of branching and recombination points, where multiple vortex sheets meet or split. In center-projected vacuum, these points act as sources and sinks of Abelian magnetic flux and appear as monopoles and anti-monopoles, reflecting genuine geometric singularities of the vortex surfaces (see the cartoon sketch in Fig.~\ref{fig:collimate}).

Strikingly, these branching regions are strongly correlated with the locations of instantons and anti-instantons. This correlation indicates that both instantons and vortex branch points shares the same geometric origin. They both arise from localized regions where the gauge field configurations exhibit nontrivial topology due to rapidly varying configurations. In this picture, center-vortex sheets provide an underlying geometric structure that organizes topological charge. When these sheets intersect, twist, or fold, they naturally generate localized topological lumps, which can be identified as instantons.

This geometric perspective unifies confinement and chiral symmetry breaking. On one hand, confinement emerges from the percolation of extended vortex surfaces, which form a dense network throughout the vacuum. As these surfaces intersect at a Wilson loop, they induce fluctuating center phases that disorder its expectation value, leading to an area-law falloff characteristic of confinement. On the other hand, chiral symmetry breaking is driven by instantons, whose zero modes produce a near-zero Dirac spectrum. If monopole worldlines and instanton-like structures originate from the same vortex geometry, these phenomena are no longer independent but instead arise as complementary manifestations of a common underlying vacuum structure.

Many lattice results~\cite{Maas:2005qt,Langfeld:1998cz,Biddle:2019gke,Biddle:2020eec,Kamleh:2023gho} as well as theoretical studies~\cite{Nguyen:2024ikq,Nguyen:2025voy,Hayashi:2024yjc,Guvendik:2024umd,Greensite:2016pfc} have supported this assertion.   

 


\subsection{Dyons (instanton-monopoles)}



Instantons in Euclidean space $R^4$ can be generalized to finite temperature with a twisted temporal boundary condition defined on a circle $R^3\times S$. This modifies Belavin-Polyakov-Schwarz-Tyupkin (BPST) instanton solutions into Kraan-van Baal-Lee-Lu (KvBLL) instantons with non-trivial holonomy \cite{Kraan:1998pm,Lee:1998bb,Kraan:1998kp}. At finite temperature in $SU(N_c)$, each KvBLL instanton (caloron) with unit topological charge are decomposed into $N_c$ self-dual dyons \cite{Zhitnitsky:2006sr,Unsal:2008ch}. Each of them carries both electric, magnetic charges, and nonzero fractional topological charge specified by nontrivial Polyakov loop. This naturally extends the zero-temperature ensemble of instanton liquid to a finite-temperature dyon plasma ensemble characterized by long-range Coulomb-like interactions, which can be quantitatively described by the dyon liquid model (DLM) \cite{Liu:2015jsa,Liu:2015ufa,Diakonov:2009jq}.  DLM has been demonstrated to support a confining phase at sufficient dyon density, offering a comprehensive explanation for the confinement–deconfinement phase transition and chiral symmetry restoration. When the temperature cools down, dyons gets denser in the vacuum and eventually recombine to instantons at zero temperature where the full 4D vacuum degrees of freedom now is described by a dilute instanton liquid (ILM) (see Fig.~\ref{fig:topo_field}). These conclusions are supported numerically by \cite{Larsen:2015vaa,Larsen:2015tso} and by mean-field analyses \cite{Liu:2015jsa,Liu:2015ufa}. For a comprehensive overview, see \cite{Shuryak:2017kct}.


\begin{figure}
    \centering
    \includegraphics[width=0.7\linewidth]{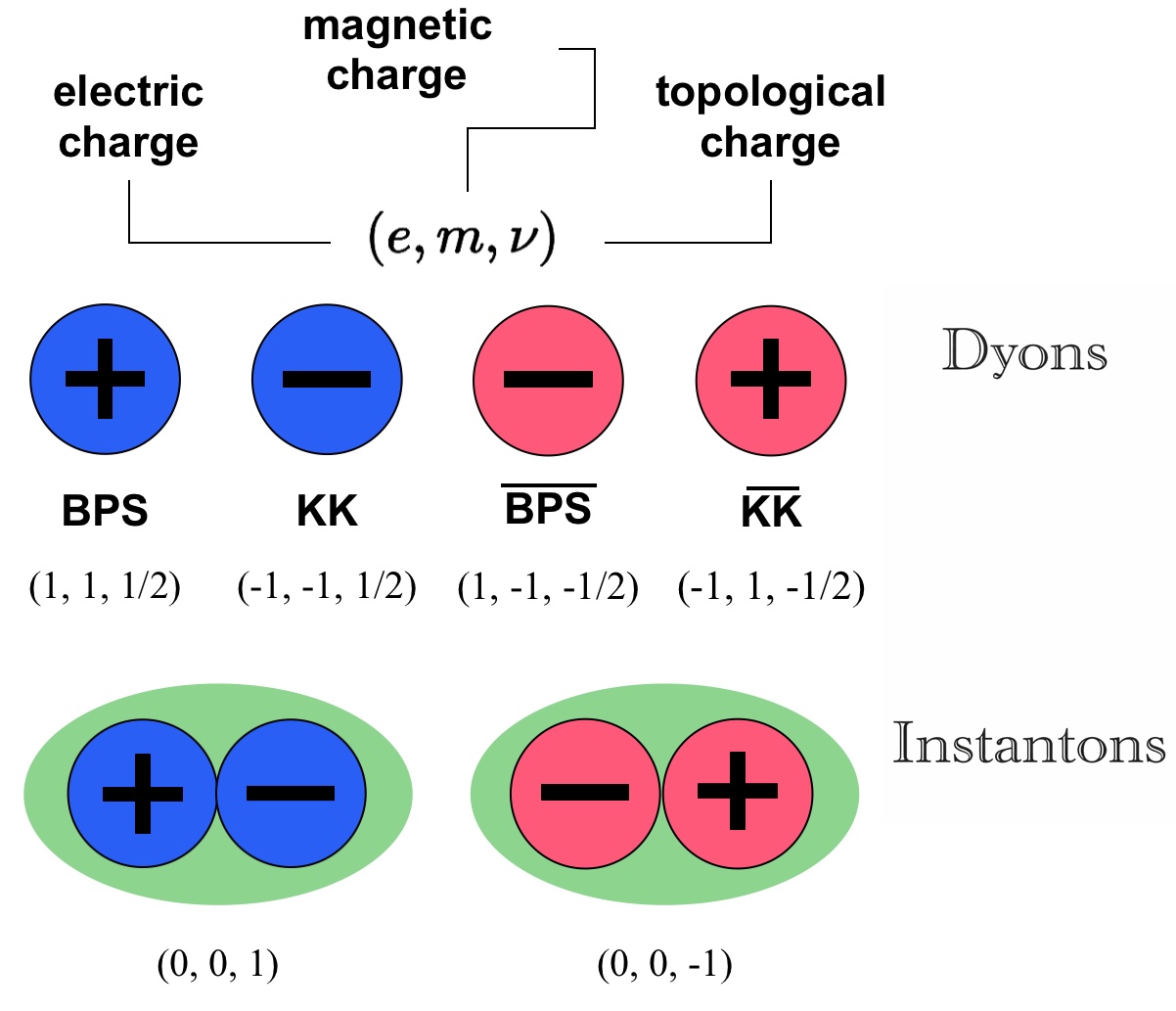}
    \caption{The constituents of instantons (calorons): BPS and KK dyons, and their anti-dyons with their electric, magnetc, and topological charges labeled}
    \label{charges}
\end{figure}

In $SU(2)$, the self-dual dyons with
electric and magnetic charges $(e, m)=(+, +)$ are called $M$, or Bogomolny–Prasad–Sommerfeld (BPS) dyons in \cite{Unsal:2007jx}. The ones with charges $(e, m)=(-, -)$ are called $L$, or Kaluza-Klein (KK) dyons. Their anti-self-duals $\bar M$ and $\bar L$ are the ones with $(e, m)=(+, -)$, and $(e, m)=(-, +)$, respectively. Their relation to instantons are illustrated in Fig.~\ref{charges}. In $SU(N_c)$, there are $N_c - 1$ BPS dyons (maximal
abelian subgroup), with charges counted from the Cartan generators, one KK dyon with the charge compensating BPS dyons to zero, and their anti-self-dual counterparts.

These dyon configurations are also closely related to monopoles, which arise as endpoints of center vortices.
They are related at finite temperature to the semiclassical description based on instanton–dyons through Poisson duality, even though monopoles themselves lack corresponding semiclassical theory description. In this dual perspective, dyon configurations can be viewed as quantum paths of moving and rotating monopoles in their collective coordinate space. Although the two descriptions are formally equivalent and encode the same underlying physics, instanton-dyon ensemble practically provide a more suitable description near and above the critical temperature $T\gtrsim T_c$ while monopole-based descriptions model the vacuum at much lower temperature $T\ll T_c$ \cite{Poppitz:2012sw,Dorey:2000qc,Poppitz:2011wy,Shuryak:2018fjr}, where $T_c$ is the critical temperature for QCD phase transition.

Consequently, at finite temperature, monopoles and instantons can be unified through a plasma picture of dyons (instanton-monopoles), which carry both topological and magnetic charges. In this picture, caloron are composed of multiple dyons, while monopoles are the dual description of them. This unified description becomes particularly relevant near the QCD phase transition, where dyon dynamics play a central role.

\subsection{Unified vacuum picture and hadron physics}

In conclusion, gradient-flow analyses based on center projection further reveal a close relation between the locations of instantons and the branching points associated with monopoles and anti-monopoles, as shown in Fig.~\ref{fig:p-v}. This connection reflects the common geometrical origin of these configurations and is consistent with their known interplay at finite temperature~\cite{Kraan:1998pm,Lee:1998bb}. 
Taken together, these observations point to an intimate relationship between chiral symmetry breaking and confinement. 

As the pseudoparticles carry much stronger chromo-electric and chromo-magnetic fields $\sqrt{E}=\sqrt{B}\approx 2.5/ \rho\approx 1.5\,{\rm GeV}$, in comparison to $\sigma_T \rho\approx 0.3\,{\rm GeV}$ associated with a center P-vortex \cite{Shuryak:2021fsu}, together with the previous observation we suggest the following interpretation. Instantons and anti-instantons dominate in low-lying hadrons, whose relatively localized worldlines are less susceptible to flux piercing. 
In contrast, monopoles and center vortices dominate in higher-lying hadrons, whose more extended worldlines enhance sensitivity to flux penetration. 
These two descriptions are likely dual aspects of the same vacuum structure, extending earlier observations in thermal QCD~\cite{Ramamurti:2018evz}.


\chapter{Instanton Liquid Model}
\label{ch:ILM}

\section{QFT in a semiclassical background field}
\label{sec:background}
In conventional perturbative QFT, the expansion is performed around the trivial vacuum. However, capturing the nontrivial structure of the vacuum requires going beyond the perturbative expansion and incorporating the full set of semiclassical field configurations. A natural way to isolate such nonperturbative contributions is to expand the theory around saddle points of the action around a nontrivial semiclassical solution. For simplicity, we begin by considering the path integral formulation of a general scalar field theory.
\begin{equation}
Z = \int \mathcal{D}\phi \, e^{-S[\phi(x)]}
\end{equation}

A naive evaluation of the path integral proceeds by expanding around the trivial saddle point, in which case the partition function reduces to the inverse determinant of the quadratic fluctuation operator obtained from the equation of motion. However, in the presence of continuous symmetries, this procedure becomes ill-defined. The fluctuation operator develops zero eigenvalues corresponding to directions in configuration space along which the action remains stationary, causing the determinant to vanish.

The origin of this problem lies in treating all fluctuations as Gaussian, including those associated with these flat directions. In reality, such modes are non-Gaussian and must be treated separately. They span a submanifold in configuration space corresponding to the orbit of the symmetry transformations. These zero modes can be parameterized by a set of collective coordinates $\gamma = (\gamma_1, \ldots, \gamma_k)$, such that the classical configurations are described by $\phi_c(x;\gamma)$. With this understanding, the field configuration should be decomposed into a semi-classical field along the orbit and perturbative fluctuations orthogonal to this submanifold.
\begin{equation}
\phi(x) = \phi_c(x;\gamma) + \delta \phi(x).
\end{equation}

To make this decomposition unique, we have to impose $k$ linear extra conditions on $\delta \phi(x)$.
\begin{equation}
\int d^4x \delta \phi(x) \psi_i(x,\gamma)=0,
\qquad i=1,\ldots,k.
\end{equation}
where the zero modes are defined as
\begin{equation}
\psi_i(x,\gamma)=\frac{\partial \phi_c(x;\gamma)}{\partial \gamma_i}
\end{equation}
These conditions fix the collective coordinates $\gamma$ by requiring that $\delta \phi(x)$ be orthogonal to the zero modes (the tangent directions of the symmetry orbit). In this way, fluctuations along the non-Gaussian zero-mode directions are excluded, while only the Gaussian fluctuations orthogonal to the orbit are retained. 

In the path integral, one can introduce the Faddeev--Popov identity to separate the Gaussian and non-Gaussian modes.
\begin{equation}
1=\int d^k\gamma\;
\delta^k\!\left(\int d^4x\delta \phi(x) \psi_i(x,\gamma)\right)
\,J(\gamma)
\end{equation}
with the Jacobian defined as a Faddeev--Popov ghost determinant
\begin{equation}
    \begin{aligned}
        J(\gamma)
=&
\det\!\left[\int d^4x\left(
\delta \phi(x)\frac{\partial \psi_i(x,\gamma)}{\partial \gamma_j}
-
\frac{\partial \phi_c(x;\gamma)}{\partial \gamma_j} \psi_i(\gamma)
\right)
\right]
    \end{aligned}
\end{equation}
Thus, the partition function becomes
\begin{equation}
Z
=
\int \mathcal{D}\delta\phi\, \int d^k\gamma\,
\delta^k\!\left(\int d^4x\delta \phi(x) \psi_i(x,\gamma)\right)J(\gamma)
e^{-S[\phi_c(x)+\delta\phi(x)]}.
\end{equation}

By rewriting the ghost determinant with a ghost action,
\begin{equation}
    \begin{aligned}
        J(\gamma)=\int d^k\eta\, d^k\bar\eta\;
e^{S_{\rm gh}[\eta,\bar\eta]}
    \end{aligned}
\end{equation}
one arrives at the Faddeev--Popov action
\begin{equation}
Z
=
\frac1{(2\pi \xi)^{k/2}}
\int\mathcal{D}\delta \phi\,\int d^k\gamma\, \delta^k\!\left(\int d^4x\delta \phi(x) \psi_i(x,\gamma)\right)J(\gamma)
e^{-S[\phi(x)_c+\delta \phi(x)]-S_{\rm gh}},
\end{equation}
with the ghost action
\begin{equation}
S_{\mathrm{gh}}
=
\bar\eta_i
\int d^4x\left(
\frac{\partial \phi_c(x;\gamma)}{\partial \gamma_j} \psi_i(x,\gamma)-\delta \phi(x)\frac{\partial \psi_i(x,\gamma)}{\partial \gamma_j}
\right)\eta_j
\end{equation}

Integrating over the Gaussian fluctuations $\delta \phi(x)$ reduces the partition function to an effective integral over the collective coordinates, which represent the contributions of the zero modes. The appearance of the ghost action can then be understood as a consequence of the orthogonal condition introduced to avoid overlaps between the Gaussian fluctuation and background zero modes.

To extend the above argument to QCD, we need to incorporate gauge symmetry properly. The presence of gauge redundancies introduces additional zero modes associated with gauge transformations, necessitating a more careful treatment of the functional integral. As a first step, we consider pure gauge theory. The YM partition function is defined as

\begin{equation}
Z = \int \mathcal{D}A\; e^{-S_{\mathrm{YM}}[A]},
\end{equation}
with
\begin{equation}
S_{\mathrm{YM}}
=
\int d^4 x\;
\frac{1}{4 g^2}
F^a_{\mu\nu}F^{a\,\mu\nu}.
\end{equation}

A similar procedure can be implemented by decomposing the field into the semi-classical solution and the small fluctuation.
\begin{equation}
A^\Omega_\mu(x) = \bar A_\mu(x;\gamma) + \delta A^\Omega_\mu(x),
\end{equation}
where $\gamma$ again denotes a $k$-dimensional set of collective coordinates that parameterize the background manifold and $\Omega$ denotes the gauge parameter in a gauge theory that defines a gauge transformation
\begin{equation}
A_\mu^\Omega
=
U A_\mu U^\dagger
+
i\,U\partial_\mu U^\dagger
\end{equation}
with $U=e^{i\Omega}\in SU(N_c)$.

Similarly, the Gaussian and non-Gaussian modes can be separated by imposing the orthogonal condition to fix the $\gamma$ degrees of freedom.
\begin{equation}
\int d^4x\; \psi_\mu^{\,i}(x,\gamma)\,\delta A_\mu(x) =0,
\qquad i=1,\ldots,4N_c
\end{equation}
where the zero modes in the gauge theory is defined as
\begin{equation}
\psi_\mu^{\,i}=
\frac{\partial \bar A_\mu}{\partial \gamma_i}, \, \, \, \, \, \, \, \bar D_\mu \psi_\mu^{\,i}=0.
\end{equation}
For simplicity, we fix the zero modes in background gauge.

In addition to the orthogonality we imposed, we also need to fix the gauge in the decomposition by introducing an extra gauge-fixing condition. Here we choose a generalized background gauge condition
\begin{equation}
\bar D_\mu\, \delta A_\mu = C(x),
\end{equation}
where the background covariant derivative is defined as
\begin{equation}
\bar D_\mu = \partial_\mu - i\,\bar A_\mu(x).
\end{equation}


The gauge theory version of the Faddeev--Popov identity is defined as
\begin{align}
1
&=
\int d^k\gamma\, \mathcal{D}\Omega\;
\delta\!\left[\bar D_\mu \delta A_\mu-C(x)\right]\,
\delta^k\!\left(\int d^4x\, \psi_\mu^{\,i}\delta A_\mu\right)
\,J(\gamma).
\end{align}

This identity gives rise to the Faddeev--Popov ghost determinant, which accounts for the nontrivial measure induced by fixing conditions. Schematically, the Jacobian reads
\begin{equation}
\begin{aligned}
J(\gamma)
=&
\det\!\Bigg[
\int d^4x\,
\Bigg(
\frac{\partial \psi_\mu^{\,i}}{\partial \gamma_j} \delta A_{\mu}
-
\psi_\mu^{\,i}\frac{\partial \bar A_\mu}{\partial \gamma_j}
\Bigg)\Bigg]
\det\big(-
\psi_\mu^{i}D_\mu
\big)\\
&\times\det\!\Bigg(
\bar{D}_\mu \frac{\partial \bar A_\mu}{\partial \gamma_j}\Bigg)
\det\!\left(\bar D_\mu D_\mu\right)
\end{aligned}
\end{equation}

In the gauge theory in a semiclassical background, this infinite-dimensional determinant can be represented by the usual ghost field $\eta(x),\bar\eta(x)$ together with extra finite-dimensional $4N_c$-ghost variables $\eta_i,\bar\eta_i$. By setting $\bar{A} = 0$, one recovers the standard Faddeev--Popov ghost determinant.

Since the partition function is independent of $C(x)$, we can further smooth the gauge fixing condition by introducing $-\frac{1}{2\xi}\int d^4 C^2(x)$ in the exponential. Eventually one arrives at
\begin{equation}
Z
=
N(\xi)
\int d^k\gamma\,
\mathcal{D}\delta A_\mu\,
\mathcal{D}\eta\,
\mathcal{D}\bar\eta\;
\delta^k\!\left(\int d^4x\, \psi_\mu^{\,i}\delta A_\mu\right)
e^{-S_{\mathrm{YM}}[\bar A+\delta A]
+
S_{\mathrm{gf}}[\bar A,\delta A]
-
S_{\mathrm{gh}}[\bar A,\delta A,\eta,\bar\eta]},
\end{equation}
where the gauge fixing term 
\begin{equation}
S_{\mathrm{gf}}
=
\frac{1}{2\xi}
\int d^4x\;
(\bar D_\mu \delta A_\mu)^2,
\end{equation}
and the ghost action
\begin{align*}
S_{\mathrm{gh}}
&=
\bar\eta_i
\left(
\int d^4x\; \psi_\mu^{\,i}(x)\,
\frac{\partial \bar A_\mu}{\partial \gamma_j}
-
\delta A_\mu \frac{\partial \psi_\mu^{\,i}}{\partial \gamma_j}
\right)\eta_j
\\
&\quad
+
\bar\eta_i \int d^4x\; \psi_\mu^{\,i} D_\mu \eta(x)
-
\int d^4x\; \bar\eta(x)\, D_\mu\frac{\partial \bar A_\mu}{\partial \gamma_i}\,\eta_i
-
\int d^4x\; \bar\eta\, \bar D_\mu D_\mu \eta(x),
\end{align*}

The integration over the gauge parameter $\Omega$ is also absorbed into the normalization factor $N(\xi)$. The field $\eta(x)$ is the usual ghost from gauge fixing and every extra orthogonal condition for the zero modes introduces an additional ghost variable $\eta_i$.

Up to quadratic order in the fluctuation $\delta A$, YM effective action reads 
\begin{align}
S_{\rm YM}
&=
S_{c}[\bar A]
+
\int d^4x\;
\frac{1}{2g^2}\,
\delta A_\mu\, \bar{D}_{\mu\nu}\, \delta A_\nu
+
\int d^4x\; \bar\eta\, \bar D^2 \eta(x)
\nonumber\\
&\quad
+
\bar\eta_i \left(\int d^4x\,\psi_\mu^{\,i}\,\frac{\partial \bar A_\mu}{\partial \gamma_j}\right)\,\eta_j
+
\int d^4x\;
\bar\eta(x)\, \bar{D}_\mu \frac{\partial \bar A_\mu}{\partial \gamma_j}\,\eta_j
+\cdots,
\end{align}
where
\begin{equation}
\bar{D}_{\mu\nu}
=
-\bar D^2 \delta_{\mu\nu}
-
2f^{abc}F^c_{\mu\nu}
+
\left(1-\frac{1}{\xi}\right)\bar{D}_\mu \bar{D}_\nu,
\end{equation}

After integrating over ghost fields and fluctuations around the saddle point, the resulting partition function can be written as
\begin{equation}
\label{eq:background}
Z
=
\int d^k\gamma\;
\det\!\left(
\psi_\mu^{\,i}\frac{\partial \bar A_\mu}{\partial \gamma_j}
\right)
\frac{\det(-\bar D^2)}{\sqrt{\det{}' D_{\mu\nu}}}
\,e^{-S_{\mathrm{c}}[\bar A]}.
\end{equation}
Here $\det{}'$ means that the zero modes are omitted from the determinant. The resulting partition function can thus be understood in three distinct contributions. The first arises from the gauge zero modes, represented by the integration over collective coordinates. The second corresponds to quantum fluctuations around the saddle point, encoded in the determinant of the fluctuation operator. The third is the semi-classical contribution, given by the action evaluated on the semiclassical field configuration.


\section{Instanton vacuum in YM}

\subsection{Instanton size distribution}
For a single instanton background, the partition function  in Eq.~\eqref{eq:background} is obtained by

\begin{equation}
\label{eq:Z_I}
Z=V\int d\rho n_0(\rho)
\end{equation}
where $d\gamma=d\rho\, d^4z\, dU$ is the conformal measure for the collective coordinates of a single (anti-)instanton (size $\rho$, center $z$, and color orientation $U$). The mean tunneling rate is
\begin{equation}
\label{eq:n}
n_0(\rho)=\frac{C_{N_c}}{\rho^5}S(\rho)^{2N_c} e^{-S(\rho)}
\end{equation}
where the inverse coupling (instanton action) is defined by $S=8\pi^2/g^2$ and $C_{N_c}$ is a constant dependent on color number $N_c$ defined as 
$$
C_{N_c} = \frac{0.466 \exp(-1.679 N_c)}{(N_c - 1)! (N_c - 2)!}
$$

In general the inverse coupling in the exponential and the prefactor runs at different loop order. This is because the exponential and prefactor in the instanton density arise from different pieces of the semiclassical expansion. The exponential $e^{-S(\rho)}$ is determined by the classical instanton action and is naturally RG-improved using the running coupling, whose scale dependence is governed by the RG and can be included to higher-loop accuracy. In contrast, the prefactor originates from the functional determinant of quantum fluctuations around the instanton background, that is, Gaussian integration over nonzero modes and zero-mode normalization. This part is only known to one-loop order~\cite{tHooft:1976snw}, and higher-order corrections remain to be computed. As a result, the two terms are generally improved by RG separately. Considering the running coupling in the exponential up to two-loop order, the vacuum tunneling rate of a single instanton reads \cite{Shuryak:1989cx,Shuryak:1987iz},
\begin{equation}
\begin{aligned}
\label{eq:dn2}
n_0(\rho)=&\frac{C_{N_c}}{\rho^5}[S_1(\rho)]^{2N_c}\exp\left[-S_2(\rho)
+ \left(2 N_c - \frac{b_1}{2 b_0}\right)
\left(\frac{b_1}{2 b_0}\right)
\frac{\ln S_1(\rho)}{S_1(\rho)}\right]
\end{aligned}
\end{equation}
where the one-loop running inverse coupling is given by $S_1=-b_0\ln(\rho\Lambda)$ and two-loop $S_2$ is given by
\begin{equation}
\label{eq:run_coup}
S_2(\rho)=S_1(\rho)
+ \frac{b_1}{2\,b_0}
\ln\!\left(
\frac{2}{b_0}\,S_1(\rho)
\right)
\end{equation}
with the beta function coefficient $b_0=(11 N_c - 2 N_f)/3$ and $b_1=(34 N_c^2 - 13 N_c N_f + 3 N_f/N_c)/3$
that appear in the Gell-Mann-Low beta function defined by
\begin{equation}
\label{eq:beta_GL}
\beta(g^2)=\mu\frac{\partial g^2}{\partial\mu}=-\frac{b_0g^4}{8\pi^2}-\frac{b_1g^6}{(8\pi^2)^2}+{\cal O}(g^8).   
\end{equation}
Here we set $N_f=0$ for YM vacuum. 

\begin{figure}
    \centering
\subfloat[\label{fig:size_distSU2}]{\includegraphics[width=0.49\linewidth]{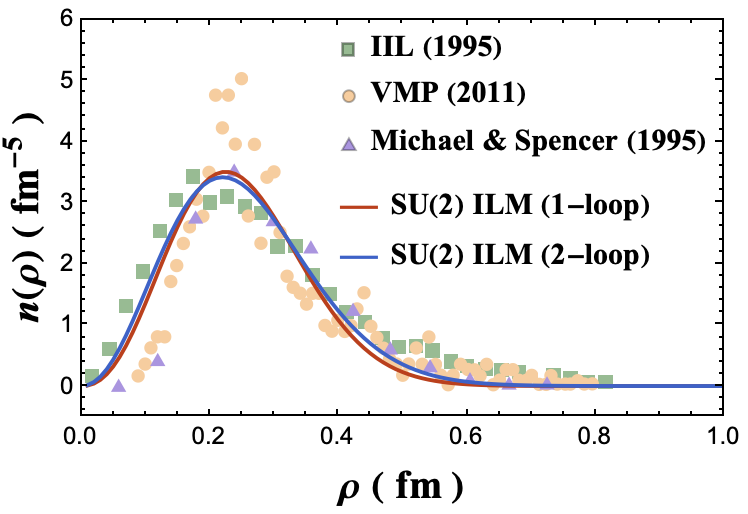}}
 \hfill
\subfloat[\label{fig:size_distSU3}]{\includegraphics[width=0.49\linewidth]{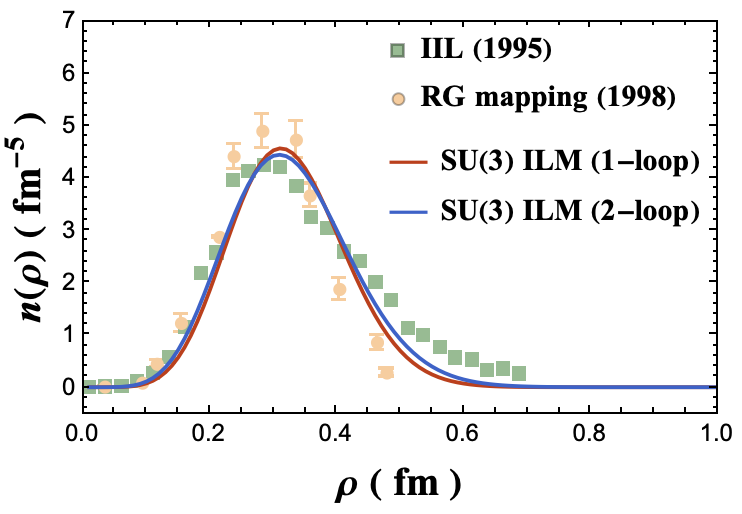}}
    \caption{(a) $SU(2)$ instanton size distribution with 1-loop (solid red) and 2-loop (solid blue) compared to lattice calculations. (b) One-loop parametrized $SU(3)$ instanton size distribution with 1-loop (solid red) and 2-loop (solid blue) compared to lattice calculations. }
    \label{fig:size_dist}
\end{figure}

In the present treatment, only the small size distribution follows from the conformal nature of the instanton moduli and perturbation theory, while the large size distribution is intrinsically nonperturbative. A cut-off scaled by $R$, the mean separation of the instantons (anti-instantons) in the vacuum must be introduced. Otherwise, the integration over the size distribution leads to IR divergences. Phenomenologically, this suppression can be attributed to a dual monopole condensate, as inferred from the flux-tube tension $\sigma=(0.44~\mathrm{GeV})^2\sim1/R^2$~\cite{Shuryak:1999fe,Dorokhov:2002qf}. With this in mind, the proposed effective size distribution reads \cite{Schafer:1996wv}

\begin{equation}
\label{eq:n3}
n(\rho)
=n_0(\rho)
\, e^{-2\pi \sigma \rho^2} \, .
\end{equation}
with $n_0$ the semiclassical instanton density defined in Eq.~\eqref{eq:n}. This quadratic $\rho$ dependence has been observed in
lattice simulations, as shown in Fig.~\ref{fig:size_dist}, where the distribution is extracted from deeply cooled gauge field configurations. 

\begin{table}
    \centering
    \begin{tabular}{|c|c|c|c|c|}
    \hline
    & $\rho$ & $n_{I+A}$ & $\sigma$ & $\Lambda$ \\
    \hline
        $SU(2)$ & $0.254(5)$ fm & $0.93(3)$ fm$^{-4}$ & $(0.271(7)\ \mathrm{GeV})^2$ & 0.3053(24) GeV \\
        $SU(3)$ & $0.328(4)$ fm & $1.07(3)$ fm$^{-4}$ & $(0.323(5)\ \mathrm{GeV})^2$ & 0.3246(25) GeV \\
    \hline
    \end{tabular}
    \caption{The instanton mean size and density obtained by fitting with the results obtained by interacting instanton liquid (IIL) ensemble \cite{Shuryak:1995pv} using 2-loop RG improved formula in Eq.~\eqref{eq:dn2}. The classical string tension $\sigma$ is estimated.}
    \label{tab:inst_dist_para}
\end{table}

Typically, the quadratic dependence at large size is related to the repulsive non-perturbative interaction among pseudoparticles \cite{Diakonov:1995qy}. For this purpose, one can also introduce the dimensionless combination $\sigma R^2$, which measures the overall repulsion among them. These interactions stabilize the ensemble by preventing uncontrolled growth in size while maintaining sufficient separation between pseudoparticles. This naturally leads to an interacting instanton liquid (IIL) ensemble, rather than a simple dilute instanton gas, as a more realistic description of the QCD vacuum.

Historically, the coefficient $\sigma$ is estimated to be $\frac1{4\pi R^2}\left(\frac{11}3N_c-4\right)(\bar\rho/R)^2\sim(0.05-0.06~\mathrm{GeV})^2$ by variational principle \cite{Diakonov:1983hh}. This indicates that semiclassical description alone are not sufficient to estimate the large-size instanton distribution. In Table~\ref{tab:inst_dist_para}, we estimate the instanton mean size $\rho$ and the density $n_{I+A}$ using \eqref{eq:n3} fitted with IIL ensemble \cite{Shuryak:1995pv}. The result is well consistent with the estimation on lattice \cite{Millo:2011zn,Hasenfratz:1998qk} for both $N_c=2$ and $N_c=3$ while the result calculated by UKQCD group indicates larger mean size $\rho=0.5$ fm \cite{Smith:1998wt}. 



In Fig.~\ref{fig:size_distSU2}, $SU(2)$ instanton size distribution in 1-loop (solid red) and 2-loop (solid blue) fitted with the result from IIL ensemble (green square) \cite{Shuryak:1995pv} using Eq.~\eqref{eq:dn2}, compared to lattice calculation using vacuum manifold projection (VMP) with lattice spacing $a=0.17$ fm on a $16^4$ lattice (orange dot) \cite{Millo:2011zn} and under-relaxed cooling algorithm with combination of $a=0.12$ fm on a $16^4$ lattice 
and $a=0.08$ fm on a $24^4$ lattice (purple triangle) \cite{Michael:1995br}. All distributions presented in Fig.~\ref{fig:size_distSU2} is normalized to the instanton density $n_{I+A}=0.93$ fm$^{-4}$ we obtain via fitting.

In Fig.~\ref{fig:size_distSU3}, $SU(3)$ instanton size distribution ($N_f=0$) in 1-loop (solid red) and 2-loop (solid blue) fitted with the result from IIL ensemble (green square) \cite{Shuryak:1995pv} using Eq.~\eqref{eq:dn2}, compared to RG mapping method on lattice (orange circle) \cite{Hasenfratz:1998qk}, and lattice results calculated by UKQCD group (orange circle) \cite{Smith:1998wt,Negele:1998ev}. All distributions presented in Fig.~\ref{fig:size_distSU3} is normalized to normalized to instanton density $n_{I+A}=1.07$ fm$^{-4}$  we obtain via fitting.

Our ILM predictions agree with lattice calculations \cite{Smith:1998wt,Millo:2011zn,Hasenfratz:1998qk} in both $N_c=2$ and $N_c=3$ case. The lattice results indicate that small instantons are more suppressed in $SU(3)$ than $SU(2)$, consistent with the ILM prediction $n(\rho) \sim \rho^{\frac{11}{3}N_c - 5}$ and instantons larger than $\rho \simeq 1/3$ fm are significantly suppressed in QCD.  The instanton size distribution obtained by lattice calculation from UKQCD group \cite{Smith:1998wt} indicates the $SU(3)$ distribution at large sizes exhibit a power-law falloff $1/\rho^{11}$, in contrast to the predicted exponential falloff, while the behavior at small sizes scales as $\rho^6$, consistent with the instanton liquid model prediction.

\subsection{Molecular configuration}
\label{sec:mole}

As discussed in Sec.~\ref{sec:vac_GF}, gauge-field configurations obtained from lattice simulations, after gradient-flow filtering, suggest a \emph{dense instanton liquid}, a dense ensemble of instanton-anti-instanton ($IA$) molecules (see Sec.~\ref{sec:DILM}). Although these additional gauge-field are not visible in configurations obtained in deep cooling as shown in Fig.~\ref{fig:p-v} since such pairs annihilate at early stages of the flow, they are essential to interpolate between the full quantum vacuum and the dilute instanton ensemble. 

The role of these molecular configurations has been explored in the context of hot \cite{Ilgenfritz:1988dh,Schafer:1994nv} and dense QCD matter \cite{Rapp:1999qa}, where they are closely tied to phase transitions. In particular, in the chiral limit, they provide the dominant surviving nonperturbative component above the critical temperature $T_c$, where chiral symmetry is restored. 

Such configurations were first studied in quantum mechanics in the context of double-well potentials and later extended to gauge theories~\cite{Balitsky:1986qn,Yung:1987zp,Verbaarschot:1991sq}. A semiclassical description is provided by minimizing the action along the streamline, leading to the notion of \emph{constrained minima}. In gauge theory, this concept arise naturally as the gradient-flow equations. The solutions interpolate between widely separated instanton--anti-instanton pairs ($R \gg \rho$) and the perturbative vacuum ($R \to 0$) and can be interpreted geometrically as Lefschetz thimbles.

Furthermore, these instanton--anti-instanton configurations are closely related to sphaleron processes.
Their streamline construction thus provides a semiclassical framework for describing sphaleron production (see Sec.~\ref{sec:landscape}). It was first applied in the electroweak context~\cite{Khoze:1991sa,Shuryak:1991pn}, and more recently revisited in QCD at LHC and RHIC collider energies
\cite{Shuryak:2021iqu}.


\subsubsection{Ratio Ansatz}
\label{sec:Ratio}
To better illustrate this, one can consider an instanton $I$ and anti-instanton $A$ of equal size $\rho$ and identical color orientation, separated by a four-distance $R$ and place their centers at $z_{I}^\mu = (0,0,0,R/2)$ and $z_{A}^{\mu} = (0,0,0,-R/2)$.

Instead of employing the streamline configurations of Verbaarschot \cite{Verbaarschot:1991sq} or Yung \cite{Yung:1987zp}, which are numerically accurate but complicated, here we adopt the ratio ansatz as a variant version,
\begin{equation}
\label{eq:Ratio}
A_\mu^a(x) 
= \frac{
\bar{\eta}_{\mu\nu}^a\, (x-z_I)^\nu \, \dfrac{\rho^2}{(x-z_I)^4}
\;+\;
\eta_{\mu\nu}^a\, (x-z_A)^\nu \, \dfrac{\rho^2}{(x-z_A)^4}
}{
1 + \dfrac{\rho^2}{(x-z_I)^2} + \dfrac{\rho^2}{(x-z_A)^2}
} \,,
\end{equation}
where the instanton-related notations such as $\eta_{\mu\nu}^a$ and $\bar{\eta}_{\mu\nu}^a$ can be found in Appendix~\ref{App:conv}.

Near one of the centers, e.g., $(x-z_I)^2 \ll (x-z_A)^2$, the corresponding term dominates, reproducing the standard singular-gauge instanton (or anti-instanton) field. For $x_4 \ll 0$, the fields are approximately anti-self-dual ($\vec{E} \approx -\vec{B}$), while for $x_4 \gg 0$ they become approximately self-dual ($\vec{E} \approx \vec{B}$), with $\vec{E}=0$ at $x_4=0$. It shows that
these configurations closely resemble sphaleron paths obtained via constrained minimization. 

\begin{figure}
    \centering
\subfloat[\label{fig:ratio_a}]{\includegraphics[height=5.5cm,width=0.48\linewidth]{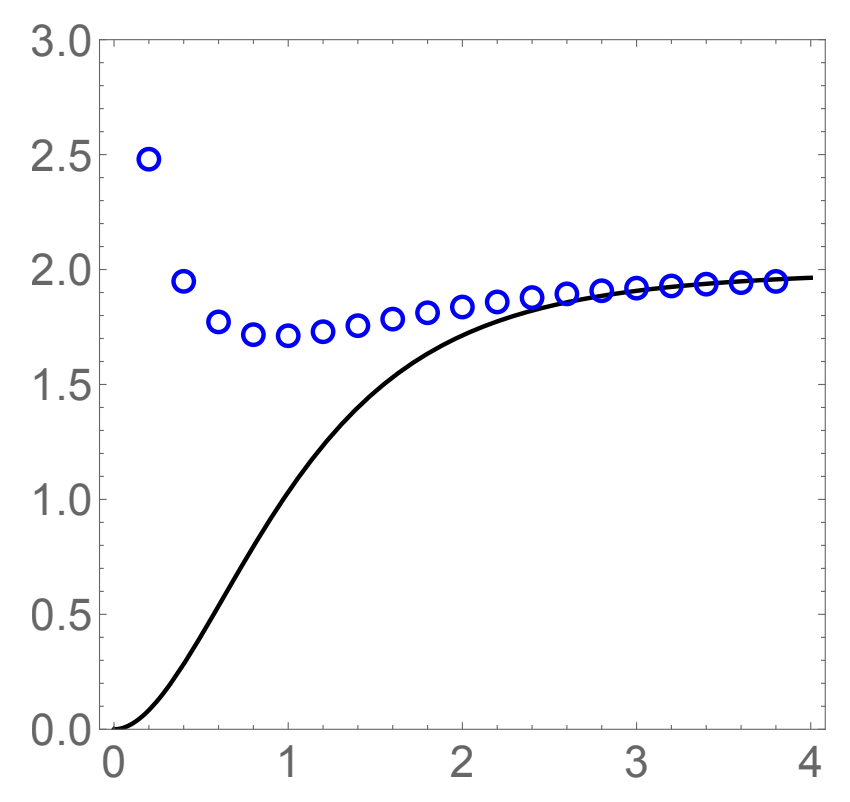}}
\hfill
\subfloat[\label{fig:ratio_b}]{\includegraphics[height=5.25cm,width=0.48\linewidth]{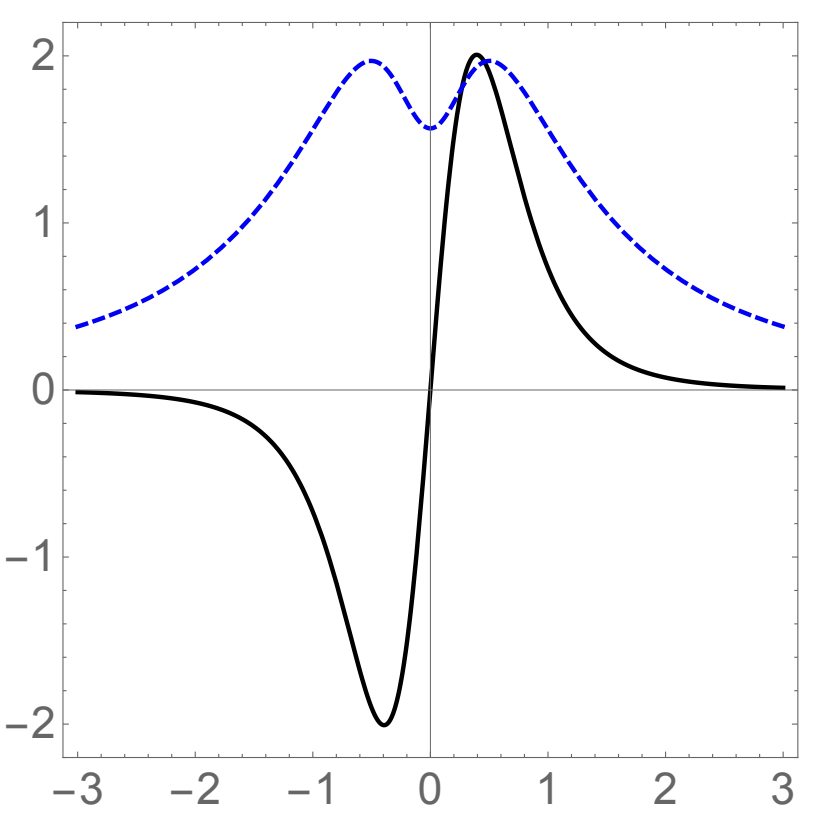}} 
    \caption{(a) Action of a $IA$ configuration (in unit of single instnaton action $8\pi^2$) in ratio ansatz vs. streamline configuration. (b) Electric field (black solid) and magnetic field (blue dashed) with $\mu= 3$, $a = 3$ as functions of $x_4/\rho$ }
    \label{fig:ratio}
\end{figure}

As illustrated in Fig.~\ref{fig:ratio_a}, the action of the $IA$ configuration, normalized to the single-instanton action $S_0 = 8\pi^2/g^2$, approaches $2$ as the separation $R \to \infty$, corresponding to two independent pseudoparticles. At small separations, however, the configuration develops a repulsive core for $R \lesssim 0.5\,\rho$, an unphysical feature induced by ratio ansatz to prevent collapse. 

Such a structure of the gauge fields reflects the underlying symmetry of the configuration. 
As presented in Fig.~\ref{fig:ratio_a}, the electric field is odd under $x^4 \to -x^4$, while the magnetic field is even. 
Similarly, $A^4$ is odd, implying that the configuration restricted to the $x^4=0$ hyperplane is purely magnetic and lies on the sphaleron path (see Sec.~\ref{sec:landscape}).  As instantons ($I$) correspond to tunneling events in Euclidean time $\tau = i x_4$, while anti-instantons ($A$) describe tunneling in the opposite direction, these molecular configurations can therefore be interpreted as forward tunneling (I) and backward tunneling (A) separated by a turning point where the electric field (the canonical field momentum) vanishes, representing incomplete tunneling events (failed tunnelings). 

Unlike isolated instantons, which are (anti-)self-dual and have vanishing stress tensor, molecular configurations are not self-dual. 
In particular, the color Poynting vector $\vec{E}^a \times \vec{B}^a$ is nonzero, indicating a nontrivial energy flow. Because the configuration is localized, the flow forms circular loops.

\subsection{Semi-classical interaction between $I$ and $A$}
\label{sec:semi-gauge-IA}
When the distance between instanton and anti-instanton is sufficiently large, these pseudoparticle develop a long-range dipole-like attractive interaction between instantons and anti-instantons. This interaction between pseudoparticles with opposite topological charges at large distances was first derived in \cite{Callan:1977gz,Forster:1977jv} by studying the interaction of an instanton with a weak, slowly varying external field strength $F_{\mu\nu}$. To illustrate this, we take the instanton field as example, the singular gauge is assumed in order to ensure that the gauge field is localized. With this in mind, one finds

\begin{equation}
\begin{aligned}
\label{inst_dipole}
     S_{\rm int}=&-i\frac{2\pi^2\rho^2}{g}\mathrm{tr}_c\left[U_I\tau^-_\mu\tau^+_\nu U_I^\dagger F_{\mu\nu}\right]\\
     =&\frac{2\pi^2\rho^2}{g}O^{ab}(U_I)\bar\eta^b_{\mu\nu}F^a_{\mu\nu}
\end{aligned}
\end{equation}
where the instanton-related notations can be found in Appendix~\ref{App:conv}. This result can be interpreted as a classical external field coupling to the color magnetic dipole moment $\frac{2\pi^2 \rho^2}{g} \bar{\eta}^a_{\mu\nu}$ of the instanton. If the external field is assumed to be an anti-instanton located at a relative distance $R$, Eq.~\eqref{inst_dipole} can be used to describe the interaction between well-separated pseudoparticles with opposite topological charges. Thus, the semi-classical gauge interaction $S_{\rm int}$ can be written as

\begin{equation}
\label{S_int}
    S_{\rm int}=\frac{32\pi^2}{g^2}\rho_I^2\rho_A^2\bar\eta_{\mu\rho}^a\eta_{\nu\rho}^bO^{ab}(U_{IA})\frac{R_\mu R_\nu}{R^6}
\end{equation}

This color force is formed by overlapping the tails of each semiclassical profiles of pseudoparticles. Since well-separated instanton–anti-instanton pairs are not significantly distorted, their interaction is well-defined semi-classically. For very close pairs, on the other hand, the instanton fields are strongly distorted. On top of that, the perturbative feature, which occurs for strongly overlapped instanton-anti-instanton pairs $R\rightarrow0$ is not included in semi-classical approximations. Thus, both the strong distortion and the perturbative feature leave the short distance description of the interaction uncertain.

It is worthing mentioning that this semi-classical gauge interaction $S_{\rm int}$ in Eq.~\eqref{S_int} can also be reproduced by the amplitude for semi-classical color force exchanges between the instantons and anti-instantons \cite{Zakharov:1992bx}

\begin{equation}
\begin{aligned}
    &\left\langle\exp\left(-\frac{2\pi^2\rho^2_I}{g}O^{ab}(U_I)\bar\eta^b_{\mu\nu}F^a_{\mu\nu}\right)\exp\left(-\frac{2\pi^2\rho^2_A}{g}O^{cd}(U_A)\eta^d_{\rho\lambda}F^c_{\rho\lambda}\right)\right\rangle\\
    =&1+\frac{4\pi^4}{g^2}\rho_I^2\rho_A^2O^{ab}(U_I)O^{cd}(U_A)\bar\eta_{\mu\nu}^b\eta_{\rho\lambda}^d\left\langle F^a_{\mu\nu}(z_I)F^c_{\rho\lambda}(z_J)\right\rangle+\cdots\\
    =&e^{-S_{\rm int}}
\end{aligned}
\end{equation}
by summing all color force exchanges with the given free propagators, 
\begin{equation}
\begin{aligned}
    \langle F^a_{\mu\nu}(x)F^b_{\rho\lambda}(0)\rangle=-\frac{2\delta^{ab}}{\pi^2x^6}\bigg[&x_\mu x_\rho\delta_{\nu\lambda}-x_\mu x_\lambda\delta_{\nu\rho}-x_\nu x_\rho\delta_{\mu\lambda}+x_\nu x_\lambda\delta_{\mu\rho}\\
    &-\frac{x^2}2\left(\delta_{\mu\rho}\delta_{\nu\lambda}-\delta_{\mu\lambda}\delta_{\nu\rho}\right)\bigg]
\end{aligned}
\end{equation}

\subsection{Instanton ensemble for YM}
\label{sec:InstYM}
Building on the previous discussion, the semi-classical partition function in Eq.~\eqref{eq:Z_I} can be generalized to the YM vacuum, which can be viewed as a statistical ensemble of multiple instantons and anti-instantons, with fluctuating pseudoparticle numbers.

This formulation provides a unified perspective on the structure of the YM vacuum, as encoded in the distributions of the action and the topological charge. These distributions are characterized by the corresponding expectation values and fluctuations, namely the mean action $\bar{S}$,  the mean topological charge $Q_{\rm top}$, the topological compressibility $\sigma_t$, and the topological susceptibility $\chi_t$. More fundamentally, the same underlying topological fluctuations produce both the trace and the axial anomalies, thereby linking the statistical description of the ensemble to essential nonperturbative features of the YM vacuum.

As a first step, we consider the canonical setting of the ensemble, in which the numbers of instantons and anti-instantons, $N_\pm$, are held fixed. In this setting, the partition function takes the form

\begin{equation}
\label{eq:Z_YM}
Z_{N_\pm}=\frac1{N_+!N_-!}\int \prod_{I=1}^{N_++N_-} d\Omega_I n(\rho_I)e^{-S_{\rm int}}
\end{equation}
where instanton tunneling rate $n(\rho)$ is defined in Eq.~\eqref{eq:dn2}, $\Omega_I$ denotes the conformal moduli space parameterized by the center, color orientation, and size $\Omega_I=(z_I,U_I,\rho_I)$ of pseudoparticles with ensemble measure defined by semi-classical interactions $S_{\rm int}$ among instantons and anti-instantons. For simplicity, only two-body interactions between instantons and anti-instantons are considered as defined in Eq.~\eqref{S_int}.

To generalize the canonical formulation to include the topological fluctuation, we introduce a auxiliary complex chemical potential, where $\theta$ is identified as the vacuum angle. 

\begin{equation}
\label{eq:grand_Z}
Z=\int\mathcal{D}A\, e^{-S_{\rm YM}[A]}\rightarrow Z(\mu,\theta)=
\sum_{N_+,N_-}
 e^{(\mu+i\theta)N_+}
 e^{(\mu-i\theta)N_-}
 Z_{N_\pm} .
\end{equation}

With the partition function in Eq.~\eqref{eq:grand_Z}, we can derive the statistical measures for the topological distributions where the pseudoparticle number $N=N_++N_-$ and the number difference $\Delta=N_+-N_-$ are allowed to fluctuate in a
grand canonical ensemble~\cite{Diakonov:1995qy,Schafer:1996wv,Zahed:2021fxk}.
\begin{equation}
\label{DISTX}
  \mathds{\mathcal P}(N_+,N_-)=\mathds P(N)\mathds Q(\Delta) 
\end{equation}
with means $\bar{N}=\langle N\rangle$ and  $Q_{\rm top}=\langle\Delta\rangle$. The corresponding measures are given by
\cite{Zahed:2021fxk,Zahed:2022wae,Diakonov:1995qy,Kacir:1996qn,Nowak:1996aj}
\begin{equation}
\label{dist}
\mathds{P}(N)=e^{\frac{bN}4 }\bigg(\frac {\bar{N}}{N}\bigg)^{\frac {bN}4 }, \qquad\, \qquad
    \mathds{Q}(\Delta)=
    \frac{1}{\sqrt{2\pi\chi_t}}\exp\left(-\frac{\Delta^2}{2\chi_t}\right)
\end{equation}
where the topological susceptibility can be related to the meson singlet mass by the Witten-Veneziano formula~\cite{Witten:1979vv,Veneziano:1979ec}
\begin{equation}
\label{CHI}
\frac{\chi_t}V=\frac{f_\pi^2}{2N_f}\left(m_{\eta^\prime}^2+m_\eta^2-2m_K^2\right)
\end{equation}
This implies that the mass of $\eta'$ is essentially due to the axial anomaly relating to non-trivial topological charge fluctuations, which can turn out to be nonzero even in the chiral limit.

To make this more explicit, we can define the relevant topological observables in YM theory. The low energy structure of the vacuum is governed by the two local gluonic operators $F^2 \equiv F_{\mu\nu}^a F^{a\,\mu\nu}$ and $F\tilde{F} \equiv F_{\mu\nu}^a \tilde{F}^{a\,\mu\nu}$. Their vacuum expectation values define the mean action $\bar S$ and the average topological charge $Q_{\rm top}$ by
\begin{equation}
\frac{g^2}{8\pi^2}\,\bar S
    =
    \frac{1}{32\pi^2}\int d^4x\, \langle F^2(x)\rangle,
    \qquad
    Q_{\rm top}
    =
    \frac{1}{32\pi^2}\int d^4x\, \langle F\tilde{F}(x)\rangle 
\end{equation}
and their connected two-point correlation functions characterize the vacuum fluctuations by
\begin{equation}
    \frac{\sigma_t}{\bar N}
    =
    \frac{1}{32\pi^2 \langle F^2\rangle}
    \int d^4x\,
    \Big(
    \langle F^2(x) F^2(0)\rangle
    -
    \langle F^2\rangle^2
    \Big),
\end{equation}
and
\begin{equation}
    \frac{\chi_t}{\bar N}
    =
    \frac{1}{32\pi^2 \langle F^2\rangle}
    \int d^4x\,
    \Big(
    \langle F\tilde{F}(x) F\tilde{F}(0)\rangle
    -
    \langle F\tilde{F}\rangle^2
    \Big).
\end{equation}

These observables can be directly related to the partition function in Eq.~\eqref{eq:grand_Z} within the grand-canonical ensemble formulation. The derivatives of the partition function with respect to the inverse coupling $8\pi^2/g^2$ yield the mean action and its fluctuations:
\begin{equation}
\label{eq:N1}
-\frac{\partial \ln Z}{\partial (8\pi^2/g^2)}
=\frac{g^2}{8\pi^2}\bar{S},
\end{equation}
\begin{equation}
\label{eq:N2}
\frac{\partial^2 \ln Z}{\partial (8\pi^2/g^2)^2}
=
\left(\frac{\sigma_t}{\bar N}\right)
\frac{V}{32\pi^2}
\langle F_{\mu\nu}^2 \rangle.
\end{equation}

For a sufficiently dilute ensemble, the mean action can be simply related to the total number of pseudoparticles under mean field,

\begin{equation}
    \bar{N}=\frac{g^2}{8\pi^2}\,\bar S
\end{equation}
thus one can further conclude 
\begin{equation}
\label{eq:N1mu}
\frac{\partial\ln Z}{\partial\mu}\bigg|_{\mu=0}=
-\frac{\partial \ln Z}{\partial (8\pi^2/g^2)}
=\bar{N},
\end{equation}
\begin{equation}
\label{eq:N2mu}
\frac{\partial^2\ln Z}{\partial\mu^2}\bigg|_{\mu=0}=\frac{\partial^2 \ln Z}{\partial (8\pi^2/g^2)^2}
=\sigma_t
\end{equation}

Similarly, for the topological sector, the derivatives with respect to the vacuum angle $\theta$ determine the topological charge and its fluctuations:
\begin{equation}
-i\,\frac{\partial \ln Z}{\partial \theta}
=
Q_{\rm top},
\end{equation}
\begin{equation}
-\frac{\partial^2 \ln Z}{\partial \theta^2}
=\chi_t.
\end{equation}

\subsubsection{Vacuum compressibility and trace anomaly}

The mean action measures the overall strength of gluon field fluctuations in the Yang–Mills vacuum, reflecting the density of gauge field configurations. This quantity is also directly tied to the vacuum energy density by trace anomaly. For a 4D Lorentz-invariant vacuum, the energy density $\epsilon_0$ is given by one quarter of the trace of the energy–momentum tensor (EMT). Through the trace anomaly, $T^\mu_{\ \mu}$ is proportional to $F^2$ with a coefficient governed by $\beta$-function. Thus one has

\begin{equation}
\epsilon_0 = \frac{1}{4}\,\langle T^\mu_{\ \mu} \rangle = 
\frac{\beta(g^2)}{4g^2}\, \langle F^2\rangle ,
\end{equation}

Due to energy--momentum conservation, the vacuum energy density carries no anomalous dimension and therefore scales by its canonical mass dimension. As a result, its RG equation is entirely fixed by dimensional analysis, 
\begin{equation}
\beta(g^2)\frac{\partial}{\partial g^2}\langle T^\mu_{\ \mu} \rangle
=
-4\,\langle T^\mu_{\ \mu} \rangle.
\end{equation}

Using the trace anomaly relation to connect $T^\mu_{\ \mu}$ and the gluonic operator $F_{\mu\nu}^2$, one can convert this RG equation into the corresponding one for the gluon condensate. This yields
\begin{equation}
\beta(g^2)\frac{\partial}{\partial g^2}\langle F^2_{\mu\nu}\rangle
=
\left[
4
-
\frac{\beta(g^2)}{g^4}
\left(
\beta(g^2)\frac{\partial}{\partial g^2}\frac{g^4}{\beta(g^2)}
\right)
\right]
\langle F^2_{\mu\nu}\rangle.
\end{equation}
This implies the gluon condensate is approximately RG-invariant as long as the strong coupling $g^2(\mu)$ is sufficiently small.

Using Eq.~\eqref{eq:N1} and \eqref{eq:N2}, the compressibility can be expressed explicitly as
\begin{equation}
\label{eq:sigma}
    \frac{\sigma_t}{\bar{N}}
    =
    \frac{1}{8\pi^2}
    \left[
    4\,\frac{g^4}{\beta(g^2)}
    -
    \left(
    \beta(g^2)\frac{\partial}{\partial g^2}\frac{g^4}{\beta(g^2)}
    \right)
    \right].
\end{equation}

In this form, it becomes clear that the compressibility is entirely controlled by the RG running of the coupling, showing that fluctuations of the gluon condensate are controlled by the scale dependence of the theory. At one-loop, the result in \eqref{eq:sigma} can be further simlified as

\begin{equation}
\label{COMP}
\frac{\sigma_t}{\bar{N}} =\frac 4b \approx \frac 1{N_c}
\end{equation}
where $b=11N_c/3$, in agreement with low-energy theorems~\cite{Novikov:1981xi}. In this formulation the cooled QCD vacuum is a quantum liquid of pseudoparticles, with a topological compressibility of about $\frac 13$ at $N_c = 3$, and 
incompressible at large $N_c$.
For a sufficiently small coupling $g^2$, this simplification leads to 

\begin{equation}
(-1)^n
\frac{d^n \ln Z}{d(8\pi^2/g^2\bigr)^n}
=
\left(\frac{4}{b}\right)^{n-1}\bar{N}.
\end{equation}
and the corresponding grand partition function as a function of $\mu$ at $\theta=0$ is given by
\begin{equation}
\label{eq:Z_grand}
Z(\mu,0)
=
\exp\!\left[
\frac{b}{4}\bar{N}\exp\!\left(\frac{4\mu}{b}\right)
\right] ,
\end{equation}
reproducing the distribution for the pseudoparticle number in Eq.~\eqref{dist}. 

\begin{figure}
    \centering
    \includegraphics[width=0.8\linewidth]{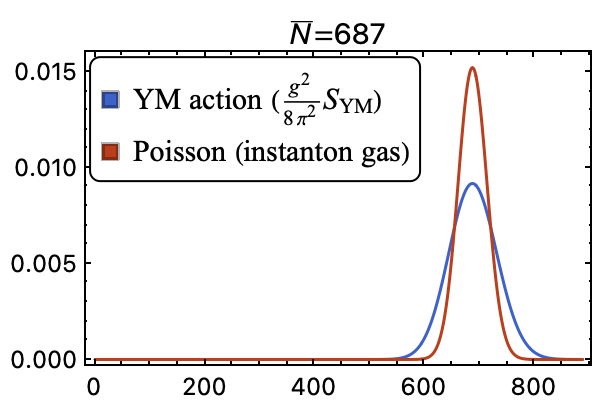}
    \caption{Mean action distribution in physical YM vacuum (blue and in ideal instanton gas (red) }
    \label{fig:mean_action}
\end{figure}

As illustrated in Fig.~\ref{fig:mean_action}, for an ideal instanton gas, the distribution of pseudoparticles is Poissonian. However, in YM theory, this simple picture is modified by interactions among pseudoparticles. The resulting deviation from the Poisson distribution is a direct consequence of renormalizability, implying nontrivial correlations among pseudoparticles to drive the distribution away from the instanton-gas limit. 

\subsubsection{Vacuum susceptibility and chiral anomaly}

The partition function in Eq.~\eqref{eq:Z_grand} can be generalized to arbitrary $\theta$ angle straightforwardly by $\bar{N}\rightarrow\langle{N}\rangle_\theta$.

\begin{equation}
Z(\mu,\theta)
=
\exp\!\left[
\frac{b}{4}\langle{N}\rangle_\theta\exp\!\left(\frac{4\mu}{b}\right)
\right]
\end{equation}
with $b=11N_c/3$.

The $\theta$ angle acts as a topological chemical potential for the number difference $\Delta$, enhancing instantons and depleting anti-instantons. The angle dependence of $\langle N\rangle_\theta$ in general is subject to the interaction between pseudoparticles in YM.

For convenience, we define $\gamma_{s,a}$ as the averaged interaction between pseudoparticle pairs with the same duality and the ones with opposite duality.

\begin{equation}
    \int d^4Rdu\, S_{\rm int}(R,u,\rho_1,\rho_2)=\frac{8\pi^2}{g^2}\rho_1^2\rho_2^2\gamma_{s,a}
\end{equation}
where the separation is defined by $R=z_1-z_2$ and the relative color orientation is $u=U^\dagger_1U_2$, whose average over the color space leads to $\gamma_{s,a}\propto N_c/(N_c^2-1)$. For the simple sum ansatz,

\begin{equation}
\label{eq:sum}
    A_\mu=\sum_IA_{I\mu}
\end{equation}
we obtain $\gamma_a=\gamma_s$ and the resulting angle dependence is given by

\begin{align}
\label{SCALE}
\langle N\rangle_\theta=\bar{N}(\cos\theta)^{4/b}
\end{align}
For $\gamma_a=0$, the angle dependence instead becomes

\begin{align}
\label{SCALE2}
\langle N\rangle_\theta=\bar{N}\cos\left(\frac4b\theta\right)
\end{align}

The general solution for quenched vacuum obtained by Feynman variational principle is \cite{Diakonov:1995qy}

\begin{equation}
\langle N\rangle_\theta
=
\cos\vartheta
\left(
\frac{
\gamma_a\cos\vartheta + \sqrt{\gamma^2_s - \gamma_a^2 \sin^2\vartheta}
}{
\gamma_s+\gamma_a
}
\right)^{\frac{4-b}{b}}
\end{equation}
where the interaction-modified angle $\vartheta$ can be controlled by
\begin{equation}
\vartheta + \frac{b-4}{4}\,
\arctan\!\left(
\frac{(\gamma_s^2-\gamma_a^2)\sin\vartheta}{\gamma_s^2\cos\vartheta + \gamma_a\sqrt{\gamma_s^2-\gamma_a^2\sin^2\vartheta}}
\right)
= \theta
\end{equation}

Using the formula above, the zero-angle topological susceptibility is obtained by
\begin{equation}
\label{chi_t}
\chi_t= \frac{4\bar N}{b-\frac{\gamma_a}{\gamma_s}(b-4)}
\end{equation}

These interaction parameters are subject to several physical constraints. First, the condition $\gamma_s + \gamma_a > 0$ ensures stability of the system by preventing the ensemble from collapsing. Second, $\gamma_s \ge 0$ is required to avoid an unphysical phase separation into two subsystems consisting purely of instantons and anti-instantons, which would correspond to maximal CP violation. Finally, the positivity of the topological susceptibility, $\chi_{\rm top} > 0$, imposes the additional constraint $\gamma_a / \gamma_s < b/(b - 4)$.

For the simple sum ansatz \eqref{eq:sum}, $\gamma_a=\gamma_s$, the topological susceptibility reduces to the average number of instantons, 

\begin{equation}
\label{eq:chi_sum}
\chi_t=\bar{N}. 
\end{equation}

For $\gamma_a=0$, Eq.~\eqref{chi_t} reduces to the result obtained by imposing exact duality, indicating the nontrivial interaction between $IA$ pairs break the duality. In general, an attractive interaction between instanton–anti-instanton ($IA$) pairs reduces the topological susceptibility, whereas a repulsive interaction enhances it.

\begin{eqnarray}
\begin{aligned}
\chi_t&<\;\frac{4}{b}\,\bar{N},\,\qquad \text{for attractive}~IA\nonumber\\
\chi_t&>\;\frac{4}{b}\,\bar{N},\,\qquad \text{for repulsive}~IA
\end{aligned}
\end{eqnarray}

A more sophisticated method using streamline ansatz \cite{1991NuPhB.362...33V} has shown that the strong $IA$ repulsion is an artifact of the simple sum ansatz. The $IA$ interaction strongly depends on the
orientation and the average repulsion is about 14 times smaller than the one obtained in \cite{Diakonov:1983hh}. Therefore, we can practically assume the average $IA$ interaction to be small such that quenched susceptibility can be approximated by

\begin{equation}
\label{ILM_chi}
    \chi_t\approx\frac4b \bar{N}
\end{equation}

\section{Instanton vacuum with quarks}
\label{sec:Inst_QCD}
The same semi-classical calculation for single instanton in Sec.~\ref{sec:background} can be generalized to QCD straightforwardly by including the quark determinant $\det(\slashed{\bar{D}}+m_f)$ for each flavor~\cite{tHooft:1976snw}.

\begin{equation}
\label{eq:Z_I_q}
Z=V\int d\rho n_0(\rho)F(m\rho)^{N_f}
\end{equation}
where the full quark contribution is approximately given by \cite{Wantz:2009qk,Dunne:2005te}
\begin{equation}
\begin{aligned}
F(m)&= \exp\Bigg(
\frac{
\frac{1}{3}\ln m
+ 0.29175
- 0.9916\, m^{2}
- 26.83\, m^{4}
- 5.292\, m^{6}
}{
1
- 3 m^{2}
+ 107.30\, m^{4}
+ 661.94\, m^{6}
+ 198.44\, m^{8}
}-\frac{1}{3}\ln m\Bigg)\\
&\xrightarrow{m\rightarrow0} 1.34\, m
\end{aligned}
\end{equation}
with the quark mass dependence induced by quark non-zero modes of single instanton \cite{Wantz:2009qk,Dunne:2005te}

Although the structure of a single instanton already breaks $U(1)$ chiral symmetry by the quark determinant. It is not sufficient to account for spontaneous chiral symmetry breaking. A single instanton amplitude has a probability proportional to very small product of Higgs-induced light quark masses $m_um_dm_s/\Lambda_{\rm QCD}^3\sim 10^{-4}$. As a result, in the chiral limit, a single-instanton configuration does not generate a nontrivial fermion determinant and thus cannot by itself induce a nonzero chiral condensate.

This limitation indicates that a description based solely on isolated instantons, as in the dilute instanton gas approximation (DIGA), is incomplete. As emphasized in \cite{Shuryak:1981ff}, the QCD vacuum is more appropriately described as an interacting liquid, in which strong correlations between instantons are mediated by light quark exchanges. In this picture, instantons and anti-instantons do not behave as independent objects, but instead form correlated clusters in the ensemble. 

Consequently, to capture the dynamics responsible for spontaneous chiral symmetry breaking, one must go beyond the single-instanton approximation and incorporate these collective effects arising from the interacting instanton ensemble.

\subsection{Spontaneous chiral symmetry breaking}

As shown in Fig.~\ref{fig:vac_q}, spontaneous chiral symmetry breaking is induced by quark propagation through the instanton ensemble via fermionic zero modes. These zero modes form a set of nearly degenerate states, allowing quarks to hop between pseudoparticles. The resulting trajectories may be either localized (forming short loops) or extended over large distances. In the thermodynamic limit $V \to \infty$, those extended loops become delocalized and give rise to an accumulation of Dirac eigenvalues near zero, $\lambda \sim 1/V$, consistent with the Banks–Casher mechanism. This mechanism provides the dynamical origin of effective quark mass from initially massless quarks and plays a central role in shaping the structure and composition of light hadrons.

The delocalization of massless quarks of definite chirality within a narrow band of near-zero modes around the Dirac virtuality $\lambda \approx 0$ is also known as the zero-mode zone (ZMZ). This structure arises from the collective mixing of quark zero modes, which generates a dense set of near-zero eigenvalues without opening a spectral gap. Instead, these eigenvalues are confined within a narrow band,
\begin{equation}
|\lambda| \sim \text{width (ZMZ)} \sim \frac{\rho^2}{R^3} \sim 50~\text{MeV},
\label{eq:ZMZ_width}
\end{equation}
as predicted in the instanton liquid model with standard phenomenological parameters.

The physical significance of this near-zero mode region has been demonstrated in lattice studies~\cite{Glozman:2012fj}, where hadronic spectra were analyzed after systematically removing Dirac eigenmodes below a cutoff $|\lambda| < \Delta$. A dramatic restructuring of the light-hadron spectrum occurs once $\Delta$ exceeds the ZMZ width. In particular, the Nambu–Goldstone modes, such as the pion, disappear entirely, demonstrating that their existence is tied to the near-zero Dirac eigenmodes responsible for spontaneous chiral symmetry breaking.

\subsection{Quark propagators in semi-classical background}
\label{sec:QP}
With the previous discussion, we introduce a systematic planar resummation to organize the quark propagation in various correlation functions in the multi-instanton background. 

For simplicity, we consider single flavor in the vacuum to begin with. The quark propagator $S(x,y)$ in the multi-instanton vacuum can be computed by the ensemble  average as \cite{Diakonov:1985eg,Pobylitsa1989TheQP}

\begin{equation}
\begin{aligned}
\label{SRIV_1}
    S=&\left\langle\frac{1}{i\slashed{\partial}+\sum_I\slashed{A}_I+im}\right\rangle\\
    =&\left\langle S_0+\sum_I(S_I-S_0)+\sum_{I\neq J}(S_I-S_0)S_0^{-1}(S_J-S_0)+\cdots\right\rangle
\end{aligned}
\end{equation}
where $S_I$ is the quark propagator with single instanton background defined as
\begin{equation}
    S_I=\frac{1}{i\slashed{\partial}+\slashed{A}_I+im}
\end{equation}
Here the ensemble average $\left\langle\cdots\right\rangle$ runs the entire instanton ensemble with sampling weighed by either the interacting between the pseudoparticles (interacting instanton liquid ensemble, IIL) or with the equal sampling (random instanton liquid ensemble, RIL) for simplicity. 


\begin{equation}
    \langle \cdots \rangle=\prod_I\int \frac{d^4z_IdU_I}{V}\cdots
\end{equation}

\begin{figure}
    \centering
    \includegraphics[width=\linewidth]{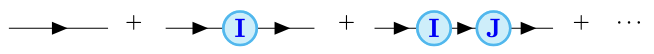}
    \caption{Quark propagator distorted by the instanton background with zero modes and non-zero modes. The zero mode contribution can be rewritten as 'tHooft vertices in \eqref{Hooft} due to the delocalization}
    \label{fig:Q_prop}
\end{figure}

This instanton expansion is presented graphically in Fig.~\ref{fig:Q_prop}. Each instanton vertex $V_I$ denoted by blue circles can be obtained by Lehmann--Symanzik--Zimmermann (LSZ) reduction presented in Eq.~\eqref{SI}, including both zero mode and non-zero mode contributions. 

\begin{equation}
\label{SI}
   V_I=S_0^{-1}(S_I-S_0)S_0^{-1} 
\end{equation}

The generalization to $N_f$ flavor vertex can be achieved straightforwardly by (see \cite{hutter2001instantonsqcdtheoryapplication} for more details)

\begin{equation}
V_I=   \prod_{f}
   [S_0^{-1}(S^f_I-S_0)S_0^{-1}]
\end{equation}

For simplicity, here our discussion will stick to the single flavor case. Large $N_c$ QCD is essentially a quenched approximation dominated by planar graphs. The same applies to its semi-classical version in terms of a random instanton vacuum. As a result, in the large $N_c$ limit, the resummation of these planar diagrams repackages the diagrams involving the same instanton at both the beginning and the end, yielding ~\cite{Pobylitsa1989TheQP,Liu:2023yuj}

\begin{equation}
\begin{aligned}
S=&\left\langle S_0+S_0\left(\sum_IM_I\right)S_0+\cdots\right\rangle=\left\langle\frac1{S_0^{-1}-\sum_IM_I}\right\rangle
\end{aligned}
\end{equation}
where the effective quark self-energy in instanton vacuum is given by the iterative equation
\begin{equation}
\begin{aligned}
\label{MqI}
    M_I
    =&V_I+V_I(S-S_0)M_I
\end{aligned}
\end{equation}


By defining the averaged total quark self-energy, 
\begin{equation}
    i\sigma=-\sum_I\int \frac{d^4z_IdU_I}{V}M_I
\end{equation}
the solution to Eq.~\eqref{MqI} yields a gap-like equation

\begin{equation}
\begin{aligned}
   &i\sigma=\frac{n_{I+A}}{N_c}\int d^4z_I\mathrm{Tr}_c\left(\slashed{A}_I\frac{1}{S_I^{-1}+i\sigma}\left(S_0^{-1}+i\sigma\right)\right)
\end{aligned}
\end{equation}
which can be solved order by order with the density expansion $\sigma=\sqrt{\frac{n_{I+A}}{N_c}}\sigma_0+\frac{n_{I+A}}{N_c}\sigma_1+\cdots$. The non-analytical dependence on the instanton density $\sqrt{n_{I+A}}$ at the leading order is the consequence of disordering in the random instanton vacuum. 

Eventually, a momentum-dependent constituent quark mass naturally emerge in instanton vacuum, determined by  \cite{Liu:2023fpj,Kock:2020frx,Pobylitsa1989TheQP}

\begin{eqnarray}
    M(k)=m+\sigma(k)
\end{eqnarray}

The nearly massless quarks acquire a substantial dynamical mass, denoted as $ M(k) $. The quark propagator can be written as
\begin{equation}
\label{running mass}
    S(x,y)=\int \frac{d^4k}{(2\pi)^4}\frac{\slashed{k}-iM(k)}{k^2+M^2(k)}e^{-ik\cdot (x-y)}
\end{equation}

\subsubsection{Zero mode delocalization in light quarks}

  
Generally, quark propagator with single instanton $S_I$ appears as a sum over zero modes and non-zero modes. The explicit solution for both zero mode and non-zero mode contributions in single instanton background can be found in Appendix~\ref{App:NZM}. In the case of light quarks, the zero modes dominates due to the nearly zero (current) mass. As a result, the propagator in the single instanton can be approximated by \cite{Diakonov:1985eg}

\begin{equation}
    S_I(x,y)\simeq\frac{\phi_I(x)\phi_I^\dagger(y)}{im}+S_0(x-y)
\end{equation} 

The non-zero mode contribution is interpolated into a free propagator $S_0$. With this interpolation, the propagator appears in the instanton resummation \eqref{SRIV_1} can be further simplified~\cite{Diakonov:1985eg,Kock:2020frx,Kock:2021spt}

\begin{equation}
\begin{aligned}
\label{eq:Q_prop0}
    S(x,y)&\simeq S_0(x-y)+\left\langle\sum_{I,J}\phi_I(x)\frac{1}{im-im D_{IJ}-T_{IJ}}\phi_J^\dagger(y)\right\rangle
\end{aligned}
\end{equation}
where the hopping integrals are defined as

\begin{equation}
\label{TIJ}
    T_{IJ}=\int d^4x\phi_I^\dagger(x) i\slashed{\partial}\phi_J(x)
\end{equation}

\begin{equation}
\label{DIJ}
    D_{IJ}=\int d^4x\phi_I^\dagger(x)\phi_J(x)-\delta_{IJ} 
\end{equation}

If we only focus on the contribution from zero modes, the iterative equation in \eqref{MqI} can be solved by ansatz assuming quark self-energy $M_I$ is of the form,

\begin{equation}
\label{self_energy_q}
M_I=\frac{1}{i\sigma_0}S_0^{-1}|\phi_I\rangle\langle\phi_I|S_0^{-1}
\end{equation}
The solution to Eq.~\eqref{MqI} can be simplified to a self-consistent condition for the disordering mass parameter $\sigma_0$.

\begin{equation}
\begin{aligned}
\label{eq:det_gap}
    &\sigma_0
    =m+8\pi^2\rho^2\int\frac{d^4k}{(2\pi)^4}\frac{\sigma(k)\left[1+\frac{m}{k^2}(m+\sigma(k))\right]}{k^2+[m+\sigma(k)]^2}\mathcal{F}(\rho k)
\end{aligned}
\end{equation}

A comparison between $M_I$ and $V_I$ reveals that the zero-mode singularity $1/m$ is regularized by the inclusion of disordering effects, leading to a finite contribution $1/\sigma_0$ even in chiral limit. By substituting the disordering $\sigma_0$ in \eqref{eq:det_gap} back into \eqref{self_energy_q} 
the constituent mass is obtained as

\begin{equation}
\label{eq:cons_m}
    M(k)\simeq m+\frac {n_{I+A}}{2N_c} \frac{4\pi^2\rho^2}{\sigma_0}\mathcal{F}(\rho k)
\end{equation} 
where $\mathcal{F}(\rho k)$ is the profile of the quark zero mode~\cite{Liu:2023yuj,Kock:2020frx} defined by
\begin{equation}
\label{ZMform}
    \sqrt{\mathcal{F}(k)}=-z\frac{d}{dz}[I_0(z)K_0(z)-I_1(z)K_1(z)]\bigg|_{z=\frac{k}{2}}
\end{equation}
The constituent mass gap in \eqref{eq:cons_m} is slightly different compared to~\cite{Kock:2020frx}.

\begin{figure}
    \centering
    \includegraphics[width=0.8\linewidth]{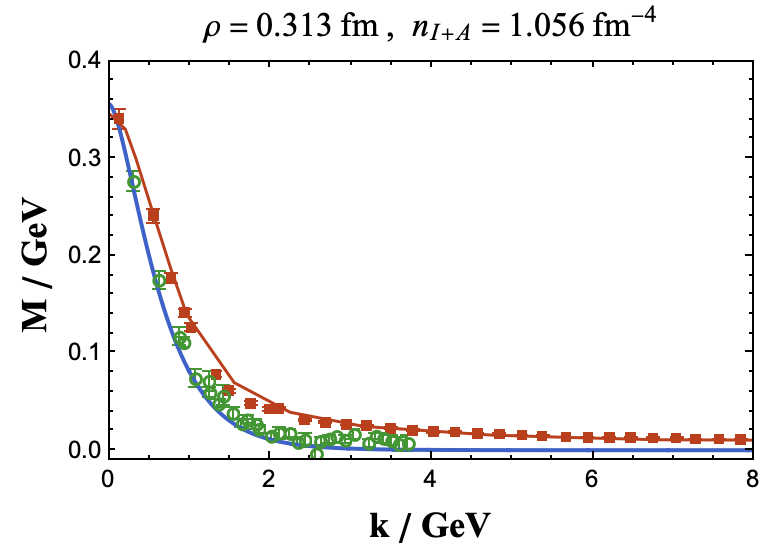}
    \caption{The constituent mass $M(k)$ running with the quark momentum $k$ with the instanton size $\rho=0.313$ fm and $n_{I+A}=1.056$ fm$^{-4}$ compared with lattice QCD using dynamical $O(a)$-improved Wilson fermions \cite{Oliveira:2018lln} (red) and result using overlap and Asqtad fermions \cite{Bowman:2004xi} in Landau gauge (green).}
    \label{fig:constituent_mass}
\end{figure}

The quark condensate in chiral limit at the leading order of instanton density reads
\begin{equation}
\begin{aligned}
\label{qq}
    \langle \bar q q\rangle=&-4N_c\int\frac{d^4k}{(2\pi)^4}\frac{M(k)}{k^2+M^2(k)}\\
    \simeq&-\frac{n_{I+A}}{\sigma_0}+\mathcal{O}(n_{I+A}^2)
\end{aligned}
\end{equation}

In Fig.~\ref{fig:constituent_mass}, we compare our result of constituent mass with the lattice QCD approach \cite{Oliveira:2018lln,Bowman:2004xi}.

This planar resummation in the multi-instanton vacuum with $1/N_c$ book-keeping can be straightforwardly generalized to any correlation functions. For more details, see \cite{Pobylitsa1989TheQP,hutter2001instantonsqcdtheoryapplication}. 

\subsubsection{Non-zero mode dominance in heavy quarks}

Since the zero modes depend inversely on the quark mass, they mostly do not contribute to heavy flavor. Instead, heavy quarks receive significant contribution from non-zero modes~\cite{Chernyshev:1995gj}. The heavy quark propagator moving in velocity $v_\mu$ can be written in the form of Wilson line as,

\begin{equation}
\begin{aligned}
S(x)=\frac{1+\slashed{v}}2\delta^3(\vec{x})\Theta(\tau)\left\langle\mathcal{P}\exp\left(i\int d\tau v_\mu A_\mu\right)\right\rangle
\end{aligned}
\end{equation}
where $\vec{x}$ denotes the transverse space perpendicular to $v_\mu$.

In this heavy mass limit, the multi-instanton contribution in propagator resums to Wilson line where rhe full instanton contribution to it is given by the exponent of the all-order single instanton result~\cite{Shuryak:2000df,Shuryak:2021hng,Dorokhov:2002qf}

\begin{equation}
\begin{aligned}
    &\left\langle\mathcal{P}\exp\left(i\int dx_\mu A_\mu\right)\right\rangle=\exp\left[\frac{1}{N_cV}\sum_I\int d^4z ~\mathrm{Tr}_c\left(W_I(\rho,z)-1\right)\right]
\end{aligned}
\end{equation}
with the single instanton inserted Wilson line
\begin{equation}
\begin{aligned}
&W_I(\rho,z)=\exp\left(i\tau^a\int_{C} dx_\mu  \frac{\bar\eta^a_{\mu\nu}(x-z)_\nu\rho^2}{(x-z)^2[(x-z)^2+\rho^2]}\right)
\end{aligned}
\end{equation}
Therefore, for the heavy quark propagator, the non-zero mode contribution thus can be studied by the straight line along $v_\mu$. The result reads

\begin{equation}
    \left\langle\mathcal{P}\exp\left(i\int_{-\infty}^{\infty} d\tau v\cdot A_I\right)\right\rangle=\cos \phi +i\frac{\vec{x}}{|\vec{x}|}\cdot\vec{\tau}\sin\phi
\end{equation}
where instanton cummulated phase is 
\begin{equation}
    \phi=\pi\left(1-\frac{|\vec{x}|}{\sqrt{|\vec{x}|^2+\rho^2}}\right)
\end{equation}

\subsection{Instanton ensemble for QCD}
\label{sec:ILM_QCD}
Building on the previous discussion, we can extend the single instanton partition function in Eq.~\eqref{eq:Z_I_q} to the QCD vacuum with $N_c$ colors and $N_f$ light quarks. This again can be formulated in a statistical ensemble of $N_+$ instantons and $N_-$ anti-instantons, with the numbers fluctuation controlled by distribution similar to Eq.~\eqref{DISTX}. 

For a canonical setting (fixed $N_\pm$), a complete quantitative framework for QCD at low resolution can be established from the partition function $Z_{N_\pm}$ given by \cite{Schafer:1996wv,Liu:2025ldh}
\begin{equation}
\begin{aligned}
\label{eq:Z_N}
   Z_{N_\pm}=&\frac{1}{N_+!N_-!}\int\prod_{I=1}^{N_++N_-}d\Omega_I n(\rho_I)e^{-S_{\rm int}}
  \prod_{f}\mathrm{Det}(\slashed{D}+m_f)
\end{aligned}
\end{equation}
where $\Omega_I$ denotes the moduli space parameterized by their size $\rho_I$, center $z_I$, and color orientation $U_I$ $\Omega_I=(z_I,U_I,\rho_I)$ for each single (anti-)instanton and $S_{\rm int}$ is the semi-classical gauge interaction among pseudoparticles. The ensemble measure is defined by the semi-classical gluonic interactions $S_{\rm int}$ among instantons and anti-instantons, sea quark determinant, and the effective tunneling rate $n(\rho)$ in Eqs.~\eqref{eq:n} and \eqref{eq:dn2} where the the running coupling now is modified by light flavor numbers. This measure depends on the relative color orientation defined as 
$U_{IJ}=U_I^\dagger U_J$ and the relative distance $R_{IJ}=z_I-z_{J}$ among each pair of instantons and anti-instantons. The details can be found in Eqs.~\eqref{eq:run_coup} and \eqref{eq:beta_GL}. See also \cite{Zakharov:1992bx,Wantz:2009qk,Shuryak:1989cx}.


The most distinctive part in QCD, compared to pure YM theory, is the presence of the quark determinant, which receives contributions from both high-momentum modes and low-momentum modes. The contribution of the higher modes are localized on the pseudoparticles and thus renormalize the mean-density rate with an additional factor of $\rho$ for each flavor. The low momentum modes, on the other hand, are approximated in the form of quasi-zero modes delocalized among the instantons and anti-instantons. Therefore, in ILM, the low-momentum part of quark determinant is usually factorized out and approximated by the determinant of the overlap matrix $T$ and $D$ in the zero mode subspace~\cite{Diakonov:1995ea, Diakonov:1995qy,Schafer:1996wv,doi:10.1142/1681}. The determinant for each quark flavor can be approximately written as~\cite{Shuryak:1989cx}

\begin{equation}
    \mathrm{det}(\slashed{D}+m)\approx \rho^{N}\mathrm{det}(
        m+mD-iT)
\end{equation}
where $T$ and $D$ are a $N\times N$ matrix with $N=N_++N_-$ and their entries $T_{IJ}$ and $D_{IJ}$ are defined by the quark zero mode overlap in Eqs.~\eqref{TIJ} and \eqref{DIJ} (see also Eq.~\eqref{eq:TIADII}). For simplicity, the matrix $D$ preceded by a small factor $m$, will often be neglected. This reflects that light quark hopping between instantons with the same duality is mostly forbidden due to the chirality. For more details, see~\cite{Diakonov:1995ea,Schafer:1996wv,doi:10.1142/1681} (reference therein). 


This approximation is equivalent to the zero-mode interpolation in Eq.~\eqref{eq:Q_prop0}. In fact, the two equations can be related through LSZ reduction. With this in mind, the effective 't Hooft vertices $\Theta_{I}$ reads

\begin{equation}
    \begin{aligned}
        \label{Hooft}
        \Theta_{I}=&\prod_f\left[m_f-\frac{4\pi^2\rho^2}{8}\bar{\psi}_fU_I\tau^\mp_\mu\tau^\pm_\nu\gamma_\mu\gamma_\nu U_I^\dagger\frac{1\mp\gamma^5}{2}\psi_f\right]
    \end{aligned}
\end{equation}
to lowest order in the current quark masses $m_f$ where
\begin{equation}
    \frac14\tau^\mp_\mu\tau^\pm_\nu\gamma_\mu\gamma_\nu=
\mathds{1}_2+\frac{1}{4}\tau^a \bar{\eta}^{a}_{\mu\nu}\sigma_{\mu\nu}
\end{equation}
The emergent vertices \eqref{Hooft} can be generalized to include non-local effect induced by finite size of the pseudoparticles. More specifically, each  quark field in the interaction vertices $\Theta_{I}$ get dressed by
\begin{align}
&\psi(k)\rightarrow\sqrt{\mathcal{F}(\rho k)}~\psi(k)
\end{align}
where the non-local quark form factor is essentially the profiling of the instanton by the quark zero mode defined in Eq.~\eqref{ZMform}.

Now the low resolution QCD path integral can be rewritten as

\begin{equation}
\begin{aligned}
\label{eq:Z_N2}
   Z_{N_\pm}=&\int \mathcal{D}\psi \mathcal{D}\bar\psi \exp\left(-\int d^4x\bar{\psi}\slashed{\partial}\psi\right)\\
   &\times\frac1{N_+!N_-!}\prod_{I=1}^{N_++N_-}\left(\int d^4z_IdU_I\int d\rho_I n(\rho_I)\rho_I^{N_f}\Theta_I(z_I) \right)e^{-S_{\rm int}}
\end{aligned}
\end{equation}

Since the size distribution is very sharp, the partition function can be further approximated by introducing a mean-field partition function for gluon sector and imposing a fixed mean instanton size $\rho$ to the remaining quark sector.
\begin{equation}
\label{eq:Zg_N2}
   Z^{(g)}_{N_\pm}=\frac{1}{N_+!N_-!}\left(\int d\rho n(\rho)V\right)^{N_++N_-} e^{-\bar{S}_{\rm int}}
\end{equation}
where $V$ is the 4-volume of the instantons live in and
$\bar{S}_{int}$ is the averaged two-body semi-classical interaction~\cite{Diakonov:1995qy,Diakonov:1983hh}.

The mean-field ILM ensemble, which we refer to as the random instanton liquid model (RILM), is thus given by

\begin{equation}
\begin{aligned}
\label{eq:Z_N3}
   Z_{N_\pm}\rightarrow&Z_{N_\pm}^{(g)}\int \mathcal{D}\psi \mathcal{D}\bar\psi e^{-\int d^4x\bar{\psi}\slashed{\partial}\psi}\prod_{I=1}^{N_++N_-}\left(\int \frac{d^4z_IdU_I}V\rho^{N_f}\Theta_I(z_I) \right)e^{-\delta S_{\rm int}(\rho)}
\end{aligned}
\end{equation}
where $\delta S_{\rm int}$ is the residual long-range semi-classical interaction among pseudoparticles which we will elaborate in Sec.~\ref{sec:tail}.

We can now generalize to a grand-canonical ensemble description of the topological distributions, in which both the total pseudoparticle number $N = N_+ + N_-$ and the number difference $\Delta = N_+ - N_-$ are allowed to fluctuate.

\begin{equation}
\label{eq:grand_ZQCD}
Z=\int\mathcal{D}\psi\mathcal{D}\bar\psi\mathcal{D}A\, e^{-S_{\rm QCD}[\psi,\bar\psi,A]}\rightarrow
\sum_{N_+,N_-}
 e^{(\mu+i\theta)N_+}
 e^{(\mu-i\theta)N_-}
 Z_{N_\pm}.
\end{equation}

This ILM ensemble formulation, among the other proposed mechanisms for describing IR QCD, successfully captures the dominant IR dynamics of the QCD vacuum via pseudoparticles (instantons and anti-instantons), which generate chirality-flipping fermionic zero modes and induce the ’t Hooft multi-fermion interaction, thereby providing a dynamical mechanism for spontaneous chiral symmetry breaking, the $U_A(1)$ anomaly, and the $\pi$-$\eta'$ mass splitting in the pseudoscalar meson sector. Lattice simulations, particularly when combined with gradient flow, provide strong support for this picture by revealing a dilute but strongly correlated ensemble of instantons and anti-instantons in smeared gauge configurations.

The resulting ensemble measure for the total number of pseudoparticles, $N$, retains the same structure, with the running coupling modified to include the effect of $N_f$. This follows from the fact that the underlying argument is governed by the trace anomaly, which is dominated by gauge-field dynamics, and by the RG running of the coupling, which is fixed by the renormalizability of QCD. In contrast, the ensemble measure for the number difference, $\Delta$, is significantly modified in the presence of dynamical quarks. This is due to topological screening induced by the strong interaction between light quarks and pseudoparticles, which sharpens the distribution of topological charge.

In (unquenched) QCD, $\chi_t$ is substantially screened by the light quarks~\cite{Diakonov:1995qy,Kacir:1996qn},
\begin{equation}
\label{SUS}
\frac{\chi_t}{\bar N}\sim 
\bigg(\frac{11}{12}N_c-\frac{n_{I+A}}{\langle\bar qq\rangle}\sum_f\frac{1}{m_f}\bigg)^{-1}
\end{equation}
This result agrees with the chiral perturbation calculation in chiral limit \cite{Leutwyler:1992yt}. As result of topological screening, the Witten-Veneziano formula \eqref{CHI} is also modified by replacing the singlet meson mass with pion mass $m_\pi^2$. 

In Table~\ref{tab:chi}, we present the susceptibility in both quenched and unquenched $N_f=2+1$ QCD vacuum obtained by Witten-Veneziano formula in \eqref{CHI} with $f_\pi=93$ MeV, $m_\pi=139$ MeV,$m_K=494$ MeV, $m_{\eta}=549$, and $m_{\eta'}=958$ MeV from Particle Data Group (PDG) \cite{ParticleDataGroup:2024cfk} and ILM prediction in two different scenario \eqref{eq:chi_sum} and \eqref{ILM_chi} using $n_{I+A}=1$ fm$^{-4}$, $m_{u}\langle\bar{u}u\rangle=m_{d}\langle\bar{d}d\rangle=-6.8\times10^{-5}$ \cite{FlavourLatticeAveragingGroupFLAG:2021npn} and $m_{s}\langle\bar{s}s\rangle=-1.7\times10^{-3}$ \cite{Harnett:2021zug}.  
\begin{table}[]
    \centering
    \begin{tabular}{|c|c|c|c|}
        \hline
        $\chi_t/V$ & WV \eqref{CHI} & sum ansatz \eqref{eq:chi_sum} & exact duality \eqref{ILM_chi} \\
        \hline
       quenched &$(180.2\,\mathrm{MeV})^4$ & $(197.3\,\mathrm{MeV})^4$ & $(153.2\,\mathrm{MeV})^4$ \\
       
      $N_f=2+1$ & $(72.6\,\mathrm{MeV})^4$ & $(75.6\,\mathrm{MeV})^4$ & $(75.1\,\mathrm{MeV})^4$ \\
        \hline
    \end{tabular}
    \caption{Topological susceptibility in YM and unquenched QCD vacuum where WV denotes the Witten-Veneziano relation}
    \label{tab:chi}
\end{table}

\begin{figure}
    \centering
\subfloat[\label{fig:topo-1}]{\includegraphics[width=0.64\linewidth]{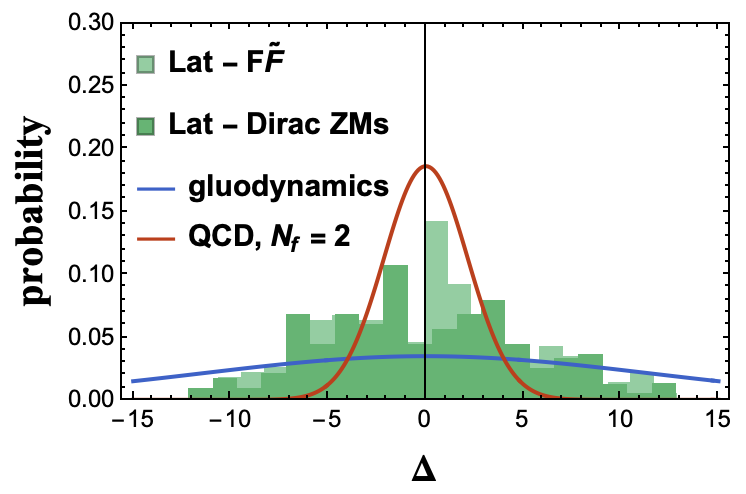}}
\hfill
\subfloat[\label{fig:topo-2}]{\includegraphics[width=0.64\linewidth]{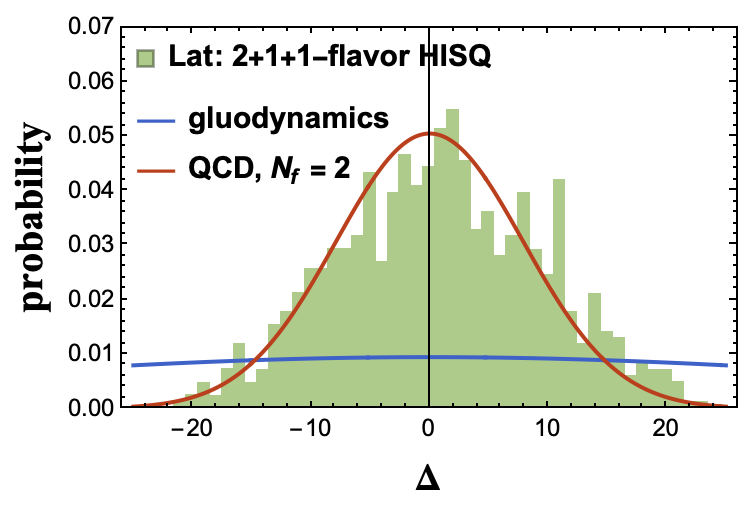}}
\hfill
\subfloat[\label{fig:topo-3}]{\includegraphics[width=0.64\linewidth]{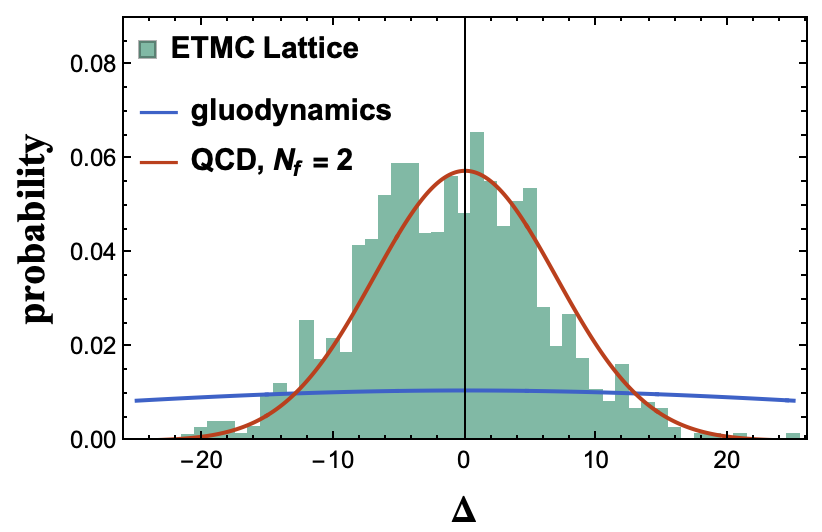}}
    \caption{ILM predicted distribution of topological charges in QCD compared to different lattice calculations}
    \label{fig:topo}
\end{figure}

In Fig.~\ref{fig:topo}, we compare our ILM prediction with the lattice calculation performed by three different lattice groups. The blue curve denotes the quenched calculation in ILM ensemble with $\chi_t/V=(197~\mathrm{MeV})^4$ and the red curve denotes the 2-flavor QCD ILM ensemble where the susceptibility is manually set to be $\chi_t/V=(85~\mathrm{MeV})^4$. Various lattice studies~\cite{Liang:2023jfj} indicate that the topological charge distribution follows a Gaussian form, in agreement with the ILM predictions.

In Fig.~\ref{fig:topo-1}, the ILM result is compared with $\chi$QCD lattice calculation using overlap fermions with the lattice size: $24^3\times64 ~a^4$ where $a=0.1105$ fm \cite{Liang:2023jfj}. The distribution presented by the light green is obtained by propagating the gluonic operator $F\tilde{F}$ with large lattice flow time $t =4a^2$ and the distribution presented by the dark green calculates the topological charges from counting the Dirac zero modes on the lattice. In Fig.~\ref{fig:topo-2}, the ILM result is compared with lattice calculation using HISQ ensemble with physical pion mass $m_\pi=135$ MeV and lattice volume: $96^3\times192 ~a^4$ where $a=0.0570$ fm \cite{Bhattacharya:2021lol}. In Fig.~\ref{fig:topo-3}, the ILM result is compared with ETMC lattice calculation using $N_f = 2+ 1+ 1$ twisted mass clover-improved fermions with the lattice size: $64^3 \times 128~a^4$ where $a=0.0801(4)$ fm and physical pion mass $m_\pi=139$ MeV \cite{Alexandrou:2020mds}.

The ILM approximates the QCD path integral by saturating it with $N_+$ instantons and $N_-$ anti-instantons, supplemented by small quantum fluctuations around those semi-classical configurations. The QCD action is expanded to quadratic order semi-classically, defining a Gaussian fluctuations that can be used in perturbation theory, while non-Gaussian fluctuations, so called zero modes, span the tangent space of a non-trivial manifold where the collective coordinates on this so-called moduli space can be interpreted as collective degrees of freedom that describes the low-energy dynamics of the underlying theory. 

\section{Canonical ensemble average}
Given a generic local QCD operator  ${\cal O}[\psi,\bar\psi,A]$ where the gluonic part is sourced by a multi-instanton vacuum~\cite{Diakonov:1995qy}, the vacuum expectation value of the ensuing operator ${\cal O}[\psi,\bar \psi,A]$ can be evaluated by

\begin{equation}
\begin{aligned}
\label{eq:avg}
\langle\mathcal{O}\rangle
    =&\frac1{Z_{N_\pm}}\int \prod_I  d\Omega_{I}e^{-S_{\rm ILM}}\mathcal{O}[S,A]
\end{aligned}
    \end{equation}
with the effective action defined by the sum of the gluonic and fermionic interaction 
\begin{equation}
\begin{aligned}
\label{eq:SILM}
    S_{\rm ILM}=&-\sum_I\ln \left[Vn(\rho_I)\rho_I^{N_f}\right]+S_{\rm int}-N_f\ln \mathrm{det}(m-iT)
\end{aligned}
\end{equation}

In Eq.~\ref{eq:avg}, all quark fields are contracted into propagators $S$ under the instanton vacuum background defined in Eq.~\eqref{eq:Q_prop0} and see~\cite{Diakonov:1985eg,Pobylitsa:1989uq,Kock:2020frx,Faccioli:2001ug} (reference therein) for more details. The first part of the effective action \eqref{eq:SILM} is governed by the instanton tunneling rate $n(\rho)$ defined in Eq.~\eqref{eq:dn2}. The second part describes the semi-classical gluonic interaction among instantons and anti-instantons and can be parameterized by \cite{Diakonov:1983hh}

\begin{equation}
    S_{\rm int}=\frac12\sum_{I\neq J}\beta_1(\sqrt{\rho_I\rho_J})u_{IJ}(R_{IJ},\rho_I,\rho_J,U_{IJ})
\end{equation}
where the gluonic interaction $u_{IJ}$ between the pair $I$ and $J$ with opposite duality is defined as \cite{Shuryak:1989cx}

\begin{equation}
\begin{aligned}
  u_{IJ}(R,\rho_I,\rho_J,U)
=&\left(
\frac{4}{\left(4 + \frac{R^2}{\rho_I\,\rho_J}\right)^2}
\left(
|u|^2
-
4
\frac{\bigl|u\cdot R\bigr|^2}
{R^2}\right)
+
\frac{16}{\left(3 + \frac{R^2}{\rho_I\,\rho_J}\right)^3}
|u|^2
\right)\\
&\times\frac{4\,\rho_I^{2}\rho_J^{2}}
{(\rho_I^{2}+\rho_J^{2})^{2}}
\end{aligned}
\end{equation}
and the gluonic interaction between the pair $I$ and $J$ with the same duality is 

\begin{equation}
    u_{IJ}(R,\rho_I,\rho_J,U)
=
\left(
\frac{1}{2}\,\mathrm{Tr}\!\left(U\mathds{1}_2 U^\dagger\mathds{1}_2\right)
-
\left|\frac{1}{2}\,\mathrm{Tr}\left(U\mathds{1}_2\right)\right|^2
\right)
\;
\frac{1.6}{\left(1+\frac{R^2}{\rho_I\rho_J}\right)^3}
\;
\frac{4\rho_I^2\rho_J^2}{\left(\rho_I^2+\rho_J^2\right)^2}
\end{equation}
with the color orientation vector defined as $u_\mu=\frac1{2i}\mathrm{tr}(U\tau_\mu^+)$. 
The last part in Eq.~\eqref{eq:SILM} is fermion induced interaction determined by the sea quark determinant with the overlap integral parameterized by \cite{Shuryak:1989cx,Shuryak:1987ja}
\begin{equation}
\begin{aligned}
T_{IJ}=&iu\cdot R\frac{4 \rho_I \rho_J}{(R^2 + 2 \rho_I\rho_J)^2}
\end{aligned}
\end{equation}

To proceed the evaluation in the QCD 
instanton vacuum, we substitute the dilute multi-instanton field with sum ansatz for simplicity.

\begin{equation}
\label{eq:gluon_field}
    A(x)=\sum_{I=1}^{N_++N_-} A_I(x)
\end{equation} 
For improved ansatz, see Sec.~\ref{sec:Ratio}.
In dilute instanton liquid ensemble, the resulting vacuum expectation value of the operator ${\cal O}$ simply splits into a sum over distinct multi-instanton background~\cite{Weiss:2021kpt}, 
With this in mind, we have
\begin{equation}
\begin{aligned}
\label{eq:gluo_op}    \langle\mathcal{O}\rangle=&\sum_{n,k}\frac{1}{n!}\binom{n}{k}\left(\frac{N_+}V\right)^k\left(\frac{N_-}V\right)^{n-k}\langle\mathcal{O}_{I_1\cdots I_n}\rangle
\end{aligned}
\end{equation}
where the effective operator in dilute $n-$instanton background $\mathcal{O}_{I_1\cdots I_n}$ is defined as

\begin{equation}
\begin{aligned}
\mathcal{O}_{I_1\cdots I_n}
    =&\mathcal{O}[S,A_{I_1},\cdots,A_{I_n}]
\end{aligned}
    \end{equation}
%



    
    
%

\subsection{ILM using Metropolis Algorithm}
\label{sec:numericalILM}
Numerically, the ensemble average in Eq.~\eqref{eq:avg} can be evaluated by Monte Carlo simulation using improved Metropolis algorithm in canonical setting~\cite{Shuryak:1988ff,Nowak:1988bh,Shuryak:1983ni} and grand canonical setting \cite{Wantz:2009qk,Wantz:2009kh,Wantz:2009mi}. With this algorithm, a set of random configuration samples $\Omega$ of the instanton ensemble is generated on a stationary Markov chain with probability given by the action of instanton liquid ensemble. In a single Metropolis step, a single instanton or anti-instanton is selected sequentially and its moduli-space variables $\Omega_I$ are proposed to change by adding an unbiased random displacements with fixed step sizes. One sweep consists of 3$N$ such steps where $N=N_++N_-$ is the total number of instantons and anti-instantons. At the end of each step, the ILM action is reevaluated to determine the Metropolis acceptance. The step size to each variable $z_I,U_I,\rho_I$ are chosen such that the acceptance rates are around 40-60\%.

In this work, we consider two types of instanton liquid ensembles: the Interacting Instanton Liquid Model (IILM) and the Random Instanton Liquid Model (RILM). These two ensembles represent a fully dynamical treatment of the instanton vacuum and a mean-field approximation, respectively.

In the RILM, the instanton ensemble is generated by fixing the density to $n_{I+A}=1~\mathrm{fm}^{-4}$ and the instanton size to its averaged value $\rho=1/3$ fm whereas the positions and color orientations of instantons are sampled independently and uniformly. 

On the other hand, the IILM incorporates the full semi-classical gluonic and fermionic interactions among pseudoparticles in the ensemble. The distribution of moduli parameters is determined dynamically through the IIL action Eq.~\eqref{eq:SILM}, leading to nontrivial correlations among pseudoparticles. 
For IILM simulations, the evaluation of the quark determinant causes the major computational cost with computational time scales as $\mathcal{O}(N^3)$ with the number of pseudoparticles $N$. However, each Monte Carlo update modifies only a single row or column of the overlap matrix. Therefore, we update the determinant using the Sherman–Morrison–Woodbury  formula~\cite{woodbury1950inverting,sherman1950adjustment}. Similar to Cholesky-based decomposition introduced in \cite{Wantz:2009qk}, our implementation updates determinant using a low-rank modification of its inverse matrix, reducing the computational complexity to $\mathcal{O}(N^2)$ per update. This approach preserves numerical stability while significantly improves large-volume simulations.

Practically, we impose periodic boundary conditions, thus instantons and anti-instantons effectively propagate on a torus. In principle, one should account for contributions from all topologically distinct paths connecting two points on the torus, rather than restricting to the shortest geodesic. However, numerical studies in \cite{Shuryak:1989cx} have shown that contributions from the those homotopy classes are negligible.

In contrast to RILM, IILM has intrinsically dependence on a low energy cut-off $\Lambda$ which signals the breakdown of the quantum mechanical fluctuations estimated by $n(\rho)$.  Similar to \cite{Schafer:1995pz,Faccioli:2001qg}, we
fix the cut-off $\Lambda$ by the instanton density $(N/L^4)\Lambda^4$ to $1$ fm$^{-4}$. 


One should keep in mind that, compared to the current-current correlation functions, more work is required to determine the partition function, which gives the overall normalization of the instanton distribution and determinantal mass. In practice, instead of direct sampling, the partition function is mostly evaluated using the thermodynamic integration method in \cite{kirkwood1935statistical,Schafer:1995pz,Schafer:1996wv} as the value is dominanted by rare events during the sampling. The result can be further improved using parallel tempering.

\subsection{RILM and effective quark theory}
\label{sec:ILMEFT}


In random instanton liquid model (RILM), the measure is reduced to 

\begin{equation}
\begin{aligned}
\label{eq:RILM}
    \prod_I \frac{d^4z_IdU_Id\rho_{I}}V e^{-S_{\rm ILM}}\rightarrow \prod_I  \frac{d^4z_{I}dU_{I}}V\prod_{f=1}^{N_f}\mathrm{det}(m-iT)\bigg|_{\mathrm{fixed}\,\rho}
\end{aligned}
\end{equation}
where we take a mean field on the gluonic interaction $S_{\rm int}$ among instantons and anti-instantons ($\delta S_{\rm int}$ in Eq.~\eqref{eq:Z_N3} has been dropped here and will be discussed later here) and approximate the resulting effective size distribution to a sharp peak around $\rho\sim0.33$ fm.

Within this mean-field framework, as we mentioned in Sec.~\ref{sec:ILM_QCD}, the zero-mode dominant quark determinant can be rewritten in terms of emergent ’t Hooft multi-fermion vertices, $\Theta_I$, using the path-integral~\cite{Liu:2025ldh}.

\begin{equation}
\begin{aligned}
\label{eq:det}
    &\prod_f\mathrm{det}(m-iT)=\int \mathcal{D}\psi \mathcal{D}\bar\psi \left(\prod_I\Theta_I\right) e^{-\int d^4x\bar{\psi}\slashed{\partial}\psi}
\end{aligned}
\end{equation}
where $\Theta_I$ is defined in Eq.~\eqref{Hooft}.

In the thermodynamic limit where the physics is insensitive to variations in $N_\pm$ for a fixed instanton density, the distance of quarks hopping through instantons and anti-instantons becomes long range, generating an effectively fully connected network that drives dynamical chiral symmetry breaking, while only small residual pairs survive as subleading correlations.


With this in mind, the ensemble average in Eq.~\eqref{eq:avg} can be rewritten as a path integral in an effective quark field theory~\cite{Liu:2024rdm}. For simplicity, we assume the size distribution is sharp around $\rho\sim0.33$ fm~\cite{Wantz:2009qk}. The averages under each $n$-cluster multi-instanton background Eq.~\eqref{eq:avg} read 

\begin{equation}
\begin{aligned}
\label{eq:op_eff}
    \langle\mathcal{O}_{I_1\cdots I_n}\rangle
    \approx&\int \mathcal{D}\psi \mathcal{D}\bar\psi e^{-\int d^4x\mathcal{L}_{\rm eff}}\left(\frac{1}{m^*}\right)^{{n\,N_f}}\\
    &\times\underbrace{\int d^4z_{I_1}dU_{I_1}\cdots d^4z_{I_n}dU_{I_n}\mathcal{O}[\psi,\bar\psi,A_{I_1},\cdots,A_{I_n}]\Theta_{I_1}\cdots\Theta_{I_n}}_{ k~\mathrm{instantons}\,(I),~n-k~\mathrm{anti-instantons}\,(A)}
\end{aligned}
\end{equation}
This framework coincides with the discussion in \cite{Diakonov:1995qy,Balla:1998rt} where only leading instanton contributions were included. The residual corrections from the inverse 't Hooft vertices $\Theta_{I}$ are given by $$\frac1{(m^*)^{N_f}}\rightarrow\frac1{(m^*)^{N_f}}\left(1-\frac{\theta_\pm-\langle\theta_\pm\rangle}{\langle\theta_\pm\rangle}\right)$$ where  $\theta_\pm=\int dU \Theta_{I,A}$ and $\langle\theta_\pm\rangle\simeq(m^*)^{N_f}$ (see Sec.~\ref{sec:det} and Appendix~\ref{app:mole}). These corrections are relevant only when computing connected averages of the effective quark operators which have a non-zero vacuum expectation value, such as the operator \(\bar\psi\psi\) and \(\bar\psi\slashed{A}\psi\). 

\begin{figure}
    \centering
    \subfloat[]{\includegraphics[width=0.35\linewidth]{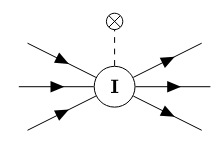}}
    \hfill
    \subfloat[]{\includegraphics[width=0.4\linewidth]{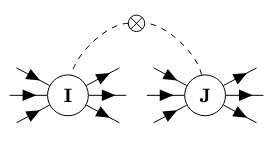}}
    \caption{Density expansion for the effective quark operators in dilute ILM ensemble}
    \label{fig:DILM_effOp}
\end{figure}
As discussed in \cite{Liu:2024rdm}, the calculations to obtain these effective multi-quark operators \eqref{eq:op_eff} in dilute multi-instanton background are illustrated by the diagrammatic contribution in~Fig.\ref{fig:DILM_effOp}, where each of the dash lines denotes the insertion of one instanton classical fields to the gluon field inside the operator. The details are presented in Appendix~\ref{App:pair}. Generally, the operators induced by the instanton clusters are non-local and usually penalized by $1/N_c$. For the quark contribution, each of the quark line passing through the pseudoparticle contributes a pair of $UU^\dagger$ in the color group integral, giving a $1/N_c$ factor, while each internal quark loop yields additional $N_c$. Hence, the $1/N_c$ counting of each operator is $1/N_c^{n_q-n_l}$ where $n_q$ is the number of external quark flavor and $n_l$ is the number of quark loops appearing in the diagrams at each order of the instanton density expansion. Similar counting can be applied to gluon contribution. Each semiclassical background field insertion represented by the dash lines yield a factor of $1/N_c$ while for gauge invariant gluonic operator, the overall color trace generates additional $N_c$.

Because instanton fields are highly localized, higher-order contributions in the instanton density to nonlocal operators are mainly generated by configurations in which several pseudoparticles form close clusters. In the dilute instanton framework, however, such configurations are strongly suppressed by the semiclassical interaction weight $e^{-S_{\rm int}}$, which favors well-separated instantons and anti-instantons. This suppression is consistent with the deeply cooled regime, where the surviving topological pseudoparticles appear as isolated semiclassical configurations. Therefore, within the dilute ensemble, higher-order density corrections are expected to be numerically small compared with the leading single-instanton contribution. See \cite{Diakonov:1995qy,Balla:1998rt,Balla:1997hf} for more details.

\begin{figure}
    \centering
\subfloat[\label{mol_1}]{\includegraphics[width=0.24\linewidth]{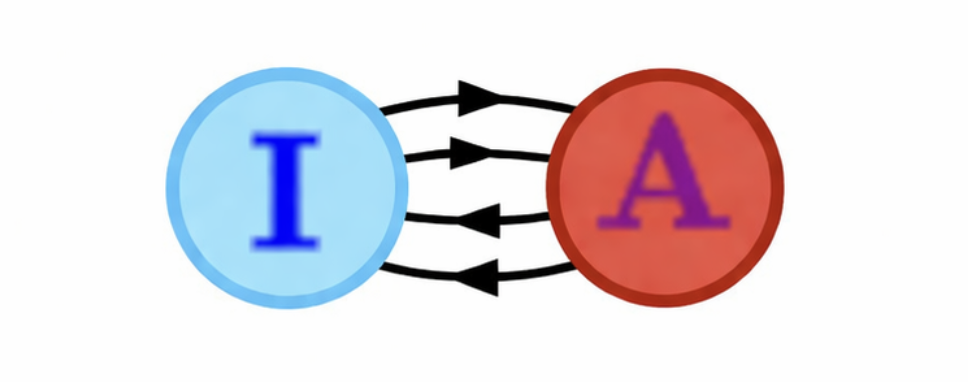}}
\hfill
\subfloat[\label{mol_2}]{\includegraphics[width=0.24\linewidth]{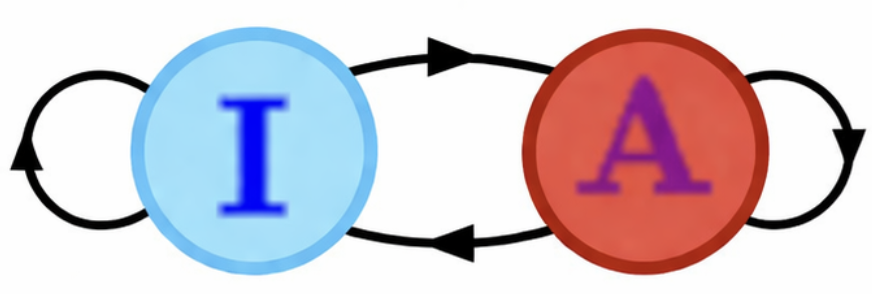}}
\hfill
\subfloat[\label{mol_3}]{\includegraphics[width=0.24\linewidth]{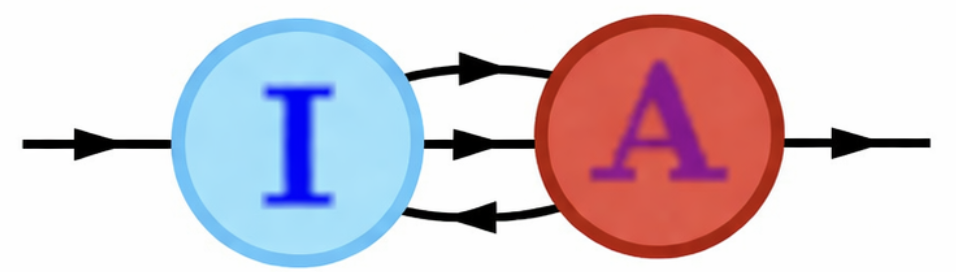}}
\hfill
\subfloat[\label{mol_4}]{\includegraphics[width=0.24\linewidth]{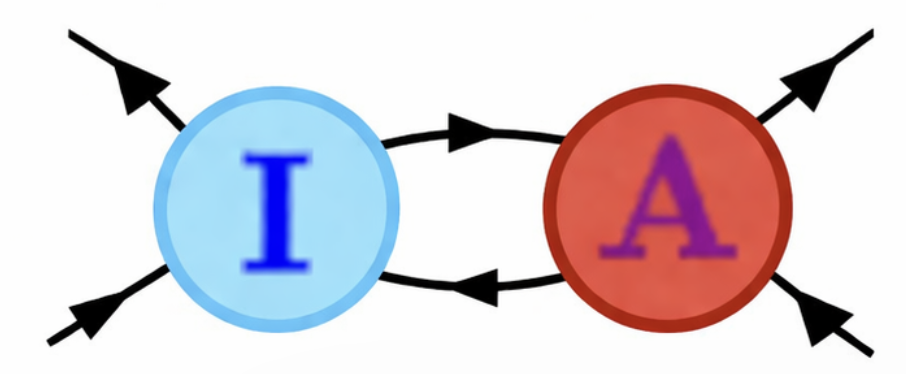}}
    \caption{Feynman diagrams for the vertices ($N_f=2$) induced by a close pair of instanton ($I$) and anti-instanton ($A$): (a) vacuum tunneling rate of a fully connected molecule where all flavors looped across the pair. (b) vacuum tunneling rate where one flavor looped across the pair. (c) one-body (two-quark) vertex induced by connecting three quark lines across $I$ and $A$, which is usually zero. (d) 
    two-body (four-quark) vertex induced by connecting two quark lines across $I$ and $A$.}
    \label{fig:mole}
\end{figure}

In addition, the ensuing effective ILM quark Lagrangian in \eqref{eq:op_eff} is defined by

\begin{equation}
\begin{aligned}
\label{eq:Leff}
    \mathcal{L}_{\rm eff}=&\bar{\psi}(\slashed{\partial}+m)\psi+\int d\rho n(\rho)\rho^{N_f}\int dU\left[\Theta_{I}+\Theta_{A}\right]\\
    &+\int d\rho_Id\rho_An(\rho_I)n(\rho_A)\rho_I^{N_f}\rho_A^{N_f}\int dud^4R\,\left[\Theta_I\Theta_A\right]_{\rm conn}+\cdots
\end{aligned}
\end{equation}
where $\psi^\dagger=i\bar\psi$ and $[\cdots]_{\rm conn}$ means the quark fields in these vertices are contracted using the free propagator, with the zero-mode contribution only associated with the instantons explicitly entering the bracket $[\cdots]_{\rm conn}$. Additional multi-instanton hoppings are not resummed into these contractions. These paired vertices are illustrated in Fig.~\ref{fig:mole}. Each flavor looping inside the pair yield a factor of $|T_{IA}|^2$. The explicit expressions of each term is discussed in Chapter~\ref{ch:low_QCD} and some useful identities for the pair calculations are presented in Appendix~\ref{App:pair}.

This separation of single instanton and molecular contribution in \eqref{eq:Leff} is only straightforwardly well-defined within a controlled power counting, such as the instanton density and $1/N_c$ resummation. Examples include the random phase approximation (RPA), or equivalently the mean-field approximation.

The explicit expressions of the 't Hooft Lagrangian is discussed in Chapter~\ref{ch:low_QCD}. In this formalism, the gluonic fields in the operator of interest are replaced by semiclassical background act as color sources that modify the color orientations of quark zero modes, thereby generating nontrivial Dirac structures. The ensemble average effectively maps QCD operators onto a set of effective quark operators with gluonic contributions systematically encoded through an expansion of instanton density $n_{I+A}$ tied to the multiple instanton and anti-instanton configurations ($I_1\cdots I_n$). Thus, Eq.~\eqref{eq:gluo_op} becomes

\begin{equation}
\begin{aligned} 
\label{eq:dilute}
\langle\mathcal{O}\rangle=&\left(\frac{n_{I+A}}2\right)\langle\mathcal{O}_{I}\rangle_{\rm eff}+\left(\frac{n_{I+A}}2\right)\langle\mathcal{O}_{A}\rangle_{\rm eff}\\
&+\frac12\left(\frac{n_{I+A}}2\right)^2\langle\mathcal{O}_{II}\rangle_{\rm eff}+\left(\frac{n_{I+A}}2\right)^2\langle\mathcal{O}_{IA}\rangle_{\rm eff}+\frac12\left(\frac{n_{I+A}}2\right)^2\langle\mathcal{O}_{AA}\rangle_{\rm eff}+\cdots
\end{aligned}
\end{equation}
where $\langle\cdots\rangle_{\rm eff}$ refers to the average evaluated using effective Lagrangian. See Appendix~\ref{App:pair} for more details. 

\subsection{Dense instanton liquid model}
\label{sec:DILM}

\begin{figure}[H]
    \centering
    \includegraphics[width=0.99\linewidth]{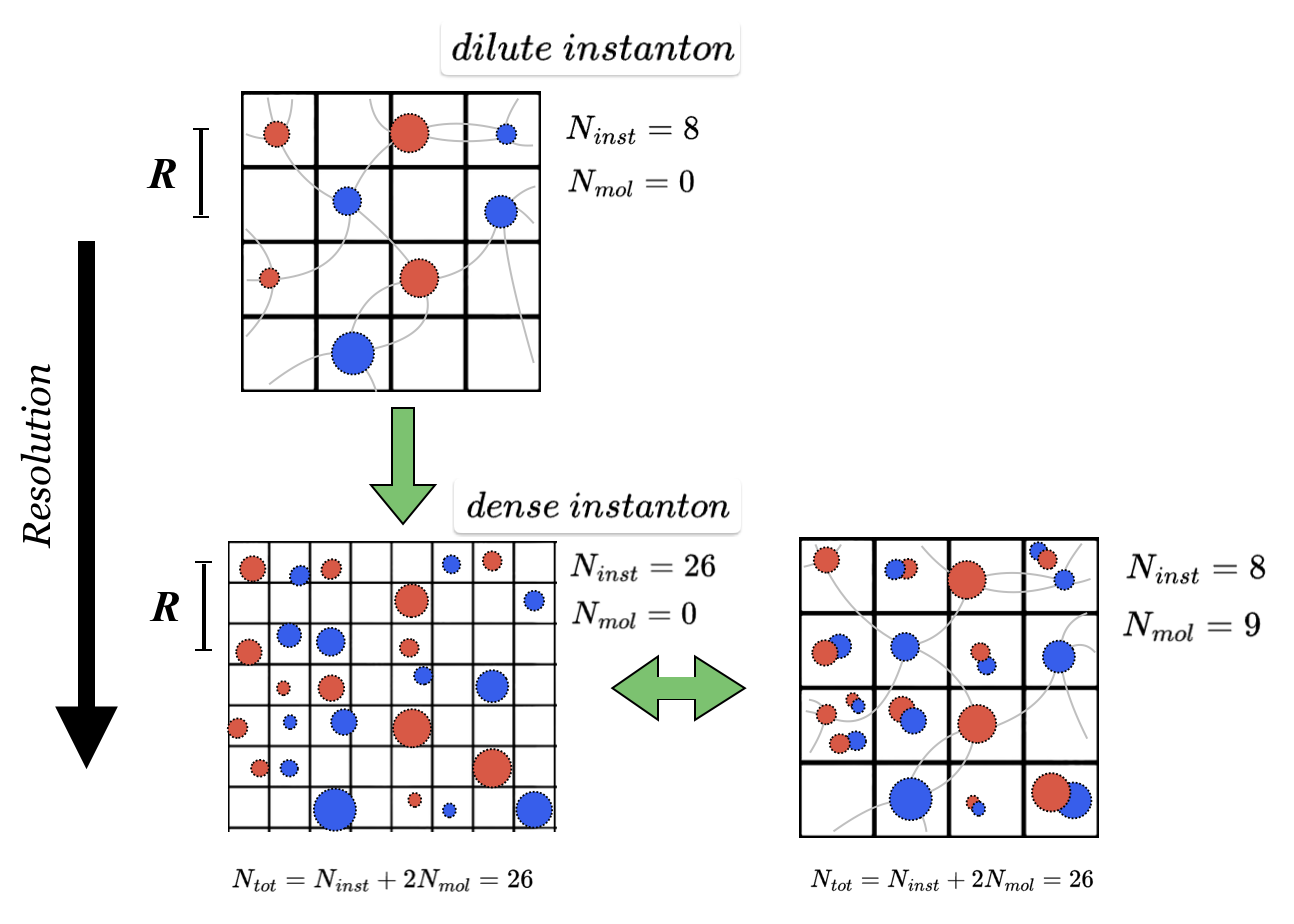}
    \caption{Red and blue circles represents the instantons and anti-instantons with gray curves represents the long range hopping among them. As the resolution is increased, short-distance correlated $(IA)$ pairs absent from the dilute ensemble are resolved as additional configurations. Their inclusion increases the total topological density of the ensemble. The emerging molecules can be cleanly separated once the resolution cut-off is introduced.}
    \label{fig:DILM}
\end{figure}

The preceding discussion in this chapter focused on the deeply cooled, dilute instanton ensemble. In this regime, the UV fluctuations have been depleted by cooling or gradient flow, and the remaining semiclassical configurations describe the long-distance topological structure of the QCD vacuum.

As discussed in Sec.~\ref{sec:vac_GF}, the QCD vacuum depends on the resolution scale. In the deeply cooled regime, UV degrees of freedom are depleted and the instanton ensemble stabilizes at a instanton density,
\begin{equation}
    n_{\rm inst}\simeq 1~{\rm fm}^{-4},
\end{equation}
which we refer to as the \emph{dilute instanton density}. As illustrated in Fig.~\ref{fig:DILM}, this dilute ensemble is characterized by a coarse resolution scale $R$, for which only isolated instantons and anti-instantons are resolved. Short-distance correlated instanton--anti-instanton pairs are not resolved as independent objects and therefore do not contribute separately to the instanton count.

When the resolution is increased, smaller-distance structures begin to emerge. In particular, instanton--anti-instanton correlated pairs, or molecules, that were previously hidden within the coarse-grained vacuum can now be resolved. This leads to a dramatic increase in the total density of topological pseudoparticles, as similarly presented in Fig.~\ref{cool2}. In this dense instanton liquid ensemble, the vacuum therefore is described by both isolated instantons and molecular pairs. Introducing a resolution cutoff $\sim R$ allows one to separate the long-distance dilute instantons from the newly resolved short-distance molecular correlations. In this sense, the dense ensemble provides an intermediate description that interpolates between the full quantum vacuum and the deeply cooled dilute instanton ensemble.

With the resolution cut-off $\bar R$, 
\begin{equation}
\langle\mathcal{O}\rangle\rightarrow\bigg[\langle\mathcal{O}\rangle_{\rm dilute}\equiv\langle\mathcal{O}\rangle-\langle\mathcal{O}\rangle|_{|z_i-z_j|\ll \bar{R}}\bigg]+n_{\rm mol}\langle\mathcal{O}\rangle_{\rm mol}
\end{equation}
where $\langle\mathcal{O}\rangle_{\rm dilute}$ is 

\begin{equation}
   \langle\mathcal{O}\rangle_{\rm dilute}=\left[\sum_{n}\frac{1}{n!}\binom{n}{k}\left(\frac{N_+}V\right)^k\left(\frac{N_-}V\right)^{n-k}\langle\mathcal{O}_{I_1\cdots I_n}\rangle\right]_{|z_i-z_j|\gg \bar{R}},
\end{equation}
as defined in \eqref{eq:gluo_op} and \eqref{eq:dilute} and the emerging molecule contribution $\langle\mathcal{O}\rangle_{\rm mol}$ is defined by

\begin{equation}
\begin{aligned}
\label{eq:op_eff_mole}
    \langle\mathcal{O}\rangle_{\rm mol}
    \approx&\int \mathcal{D}\psi \mathcal{D}\bar\psi e^{-\int d^4x\mathcal{L}_{\rm eff}}\left(\frac{1}{m^*}\right)^{{2\,N_f}}\\
    &\times\frac{1}{\bar{R}^4}\int d^4z_{I}dU_{I}d^4z_{A}dU_{A}\mathcal{O}[\psi,\bar\psi,A_{I},A_{A}][\Theta_{I}\Theta_{A}]_{\rm conn}\bigg|_{|z_I-z_A|\ll \bar R\sim\frac1{\sqrt[4]{n_{\rm inst}}}}
\end{aligned}
\end{equation}

Thus, in this dense formalism, Eq.~\eqref{eq:gluo_op} becomes

\begin{equation}
\begin{aligned}
\label{vac_OPE0}
    \mathcal{O}_{\mathrm{QCD}}[\psi,\bar \psi,A]\approx&\, \frac{n_{\rm inst}}{2}\sum_{n=1}^{N_f}\left(\frac{4\pi^2\rho^2}{m^*}\right)^n\bigg(\beta^{(I)}\mathcal{O}^{(I)}_n[\psi,\bar \psi]+\beta^{(A)}\mathcal{O}^{(A)}_n[\psi,\bar \psi]\bigg)\\
&+n_{\rm mol}\sum_{n=1}^{N_f}\bigg(\gamma^{(n)}_{IA}\beta^{(IA)}(\rho q)\mathcal{O}_n^{(IA)}[\psi,\bar \psi]\bigg)\bigg|_{R\ll\bar{R}}+\cdots
\end{aligned}
\end{equation}
with a clear RG implication on the emerging molecular contribution. The dense ILM vacuum OPE in \eqref{vac_OPE0} is controlled by the vacuum densities $\{n_{\rm inst}=n_{I+A},~n_{\rm mol}=n_{IA}\}$, the determinantal mass $m^*$ and vacuum hopping per density $\gamma_{IA}$ where $n_{I+A}$ is the single instanton density (dilute density) and $n_{IA}$ is the pair density in \eqref{pair_densityIA}. $\beta(\rho q)$ is the pseudoparticle profiles of each configurations (single-instantons and molecules) probed by the external momentum transfer $q$, which is zero for vacuum expectation but can be non-zero when hadron sources present (form factors). 
Each matrix element in Eqs.~\eqref{eq:gluo_op} and \eqref{vac_OPE0} can be either evaluated by statistical ensemble numerically (see Sec.~\ref{sec:numericalILM}) or analytically by the effective Lagrangian \eqref{eq:Leff} (see also \cite{Liu:2024rdm,Liu:2024jno,Liu:2024yqa}).

\subsection{Vacuum resummation}

The vacuum tunneling amplitudes are defined by the diagrams appearing in $\langle\mathcal{L}_{\rm eff}\rangle$, which are related to the determinantal mass. For the single-instanton contribution, $\mathcal{L}_I$, the average simply produces the overall density $n_{I+A}$. For the molecular contribution, $\mathcal{L}_{IA}$, the vacuum expectation value of the corresponding pair vertices is given by

\begin{equation}
\begin{aligned}
\label{pair_densityIA}
    \langle\mathcal{L}_{IA}\rangle=&\int d\rho_Id\rho_{A}n(\rho_I)n(\rho_{A})\rho_I^{N_f}\rho_{A}^{N_f}\int d^4Rdu\sum_{n=1}^{N_f}\binom{N_f}{n}\left(\frac{|T_{IA}(u,R)|}{m^*}\right)^{2n}\\    =&\left(\frac{n_{I+A}}2\right)^2\int d^4Rdu\sum_{n=1}^{N_f}\binom{N_f}{n}\left(\frac{|T_{IA}|}{m^*}\right)^{2n}
\end{aligned}
\end{equation}

\begin{figure}
    \centering
    \includegraphics[width=0.75\linewidth]{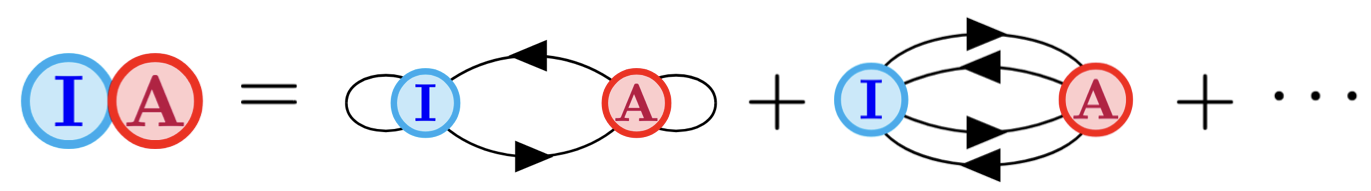}
    \caption{Vacuum resummation in the $IA$ pair. Those internal quark loops are often omitted in diagrams for simplicity}
    \label{fig:IA_sum}
\end{figure}

The first bracket is counting the pair density of pseudoparticles in the vacuum and the second is the light quark hopping between the pseudoparticles. The summation over all disconnected contractions in $[\Theta_I \Theta_A]_{\rm conn}$ are shown in Fig.~\ref{fig:IA_sum}. For simplicity, we will all omit the quark lines inside molecules.

\subsection{Determinantal mass}
\label{sec:det}
The concept of the determinantal mass $m^*$ first emerge as the ensemble average of the emergent vertices $\Theta_I$ \eqref{Hooft},

\begin{equation}
\label{eq:det_def}
    \left\langle 
    \prod_f\rho^{N}\mathrm{Det}(\slashed{D}+m)_{\mathrm{ZM}}\right\rangle\simeq\left\langle\rho^{NN_f} \prod_I\Theta_I\right\rangle=(\rho m^*)^{NN_f}
\end{equation}
which measures the geometrical mean of eigenvalues near the quasi-zero modes per flavor averaged over the moduli space of topological vacuum fluctuations~\cite{Liu:2025ldh,Faccioli:2001ug} (reference therein). At leading order of instanton density, $\langle\theta_\pm\rangle\approx(m^*)^{N_f}$. The definition can be further simplified in RILM

\begin{equation}
\label{eq:mdet}
    (m^*)^{N_f}=\left(\overline{\mathrm{det}(
        m-iT)]^{N_f}}\right)^{1/N}
\end{equation}
where the overline denotes the average over the moduli space in canonical RILM. An approximate estimation can also be made by the vacuum expectation value of 't Hooft vertex Eq.~\eqref{Hooft} (leading order of $1/N_c$ expansion). The resulting mass gap equation for each flavor is \cite{Liu:2024rdm,Schafer:1996wv}

\begin{equation}
\label{eq:det_gap_2}
    m^*\simeq m-\frac{2\pi^2\rho^2}{N_c}\langle\bar q\mathcal{F}(i\rho \partial)q\rangle
\end{equation}
with typically $\langle\bar q\mathcal{F}(i\rho \partial)q\rangle/\langle\bar qq\rangle= 0.27$ -- $0.3$. The additional suppression on quark condensate is induced by the instanton size. Using $n_{I+A}=1$ fm$^{-4}$, $\rho=0.313$ fm, quark current mass $m=4$ MeV, and $\langle \bar q\mathcal{F}(i\rho \partial)q\rangle=-0.286\times(276~\mathrm{MeV})^3$ where quark condensate is estimated using chiral Gell-Mann–Oakes–Renner (GOR) relation with $m_\pi=139$ MeV and $f_\pi=93$ MeV, Eq.~\eqref{eq:det_gap_2} gives $m^*=103.6$ MeV, which is consistent with the result $m^*\sim 103\,\rm  MeV$~\cite{Faccioli:2001ug}, following from numerical simulations of ensembles of interacting pseudoparticles.

The calculation of determinantal mass can also be systematically corrected using effective quark theory in \eqref{eq:Leff} by
\begin{equation}
\begin{aligned}
\label{eq:det_m_full}
    \left(m^*\right)^{NN_f}=&\langle \theta_+\rangle^{N_+} _{\rm eff}\langle \theta_- \rangle _{\rm eff}^{N_-}+N_+N_-\langle\theta_+\theta_-\rangle _{\rm eff}\langle \theta_+ \rangle _{\rm eff}^{N_+-1}\langle \theta_- \rangle _{\rm eff}^{N_--1}+\cdots
\end{aligned}
\end{equation}
where the first term gives the leading single-instanton contribution, while the connected instanton--anti-instanton molecule diagrams generate subleading corrections to the determinantal mass.

Furthermore, their values quantify the extent to which the presence of fermions suppresses the instanton density relative to the corresponding quenched ensemble. 

In the instanton liquid model (ILM), the vacuum transition amplitude is given by
\begin{equation}
    \frac{n_{I+A}}{2}
    =
    \int d\rho \, n(\rho)\,
    \prod_{f=1}^{N_f} \left(m_f^* \rho\right),
\end{equation}
where $n_{I+A}$ denotes the total instanton density and $n(\rho)$ is the instanton size distribution introduced in Eq.~\eqref{eq:dn2}. The determinantal mass $m_f^*$ encodes the effect of the fermion determinant and acts to suppress vacuum tunneling. Using the estimated determinantal mass $m^*=103.6$ MeV, the effective instanton-quark coupling in \eqref{eq:Leff} can be estimated by
\begin{equation}
    \frac{n_{I+A}}{2}
    \left(\frac{4\pi^2\rho^2}{m^*}\right)^{N_f}\approx696.7~\mathrm{GeV}^{-2}~(N_f=2)
\end{equation}

\subsection{Semi-classical soft gluons}
\label{sec:tail}
\begin{figure}
\centering
\subfloat[\label{fig:tHooft_3}]{\includegraphics[width=0.33\linewidth]{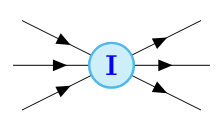}}
\hfill
\subfloat[\label{fig:tHooft_2}]{\includegraphics[width=0.33\linewidth]{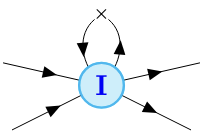}}
\hfill
\subfloat[\label{fig:tHooft_1}]{\includegraphics[width=0.33\linewidth]{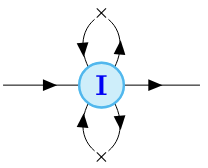}}
\caption{Feynman diagrams in the RILM effecitve quark theory. (a) $2N_f$-quark 't Hooft vertices induced by zero modes. (b) one-flavor reduction by mean field mass insertion $m^*$. (c) two-flavor reduction.}
\label{fig:ILM_Feyn}
\end{figure}

In RILM as discussed in Sec.~\ref{sec:ILM_QCD}, the quark degrees of freedom and their dynamics are mostly encoded in the interaction vertices $\Theta_{I,A}$~\eqref{Hooft} as presented in Fig.~\ref{fig:ILM_Feyn}. Due to Pauli exclusion, only different flavor can pass through the same instanton simultaneously. This results in a very strong flavor dependence in various dynamical processes. First, in Fig.~\ref{fig:tHooft_3}, the Feynman diagram denotes typical 't Hooft determinantal interaction involving three flavors.  Following in Fig.~\ref{fig:tHooft_2} and \ref{fig:tHooft_1}, the flavor reduction of the determinantal interaction can be achieved by looping up some of the flavors with the insertion of determinantal quark mass $m^*$. 

We now turn to the residual semiclassical gluonic interactions appearing in Eq.~\eqref{eq:Z_N3}. In this RILM framework, the soft gluonic degrees of freedom are described semiclassically. At large distances, this field is determined by its field strength $F_{\mu\nu}$ and propagates through the free gluon propagator, giving rise to a dipole-like tail.

\begin{equation}
\begin{aligned}
\label{LSZ_g}
    A_{I}(x-z_I)&=\int \frac{d^4k}{(2\pi)^4}e^{ik\cdot (x-z_I)} \left[-i\frac{4\pi^2\rho^2}g\bar\eta^a_{\mu\nu}\frac{k_\nu}{k^2}\mathcal{F}_g(\rho k)\right]\\
    &=\frac{2\pi^2\rho^2}{g}\bar{\eta}_{\mu\nu}^a\mathcal{F}_g(i\rho\partial_{z_I}) \langle F^a_{\mu\nu}(z_I) A(x)\rangle_{\mathrm{free}}\\
    &\xrightarrow{\rho\rightarrow0}\frac{2\pi^2\rho^2}{g}\bar{\eta}_{\mu\nu}^a\langle F^a_{\mu\nu}(z_I) A(x)\rangle_{\mathrm{free}}
\end{aligned}
\end{equation}

This allows us to include the soft gluon degrees of freedom as semiclassical gauge interactions in our effective RILM Lagrangian in Eq.~\eqref{eq:Leff}. The linearized interaction Eq.~\eqref{inst_dipole} is attached to the instanton vertex $\Theta_{I}(x)$~\eqref{Hooft} exponentially, resulting in \cite{Vainshtein:1981wh}

\begin{align}
\label{eq:tHooft_g}
\Theta_{I}(x)\rightarrow \Theta_{I}(x)e^{i\frac{2\pi^2\rho^2}{g}\mathrm{tr}_c\left[U_I\tau^\mp_\mu\tau^\pm_\nu U_I^\dagger F_{\mu\nu}\right]} 
\end{align}
The finite size effect of instanton color-magnetic moment can be straightforwardly recovered by 
\begin{align}
\frac{2\pi^2\rho^2}{g}\rightarrow \frac{2\pi^2\rho^2}{g}\mathcal{F}_g(\rho q)
\end{align}
where the finite-sized color-magnetic moment profile is defined as
\begin{equation}
\label{eq:g_form}
    \mathcal{F}_g(q)=\frac{4}{q^2}-2K_2(q)
\end{equation}

The color field strength $F_{\mu\nu}$ follows from the LSZ reduction of pseudoparticle field profile, and is coupled to the color-magnetic moment of individual instantons~\cite{Kochelev:1996pv,Qian:2015wyq,Diakonov:2002fq}. It can be interpreted as color field sourcing the tail of the instanton profiles.

\begin{figure}
\centering
\subfloat[\label{fig:tHooft_g}]{\includegraphics[width=.95\linewidth]{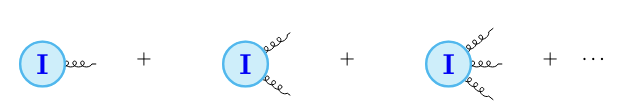}}
\hfill
\subfloat[\label{fig:tHooft_qg1}]{\includegraphics[width=.3\linewidth]{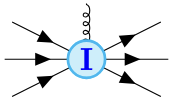}}
\hfill
\subfloat[\label{fig:tHooft_qg2}]{\includegraphics[width=.3\linewidth]{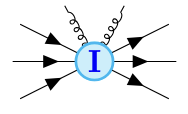}}
\caption{Feynman diagrams in the RILM effecitve Lagrangian. (a) semi-classical emission of gluon plane wave from instanton vacuum. (b), (c) quark 't Hooft vertices combined with semi-classical gluon emission }
\label{fig:ILM2}
\end{figure}


After averaging over the color orientation, the full effective Lagrangian $\mathcal{L}_{\rm eff}^{q+g}$ following from \eqref{eq:tHooft_g} yield additional semi-classical soft gluon vertices sourced by external color fields on top of the emergent multi-flavor vertices. The additional vertices are graphically presented in Fig.~\ref{fig:ILM2}. 

When the color sourcing field $F_{\mu\nu}$ contracts with the outgoing gluon states, it can also be viewed as gluon emission from the failed vacuum tunneling as illustrated in Fig.~\ref{fig:tHooft_g} (see Sec.~\ref{sec:mole}). Another important feature introduced by these vertices is the anomalous quark chromomagnetic moment \cite{Kochelev:1996pv} as illustrated in Fig.~\ref{fig:tHooft_qg1}, which has significant applications, including the study of gluon distributions in hadrons \cite{Kochelev:2015pqd}, the odderon \cite{Kochelev:2013csa}, the Pauli form factor \cite{Zhang:2017zpi}, and spin physics \cite{Cherednikov:2006zn,Kochelev:2008fy}. 

These interactions in $\mathcal{L}^{q+g}_{\rm eff}$ have similar book-keeping in $1/N_c$. For a single instanton with $n_q$ open quark flavors and $n_g$ gluons, the vertices in Eq.~\eqref{Hooft} and Eq.~\eqref{eq:tHooft_g} give rise to an effective 't Hooft interaction coupling. 
$$
G_I^{q+g}\sim\frac{n_{I+A}}{2}\frac{1}{N_c^{n_q+n_g}}\left(\frac{4\pi^2\rho^2}{m^*}\right)^{n_q}\left(\frac{2\pi^2\rho^2}{g}\right)^{n_g}
$$
The emergent couplings with constituent quarks and gluons are determined through color averaging, with each $UU^\dagger$ pair contributing a $1/N_c$ factor in the large $N_c$ limit.
Note that each gluon emission from the instanton is further suppressed by the instanton size. Now the effective interactions are given by \cite{Liu:2024rdm}.

\begin{equation}
\begin{aligned}
\label{LEFFSIA}
\mathcal{L}^{q+g}_{\mathrm{eff}}=&\bar{\psi}(i\slashed{\partial}-m)\psi+\frac{n_{I+A}}{2}\mathcal{V}_{n_q=0}+\frac{n_{I+A}}{2}\left(\frac{4\pi^2\rho^2}{m^*}\right)\mathcal{V}_{n_q=1}\\
&+\frac{n_{I+A}}{2}\left(\frac{4\pi^2\rho^2}{m^*}\right)^2\mathcal{V}_{n_q=2}+\mathcal{O}\left(\frac{n_{I+A}}{2}\left(\frac{4\pi^2\rho^2}{m^*}\right)^3\right)
\end{aligned}
\end{equation}
where the zero-body (gluodynamic) interaction (as presented in Fig.~\ref{fig:tHooft_g}) is defined as
\begin{equation}
\begin{aligned}
    \mathcal{V}_{n_q=0}=&\frac{1}{N^2_c-1}\left(\frac{2\pi^2\rho^2}{g}\right)^2(F^a_{\mu\nu})^2+\frac{4}{3N_c(N^2_c-1)}\left(\frac{2\pi^2\rho^2}{g}\right)^3f^{abc}F^a_{\mu\nu}F^b_{\mu\lambda}F^c_{\nu\lambda}\\
    &+\mathcal{O}\left(\left(\frac{2\pi^2\rho^2}{g}\right)^4\right)
\end{aligned}
\end{equation}
and the one-body interaction  (as presented in Figs.~\ref{fig:ILM_Feyn}, \ref{fig:tHooft_qg1} and \ref{fig:tHooft_qg2}) is defined as
\begin{equation}
\begin{aligned}
   \mathcal{V}_{n_q=1}=& -\frac{1}{N_c}\bar{\psi}\psi+\frac{1}{N^2_c-1}\left(\frac{2\pi^2\rho^2}{g}\right)\bar{\psi}\sigma_{\mu\nu}\frac{\lambda^a}2\psi F^a_{\mu\nu}\\
   &-\frac{1}{N_c(N^2_c-1)}\left(\frac{2\pi^2\rho^2}{g}\right)^2f^{abc}\bar{\psi}\sigma_{\mu\nu}\lambda^a\psi F^b_{\mu\rho}F^c_{\nu\rho}\\
&-\frac{1}{N_c(N^2_c-1)}\left(\frac{2\pi^2\rho^2}{g}\right)^2\left(\delta^{bc}\bar{\psi}\psi+\frac{N_c}{4(N_c+2)}d^{abc}\bar{\psi}\lambda^a\psi\right) F^b_{\mu\nu}F^c_{\mu\nu}\\
&-\frac{1}{N_c(N^2_c-1)}\left(\frac{2\pi^2\rho^2}{g}\right)^2\left(\delta^{bc}\bar{\psi}\gamma^5\psi+\frac{N_c}{4(N_c+2)}d^{abc}\bar{\psi}\lambda^a\gamma^5\psi\right)F^b_{\mu\nu}\tilde{F}^c_{\mu\nu}\\
&+\mathcal{O}\left(\left(\frac{2\pi^2\rho^2}{g}\right)^3\right)\\
\end{aligned}
\end{equation}
and the two-body interaction  (as presented in Figs.~\ref{fig:ILM_Feyn}, \ref{fig:tHooft_qg1} and \ref{fig:tHooft_qg2}) is defined as
\begin{equation}
\begin{aligned}
    \mathcal{V}_{n_q=2}=&\frac{2N_c-1}{16N_c(N^2_c-1)}\left[(\bar{\psi}\psi)^2-(\bar{\psi}\tau^a\psi)^2-(\bar{\psi}i\gamma^5\psi)^2+(\bar{\psi}i\gamma^5\tau^a\psi)^2\right]\\
    &+\frac{1}{32N_c(N^2_c-1)}\left[\left(\bar{\psi}\sigma_{\mu\nu}\psi\right)^2-\left(\bar{\psi}\sigma_{\mu\nu}\tau^a\psi\right)^2\right]\\
    &+\mathcal{O}\left(\left(\frac{2\pi^2\rho^2}{g}\right)\right)\\
\end{aligned}
\end{equation}

Here we present pure color sources, one-body interactions with up to three semi-classical gluons, and two-body interactions with one color sourcing field. Higher-order interactions follows similar scaling but increases in complexity. The $1/N_c$ counting of each vertex 
implies the more quarks and gluons involved in one instanton, the more $1/N_c$ suppression.

The use of the gluonic vertices in Eq.~\eqref{eq:tHooft_g} is justified in momentum space diagrams, when the exchanging semi-classical gluons carry energies below the sphaleron mass
(the top of the tunneling barrier)
\begin{equation}
M_S=\int d^3x \frac 18{F^2_{\mu\nu}(0,\vec x)}
=\frac{3\pi}{4\alpha_s\rho}
\end{equation}
The sphaleron mass is given by $M_S\sim 2.5$ GeV, for $\alpha_s(1/\rho)\sim 0.42$--$0.7$~\cite{Schafer:1996wv}.

\section{Grand canonical ensemble average }

In a realistic QCD vacuum, topological fluctuations must be taken into account. This can be achieved by generalizing the current ILM to a grand-canonical ensemble, in which both the total pseudoparticle number $N = N_+ + N_-$ and the number difference $\Delta = N_+ - N_-$ are allowed to fluctuate. Within this ensemble, the expectation value of a generic QCD operator can be computed in
\begin{equation}
\begin{aligned}
\label{grand_canonical}
    \langle \mathcal{O}_{\mathrm{QCD}}\rangle=&\sum_{N_+,N_-}\mathcal {P}(N_+,N_-)\langle \mathcal{O}_{\mathrm{QCD}}\rangle_{N_\pm}
\equiv\overline{\langle\mathcal{O}_{\mathrm{QCD}}\rangle}_{N_\pm} 
\end{aligned}
\end{equation}
The  averaging is carried  over the configurations with fixed $N_\pm$ (canonical ensemble average),  followed by an averaging over the distribution (\ref{DISTX}), where the corresponding $\sigma_t$ and $\chi_t$ are modified by the presence of light quark accordingly.

The simplest example is the calculation of $F^2$ and $F\tilde{F}$. In the canonical ensemble, the vacuum average of the gluonic scalar operator is directly proportional to the total number of instantons, $N$. 
\begin{equation}
    \frac1{32\pi^2}\langle F^2_{\mu\nu}\rangle_{N_\pm}= \frac NV
\end{equation}

After taking the fluctuations into account, the VEV of gluonic scalar operator, namely gluon condensate, corresponds to the instanton density $n_{I+A}$. 

The same calculation applies to the gluonic pseudoscalar operator. Its vacuum average in canonical ensemble is proportional to the total instanton number difference $\Delta$.

\begin{equation}
\frac1{32\pi^2}\langle F_{\mu\nu}\tilde{F}_{\mu\nu}\rangle_{N_\pm}= \frac \Delta V
\end{equation}

After taking the fluctuations into account, the VEV of gluonic pseudoscalar operator becomes zero, indicating that the QCD vacuum is topologically neutral.

\subsection{Hadronic matrix element in grand canonical ensemble}
\label{sec:had_grand}

The similar evaluation can be extended to the off-foward hadronic matrix elements in a numerical ensemble setting as well. Yet the calculation is more involved. First, the matrix element can be formally written as a large time reduction of a  3-point function.
\begin{equation}
    \frac{\langle h|\mathcal{O}_{\mathrm{QCD}}|h\rangle}{\langle h|h\rangle}=\lim_{t\rightarrow\infty}\frac{\langle J^\dagger_h(t/2)\mathcal{O}_{\mathrm{QCD}} J_h(-t/2)\rangle}{\langle J^\dagger_h(t/2)J_h(-t/2)\rangle}
\end{equation}
where $J_h(t)$ is a pertinent source for  the hadronic state $h$ defined by

\begin{equation}
    J_h(p,t)=\int d^3\vec{x} e^{-i\vec{p}\cdot \vec{x}}J_h(x)
\end{equation}


In this setting, the diagrams leading in $1/N_c$ counting are usually the diagrams disconnected to the hadron sources, resulting in no contribution to the hadron matrix element for fixed $N_\pm$.

\begin{equation}
\begin{aligned}
    &\langle J^\dagger_h(t/2)\mathcal{O}_{\rm QCD}J_h(-t/2)\rangle_{N_\pm}=\langle J^\dagger_h(t/2)J_h(-t/2)\rangle_{N_\pm}\langle\mathcal{O}_{\rm QCD}\rangle_{N_\pm}\left(1+\mathcal{O}(1/N_c)\right)
\end{aligned}
\end{equation}

However, these matrix elements are also subject to the vacuum fluctuation of the topological pseudoparticles when external hadron sources present, which is not included in canonical ensemble. As a result, the canonical ensemble formulation has to be generalized to the grand canonical ensemble with varying $N_\pm$. By extending the calculation to the grand canonical framework, the ensuing 3-point correlation function is carried out by



\begin{equation}
\begin{aligned}
&\langle J^\dagger_h(t/2)\mathcal{O}_{\mathrm{QCD}} J_h(-t/2)\rangle\\
=&\sum_{N_+,N_-}\mathcal{P}(N_+,N_-)\left[\langle\mathcal{O}_{\rm QCD}\rangle_{N_\pm}-\overline{\langle\mathcal{O}_{\rm QCD}\rangle}_{N_\pm}\right]\langle J_h^\dagger(t/2) J_h(t/2)\rangle_{N_\pm}
\end{aligned}    
\end{equation}

By expanding the fluctuation to the leading order and implementing the asymptotic Euclidean time $t$ limit,
$$\lim_{t\rightarrow\infty}\left\langle J^\dagger_h(t/2)J_h(-t/2)\right\rangle_{N_\pm}\rightarrow e^{-m_h(N_+,N_-)t}$$
the hadronic matrix element reads
%
\begin{equation}
\begin{aligned}
    \frac{\langle h|\mathcal{O}|h\rangle}{V}
    =&-\overline{\left[\langle\mathcal{O}_{\rm QCD}\rangle_{N_\pm}-\overline{\langle\mathcal{O}_{\rm QCD}\rangle}_{N_\pm}\right](N-\bar{N})}\left(\frac{\partial m^2_h}{\partial N}\right)\bigg|_{\substack{N=\bar{N}\\ \Delta =0}}\\
    &-\overline{\left[\langle\mathcal{O}_{\rm QCD}\rangle_{N_\pm}-\overline{\langle\mathcal{O}_{\rm QCD}\rangle_{N_\pm}}\right]\Delta}\left(\frac{\partial m^2_h}{\partial \Delta}\right)\bigg|_{\substack{N=\bar{N}\\ \Delta =0}}
\end{aligned}
\end{equation}
where the overline average is defined by
\begin{equation}
\begin{aligned}
\overline{X}\equiv\sum_{N_+,N_-}\mathcal {P}(N_+,N_-)X
\end{aligned}
\end{equation}
 
We can directly apply this calculation to the matrix element of gluonic scalar and pseudoscalar operators at the leading $1/N_c$. The scalar gluon matrix elemenet is tied to the topological compressibility,
\begin{equation}
\begin{aligned}
    \frac1{32\pi^2}\langle h|F^2_{\mu\nu}|h\rangle
    =-2m^2_h \sigma_t
    \frac{\partial\ln m_h}{\partial N}\bigg|_{\substack{N=\bar{N}\\ \Delta =0}}
\end{aligned}
\end{equation}
and the matrix element of pseudoscalar gluon at the leading $1/N_c$ is tied to the topological susceptibility. 
\begin{equation}
\begin{aligned}
    \frac1{32\pi^2}\langle h|F_{\mu\nu}\tilde{F}_{\mu\nu}|h\rangle
    =-2m^2_h \chi_t
    \frac{\partial\ln m_h}{\partial \Delta}\bigg|_{\substack{N=\bar{N}\\ \Delta =0}}
\end{aligned}
\end{equation}

This general framework can provide a robust framework for vast applications in calculations of various hadronic matrix element \cite{Liu:2024rdm,Liu:2024jno,Liu:2024sqj,Liu:2024vkj,Weiss:2021kpt,Diakonov:1995qy,Kim:2023pll}.

\chapter{Low energy QCD}
\label{ch:low_QCD}

\begin{figure}
    \centering
    \includegraphics[width=1\linewidth]{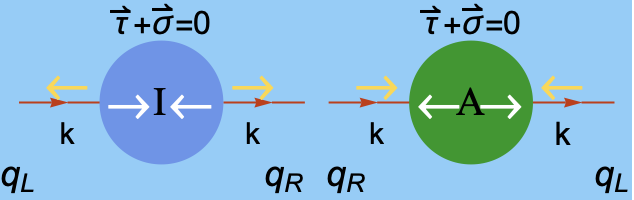}
    \caption{Light quarks flip their chirality when passing through the instanton (left) and anti-instanton (right)}
    \label{fig:inst-q}
\end{figure}

As mentioned earlier in Ch.~\ref{ch:ILM}, the ILM provides not only qualitative insight but also a quantitatively reasonable description of chiral physics. At low resolution, the QCD vacuum is predominantly populated by topologically active instantons and anti-instantons, which are Euclidean tunneling configurations between vacua with different winding numbers. As shown in Fig.~\ref{fig:inst-q}, light quarks interacting with these topological configurations develop zero modes with fixed handedness. For instance, a massless left-handed quark tunneling through an instanton can appear as a right-handed massless quark. The same scenario happens at an anti-instanton with the handedness of the quark flipped. Therefore, for temperatures $T < T_c$, the $SU(N_f)$ chiral symmetry is spontaneously broken, and the quark--antiquark pairs can be replaced by their condensate. This leads to effective two-quark and four-quark operators, both of which play a central role in hadronic physics. The two-quark operator generates a nonzero constituent quark mass, while the resulting four-fermion operators induce a variety of additional flavor dependent effects.

With this in mind, we highlight several of its key implications for hadronic spectroscopy. First, it establishes a quantitative relation between the quark condensate and the instanton density. Furthermore, the lightest chiral multiplet, consisting of the $\sigma$ and $\pi$ mesons, is dominated by so-called ``instanton bubble chain'' diagrams involving the ’t Hooft interaction as shown in Fig.~\ref{fig:bubble}. These diagrams can be resummed using the Bethe--Salpeter (BS) equation, allowing one to relate all chiral parameters to the key vacuum properties, such as $n_{I+A}$ and $\rho$.

Unlike superconductors or the Fermi surface in metals, the surface of the massless Dirac sea becomes gapped only if the attractive interaction exceeds a critical strength. This phenomenon, first observed by Nambu and Jona-Lasinio, also applies to the ’t Hooft interaction. 

Other mesons, such as vector mesons $\rho$, $\omega$, axial mesons $f_1$, $a_1$, tensor mesons $f_2$, $a_2$, and other higher-spin states, do not couple directly to instantons, as indicated by analyses of empirical correlation functions~\cite{Shuryak:1993kg}. Chiral channels of the $LL + RR$ type, however, are sensitive to interactions induced by instanton--anti-instanton ($IA$) molecules. These effects will be discussed in this chapter as well.

The simplest baryons in the flavor $SU(3)_f$ decuplet are $\Delta$, $\Sigma$, $\Xi$, and $\Omega$. These states can be viewed as three constituent quarks bound primarily by confinement. The corresponding nonrelativistic potentials, obtained by averaging three Wilson lines in the background field of $IA$ molecules, have been widely used \cite{Shuryak:2022thi}.

\section{'t Hooft Lagrangian}
\label{tHooft_L}

The important role of topological configurations (instantons and anti-instantons) to the quark sector is technically understood through the so-called ’t Hooft effective Lagrangian, derived from Eqs.~\eqref{Hooft} and \eqref{eq:Leff}. Since its explicit form is rather involved, we extend the discussion in this section.

The ’t Hooft effective Lagrangian incorporates all fermion flavors in a single nonlocal vertex. In QCD, this involves only the light quarks $(u,d,s)$, and the interaction takes the form of a six-fermion operator. The chiralities of quark and antiquark lines are always opposite, so any attempt to close these lines into loops vanishes for massless quarks. This interaction explicitly violates the $U_A(1)$ chiral symmetry, as it changes the chiral charge $Q_5$ of a state.

\subsection{Two-flavor case}

By explicitly carrying out the color average in the effective Lagrangian in Eq.~\eqref{eq:Leff} for $N_f=2$, the induced interactions for the light quarks from single instanton plus anti-instanton give \cite{Liu:2023fpj,Rapp:1999qa,Shuryak:2021yif}

\begin{equation}
\label{THOOFT1}
\begin{aligned}
    \mathcal{L}_I=\frac{G_I}{8(N^2_c-1)}&\Big\{\frac{2N_c-1}{2N_c}\left[(\bar{\psi}\psi)^2-(\bar{\psi}\tau^a\psi)^2-(\bar{\psi}i\gamma^5\psi)^2+(\bar{\psi}i\gamma^5\tau^a\psi)^2\right]\\
    &-\frac{1}{4N_c}\left[\left(\bar{\psi}\sigma_{\mu\nu}\psi\right)^2-\left(\bar{\psi}\sigma_{\mu\nu}\tau^a\psi\right)^2\right]\Big\}\\
\end{aligned}
\end{equation}
which are seen to mix $LR$ chiralities. The effective coupling is given by
\begin{equation}
\label{eq:GI}
    G_I=\int d\rho n(\rho)\rho^{N_f}(4\pi^2\rho^2)^{N_f}\simeq\frac{n_{I+A}}{2}
    \left(\frac{4\pi^2\rho^2}{m^*}\right)^{N_f}
\end{equation}
and fixed by the mean-instanton density and $m^*_f$ the induced determinantal mass~\cite{Vainshtein:1981wh,Liu:2025ldh}. Note that there is no vector or axial vector channel in Eq.~\eqref{THOOFT1}. To obtain vector bound states, it is necessary to go beyond the single instanton-induced interaction. In the interacting instanton vacuum, additional interactions involve instanton clusters, as shown in Fig.~\ref{fig:vac_q}. The simplest and relevant cluster configurations are molecules (see Sec.~\ref{sec:mole} and Refs.~\cite{Liu:2023fpj,Rapp:1999qa,Schafer:1994nv}). From Eq.~\eqref{eq:Leff}, the corresponding Lagrangian reads

\begin{equation}
\label{THOOFT2}
 \begin{aligned}
\mathcal{L}_{IA}=&G_{IA}\bigg\{\frac{1}{N_c(N_c-1)}\left[(\bar{\psi}\gamma^\mu\psi)^2+(\bar{\psi}\gamma^\mu\gamma^5\psi)^2\right]\\
&-\frac{N_c-2}{N_c(N_c^2-1)}\left[(\bar{\psi}\gamma^\mu\psi)^2- (\bar{\psi}\gamma^\mu\gamma^5\psi)^2\right]\\
    &+\frac{2N_c-1}{N_c(N_c^2-1)}\left[(\bar{\psi}\psi)^2+(\bar{\psi}\tau^a\psi)^2+(\bar{\psi}i\gamma^5\psi)^2+(\bar{\psi}i\gamma^5\tau^a\psi)^2\right]\\
     &-\frac{1}{2N_c(N_c-1)}\left[(\bar{\psi}\gamma^\mu\psi)^2+(\bar{\psi}\tau^a\gamma^\mu\psi)^2+(\bar{\psi}\gamma^\mu\gamma^5\psi)^2+(\bar{\psi}\tau^a\gamma^\mu\gamma^5\psi)^2\right]\bigg\}  
\end{aligned}
\end{equation}
which are $LL$ and $RR$ chirality preserving, in contrast to (\ref{THOOFT1}). The effective molecule-induced coupling is defined as
 \begin{equation}
 \begin{aligned}
 \label{MOLX}
     G_{IA}=&\int d\rho_I d\rho_{A}\int dud^4R ~ \frac1{8}(4\pi^2\rho^2_I)(4\pi^2\rho^2_{A})n(\rho_I)n(\rho_{A})\rho_I^{N_f}\rho_{A}^{N_f}|T_{IA}(u,R)|^{2N_f-2}
\end{aligned}
\end{equation}
 
Here $R=z_I-z_{A}$ is the relative molecular separation, $u=U^\dagger_IU_{A}$ is the relative
molecular color orientation, and $T_{IA}$, also defined in Eq.~\eqref{TIJ}, is the hopping quark matrix jumping from anti-instanton to instanton and its conjugate defines the reverse process.

\begin{equation}
\begin{aligned}
\label{TIA}
    T_{IA}(u,R)&=\int d^4x\phi_I^\dagger(x-z_I) i\slashed{\partial}\phi_A(x-z_A)=-4\pi^2\rho^2\mathrm{tr}(\tau_\mu^+u)\frac{R_\mu}{R}\frac{dT(R)}{dR}
\end{aligned}
\end{equation}
with the hopping integral $T(R)$ defined as
\begin{equation}
\begin{aligned}
\label{eq:hop}
    T(R)=&\frac{1}{2\pi^2R}\int_0^\infty dk\mathcal{F}(\rho k)J_1(k R)
\end{aligned}
\end{equation}
and $J_n$ Bessel functions of the first kind.

The pair-induced coupling in (\ref{MOLX}) is understood as the failed tunneling rate for a molecular configuration, whereby each of quark lines hopping inside the pair is removed by each division of $T_{IA}$, the overlap of the zero mode between instanton and anti-instanton. The strength of the induced molecular coupling $G_{IA}$ to the single coupling $G_I$ can be parameterized as

\begin{equation}
\label{eq:IA_corr}
G_{IA}=\frac18G_I^2\int dud^4R\left(\frac{T_{IA}(u,R)}{4\pi^2\rho^2}\right)^{2N_f-2}
\end{equation}
where the parameter $T_{IA}$ measures the correlation between the instanton and anti-instantons induced by quarks.



Low-lying meson dynamics at the low energy can be completely described by the instanton-induced interaction \cite{Liu:2023fpj,Shuryak:2021fsu}. To have physical mass spectrum of light mesons consistent with the experiments, the parameters in the 't Hooft Lagrangian has to be fixed by specific values. 
These values are subject to the values of instanton size $\rho$ and instanton density $n_{I+A}$. 

In Table~\ref{tab:parameters_ILM}, we present the fitted parameters in ILM using instanton size $\rho=0.33(2)$ fm and constituent mass $M=395(3)$ MeV with fixed pion mass $m_\pi=139.4$ MeV and rho meson mass $m_\rho=785(6)$ MeV. One sample of the data sets is presented in the first row to illustrate the numerical relations among the parameters. The last row labeled by asterisk $*$ shows the mean values and boostrap uncertainties to illustrate the sensitivity with respect to the uncertainty in instanton size $\rho$ and constituent mass $M$.

\begin{table*}
    \centering
\begin{tabular}{|c|c|c|c|c|c|c|}
   \hline
    & $n_{I+A}$ & $\rho$ & $M$ \\
    \hline
  ILM & $0.85~\mathrm{fm}^{-4}$ & $0.33$ fm & $395.2$ MeV \\
    \hline
  ILM* & $0.88(30)~\mathrm{fm}^{-4}$ & $0.33(2)$ fm & $395(3)$ MeV \\
   \hline
\end{tabular}
\hfill \\[10pt]
\begin{tabular}{|c|c|c|c|c|c|c|}
   \hline
    & $G_I$  & $G_{IA}$ & $\sqrt{\langle T_{IA}^2\rangle}$ \\
    \hline
  ILM & $658.7~\mathrm{GeV}^{-2}$ & $67.8~\mathrm{GeV}^{-2}$ & $3.9$ GeV$^{-1}$ \\
    \hline
  ILM* & $683\pm157~\mathrm{GeV}^{-2}$ & $67\pm11~\mathrm{GeV}^{-2}$ & $3.7(1.0)$ GeV$^{-1}$ \\
   \hline
\end{tabular}
    \caption{The fitted parameters in ILM using instanton size $\rho=0.33(2)$ fm and constituent mass $M=395(3)$ MeV with fixed pion mass $m_\pi=139.4$ MeV and rho meson mass $m_\rho=785(6)$ MeV.}
    \label{tab:parameters_ILM}
\end{table*}

\subsection{Three-flavor case}
The same color average in Eq.~\eqref{eq:Leff} for $N_f=3$ carries out the induced interactions for the light quarks $u,d,s$. The interaction reads

\begin{align}
\label{eq:Hoot_3f}
\mathcal{L}^{N_f=3}_{I}
&= -  
\frac{G_I}{N_c (N_c^2 - 1)}
\left(
\frac{2N_c + 1}{2N_c + 4}\mathrm{det}\bar{\psi}_R\psi_L
+ \frac{3}{8 (N_c + 1)}\mathrm{det}\bar{\psi}_R\sigma_{\mu\nu}\psi_L
\right)
+ (L \leftrightarrow R)
\end{align}
where the typical 't Hooft determinantal interaction in $N_f=3$ is defined as,

\begin{equation}
\mathrm{det}\bar{\psi}_R\psi_L=\frac16\epsilon_{ijk}\epsilon_{lmn}\bar{\psi}_{R i}\psi_{L l}\,
\bar{\psi}_{R j}\psi_{L m}\,
\bar{\psi}_{R k}\psi_{L n}=\left|\begin{array}{ccc}
\bar{u}_Ru_L & \bar{u}_Rd_L & \bar{u}_Rs_L \\
\bar{d}_Ru_L & \bar{d}_Rd_L & \bar{d}_Rs_L \\
\bar{s}_Ru_L & \bar{s}_Rd_L & \bar{s}_Rs_L \end{array}\right|
\end{equation}
and
\begin{equation}
\begin{aligned}
\mathrm{det}\bar{\psi}_R\sigma_{\mu\nu}\psi_L&=   \frac16\epsilon_{ijk}\epsilon_{lmn} \,
\bar{\psi}_{R i}\psi_{L l}\,
\bar{\psi}_{R j}\sigma_{\mu\nu}\psi_{L m}\,
\bar{\psi}_{R k}\sigma_{\mu\nu}\psi_{L n}\\
&=\bar u_{R}u_{L}\left|\begin{array}{cc}
\bar d_{R}\sigma_{\mu\nu}d_{L} & \bar d_{R}\sigma_{\mu\nu}s_{L} \\
\bar s_{R}\sigma_{\mu\nu}d_{L} & \bar s_{R}\sigma_{\mu\nu}s_{L}  
\end{array}\right|
-\bar u_{R}d_{L}\left|\begin{array}{cc}
\bar d_{R}\sigma_{\mu\nu}u_{L} & \bar d_{R}\sigma_{\mu\nu}s_{L} \\
\bar s_{R}\sigma_{\mu\nu}u_{L} & \bar s_{R}\sigma_{\mu\nu}s_{L}  
\end{array}\right|\\
&+\bar u_{R}s_{L}\left|\begin{array}{cc}
\bar d_{R}\sigma_{\mu\nu}u_{L} & \bar d_{R}\sigma_{\mu\nu}d_{L} \\
\bar s_{R}\sigma_{\mu\nu}u_{L} & \bar s_{R}\sigma_{\mu\nu}d_{L}  
\end{array}\right|
-\bar d_{R}u_{L}\left|\begin{array}{cc}
\bar u_{R}\sigma_{\mu\nu}d_{L} & \bar u_{R}\sigma_{\mu\nu}s_{L} \\
\bar s_{R}\sigma_{\mu\nu}d_{L} & \bar s_{R}\sigma_{\mu\nu}s_{L}  
\end{array}\right|\\
&+\bar d_{R}d_{L}\left|\begin{array}{cc}
\bar u_{R}\sigma_{\mu\nu}u_{L} & \bar u_{R}\sigma_{\mu\nu}s_{L} \\
\bar s_{R}\sigma_{\mu\nu}u_{L} & \bar s_{R}\sigma_{\mu\nu}s_{L}  
\end{array}\right|
-\bar d_{R}s_{L}\left|\begin{array}{cc}
\bar u_{R}\sigma_{\mu\nu}u_{L} & \bar u_{R}\sigma_{\mu\nu}d_{L} \\
\bar s_{R}\sigma_{\mu\nu}u_{L} & \bar s_{R}\sigma_{\mu\nu}d_{L}  
\end{array}\right|\\
&+\bar s_{R}u_{L}\left|\begin{array}{cc}
\bar u_{R}\sigma_{\mu\nu}d_{L} & \bar u_{R}\sigma_{\mu\nu}s_{L} \\
\bar d_{R}\sigma_{\mu\nu}d_{L} & \bar d_{R}\sigma_{\mu\nu}s_{L}  
\end{array}\right|
-\bar s_{R}d_{L}\left|\begin{array}{cc}
\bar u_{R}\sigma_{\mu\nu}u_{L} & \bar u_{R}\sigma_{\mu\nu}s_{L} \\
\bar d_{R}\sigma_{\mu\nu}u_{L} & \bar d_{R}\sigma_{\mu\nu}s_{L}  
\end{array}\right|\\
&+\bar s_{R}s_{L}\left|\begin{array}{cc}
\bar u_{R}\sigma_{\mu\nu}u_{L} & \bar u_{R}\sigma_{\mu\nu}d_{L} \\
\bar d_{R}\sigma_{\mu\nu}u_{L} & \bar d_{R}\sigma_{\mu\nu}d_{L}  
\end{array}\right|
\end{aligned}
\end{equation}

\section{Bosonization}
\label{sec:bos}
By averaging over the color orientation of instantons using $1/N_c$ as a book-keeping argument, the leading order of the single-instanton 't Hooft effective Lagrangian in Eq.~\eqref{eq:Leff} reads

\begin{equation}
\label{tHooft}
\mathcal{L}_{\mathrm{eff}}=\bar{\psi}\left(i\slashed{\partial}-m\right)\psi-\frac{G_I}{N_c^{N_f}}\left(\mathrm{det}\bar{\psi}_L\psi_R+\mathrm{det}\bar{\psi}_R\psi_L\right)
\end{equation}


This framework classifies QCD degrees of freedom into two categories: (i) those with masses \(\geq 1/\rho\) and (ii) those with masses \(\ll 1/\rho\). In low-energy strong interactions, where momenta are much smaller than \(1/\rho \simeq 600\) MeV, the heavy modes can be neglected, focusing only on the light degrees of freedom. The only relevant low-mass states are Goldstone pseudoscalar mesons and quarks, which acquire a dynamically generated mass \(M \simeq 300\)–\(400\) MeV \(\ll 1/\rho\). Consequently, for momenta \(k \ll 1/\rho\), QCD reduces to a simpler yet nontrivial theory of massive quarks interacting with nearly massless pions. With this in mind, at lower energy scale $\mu\sim300-400$ MeV, the Lagrangian in Eq.~\eqref{tHooft} can be approximately bosonized by introducing \(N_f \times N_f\) auxiliary fields, as detailed in \cite{Kacir:1996qn}.

\begin{equation}
\label{semi-bos}
    \mathcal{L}_{bos}=\bar{\psi}(i\slashed{\partial}-m)\psi+\frac{2\pi^2\rho^2}{N_c}\bar{\sigma}\bar\psi\left[\frac{1-\gamma^5}{2}U+\frac{1+\gamma^5}{2}U^\dagger\right]\psi+\frac12\mathrm{Tr}\left[m\bar{\sigma}(U+U^\dagger)\right]
\end{equation}
where $N_f\times N_f$ auxiliary bosonic field is defined as
\begin{equation}
U=\exp\left(i\pi^a\tau^a/F_\pi\right)
\end{equation}

The second term in Eq.~\eqref{semi-bos} represents the quark-meson effective interaction with Goldberger-Treiman (GT) relation manifested.

\begin{equation}
    g_{\pi qq}=\frac{2\pi^2\rho^2}{N_c}\frac{\bar\sigma}{F_\pi}=\frac{M}{F_\pi}
\end{equation}
where $\bar\sigma=-\langle\bar q q\rangle$ with the identification of the constituent mass obtained by Eq.~\eqref{eq:cons_m} and quark condensate approximated by Eq.~\eqref{qq}. The last term determines the mass of the (pseudo) Goldstone boson by GOR relation.

\begin{equation}
    m^2_\pi=\frac{2m\bar\sigma}{F_\pi^2}
\end{equation}

More specifically, the semi-bosonized Lagrangian in Eq.~\eqref{semi-bos} can be rewritten as \cite{Diakonov:1995ea}

\begin{equation}
\label{chiral}
    \mathcal{L}_{bos}=\bar{\psi}(i\slashed{\partial}-MU^{\gamma^5})\psi
\end{equation}
by using the identity

\begin{equation}
    \frac{1-\gamma^5}{2}U+\frac{1+\gamma^5}{2}U^\dagger=U^{\gamma^5}
\end{equation}
where the pseudoscalar Goldstone modes are manifested by
\begin{equation}
U^{\gamma^5}=\exp\left(i\pi^a\tau^a\gamma^5/F_\pi\right)
\end{equation}

By including the full RILM effective quark Langrangian with single instantons and molecules, the effective quark-meson chiral theory can be generalized in a semi-bosonization form \cite{Kacir:1996qn,Osipov:2003xu,Liu:2023yuj,Liu:2023fpj}. In the case of $N_f=2$, we have

\begin{equation}
\begin{aligned}
\label{eq:mes}
    \mathcal{L}_{bos}=&\bar{\psi}(i\slashed{\partial}-m)\psi+\bar{\psi}\bigg(\sigma+a^a_0\tau^a+i\gamma^5\eta'+i\gamma^5\tau^a\pi^a\\
    &-\gamma^\mu\omega_\mu-\gamma^\mu\tau^a\rho^a_\mu-\gamma^\mu\gamma^5 f_{1\mu}-\gamma^\mu\gamma^5\tau^aa^a_{1\mu}\bigg)\psi\\
    &-\frac{N_c}{2g_\sigma}\sigma^2-\frac{N_c}{2g_{a_0}}a_0^2-\frac{N_c}{2g_{\eta'}}(\eta')^2-\frac{N_c}{2g_\pi}\pi^2\\
    &+\frac{N_c}{2g_\omega}\omega_\mu^2+\frac{N_c}{2g_\rho}\rho_\mu^2+\frac{N_c}{2g_{f_1}}f_{1\mu}^2+\frac{N_c}{2g_{a_1}}a_{1\mu}^2
 \end{aligned}
\end{equation}
where the couplings in each channel are defined in Eq.~\eqref{tHooft_couplings}.

This formulation resembles the nonlinear sigma model and 
By choosing an nontrivial vacuum in chiral broken phase, the scalar field is shifted by an notrivial expectation value $\sigma\rightarrow \langle\sigma\rangle+\sigma$ which breaks the chiral symmetry spontaneously. In this phase, the quark constituent mass can determined by the gap equation.

\begin{equation}
\label{eq:Ch_gap}
M=m-\langle\sigma\rangle
\end{equation}

This equation is equivalent to the gap equation presented in Eq.~\eqref{eq:gap} with the identification $\langle\sigma\rangle=g_\sigma\langle\bar\psi\psi\rangle/N_c$ where $\langle\bar\psi\psi\rangle=\langle\bar uu\rangle+\langle\bar dd\rangle$ in two flavor case.

\section{Chiral Lagrangian}
\label{sec:chiral}
If one integrates off the quark fields in Eq.~\eqref{chiral}, one gets the effective chiral
Lagrangian. This idea also is also known as chiral-quark-soliton model \cite{Diakonov:1997sj}.
\begin{equation}
\begin{aligned}
\label{ChPT}
&S_{\mathrm{\chi PT}}=\frac{F_\pi^2}{4}\int d^4x\mathrm{Tr}\left(L_\mu L_\mu\right)\\
&-\frac{N_c}{192\pi^2}\int d^4x\left[2\mathrm{Tr}(\partial_\mu L_\mu)^2+\mathrm{Tr} (L_\mu L_\nu L_\mu L_\nu)\right]\\
&+\frac{N_c}{240\pi^2}\int d^5x\epsilon_{\mu\nu\rho\lambda\sigma}\mathrm{Tr}(L_\mu L_\nu L_\rho L_\lambda L_\sigma)
\end{aligned}
\end{equation}
where the chiral field is defined as
\begin{align}
    L_\mu=iU^\dagger\partial_\mu U
\end{align}

The first term here is the old Weinberg chiral lagrangian \cite{Weinberg:1991um} with
\begin{equation}
    F^2_\pi=4N_c\int \frac{d^4k}{(2\pi)^4}\frac{M^2(k)}{[k^2+M^2(k)]^2}
\end{equation}
which has also been observed in various formulation of instanton model \cite{Liu:2021evw}. The second term are the four-derivative Gasser–Leutwyler terms \cite{Gasser:1983yg,Scherer:2002tk} (with coefficients which turn out to agree with those following from the analysis of the data); the last term is the so-called Wess–Zumino term \cite{Lee:2020ojw}. Note that the $F_\pi$ constant originally diverges logarithmically at large momenta but is smoothly cut in ILM by the momentum dependent mass at $k \sim 1/\rho$ as a result of the finite instanton size.

Similarly, the bosonization procedure can be extended to Eq.~\eqref{eq:mes}. The saddle point expansion of the result gives the effective kinetic terms for the mesons (mesonic propagators) and the higher order expansion corresponds to the meson-meson interactions. The pseudoscalar mesons and axial vector mesons are mixed. The mixing angle between the pseudoscalar and axial vector channel can be determined by the diagonalization. 

\section{Schwinger--Dyson and Bethe--Salpeter--Faddeev equations}

In this section, we briefly summarize the connection between the Green’s function, the transfer matrix, and the Schwinger-Dyson (SD) equation, in close analogy with the relation between the covariant bound-state wave function and the Bethe--Salpeter--Faddeev (BSF) equation. For brevity, the equations are presented in quantum operator formalism in Hilbert space.

The full Green’s function $G$ is related to the free propagator $G_0$ and the transfer matrix $T$ through
\begin{equation}
    G = G_0 + G_0 \, T \, G_0 \, .
\end{equation}
The corresponding SD equation for the scattering matrix is
\begin{equation}
    T = K + K \, G_0 \, T \, ,
\end{equation}
where $K$ denotes the irreducible interaction kernel. In ILM, this kernel is specified by the effective multi-quark interaction induced by the instantons and anti-instantons in the vacuum.

When the interaction is sufficiently attractive to generate a bound state, the transfer matrix develops a pole at the mass of a hadron bound-state $X$,
\begin{equation}
    T \;\xrightarrow[P^2\to m_X^2]{}\; \frac{-i\,\Gamma\,\bar\Gamma}{P^2 - m_X^2} \, ,
\end{equation}
where $\bar\Gamma=\gamma^0\Gamma^\dagger\gamma^0$ with total momentum $P$, the sum of all incoming external momenta. Here $\Gamma$ is the amputated BSF vertex function, while the corresponding covariant BSF wave function is obtained by attaching the constituent free propagators,
\begin{equation}
    \Psi = G_0 \, \Gamma \, ,
\end{equation}

Substituting the pole structure into the SD equation yields the homogeneous bound-state equation
\begin{equation}
    \Gamma = K \, G_0 \, \Gamma \, ,
\end{equation}
which is the BSF equation for the hadron vertex. Solving the equation simultaneously produces a gap-like equation to determine the hadron mass and the form of the vertex function in terms of Dirac matrices.

For an $n$-body Green's function where $n>2$, the dynamics can be reduced to a hierarchy of coupled two-body equations if the interaction kernel $K$ contains only two-body interactions. In this case, repeated interactions within a given pair of any two quark lines can be resummed into an effective two-body transfer matrix that satisfys its own SD equation. The full transfer matrix can then be expressed in terms of these reduced matrix with quark-exchange kernel that permutes quark lines between successive pairs. For more details, see \cite{Ishii:1995bu} (references therein)

\section{Mesons in $N_f=2$}
\label{sec:mes}

\subsection{Hartree-Fock construction}

\begin{figure}
    \centering
    \includegraphics[width=\linewidth]{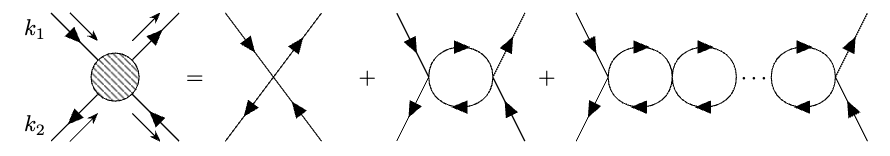}
    \caption{Quark-anti-quark bubble chain in meson channels}
    \label{fig:bubble}
\end{figure}

In order to incorporate the quark–antiquark bubble resummation as shown in Fig.~\ref{fig:bubble}, also known as RPA, one constructs effective interaction potentials in the mesonic channels by combining the direct and exchange contributions. This is achieved by performing a Fierz transformation on each four-quark vertex in Eqs.~\eqref{THOOFT1} and \eqref{THOOFT2}. Thus the resulting effective Hartree-Fock potential $\mathcal{L}^{(\mathrm{mes})}$ for the mesonic  channel can be constructed by combining the original interaction (direct) and its Fierz transform (exchange) contribution. The overall minus sign in the exchange channel is generated by anticommutativity of the quark fields. One should expect this potential reproduces the result of a full calculation by evaluating the direct fermionic bubble only (Hartree term). With this in mind, the effective meson Lagrangian reads

\begin{equation}
\begin{aligned}
\label{eq:mes_LI}
    \mathcal L^{(\mathrm{mes})}_{I}
    =&\frac{G_I}{8N_c^2}\bigg\{\left[(\bar{\psi}\psi)^2-(\bar{\psi}\tau^a\psi)^2-(\bar{\psi}i\gamma^5\psi)^2+(\bar{\psi}i\gamma^5\tau^a\psi)^2\right]\\
    &+\frac{N_c-2}{2(N^2_c-1)}\left[(\bar{\psi}\lambda^\alpha\psi)^2-(\bar{\psi}\lambda^\alpha\tau^a\psi)^2-(\bar{\psi}i\gamma^5\lambda^\alpha\psi)^2+(\bar{\psi}i\gamma^5\lambda^\alpha\tau^a\psi)^2\right]\\
    &+\frac{N_c}{4(N^2_c-1)}\left[\left(\bar{\psi}\sigma_{\mu\nu}\lambda^\alpha\psi\right)^2-\left(\bar{\psi}\sigma_{\mu\nu}\lambda^\alpha\tau^a\psi\right)^2\right]\bigg\}\\
\end{aligned}
\end{equation}

\begin{equation}
\begin{aligned}
\label{eq:mes_LIA}
      &\mathcal L^{(\mathrm{mes})}_{IA}=-\frac{G_{IA}}{2N_c^2}\bigg\{\left[(\bar{\psi}\gamma^\mu\psi)^2+(\bar{\psi}\tau^a\gamma^\mu\psi)^2+(\bar{\psi}\gamma^\mu\gamma^5\psi)^2+(\bar{\psi}\tau^a\gamma^\mu\gamma^5\psi)^2\right]\\
      &-4(\bar{\psi}\gamma^\mu\gamma^5\psi)^2-4\left[(\bar{\psi}\psi)^2+(\bar{\psi}\tau^a\psi)^2+(\bar{\psi}i\gamma^5\psi)^2+(\bar{\psi}i\gamma^5\tau^a\psi)^2\right]\\
     &+\frac{2}{N_c-1}\left[(\bar{\psi}\gamma^\mu\lambda^\alpha\psi)^2+(\bar{\psi}\gamma^\mu\gamma^5\lambda^\alpha\psi)^2\right]+\frac{2(2N_c-1)}{N_c^2-1}\left[(\bar{\psi}\gamma^\mu\lambda^\alpha\psi)^2-(\bar{\psi}\gamma^\mu\gamma^5\lambda^\alpha\psi)^2\right]\\
     &-\frac{2(N_c-2)}{N_c^2-1}\left[(\bar{\psi}\lambda^\alpha\psi)^2+(\bar{\psi}\lambda^\alpha\tau^a\psi)^2+(\bar{\psi}i\gamma^5\lambda^\alpha\psi)^2+(\bar{\psi}i\gamma^5\lambda^\alpha\tau^a\psi)^2\right]\\
     &-\frac{1}{N_c-1}\left[(\bar{\psi}\gamma^\mu\lambda^\alpha\psi)^2+(\bar{\psi}\gamma^\mu\lambda^\alpha\tau^a\psi)^2+(\bar{\psi}\gamma^\mu\gamma^5\lambda^\alpha\psi)^2+(\bar{\psi}\gamma^\mu\gamma^5\lambda^\alpha\tau^a\psi)^2\right]
     \bigg\}
\end{aligned}
\end{equation}

The interactions consist of the color singlet and octet channels. For the convenience of the $1/N_c$ counting, we normalized the $\mathrm{tr}(\lambda^\alpha\lambda^\beta)=N_c\delta^{\alpha\beta}$.The additional $-4(\bar{\psi}\gamma^\mu\gamma^5\psi)^2$ term in the first line of Eq.~\eqref{eq:mes_LIA} is the anomalous contribution from the instanton molecule contribution, breaking the $U(1)_A$ symmetry. For the meson bound states, only the color singlet interaction is relevant. Thus, the resulting effective Lagrangian for mesons can be written as


\begin{equation}
\begin{aligned}
\label{L_mes}
\mathcal{L}=&\bar{\psi}(i\slashed{\partial}-M)\psi+\frac{g_\sigma}{2N_c}(\bar{\psi}\psi)^2+\frac{g_{a_0}}{2N_c}(\bar{\psi}\tau^a\psi)^2+\frac{g_{\eta'}}{2N_c}(\bar{\psi}i\gamma^5\psi)^2+\frac{g_\pi}{2N_c}(\bar{\psi}i\gamma^5\tau^a\psi)^2\\
 &-\frac{g_\omega}{2N_c}(\bar{\psi}\gamma_\mu\psi)^2-\frac{g_\rho}{2N_c}(\bar{\psi}\gamma_\mu\tau^a\psi)^2-\frac{g_{f_1}}{2N_c}(\bar{\psi}\gamma_\mu\gamma^5\psi)^2-\frac{g_{a_1}}{2N_c}(\bar{\psi}\gamma_\mu\gamma^5\tau^a\psi)^2 
\end{aligned}
\end{equation}
where $M$ is the constituent quark mass which we will explain in Sec.~\ref{sec:gap}. The effective couplings of each channel are fixed by single instanton induced coupling $G_{I}$ and instanton-anti-instanton pair induced coupling $G_{IA}$ from the instanton vacuum.

\begin{align}
\label{tHooft_couplings}
&g_\sigma=g_\pi=\frac{G_{I}}{4N_c}+\frac{4G_{IA}}{N_c}  \nonumber\\
&g_{a_0}=g_{\eta'}=-\frac{G_{I}}{4N_c}+\frac{4G_{IA}}{N_c} \nonumber\\
&g_\omega=g_{\rho}=g_{a_1}=\frac{G_{IA}}{N_c} \nonumber\\
&g_{f_1}=-3\frac{G_{IA}}{N_c} 
\end{align}

Those quark fields in the emergent vertices in (\ref{L_mes}) are modified by finite size of the pseudoparticles. More specifically, each interacting quark field get dressed by
\begin{align}
&\psi(k)\rightarrow\sqrt{\mathcal{F}(\rho k)}~\psi(k)
\end{align}
with non-local quark form factor $\mathcal{F}(k)$ which is essentially the profile of quark zero mode  defined in \eqref{ZMform}. 

\subsection{Contituent quark mass and gap equation}
\label{sec:gap}

\begin{figure}
    \centering
\subfloat[]{\includegraphics[width=0.5\linewidth]{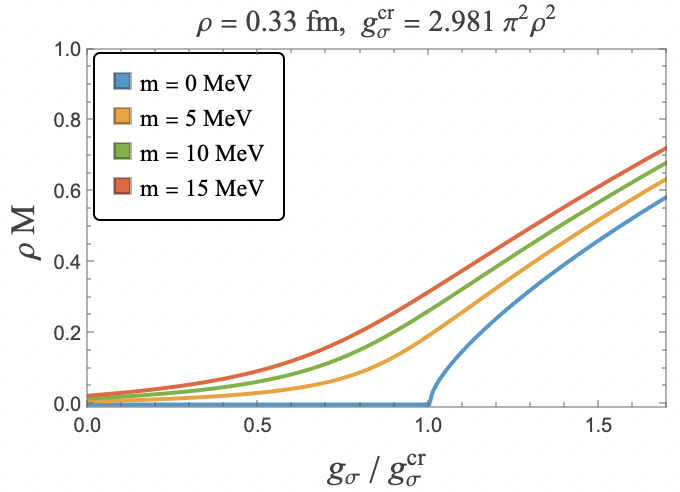}}
\hfill
\subfloat[]{\includegraphics[width=0.5\linewidth]{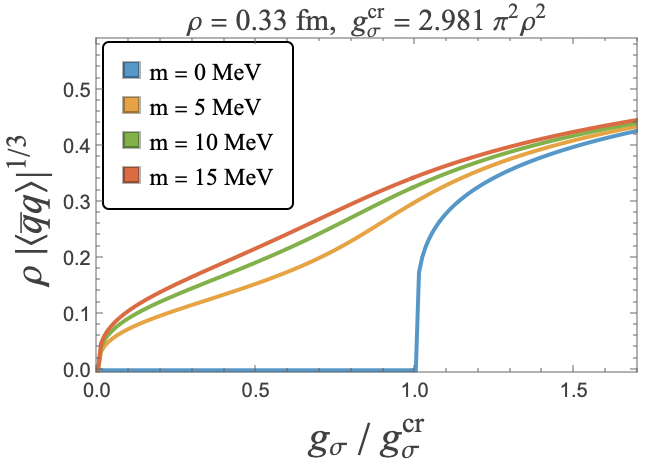}}
    \caption{(a) The plot for the $N_f=2$ gap equation in Eq.~\eqref{eq:gap}. Quark constituent mass vs. the 't Hooft coupling $g_\sigma$ for increasing current quark mass (bottom-up). (b) The plot for the quark condensate with $N_f=2$ dynamical quark in the instanton vacuum. Quark condensate vs. 't Hooft coupling $g_\sigma$ in $\sigma$ channel for increasing current quark mass (bottom-up).}
    \label{fig:chiral}
\end{figure}

Constituent quark mass signals the dynamical breaking of chiral symmetry. In Fig.~\ref{fig:chiral}, a clear second-order phase transition for both constituent quark mass and quark condensate is observed in the chiral limit for different values of $g_\sigma$, reflecting the fact that the emergence of chiral broken phases is induced by the vacuum density $n_{I+A}$. By choosing a chiral broken vacuum, quarks obtain a dynamical constant constituent mass defined by $$M(k)=m+(M-m)\mathcal{F}(k)$$ where the zero mode profile is defined in \eqref{ZMform}. Its value is determined by the gap equation through quark mass $m$, instanton size $\rho$, and density of pseudoparticles $n_{I+A}$ where the last parameter is hidden in the effective coupling $g_\sigma$ in the flavor-singlet scalar channel.

\begin{equation}
\label{eq:gap}
    M=m+2g_\sigma\int \frac{d^4k}{(2\pi)^4}
\frac{4M(k)\mathcal{F}(k)}{k^2 + M^2}
\end{equation}
where we have approximated $M(k)$ in the denominator to simply $M$. This form of the gap equation closely resembles Eqs.~\eqref{eq:cons_m} and \eqref{eq:Ch_gap}. The primary distinction lies in the treatment of the dynamical mass term, which is implemented through different approximations or resummation schemes. Despite these methodological differences, all formulations consistently capture the leading-order contribution in the $1/N_c$ expansion and are governed by the same underlying quark condensate. Consequently, they are parametrically equivalent, though they may differ quantitatively due to subleading corrections.

We can also define quark condensate directly as the order parameter of the chiral phase transition.

\begin{equation}
\label{eq:cond}
    \langle\bar qq\rangle=-N_c\int \frac{d^4k}{(2\pi)^4}\mathrm{Tr}[S(k)-S_0(k)]
\end{equation}
where $S_0(k)=\frac{i(\slashed{k}+m)}{k^2-m^2}$ is the free quark propagator and the quark propagator in the instanton vacuum is defined by
\begin{equation}
    S(k)=\frac{i(\slashed{k}+M(k))}{k^2-M^2(k)}
\end{equation}
with the momentum dependent constituent mass $M(k)$ determined by the gap equation in \eqref{eq:gap}.

\subsection{Meson spectrum}

The meson structures in a covariant formalism can be investigated by Bethe-Salpeter equation in a systematic $1/N_c$ expasion. The solution can be obtained by resumming the quark bubble diagrams in Fig.~\ref{fig:bubble}. The master integrals for each quark loop is evaluated in Appendix \ref{app:vac_pol}. The higher order $1/N_c$ correction can be found in \cite{Oertel:2000sr}. 



%

With this in mind, the meson mass can be determined by the gap-like equations expressed in terms of quark bubble functions:

\begin{equation}
\label{bound_eq_omega}
    g_{\omega,\rho}\Pi_{VV}(m_{\omega,\rho}^2)=1
\end{equation}

\begin{equation}
\label{bound_eq_2}
    g_{\sigma,a_0}\Pi_{SS}(m_{\sigma,a_0}^2)=1
\end{equation}

\begin{equation}
\label{bound_eq_3}
    g_{f_1,a_1}\Pi^{(t)}_{AA}(m_{f_1,a_1}^2)=1
\end{equation}

The pseudoscalar mesons channel will mix with the longitudianl axial mesons. By taking the mixing effect into account, the pseudoscalar mass pole can be determined by

\begin{equation}
\begin{aligned}
\label{bound_eq_4}
    &g_{\eta',\pi}\Pi_{PP}(m^2_{\eta',\pi})+g_{f_1,a_1}\Pi^{(l)}_{AA}(m^2_{\eta',\pi})\\
    &-g_{\pi,\eta'}g_{f_1,a_1}\left[\Pi_{PP}(m^2_{\eta',\pi})\Pi^{(l)}_{AA}(m^2_\pi)+\Pi^2_{PA}(m^2_{\eta',\pi})\right]=1
\end{aligned}
\end{equation}

To have physical pion and rho meson masses consistent with the experiments, the parameters in the 't Hooft Lagrangian in \eqref{L_mes} has to be fixed by specific values. These values are subject to the values of vacuum properties such as instanton size $\rho$ and instanton density $n_{I+A}$. As shown in Table~\ref{tab:parameters_ILM}, we present one sample of the data sets to illustrate the numerical relations among the parameters. The parameters labeled by asterisk $*$ shows overal mean values and boostrap uncertainties to illustrate the sensitivity among the parameters.

With the values of $G_I$ and $G_{IA}$ given in Table~\ref{tab:parameters_ILM}, the couplings in each channel are presented in Table~\ref{tab:g} with the uncertainties of each coupling to illustrate the sensitivity with respect to $\rho$ and $M$.
The couplings in all channels except for $f_1$ are attractive after the contributions from instanton-anti-instanton molecules are included in 't Hooft Lagrangian. 

\begin{table}
    \centering
    \begin{tabular}{|c|c|c|c|c|}
    \hline
        & $g_{\sigma,\pi}$ & $g_{a_0,\eta'}$ & $g_{\omega,\rho,a_1}$ & $g_{f_1}$ \\
         \hline
        ILM & 145.3 & 35.5 & 22.6 & $-67.8$ \\
        \hline 
       ILM* & $146(18)$ & $32(21)$ & $22(4)$ & $-67(11)$ \\
    \hline
    \end{tabular}
    \caption{The values are in $\mathrm{GeV}^{-2}$}
    \label{tab:g}
\end{table}

Given the coupling strengths in each channel, the Bethe-Salpeter equations in Eqs.~\eqref{bound_eq_1}, \eqref{bound_eq_2},  \eqref{bound_eq_3}, and \eqref{bound_eq_4} can determine the bound state mass for $\sigma$, $\eta'$.  However, in $a_0$ and $a_1$ channel, the interaction remains too weak to produce a bound state below the two–quark threshold $2M$.

\begin{table*}
    \centering
    \begin{tabular}{|c|c|c|c|c|c|c|c|}
    \hline
    Model & $m_{\sigma}$ (MeV) 
    &$m_{\pi^0}$ (MeV) & $m_{\pi^\pm}$ (MeV) & $m_{\eta'}$ (MeV)\\
    \hline
     ILM & $682.1$ & $139.4$ & $139.4$ & $640.3$  \\
     ILM* & $682(5)$ & $139.4$ & $139.4$ & $650(70)$  \\
     IL(spec)\cite{Shuryak:1992jz,Shuryak:1992ke,hutter2001instantonsqcdtheoryapplication} & $543$ & $142\pm14$ & $142\pm14$ & $-$ \\
     PDG \cite{ParticleDataGroup:2014cgo,ParticleDataGroup:2016lqr,ParticleDataGroup:2018ovx,ParticleDataGroup:2024cfk} & $400-800$ & $134.9766(6)$& $139.57018(35)$ & $957.78(6)$ \\
     \hline
    \end{tabular}
\hfill\\[10pt]
    \begin{tabular}{|c|c|c|c|c|c|c|c|}
    \hline
    Model & $m_{\omega}$ (MeV) & $m_{\rho}$ (MeV) \\
    \hline
     ILM & $785$ & $785$  \\
     ILM* & $785(6)$ & $785(6)$  \\
     IL(spec)\cite{Shuryak:1992jz,Shuryak:1992ke,hutter2001instantonsqcdtheoryapplication} & $950\pm100$ &  $950\pm100$ \\
     PDG \cite{ParticleDataGroup:2014cgo,ParticleDataGroup:2016lqr,ParticleDataGroup:2018ovx,ParticleDataGroup:2024cfk} & $782.65(12)$ & $775.26(25)$ \\
     \hline
    \end{tabular}
    \caption{Meson spectrum in two-flavor instanton liquid model (ILM) using Bethe-Salpeter equation solved by quark bubble summation in large $N_c$ with mean values and boostrap uncertainties labeled by asterisk $*$. The results are compared to the similar ILM framework ($N_f=3$) using spectral fitting \cite{Shuryak:1992jz,Shuryak:1992ke,hutter2001instantonsqcdtheoryapplication} and PDG.}
    \label{tab:mes0_spec}
\end{table*}

The resulting meson spectrum is presented in Table~\ref{tab:mes0_spec}. 
Without instanton-anti-instanton molecules, isovector scalar $a_0$ and isoscalar pseudoscalar $\eta'$ channel are purely repulsive and impossible to form bound states. However, when instanton–anti-instanton pair contribution is included, these channels become weakly attractive. Compared to the findings in~\cite{Shuryak:1992jz,Shuryak:1992ke,hutter2001instantonsqcdtheoryapplication}, no bound states were obtained since the molecular contribution is not sufficiently strong to overcome the repulsion from single-instanton effects. In spin-$1$ sector, due to the axial anomaly, $g_{f_1}$ coupling is strongly repulsive while the other vector coupling remains the same. As a result, the isoscalar axial meson $f_1$ is not bounded by instanton vacuum while the isovector axial meson $a_1$ exhibits a resonance above the threshold.

In Table~\ref{tab:parameters_ILM_2}, we compare the ILM prediction for current quark mass (Higgs-induced mass) and quark condensate to the Flavour Lattice Averaging Group (FLAG) \cite{FlavourLatticeAveragingGroupFLAG:2021npn,FlavourLatticeAveragingGroupFLAG:2024oxs} at the resolution $\mu=$2 GeV. The last column is the RG invariant combination in GeV$^4$. Although the value of the RG invariant combination predicted by ILM only approximately aligned with FLAG. It successfully generate the low energy GOR relation, indicating the correct chiral picture and the pion decay constant is given by $f_\pi=82.4$ MeV with the pion mass given in Table~\ref{tab:mes0_spec}.

\begin{table*}
    \centering
    \begin{tabular}{|c|c|c|c|}
    \hline
     & $m$ &  $|\langle\bar{q}q\rangle|$ & $m|\langle\bar{q}q\rangle|$   \\[5pt] 
    \hline
    ILM & $3.585$ MeV & ($264$ MeV)$^3$ & $6.57\times10^{-5}$ \\
    \hline
    ILM* & $3.60(24)$ MeV & ($264(12)$ MeV)$^3$ & $6.6(1.0)\times10^{-5}$ \\
     \hline
    FLAG \cite{FlavourLatticeAveragingGroupFLAG:2021npn,FlavourLatticeAveragingGroupFLAG:2024oxs} & $3.381(40)$ MeV & ($272(5)$ MeV)$^3$ & $6.8(4)\times10^{-5}$ \\
   \hline
\end{tabular}
    \caption{Quark mass vs. chiral condensate in the ILM using Eqs.~\eqref{eq:gap} and \eqref{eq:cond} with the parameters listed in Table~\ref{tab:parameters_ILM} at the resolution
    $\mu=1/\rho\approx600$ MeV. The last column presents the RG invariant combination in GeV$^4$}
    \label{tab:parameters_ILM_2}
\end{table*}

\subsection{BS wave functions}

\begin{figure}
    \centering
\subfloat[]{\includegraphics[width=0.47\linewidth]{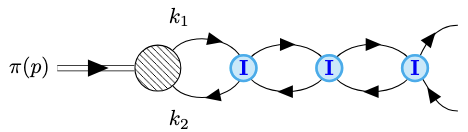}}  
\hfill
\subfloat[]{\includegraphics[width=0.47\linewidth]{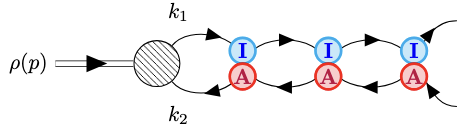}} 
    \caption{Pion and rho meson bound state in ILM}
    \label{fig:BS_mes}
\end{figure}

The meson wave function in two-body Fock space is related to the Bethe-Salpeter (BS) wave functions defined by

\begin{equation}
\begin{aligned}
    &i(2\pi)^4\delta^4(p-k_1-k_2)\Psi_{X,ij}(k_1,k_2;P\lambda)\delta_{\alpha\beta}\\
    =&\int d^4x_1d^4x_2 e^{ik_1\cdot x_1+ik_2\cdot x_2}\langle0|\psi_{\alpha i}(x_1)\bar{\psi}_{\beta j}(x_2)|X(P,\lambda)\rangle    
\end{aligned}
\end{equation}
where $\alpha,\beta$ denote the color indices and $i,j$ denote the flavor-Dirac indices carried by the quark fields $\psi$. 

As shown in Fig.~\ref{fig:BS_mes}, the solution for meson BS equation can be parameterized by

\begin{equation}
    \begin{aligned}
    \label{eq:mes_wf1}
        &\Psi_\sigma(k_1,k_2;P)=-\sqrt{Z_{\sigma}} \sqrt{\mathcal{F}(k_1)\mathcal{F}(k_2)}\, S(k_1)S(-k_2) \\[8pt]
    &\Psi_{a_0}(k_1,k_2;P)=-\sqrt{Z_{a_0}}\sqrt{\mathcal{F}(k_1)\mathcal{F}(k_2)}\, S(k_1)\tau^a S(-k_2) 
    \end{aligned}
\end{equation}

\begin{equation}
\begin{aligned}
\label{eq:mes_wf2}
    &\Psi_{\eta'}(k_1,k_2;P)=-\sqrt{Z_{\eta'}} \sqrt{\mathcal{F}(k_1)\mathcal{F}(k_2)}\, S(k_1)\left(i\gamma^5\cos\theta_{\eta'f_1}+i\hat{\slashed{P}}\gamma^5\sin\theta_{\eta'f_1}\right)  S(-k_2) \\[8pt]
    &\Psi_\pi(k_1,k_2;P)=-\sqrt{Z_{\pi}} \sqrt{\mathcal{F}(k_1)\mathcal{F}(k_2)}\, S(k_1)\left(i\gamma^5\cos\theta_{\pi a_1}+i\hat{\slashed{P}}\gamma^5\sin\theta_{\pi a_1}\right)\tau^a S(-k_2)
\end{aligned}
\end{equation}

Similarly for the spin-$1$ meson, we have

\begin{equation}
\begin{aligned}
\label{eq:mes_wf3}
    &\Psi_\omega(k_1,k_2;P\lambda)=-\sqrt{Z_{\omega}} \sqrt{\mathcal{F}(k_1)\mathcal{F}(k_2)}\, \epsilon_\lambda^\mu(P)S(k_1)\gamma_\mu S(-k_2)\\[8pt]
    &\Psi_{\rho}(k_1,k_2;P\lambda)=-\sqrt{Z_{\rho}} \sqrt{\mathcal{F}(k_1)\mathcal{F}(k_2)}\, \epsilon_\lambda^\mu(P)S(k_1)\gamma_\mu\tau^a S(-k_2)\\[8pt]
    &\Psi_{f_1}(k_1,k_2;P\lambda)=-\sqrt{Z_{f_1}} \sqrt{\mathcal{F}(k_1)\mathcal{F}(k_2)}\, \epsilon_\lambda^\mu(P)S(k_1)\gamma_\mu\gamma^5 S(-k_2)\\[8pt]
    &\Psi_{a_1}(k_1,k_2;P\lambda)=-\sqrt{Z_{a_1}} \sqrt{\mathcal{F}(k_1)\mathcal{F}(k_2)}\, \epsilon_\lambda^\mu(P)S(k_1)\gamma_\mu\gamma^5\tau^a S(-k_2)
\end{aligned}
\end{equation}
where the wave function renormalization constants $Z_{X}$ can be determined by

\begin{equation}
 Z_{\sigma,a_0}=\frac1{N_c}\left(\frac{\partial \Pi_{SS}(P^2)}{\partial P^2}\right)^{-1}\Big|_{P^2=m_\sigma^2} 
\end{equation}

\begin{equation}
   Z_{\omega,\rho}=\frac1{N_c}\left(\frac{\partial \Pi_{VV}(P^2)}{\partial P^2}\right)^{-1}\Big|_{P^2=m_{\omega,\rho}^2}  
\end{equation}

\begin{equation}
 Z_{f_1,a_1}=\frac1{N_c}\left(\frac{\partial \Pi^{(t)}_{AA}(P^2)}{\partial P^2}\right)^{-1}\Big|_{P^2=m_{f_1,a_1}^2}
\end{equation}

\begin{equation}
\begin{aligned}
&Z_{\eta',\pi}=\\
&\frac{\frac1{N_c}\left(g_{\eta',\pi}-g_{f_1,a_1}+g_{\eta',\pi}g_{f_1,a_1}(\Pi_{PP}-\Pi^{(l)}_{AA})\right)}{g_{\eta',\pi}(1-g_{f_1,a_1}\Pi^{(l)}_{AA})\frac{\partial \Pi_{PP}}{\partial P^2}+g_{f_1,a_1}(1-g_{\eta',\pi}\Pi_{PP})\frac{\partial \Pi^{(l)}_{AA}}{\partial P^2}-2g_{\eta',\pi} g_{f_1,a_1}\Pi_{PA}\frac{\partial \Pi_{PA}}{\partial P^2}}\Bigg|_{P^2=m_{\eta',\pi}^2}
\end{aligned}
\end{equation}

The pseudoscalar channel is mixed with time-component of axial vector channel. The mixing angle $\theta_{\eta'f_1}$  and the mixing angle $\theta_{\pi a_1}$ are obtained by the diagonalization of the pseudoscalar and time-component axial vector channels. The mixing angles can also be determined by the quark bubble functions defined in Appendix~\ref{app:vac_pol}.

\begin{equation}
\begin{aligned}
\tan\theta_{\eta'f_1, \pi a_1}
=&\frac{g_{f_1,a_1}\Pi_{PA}(m^2_{\eta',\pi})}{1-g_{f_1,a_1}\Pi^{(l)}_{AA}(m^2_{\eta',\pi})}=\frac{1-g_{\eta',\pi}\Pi_{PP}(m^2_{\eta',\pi})}{-g_{\eta',\pi}\Pi_{PA}(m^2_{\eta',\pi})}
\end{aligned}
\end{equation}
where the range of mixing angle is within $-\pi/2<\theta_{\eta'f_1, \pi a_1}<\pi/2$. 

The meson wave function renormalization constant are estimated in Table~\ref{tab:renorm_mes} and the mixing angles between pseudoscalar and axial channel are also presented in Table~\ref{tab:PA_mixing}.

\begin{table}
\begin{center}
\begin{tabular}{|c|c|c|c|c|c|}
   \hline
     &  $\sqrt{Z_{\sigma}}$ & $\sqrt{Z_{\pi}}$ & $\sqrt{Z_{\omega,\rho}}$ & $\sqrt{Z_{\eta'}}$\\
   \hline
   ILM  & $3.94$ & $5.09$ & $0.746$ & $1.94$ \\
   \hline
   ILM*  & $3.95(15)$ & $5.11(23)$ & $0.731(14)$ & $1.94(30)$ \\
   \hline
\end{tabular}
\end{center}
    \caption{Meson wave function renormalization constants, or effective meson-quark coupling}
    \label{tab:renorm_mes}
\end{table}

\begin{table}
\begin{center}
\begin{tabular}{|c|c|c|}
\hline
Model &$\theta_{\eta' f_1}$ & $\theta_{\pi a_1}$ \\
\hline
ILM & $-51^\circ$ & $0.955^\circ$\\
\hline
ILM* & $-51(25)^\circ$ & $0.88(34)^\circ$\\
\hline
\end{tabular}
\end{center}
    \caption{Pseudoscalar-axial-vector mixing angle}
    \label{tab:PA_mixing}
\end{table}

\subsection{Light Front Wave Functions}
\label{sec:LFWF}
The light front wave function (LFWF) is defined by spanning in the Fock space. The $2$-body meson LFWF at leading Fock space reads

\begin{equation}
    \ket{X}\approx\sum_{s_1,s_2}\int d[12]\frac{\delta_{\alpha\beta}}{\sqrt{N_c}}\Phi_X(k_1,k_2,s_1,s_2)b_{s_1,\alpha}^\dagger(k_1) c^\dagger_{s_2,\beta}(k_2)\ket{0}
\end{equation}

where the $2$-body phase space integral is defined as

\begin{equation}
\begin{aligned}
\label{eq:n-body}
    &d[12]
    =\prod_{i=1}^2\frac{dx_{i}d^2k_{i\perp}}{(2\pi)^32x_i}(2\pi)^32\delta\left(1-\sum_{i=1}^2x_{i}\right)\delta^2\left(p_\perp-\sum_{i=1}^2k_{i\perp}\right)
\end{aligned}
\end{equation}
The solutions in light front formalism should be equivalent to the Lorentz covariant formalism. In our method we solve $2$-body BS equation by summing over all planar quark bubble diagrams at $1/N_c$ limit. It is easy to show it is equivalent to leading Fock LFWF by directly integrating the BS wave function over the $k_{1,2}^-$ component of the momentum integrals within the physical region $\delta(k_1^-+k_2^--p^-)$. The higher Fock state LFWF correspond to the diagrams that is sub-leading in $1/N_c$. 
By projecting out the bounded quark spins
\begin{equation}
\begin{aligned}
    &\frac{1}{2x\bar{x}}\Phi_{X}(x,k_\perp,s_1,s_2)=&i\sqrt{N_c}P^+\int_{-\infty}^\infty \frac{dk_1^-}{2\pi}\frac{\bar{u}_{s_1}\gamma^+}{2k_1^+}\Psi_X(k_1,k_2;P)\frac{\gamma^+v_{s_2}}{2k_2^+}\bigg|_{k_{1}^+=xP^+}\\
\end{aligned}
\end{equation}

Using the light front integral in \cite{Liu:2023fpj} with $P^2=m_X^2$, 
\begin{equation}
\begin{aligned}
    &iP^+\int_{-\infty}^{\infty
    }\frac{dk^-}{2\pi}\frac{\sqrt{\mathcal{F}\left(k\right)\mathcal{F}\left(P-k\right)}}{(k^2-M^2)[(P-k)^2-M^2]}\\
    &\rightarrow\frac{\theta(x\bar{x})}{2x\bar{x}}\frac{1}{m_X^2-\frac{k_\perp^2+M^2}{x\bar{x}}}\mathcal{F}\left(\frac{k_\perp}{\lambda_X\sqrt{x\bar{x}}}\right)
\end{aligned}
\end{equation}

Therefore, the meson light front wavefunction for $N_f=2$ would be

\begin{equation}
\begin{aligned}
\label{WFX}
\Phi_X=&\phi_X(x_1,x_2,k_{1\perp},k_{2\perp})
\bar u_{s_1}\Gamma_Xv_{s_2}
\end{aligned}
\end{equation}
with $\Gamma_X$ the BS vertex function.
For spin-$0$ mesons, $\Gamma_X=\frac{1}{\sqrt{2}}\{1,\tau^a,i\gamma^5,i\gamma^5\tau^a\}$ for $\sigma$, $a_0$, $\eta'$, $\pi$ mesons and for spin-$1$ mesons, we introduce the polarization vector $\varepsilon^\mu_\lambda(p)$ to carry the vector polarizations. Thus, $\Gamma_X=\frac{1}{\sqrt{2}}\varepsilon^\mu_\lambda(p)\{\gamma_\mu,\gamma_\mu\tau^a,\gamma_\mu\gamma^5,\gamma_\mu\gamma^5\tau^a\}$ for $\omega$, $\rho$, $f_1$, $a_1$ mesons. The momentum wave function for meson $X$ in \eqref{WFX} is defined by \cite{Liu:2023fpj,Liu:2023yuj} with $x_1=1-x_2=x$ and $k_{1\perp}=-k_{2\perp}=k_{\perp}$,
\begin{equation}
\begin{aligned}
\label{WFX2}
&\phi_X(x,k_{\perp})=-\frac{\sqrt{N_cZ_X}}{m_X^2-\frac{k_{\perp}^2+M^2}{x\bar{x}}}\,\mathcal{F}\left(\frac{k_{\perp}^2}{\lambda_X x\bar{x}}\right)
\end{aligned}
\end{equation}
with $\sqrt{Z_X}$ the meson wave function renormalization constant and $\lambda_X$ the parameter that makes sure the wave function in Eqs. \eqref{eq:mes_wf1}, \eqref{eq:mes_wf2}, and \eqref{eq:mes_wf3} are properly normalized in lightfront setting~\cite{Kock:2020frx,Liu:2023fpj}.

For completeness, the explicit expressions of the light-cone wave functions in each meson channel are presented in Appendix~\ref{app:LCWF}, following the classification of \cite{Ji:2003yj}.

\section{Mesons in $N_f=3$}
\label{sec:mes_3}

\begin{table}
    \centering
    \begin{tabular}{|c|c|c|c|c|c|c|}
   \hline
 $G_I$  & $G_{IA}$ & $\sqrt{\langle T_{IA}^2\rangle}$ & $\sqrt[4]{\langle T_{IA}^4\rangle}$ & $M_u$ & $M_s$   \\ 
    \hline
   $182.4$~$\mathrm{fm}^{5}$ & 64.8 GeV$^{-2}$ & 3.57 GeV$^{-1}$ & 0.675 & $395.2$ MeV & $591.7$ MeV  \\
   \hline
\end{tabular}
    \caption{The fitted parameters in ILM ($N_f=3$) using instanton size $\rho=0.33$ fm and constituent mass $M_u=395.2$ MeV with fixed pion mass $m_\pi=139.4$ MeV, kaon mass $m_K=473$ MeV, and rho meson mass $m_\rho=787$ MeV. $G_I$ listed here denotes the 3-flavor single instanton coupling (see Eq.~\eqref{eq:inst_couple_3Nf}).}
    \label{tab:parameters_ILM_Nf3}
\end{table}


\begin{table}
    \centering
    \begin{tabular}{|c|c|c|c|c|}
   \hline
    & $m_u$ & $m_s$ & $\langle\bar{u}u\rangle$ &  $\langle\bar{s}s\rangle$   \\ 
    \hline
  ILM & $3.80$ & $100.6$ & $-(263.6)^3$ & $-(267.7)^3$ \\
   \hline
   Lattice \cite{McNeile:2012xh} & $3.36(6)$ & $ 92.2(1.0)$ & $-(283(2))^3$ & $-(290(15))^3$ \\
   \hline
\end{tabular}
    \caption{Quark mass (MeV) vs. chiral condensate (MeV$^3$) in the ILM using Eqs.~\eqref{eq:cond} and \eqref{gap} with the parameters listed in Table~\ref{tab:parameters_ILM_Nf3} at the resolution $\mu\approx1/\rho$ where $\rho=0.33$~fm. The estimate is compared with lattice calculation using Highly Improved Staggered Quarks (HISQ).}
    \label{tab:condensate_ILM}
\end{table}

In the case of $N_f=3$, the 't Hooft interaction emerging from the QCD instanton vacuum is typically a six-quark interaction. In $1/N_c$ expansion, the six-quark interaction can be dimensionally reduced to four-quark 't Hooft interaction that resembles Eq.~\eqref{THOOFT1} by looping one of the three flavors into corresponding condensate. The effective coupling constant for the remaining two flavor thus reads

\begin{equation}
\label{eq:inst_couple_3Nf}
G_I(N_f=2)=-\frac{G_I(N_f=3)}{2N_c}\left(\frac{N_cm_q^*}{2\pi^2\rho^2}\right)
\end{equation}
with value $G_I=182.4\,\mathrm{fm}^5=6.1\times10^5\,\mathrm{GeV}^{-5}$ fixed by pion, kaon, and rho meson masses. $m_{u,d}^*$ is proportional to quark condensate $2\pi^2\rho^2\langle\bar uu\rangle/N_c$ for degenerate $u$ and $d$ due to the relatively light quark mass, while the strange quark mass is not negligible in $m^*_s$.

With this in mind, in the mean field approximation along with the $1/N_c$ expansion, the effective 't Hooft Lagrangian for $N_f=3$ mesons reads

\begin{equation}
\begin{aligned}
\mathcal{L}_I=\sum_{q=u,d,s}\frac{g_q}{2N_c}\left[(\bar{\psi}\mathds{1}_q\psi)^2-(\bar{\psi}i\gamma^5\mathds{1}_q\psi)^2-(\bar{\psi}\tau^a_q\psi)^2+(\bar{\psi}i\gamma^5\tau^a_q\psi)^2\right]\\
\end{aligned}
\end{equation}

In contrast, the additional vertices induced by instanton–anti-instanton pairs contribute a hierarchy of multi-fermion interactions in three flavor as multiple quark flavors can be simultaneously ``locked'' within a correlated instanton–anti-instanton configuration, giving rise to vertices with varying numbers of external quark legs (see Sec.~\ref{sec:tail})

Among these molecular contributions, the two-body (four-quark) interactions are particularly prominent, give rise to the same form in \eqref{THOOFT2}~\cite{Rapp:1999qa}. This is a direct consequence of the $1/N_c$ counting: vertices with a larger number of quark legs are suppressed by higher powers of $1/N_c$ (see Sec.~\ref{sec:tail}). As a result, the four-quark interactions provide the leading molecular contribution, while higher-body operators enter only as subleading corrections. 

At leading order in the $1/N_c$ expansion, the dominant contribution is governed by two-body operators, corresponding to effective four-quark interactions. These terms provide the primary mechanism for dynamical chiral symmetry breaking, while higher-body operators are suppressed by additional powers of $1/N_c$.

\begin{equation}
\begin{aligned}
\mathcal{L}_{IA}=&-\frac{g_{IA}}{2N_c}\left[\frac23(\bar{\psi}\gamma^\mu\psi)^2+(\bar{\psi}\gamma^\mu\lambda^a\psi)^2+\frac23(\bar{\psi}\gamma^\mu\gamma^5\psi)^2+(\bar{\psi}\gamma^\mu\gamma^5\lambda^a\psi)^2\right]\\
&+\frac{2g_{IA}}{N_c}(\bar{\psi}\gamma^\mu\gamma^5\psi)^2+\frac{2g_{IA}}{N_c}\left[\frac{2}{3}(\bar{\psi}\psi)^2+(\bar{\psi}\lambda^a\psi)^2+\frac23(\bar{\psi}i\gamma^5\psi)^2+(\bar{\psi}i\gamma^5\lambda^a\psi)^2\right]
\end{aligned}
\end{equation}

\begin{align}
    g_{q}=&\frac{G_I(N_f=3)}{4N_c}\left(\frac{m_q^*}{4\pi^2\rho^2}\right) &
    g_{IA}=&\frac{G_{IA}}{N_c}
\end{align}

The 't Hooft Lagrangian is divided into three sectors with three different $SU(2)_{\mathrm{flavor}}$ doublet in the $SU(3)_{\mathrm{flavor}}$ flavor space with the corresponding effective $4$-quark coupling. Each sector is generated by looping one of the flavor $q$ to form a quark condensate $\langle\bar{q}q\rangle$. The Pauli matrices $\tau^a_q$ and identity matrices $\mathds{1}_q$ corresponds to the  generators for each $SU(2)_{\mathrm{flavor}}$ sector. 

\begin{itemize}
    \item For $q=s$, the isospin matrices can be defined as, 
\begin{center}
    $\mathds{1}_s=\frac{2}{3}\mathds{1}+\frac{\sqrt{3}\lambda^8}{3}$, $\tau_s^\pm=(\lambda^1\pm i\lambda^2)/\sqrt{2}$ and $\tau_s^3=\lambda^3$
\end{center}

\item For $q=d$, the $U$-spin matrices can be defined as

\begin{center}
$\mathds{1}_d=\frac{2}{3}\mathds{1}+\frac{\lambda^3}{2}-\frac{\sqrt{3}\lambda^8}{6}$, $\tau_d^\pm=(\lambda^4\pm i\lambda^5))/\sqrt{2}$ and $\tau_d^3=(\sqrt{3}\lambda^8+\lambda^3)/2$
\end{center}

\item For $q=u$, the matrices for V-spin are,
\begin{center}
$\mathds{1}_u=\frac{2}{3}\mathds{1}-\frac{\lambda^3}{2}-\frac{\sqrt{3}\lambda^8}{6}$, $\tau_u^\pm=(\lambda^6\pm i\lambda^7))/\sqrt{2}$ and $\tau_u^3=(\sqrt{3}\lambda^8-\lambda^3)/2$
\end{center}
\end{itemize}
where $\mathds{1}$ is the $3\times3$ identity matrix and $\lambda^a$ is the Gell-Mann matrices in flavor space.

Now with this in mind, the Lagrangian reads
\begin{equation}
\begin{aligned}
\mathcal{L}=&\bar{\psi}\left(i\slashed{\partial}-M\right)\psi+\mathcal{L}_S +\mathcal{L}_P +\mathcal{L}_V +\mathcal{L}_A
\end{aligned}
\end{equation}

The quark constituent mass $M$ can be expressed in terms of Gell-Mann matrices in flavor space as well.

\begin{equation}
    M=M_u\left(\frac{1}{3}\mathds{1}+\frac{\lambda^3}{2}+\frac{\sqrt{3}}{6}\lambda^8\right)+M_d\left(\frac{1}{3}\mathds{1}-\frac{\lambda^3}{2}+\frac{\sqrt{3}}{6}\lambda^8\right)+M_s\left(\frac{1}{3}\mathds{1}-\frac{\sqrt{3}}{3}\lambda^8\right)
\end{equation}
with the gap equations given by 

\begin{eqnarray}
\label{gap}
M_u=&m_u+\frac{G_I}{4N_c}\sigma_d\sigma_s+\frac{8G_{IA}}{N_c}\sigma_u \nonumber\\[8pt]
M_d=&m_d+\frac{G_I}{4N_c}\sigma_u\sigma_s+\frac{8G_{IA}}{N_c}\sigma_d \nonumber\\[8pt]
M_s=&m_s+\frac{G_I}{4N_c}\sigma_u\sigma_d+\frac{8G_{IA}}{N_c}\sigma_s 
\end{eqnarray}
where for $u$, $d$ and $s$, the mass gap parameter reads
\begin{equation}
    \sigma_q\equiv\frac{m_q^*}{2\pi^2\rho^2}=4\int\frac{d^4k}{(2\pi)^4}\frac{m_q+(M_q-m_q)\mathcal{F}(k)}{k^2+[m_q+(M_q-m_q)\mathcal{F}(k)]^2}\mathcal{F}(k)
\end{equation}

The dynamical mass of the $u$ quark receives contributions not only from its own channel but also from the $d$ and $s$ quark condensates through flavor-mixing interactions. An analogous pattern holds for the $d$ and $s$ quarks. Consequently, even in the chiral limit, a nonvanishing mass gap occurs for all flavors, determined self-consistently by the system of gap equations.


\subsection{Scalar meson spectrum}

In the scalar meson $0^+$ channel, the effective interaction reads
\begin{equation}
\begin{aligned}
    \mathcal{L}_{S}=&\frac{g_{\sigma}}{2N_c}\left(\bar{\psi}\lambda^0\psi\cos\theta_{\sigma f_0}+\bar{\psi}\lambda^8\psi\sin\theta_{\sigma f_0}\right)^2+\frac{g_{f_0}}{2N_c}\left(\bar{\psi}\lambda^8\psi\cos\theta_{\sigma f_0}-\bar{\psi}\lambda^0\psi\sin\theta_{\sigma f_0}\right)^2\\
    &+\frac{g_{a_0}}{2N_c}(\bar{\psi}\lambda^3\psi)^2+\frac{g_{a_0}}{2N_c}(\bar{\psi}\tau_I^a\psi)^2+\frac{g_{K^*_0}}{2N_c}(\bar{\psi}\tau_U^a\psi)^2+\frac{g_{K^*_0}}{2N_c}(\bar{\psi}\tau_V^a\psi)^2
\end{aligned}
\end{equation}
where $\lambda^0\equiv\sqrt{\tfrac23}\mathds{1}$ and the mixing angle is determined by \eqref{S_mixing} and will be discussed later in this section.
In flavored channels, the couplings are 
\begin{equation}
g_{a_0} = -g_s + 4g_{IA}, \qquad
g_{K^*_0} = -g_u + 4g_{IA}
\end{equation}

With only single-instanton contributions, the isovector scalar $a_0$ and strange scalar $K_0^*$ channels are purely repulsive and do not support bound states. This situation changes once correlated instanton–anti-instanton pairs (``molecules'') are included. Although such configurations carry zero net topological charge, they induce additional interactions that partially compensate the repulsion from single instantons. As a result, the net interaction in these channels becomes weakly attractive, allowing for bound-state formation when the molecular density is sufficiently large.

In the unflavored channel, the couplings are more involved due to the flavor mixing, given by
\begin{equation}
\begin{aligned}
\label{eq:sca_mix}
g_{\sigma,f_0}=\frac12 \left(g_s+8 g_{IA} \pm \sqrt{8 g_u^2+g_s^2}\right)
\end{aligned}
\end{equation}

The mixing is induced by the breaking of the flavor symmetry $SU(N_f)$. In the limit of exact $SU(3)$ flavor symmetry $g_s=g_u$, the coupling in \eqref{eq:sca_mix} is reduced to

\begin{equation}
g_{\sigma} = 2g_s + 4g_{IA}
, \qquad
g_{f_0} = -g_s + 4g_{IA}
\end{equation}


\subsubsection{Mass spectrum}

The mass spectrum of light scalar mesons ($N_f=3$) can be obtained by the Bethe–Salpeter resummation, as illustrated in Fig.~\ref{fig:bubble}. The mass of the isovector scalar $a_0$ is determined by

\begin{equation}
\label{bound_eq_a0}
    g_{a_0}\Pi_{a_0a_0}(m_{a_0}^2)=1
\end{equation}
where the bubble diagram in $a_0$ channel reads
\begin{equation}
    \Pi_{a_0a_0}=2\Pi^{ud}_{SS}=\Pi^{uu}_{SS}+\Pi^{dd}_{SS}
\end{equation}
with each quark bubble functions $\Pi^{qq'}_{SS}$ defined in Appendix \ref{app:vac_pol}. The mass of the scalar partner of kaon can also be determined by
\begin{equation}
\label{bound_eq_Kstar}
    g_{K^*_0}\Pi_{K^*_0K^*_0}(m_{K^*_0}^2)=1
\end{equation}
with the bubble diagram in $K^*_0$ channel is given by
\begin{equation}
    \Pi_{K^*_0K^*_0}=2\Pi^{us}_{SS}=2\Pi^{ds}_{SS}
\end{equation}

For the unflavored channel, the quark bubble functions are mixed between the $\sigma$ and $f_0$ channel.

\begin{equation}
\begin{aligned}
\Pi_{\sigma\sigma}=&\left(\Pi_{PP}^{uu}+\Pi_{PP}^{dd}\right)\left(\frac23\cos^2\theta_{\sigma f_0}+\frac{2\sqrt{2}}{3}\cos\theta_{\sigma f_0} \sin\theta_{\sigma f_0}+\frac13\sin^2\theta_{\sigma f_0}\right)\\
&+\Pi_{PP}^{ss}\left(\frac23\cos^2\theta_{\sigma f_0}-\frac{4\sqrt{2}}{3}\cos\theta_{\sigma f_0} \sin\theta_{\sigma f_0}+\frac43\sin^2\theta_{\sigma f_0}\right)
\end{aligned}
\end{equation}

\begin{equation}
\begin{aligned}
\Pi_{\sigma f_0}=&\left(\Pi_{PP}^{uu}+\Pi_{PP}^{dd}\right)\left(\frac{1}{\sqrt{3}}\cos^2\theta_{\sigma f_0}-\frac{1}{\sqrt{3}}\sin^2\theta_{\sigma f_0}-\frac23\cos\theta_{\sigma f_0} \sin\theta_{\sigma f_0}\right)\\
     &-\Pi_{PP}^{ss} \left(\frac2{\sqrt{3}}\cos^2\theta_{\sigma f_0}-\frac2{\sqrt{3}}\sin^2\theta_{\sigma f_0}-\frac13\cos\theta_{\sigma f_0} \sin\theta_{\sigma f_0}\right)
\end{aligned}
\end{equation}

\begin{equation}
\begin{aligned}
\Pi_{f_0f_0}=&\left(\Pi_{PP}^{uu}+\Pi_{PP}^{dd}\right)\left(\frac23\sin^2\theta_{\sigma f_0}-\frac{2\sqrt{2}}{3}\cos\theta_{\sigma f_0} \sin\theta_{\sigma f_0}+\frac13\cos^2\theta_{\sigma f_0}\right)\\
&+\Pi_{PP}^{ss}\left(\frac23\sin^2\theta_{\sigma f_0}+\frac{4\sqrt{2}}{3}\cos\theta_{\sigma f_0} \sin\theta_{\sigma f_0}+\frac43\cos^2\theta_{\sigma f_0}\right)
\end{aligned}
\end{equation}
where at tree level, the mixing angle $\theta_{\sigma f_0}$ ($0<\theta_{\sigma f_0}<\pi/2$) induced by simply the couplings is given by

\begin{align}
\label{S_mixing}
\tan2\theta_{\sigma f_0}=\frac{2\sqrt{2}(g_s-g_u)}{g_s+8g_u}\propto m^*_s-m^*_u,
\end{align}

Beyond tree level, the mixing angle is shifted by bubble (RPA) resummation. The resulting shift $\delta\theta_{\sigma f_0}$ is determined by
\begin{equation}
\begin{aligned}
\label{S_mixing_2}
\tan\delta\theta_{\sigma f_0}=&\frac{g_{f_0}\Pi_{\sigma f_0}(m^2_{\sigma})}{1-g_{f_0}\Pi_{f_0f_0}(m^2_{\sigma})}=\frac{1-g_{\sigma}\Pi_{\sigma\sigma}(m^2_{\sigma})}{g_{\sigma}\Pi_{\sigma f_0}(m^2_{\sigma})}
\end{aligned}
\end{equation}

The mass of the resulting $\sigma$ bound state is then determined by 
\begin{equation}
    g_\sigma\Pi_{\sigma\sigma}(m^2_{\sigma})+g_{f_0}\Pi_{f_0f_0}(m^2_{\sigma})-g_\sigma g_{f_0}\left[\Pi_{\sigma\sigma}(m^2_{\sigma})\Pi_{f_0f_0}(m^2_{\sigma})-\Pi^2_{\sigma f_0}(m^2_{\sigma})\right]=1
\end{equation}
The physical $\sigma$ solution is selected by the condition
\begin{equation}
g_\sigma + g_{f_0}
-g_\sigma g_{f_0}
\left[
\Pi_{\sigma\sigma}(m^2_{\sigma})
+\Pi_{f_0f_0}(m^2_{\sigma})
\right]>0 .
\end{equation}
The opposite condition corresponds to the orthogonal eigenchannel and would instead identify the $f_0$ bound-state solution.

With the values of $G_I$ and $G_{IA}$ fixed in Table~\ref{tab:parameters_ILM_Nf3}, the couplings in each channel now read

\begin{center}
\begin{align*}
&g_{a_0}=27.76~ \mathrm{GeV}^{-2} && g_{K_0^*} =52.09~ \mathrm{GeV}^{-2}  \\
&g_\sigma=172.39
~ \mathrm{GeV}^{-2} 
&&g_{f_0}=59.03
~ \mathrm{GeV}^{-2}
\end{align*}
\end{center}

Compared with the two-flavor couplings in Table~\ref{tab:g}, the inclusion of the strange quark enhance the coupling strength in the isoscalar scalar $\sigma$ channel. Consequently, the pion and $\sigma$ meson no longer share the same coupling, reflecting the enhanced flavor mixing induced by the presence of strangeness. In contrast, the coupling in the isovector scalar channel remains of the same order, indicating that it is comparatively less sensitive to strange-quark effects.

\begin{table}[]
    \centering
    \begin{tabular}{|c|c|c|c|c|c|}
    \hline
    Model & $m_{\sigma}$ (MeV) & $m_{f_0}$ (MeV) & $m_{a_0}$ (MeV) & $m_{K^*_0}$ (MeV) \\
    \hline
     ILM & $677.36$ & $-$ & $-$ & $-$  \\
     PDG \cite{ParticleDataGroup:2014cgo,ParticleDataGroup:2016lqr,ParticleDataGroup:2018ovx,ParticleDataGroup:2024cfk} & $400\sim800$ & $990\pm20$ & $980\pm20$ & $845\pm17$ \\
     \hline
\end{tabular}   
    \caption{Mass spectrum of scalar mesons ($N_f=3$), where `` $–$ '' indicates that the state lies above lowest threshold $2M_u$ and is not stably bound.}
    \label{tab:s_meson}
\end{table}
The scalar meson masses are presented in Table~\ref{tab:s_meson}. Only the $\sigma$ meson is bound by the instanton-induced interaction. The remaining scalar mesons lie above the corresponding mass thresholds, indicating that the instanton interaction alone is not sufficiently strong to bind them.

The mass is satisfied with the linear Gell-Mann-Okubo mass relation with the mixing angle given in Table~\ref{tab:s_meson_mixing}:
\begin{equation}
    4m_{K^*_0}-m_{a_0}=3(m_{f_0}\cos^2\theta_S+m_\sigma\sin^2\theta_S)
\end{equation}

\begin{table}
    \centering
\begin{tabular}{|c|c|}
    \hline
    Model  & $\theta_S$  \\
    \hline
     ILM & $15.13^\circ$  \\
     Experimental analysis \cite{Oller:2003vf}  & $ 19^\circ \pm 5^\circ$ \\
     Diquark exchange model \cite{Agaev:2017cfz}  & $-33^\circ.00 \pm 1^\circ.17$ \\
     ChPT \cite{Napsuciale:1998ip}   & $-65^\circ \sim -79^\circ$ \\
     \hline
\end{tabular}   
\caption{Mixing angle of $\sigma$ and $f_0$ meson in ILM with $\rho=0.33$ fm. The result is compared to the conventional mixing angle $\theta_S$ in the normalized singlet-octet basis}
    \label{tab:s_meson_mixing}
\end{table}
The scalar mixing angle in our model basis ($\mathds{1}$, $\lambda^8$) is found to be $\theta_{\sigma f_0}+\delta\theta_{\sigma f_0}=15.13^\circ$ in total, with the leading-order angle in \eqref{S_mixing} $\theta_{\sigma f_0}=
5.837^\circ$, and the increasing of $\delta\theta_{\sigma f_0}=9.292^\circ$ after resumming the bubble diagrams in \eqref{S_mixing_2}. Our model basis can be straightforwardly related to the standard normalized singlet-octet basis with the mixing angles are directly related by $\theta_S=\theta_{\sigma f_0}+\delta\theta_{\sigma f_0}$. 

The comparison is listed in Table~\ref{tab:s_meson_mixing}. The mixing angle is consistent with the experimental analysis in \cite{Oller:2003vf} showing that the $f_0(980)$ meson is mostly isosinglet octet and the $\sigma$ is predominantly singlet. 

The mixing of singlet and octet current to form the physical mesons implies that all of these decays can run through the
superallowed Okubo–Zweig–Iizuka (OZI) mechanism. Without the
mixing, the decay $f_0(980) \rightarrow \pi \pi$ can proceed due to one gluon exchange, whereas $f_0(980) \rightarrow KK$ is still OZI superallowed process. 


Similarly, the wave function renormalization constants $Z_{X}$, or effecitve meson-quark couplings are given by

\begin{equation}
 Z_{a_0}=\frac1{N_c}\left(\frac{\partial \Pi_{a_0a_0}(P^2)}{\partial P^2}\right)^{-1}\Big|_{P^2=m_{a_0}^2} 
\end{equation}

\begin{equation}
Z_{K^*_0}=\frac1{N_c}\left(\frac{\partial \Pi_{K^*K^*}(P^2)}{\partial P^2}\right)^{-1}\Big|_{P^2=m_{K^*_0}^2} 
\end{equation}

\begin{equation}
Z_{\sigma}=\frac{1}{N_c}\frac{g_{\sigma}+g_{f_0}-g_{\sigma}g_{f_0}(\Pi_{\sigma\sigma}+\Pi_{f_0 f_0})}{g_{\sigma}(1-g_{f_0}\Pi_{f_0f_0})\frac{\partial \Pi_{\sigma\sigma}}{\partial P^2}+g_{f_0}(1-g_{\sigma}\Pi_{\sigma\sigma})\frac{\partial \Pi_{f_0 f_0}}{\partial P^2}+2g_{f_0} g_{f_0}\Pi_{\sigma f_0}\frac{\partial}{\partial P^2}\Pi_{\sigma f_0}}\Bigg|_{P^2=m_{\sigma}^2}
\end{equation}

\begin{table}[H]
\begin{center}
\begin{tabular}{|c|c|c|c|c|}
   \hline
     &  $\sqrt{Z_{\sigma}}$ & $\sqrt{Z_{a_0}}$ & $\sqrt{Z_{K_0^*}}$ & $\sqrt{Z_{f_0}}$ \\
   \hline
   ILM  & $4.224$ & $-$ & $-$ & $-$ \\
   \hline
\end{tabular}
\end{center}
    \caption{Effective meson-quark coupling in scalar channels with parameters given in Table~\ref{tab:parameters_ILM_Nf3}}
\end{table}

\subsection{Pseudoscalar meson spectrum}

In general, the pseudoscalar channels also mix with the axial-vector channels. For simplicity, we neglect this mixing here and focus on the mixing effects induced by strangeness. In the pseudoscalar meson $0^-$ channel, the Hatree-Fock effective meson interaction is given by
\begin{equation}
\begin{aligned}
    \mathcal{L}_{P}=&\frac{g_{\eta'}}{2N_c}\left(\bar{\psi}i\gamma^5\lambda^0\psi\cos\theta_{\eta\eta'}+\bar{\psi}i\gamma^5\lambda^8\psi\sin\theta_{\eta\eta'}\right)^2\\
    &+\frac{g_\eta}{2N_c}\left(\bar{\psi}i\gamma^5\lambda^8\psi\cos\theta_{\eta\eta'}-\bar{\psi}i\gamma^5\lambda^0\psi\sin\theta_{\eta\eta'}\right)^2\\
    &+\frac{g_{\pi}}{2N_c}(\bar{\psi}i\gamma^5\lambda^3\psi)^2+\frac{g_\pi}{2N_c}(\bar{\psi}i\gamma^5\tau_I^a\psi)^2+\frac{g_{K}}{2N_c}(\bar{\psi}i\gamma^5\tau_U^a\psi)^2+\frac{g_{K}}{2N_c}(\bar{\psi}i\gamma^5\tau_V^a\psi)^2\\
\end{aligned}
\end{equation}
with $\lambda^0\equiv\sqrt{\tfrac23}\mathds{1}$ and the mixing angle determined by \eqref{P_mixing}.

In the flavored channels, the couplings are

\begin{equation}
\label{tHooft_couplings_3}
g_\pi = g_s + 4g_{IA}, \qquad
g_K = g_u + 4g_{IA}
\end{equation}

In the flavor-singlet channels, the singlet and octet components mix, leading to the following effective couplings
\begin{equation}
\begin{aligned}
\label{eq:mix_pseudo}
g_{\eta,\eta'}=\frac12 \left(-g_s+8 g_{IA} \pm \sqrt{8 g_u^2+g_s^2}\right)
\end{aligned}
\end{equation}
In the exact flavor $SU(3)$ limit $g_s=g_u$, the couplings in \eqref{eq:mix_pseudo} are reduced to

\begin{equation}
g_\eta = g_s + 4g_{IA}, \qquad
g_{\eta'} = -2g_s + 4g_{IA} 
\end{equation}

\subsubsection{Mass spectrum}

The mass spectrum of light scalar mesons ($N_f=3$) is obtained via BS resummation of the reduced four-quark interaction in the RPA ($1/N_c$) approximation, as illustrated in Fig.~\ref{fig:bubble}. In this framework, the mass of a pion is determined by

\begin{equation}
\label{bound_eq_1}
    g_{\pi}\Pi_{\pi\pi}(m_{\pi}^2)=1
\end{equation}
where the bubble diagram in $\pi$ channel is defined by 
\begin{equation}
    \Pi_{\pi\pi}=2\Pi^{ud}_{PP}=\Pi^{uu}_{PP}+\Pi^{dd}_{PP}
\end{equation} 
with the quark bubble functions $\Pi^{qq'}_{PP}$ defined in Appendix \ref{app:vac_pol}. The mass of kaon can be determined by
\begin{equation}
\label{bound_eq_K}
    g_{K}\Pi_{KK}(m_{K}^2)=1
\end{equation}
with the quark bubbles given by
\begin{equation}
    \Pi_{KK}=2\Pi^{us}_{PP}=2\Pi^{ds}_{PP}
\end{equation}

For $\eta$ and $\eta'$ meson, the gap-like equations are more involved due to the mixing. The quark bubble functions in $\eta$ and $\eta'$ channels are given by

\begin{equation}
\begin{aligned}
\Pi_{\eta'\eta'}=&\left(\Pi_{PP}^{uu}+\Pi_{PP}^{dd}\right)\left(\frac23\cos^2\theta_{\eta\eta'}+\frac{2\sqrt{2}}{3}\cos\theta_{\eta\eta'} \sin\theta_{\eta\eta'}+\frac13\sin^2\theta_{\eta\eta'}\right)\\
&+\Pi_{PP}^{ss}\left(\frac23\cos^2\theta_{\eta\eta'}-\frac{4\sqrt{2}}{3}\cos\theta_{\eta\eta'} \sin\theta_{\eta\eta'}+\frac43\sin^2\theta_{\eta\eta'}\right)
\end{aligned}
\end{equation}

\begin{equation}
\begin{aligned}
     \Pi_{\eta'\eta}=&\left(\Pi_{PP}^{uu}+\Pi_{PP}^{dd}\right)\left(\frac{1}{\sqrt{3}}\cos^2\theta_{\eta\eta'}-\frac{1}{\sqrt{3}}\sin^2\theta_{\eta\eta'}-\frac23\cos\theta_{\eta\eta'} \sin\theta_{\eta\eta'}\right)\\
     &-\Pi_{PP}^{ss} \left(\frac2{\sqrt{3}}\cos^2\theta_{\eta\eta'}-\frac2{\sqrt{3}}\sin^2\theta_{\eta\eta'}-\frac13\cos\theta_{\eta\eta'} \sin\theta_{\eta\eta'}\right)
\end{aligned}
\end{equation}

\begin{equation}
\begin{aligned}
\Pi_{\eta\eta}=&\left(\Pi_{PP}^{uu}+\Pi_{PP}^{dd}\right)\left(\frac23\sin^2\theta_{\eta\eta'}-\frac{2\sqrt{2}}{3}\cos\theta_{\eta\eta'} \sin\theta_{\eta\eta'}+\frac13\cos^2\theta_{\eta\eta'}\right)\\
&+\Pi_{PP}^{ss}\left(\frac23\sin^2\theta_{\eta\eta'}+\frac{4\sqrt{2}}{3}\cos\theta_{\eta\eta'} \sin\theta_{\eta\eta'}+\frac43\cos^2\theta_{\eta\eta'}\right)
\end{aligned}
\end{equation}
where the tree-level mixing angle $\theta_{\eta \eta'}$ are determined by 
\begin{align}
\label{P_mixing}
\tan2\theta_{\eta\eta'}=\frac{2\sqrt{2}(g_s-g_u)}{g_s+8g_u}\propto m_s^*-m_u^*
\end{align}
RPA resummation induces a modest shift of this angle and enhances the mixing between the $\eta$ and $\eta'$ channels, where the shift of the mixing angle is given by

\begin{equation}
\begin{aligned}
\tan\delta\theta_{\eta\eta'}=&\frac{g_{\eta'}\Pi_{\eta'\eta}(m^2_{\eta})}{1-g_{\eta'}\Pi_{\eta'\eta'}(m^2_{\eta})}=\frac{1-g_{\eta}\Pi_{\eta\eta}(m^2_{\eta})}{g_{\eta}\Pi_{\eta'\eta}(m^2_{\eta})}
\end{aligned}
\end{equation}
With this in mind, the $\eta$ bound-state mass is determined by
\begin{equation}
g_\eta\Pi_{\eta\eta}(m^2_{\eta})
+g_{\eta'}\Pi_{\eta'\eta'}(m^2_{\eta})
-g_\eta g_{\eta'}
\left[
\Pi_{\eta\eta}(m^2_{\eta})\Pi_{\eta'\eta'}(m^2_{\eta})
-\Pi^2_{\eta\eta'}(m^2_{\eta})
\right]
=1 ,
\end{equation}
with the physical solution selected by the condition
\begin{equation}
    g_\eta + g_{\eta'}-g_\eta g_{\eta'}\left[\Pi_{\eta\eta}(m^2_{\eta})+\Pi_{\eta'\eta'}(m^2_{\eta})\right]>0,
\end{equation}
which identifies the $\eta$ state, while the opposite condition corresponds to the orthogonal combination associated with the $\eta'$ channel.

With the values of $G_I$ and $G_{IA}$ fixed in Table~\ref{tab:parameters_ILM_Nf3}, the couplings in each channel are given by

\begin{center}
\begin{align*}
&g_\pi=145.18~ \mathrm{GeV}^{-2} &&g_{K}=120.69~ \mathrm{GeV}^{-2} \\
&g_{\eta}=113.75~ \mathrm{GeV}^{-2} && g_{\eta'}=0.396~ \mathrm{GeV}^{-2}  
\end{align*}
\end{center}

Compared with the two-flavor couplings in Table~\ref{tab:g}, the inclusion of the strange quark drastically splits the coupling strengths in the $\eta$ and $\eta'$ channels. Consequently, the $\eta$ mass is further reduced by about 100 MeV (see Table~\ref{tab:p_meson_spec}), while the $\eta'$ channel is no longer bound, indicating that $\eta$–$\eta'$ mixing is driven by the presence of strangeness. With the chosen parameters, the $\eta'$ channel is slightly attractive, although this behavior is sensitive to the values of input parameters and may become weakly repulsive for different choices. In general, the $\eta'$ does not couple to the quark zero modes, indicating that its mass is not primarily generated by spontaneous chiral symmetry breaking, but is instead driven by the axial anomaly.

\begin{table}
    \centering
    \begin{tabular}{|c|c|c|c|c|c|c|}
    \hline
    Model & $m_{\eta}$ (MeV) & $m_{\eta'}$ (MeV) &$m_{\pi^0}$ (MeV)   \\
    \hline
     ILM & $537.2$ & $-$ & $139.4$  \\
     PDG \cite{ParticleDataGroup:2014cgo,ParticleDataGroup:2016lqr,ParticleDataGroup:2018ovx,ParticleDataGroup:2024cfk} & $547.862(17)$ & $782.65(12)$ & $134.9766(6)$  \\
     \hline
    \end{tabular}
    \hfill
    \begin{tabular}{|c|c|c|c|c|c|c|}
    \hline
    Model &$m_{\pi^\pm}$ (MeV)   & $m_{K^\pm}$ (MeV) & $m_{K^0,\bar K^0}$ (MeV)   \\
    \hline
     ILM & $139.4$ & $474.0$ & $474.0$  \\
     PDG \cite{ParticleDataGroup:2014cgo,ParticleDataGroup:2016lqr,ParticleDataGroup:2018ovx,ParticleDataGroup:2024cfk} & $139.57018(35)$ & $493.677(13)$ & $497.611(13)$  \\
     \hline
    \end{tabular}
    \caption{Mass spectrum of pseudoscalar mesons ($N_f=3$), where `` $–$ '' indicates that the state lies above lowest threshold $2M_u$ and is not stably bound.}
    \label{tab:p_meson_spec}
\end{table}

\begin{table}
    \centering
\begin{tabular}{|c|c|}
    \hline
    Model & $~\theta_{P}~$  \\
    \hline
     ILM & $ 5.887^\circ $   \\
     Experimental analysis \cite{Bramon:1997va} & $-15.5^\circ \pm 1.3^\circ$   \\
    Experimental analysis \cite{Bramon:2000fr} & $-17.0^\circ\pm2.4^\circ$ \\
     \hline
\end{tabular}   
\caption{Mixing angle of $\eta$ and $\eta'$ meson in ILM with parameters given in Table~\ref{tab:parameters_ILM_Nf3}, compared to the experimental analysis.}
    \label{tab:p_meson_mixing}
\end{table}

The pseudoscalar mixing angle in our model basis $(\sqrt{\tfrac23}\mathds{1},\lambda^8)$ is given by $\theta_{\eta \eta'}+\delta\theta_{\eta \eta'}=5.887^\circ$ in total, with the Lagrangian-induced angle in \eqref{S_mixing} $\theta_{\eta \eta'}=5.837^\circ$, and the loop-level $\delta\theta_{\eta\eta'}=0.05014^\circ$ after RPA bubble resummation in \eqref{S_mixing_2}. Our model basis can be straightforwardly related to the standard normalized singlet--octet state basis with the mixing angle related by $\theta_P=\theta_{\eta\eta'}+\delta\theta_{\eta\eta'}$.  


The comparison is listed in Table~\ref{tab:p_meson_mixing}. The mixing angle is consistent with the experimental analysis in \cite{Bramon:1997va} showing that the $\eta'$ meson is mostly isosinglet octet and the $\eta$ is predominantly singlet. The review can be found in \cite{DiDonato:2011kr}


In the pseudoscalar channels, the effective pseudoscalar meson-quark couplings are given by

\begin{equation}
  Z_{\pi}=\frac1{N_c}\left(\frac{\partial \Pi_{\pi\pi}(P^2)}{\partial P^2}\right)^{-1}\Big|_{P^2=m_{\pi}^2}  
\end{equation}

\begin{equation}
  Z_{K}=\frac1{N_c}\left(\frac{\partial \Pi_{KK}(P^2)}{\partial P^2}\right)^{-1}\Big|_{P^2=m_{K}^2}  
\end{equation}

\begin{equation}
Z_{\eta}=\frac{1}{N_c}\frac{g_{\eta'}+g_{\eta}-g_{\eta'}g_\eta(\Pi_{\eta'\eta'}+\Pi_{\eta\eta})}{g_{\eta'}(1-g_\eta\Pi_{\eta\eta})\frac{\partial \Pi_{\eta'\eta'}}{\partial P^2}+g_\eta(1-g_{\eta'}\Pi_{\eta'\eta'})\frac{\partial \Pi_{\eta\eta}}{\partial P^2}+2g_{\eta'} g_\eta\Pi_{\eta'\eta}\frac{\partial \Pi_{\eta'\eta}}{\partial P^2}}\Bigg|_{P^2=m_{\eta',\pi}^2}
\end{equation}

\begin{table}[H]
\begin{center}
\begin{tabular}{|c|c|c|c|c|}
   \hline
     &  $\sqrt{Z_{\pi}}$ & $\sqrt{Z_{K}}$ & $\sqrt{Z_{\eta}}$ & $\sqrt{Z_{\eta'}}$ \\
   \hline
   ILM  & $4.939$ & $4.588$ & $4.379$ & $-$ \\
   \hline
\end{tabular}
\end{center}
    \caption{Effective meson-quark couplings in the pseudoscalar channels with parameters given in Table~\ref{tab:parameters_ILM_Nf3}}
\end{table}

\subsection{Vector meson spectrum}
In the vector meson $1^-$ channel, the interaction is of form
\begin{equation}
\begin{aligned}
    \mathcal{L}_{V}=&-\frac{g_\omega}{2N_c}\left(\bar{\psi}\gamma_\mu\lambda^0\psi\right)^2-\frac{g_\phi}{2N_c}(\bar{\psi}\gamma_\mu\lambda^8\psi)^2-\frac{g_\rho}{2N_c}(\bar{\psi}\gamma_\mu\lambda^3\psi)^2\\
&-\frac{g_\rho}{2N_c}(\bar{\psi}\gamma_\mu\tau^a_I\psi)^2-\frac{g_{K^*}}{2N_c}(\bar{\psi}\gamma_\mu\tau_U^a\psi)^2-\frac{g_{K^*}}{2N_c}(\bar{\psi}\gamma_\mu\tau_V^a\psi)^2\\
\end{aligned}
\end{equation}
where the couplings are given by
\begin{equation}
g_{\rho} = g_{K^*} = g_{\phi} = g_{\omega} = g_{IA}
\end{equation}

One should note that the molecule-induced coupling is universal across the vector meson channels. As a consequence, at leading order the $\omega$ and $\phi$ mesons do not mix in the singlet–octet basis. The $\omega$--$\phi$ mixing emerges dynamically through bubble resummation, leading to the decoupling of strangeness from the isospin sector.

\subsubsection{Mass spectrum}

The mass spectrum for the vector light mesons and their effective coupling with quarks can be obtained using the standard RPA resummation in each channl, as in Fig.~\ref{fig:bubble}. For flavored channels, the mass of isospin vector $\rho$ can be determined by

\begin{equation}
\label{bound_eq_rho}
    g_{\rho}\Pi_{\rho\rho}(m_{\rho}^2)=1
\end{equation}
where the vacuum bubbles in $\rho$ channel are defined by
\begin{equation}
    \Pi_{\rho\rho}=2\Pi^{ud}_{VV}=\Pi^{uu}_{VV}+\Pi^{dd}_{VV}
\end{equation}
with $\Pi^{qq'}_{VV}$ defined in Appendix \ref{app:vac_pol}. The mass of kaon can be determined by
\begin{equation}
\label{bound_eq_Ks}
    g_{K^*}\Pi_{K^*K^*}(m_{K^*}^2)=1
\end{equation}
with the vacuum quark exchange in $K^*$ channel defined by
\begin{equation}
    \Pi_{K^*K^*}=2\Pi^{us}_{VV}=2\Pi^{ds}_{VV}
\end{equation}

For the unflavored channel, the quark bubble functions are given by

\begin{equation}
\Pi_{\omega\omega}=\frac23\left(\Pi_{VV}^{uu}+\Pi_{VV}^{dd}+\Pi_{VV}^{ss}\right)
\end{equation}

\begin{equation}
    \Pi_{\omega\phi}=\frac{\sqrt{2}}{3}\left(\Pi_{VV}^{uu}+\Pi_{VV}^{dd}-2\Pi_{VV}^{ss}\right)
\end{equation}

\begin{equation}
\Pi_{\phi\phi}=\frac13\left(\Pi_{VV}^{uu}+\Pi_{VV}^{dd}+4\Pi_{VV}^{ss}\right)
\end{equation}
written in flavor basis $u$, $d$, and $s$.

The RPA resummation will mix the $\omega$ and $\phi$ with the mixing angle determined by

\begin{equation}
\begin{aligned}
\label{V_mixing_2}
\tan\delta\theta_{\omega\phi}=&\frac{g_{\phi}\Pi_{\omega\phi}(m^2_{\omega})}{1-g_{\phi}\Pi_{\phi\phi}(m^2_{\omega})}=\frac{1-g_{\omega}\Pi_{\omega\omega}(m^2_{\omega})}{g_{\omega}\Pi_{\omega\phi}(m^2_{\omega})}
\end{aligned}
\end{equation}  
 
The $\omega$ and $\phi$ meson bound state mass can be determined by

\begin{equation}
    g_\omega\Pi_{\omega\omega}(m^2_{\omega})+g_{\phi}\Pi_{\phi\phi}(m^2_{\omega})-g_\omega g_{\phi}\left[\Pi_{\omega\omega}(m^2_{\omega})\Pi_{\phi\phi}(m^2_{\omega})-\Pi^2_{\omega\phi}(m^2_{\omega})\right]=1
\end{equation}
Using the relation $g_\omega=g_{\phi}$, this equation can be further reduced to
\begin{equation}
\left(1 - 2g_{\phi}\,\Pi^{uu}_{VV}\right)
\left(1 - 2g_{\phi}\,\Pi^{ss}_{VV}\right)=0 .
\end{equation}
This indicates that, once strangeness is included, $\omega$–$\phi$ mixing effectively decouples the strange sector from the isospin sector. The $\omega$ meson remains degenerate with the $\rho$ meson and retains a purely light-quark composition, while the $\phi$ meson is composed predominantly of a pure strange bound state, consistent with the strict OZI limit, $\omega \sim \bar u u + \bar d d$ and $\phi \sim \bar s s$. This suggests that, in the vector channel, molecule-induced interactions do not lead to OZI violation. 

This mixing between the $\omega$ and $\phi$ mesons plays an important role in understanding OZI violation and the explicit breaking of flavor $SU(3)$ symmetry in QCD. In QCD, such mixing arises primarily from the mass difference between light and strange quarks \cite{Kucukarslan:2006wk}. 

With the values of $G_{IA}$ fixed in Table~\ref{tab:parameters_ILM_Nf3}, the couplings in each vector channel are given by

\begin{center}
\begin{align*}
&g_\rho=g_{K^*}=g_{\phi}=g_{\omega}=21.60~ \mathrm{GeV}^{-2} 
\end{align*}
\end{center}

Compared to the two-flavor couplings in Table~\ref{tab:g}, the inclusion of the strange quark does not significantly affect the $\omega$ and $\rho$ vector channels. The couplings remain of the same order. This insensitivity to flavor-symmetry breaking suggests that the molecule-induced interaction is largely flavor blind and generates only weak flavor correlations.

\begin{table}[]
    \centering
    \begin{tabular}{|c|c|c|c|c|}
    \hline
    Model & $m_{\omega}$ (MeV) & $m_{\rho}$ (MeV)  & $m_{\phi}$ (MeV) & $m_{K^{*}}$ (MeV) \\
    \hline
     ILM & $787$ & $787$ & $1093.3$ & $-$  \\
     PDG \cite{ParticleDataGroup:2014cgo,ParticleDataGroup:2016lqr,ParticleDataGroup:2018ovx,ParticleDataGroup:2024cfk} & $782.66(13)$ & $775.26(25)$ & $1019.461(16)$ & $891.67(26)$  \\
    \hline
    \end{tabular}  
    \caption{Mass spectrum of vector meson in ILM with $\rho=0.33$ fm. Due to its decoupling from the isospin sector, the effective lowest mass threshold for $\phi$ becomes $2M_s\approx1183.4$ MeV}
    \label{tab:v_meson_spec}
\end{table}

\begin{table}[]
\centering
\begin{tabular}{|c|c|}
    \hline
    Model & $~\theta_{V}~$  \\
    \hline
     ILM & $35.26^\circ$  \\
     OZI & $35.27^\circ$ \\
     Experimental analysis \cite{Bramon:2000fr} & $38.6^\circ\pm0.2^\circ$  \\
     \hline
\end{tabular}   
    \caption{Mixing angle of $\omega$ and $\phi$ meson in ILM with $\rho=0.33$ fm. This value indicates that the $\phi$ meson is predominantly composed of strange quarks, while the $\omega$ meson remains largely within the isospin sector.}
    \label{tab:v_meson_mixing}
\end{table}

The vector mixing angle in our model basis ($\sqrt{\tfrac23}\mathds{1}$, $\lambda^8$) is given by $\delta\theta_{\omega\phi}=35.26^\circ$ in total, with the contribution from the resummation of the vacuum polarizations in \eqref{V_mixing_2}. This basis can be straightforwardly identified as the standard normalized singlet-octet basis employed in various phenomenological analysis. As a result, the mixing angles are connected to the phenomenological mixing angle $\theta_V=\delta\theta_{\omega\phi}$.  

We can also compute the effective meson-quark couplings (the meson wave function renormalization) by

\begin{equation}
  Z_{\rho}=\frac1{N_c}\left(\frac{\partial \Pi_{\rho\rho}(P^2)}{\partial P^2}\right)^{-1}\Big|_{P^2=m_{\rho}^2}  
\end{equation}

\begin{equation}
  Z_{K^*}=\frac1{N_c}\left(\frac{\partial \Pi_{K^*K^*}(P^2)}{\partial P^2}\right)^{-1}\Big|_{P^2=m_{K^*}^2}  
\end{equation}

\begin{equation}
Z_{\omega,\phi}=\frac{1}{N_c}\frac{g_{\omega}+g_{\phi}-g_{\omega}g_\phi(\Pi_{\omega\omega}+\Pi_{\phi\phi})}{g_{\omega}(1-g_\phi\Pi_{\phi\phi})\frac{\partial \Pi_{\omega\omega}}{\partial P^2}+g_\phi(1-g_{\omega}\Pi_{\omega\omega})\frac{\partial \Pi_{\phi\phi}}{\partial P^2}+2g_{\omega} g_\phi\Pi_{\omega\phi}\frac{\partial \Pi_{\omega\phi}}{\partial P^2}}\Bigg|_{P^2=m_{\omega}^2}
\end{equation}

\begin{table}[H]
\begin{center}
\begin{tabular}{|c|c|c|c|c|}
   \hline
     &  $\sqrt{Z_{\rho}}$ & $\sqrt{Z_{K^*}}$ & $\sqrt{Z_{\omega}}$ & $\sqrt{Z_{\phi}}$ \\
   \hline
   ILM  & $0.655$ & $-$ & $0.655$ & $-$ \\
   \hline
\end{tabular}
\end{center}
    \caption{Effective meson-quark coupling with parameters given in Table~\ref{tab:parameters_ILM_Nf3}}
\end{table}

\subsection{Axial vector meson spectrum}
In the axial vector meson $1^+$ channel, the effective meson interaction is given by
\begin{equation}
\begin{aligned}
    \mathcal{L}_{A}=&-\frac{g_{f_1}}{2N_c}\left(\bar{\psi}\gamma_\mu\lambda^0\gamma^5\psi\right)^2-\frac{g_{f_1'}}{2N_c}(\bar{\psi}\gamma_\mu\gamma^5\lambda^8\psi)^2-\frac{g_{a_1}}{2N_c}(\bar{\psi}\gamma_\mu\gamma^5\lambda^3\psi)^2\\
&-\frac{g_{a_1}}{2N_c}(\bar{\psi}\gamma_\mu\gamma^5\tau^a_I\psi)^2-\frac{g_{K_1}}{2N_c}(\bar{\psi}\gamma_\mu\gamma^5\tau_U^a\psi)^2-\frac{g_{K_1}}{2N_c}(\bar{\psi}\gamma_\mu\gamma^5\tau_V^a\psi)^2\\
\end{aligned}
\end{equation}
where the couplings are given by
\begin{equation}
g_{a_1} = g_{K_1} = g_{f_1'} = g_{IA}, \qquad
g_{f_1} = -5 g_{IA}
\end{equation}

\subsubsection{Mass spectrum}
Resumming the quark bubble diagrams in RPA as in Fig.~\ref{fig:bubble}, the leading $1/N_c$ contributions determine the masses in the axial vector channel. The mass of isospin axial vector $a_1$ is given by 
\begin{equation}
\label{bound_eq_a1}
    g_{a_1}\Pi_{a_1a_1}(m_{a_1}^2)=1
\end{equation}
where the bubble diagrams in $a_1$ channel is defined by
\begin{equation}
    \Pi_{a_1a_1}=2\Pi^{ud}_{AA,(t)}=\Pi^{uu}_{AA,(t)}+\Pi^{dd}_{AA,(t)}
\end{equation}

The mass of strange axial vector $K_1$ can be determined by
\begin{equation}
\label{bound_eq_K1}
    g_{K_1}\Pi_{K_1K_1}(m_{K_1}^2)=1
\end{equation}
with the bubble diagrams in $K_1$ channel defined by
\begin{equation}
    \Pi_{K_1K_1}=2\Pi^{us}_{AA,(t)}=2\Pi^{ds}_{AA,(t)}
\end{equation}

For the unflavored channel, the vacuum bubble diagrams are mixed between the $f_1$ and $f_1'$ channels, given by

\begin{equation}
\Pi_{f_1f_1}=\frac23\left(\Pi_{AA,(t)}^{uu}+\Pi_{AA,(t)}^{dd}+\Pi_{AA,(t)}^{ss}\right)
\end{equation}

\begin{equation}
    \Pi_{f_1f_1'}=\frac{\sqrt{2}}{3}\left(\Pi_{AA,(t)}^{uu}+\Pi_{AA,(t)}^{dd}-2\Pi_{AA,(t)}^{ss}\right)
\end{equation}

\begin{equation}
\Pi_{f_1'f_1'}=\frac13\left(\Pi_{AA,(t)}^{uu}+\Pi_{AA,(t)}^{dd}+4\Pi_{AA,(t)}^{ss}\right)
\end{equation}

Similar to vector channels, at tree level, $f_1$ and $f_1'$ meson does not mix in singlet and octet basis due to the nature of the instanton-anti-instanton-molecule-induced interaction. Beyond tree level, the resummation of the bubble diagrams mixes the $f_1$ and $f_1'$ with a mixing angle determined by

\begin{equation}
\begin{aligned}
\tan\delta\theta_{f_1 f_1'}=&\frac{g_{f_1'}\Pi_{f_1 f_1'}(m^2_{f_1})}{1-g_{f_1'}\Pi_{f_1'f_1'}(m^2_{f_1})}=\frac{1-g_{f_1}\Pi_{f_1f_1}(m^2_{f_1})}{g_{f_1}\Pi_{f_1 f_1'}(m^2_{f_1})}
\end{aligned}
\end{equation}
where the mixing angles are directly related to the phenomenological mixing angle by $\theta_A=\delta\theta_{f_1 f_1'}$.

The $f_1$ and $f_1'$ bound state mass can be determined by  

\begin{equation}
    g_{f_1}\Pi_{f_1f_1}(m^2_{f_1})+g_{f_1'}\Pi_{f_1'f_1'}(m^2_{f_1})-g_{f_1} g_{f_1'}\left[\Pi_{f_1f_1}(m^2_{f_1})\Pi_{f_1'f_1'}(m^2_{f_1})-\Pi^2_{f_1f_1'}(m^2_{f_1})\right]=1
\end{equation}

With the values of $G_I$ and $G_{IA}$ fixed in Table~\ref{tab:parameters_ILM_Nf3}, the couplings in each channel now read

\begin{center}
\begin{align*}
&g_{a_1}=g_{K_1}=g_{f_1'}=21.60~ \mathrm{GeV}^{-2} && g_{f_1}=-107.99~ \mathrm{GeV}^{-2}  \\
\end{align*}
\end{center}

\begin{table}[]
    \centering
    \begin{tabular}{|c|c|c|c|c|}
    \hline
    Model & $m_{a_1}$ (MeV) & $m_{K_1}$ (MeV)  & $m_{f_1}$ (MeV) & $m_{f_1'}$ (MeV) \\
    \hline
     ILM & $-$ & $-$ & $-$ & $-$  \\
     PDG \cite{ParticleDataGroup:2014cgo,ParticleDataGroup:2016lqr,ParticleDataGroup:2018ovx,ParticleDataGroup:2024cfk}  & $1230(40)$ & $1272(7)$ & $1281.9(5)$ & $ 1426.3(9)$   \\
     \hline
\end{tabular}   
    \caption{Mass spectrum of axial vector mesons in ILM with parameters given in Table~\ref{tab:parameters_ILM_Nf3}}
    \label{tab:a_meson_spec}
\end{table}

The $f_1$ channel is strongly repulsive. Compared to the two-flavor couplings in Table~\ref{tab:g}, the inclusion of the strange quark enhances the repulsion in the singlet component of the axial-vector channel. In contrast, the coupling in the isovector $a_1$ channel remains of the same order, indicating that it is comparatively less sensitive to strange-quark effects.

As indicated in Table~\ref{tab:a_meson_spec}, the PDG data show that axial-vector mesons lie above the thresholds $2M_u$ and $2M_s$, implying that their binding is not primarily driven by chiral symmetry breaking.


\section{Diquarks in $N_f=2$}

The spectrum and low-energy dynamics of diquarks are governed by the instanton-induced interaction. In the case of $N_c=2$, diquarks can be color-singlet and thus coincide with baryonic states. An extended $SU(4)$ supersymmetry, Pauli–G\"{u}rsey symmetry emerges in two-color QCD with two light flavors in the chiral limit, unifying massless baryons and mesons within the same multiplet. For physical QCD with $N_c=3$, although diquarks are no longer color-singlet asymptotic states, they remain essential building blocks in both light and heavy–light baryons \cite{Shuryak:2022wtk,Schafer:1993ra}. The form of the diquark bound states is constrained by color, isospin, and Dirac matrices $C\Gamma$ where $C=i\gamma^0\gamma^2$ is the charge conjugation matrix. In the three-color QCD, the color matrices are divided into two irreducible representations for diquarks,
$$\bf3\otimes\bf3=\bf{6}\oplus\bar{\bf{3}}$$ 
where $\bf6$ is color-symmetric and  $\bar{\bf3}$ is color-antisymmetric.
The isospin matrix is given by $\tau^2$ for an isosinglet diquark and by $\tau^2\tau^a$ for an isovector diquark. Due to the overall antisymmetric characteristic regarding exchanging two quarks, the diquark channels associated with $\Gamma=1,\gamma^5,\gamma_\mu\gamma^5$ are constrained to be isoscalar diquarks whereas $\Gamma=\gamma_\mu, \sigma_{\mu\nu}$ gives isovector diquarks. Note that the parity of the Dirac matrix $C\Gamma$ is opposite to that of the matrix $\Gamma$ as the internal parity of a $qq$ pair is positive while the one of the $\bar qq$ system is negative.

\subsection{Hartree-Fock construction}

The effective diquark interaction can be constructed by transposing the ’t Hooft vertices derived in Eq.~\eqref{THOOFT1} and \eqref{THOOFT2} \cite{Liu:2025ldh,Liu:2023fpj,Schafer:1996wv}, followed by a Fierz transformation (see Appendix~\ref{App:Fierz}). The resulting effective vertices, restricted to direct fermionic bubble diagrams, fully reproduce the contributions obtained from the original instanton-induced 't Hooft Lagrangian. The leading contribution comes from single instanton, corresponding to the effective Lagrangian given by


\begin{equation}
\begin{aligned}
\label{eq:diq}
    \mathcal L^{(\mathrm{diq})}_{I}
    =\frac{G_I}{8N_c^2}&\bigg\{(\bar{\psi}\tau^2\beta_{A}^\alpha i\gamma^5C\bar{\psi}^T)(\psi^TC\tau^2\beta_{A}^\alpha i\gamma^5\psi)\\
    &-\frac{1}{N_c-1}(\bar{\psi}\tau^2\beta_{A}^\alpha C\bar{\psi}^T)(\psi^TC\tau^2\beta_{A}^\alpha\psi)\\
    &-\frac{1}{2(N_c+1)}(\bar{\psi}\tau^2\beta^\alpha_{S}\sigma_{\mu\nu}C\bar{\psi}^T)(\psi^TC\tau^2\beta_{S}^\alpha\sigma_{\mu\nu}\psi)\bigg\}\\
\end{aligned}
\end{equation}
where the effective coupling for single instanton interaction is defined in Eq.~\eqref{eq:GI} and fixed by pion mass (see Sec.~\ref{sec:mes}). The color matrices $\beta^a_{S,A}$ correspond to the two irreducible representations in $SU(3)$ for diquark bound states. In color-antisymmetric representation $\bar{\bf3}$, the three generators are defined by $(\beta_A^\alpha)_{\beta\gamma}=-i\sqrt{N_c/2}\ \epsilon_{\alpha\beta\gamma}$, with the normalization $\mathrm{tr}(\beta^\alpha\beta^\beta)=N_c\delta^{\alpha\beta}$ for $1/N_c$ counting, associated with the antisymmetric Gell-Mann matrices

\begin{align}
    \beta^3_{A}=\sqrt{\frac{N_c}2}\lambda^2, \qquad 
    \beta^2_{A}=-\sqrt{\frac{N_c}2}\lambda^5, \qquad \beta^1_{A}=\sqrt{\frac{N_c}2}\lambda^7
\end{align}
In the color-symmetric representation $\bf{6}$, the six generators $\beta^\alpha_{S}$ are defined by $\beta^1_S=\sqrt{N_c}\delta_{1\alpha}\delta_{1\beta}$, $\beta^2_S=\sqrt{N_c}\delta_{2\alpha}\delta_{2\beta}$, $\beta^3_S=\sqrt{N_c}\delta_{3\alpha}\delta_{3\beta}$ for the diagonal color matrices and $\beta^{4}_{S}=\sqrt{N_c/2}\left(\delta_{1\alpha}\delta_{2\beta}+\delta_{2\alpha}\delta_{1\beta}\right)$, $\beta^{5}_{S}=\sqrt{N_c/2}\left(\delta_{1\alpha}\delta_{3\beta}+\delta_{3\alpha}\delta_{1\beta}\right)$, $\beta^{6}_{S}=\sqrt{N_c/2}\left(\delta_{2\alpha}\delta_{3\beta}+\delta_{3\alpha}\delta_{2\beta}\right)$ for the non-diagonal color matrices. The former three are the linear combinations of the diagonal Gell-Mann matrices $\lambda^{8}$, $\lambda^{8}$ and the color identity matrix $\mathds{1}_c$ while the latter three correspond to non-diagonal symmetric Gell-Mann matrices $\lambda^{1}$, $\lambda^{4}$, $\lambda^{6}$.

\begin{align}
    \beta^1_S&=\sqrt{N_c}\left(\frac{1}{3}\mathds{1}_c+\frac{\lambda^3}{2}+\frac{\sqrt{3}}{6}\lambda^8\right) & \beta^{4}_{S}&=\sqrt{\frac{N_c}{2}}\lambda^{1} \nonumber\\
    \beta^2_S&=\sqrt{N_c}\left(\frac{1}{3}\mathds{1}_c-\frac{\lambda^3}{2}+\frac{\sqrt{3}}{6}\lambda^8\right) & \beta^{5}_{S}&=\sqrt{\frac{N_c}{2}}\lambda^{4} \nonumber\\
    \beta^3_S&=\sqrt{N_c}\left(\frac{1}{3}\mathds{1}_c-\frac{\sqrt{3}}{3}\lambda^8\right) & \beta^{6}_{S}&=\sqrt{\frac{N_c}{2}}\lambda^{6}
\end{align}

The color-antisymmetric $\bar{\bf{3}}$ channel induces a strong attraction for isosinglet scalar diquark $(\psi^T Ci\gamma^5\psi)$ and a strong repulsion for isosinglet pseudoscalar diqaurk $(\psi^T C\psi)$ while color-symmetric $\bf{6}$ channel produces a weakly attraction for isovector tensor diquark.

We note that no vector or axial vector channels are included in \eqref{eq:diq}. The formation of spin-1 diquark bound states therefore requires interactions beyond the single-instanton approximation. Correlated clusters formed by a few nearby instantons and anti-instantons become relevant. The next-to-leading corrections beyond the single-instanton contribution arise from molecular configurations, namely correlated instanton–anti-instanton pairs \cite{Schafer:1994nv}




To obtain spin-1 diquark bound states, the additional effective Lagrangian from the instanton-anti-instanton pair reads

\begin{equation}
\begin{aligned}
\label{eq:diq_2}
    \mathcal L^{(\mathrm{diq})}_{IA}
    =\frac{G_{IA}}{2N_c^2}&\bigg\{-\frac{1}{N_c-1}\bar{\psi}\gamma^\mu\gamma^5\tau^2\beta^\alpha_AC\bar{\psi}^T\psi^TC\gamma_\mu\gamma^5\tau^2\beta^\alpha_A\psi\\
    &-\frac{1}{N_c-1}\bar{\psi}\gamma^\mu\tau^2\tau^a\beta^\alpha_A C\bar{\psi}^T\psi^TC\gamma_\mu\tau^a\tau^2\beta^\alpha_A\psi \\
       &+\frac{4}{N_c-1}\left(\bar{\psi}i\gamma^5\tau^2\beta^\alpha_AC\bar{\psi}^T\psi^TCi\gamma^5\tau^2\beta^\alpha_A\psi+\bar{\psi}\tau^2\beta^\alpha_AC\bar{\psi}^T\psi^TC\tau^2\beta^\alpha_A\psi\right)\\
       &+\frac{3}{N_c+1}\bar{\psi}\gamma^\mu\tau^2\beta^\alpha_S C\bar{\psi}^T\psi^TC\gamma_\mu\tau^2\beta^\alpha_S\psi\\
       &+\frac{3}{N_c+1}\bar{\psi}\gamma^\mu\gamma^5\tau^2\tau^a\beta^\alpha_SC\bar{\psi}^T\psi^TC\gamma_\mu\gamma^5\tau^a\tau^2\beta^\alpha_S\psi\bigg\}\\
\end{aligned}
\end{equation}
The effective molecule-induced coupling is defined in \eqref{MOLX}. The pair-induced coupling in (\ref{MOLX}) is understood as the failed tunneling rate for a molecular configuration, whereby each of quark lines hopping inside the pair is removed by each division of $T_{IA}$, the overlap of the zero mode between instanton and anti-instanton.

The instanton-anti-instanton molecules in color-antisymmetric $\bar{\bf{3}}$ channel attracts for both isosinglet scalar and pseudoscalar diqaurk, particularly turning the pseudoscalar channel from repulsive to attractive, thus demonstrating the importance of molecular configurations. The vector diquark and axial vector diquark channels are also weakly attracted in color-symmetric $\bar{\bf3}$ channel while in color-symmetric $\bf{6}$ channel, vector and axial vector diquark are weakly repulsive.

Those quark fields in the emergent vertices in \eqref{eq:diq} and \eqref{eq:diq_2} are modified by finite size effects of the pseudoparticles (see Sec.~\ref{sec:mes}). Now the resulting full Lagrangian including single instanton and instanton molecules reads,

\begin{equation}
\begin{aligned}
\mathcal{L}=&\bar{\psi}(i\slashed{\partial}-M)\psi+\frac{g_{\bar3_S}}{2N_c}\bar{\psi}i\gamma^5\tau^2\lambda_{A}^\alpha C\bar{\psi}^T\psi^TCi\gamma^5\tau^2\lambda_{A}^\alpha\psi+\frac{g_{\bar3_P}}{2N_c}\bar{\psi}\tau^2\lambda_{A}^\alpha C\bar{\psi}^T\psi^TC\tau^2\lambda_{A}^\alpha\psi\\
 &-\frac{g_{\bar3_V}}{2N_c}\bar{\psi}\gamma^\mu\gamma^5\tau^2\lambda^\alpha_AC\bar{\psi}^T\psi^TC\gamma_\mu\gamma^5\tau^2\lambda^\alpha_A\psi-\frac{g_{\bar3_A}}{2N_c}\bar{\psi}\gamma^\mu\tau^2\tau^a\lambda^\alpha_A C\bar{\psi}^T\psi^TC\gamma_\mu\tau^2\tau^a\lambda^\alpha_A\psi \\
 &-\frac{g_{6_V}}{2N_c}\bar{\psi}\gamma^\mu\gamma^5\tau^2\tau^a\lambda^\alpha_SC\bar{\psi}^T\psi^TC\gamma_\mu\gamma^5\tau^2\tau^a\lambda^\alpha_S\psi-\frac{g_{6_A}}{2N_c}\bar{\psi}\gamma^\mu\tau^2\lambda^\alpha_S C\bar{\psi}^T\psi^TC\gamma_\mu\tau^2\lambda^\alpha_S\psi\\
 &-\frac{g_{6_T}}{2N_c}\bar{\psi}\tau^2\lambda^\alpha_{S}\sigma_{\mu\nu}C\bar{\psi^T}\psi^TC\tau^2\lambda_{S}^\alpha\sigma_{\mu\nu}\psi
\end{aligned}
\end{equation}
where the coupling constants of each channel are fixed by two parameters $G_{I}$ and $G_{IA}$ from the instanton vacuum.

\begin{align}
\label{diqaurk_tHooft_couplings}
&g_{\bar3_S}=\left(\frac1{N_c-1}\right)\left(\frac{G_{I}}{4N_c}+\frac{4G_{IA}}{N_c}\right) && g_{\bar3_P}=\left(\frac1{N_c-1}\right)\left(-\frac{G_{I}}{4N_c}+\frac{4G_{IA}}{N_c}\right) \nonumber\\
&g_{\bar3_V}=\left(\frac1{N_c-1}\right)\frac{G_{IA}}{N_c} \nonumber && g_{\bar3_A}=\left(\frac1{N_c-1}\right)\frac{G_{IA}}{N_c} \nonumber\\
&g_{6_V}=-\left(\frac1{N_c+1}\right)\frac{3G_{IA}}{N_c} && g_{6_A}=-\left(\frac1{N_c+1}\right)\frac{3G_{IA}}{N_c} \nonumber\\
&g_{6_T}=\left(\frac1{N_c+1}\right)\frac{G_{I}}{8N_c}
\end{align}

With given constituent mass $M=395.17$ MeV~\cite{Liu:2023fpj} and phenomenological value of instanton size $\rho=0.33$ fm~\cite{Schafer:1996wv}, the coupling strength $G_I$ and $G_{IA}$ is fixed by physical pion mass ($m_\pi=139.3$ MeV) and rho mass ($m_\rho=780$ MeV). The values of those three parameters are presented in Table~\ref{tab:parameters_ILM_diq}. 

The couplings in each diquark channel can also be determined from the values of $G_I$ and $G_{IA}$ given in Table~\ref{tab:parameters_ILM_diq}

\begin{center}
\begin{align*}
&g_{\bar3_S}=74.41~ \mathrm{GeV}^{-2}  \\
&g_{\bar3_V}=g_{\bar3_A}=12.40~ \mathrm{GeV}^{-2} \\
&g_{\bar3_P}=24.75~ \mathrm{GeV}^{-2}  \\
\end{align*}
\end{center}

\begin{table}
    \centering
    \begin{tabular}{|c|c|c|c|c|}
   \hline
    & $G_I$  & $G_{IA}$ & $M$   \\ 
    \hline
  ILM & 596.02~$\mathrm{GeV}^{-2}$ & 74.37 $\mathrm{GeV}^{-2}$ &
  $395.17$ MeV  \\
   \hline
\end{tabular}
    \caption{Parameters in instanton liquid model with $\rho=0.33$ fm with physical pion mass ($m_\pi=139.3$ MeV) and rho mass ($m_\rho=780$ MeV)}
    \label{tab:parameters_ILM_diq}
\end{table}

For completeness, we also fix the quark mass $m$ and quark condensate $\langle\bar\psi\psi\rangle=\langle\bar u u\rangle=\langle\bar d d\rangle$ by the gap equation \eqref{eq:gap} induced by chiral symmetry breaking~\cite{Liu:2025ldh}. The values are compared to the calculation from Flavour Lattice Averaging Group (FLAG)~\cite{FlavourLatticeAveragingGroupFLAG:2021npn}


The correlation between instanton and anti-instantons can be parameterized by dimensionless $\langle(\rho T_{IA})^{2N_f-2}\rangle$ and measured by the strength of the induced molecular coupling $G_{IA}$ to the single coupling $G_I$ using Eq.~\eqref{eq:IA_corr}. With the values in Table~\ref{tab:parameters_ILM_diq}, we have $\langle(\rho T_{IA})^{2}\rangle=60.34$.   

For general $N_c$, the diqaurk couplings are down by the additional universal color factor $1/(N_c - 1)$ relative to the couplings of the corresponding meson partner \cite{Rapp:1999qa,Shuryak:2022wtk}. In the case of two-color QCD, the color factor gives a unity, implying additional Pauli-G$\ddot{\mathrm{u}}$rsey symmetry that mixes quarks with antiquarks. Therefore, diquarks (the baryons of the two-color QCD) are degenerate with the corresponding mesons, forming a supersymmetric pair. Since our focus is on the diquark states relevant for nucleon structure, we restrict the follwoing discussion to the color–antisymmetric $\bar{\mathbf{3}}$ channel.  

\subsection{Diquark spectrum}

\begin{figure}
    \centering
    \includegraphics[width=\linewidth]{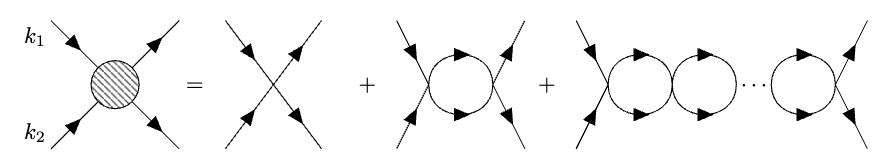}
    \caption{The Feynman diagrams for the quark bubble resummation in the diquark channels.}
    \label{fig:diquark_bubble}
\end{figure}

The diquark correlation functions in each channel can be calculated by suming the leading $1/N_c$ $s$-channel bubble diagrams presented in Fig.~\ref{fig:diquark_bubble}. This procedure is also known as the leading $1/N_c$ order RPA, or Hartree-Fock approximation. The loop integrals for each bubble diagrams are presented in Appendix~\ref{app:vac_pol}. 

In the scalar diquark channel, the attraction is strong enough to generate a diquark bound state, implying a mass pole below the two–quark threshold. The corresponding mass can be obtained from the pole condition of the resummed quark–quark correlation function, leading to a gap-like equation for the diquark mass. Since the isosinglet scalar diquark channel mixes with the longitudinal isosinglet vector diquark, by taking the mixing effect into account, the scalar diquark mass can be determined by

\begin{equation}
\label{bound_eq_diS}
    g_{\bar3_S}\Pi_{PP}(m^2_{qq_S})+g_{\bar3_V}\Pi^{(l)}_{AA}(m^2_{qq_S})-g_{\bar3_S}g_{\bar3_V}\left[\Pi_{PP}(m^2_{qq_S})\Pi^{(l)}_{AA}(m^2_{qq_S})+\Pi^2_{PA}(m^2_{qq_S})\right]=1
\end{equation}

While at leading order, the single-instanton contribution leaves no interaction in the vector and axial-vector diquark channels and induces strong repulsion in pseudoscalar channel, the situation changes once molecular configurations are taken into account. The instanton–anti-instanton molecules induce overall attraction to each channel and modify the diquark dynamics. Thus, for the pseudoscalar, vector and axial diquark, similar gap-like equation can be derived from the bubble summation.
\begin{equation}
\label{bound_eq_diqP}
    g_{\bar3_P}\Pi_{SS}(m_{qq_P}^2)=1
\end{equation}

\begin{equation}
\label{bound_eq_diqV}
    g_{\bar3_V}\Pi^{(t)}_{AA}(m_{qq_V}^2)=1
\end{equation}

\begin{equation}
\label{bound_eq_diqA}
    g_{\bar3_A}\Pi_{VV}(m_{qq_A}^2)=1
\end{equation}

However, these equations do not have solutions below the threshold $2M$ for the present coupling fixed by physical pion and rho meson mass. The interaction in pseudoscalar, vector, and axial channels remains too weak to produce a diquark bound state below the two–quark threshold. Instead, the corresponding diquark scattering amplitude exhibit a resonance in the continuum region above threshold. We therefore determine the resonance by performing a spectral fit in space-like region, using spectral density $\rho_{qq}(s)$ with phenomenological two-body continuum spectral density and an additional single pole structure.

%
\begin{equation}
\begin{aligned}
\label{eq:spec3}
    \frac{2g_{\bar3}/N_c}{1-g_{\bar3}\Pi(-Q^2)}=&\frac{2g_{\bar3}/N_c}{1-g_{\bar3}\Pi(0)}-\int_{0}^\infty ds\frac{Q^2\rho_{qq}(s)}{(s+Q^2)s}
\end{aligned}
\end{equation}
where the spectral density $\rho_{qq}(s)$ is parameterized by

\begin{equation}
    \rho_{qq}(s)=\lambda^2_{qq}\delta(s-m_{qq}^2)-\gamma_{qq}\sqrt{1-\frac{4M^2}s}\theta(s-4M^2)
\end{equation}
with $m_{qq}$ and $\lambda_{qq}$ the diquark mass and effective quark-diquark coupling strength in the corresponding diquark channel and $\gamma_{qq}$ the phenomenological constant that denotes the coupling to the continuum states. This parameterization can determine the resonance mass and diquark coupling quite robustly. The corresponding decay width $\Gamma_{qq}$ can be included by smearing the resonance pole structure. 

\begin{equation}
\label{eq:decay}
    \delta(s-m_{qq}^2)\rightarrow\frac{1}{\pi}\frac{m_{qq}\Gamma_{qq}}{(s-m_{qq}^2)^2+m_{qq}^2\Gamma^2_{qq}}
\end{equation}

\begin{table*}
    \centering
\begin{tabular}{|c|c|c|c|c|c|c|c|}
    \hline
    Model & $m_{qq_S}$ (MeV) & $m_{qq_P}$ (MeV) &$m_{qq_V}$ (MeV) & $m_{qq_A}$ (MeV) \\
    \hline
     ILM & $575.1$ & $891.7$ & $807.0$ & $818.4$ \\
     RL-DSE \cite{Roberts:2011cf} & $780$ & $1060$ & $930$ & $1160$ \\
      Lattice\cite{Bi:2015ifa} & $690(47)$ & $-$ & $-$ & $990(60)$ \\
      ILM \cite{Schafer:1993ra} & $420 \pm 30$ & $-$ & $940 \pm 20$ & $940 \pm 20$  \\ 
     \hline
\end{tabular}   
    \caption{The mass spectrum on color antisymmetric diquark bound states. The result is compared to the calculation in \cite{Roberts:2011cf} using RL-truncated DSE and lattice calculation \cite{Bi:2015ifa} using Laudau gauge  with a slight larger constituent quark mass $M_q(\mathrm{lattice})=427(25)$ MeV compared to our values in Table~\ref{tab:parameters_ILM}.}
    \label{tab:diquark_m}
\end{table*}

With the parameters given in Table~\ref{tab:parameters_ILM_diq} together with \eqref{bound_eq_diS} and \eqref{eq:spec3}, the diquark mass spectrum are shown in Table~\ref{tab:diq-q_couple}. The pseudoscalar $[qq]_{P}$, vector $[qq]_{V}$, and axial vector $[qq]_{A}$ channels are found above the threshold mass, $2M = 790$ MeV. The only channel bounded below the threshold is the scalar $[qq]_{S}$. The diquark masses are compared to the calculation in \cite{Roberts:2011cf} using rainbow ladder (RL) truncated Dyson-Schwinger equation (DSE) and lattice calculation \cite{Bi:2015ifa}. A lighter mass is expected as the instanton vacuum does not fully cover long-range confinement string contribution.

The color factor $1/(N_c - 1)$ in \eqref{diqaurk_tHooft_couplings} implies the real QCD with $N_c = 3$ is half-way between $N_c = 2$ (the “small $N_c$ limit” of QCD) with a relative weight of $1$, and $N_c = \infty$ (the “large $N_c$ limit” of QCD) with relative weight $0$ \cite{Shuryak:2005pk}. Therefore, roughly speaking, the scalar-isoscalar diquarks are half Goldstone bosons with a binding energy of $2M-m_{qq_S}\approx200$ MeV, which is about half of that for pion, $(2M-m_\pi)/2\approx315$ MeV.

We also compute the effective scalar quark-diquark couplings using 





\begin{equation}
\label{eq:sca_di}
\lambda^2_{qq_S}=\frac{\frac2{N_c}\left(g_{\bar3_S}-g_{\bar3_V}+g_{\bar3_S}g_{\bar3_V}(\Pi_{PP}-\Pi^{(l)}_{AA})\right)}{g_{\bar3_S}(1-g_{\bar3_V}\Pi^{(l)}_{AA})\frac{\partial \Pi_{PP}}{\partial P^2}+g_{\bar3_V}(1-g_{\bar3_S}\Pi_{PP})\frac{\partial \Pi^{(l)}_{AA}}{\partial P^2}-2g_{\bar3_S} g_{\bar3_V}\Pi_{PA}\frac{\partial\Pi_{PA}}{\partial P^2}}\Bigg|_{P^2=m_{qq_S}^2}
\end{equation}


As the scalar diquark is bounded below the threshold. For the pseudoscalar, vector, and axial-vector diquark channels, which correspond to states above the two-particle threshold, we employ spectral fits to extract their effective diquark–quark couplings. 

The mixing angle $\theta_{SV}$ between the $[qq]_{S}$ and $[qq]_{V}$ channels can be obtained by the diagonalization of the $[qq]_S$ and $[qq]_V$ bubble diagrams. The scalar-vector mixing angle would be determined by

\begin{equation}
\begin{aligned}
\tan\theta_{SV}=&\frac{g_{\bar3_V}\Pi_{PA}(m^2_{qq_S})}{1-g_{\bar3_V}\Pi^{(l)}_{AA}(m^2_{qq_S})}=\frac{1-g_{\bar3_S}\Pi_{PP}(m^2_{qq_S})}{-g_{\bar3_S}\Pi_{PA}(m^2_{qq_S})}
\end{aligned}
\end{equation}

With the parameters in Table~\ref{tab:parameters_ILM}, the mixing angle between the scalar and vector diqaurk channel is $\theta_{SV}=4.17^\circ$, indicating a small mixing effect from the time-component of vector diquark channel. Thus in the followinf analysis, we will neglect the mixing.

\begin{table}
\begin{center}
\begin{tabular}{|c|c|c|c|c|}
   \hline
     &  $\lambda_{qq_S}$ & $\lambda_{qq_P}$ & $\lambda_{qq_V}$ & $\lambda_{qq_A}$\\ 
   \hline
   ILM & $4.26$ & $1.56$ & $0.104$ & $0.686$ \\
   \hline
\end{tabular}
\end{center}
    \caption{Effective diquark-quark couplings. The couplings in the pseudoscalar, vector, and axial vector diquark channel are determined by spectral fit in \eqref{eq:spec3} while the coupling in scalar diquark channel is determined by \eqref{eq:sca_di}.}
    \label{tab:diq-q_couple}
\end{table}


\section{Baryons in $N_f=2$}
\label{sec:bary}

\begin{figure}
    \centering
    \subfloat[]{\includegraphics[width=0.5\linewidth]{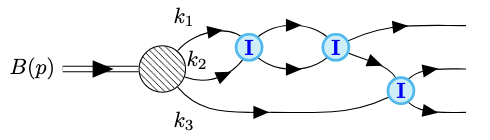}}
    \hfill
    \subfloat[]{\includegraphics[width=0.5\linewidth]{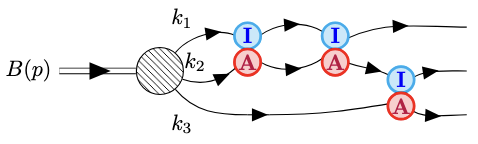}}
    \caption{Baryon bound state in ILM. Nucleon has contribution from both (a) and (b) while $\Delta$ only receives contribution from (b).}
    \label{fig:WF_b}
\end{figure}

\subsection{BSF wave functions}

The covariant BSF nucleon wave function can be defined as

\begin{equation}
\begin{aligned}
\label{NBS}
&(2\pi)^4\delta^4(p-k_1-k_2-k_3)\Psi_{B,ijk}(k_1,k_2,k_3;ps)\frac1{\sqrt{2N_c}}\epsilon_{\alpha\beta\gamma}\\
=&-i\int d^4x_1d^4x_2d^4x_3 e^{ik_1x_1+ik_2\cdot x_2+ik_3\cdot x_3}\frac1{3!}\langle0|\psi_{\alpha i}(x_1)\psi_{\beta j}(x_2)\psi_{\gamma k}(x_3)|B(PS)\rangle    
\end{aligned}
\end{equation}
where $\alpha,\beta, \gamma$ denote the color indices and $i,j,k$ denote the flavor-Dirac indices.

The Pauli principle requires the full Dirac-flavor-color
structure of the BSF wave function to be antisymmetric under exchange of any two quark legs, i.e. under a combined permutation of Dirac, flavor and color indices and the corresponding exchange of momenta. The color wave function in is an antisymmetric singlet, denoted by $\epsilon_{\alpha\beta\gamma}$ in \eqref{NBS}. The transformation
properties of the flavor and Dirac structure follow from Fierz identities. With the constraint of a fully symmetric Dirac-flavor part, one thus obtains the transformation behavior of the Dirac amplitudes. 

As shown in Fig.~\ref{fig:WF_b}, in our case, only two-body interactions are involved in the interaction kernel and thus we can futher reduced the BSF equation into quark-diquark form.

\begin{figure}
    \centering
    \includegraphics[width=1\linewidth]{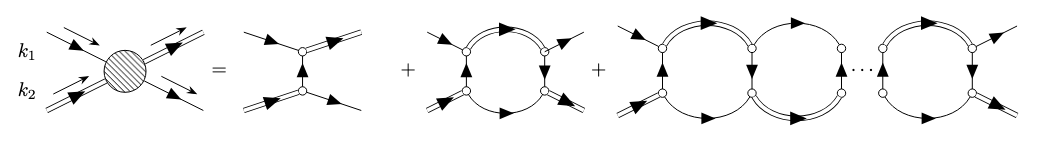}
    \caption{The Feynman diagrams for the quark-diquark bubble resummation in the baryon channels.}
    \label{fig:nucleon_bubble}
\end{figure}

The nucleon structure as a quark-diquark bound state has been discussed in many literatures \cite{Ishii:1995bu,Cloet:2005pp,Oettel:1998bk,Buck:1992wz,Cloet:2014rja,Rezaeian:2003kq,Oettel:2000jj,Hellstern:1997pg,Mineo:2002bg,Mineo:1999eq,Bloch:1999ke}. It is well established that the instanton vacuum induces strong attractive interactions in the diquark channels through the effective Lagrangian in \eqref{eq:diq} and \eqref{eq:diq_2}. As a result, a pair of tightly correlated quarks are dynamically bounded as a diquark. These diquarks subsequently interact with an additional quark via quark exchange processes as shown in Fig.~\ref{fig:nucleon_bubble}, forming a baryon bound state.



Here in this section, we construct the nucleon as a quark-diquark bound state. The color structure can be determined by projection $\mathbf{3}\otimes\bar{\mathbf{3}}=\mathbf{8}\oplus\mathbf{1}$ as the baryon states are color-singlet. For the isospin part, as nucleon is an isospin doublet state with spin-$1/2$, it mixes the scalar diquark and axial diquark whereas the $\Delta$ baryon only consists of axial diquark due to the spin-$3/2$ and isospin quadrplet nature. 

The nucleon $N$ and $\Delta$ masses and their corresponding Bethe-Salpeter wave functions are calculated in the large $N_c$ approximation to the quark exchange between a scalar or axial vector diquark and a constituent quark. 
 
To construct the nucleon as a $3$-quark
bound state, we adopt a simple static approximation \cite{Mineo:2002bg} to the full Faddeev equation in the instanton liquid model, including correlations in the scalar and axial vector diquark channels. The quark exchange matrix for nucleons mixes the scalar diquarks and axial diquarks. 

\begin{equation}
\label{quark_exchange}
\begin{aligned}
    iG_{N}=&(\beta^\alpha_A\beta^{\alpha'}_A)
    \begin{pmatrix}
        i\gamma^5C\tau^2S^T(k)Ci\gamma^5\tau^2 & \gamma^{\mu}C\tau^2\tau^a S^T(k)Ci\gamma^5\tau^2 \\
        i\gamma^5C\tau^2S^T(k)C\gamma^{\mu'}\tau^{a'}\tau^2& \tau^2\tau^a\gamma^{\mu}C S^T(k)C\gamma^{\mu'}\tau^{a'}\tau^2 \\
    \end{pmatrix}\\
    \rightarrow&(\beta^\alpha_A\beta^{\alpha'}_A)
    \begin{pmatrix}
        \tau^2\tau^2 & \tau^2\tau^a\tau^2 \\
        \tau^2\tau^{a'}\tau^2& \tau^2\tau^a\tau^{a'}\tau^2 \\
    \end{pmatrix}\frac{i}{M}
    \begin{pmatrix}
        -1 & i\gamma^{\mu}\gamma^5 \\
        i\gamma^5\gamma^{\mu'}& \gamma^{\mu} \gamma^{\mu'} \\
    \end{pmatrix}
\end{aligned}
\end{equation}

For the $\Delta$, the isospin-$3/2$ and spin-$3/2$ baryon, can only be formed by the isovector and spin-$1$ diquark. Therefore, the quark exchange kernel reads

\begin{equation}
    iG_{\Delta}=(\beta^\alpha_A\beta^{\alpha'}_A)_{\beta'\beta}
     \tau^2\tau^a\gamma^{\mu}c S^T(k)C\gamma^{\mu'}\tau^{a'}\tau^2 \rightarrow(\beta^\alpha_A\beta^{\alpha'}_A)_{\beta'\beta}(\tau^2\tau^a\tau^{a'}\tau^2)\frac{i}{M}\gamma^{\mu} \gamma^{\mu'} 
\end{equation}

As nucleon is color-singlet, the color matrix in the quark exchange kernel in \eqref{quark_exchange} is projected onto a color-singlet subspace and reduced to $-3$  

\begin{equation}
(\beta^\alpha_A\beta^{\alpha'}_A)_{\beta'\beta}\rightarrow-3
\end{equation}

In isospin space, the isoscalar diquark (scalar) only contributes to $I=1/2$ baryon as $\frac12\otimes0=\frac12$ while the isovector diquark (axial vector) contributes to both $I=3/2$ and $I=1/2$ baryon as $\frac12\otimes1=\frac32\oplus\frac12$. Now with this in mind, the quark exchange kernel for nucleon with color and isospin projected yields a 5 by 5 matrix

\begin{equation}
\label{G_N}
    G_N\rightarrow-\frac{3}{M}
    \begin{pmatrix}
        -1 & \sqrt{3}i\gamma^\mu \gamma^5 \\
        \sqrt{3}i\gamma^5\gamma^{\mu'}& -\gamma^\mu\gamma^{\mu'}\\
    \end{pmatrix}
\end{equation}
For $\Delta$ baryon, we have a 4 by 4 matrix
\begin{equation}
\label{G_Delta}
    G^{\mu'\mu}_\Delta\rightarrow-2\times\frac{3}{M}\gamma^\mu \gamma^{\mu'}
\end{equation}

The remaining structure to deal with in the quark exchange kernel is the spin. The quark exchange kernel for nucleon is a tensor product of $5\times5$ mixing matrix between scalar and axial vector diquarks and $4\times4$ Dirac structure. For the $\Delta$-baryon, the quark exchange kernel is a tensor product of $4\times4$ matrix from the axial vector diquark and $4\times4$ Dirac structure. (see details in \cite{Ishii:1995bu} and Appendix~\ref{app:Baryon})

\begin{figure}
    \subfloat[\label{color}]{\includegraphics[width=0.5\linewidth]{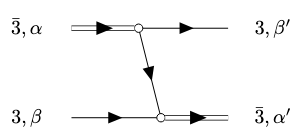}}
\hfill
    \subfloat[\label{isospin}]{\includegraphics[width=0.5\linewidth]{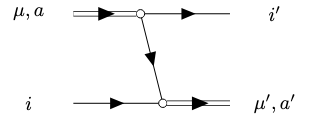}}
\caption{The quark exchange matrix in color space (a) and in flavor-spin space (b)}
\end{figure}

By resuming the all the quark-diquark bubble diagrams in Fig.~\ref{fig:nucleon_bubble}, the BSF resummation gives the effective propagator for nucleon and $\Delta$
\begin{equation}
\label{eq:S_N}
    S_{N,\Delta}(P)=\left[1-G_{N,\Delta}\Pi_{N,\Delta}(P)\right]^{-1}iG_{N,\Delta}
\end{equation}

Each quark-diquark bubble diagram are defined as

\begin{equation}
\label{Pi_N}
    \Pi_N(P)
    =\begin{pmatrix}
        \Pi_{s}(P) & 0 \\
        0 & \Pi^{\mu\nu}_{a}(P)
    \end{pmatrix}
\end{equation}
and
\begin{equation}
\label{Pi_Delta}
    \Pi_\Delta(P)=\Pi^{\mu\nu}_{a}(P)
\end{equation}
where the loop integrals are defined by
\begin{equation}
    \Pi_{s}(P)=i\int \frac{d^4k}{(2\pi)^4}S(k)D_{s}(P-k)\mathcal{F}(k)
\end{equation}
and
\begin{equation}
    \Pi^{\mu\nu}_{a}(P)=i\int \frac{d^4k}{(2\pi)^4}S(k)D^{\mu\nu}_{a}(P-k)\mathcal{F}(k)
\end{equation}
with the scalar diquark propagator defined as

\begin{equation}
\begin{aligned}
    D_{s}(k)=&\frac2{N_c}\frac{i\left[g_{\bar3_S}-g_{\bar3_V}+g_{\bar3_S}g_{\bar3_V}\left(\Pi_{PP}-\Pi^{(l)}_{AA}\right)\right]}{(1-g_{\bar3_S}\Pi_{PP})(1-g_{\bar3_V}\Pi^{(l)}_{AA})+g_{\bar3_S}g_{qq_V}\Pi^2_{PA}}
    \rightarrow \frac{i\lambda^2_{qq_S}}{P^2-m_{qq_S}^2}
\end{aligned}
\end{equation}
and the axial vector diquark propagator defined as
\begin{equation}
\begin{aligned}
    D_{a}^{\mu\nu}(k)=&\frac{2}{N_c}\frac{-ig_{qq_A}}{1-g_{qq_A}\Pi_{VV}}\left(g^{\mu\nu}-g_{qq_A}\Pi_{VV}\frac{k^\mu k^\nu}{k^2}\right)\rightarrow \frac{i\lambda^2_{qq_A}}{P^2-m_{qq_A}^2}
\end{aligned}
\end{equation}

To make the contribution from scalar diquark and axial diquark trackable, we adapt the pole approximation to the diquark propagators with diquark masses and quark-diquark effective couplings also fixed two-body calculation in the last section.

Around the baryon mass pole, Eq.~\eqref{eq:S_N} becomes
\begin{equation}
    S_{N,\Delta}(P)\rightarrow\frac{-iZ_{N,\Delta}}{P^2-m_{N,\Delta}^2}\sum_S\Gamma_{N,\Delta}(P,S)\bar\Gamma_{N,\Delta}(P,S)
\end{equation}
where the nucleon vertex function is
\begin{equation}
\begin{aligned}
\label{vertex_N}
    &\Gamma_N(P,S)=\begin{pmatrix}
        -i\Gamma_s \\
        \Gamma_a^\mu
    \end{pmatrix}u_N(P,S)=
    \begin{pmatrix}
        -i\alpha_1 \\
        \left(\alpha_2-\frac{\alpha_3}{\sqrt{3}}\right) \frac{P^\mu}{m_N}\gamma^5-\frac{\alpha_3}{\sqrt{3}}\gamma^\mu\gamma^5
    \end{pmatrix}u_N(P,S)
\end{aligned}
\end{equation}
and the one for delta baryon is
\begin{equation}
\label{vertex_Delta}
    \Gamma_\Delta(P,S)=u^\mu_\Delta(P,S)
\end{equation}

With proper choice of basis and the positive energy projection, the BSF equation is eventually reduced to corresponding gap-like equations that determine the baryon mass in each channel~\cite{Ishii:1995bu}.
\begin{equation}
\label{baryon_eq}
    \Gamma_{N,\Delta}(P,S)=G_{N,\Delta}\Pi_{N,\Delta}\Gamma_{N,\Delta}(P,S)
\end{equation}
The details for solving the Faddeev equation can be found in Appendix \ref{app:Baryon}. 

The nucleon and $\Delta$ masses are presented in Table~\ref{tab:baryon_spec}. In the absence of axial diquark contributions, the nucleon mass is $929$ MeV, roughly $10$ MeV heavier than the one with the axial diquark  included.

The contribution of the scalar and axial diquark channels can be quantified by the inverse wave function renormalization $Z^{-1}_N=Z_{N_s}^{-1}+Z_{N_a}^{-1}$ of nucleon channel,

\begin{equation}
\begin{aligned}
   Z_{N_s}^{-1}=& -\,\bar{\Gamma}_s(p)\,
\frac{\partial \Pi_{s}(p)}{\partial p^2}\,
\Gamma_s(p)\\
Z_{N_a}^{-1}=& -\,\bar{\Gamma}^\mu_a(p)\,
\frac{\partial \Pi^{\mu\nu}_{a}(p)}{\partial p^2}\,
\Gamma^\nu_a(p)\\
\end{aligned}
\end{equation}

With the coefficients in \eqref{vertex_N} obtained by solving BSF equation, $\alpha_1=-0.292$, $\alpha_2=0.488$, $\alpha_3=0.822 $, the contribution from each diquark reads

\begin{align}
   \frac{Z_{N_s}^{-1}}{Z_N^{-1}}&=97.7\% & \frac{Z_{N_a}^{-1}}{Z_N^{-1}}&=2.3\% 
\end{align}

\begin{table}[]
    \centering
    \begin{tabular}{|c|c|c|}
    \hline
    Model & $m_{N}$ (MeV) & $m_{\Delta}$ (MeV) \\
    \hline
     ILM & $919.47$ & $1349.2$ \\
     PDG \cite{ParticleDataGroup:2014cgo,ParticleDataGroup:2016lqr,ParticleDataGroup:2018ovx,ParticleDataGroup:2024cfk} & $938.27$ & $1232$ \\
     \hline
    \end{tabular}
    \caption{The mass for nucleon and $\Delta$ using the ILM parameters in Table~\ref{tab:parameters_ILM} fixed by physical pion and rho mass}
    \label{tab:baryon_spec}
\end{table}

For completeness, we compare our diquark admixture estimates with the NJL result in~\cite{Mineo:2002bg}. They concluded that the reasonable range for the axial-diquark admixture is 2\%–10\%. \cite{Shuryak:2022wtk} similarly point to a strong scalar-diquark dominance, with a scalar-diquark contribution of roughly 
70\%–90\% in the nucleon.

Now the nucleon and delta baryon BSF wave functions reads

\begin{equation}
\begin{aligned}
\label{eq:BSFWF_N}
&\Psi_N(k_1,k_2,k_3;PS)=\\
&\frac1{\sqrt{6}}\sqrt{Z_{N}}\sqrt{\mathcal{F}(k_1)\mathcal{F}(k_2)\mathcal{F}(k_3)}
       [S(k_1)\gamma^5C\tau^2S^T(k_2)]D(k_1+k_2)S(k_3)\Gamma_su_N\\
      &+\frac1{\sqrt{6}}\sqrt{Z_{N}}\sqrt{\mathcal{F}(k_1)\mathcal{F}(k_2)\mathcal{F}(k_3)}[S(k_1) \gamma^\mu C\tau^2\tau^aS^T(k_2)]D_{\mu\nu}(k_1+k_2)S(k_3)\tau^a\Gamma^\nu_au_N\\
&+ S_3\mathrm{~permutation}~(1\leftrightarrow2\leftrightarrow3)
\end{aligned}
\end{equation}
and
\begin{equation}
\begin{aligned}
\label{eq:BSFWF_D}
   &\Psi_\Delta(k_1,k_2,k_3;PS)=\\
   & \frac1{\sqrt{6}}\sqrt{Z_{\Delta}}\sqrt{\mathcal{F}(k_1)\mathcal{F}(k_2)\mathcal{F}(k_3)} \left[S(k_1)\gamma^\mu C\tau^2\tau^a S^T(k_2)\right]
       D_{\mu\nu}(k_1+k_2)S(k_3)\xi^a_{\frac32}u^{\nu}_{\Delta}\\
    &+ S_3\mathrm{~permutation}~(1\leftrightarrow2\leftrightarrow3)
\end{aligned}
\end{equation}
with $k_1$, $k_2$, $k_3$ the constituent quark momenta and $P$ the total momenta of the baryon and $1/\sqrt{6}$ the symmetrization factor among permutations. $u^{\nu}_{\Delta}$ is the Rita-Schwinger spinor defined in \eqref{eq:RS_spinor} with the flavor projection $\xi^a_{\frac32}$ to isospin-$3/2$ defined by

\begin{equation}
(\xi^a_{\frac32})_{i',i}=\left\langle1 a;\frac12 i'\big|\frac32i\right\rangle 
\end{equation}

The effective quark-diquark spinor parameterization naturally emerges to characterize the scalar and axial diquark correlations.

\subsection{Light Front Wave Functions}

The light front wave function is defined by spanning in the Fock space. The $3$-body baryon light front wave function at leading Fock space reads

\begin{equation}
    \ket{B}\approx\sum_{s_1,s_2,s_3}\int d[123]\frac1{\sqrt{6}}\Phi_B(k_1,k_2,k_3,s_1,s_2,s_3)\frac{\epsilon_{\alpha\beta\gamma}}{\sqrt{2N_c}}b_{s_1,\alpha}^\dagger(k_1) b^\dagger_{s_2,\beta}(k_2)b^\dagger_{s_3,\gamma}(k_3)\ket{0}
\end{equation}
where $\frac1{\sqrt{6}}$ is the symmetrization factor and the $3$-body phase space integral is defined as

\begin{equation}
\begin{aligned}
    &d[123]
    =\prod_{i=1}^3\frac{dx_{i}d^2k_{i\perp}}{(2\pi)^32x_i}(2\pi)^32\delta\left(1-\sum_{i=1}^3x_{i}\right)\delta^2\left(p_\perp-\sum_{i=1}^3k_{i\perp}\right)
\end{aligned}
\end{equation}
The solutions in light front formalism should be equivalent to the Lorentz covariant formalism. In our method we solve $3$-body Faddeev equation by summing over all planar quark bubble diagrams at $1/N_c$ limit. It is easy to show it is equivalent to leading Fock light front wave function by directly integrating the Faddeev wave function over the $k_{1,2,3}^-$ component of the momentum integrals within the physical region $\delta(k_1^-+k_2^-+k_3^-p^-)$. The higher Fock state light front wave function correspond to the diagrams that is sub-leading in $1/N_c$.

\begin{equation}
\begin{aligned}
    &\frac{1}{4x_1x_2x_3}\Phi_{B}(x_1,x_2,x_3,k_{1\perp},k_{2\perp},k_{3\perp},s_1,s_2,s_3)=\\
    &-i(P^+)^2\int_{-\infty}^{\infty
    }\frac{dk_1^-dk_2^-}{(2\pi)^2}\frac{[\bar{u}_{s_1}\gamma^+]_i}{2k_1^+}\frac{[\bar{u}_{s_2}\gamma^+]_j}{2k_2^+}\frac{[\bar{u}_{s_3}\gamma^+]_k}{2k_3^+}\Psi_{B,ijk}(k_1,k_2,k_3;ps)\\
\end{aligned}
\end{equation}
where $k_{1,2,3}^+=x_{1,2,3}P^+$.

Using the light front integral in \cite{Liu:2023fpj}, 
\begin{equation}
\begin{aligned}
    &(iP^+)^2\int_{-\infty}^{\infty
    }\frac{dk_1^-dk_2^-}{(2\pi)^2}\frac{\sqrt{\mathcal{F}(k_1)\mathcal{F}(k_2)\mathcal{F}(k_3)}}{(k_1^2-M^2)(k_2^2-M^2)(k_3^2-M^2)}\frac{1}{(k_1+k_2)^2-m_{qq}^2}\bigg|_{P^2=m_N^2}\\
    \rightarrow&\frac{\theta(x_1x_3)\theta(x_2x_3)}{4x_1x_2x_3}\\
    &\times\frac{\mathcal{F}\left(\frac{k_{1\perp}^2}{\lambda x_1}+\frac{k_{2\perp}^2}{\lambda x_2}+\frac{k_{3\perp}^2}{\lambda x_3}\right)}{\bar{x}_3\left(m_B^2-\frac{k_{1\perp}^2+M^2}{x_1}-\frac{k_{2\perp}^2+M^2}{x_2}-\frac{k_{3\perp}^2+M^2}{x_3}\right)\left(\frac{k_{1\perp}^2+M^2}{x_1}+\frac{k_{2\perp}^2+M^2}{x_2}-\frac{k_{3\perp}^2+m_{qq}^2}{\bar{x}_3}\right)}
\end{aligned}
\end{equation}
where $x_1+x_2+x_3=1$ and $k_{1\perp}+k_{2\perp}+k_{3\perp}=0$. The parameter $\lambda_B$ is of order unity and fixed such that the light-cone wave functions are properly normalized with the normalization condition in both the nucleon and delta channels,

\begin{equation}
    \sum_{s_1s_2s_3}\int d[123] \left|\Phi_B(1,2,3,s_1,s_2,s_3)\right|^2=1
\end{equation}
where $(1,2,3)$ denotes the 3 constituent quark phase space $x_1$, $x_2$, $x_3$, $k_{1\perp}$, $k_{2\perp}$, $k_{3\perp}$.

Therefore, the nucleon and delta baryon light front wavefunction would be

\begin{equation}
\begin{aligned}
\label{eq:LCWF_N}
&\Phi_N=\\
&\frac1{\sqrt{6}}\phi_{N_s}(x_1,x_2,x_3,k_{1\perp},k_{2\perp},k_{3\perp})\left(\bar u_{s_1}\gamma^5C i\tau^2\bar u_{s_2}^T\right)\bar u_{s_3}\Gamma_su_N\\
&+\frac1{\sqrt{6}}\phi_{N_a}(x_1,x_2,x_3,k_{1\perp},k_{2\perp},k_{3\perp})\left(g_{\mu\nu}-\frac{k_{12\mu} k_{12\nu}}{m_{qq_A}^2}\right)\left(\bar u_{s_1}\gamma^\mu Ci\tau^2\tau^a\bar u_{s_2}^T\right)\bar u_{s_3}\tau^a\Gamma_a^\nu u_N\\
&+ S_3\mathrm{~permutation}~(1\leftrightarrow2\leftrightarrow3)
\end{aligned}
\end{equation}

\begin{equation}
\begin{aligned}
\label{eq:LCWF_del}
\Phi_\Delta=&\frac1{\sqrt{6}}\phi_{\Delta}(x_1,x_2,x_3,k_{1\perp},k_{2\perp},k_{3\perp})\left(g_{\mu\nu}-\frac{k_{12\mu} k_{12\nu}}{m_{qq_A}^2}\right)\left(\bar u_{s_1}\gamma^\mu Ci\tau^2\tau^a\bar u_{s_2}^T\right)\bar u_{s_3} u^\nu_\Delta\\
&+ S_3\mathrm{~permutation}~(1\leftrightarrow2\leftrightarrow3)
\end{aligned}
\end{equation}
where $k_{12}=k_1+k_2$ abd $u_{s_{1,2,3}}$ denotes the light front quark $4-$spinor (see Appendix~\ref{Appx:LFspinor}) and the momentum wave function is defined by
\begin{equation}
\begin{aligned}
\label{eq:WFNs}
&\phi_{N_s}(x_1,x_2,x_3,k_{1\perp},k_{2\perp},k_{3\perp})=\\
&\frac{\sqrt{Z_{N}}\mathcal{F}\left(\frac{k_{1\perp}^2}{\lambda_N x_1}+\frac{k_{2\perp}^2}{\lambda_N x_2}+\frac{k_{3\perp}^2}{\lambda_N x_3}\right)}{\bar{x}_3\left(\frac{k_{1\perp}^2+M^2}{x_1}+\frac{k_{2\perp}^2+M^2}{x_2}-\frac{k_{3\perp}^2+m_{qq_S}^2}{\bar{x}_3}\right)\left(m_N^2-\frac{k_{1\perp}^2+M^2}{x_1}-\frac{k_{2\perp}^2+M^2}{x_2}-\frac{k_{3\perp}^2+M^2}{x_3}\right)}
\end{aligned}
\end{equation}
and
\begin{equation}
\begin{aligned}
\label{eq:WFNa}
&\phi_{N_a}(x_1,x_2,x_3,k_{1\perp},k_{2\perp},k_{3\perp})=\\
&\frac{-\sqrt{Z_{N}}\mathcal{F}\left(\frac{k_{1\perp}^2}{\lambda_N x_1}+\frac{k_{2\perp}^2}{\lambda_N x_2}+\frac{k_{3\perp}^2}{\lambda_N x_3}\right)}{\bar{x}_3\left(\frac{k_{1\perp}^2+M^2}{x_1}+\frac{k_{2\perp}^2+M^2}{x_2}-\frac{k_{3\perp}^2+m_{qq_A}^2-im_{qq_A}\Gamma_{qq_A}}{\bar{x}_3}\right)\left(m_N^2-\frac{k_{1\perp}^2+M^2}{x_1}-\frac{k_{2\perp}^2+M^2}{x_2}-\frac{k_{3\perp}^2+M^2}{x_3}\right)}
\end{aligned}
\end{equation}
and
\begin{equation}
\begin{aligned}
\label{eq:WFD}
&\phi_{\Delta}(x_1,x_2,x_3,k_{1\perp},k_{2\perp},k_{3\perp})=\\
&\frac{-\sqrt{Z_{\Delta}}\mathcal{F}\left(\frac{k_{1\perp}^2}{\lambda_\Delta x_1}+\frac{k_{2\perp}^2}{\lambda_\Delta x_2}+\frac{k_{3\perp}^2}{\lambda_\Delta x_3}\right)}{\bar{x}_3\left(\frac{k_{1\perp}^2+M^2}{x_1}+\frac{k_{2\perp}^2+M^2}{x_2}-\frac{k_{3\perp}^2+m_{qq_A}^2-im_{qq_A}\Gamma_{qq_A}}{\bar{x}_3}\right)\left(m_\Delta^2-\frac{k_{1\perp}^2+M^2}{x_1}-\frac{k_{2\perp}^2+M^2}{x_2}-\frac{k_{3\perp}^2+M^2}{x_3}\right)}
\end{aligned}
\end{equation}
with the normalization constant $\sqrt{Z_N}$ and $\sqrt{Z_\Delta}$ related to the wave function renormalization constant in \eqref{eq:BSFWF_N} and \eqref{eq:BSFWF_D}.

The explicit wave functions written in the conventional basis of light front wave functions in \cite{Ji:2003yj} can be found in Appendix~\ref{app:LCWF}.

The calculation demonstrates that in the ILM, the proton and neutron are strongly correlated quark-diquark states, with a  tight scalar-iso-scalar diquark $[ud]_{S}$ and weaker  axial-vector flavor-triplet  diquark $[ud]_{A}$~\cite{Schafer:1996wv}. This strong correlation in the scalar channel follows from the particle-anti-particle symmetry between the spin-$0$ pion and the spin-$0$ diquark.

\begin{table}
\begin{center}
\begin{tabular}{|c|c|c|}
   \hline
     &  $\sqrt{Z_{N}}$ & $\sqrt{Z_{\Delta}}$ \\
   \hline
   ILM & $3.69$ & $3.31$ \\
   \hline
\end{tabular}
\end{center}
    \caption{The wave function renormalization of nucleon and delta baryon wave function where we have normalize the nucleon vertex mixing coefficients $\alpha_{1,2,3}$ to 1 (see Appendix~\ref{app:Baryon}).}
    \label{tab:WFren}
\end{table}

\subsection{SU(6) LFWF}

Since the conventional baryon $SU(6)$ wave functions can be recast in a quark–diquark formulation~\cite{Anselmino:1992vg} (and references therein), we can also construct a baryon $SU(6)$ wave functions by extracting the $SU(6)$ components from the full light-cone wave functions in \eqref{eq:LCWF_N} and \eqref{eq:LCWF_del}. We will further use this to assess of the accuracy of the $SU(6)$ approximation.

The $SU(6)$ nucleon favor-spin wave functions can be obtained by combining the mixed anti-symmetric basis and symmetric basis equally.

\begin{equation}
\begin{aligned}
\label{SU6}
&|p,+1/2\rangle=&\frac 1{\sqrt {18}}
 \bigg(3[ud]_S u_\uparrow + 2[uu]^+_A d_\downarrow 
 -\sqrt 2[uu]^0_A d_\uparrow-\sqrt 2[ud]^+_A u_\downarrow + [ud]^0_A u_\uparrow\bigg)\nonumber\\
&|n,+1/2\rangle=&\frac 1{\sqrt {18}}
 \bigg(3[ud]_S d_\uparrow + 2[dd]^+_A u_\downarrow 
 -\sqrt 2[dd]^0_A u_\uparrow-\sqrt 2[ud]^+_A d_\downarrow + [ud]^0_A d_\uparrow\bigg)
\end{aligned}
\end{equation}
The diquark states appearing above are defined in terms of two-quark combinations with definite spin and isospin. The scalar (isoscalar, spin-0) diquark is antisymmetric in both flavor and spin,
\begin{equation}
[ud]_S \equiv \frac{1}{\sqrt{2}}\left( u^\uparrow d^\downarrow - u^\downarrow d^\uparrow - d^\uparrow u^\downarrow + d^\downarrow u^\uparrow \right),
\end{equation}
which can be written compactly as an antisymmetric combination in flavor and spin.
The axial-vector (isovector, spin-1) diquarks are symmetric in spin and carry isospin components labeled by the superscript $+ , 0 , -$. They are given by
\begin{align}
[uu]^+_A &\equiv u^\uparrow u^\uparrow, \\
[dd]^+_A &\equiv d^\uparrow d^\uparrow, \\
[uu]^0_A &\equiv \frac{1}{\sqrt{2}}\left( u^\uparrow u^\downarrow + u^\downarrow u^\uparrow \right), \\
[dd]^0_A &\equiv \frac{1}{\sqrt{2}}\left( d^\uparrow d^\downarrow + d^\downarrow d^\uparrow \right), \\
[ud]^+_A &\equiv \frac{1}{\sqrt{2}}\left( u^\uparrow d^\uparrow + d^\uparrow u^\uparrow \right), \\
[ud]^0_A &\equiv \frac{1}{2}\left( u^\uparrow d^\downarrow + u^\downarrow d^\uparrow + d^\uparrow u^\downarrow + d^\downarrow u^\uparrow \right), \\
[ud]^-_A &\equiv \frac{1}{\sqrt{2}}\left( u^\downarrow d^\downarrow + d^\downarrow u^\downarrow \right).
\end{align}
These definitions ensure that the scalar diquark is fully antisymmetric, while the axial-vector diquarks form a symmetric spin-triplet and isospin-triplet representation, consistent with the Pauli principle when combined with the antisymmetric color structure.

 Similarly for delta, we can construct the flavor-spin wave functions in quark-diquark form. The $SU(6)$ quark-diquark wave functions for a helcity $\lambda=+1/2$ state thus read
\begin{equation}
    \begin{aligned}
\label{SU6_del}
        & |\Delta^{++},+1/2\rangle=\frac1{\sqrt{3}}\bigg([uu]_A^+ u_\downarrow+\sqrt{2}[uu]_A^0 u_\uparrow\bigg)\\
& |\Delta^{+},+1/2\rangle=\frac{1}{\sqrt{9}}\bigg([uu]_A^+ d_\downarrow+\sqrt{2}[uu]_A^0 d_\uparrow+\sqrt{2}[ud]_A^+ u_\downarrow+2[ud]_A^0 u_\uparrow\bigg)\\
& |\Delta^{0},+1/2\rangle=\frac{1}{\sqrt{9}}\bigg([dd]_A^+ u_\downarrow+\sqrt{2}[dd]_A^0 u_\uparrow+\sqrt{2}[ud]_A^+ d_\downarrow+2[ud]_A^0 d_\uparrow\bigg)\\
& |\Delta^{-},+1/2\rangle=\frac1{\sqrt{3}}\bigg([dd]_A^+ d_\downarrow+\sqrt{2}[dd]_A^0 d_\uparrow\bigg)
    \end{aligned}
\end{equation}

and for helicity $\lambda=+3/2$ state
\begin{equation}
    \begin{aligned}
\label{SU6_1}
        & |\Delta^{++},+3/2\rangle=[uu]_A^+ u_\uparrow
    \end{aligned}
\end{equation}

\begin{equation}
    \begin{aligned}
\label{SU6_2}
& |\Delta^{-},+3/2\rangle=[dd]_A^+ d_\uparrow
    \end{aligned}
\end{equation}

\begin{equation}
    \begin{aligned}
\label{SU6_3}
& |\Delta^{+},+3/2\rangle=\frac{1}{\sqrt{3}}\bigg([uu]_A^+ u_\uparrow+\sqrt{2}[ud]_A^+ u_\uparrow\bigg)
    \end{aligned}
\end{equation}

\begin{equation}
    \begin{aligned}
\label{SU6_4}
& |\Delta^{0},+3/2\rangle=\frac{1}{\sqrt{3}}\bigg([dd]_A^+ u_\uparrow+\sqrt{2}[ud]_A^+ d_\uparrow\bigg)
    \end{aligned}
\end{equation}

These baryon flavor and spin basis can be rewritten as covariant quark-diquark 4-spinor form after Lorentz boost. For the mixed antisymmetric flavor-spin combination, the result reads

\begin{equation}
\begin{aligned}
    &\begin{pmatrix}
   |udu\rangle-|duu\rangle \\
        |udd\rangle-|dud\rangle
    \end{pmatrix}\rightarrow i\tau^2\otimes 1
\end{aligned}
\end{equation}

and
\begin{equation}
    \begin{pmatrix}
   |\uparrow\downarrow\uparrow\rangle-|\downarrow\uparrow\uparrow\rangle \\
        |\uparrow\downarrow\downarrow\rangle-|\downarrow\uparrow\downarrow\rangle
    \end{pmatrix}\rightarrow \frac1{2M^{3/2}}(u_{s_1}^TC\gamma^5u_{s_2})u_{s_3}
\end{equation}

For the mixed symmetric combination, the result reads

\begin{equation}
\begin{aligned}
    &\begin{pmatrix}
   2|uud\rangle-|udu\rangle-|duu\rangle \\
    -2|ddu\rangle+|dud\rangle+|udd\rangle
    \end{pmatrix}\rightarrow i\tau^2\tau^a\otimes \tau^a
\end{aligned}
\end{equation}

and
\begin{equation}
\begin{aligned}
    &\begin{pmatrix}
   2|\uparrow\uparrow\downarrow\rangle-|\uparrow\downarrow\uparrow\rangle-|\downarrow\uparrow\uparrow\rangle \\
        -2|\downarrow\downarrow\uparrow\rangle+|\downarrow\uparrow\downarrow\rangle+|\uparrow\downarrow\downarrow\rangle
    \end{pmatrix}\rightarrow \frac{-1}{2M^{3/2}}(u_{s_1}^TC\gamma^\mu u_{s_2})\left(\gamma_\mu+\frac{P_\mu}{m_N}\right)\gamma^5 u_{s_3}
\end{aligned}
\end{equation}
where the 4-spinor normalization is $\bar{u}_su_s=2M$.

%

%
Using these identities, the 
$SU(6)$ wave functions are constructed from the full light-cone wave functions in Eqs.~\eqref{eq:LCWF_N} and \eqref{eq:LCWF_del}, allowing us to assess the validity of the $SU(6)$ approximation for baryons.
 
\begin{equation}
\begin{aligned}
\Phi_N^{SU(6)}&=\frac1{\sqrt{6}}\phi^{SU(6)}_N(1,2,3)\left(\bar u_{s_1}\gamma^5C i\tau^2\bar u_{s_2}^T\right)\bar u_{s_3}u_N\\
&-\frac1{3\sqrt{6}}\phi^{SU(6)}_N(1,2,3)\left(\bar u_{s_1}\gamma^\mu Ci\tau^2\tau^a\bar u_{s_2}^T\right)\bar u_{s_3}\tau^a\left(\gamma_\mu+\frac{P_\mu}{m_N}\right)\gamma^5 u_N\\
&+S_3~\mathrm{permutation}
\end{aligned}
\end{equation}
where \begin{equation}
\phi^{SU(6)}_N(1,2,3)=\frac12\left[\alpha_1\phi_s(1,2,3)+\sqrt{3}\alpha_3\phi_a(1,2,3)\right]
\end{equation}



\chapter{Form factors}
\label{ch:FF}

\begin{figure}
    \centering
    \includegraphics[width=0.7\linewidth]{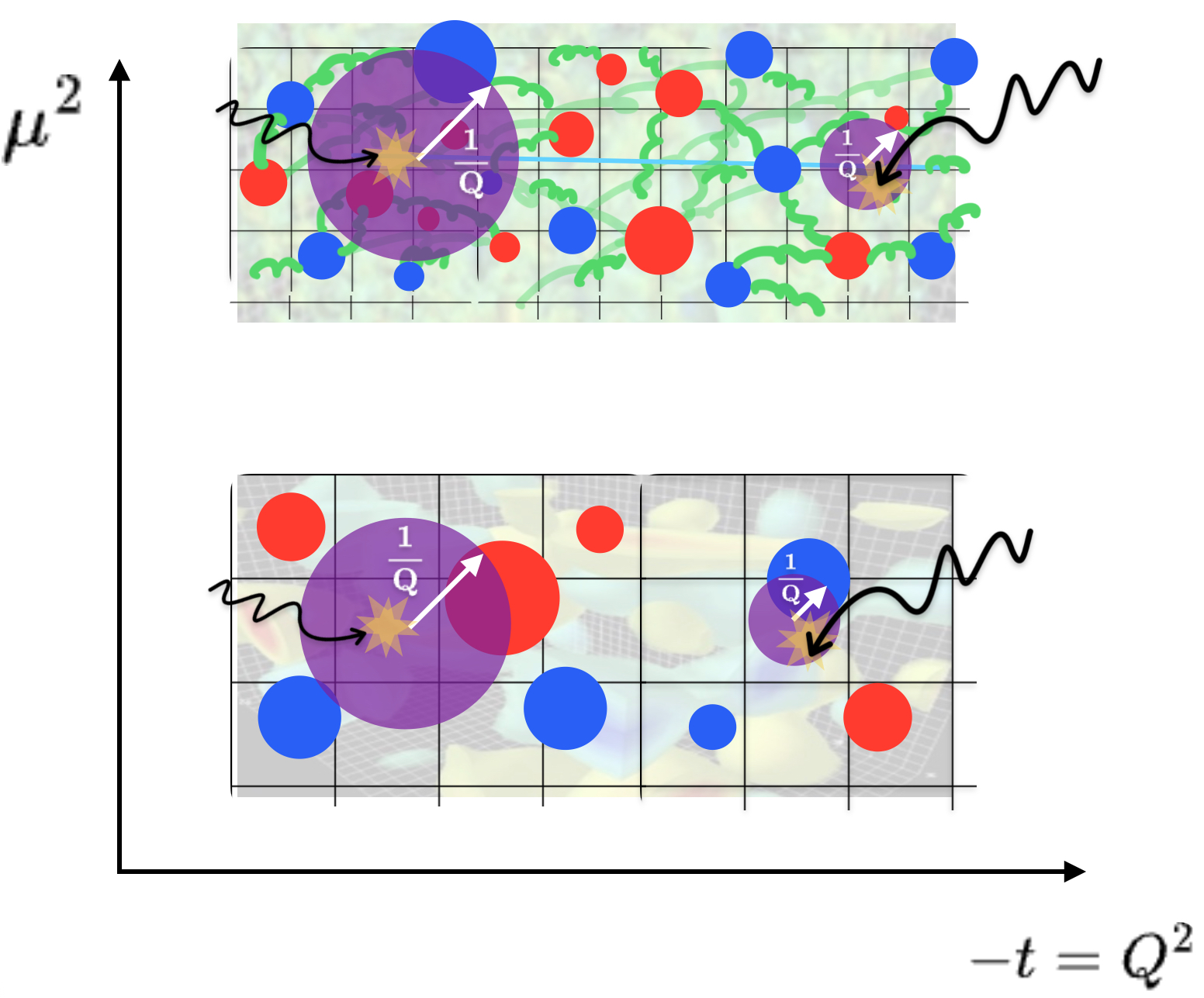}
    \caption{Illustration for QCD operator probing the vacuum with hadronic sources presented. The form factors in general are determined by renormalization scale $\mu$ and hadron kinematic momentum transfer $t=-Q^2$.}
    \label{fig:FF_vac}
\end{figure}

In general, the form factor can be described by the hadronic states with the probe defined by QCD operators $\mathcal{O}$. Therefore, form factors are associated with two scales $\mu$ and $Q$ as illustrated in Fig.~\ref{fig:FF_vac}. To calculate the form factor, we need the information of hadron states and the probe. 

In this chapter, we employ the standard framework of ILM to evaluate light hadron form factors of quark and gluonic origin in the QCD vacuum.  At large momentum transfer, we adopt a RG scheme with $Q^2=\mu^2$ to avoid large logarithms in perturbative QCD (pQCD) calculation.  At the low-to-intermediate region ($Q^2 \lesssim 10~\mathrm{GeV}^2$), the form factors are evaluated at a resolution scale set by the inverse instanton size $1/\rho$, where the gauge fields are dominated by localized configurations of single instantons and correlated instanton–anti-instanton pairs in a dilute liquid ensemble. We compute gluonic form factors associated with the scalar, pseudoscalar, and EMT operators, incorporating both isolated instantons and molecular configurations. The EMT results are then used to quantify the gluon contributions to Ji’s mass and spin sum rules. As outlined in Sec.~\ref{sec:vac_GF} and Ch.~\ref{ch:had-to-par}, these result can be evolved to higher resolution through gradient flow scheme, yielding nucleon mass and spin decompositions that are in good agreement with recent lattice QCD determinations.

\section{Form factors and factorization}

At large momentum transfer $Q\gg 1$ GeV, or the hard region, the meson form factors are fixed by factorization and the perturbative hard scattering. The factorization can be naturally described by Breit (brick-wall) frame,

\begin{align*}
    p_\mu=&(Q/\sqrt{2},0^-,0_\perp) &
    p'_\mu=&(0^+,Q/\sqrt{2},0_\perp)
\end{align*}
where momentum $p$ recoils as $p'$ in the opposite direction under the probe $Q$. This particular momentum configuration presents the fast moving hadron in large $Q^2$, allowing the factorization to be formulated in terms of perturbative hard kernels, $ T(x,x',\mu/Q, Q^2)$, and light cone observables, namely the nonperturbative hadron distribution amplitudes (DAs), $\varphi(x, \mu) $ (longitudinal momentum light-cone wave functions), at a specified renormalization scale $\mu $, where $ Q \gtrsim \mu \gg \Lambda_{\rm QCD}$.

\begin{figure}
    \centering
\subfloat[\label{fig:m_FF}]{\includegraphics[width=0.5\linewidth]{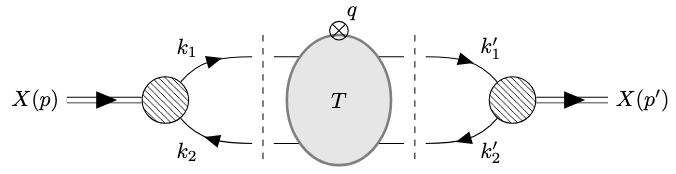}}
\hfill
\subfloat[\label{fig:b_FF}]{\includegraphics[width=0.5\linewidth]{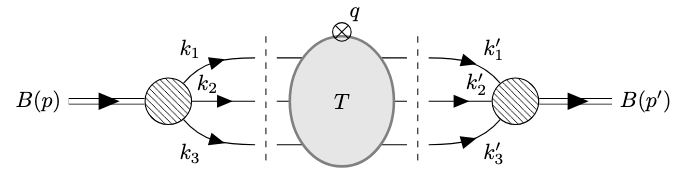}}
    \caption{The factorization of (a) meson and (b) baryon form factors at large $Q^2$ ($Q^2\gg 10$ GeV$^2$)}
\label{fig:FF}
\end{figure}

\begin{figure}
    \centering
\subfloat[\label{T1}]{\includegraphics[width=0.5\linewidth]{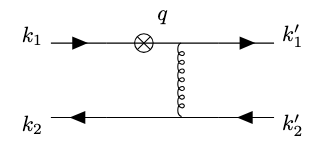}}
\hfill
\subfloat[\label{T3}]{\includegraphics[width=0.5\linewidth]{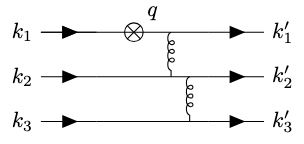}} 
    \caption{(a) and (b) are the perturbative gluon exchange diagrams at the leading order of $\alpha_s(Q^2)$.}
    \label{fig:T}
\end{figure}

As illustrated in Fig.~\ref{fig:FF}, the hadron form factor reads  \cite{Braaten:1987yy,Sterman:1997sx,Dittes:1983dy},

\begin{equation}
\begin{aligned}
&\langle p' |\mathcal{O}_{\rm QCD}| p\rangle=\int dx\int dx'\varphi^*(x,\mu) T(x,x',\mu/Q,Q^2)\varphi(x',\mu)
\end{aligned}
\end{equation}
where the integration over the longitudinal momentum fraction is defined as $$dx=dx_1dx_2\delta(1-x_1-x_2)$$ for meson and $$dx=dx_1dx_2dx_3\delta(1-x_1-x_2-x_3)$$ for baryon. In Fig.~\ref{fig:T}, we show the hard kernel diagrams with perturbative gluon exchange at the leading order of $\alpha_s$.

Incorporating twist expansion of the DAs, this factorization formulation can be systematically extended to include power corrections suppressed by $1/Q^2$, thereby enhancing the accuracy by including subleading power corrected contributions~\cite{Shuryak:2020ktq,Liu:2024vkj}. Higher-order perturbative corrections often results in a logarithmic scale dependence of the form $\ln^n(\mu^2/Q^2)$ in the hard kernel, arising from the truncation at finite order. To avoid large logarithmic from comprimising the validity of factorization, a renormalization scheme such that $\mu\sim Q$ is often assumed.

At asymptotic limit ($Q^2\rightarrow\infty$), the form factor follows $n-$pole form by Brodsky-Farrar (constituent quark) counting rule \cite{Brodsky:1973kr} where $n$ is the minimum number of the parton spectators. In the case of electromagnetic form factor, it is monopole for meson and dipole for baryon.
\begin{equation}
\langle p' |\mathcal{O}_{\rm QCD}| p\rangle\sim\frac1{(Q^2)^n}
\end{equation}

\begin{figure}
    \centering
\includegraphics[width=0.8\linewidth]{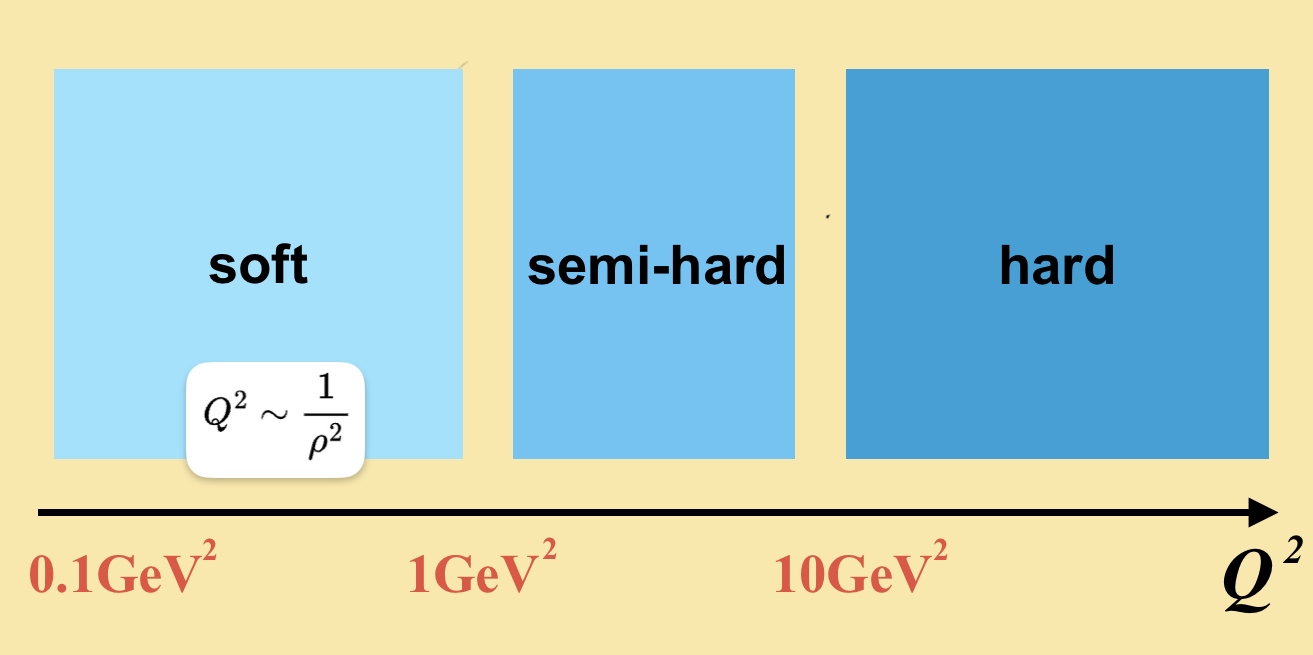}
    \caption{The soft, semi-hard, and hard $Q^2$ energy regions characterizing different regimes of the hadronic form factors}
    \label{fig:scales}
\end{figure}

\begin{figure}
    \centering
\subfloat[]{\includegraphics[width=0.5\linewidth]{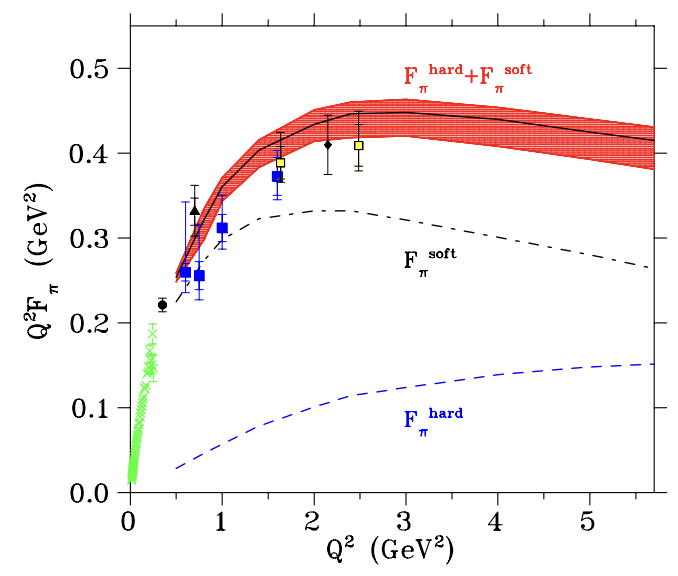}}
\hfill
\subfloat[]{\includegraphics[width=0.5\linewidth]{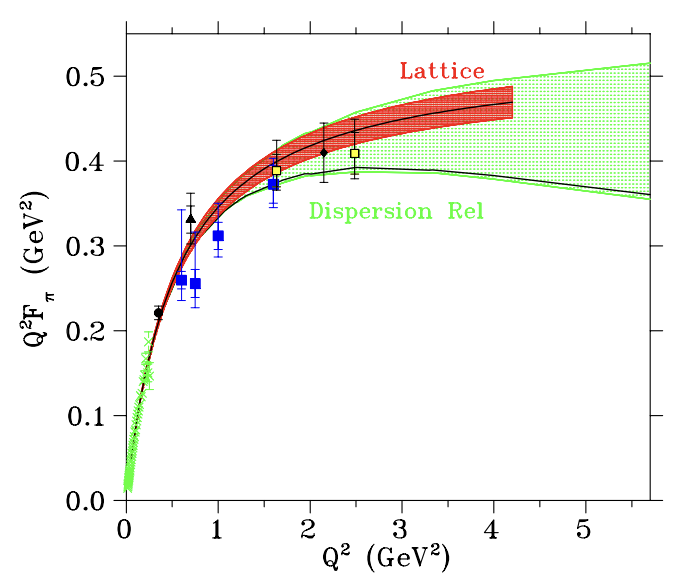}}
    \caption{(a) Pion form factor $F_\pi$ data \cite{JeffersonLab:2008jve} compared with LO+NLO hard contributions \cite{Bakulev:2004cu} based on an analysis of the pion-photon transition form factor from CLEO \cite{CLEO:1997fho} and CELLO \cite{CELLO:1990klc}. The supplemented soft component is estimated from local quark–hadron duality. (b) Comparison with dispersion relations \cite{Geshkenbein:1998gu} and lattice QCD \cite{QCDSFUKQCD:2006gmg} where the lattice band includes statistical and chiral extrapolation uncertainties.}
    \label{fig:JLab}
\end{figure}

Unfortunately, there are big discrepancies between such an asymptotic theory and existing experimental data, which remain in the semi-hard regime, as illustrated in Fig.~\ref{fig:scales}. This is particularly clear from current JLab data on the pion electromagnetic form factors~\cite{JeffersonLab:2008jve} and lattice simulations~\cite{QCDSFUKQCD:2006gmg}, as shown in Fig.~\ref{fig:JLab}. The measured meson form factors are well above the predictions of the perturbative QCD scaling laws, even when twist corrections are included.  This is not surprising since there is a drastic difference between the scales in jet physics and exclusive processes. The former are well defined above the scale $\mu^2 \sim 100\,{\rm GeV}^2$ (hard regime), while the exclusive processes are defined at much smaller scale, within $1-10\,{\rm GeV}^2$ (semi-hard regime). 

As a result, the study of form factors in the intermediate regime is therefore both crucial and increasingly relevant.

\section{Form factors in ILM}

At intermediate momentum transfer, $\mu^2 \sim Q^2 \lesssim 10~\mathrm{GeV}^2$, perturbative factorization becomes unreliable as nonperturbative dynamics grow increasingly important. In this regime, an ILM-based description provides a natural framework. The transition from a perturbative factorization picture to this framework involves two key modifications. First, topological gauge-field configurations alter the effective quark–gluon interaction, dressing the point-like probe vertex. Second, the light-front partonic picture at high resolution $\mu$ transitions into a constituent quark description at intermediate scales. (see Chs.~\ref{ch:vac} and \ref{ch:had-to-par})

As discussed in Ch.~\ref{ch:ILM}, the averages over the QCD operators convert the gluonic part into the corresponding effective quark degrees of freedom. The gluonic part of the operator distorts the color orientation of 't Hooft vertices, producing various quark spin structure, mapping the operators into numerous effective quark operators that can be evaluated further by the effective quark Lagrangian \eqref{eq:Leff} in the instanton vacuum (see Secs.~\ref{sec:ILMEFT} and \ref{sec:DILM}).


        \begin{figure}
        \centering
        \subfloat[]{\includegraphics[width=0.09\linewidth]{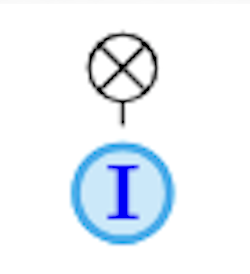}}
        \hfill
        \subfloat[]{
        \includegraphics[width=0.22\linewidth]{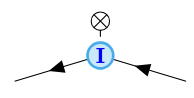}}
        \hfill
        \subfloat[]{
        \includegraphics[width=0.21\linewidth]{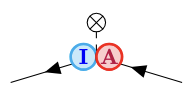}}
        \hfill
        \subfloat[]{
        \includegraphics[width=0.22\linewidth]{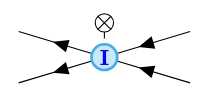}}
        \caption{(a) Disconnected contribution to the gluonic operators. (b) One-body operator arising from single (anti-)instantons. (c) One-body operator arising from close instanton--anti-instanton molecular configurations. (d) Two-body operator arising from (anti-)instantons.}
    \label{fig:ILM_exp1}
\end{figure}

We present the first few terms in the expansion \eqref{vac_OPE0}, which can also be illustrated diagrammatically in Fig.~\ref{fig:ILM_exp1}. Diagram (a) represents the disconnected contribution, which becomes nonzero in the presence of hadron sources due to fluctuations. Diagrams (b) and (c) correspond to one-body operators: diagram (b) arises from a single-instanton contribution, while diagrams (c) originate from correlated $IA$ pair configurations, respectively (see Sec.~\ref{sec:ILMEFT} and Appendix~\ref{App:pair}). Diagram (d) corresponds to two-body operators. These many-body contributions are omitted, as they are suppressed by $1/N_c$ and typically yield small matrix elements in hadronic observables.

When employing the effective Lagrangian to compute the resulting quark operators in \eqref{vac_OPE0}, one can also include the non-zero mode contributions which introduce additional structures. The nonzero-mode contributions alone generate a chirality-preserving effective vertex, and their interference with the zero modes produces a chirality-flipping effective vertex. As a result, the point-like probe vertex is significantly modified by those quark zero modes and quark nonzero modes induced by the topological pseudoparticles in the vacuum.

The arguments for vacuum averages can be extended to hadronic matrix elements and form factors. Given the comparable size of instantons and light hadrons, the off-forward  matrix element of the operator $\mathcal{O}$ in a hadron state can be expressed as an ensemble average, similar to the expansion \eqref{eq:gluo_op}, with vacuum bra-ket replaced by in-out on-shell hadronic states 
\begin{equation}
\label{eq:o_had_exp}
    \langle \mathcal{O}\rangle_{N_\pm}\rightarrow \langle h(p')| \mathcal{O}|h(p)\rangle_{N_\pm},
\end{equation}
which generalizes the arguments in~\cite{Weiss:2021kpt} to off-forward and multi-instanton contributions. Translational symmetry relates the hadronic matrix element of $\mathcal{O}$ to the momentum transfer  between the hadronic states, 
\begin{equation}
     \langle p'|\mathcal{O}|p\rangle = \frac{1}{V}\int d^4x\langle p'|\mathcal{O}(x)|p\rangle e^{-iq\cdot x}
\end{equation}
The recoiling hadron momentum is defined as $p'=p+q$, and the forward limit follows from $q\rightarrow0$.  The extension to hadron matrix element in \eqref{eq:o_had_exp} is highly nontrivial. Especially at nonzero momentum transfer $q\neq0$, the resummation of instanton-induced interactions among quarks typically leads to meson- or glueball-dominated exchanges and a pion cloud surrounding hadrons~\cite{Liu:2024rdm,Liu:2023fpj}, thus obscuring a clean separation between the source and the probe vertex.

\begin{figure*}
    \centering
\subfloat[\label{fig:hard_m}]{\includegraphics[width=.49\linewidth]{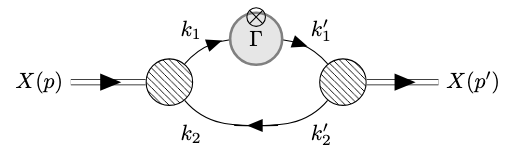}}
\hfill
\subfloat[\label{fig:hard_b}]{\includegraphics[width=.49\linewidth]{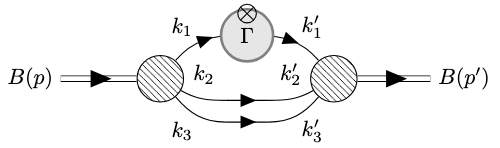}}
    \caption{The form factors of (a) meson $X$ and (b) baryon $B$ at small $Q^2$ with hadron state addressed by the BS wave function. The probe denoted by the cross dot is dressed by the instanton vacuum (See text). The momentum conservation requires $k_1'=k_1+q$ and $k_{2,3}'=k_{2,3}$. }
    \label{fig:WF_FF}
\end{figure*}

For the external hadron states, at the renormalization scale $\mu \sim Q \sim 1/\rho$ in this regime, hadronic dynamics are primarily governed by the emergent ’t Hooft interactions~\eqref{Hooft} and \eqref{eq:Leff}. These interactions can be systematically organized by resumming the leading $1/N_c$ $s$-channel bubble diagrams (RPA) among constituent quarks, providing a transparent dynamical picture that leads to the hadronic BSF wave function $\Psi_h(k; p)$, as illustrated in Figs.~\ref{fig:BS_mes} and \ref{fig:WF_b} (see Secs.~\ref{sec:mes}, \ref{sec:mes_3}, and \ref{sec:bary}). With these wave functions, together with the effective operators derived in Eq.~\eqref{vac_OPE0}, the hadron form factors can often be computed directly, as illustrated in Figs.~\ref{fig:hard_m} and \ref{fig:hard_b}.

\section{Scalar form factors and trace anomaly}

In QCD, the  breaking of conformal symmetry is captured by the trace anomaly

\begin{equation}
\label{3X}
T^\mu{}_\mu= \left(\frac{\beta(g^2)}{4g^2}F^2_{\mu\nu}+\gamma_mm\overline\psi\psi\right)+m\overline\psi\psi
\end{equation}
where $\gamma_m$ is the quark mass anomalous dimension generated by the renormalization of $F^2$ \cite{Tarrach:1981bi,Nielsen:1977sy,Collins:1976yq}, indicating its deep connection to the quark running mass. The two-loop results are given by \cite{Tarrach:1981bi}

\begin{equation}
    \gamma_{m}(g^2) = \frac{g^2}{8\pi^2}3\, C_F+\left(\frac{g^2}{8\pi^2}\right)^2\left(\frac{3}{4} C_F^2 + \frac{97}{12} C_F C_A - \frac{5}{6} C_F N_f\right)
\end{equation}

Strictly speaking, only $m\bar\psi\psi$ is RG-invariant while $F^2$ is only RG-invariant at 1-loop in chiral limit.

By Poincare symmetry, the hadronic mass can be decomposed in a RG-invariant (resolution independent) way in term of the ``invariant'' mass (fixed by $\Lambda_{\rm QCD}$) and the chiral breaking mass  (sigma term) at some resolution. 

\begin{equation}
\begin{aligned}
    m_h=m_h^{(\mathrm{inv})}+\sigma_{\pi h}
\end{aligned}
\end{equation}
which is distinct from the mass budgeting sum rule to be discussed below.

The trace identity Eq.~\eqref{3X} reflects on the general fact that all hadron masses in QCD are tied to the quantum breaking of conformal symmetry, and should be enforced by any non-perturbative QCD description. 

\begin{figure}
    \centering
\includegraphics[width=.75\linewidth]{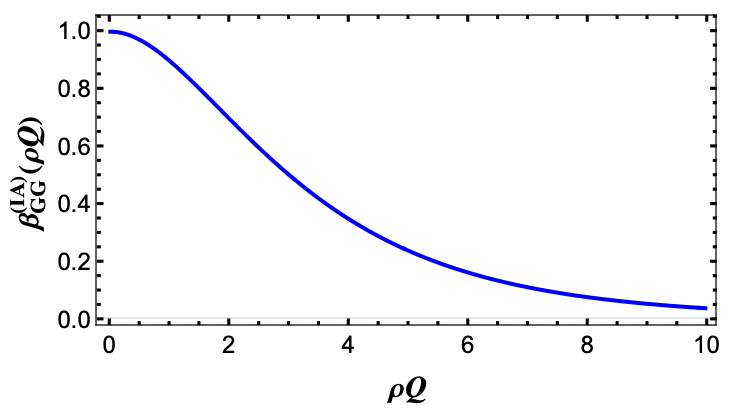}
    \caption{The instanton form factors induced by molecular profile for gluonic scalar operators}
\label{fig:betaGG}
\end{figure}

In this section, we will discuss the quark and gluon scalar form factors for pions, nucleons, and rho mesons. The gluonic scalar operator $F^2$ is mapped to effective quark operators by Eqs.~\eqref{eq:gluo_op} and \eqref{eq:op_eff} up to a pair of pseudoparticles using simply sum ansatz (see Ch.~\ref{ch:ILM}). The details of calculation can be found in \cite{Liu:2024rdm}. Here we explicitly show the result

\begin{equation}
\begin{aligned}
\label{eq:GG0}
    \langle h'| \frac{g^2}{32\pi^2}F^2|h\rangle=&-2m_h^2\left(\bar N\frac{\partial\ln m_h}{\partial N}\right)\left(\frac{\sigma_t}{\bar N}\right)\frac{(2\pi)^4\delta^4(q)}V\\
    &+\frac{n_{IA}\gamma_{IA}}{2N_c(N_c^2-1)}\frac{1}{18}\beta_{GG}^{(IA)}(\rho q)\rho^2q_\mu q_\nu\langle h'|\bar{\psi}\left(\gamma_{(\mu} i\overleftrightarrow{\partial}_{\nu)}-\frac14g_{\rho\lambda}i\slashed{\partial}\right)\psi|h\rangle\\
    &+\cdots
\end{aligned}
\end{equation}
where the instanton profile form factors are defined as

\begin{align}
    \beta^{(IA)}_{GG}(q)=&\frac{1}{q}\int_0^\infty dx\frac{576 x^2}{(1+x^2)^4}\frac{J_3(qx)}{q^2x^2}\nonumber\\
    =&\frac{576}{q^6}(-32 + q^2)  +\frac{144}{q^5}(128 + 28 q^2 + q^4)K_1(q) \nonumber\\
    &+\frac{12}{q^4} (768 + 72 q^2 + q^4) K_0( q)
\end{align}
and illustrated in Fig.~\ref{fig:betaGG}

$\gamma_{IA}$ is the typical coupling in one-body operators arise from $IA$ pair defined by (see Appendix~\ref{App:pair})

\begin{equation}
\begin{aligned}
\gamma_{IA}=&\left\langle\sum_{n=0}^{N_f-1}\binom{N_f-1}{n}\left(\frac{|T_{IA}|}{m^*}\right)^{2n}\frac{(4\pi^2\rho^2)^2}{(m^*)^2}\left(\frac{-1}4R\frac{dT(R)}{dR}\right)\right\rangle
\end{aligned}
\end{equation}


The leading $1/N_c$ contribution in \eqref{eq:GG0} arises from diagrams that are disconnected from the hadronic source in a fixed-$N_\pm$ configuration \cite{Diakonov:1995qy}, where nontrivial values emerge due to fluctuations in the instanton number (see Sec.~\ref{sec:had_grand}).

The $\delta^4(q)$ term reflects the assumption of a homogeneously distributed vacuum, where scalar glueball probe is disconnected from the hadronic source and exclusively sourced by equally distributed semiclassical instanton and anti-instanton fields (mean field). To account for inhomogeneous structure in the instanton vacuum, the momentum dependence can be generalized to a correlation function of scalar glueballs.

\begin{equation}
\frac{\sigma_t}{\bar{N}}\frac{(2\pi)^4}{V}\delta^4(q)\rightarrow
\frac{\sigma_t(Q^2)}{\bar N}\equiv\frac{1}{32\pi^2\langle F^2\rangle}\int d^4xe^{-iq\cdot x}\langle F^2(x)F^2(0)\rangle_c
\end{equation}
where the topological compressibility is given by $\sigma_t=\frac{4}b \bar{N}$ with $b=\frac{11}3N_c-\frac23N_f$, the one loop coefficient for the beta function (see Secs.~\ref{sec:InstYM} and~\ref{sec:ILM_QCD}).

This gluonic scalar vacuum form factor $\sigma_t$ captures important aspects of the
gluonic correlation function in the QCD vacuum, discussed in the bosonized ILM in~\cite{Kacir:1996qn} and in the full ILM in~\cite{Schafer:1996wv}. More specifically, we define the point-to-point correlator of the gluonic scalar source $F^2$ as
\begin{equation}
    \Pi_{GG}(x)=\langle F^2(x)F^2(0)\rangle_c
\end{equation}

\begin{figure}
    \centering
       \includegraphics[width=\linewidth]{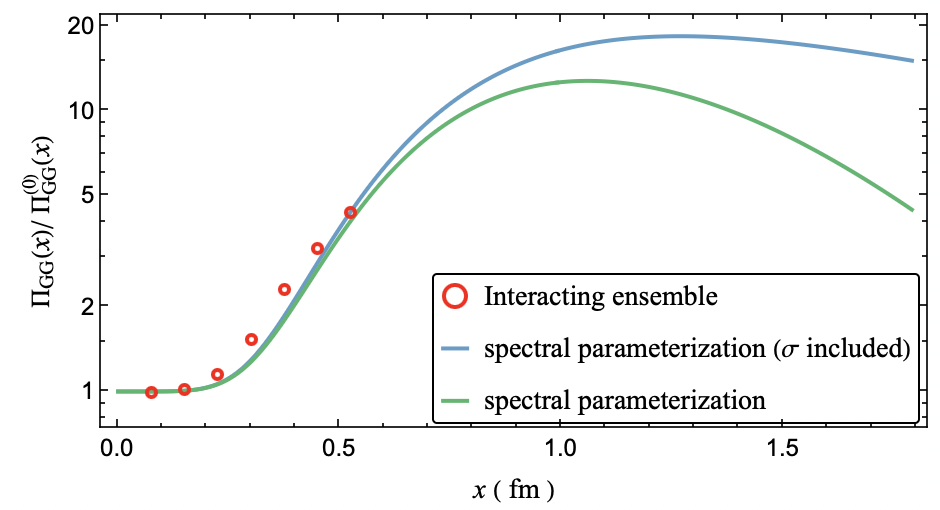}
    \caption{ Spectral function from the ILM (open red circles)~\cite{Schafer:1994fd}, 
    normalized by the perturbative contribution in coordinate space. The
    comparison is to the empirical spectral parametrization  (\ref{eq:traceFF2}) with the scalar meson $\sigma$ (blue line) and without the scalar meson $\sigma$ (green line).}
    \label{fig_G2_corr_vac}
\end{figure}

In Fig.~\ref{fig_G2_corr_vac} we show the $x$-space point-to-point correlator normalized to the perturbative (two gluon) result, given by

\begin{equation}
    \Pi_{GG}^{(0)}(x)=\frac{384\,g^4}{\pi^4 x^8}
\end{equation}

The $x$-space spectral function (dashed line) accounts for the scalar-sigma,
plus the scalar $0^{++}$ glueball, plus the soft 2-pion cut  and the hard 2-parton cut,
\begin{equation}
\begin{aligned}
\label{eq:traceFF2}
&\Pi_{GG}(x)=\lambda_{0^{++}}^2 D(m_{0^{++}},x)+\lambda_{\sigma}^2 D(m_{\sigma},x)\\
&+\int_{4m_\pi^2}^{\Lambda_\chi^2}ds\,\frac{3}{64\pi^2}\sqrt{1-\frac{4m_\pi^2}{s}}(s-2m_\pi^2)^2D(\sqrt{s},x)+\frac{2g^4}{\pi^2}\int_{s_0}^{\infty}ds\,s^2D(\sqrt{s},x)
\end{aligned}
\end{equation}
where $D(m,x)=\frac{m}{4\pi^2x}K_1(mx)$. The parameters used in the spectral parameterization are as  $\lambda_{0^{++}}=15.6$ \rm GeV$^{3}$, $\lambda_{\sigma}=2.6$ GeV$^{3}$, $m_{0^{++}}=1.2$ GeV, and $m_\sigma=680$ MeV to be consistent with the interacting ensemble calculation \cite{Schafer:1994fd} and ILM result in Table~\ref{tab:mes0_spec} and \ref{tab:s_meson}. For the two-pion threshold, we use $m_\pi=138$ MeV given by Table~\ref{tab:mes0_spec} and \ref{tab:p_meson_spec} with the pion decay constant $f_\pi=93$ MeV and the chiral symmetry breaking scale $\Lambda_\chi=1.1$ GeV. Note that we have re-instated the gauge coupling with a value $\alpha_s=0.315$, when accounting for the perturbative contributions with a perturbative threshold set to $s_0=2.4$ GeV~\cite{Schafer:1994fd}. The scalar glueball mass is slightly lighter compared to the results given by lattice calculations~\cite{Schafer:1994fd,Chen:2005mg,Sun:2017ipk}.

In this section, since the gluonic scalar correlation function is dominated by scalar glueball exchange, we simply parameterize the correlation function by dipole form with $m_{0^{++}}=1.5-1.7$ GeV given by lattice calculations~\cite{Schafer:1994fd,Chen:2005mg,Sun:2017ipk}.

\begin{equation}
\label{eq:traceFF}
    \sigma_t(Q^2)=\frac{\sigma_t}{(1+Q^2/m_{0^{++}}^2)^2}
\end{equation}

At zero momentum transfer in Eq.~\eqref{eq:GG0}, the matrix element is dominated by density fluctuations, corresponding to the disconnected diagrams. Additional corrections arise from quark mixing in the trace anomaly, which is manifested as instanton-quark interaction at medium resolution in ILM

\begin{equation}
\begin{aligned}
    \frac{1}{2m_h}\langle h| \frac{\beta(g^2)}{4g^2}F^2|h\rangle=&-\left(\bar N\frac{\partial m_h}{\partial N}\right)\frac{8\pi^2\beta(g^2)}{g^4}\left(\frac{\sigma_t}{\bar N}\right)\\
\end{aligned}
\end{equation}
where $\sigma_{\pi h}$ is the pion-hadron sigma term.
Since the full trace anomaly is RG invariant, the gluonic matrix element together with the quark RG mixing leads to the definition of an RG-invariant (effective) mass $m_h^{(\rm inv)}$. As a result, one can define the invariant mass as

\begin{equation}
\begin{aligned}
\label{eq:GG}
m^{(\rm inv)}_h\equiv&\frac1{2m_h}\langle h| \left(\frac{\beta(g^2)}{4g^2}F^2_{\mu\nu}+\gamma_mm\overline\psi\psi\right)|h\rangle\\
=&-\left(\bar N\frac{\partial m_h}{\partial N}\right)\frac{8\pi^2\beta(g^2)}{g^4}\left(\frac{\sigma_t}{\bar N}\right)+\gamma_m\sigma_{\pi h}\\
\approx&\,4\left(\bar N\frac{\partial m_h}{\partial N}\right)
+\mathcal{O}(g^2)
\end{aligned}
\end{equation}

At non-zero momentum transfer, the scalar gluonic form factor shows that the $F^2$ in a hadron couples primarily to disconnected scalar glueball exchange with mixing of the scalar meson ($\sigma$) \cite{Kacir:1996qn}. In the following calculations for pions and nucleons, since the full result is not yet available, we model the contribution using a meson-dominance exchange, motivated by the ILM observations in \cite{Kacir:1996qn} and in Eqs.~\eqref{eq:GG0} and \eqref{eq:traceFF}.

\subsection{Pion}

The pion scalar form factors are defined by

\begin{align}
\label{pi_scalar}
&\langle \pi'|\frac{\beta(g^2)}{4g^2}F^2|\pi\rangle=2m_\pi^2G_\pi(Q^2)\\
&\langle \pi'|m\bar\psi\psi|\pi\rangle=2m^2_\pi\sigma_\pi(Q^2)
\end{align}

The parameterization of gluonic scalar form factor is motivated by \cite{Kacir:1996qn} and Eq.~\eqref{eq:GG0}. The scalar glueball and $\sigma$ meson mixing is represented by $\sigma_t$ multiplied by $\sigma$ exchange with $\mathcal{O}(\rho Q^2)$ correction introduced by instanton profiles with finite size $\rho=0.313$ fm.

\begin{equation}
\begin{aligned}
    G_\pi(Q^2)=&\left(1-\frac{\sigma_{\pi \pi}}{m_\pi}\right)\frac{\sigma_t(Q^2)}{\sigma_t}\left(\frac{1+\left(2.52 - 38.6\tfrac{\sigma_{\pi \pi}}{m_\pi}\right) \rho^2Q^2}{1+Q^2/m^2_\sigma}\right)
\end{aligned}
\end{equation}

\begin{figure}
    \centering
\subfloat[\label{fig:scalarFF_pi}]{\includegraphics[width=0.5\linewidth]{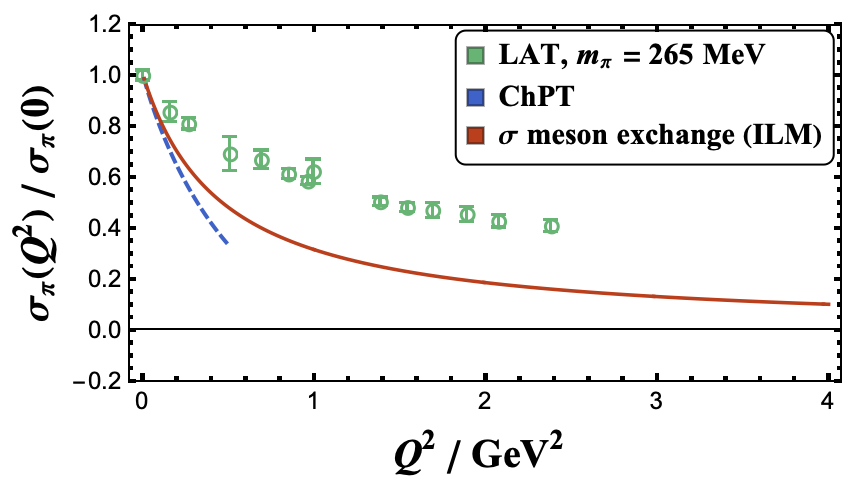}}
\hfill
\subfloat[\label{fig:scalarFF_nuc}]{\includegraphics[width=0.5\linewidth]{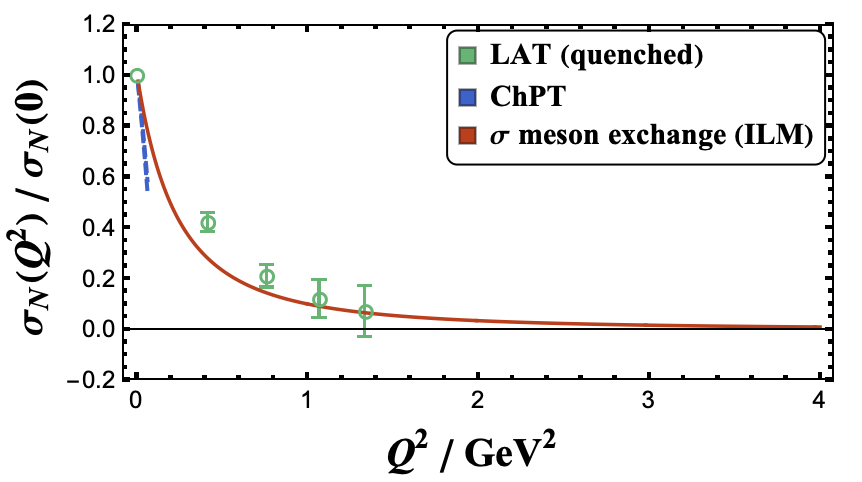}}
    \caption{(a) Normalized quark scalar form factors in pion using $\sigma$ exchange monopole form in ILM \eqref{eq:sca_pi} (red solid) compared with ChPT (black dashed curve) and lattice calculation (green circle) from ETMC. (b) Normalized quark scalar form factors in nucleon using $\sigma$ exchange dipole form in ILM \eqref{eq:sca_nuc} (red solid) compared with ChPT (black dashed curve) and lattice calculations  (green circle) using quenched ensemble with heavy quark mass $~120$ MeV }
    \label{fig:scalarFF}
\end{figure}

For the quark scalar form factor, we assume the scalar meson dominance and thus the quark scalar form factor is estimated by

\begin{equation}
\label{eq:sca_pi}
\sigma_\pi(Q^2)=\frac{\sigma_{\pi\pi}/m_\pi}{1+Q^2/m^2_\sigma}
\end{equation}
with scalar meson mass $m_\sigma=683$ MeV (see Sec.~\ref{sec:mes}) and the pion-pion sigma term $\sigma_{\pi\pi}=0.484m_\pi$ \cite{Bijnens:1998fm,Bijnens:1997vq}. The curve is numerically compared with chiral perturbation theory (ChPT) in \cite{Bijnens:1998fm,Bijnens:1997vq} at small $Q^2$ and lattice calculation from ETMC \cite{Alexandrou:2021ztx} using an $N_f=2+1+1$ ensemble with clover improved maximally twisted mass fermions and slightly larger pion mass $m_\pi=260$ MeV, as illustrated in Fig.~\ref{fig:scalarFF_pi}

\begin{figure}
    \centering
    \includegraphics[width=1\linewidth]{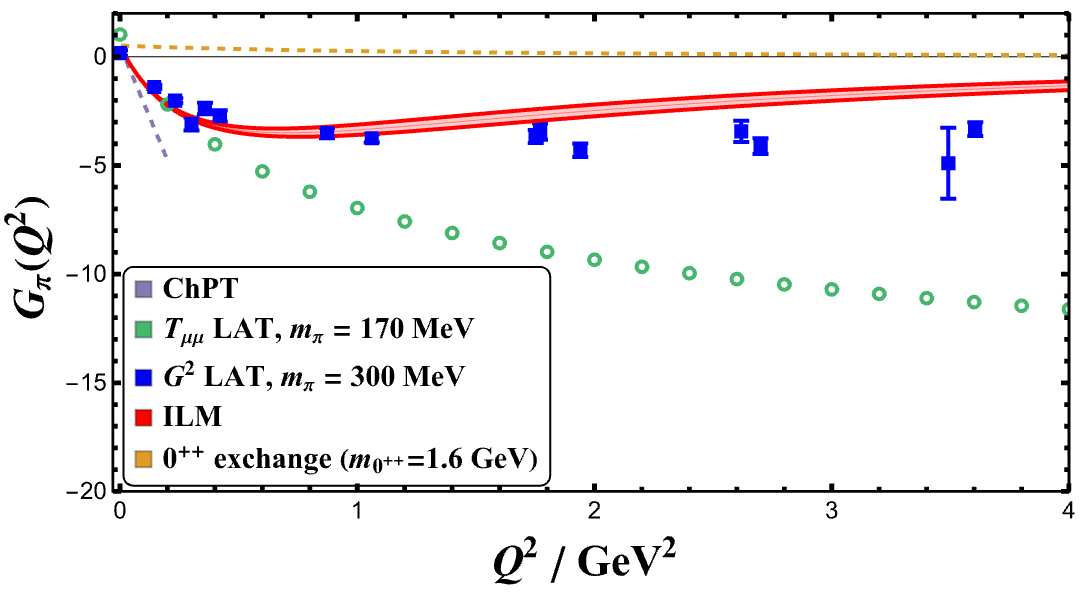}
    \caption{Gluonic scalar form factor of pion (red solid line) together with the $0^{++}$ glueball correlation function \eqref{eq:traceFF} (orange dashed line). Blue square represents the lattice from \cite{Wang:2024lrm} with pion mass $m_\pi=300$ MeV, the purple dashed curve represents the ChPT prediction. The red band reflects the range of scalar glueball mass $m_{0^{++}}=1.5-1.7$ GeV used in scalar form factor. The green circle represents the lattice dipole fit of EMT trace form factor (see text).}
    \label{fig:G_pi}
\end{figure}

In Fig.~\ref{fig:G_pi},  we show our result for the behavior of the gluonic scalar form factor $G_\pi(Q^2)$ in the ILM with pion mass 138 MeV and a range of scalar glueball mass $m_{0^{++}}=1.5-1.7$ GeV (red-spread). The lattice results of $F^2$ form factor (blue squares)~\cite{wang2024trace} are shown along with the lattice monopole fit of $T^{\mu}{}_\mu$ form factor (green circle)~\cite{Hackett:2023nkr} and the result from ChPT (purple-dashed line). The resulting form factor, as indicated in Fig.~\ref{fig:G_pi}, is weakly dependent on both the glueball mass and pion sigma term $\sigma_{\pi\pi}$, which is fixed by $0.484m_\pi$.

\begin{table*}
    \centering
    \begin{tabular}{|c|c|c|c|c|c|c|c|}
   \hline
     & $m_\sigma$ & $m_{\eta'}$ & $m_{f_2}$ \\ 
    \hline
   ILM & 683 MeV & 682 MeV & 1.275 GeV   \\
   PDG \cite{ParticleDataGroup:2024cfk,ParticleDataGroup:2010dbb} & 400--550 MeV & 957.78(6) MeV & 1.2751(12) GeV \\
   \hline
\end{tabular}
\begin{tabular}{|c|c|c|c|c|c|c|c|}
   \hline
     & $m_{0^{++}}$ & $m_{0^{-+}}$ \\ 
    \hline
   ILM & 1.5 -- 1.7 GeV &  2.5 GeV  \\
   Lattice \cite{Chen:2005mg} & 1.710(50)(80) GeV & 2.560(35)(120) GeV \\
   \hline
\end{tabular}
    \caption{Meson ($\sigma$, $\eta'$, $f_2$) and glueball ($0^{++}$, $0^{-+}$) masses entering the form factor calculations within the ILM framework in this section}
    \label{tab:mass_ILM}
\end{table*}

\subsection{Nucleon}

\begin{figure}
    \centering
    \includegraphics[width=
    \linewidth]{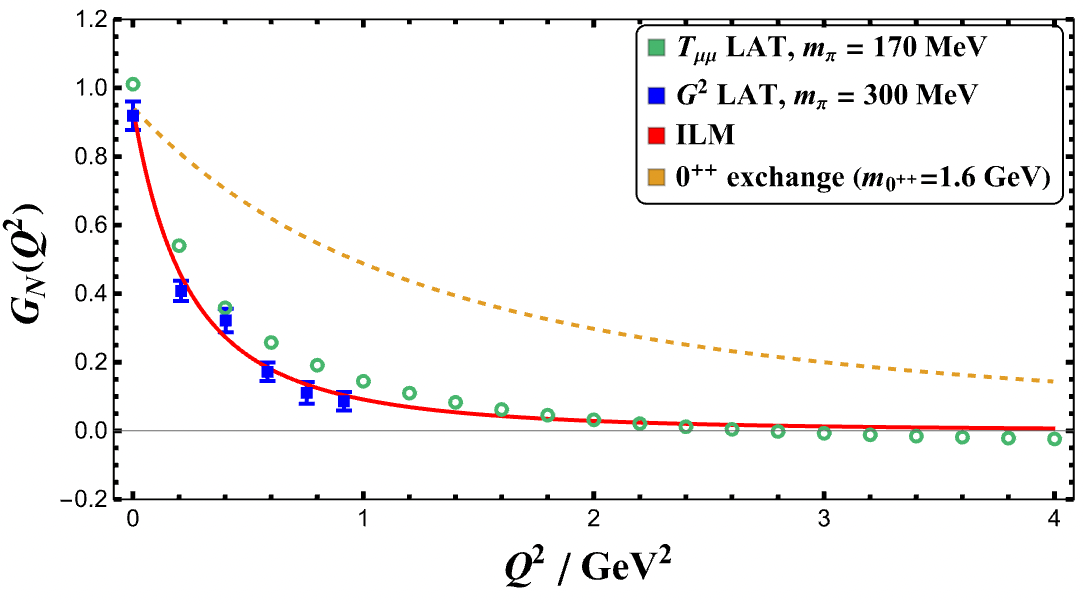}
    \caption{Nucleon gluonic scalar form factor \eqref{GN} (red solid line) together with the $0^{++}$ glueball correlation function \eqref{eq:traceFF} using scalar glueball mass $m_{0^{++}}=1.6$ GeV (orange dashed line are presented. The red band reflects the sensitivity of the form factor to scalar glueball mass $m_{0^{++}}=1.5-1.7$ GeV. Results are compared with lattice data (blue points)~\cite{Wang:2024lrm} and the nucleon trace form factor reconstructed from lattice dipole fits of the EMT form factors~\cite{Hackett:2023rif} (green open circles).}
    \label{fig:G_N}
\end{figure}

The nucleon scalar form factors are defined by
\begin{align}
\label{N_scalar}
&\langle N'|\frac{\beta(g^2)}{4g^2}F^2|N\rangle=m_NG_N(Q^2)\bar{u}_{s'}(p')u_s(p)\nonumber\\
&\langle N'|m\bar\psi\psi|N\rangle=m_N\sigma_N(Q^2)\bar{u}_{s'}(p')u_s(p)
\end{align}

For the quark scalar form factor, the scalar form factor of quark content is estimated by scalar meson dominance

\begin{equation}
\label{eq:sca_nuc}
    \sigma_N(Q^2)=\frac{\sigma_{\pi N}/m_N}{(1+Q^2/m^2_\sigma)^2}
\end{equation}
with scalar meson mass $m_\sigma=683$ MeV (see Sec.~\ref{sec:mes}). We also chose a reasonable value for the pion-nucleon sigma term $\sigma_{\pi N}=53$ MeV indicated in \cite{Schweitzer:2003sb,Hoferichter:2023ptl,FlavourLatticeAveragingGroupFLAG:2024oxs} (references therein). The curve is numerically compared with dispersively improved ChPT in \cite{Alarcon:2017ivh} at small $Q^2$ and lattice calculation from \cite{Dong:1995ec}, as illustrated in Fig.~\ref{fig:scalarFF_nuc}. This lattice calculation used quenched ensemble with large unphysical quark mass $m\sim120$ MeV, resulting in $\sigma_{\pi N}=49.7\pm2.6$ MeV.


The nucleon scalar form factor of gluonic content is parameterized by

\begin{equation}
\begin{aligned}
\label{GN}
    &G_N(Q^2)=\left(1-\frac{\sigma_{\pi N}}{m_N}\right)\frac{\sigma_t(Q^2)}{\sigma_t}\left(\frac{1+\left(2.52 - 38.6\tfrac{\sigma_{\pi N}}{m_N}\right)\rho^2Q^2}{(1+Q^2/m^2_\sigma)^2}\right)
\end{aligned}
\end{equation}
The scalar glueball and $\sigma$ meson mixing is represented by $\sigma_t$ multiplied by $\sigma$ exchange with $\mathcal{O}(\rho Q^2)$ correction introduced by instanton finite size $\rho=0.313$ fm.

In Fig.~\ref{fig:G_N}, we show the result for the anomalous gluonic contribution to the trace of the EMT in the nucleon Eq.~\eqref{3X} (red-solid line) compared to the recent lattice results of $F^2$
(blue squares) from~\cite{Wang:2024lrm}. The trace of the EMT in the nucleon  (green-open circles) is reconstructed by the lattice dipole fit of nucleon $A$- and $D$-form factors from~\cite{Hackett:2023rif}. As expected, our results (red-solid line) including connected contribution supplemented with the disconnected $0^{++}$ glueball exchange contribution (dashed-orange line) yields a good account of the reported lattice results in $Q^2\lesssim1/\rho^2$. The resulting form factor is weakly sensitive to the glueball mass and nucleon sigma term $\sigma_{\pi N}$, which is set to be 53 MeV.

Overall, the gluonic component of the nucleon scalar form factor is dominated by the connected instanton contribution and the disconnected scalar glueball exchange (orange curve in Figs.~\ref{fig:G_N}). In contrast to nucleon, a comparison with Figs.~\ref{fig:G_pi} and \ref{fig:G_N} shows that the opposite pattern emerges for the pion~\cite{Liu:2024rdm}, consistent with lattice results~\cite{Wang:2024lrm}. This difference can be traced to the suppression of the glueball-exchange contribution in the forward limit for the pion, which follows from its Goldstone nature.

\section{Gravitational form factor}
\label{sec:GFF}
Another gluonic operator of interest is the gluonic tensor tied to the QCD energy-momentum tensor (EMT).  To evaluate the QCD energy-momentum tensor using local or non-local  effective 
formulations is subtle, for a recent discussion see~\cite{Freese:2019bhb}. In the present approach, it follows the same reasoning as that for the scalar operators discussed above.

With this in mind, the EMT in QCD is given by
\begin{equation}
    T_{\mu\nu}=T^g_{\mu\nu}+T^q_{\mu\nu}
\end{equation}
where $T^g_{\mu\nu}$ is the gluonic EMT and $T^q_{\mu\nu}$ is the quark EMT, given by~\cite{Polyakov:2018exb}
\begin{equation}
\begin{aligned}
T^q_{\mu\nu}
&=
\bar\psi
\gamma^{(\mu} i\overleftrightarrow{D}^{\nu)}\psi
-g^{\mu\nu}\bar\psi
\left(
i\overleftrightarrow{\slashed{D}}
-m
\right)\psi ,
\\[6pt]
T^g_{\mu\nu}
&=
-F^{a\mu\lambda}F^{a\nu}{}_{\lambda}
+
\frac{1}{4}g^{\mu\nu}
F^2_{\lambda\sigma}
\end{aligned}
\end{equation}

All issues of operator renormalization
are understood in the sense of a gradient flow cooling to the
semiclassical point with fixed topological charge per 4-volume, with the instanton density as the sole scale.
The energy-momentum tensor can be decomposed as the sum of a traceless and traceful part~\cite{Ji:2021mtz,Ji:1995sv},
\begin{equation}
\label{EMT_decop}
T^{q,g}_{\mu\nu}=\bar{T}^{q,g}_{\mu\nu}+\frac{1}{4}g_{\mu\nu}T^{q,g}_{\alpha\alpha}
\end{equation}
where the traceless part includes the gluonic tensor
\begin{equation}
\begin{aligned}
\label{eq:EMTg}
\bar{T}^g_{\mu\nu}=\frac{1}{4}g_{\mu\nu}F^2-F^{a}_{\mu\lambda}F^{a}_{\nu\lambda}
\end{aligned}
\end{equation}
and
\begin{equation}
    \bar{T}^q_{\mu\nu}=\bar{\psi}\left(\gamma_{(\mu} i\overleftrightarrow{D}_{\nu)}-\frac{1}{4}g_{\mu\nu}i\overleftrightarrow{\slashed{D}}\right)\psi
\end{equation}
and the traceful part $T^{\alpha}{}_{\alpha}$ is given by the trace anomaly in Eq.~\eqref{3X}. Both contributions belong to different irreducible representations of the Lorentz group. Thus, they do not mix in the renormalization.

\begin{figure}
\centering
\begin{minipage}{0.48\linewidth}
    \centering
    \subfloat[\label{fig:ILM_exp2q}]{
        \includegraphics[width=0.5\linewidth]{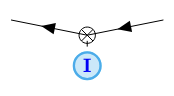}
    }
\end{minipage}
\hfill
\begin{minipage}{0.48\linewidth}
    \centering
    \subfloat[\label{fig:ILM_exp2g}]{
        \includegraphics[width=0.5\linewidth]{figures/v3.png}
    }
\end{minipage}

\caption{(a) One-body operator arising from single (anti-)instantons in the gauge potential $\bar\psi A\psi$ in quark EMT (b) One-body operator arising from instanton--anti-instanton molecular configurations in gluon EMT.}
\label{fig:ILM_exp2}
\end{figure}

\begin{figure}
    \centering
\subfloat[\label{fig:betaqAq}]{\includegraphics[width=.5\linewidth]{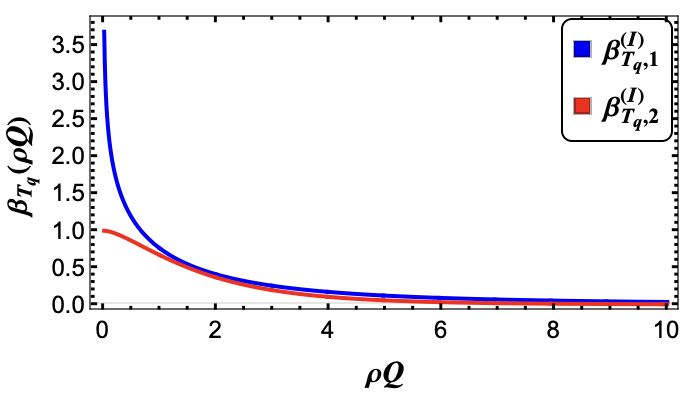}}
\hfill
\subfloat[\label{fig:betaTg}]{\includegraphics[width=.5\linewidth]{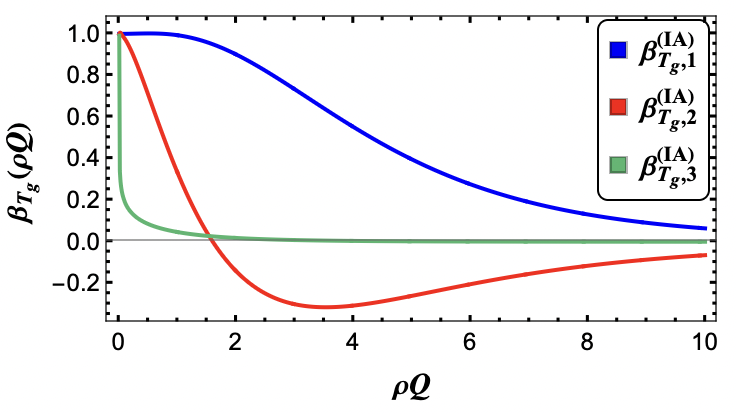}}
    \caption{The instanton form factors induced by single pseudoparticle in quark EMT and molecules in gluon EMT operators.}
    \label{fig:betaForm}
\end{figure}

For the quark EMT, as illustrated in Fig.~\ref{fig:betaqAq}, the hadronic matrix elements of the traceless and tracefull part are given by

\begin{equation}
\begin{aligned}
\label{eq:qEMT1}
   \langle h'|\bar{T}^q_{\mu\nu}|h\rangle=&\langle h'|\bar{\psi}\left(\gamma_{(\mu} i\overleftrightarrow{\partial}_{\nu)}-\frac{1}{4}g_{\mu\nu}i\overleftrightarrow{\slashed{\partial}}\right)\psi|h\rangle\\
    &-\left(\frac{n_{I+A}}2\right)\frac{1}{8N_c}\left(\frac{4\pi^2\rho^2}{m^*}\right)\beta^{(I)}_{T_q,1}(\rho q)\rho^2\left(q_\mu q_\nu-\frac14 g_{\mu\nu}q^2\right)\langle h'|\bar\psi\psi|h\rangle\\
\end{aligned}
\end{equation}
and
\begin{equation}
\begin{aligned}
\label{eq:qEMT2}
   \langle h'|T^q_{\mu\mu}|h\rangle=&\langle h'|\bar{\psi}i\overleftrightarrow{\slashed{\partial}}\psi|h\rangle
    -\left(\frac{n_{I+A}}2\right)\frac{1}{N_c}\left(\frac{4\pi^2\rho^2}{m^*}\right)\left(2\beta^{(I)}_{T_q,2}(\rho q)-1\right)\langle h'|\bar\psi\psi|h\rangle\\
\end{aligned}
\end{equation}
where the single instanton profiling form factors are defined as


\begin{equation}
\begin{aligned}
    \beta^{(I)}_{T_q,1}(q)=&\frac{1}{q}\int_0^\infty dxK_D(x)\frac{32}{1+x^2}\frac{J_3(qx)}{q^2x^2}
\end{aligned}
\end{equation}

\begin{equation}
\begin{aligned}
    \beta^{(I)}_{T_q,2}(q)=&\frac{1}{q}\int_0^\infty dxK_D(x)\frac{6}{x^2(1+x^2)}J_1(qx)
\end{aligned}
\end{equation}
At $q\rightarrow0$, we have 
$\beta^{(I)}_{T_q,1}(q\rightarrow0)=-\frac{5}{18}
-\frac{2\gamma_E}{3}
+\frac{4}{3}\ln 2
-\frac{2}{3}\ln q$ and $\beta^{(I)}_{T_q,2}(0)=1$.

If we ignore the $\mathcal{O}(\rho^2q^2)$ correction, the matrix element of full effective quark EMT by combining \eqref{eq:qEMT1} and \eqref{eq:qEMT2} resembles the EMT matrix element derived from effective 't Hooft Lagrangian \eqref{THOOFT1}, which is the leading single instanton part in \eqref{eq:Leff}. 
More specifically, the full quark EMT is given by

\begin{equation}
\begin{aligned}
   \langle h'|T^{q}_{\mu\nu}|h\rangle=&\langle N'|\bar{\psi}\gamma_{(\mu} i\overleftrightarrow{\partial}_{\nu)}\psi|N\rangle-g_{\mu\nu}\left(\frac{n_{I+A}}2\right)\frac{1}{4N_c}\left(\frac{4\pi^2\rho^2}{m^*}\right)\langle N'|\bar\psi\psi|N\rangle\\
   &-\left(\frac{n_{I+A}}2\right)\frac{1}{8N_c}\left(\frac{4\pi^2\rho^2}{m^*}\right)\beta_{T_q,1}(\rho q)\rho^2\left(q_\mu q_\nu-\frac14 g_{\mu\nu}q^2\right)\langle N'|\bar\psi\psi|N\rangle\\
    &-\left(\frac{n_{I+A}}2\right)\frac{1}{2N_c}\left(\frac{4\pi^2\rho^2}{m^*}\right)\left(\beta_{T_q,2}(\rho q)-1\right)g_{\mu\nu}\langle N'|\bar\psi\psi|N\rangle
\end{aligned}
\end{equation}

This reflects the reduction of the full QCD equations of motion to those of the effective 't Hooft Lagrangian at low resolution, corresponding to the deeply cooled regime in which dynamical gluon modes are depleted.

\begin{equation}
\label{eq:EOM}
    \langle h|\bar\psi\left(i\slashed{D}-m\right)
\psi|h\rangle\rightarrow\langle h|\bar\psi\left(i\slashed{\partial}-m+ \sum_{a}[G_{a}
\bar\psi(x)\Gamma_a\psi(x)]\, \Gamma_a\right)
\psi|h\rangle=0,
\end{equation}
where $\Gamma_a$ here denotes the spin-flavor structure induced by instantons and $G_a$ is the coupling in the corresponding channel. With Dyson-Schwinger relation given by \eqref{eq:EOM}, the hadronic matrix element of 't Hooft EMT is equivalent to the quark EMT in ILM.

\begin{equation}
\begin{aligned}
\langle h'|T^{q}_{\mu\nu}|h\rangle=\langle h'|T^{\rm 'tHooft}_{\mu\nu}|h\rangle+\mathcal{O}(\rho^2q^2)
\end{aligned}
\end{equation}
where the 't Hooft EMT is defined as
\begin{equation}
\begin{aligned}
T^{\rm 'tHooft}_{\mu\nu}(x)
&=
\bar\psi(x)\,
i\gamma_{(\mu}\overleftrightarrow{\partial}_{\nu)}
\psi(x)
-
g_{\mu\nu}
\bar\psi(x)
\left(i\overleftrightarrow{\slashed{\partial}}-m\right)
\psi(x)
\\[4pt]
&\quad
-
\frac12 g_{\mu\nu}
\sum_a G_a
\left(\bar\psi(x)\Gamma_a\psi(x)\right)^2 .
\end{aligned}
\end{equation}

As discussed in Sec.~\ref{sec:GFF}, the connection to the UV dynamics is restored through the molecular contribution where $T_{\mu\nu}^g$ starts to have nontrivial values. As illustrated in Fig.~\ref{fig:ILM_exp2g}, for the traceless gluonic EMT, the hadronic matrix element receives contributions solely from instanton–anti-instanton molecular interactions.

\begin{equation}
\begin{aligned}
\label{eq:EMTFF}
   &\langle h'|g^2\bar{T}^g_{\mu\nu}|h\rangle=
  \frac{n_{IA}\gamma_{IA}}{2N_c(N_c^2-1)}\Bigg\{\frac{4\pi^2}{3}\beta^{(IA)}_{T_g,1}(\rho q)\left(g_{\mu\rho} g_{\nu\lambda}+g_{\nu\rho} g_{\mu\lambda}\right)\\
  &-\frac{2\pi^2\rho^2}{9}\beta^{(IA)}_{T_g,2}(\rho q)\left(q_\mu q_\rho g_{\nu\lambda}+q_\nu q_\rho g_{\mu\lambda}-\frac{1}{2} g_{\mu\nu}q_\rho q_\lambda\right)\\
    &\quad-2\pi^2\rho^4\beta^{(IA)}_{T_g,3}(\rho q)\left(q_\mu q_\nu-\frac{1}{4}q^2g_{\mu\nu}\right)q_\rho q_\lambda\Bigg\}\langle h'|\bar{\psi}\left(\gamma_{(\rho} i\overleftrightarrow{\partial}_{\lambda)}-\frac{1}{4}g_{\rho\lambda}i\overleftrightarrow{\slashed{\partial}}\right)\psi|h\rangle
\end{aligned}
\end{equation}
where the instanton profile form factors are given by

\begin{equation}
\label{BETAEMT}
    \beta^{(IA)}_{T_g,1}(q)=\frac{1}{q}\int_0^\infty dx\left[\frac{24}{(1+x^2)^4}J_1(qx)+\frac{24x^2}{(1+x^2)^4}J_3(qx)-\frac{192}{(1+x^2)^3}\frac{J_3(qx)}{q^2x^2}\right]
\end{equation}

\begin{equation}
    \beta^{(IA)}_{T_g,2}(q)=\frac{1}{q}\int_0^\infty dx9x^2\left[\frac{128x^2}{(1+x^2)^4}\frac{J_3(qx)}{q^2x^2}-\frac{512}{(1+x^2)^3}\frac{J_4(qx)}{q^3x^3}\right]
\end{equation}

\begin{equation}
    \beta^{(IA)}_{T_g,3}(q)=\frac{1}{q}\int_0^\infty dx\frac{256x^4}{(1+x^2)^3}\frac{J_5(qx)}{q^4x^4}
\end{equation}
They are normalized to 1 and their momentum dependence is illustrated in Fig.~\ref{fig:betaTg}.

\begin{table*}
    \centering
    \begin{tabular}{|c|c|c|c|c|c|c|}
   \hline
      $n_{I+A}$ & $\rho$ & $m^*$ & $M$ & $n_{IA}$ & $\gamma_{IA}$  \\ 
    \hline
    1.0 fm$^{-4}$ & $0.313$ fm & $100.70$ MeV & $395.17$ MeV & 0.0127 fm$^{-4}$ & $41.13$ fm$^4$ \\
   \hline
\end{tabular}
    \caption{ILM parameters used in this section for gluonic EMT form factors with experimental inputs $m_\pi=138$ MeV, $m_N=938$ MeV.}
    \label{tab:parameters_ILM_FF}
\end{table*}

\subsubsection{Pion}

For pion, the gravitational4 form factors are defined by

\begin{align}
\label{THETA123}
&\langle\pi'|T^{q,g}_{\mu\nu}|\pi\rangle=2 {\bar{p}^\mu \bar{p}^\nu}\,A^{q,g}_\pi(Q^2)+
\frac 12 ({q^\mu q^\nu}-g^{\mu\nu}q^2)\,D^{q,g}_\pi(Q^2)
\end{align}
where we have omitted the anomalous trace term $\propto g^{\mu\nu}$ which will be dropped immediately later when we consider traceless part only.

The $A$ form factor, or the mass distribution, is the component of the EMT related to the mass/energy
distribution and $D$ is related to the mechanical property (shear and pressure) inside the hadron.
The trace form factor for pion is defined as
\begin{align}
\label{pi_scalar_tr}
&\langle \pi'|T_{\mu\mu}|\pi\rangle=2m_\pi^2T_\pi(Q^2)
\end{align}

The trace form factor can be related to the scalar form factors and gravitational form factors
\begin{equation}
\begin{aligned}
    T_\pi(Q^2)=&\left(1+\frac{Q^2}{4m_\pi^2}\right)A_\pi(Q^2)+\frac{3Q^2}{4m_\pi^2}D_\pi(Q^2)\\
    \approx&\,G_\pi(Q^2)+\sigma_\pi(Q^2)
\end{aligned}
\end{equation}
One may define $T_\pi^q$ and $T_\pi^g$ in terms of the separated quark and gluon gravitational form factors. However, this decomposition is subtle. In general, quark and gluon EMT operators mix under renormalization, and therefore such quantities cannot be unambiguously separated or directly mapped onto one another, especially in UV region \cite{Hatta:2018sqd}.
\begin{figure}
    \centering
    \includegraphics[width=1\linewidth]{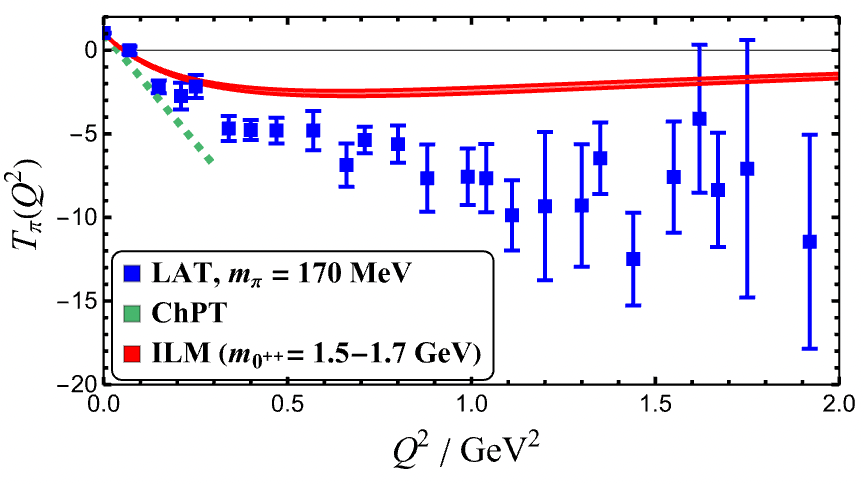}
    \caption{Pion trace form factor. Blue square represents the reconstruction of the lattice results for $T_\pi(q)$ using pion gravitational $A,D$ form factors from~\cite{Hackett:2023nkr} with pion mass $m_\pi=170$ MeV, the green dashed curve represents the ChPT prediction in \cite{Chen:1997zza,Novikov:1980fa}, and the red curve represents the ILM result with the band indicating the range of scalar glueball mass $1.5-1.7$ GeV }
    \label{fig:T_pi}
\end{figure}

In Fig.~\ref{fig:T_pi},  we show our result for the trace of the EMT $T_\pi(q)$ in the ILM (red-solid line), versus the reconstructed lattice results (blue squares) from
\cite{Hackett:2023nkr}, and the result of ChPT (green-dashed line). In the reconstruction of the lattice results for $T_\pi(q)$, we made use of the lattice calculations of pion gravitational $A,D$ form factors from~\cite{Hackett:2023nkr} where the errors were generated by error propagation. The resulting form factor is weakly dependent on both the glueball mass and sigma term $\sigma_{\pi\pi}$, as indicated in Fig.~\ref{fig:T_pi}.

To proceed the evaluation of pion EMT form factors $A_\pi$ and $D_\pi$, the mass form factor $A_\pi(Q^2)$ is estimated by isoscalar tensor meson dominance with mass $m_{f_2}=1.275$ GeV.

\begin{equation}
    A_\pi(Q^2)=\frac{1}{1+Q^2/m^2_{f_2}}
\end{equation}
and the internal force form factor $D_\pi(Q^2)$ is estimated by mixing between the isoscalar tensor meson and scalar meson exchange given by $G_\pi(Q^2)$, $\sigma_\pi(Q^2)$ and $A_\pi(Q^2)$ due to trace identity.

\begin{equation}
\begin{aligned}
    D_\pi(Q^2)=\frac{4m_\pi^2}{3Q^2}\left[G_\pi(Q^2)+\sigma_\pi(Q^2)-\left(1+\frac{Q^2}{4m_\pi^2}\right)A_\pi(Q^2)\right]
\end{aligned}
\end{equation}

The pion gluonic EMT form factor is evaluated by

\begin{equation}
\begin{aligned}
\label{eq:A_pi}
   &A_\pi^g(Q^2)=
   \frac{n_{IA}}{2N_c(N_c^2-1)}\gamma_{IA}\frac{8\pi^2}{3}\beta^{(IA)}_{T_g,1}(\rho q)A_\pi^q(Q^2)
\end{aligned}
\end{equation}

\begin{equation}
\begin{aligned}
\label{eq:D_pi}
   &D_\pi^g(Q^2)=\\
   &\frac{n_{IA}\gamma_{IA}}{2N_c(N_c^2-1)}\Bigg\{\Bigg[\frac{8\pi^2}{3}\beta^{(IA)}_{T_g,1}(\rho q)+\frac{\pi^2}{3}\rho^2Q^2\beta^{(IA)}_{T_g,2}(\rho q)-\frac{3\pi^2}2\rho^4Q^4\beta^{(IA)}_{T_g,3}(\rho q)\Bigg]D_\pi^q(Q^2)\\
    &\qquad\qquad+\frac{4\pi^2}9\rho^2m_\pi^2\left(\beta^{(IA)}_{T_g,2}(\rho q)-\frac92\rho^2Q^2\beta^{(IA)}_{T_g,3}(\rho q)\right)\left(1+\frac{Q^2}{4m_\pi^2}\right)A_\pi^q(Q^2)\Bigg\}
\end{aligned}
\end{equation}

At non-zero momentum transfer, Eq.~\eqref{eq:A_pi} show that the gluonic gravitational form factor in a hadron is dominated by tensor meson ($f_2$) exchanges, while Eq.~\eqref{eq:D_pi} mixes with the scalar meson ($\sigma$) and tensor meson ($f_2$).

\begin{figure}
    \centering
\subfloat[\label{fig:A_p}]{\includegraphics[width=.5\linewidth]{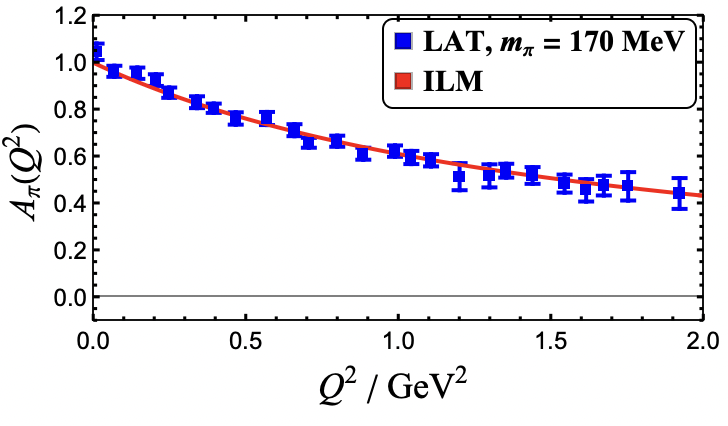}}
\hfill
\subfloat[\label{fig:D_p}]{\includegraphics[width=.5\linewidth]{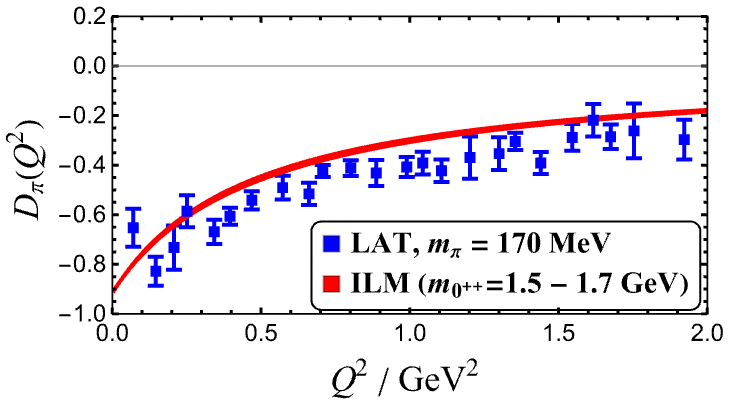}}
    \caption{Pion form factors $A_\pi(Q^2)$ (left red line) and $D_\pi(Q^2)$
(right red line) in the ILM, compared to the lattice results (blue square)~\cite{Hackett:2023nkr}}
    \label{fig:GFF_pi}
\end{figure}

In Fig.~\ref{fig:A_p}, we show  our result for the pion gravitational form factor $A_\pi$ in the ILM (red-solid line) in comparison to the lattice results (blue squares)~\cite{Hackett:2023nkr}. We have used the standard ILM parameters given in Table~\ref{tab:parameters_ILM_FF} and \ref{tab:mass_ILM}, with a mean instanton size $\rho=0.313\,\rm fm$ and determinantal mass $m^*=100.7\,\rm MeV$. The agreement of the ILM and lattice results is excellent.

Due to mixing with the trace anomaly, the result depends on the scalar glueball mass where we take $m_{0^{++}}=1.5-1.7$ GeV and $\sigma_{\pi\pi}=0.484m_\pi$, as indicated by the red band. At zero momentum transfer, $D_\pi=-0.915$ (for $m_{0^{++}}=1.6$ GeV) in consistency with the ChPT prediction $D_\pi=-1$. The resulting form factor is weakly dependent on both the glueball mass and sigma term $\sigma_{\pi\pi}$.

\begin{figure}
    \centering
\subfloat[\label{fig:A_qg_pi}]{\includegraphics[width=.5\linewidth]{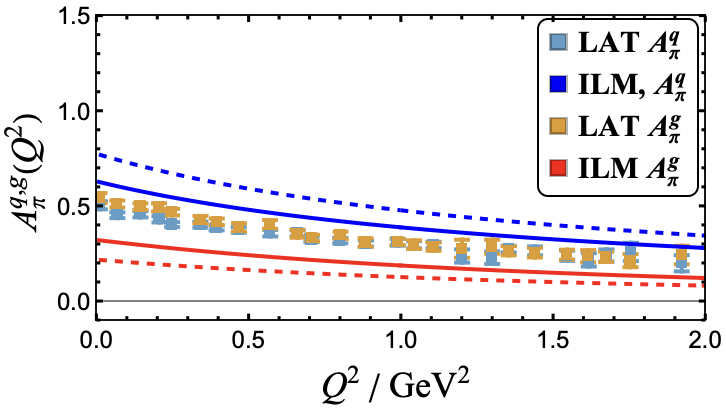}}
\hfill
\subfloat[\label{fig:D_qg_pi}]{\includegraphics[width=.5\linewidth]{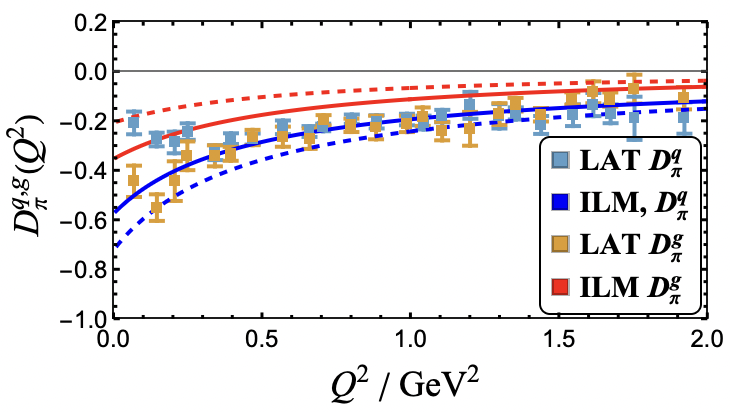}}
    \caption{(a) Quark contribution $A^q_\pi(Q^2)$ (blue) and gluon contribution $A^g_\pi(Q^2)$ (red) to the pion gravitational form factor in the ILM at a resolution $\mu=1/\rho$ (dashed lines) and at a resolution $\mu=2\,\rm GeV$ (solid lines). The lattice results are from~\cite{Hackett:2023nkr}. 
  (b) Quark contribution $D^q_\pi(Q^2)$ (blue) and gluon contribution $D^g_\pi(Q^2)$ (red) to the pion gravitational form factor in the ILM at a resolution $\mu=1/\rho$ (dashed lines) and at a resolution $\mu=2\,\rm GeV$ (solid lines). The lattice results are from~\cite{Hackett:2023nkr}.}
    \label{fig:GFF_qg_pi}
\end{figure}

For the separate gravitational form factors, quark and gluon contributions are scale-dependent.  In Fig.~\ref{fig:A_qg_pi}, we show the separate quark and gluon contributions to the pion gravitational form factor $A$ in the ILM. At the resolution $\mu=1/\rho$, the quark contribution $A_\pi^q(Q^2)$ (blue-dashed line) overshadows the gluon contribution $A_\pi^g(Q^2)$ (orange-dashed line). Under 
QCD evolution, the charges are evolved to the resolution
$\mu=2\,\rm GeV$ where the quark contribution $A_\pi^q(Q^2)$ 
(blue solid line) becomes comparable to the 
gluon contribution $A_\pi^g(Q^2)$ (orange solid line). They
are in good agreement with the 
lattice results for $A_\pi^q(Q^2)$ (blue squares) and 
$A_\pi^g(Q^2)$ (orange squares) at the scale $\mu=2\,\rm GeV$~\cite{Hackett:2023nkr}. The same
results and comparison for the quark and gluon contributions to the pion $D$ form factor are shown in Fig.~\ref{fig:D_qg_pi}.

At zero momentum transfer, the gravitational form factors $A_h^{q,g}(0)$ correspond to the momentum fractions carried by quarks and gluons inside a hadron. In ILM, these quantities are related by the molecular configurations and constrained by momentum conservation. Since RG evolution is independent of the hadronic state, $A^{q,g}(0)$ is universal for different hadrons, such as the pion and the nucleon. The distinction between them arises only through their momentum-transfer dependence, thus the gravitational $A$ form factor radius.

\begin{table}
    \centering
    \begin{tabular}{|c|c|c|c|c|}
   \hline
      & $A_\pi^q(0)$ & $A_\pi^g(0)$  & $D_\pi^q(0)$  & $D_\pi^g(0)$ \\ 
    \hline
   ILM ($0.6$ GeV) & 0.714 & 0.286 & $-0.713$ & $-0.202$    \\
   \hline
   ILM ($2$ GeV) & 0.581 & 0.419 & $-0.567$ & $-0.348$ \\
   \hline
   MIT ($m_\pi=450$ MeV, 2018) \cite{Shanahan:2018pib} & -- & 0.58(5) & -- & $-1.2(1)$ \\
   \hline
\end{tabular}
    \caption{The $A$ charge (quark and gluon momentum fraction) and $D$ charge using parameters in Table~\ref{tab:parameters_ILM_FF} with $m_{0^{++}}=1.6$ GeV and $\sigma_{\pi\pi}=0.484m_\pi$ are evolved to $2$ GeV and compared to lattice calculation using a clover improved quark action and Lüscher-Weisz gauge action in a $L^3 \times T= 323 \times 96$, with large pion mass $m_\pi=450$ MeV and lattice spacing $a=0.1167(16)$ fm}
    \label{tab:GFF_pi}
\end{table}

\subsubsection{Nucleon}

For nucleon (spin-$1/2$), the gravitational form factors are defined by

\begin{equation}
\begin{aligned}
\label{T_traceless}
    \langle N'|T^{q,g}_{\mu\nu}|N\rangle=&\bar{u}_{s'}(p')\bigg(A_N^{q,g}(Q^2)\frac{\bar{p}^{\mu} \bar{p}^{\nu}}{m_N}+J_N^{q,g}(Q^2)\frac{i\bar{p}^{(\mu}\sigma^{\nu)\alpha}q_\alpha}{m_N}\\
    &+D_N^{q,g}(Q^2)\frac{1}{4m_N}\left(q^{\mu}q^{\nu}-g^{\mu\nu}q^2\right)\bigg)u_{s}(p)
    \end{aligned}
\end{equation}
where we have omitted the anomalous trace term $\propto g^{\mu\nu}$ which will be dropped immediately later when we consider traceless part only. The new component $J$ form factor is related to the angular momentum distribution inside the nucleon. 

The trace form factor for a nucleon is defined as
\begin{align}
\label{N_scalar_trace}
&\langle N'|T^{\mu}{}_{\mu}|N\rangle=m_NT_N(Q^2)\bar{u}_{s'}(p')u_{s}(p)
\end{align}
Using the trace identity in Eq.~\eqref{3X}, the trace form factor relates the scalar form factors and gravitational form factors by
\begin{equation}
\begin{aligned}
    T_N(Q^2)
    \equiv&\left(1+\frac{Q^2}{4m_N^2}\right)A_N(Q^2)-\frac{Q^2}{2m_N^2}J_N(Q^2)+\frac{3Q^2}{4m_N^2}D_N(Q^2)\\
    \approx&G_N(Q^2)+\sigma_N(Q^2)
\end{aligned}
\end{equation}

One may define $T_N^q$ and $T_N^g$ in terms of the separated quark and gluon gravitational form factors. However, one should keep in mind that comparing these individual contributions with the trace-anomaly or scalar form factors is incorrect. In general, quark and gluon EMT operators mix under renormalization, and therefore such quantities cannot be unambiguously separated or directly mapped onto one another.

\begin{figure}
    \centering
    \includegraphics[width=1\linewidth]{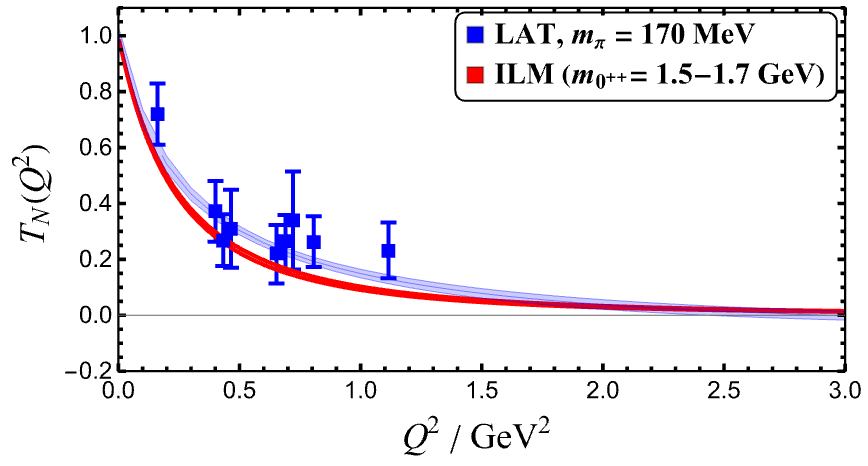}
    \caption{Nucleon trace form factor in \eqref{N_scalar} with $\sigma_{\pi N}=53$ MeV where the red band indicates the range of $m_{0^{++}}=1.5-1.7$ GeV.}
    \label{fig:T_N}
\end{figure}

In Fig.~\ref{fig:T_N}, we show our result for the trace of the EMT $T_N(q)$ in the ILM (red-solid line), versus the reconstructed lattice results combining nearest data point (blue squares) and dipole fit (blue band) from
\cite{Hackett:2023rif}. In the reconstruction of the lattice results for $T_N(q)$, we made use of the lattice calculations of nucleon gravitational $A,D$ form factors from~\cite{Hackett:2023rif} where the errors were generated by error propagation in blue squares and the band in the blue curve indicate the uncertainty from dipole fitting. The resulting form factor is weakly dependent on both the glueball mass and sigma term $\sigma_{\pi N}$.

To proceed the evaluation of nucleon EMT form factors $A_N$ and $D_N$, the mass form factor $A_N(Q^2)$ is estimated by isoscalar tensor meson dominance with mass $m_{f_2}=1.275$ GeV.

\begin{equation}
    A_N(Q^2)=\frac{1}{(1+Q^2/m^2_{f_2})^2}
\end{equation}
and the internal force form factor $D_N(Q^2)$ is estimated by mixing between the isoscalar tensor meson, scalar meson and scalar glueball exchange given by $A_N(Q^2)$, $J_N(Q^2)$, $\sigma_N(Q^2)$, and $G_N(Q^2)$ due to trace identity.

\begin{equation}
\begin{aligned}
    D_N(Q^2)=\frac{4m_N^2}{3Q^2}\left[G_N(Q^2)+\sigma_N(Q^2)-\left(1+\frac{Q^2}{4m_N^2}\right)A_N(Q^2)+\frac{Q^2}{2m_N^2}J_N(Q^2)\right]
\end{aligned}
\end{equation}

Using the result in Eq.~\eqref{eq:EMTFF}, the nucleon gravitational form factor is given by

\begin{equation}
\begin{aligned}
   &A_N^g(Q^2)=
   \frac{n_{IA}}{2N_c(N_c^2-1)}\gamma_{IA}\frac{8\pi^2}{3}\beta^{(IA)}_{T_g,1}(\rho q)A_N^q(Q^2)
\end{aligned}
\end{equation}

\begin{equation}
\begin{aligned}
   &J_N^g(Q^2)=
   \frac{n_{IA}}{2N_c(N_c^2-1)}\gamma_{IA}\frac{8\pi^2}{3}\beta^{(IA)}_{T_g,1}(\rho q)J_N^q(Q^2)
\end{aligned}
\end{equation}

\begin{equation}
\begin{aligned}
   &D_N^g(Q^2)=\\
   &\frac{n_{IA}\gamma_{IA}}{2N_c(N_c^2-1)}\Bigg\{\Bigg[\frac{8\pi^2}{3}\beta^{(IA)}_{T_g,1}(\rho q)+\frac{\pi^2}{3}\rho^2Q^2\beta^{(IA)}_{T_g,2}(\rho q)-\frac{3\pi^2}2\rho^4Q^4\beta^{(IA)}_{T_g,3}(\rho q)\Bigg]D_N^q(Q^2)\\
    &+\frac{4\pi^2}9\rho^2m_N^2\left(\beta^{(IA)}_{T_g,2}(\rho q)-\frac92\rho^2Q^2\beta^{(IA)}_{T_g,3}(\rho q)\right)\left(A_N^q(Q^2)-\frac{Q^2}{4m_N^2}B_N^q(Q^2)\right)\Bigg\}
\end{aligned}
\end{equation}

\begin{figure}
    \centering
\subfloat[\label{fig:A_N}]{\includegraphics[width=.5\linewidth]{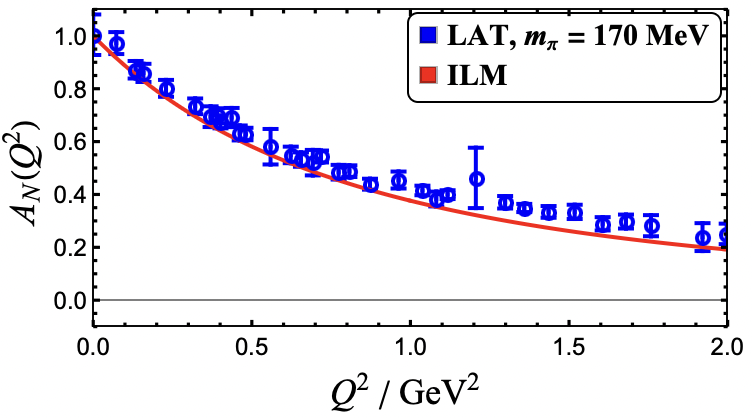}}
\hfill
\subfloat[\label{fig:D_N}]{\includegraphics[width=.5\linewidth]{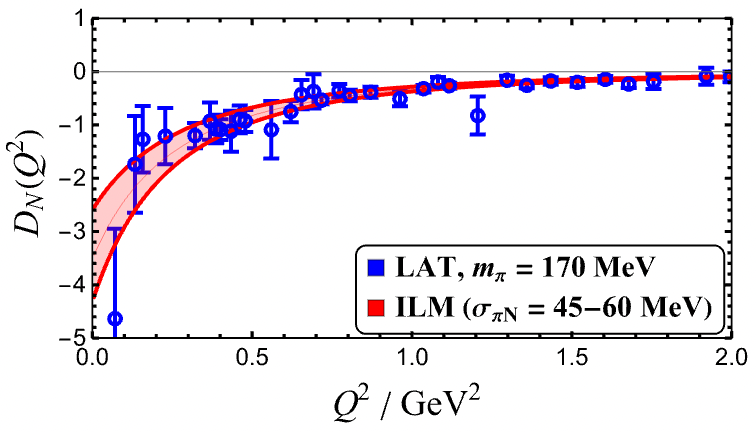}}
    \caption{Nucleon gravitational form factors $A_N(Q^2)$ and $D_N(Q^2)$ in the ILM with red bands indicating the range of $\sigma_{\pi N}=45-60$ MeV, compared to the fitted lattice results~\cite{Hackett:2023nkr} using dipole form.}
\label{fig:GFF_N}
\end{figure}

In Fig.~\ref{fig:A_N}, we show  our result for the nucleon gravitational form factor $A_N$ in the ILM (red-solid line) in comparison to the lattice results (blue squares)~\cite{Hackett:2023nkr}. We have used the standard ILM parameters given in Table~\ref{tab:parameters_ILM_FF} and \ref{tab:mass_ILM}, with a mean instanton size $\rho=0.313\,\rm fm$ and determinantal mass $m^*=100.7\,\rm MeV$. The agreement of the ILM and lattice results is excellent. 

In Fig.~\ref{fig:D_N}, we show our result
for the $D_N$ (red curve) versus the lattice results (blue squares)~\cite{Hackett:2023nkr} where the red band represents the sensitivity of $D_N$ to pion-nucleon sigma term in the range $45-60$ MeV with $m_{0^{++}}=1.6$ GeV. At zero momentum transfer, the nucleon $D$ charge is given by $D_N=-3.51$ using $\sigma_{\pi N}=53$ MeV and $m_{0^{++}}=1.6$ GeV. Its absolute value is of the same order as the lattice result, $D_N^{\rm lat} = -3.872(97)$~\cite{Hackett:2023nkr}, as well as predictions from various low-energy models. The discrepancy is partly driven by the sensitivity to the pion–nucleon sigma term $\sigma_{\pi N}$. For $\sigma_{\pi N} = 45$ MeV, we obtain $D_N = -2.58$, while for $\sigma_{\pi N} = 60$ MeV, the value shifts to $D_N = -4.31$.

In Table~\ref{tab:GFF}, we obtained the nucleon $A$ and $D$ charge using ILM parameters in Table~\ref{tab:parameters_ILM_FF} with $m_{0^{++}}=1.6$ GeV and $\sigma_{\pi\pi}=0.484m_\pi$, compared with various lattice calculations and phenomenological analysis. In \cite{Shanahan:2018pib} and \cite{Pefkou:2021fni}, they presents lattice calcualtion for gluonic gravitational form factor for pion, nucleon, and rho meson using a large quark mass corresponding to $m_\pi\sim450$ MeV, while in \cite{Hackett:2023rif} and \cite{Alexandrou:2020sml}, the results are obtained at physical pion mass.

\begin{table}
    \centering
    \begin{tabular}{|c|c|c|c|c|}
   \hline
      & $A_N^q(0)$ & $A_N^g(0)$  & $D_N^q(0)/4$  & $D_N^g(0)/4$ \\ 
    \hline
   ILM ($0.6$ GeV) & 0.714 & 0.286 & $-0.703$ & $-0.180$   \\
    \hline
   ILM ($2$ GeV) & 0.581 & 0.419 & $-0.553$ & $-0.325$ \\
    Asymptotic \cite{Gross:1974cs,Politzer:1974sm} & $0.529$ & $0.471$ & -- & -- \\
    MIT (2018) \cite{Shanahan:2018pib} & -- & 0.54(8) & -- & $-2.50(75)$ \\
    MIT (2022) \cite{Pefkou:2021fni} & 0.57(4) & 0.429(39) & -- & $-0.485$ \\
    MIT (2023) \cite{Hackett:2023rif} & $0.510(25)$ & $0.501(27)$ & $-0.325(12)$ & $-0.643(21)$\\
    ETMC \cite{Alexandrou:2020sml} & 0.599(60) & 0.427(92) & & \\
    Global Analysis \cite{Hou:2019efy} & $0.58(1)$ & $0.414(8)$ & -- & -- \\
    DVCS \cite{Burkert:2018bqq} &-- & -- & $-0.408(28)(33)$ & --  \\
    \hline
\end{tabular}
    \caption{Nucleon gravitational charge $A$ and $D$ estimated using $\sigma_{\pi N}=53$ MeV and $m_{0^{++}}=1.6$ GeV, compared with various lattice calculations and global analyses.}
    \label{tab:GFF}
\end{table}

\begin{table}
    \centering
    \begin{tabular}{|c|c|c|c|c|c|c|}
   \hline
     & ILM  & CSM \cite{Jung:2013bya} & Skyrme \cite{Cebulla:2007ei} & $\chi$QSM \cite{Wakamatsu:2007uc} & MIT \cite{Hackett:2023rif} \\ 
    \hline
   $D_N(0)$  & $-3.51$ & $-5.03$ & $-4.48$ &  $-2.35$ & $-3.872(97)$  \\
    \hline
\end{tabular}
    \caption{Nucleon $D$ charge in ILM ($N_f=2$) using $\sigma_{\pi N}=53$ MeV and $m_{0^{++}}=1.6$ GeV, compared with various chiral models and lattice calucation from MIT group \cite{Hackett:2023rif}.}
    \label{tab:GFF2}
\end{table}

\begin{figure}
    \centering
\subfloat[\label{fig:A_qg_N}]{\includegraphics[width=.5\linewidth]{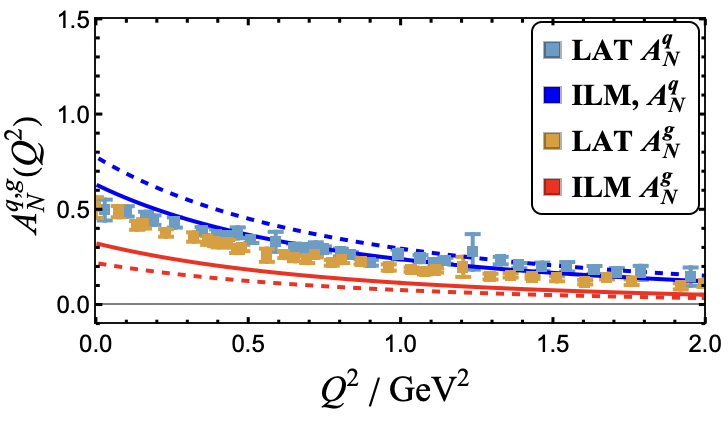}}
\hfill
\subfloat[\label{fig:D_qg_N}]{\includegraphics[width=.5\linewidth]{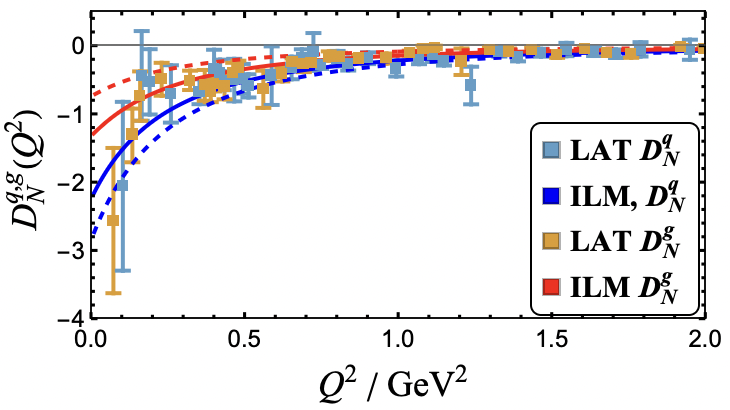}}
    \caption{(a) Quark contribution $A^q_N(Q^2)$ (blue) and gluon contribution $A^g_N(Q^2)$ (orange) to the pion gravitational form factor in the ILM at a resolution $\mu=1/\rho$ (dashed lines) and at a resolution $\mu=2\,\rm GeV$ (solid lines). The lattice results are from~\cite{Hackett:2023nkr}. 
  (b) Quark contribution $D^q_N(Q^2)$ (blue) and gluon contribution $D^g_N(Q^2)$ (orange) to the pion gravitational form factor in the ILM at a resolution $\mu=1/\rho$ (dashed lines) and at a resolution $\mu=2\,\rm GeV$ (solid lines). The lattice results are from~\cite{Hackett:2023nkr}.}
    \label{fig:GFF_qg_N}
\end{figure}

\section{Pseudoscalar and Axial form factor}

Similar to the gluonic scalar operator, pseudoscalar $F\tilde F$ is also dominated by vacuum fluctuation in the disconnected diagram, as illustrated in Fig.~\ref{fig:ILM_exp1}.

\begin{equation}
\begin{aligned}
\label{eq:GGtilde}
    \langle h'| \frac{g^2}{32\pi^2}F\tilde{F}|h\rangle&=-2m_h^2 \left(\bar{N}\frac{\partial \ln m_h}{\partial \Delta}\right)\left(\frac{\chi_t}{\bar N}\right)\frac{(2\pi)^4\delta^4(q)}{V}\\
    &-\frac{n_{IA}\gamma_{IA}}{9N_c(N_c^2-1)}\beta^{(IA)}_{G\tilde{G}}(\rho q)\rho^2q_\mu q_\nu\langle h'|\bar{\psi}\left(\gamma_{(\mu} i\overleftrightarrow{\partial}_{\nu)}-\frac{1}{4}g_{\mu\nu}i\overleftrightarrow{\slashed{\partial}}\right)\gamma^5\psi|h\rangle
\end{aligned}
\end{equation}
where the instanton form factors are defined as

\begin{align}
\label{BETAGGTILDE}
\beta^{(IA)}_{G\tilde{G}}(q)
&=
\frac{1}{q}\int_0^\infty dx\,\frac{576x^2}{(1+x^2)^4}\frac{J_3(qx)}{q^2x^2}=\beta^{(IA)}_{GG}(q)
\end{align}
For a nucleon state, the last matrix element is zero by construction. 

It is clear that the universal fluctuation dominates the gluonic pseudoscalar form factor at the leading order of the $1/N_c$ expansion. The $\delta^4(q)$ contribution reflects on the homogeneous correlation between pseudoscalar glueball, where the pseudoscalar probe is disconnected from the hadronic source and exclusively sourced by equally distributed semiclassical instanton and
anti-instanton fields (mean field). Similarly, one can smear the momentum dependence to a connected correlation function

\begin{equation}
\frac{\chi_t}{\bar{N}}\frac{(2\pi)^4}{V}\delta^4(q)\rightarrow
\frac{\chi_t(Q^2)}{\bar N}\equiv\frac{1}{32\pi^2\langle F^2\rangle}\int d^4xe^{-iq\cdot x}\langle F\tilde{F}(x)F\tilde{F}(0)\rangle_c
\end{equation}
where in quenched vacuum, the topological susceptibility is of the order of the instanton density (see Sec.~\ref{sec:ILM_QCD})

\begin{equation}
\chi_t\simeq\bar{N}
\end{equation}

This quenched topological susceptibility can be related to the meson singlet mass by the Witten-Veneziano formula~\cite{Witten:1979vv,Veneziano:1979ec}
\begin{equation}
\frac{\chi_t}V=\frac{f_\pi^2}{2N_f}\left(m_{\eta^\prime}^2+m_\eta^2-2m_K^2\right)
\end{equation}
This implies that the mass of $\eta'$ is essentially due to the axial anomaly relating to non-trivial topological charge fluctuations, which can turn out to be nonzero even in the chiral limit. However, in (unquenched) QCD, $\chi_t$ is substantially screened by the light quarks~\cite{Diakonov:1995qy,Kacir:1996qn} (see Sec.~\ref{sec:ILM_QCD})
and Witten-Veneziano formula is modified by $m_{\eta^\prime}^2+m_\eta^2-2m_K^2\rightarrow m_\pi^2$.

At zero momentum transfer in Eq.~\eqref{eq:GGtilde}, the anomalous gluonic contribution to the intrinsic spin is corrected by quark mixing induced 'tHooft type quark-instanton interaction. The result reads

\begin{equation}
\begin{aligned}
\label{eq:GGtil}
\Delta \tilde{g}\equiv\frac{\langle h| \frac{g^2}{32\pi^2}F\tilde{F}|h\rangle}{im_hq\cdot S}=& -\left(\frac{\chi_t}{\bar N}\right)\frac{2m_h}{iq\cdot S}\left(\bar N\frac{\partial m_h}{\partial \Delta}\right)
\approx&\frac{M}{m_h}\frac{\chi_t}{\bar N}
\end{aligned}
\end{equation}
where $iq\cdot S=\bar{u}_s(p')i\gamma^5u_s(p)$ is the pseudoscalar spinor product, $M$ is the constituent mass \eqref{eq:cons_m} and $\Delta\tilde{q}$ is the pseudoscalar charge of nucleon defined in Eq.~\eqref{eq:ABJ}. The leading contribution arises from vacuum fluctuations (disconnected diagrams), with a subleading quark mixing correction induced by the emergent ’t Hooft-type quark–instanton interaction. 

In quenched limit, the anomalous spin $\Delta\tilde{g}$ is purely determined by topological fluctuation. With $M=395$ MeV, $m_N=938$ MeV, and topological susceptibility $(\chi_t/V)^{1/4} \simeq 153$--$180~\mathrm{MeV}$ as indicated in Table~\ref{tab:chi}, the quenched anomalous spin reads

\begin{equation}
    \Delta\tilde{g}^{(\rm quench)}=0.152 - 0.292
\end{equation}

However, the story changes drastically in the unquenched vacuum. The topological susceptibility is strongly suppressed by quark-mass screening, leaving $(\chi_t/V)^{1/4} \simeq 73$ -- $76~\mathrm{MeV}$. Thus, in unquenched vacuum, 

\begin{equation}
\label{eq:unq}
    \Delta\tilde{g}^{(\rm unquench)}\simeq(7.89\,\mathrm{-}\,9.27)\times10^{-3}.
\end{equation}
This indicates that the reduction induced by sea quark is crucial in determining $\Delta \tilde{g}$. 

For the pseudoscalar charge, using $m_N=938$ MeV, $m_{\eta'}=958$ MeV \cite{ParticleDataGroup:2024cfk}, phenomenologically estimated  nucleon-$\eta'$ coupling $g_{NN\eta'}= 1.4 \pm 1.1$ \cite{Singh:2010wd,Nasrallah:2005hn,Feldmann:1999uf}, and pseudoscalar current coupling in $\eta'$ meson $h^{u+d+s}_{\eta'}=0.0347(29)$ GeV$^3$ \cite{Beneke:2002jn} (mostly dominated by strange quark), ILM simple estimation on the pseudoscalar current density $\Delta \tilde{q}$ for $N_f=3$ is obtained by
\begin{equation}
\begin{aligned}
\label{eq:dtq}
    \Delta\tilde{q}=\frac{g_{NN\eta'}h^{u+d+s}_{\eta'}}{m_{\eta'}^2m_N}=0.06\pm0.04
\end{aligned}
\end{equation}

 

At non-zero momentum transfer, Eq.~\eqref{eq:GGtilde} shows that the pseudoscalar gluonic $F\tilde F$ in a hadron couples primarily to pseudoscalar glueball exchange with mixing of the isoscalar pseudoscalar meson ($\eta'$) and isoscalar spin-2 meson with negative parity ($\pi_2$) channel

Without loss of generality, we will discuss nucleon. The nucleon pseudoscalar form factors are defined by
\begin{align}
\label{N_pseuoscalar}
&\langle N'|\frac{g^2}{32\pi^2}F\tilde{F}|N\rangle=m_N\tilde{G}_N(Q^2)\bar{u}_{s'}(p')i\gamma^5u_s(p)\nonumber\\[5pt]
&\langle N'|m\bar\psi i\gamma^5\psi|N\rangle=m_N\tilde{G}_P(Q^2)\bar{u}_{s'}(p')i\gamma^5u_s(p)
\end{align}
and the axial form factor along with the induced pseudoscalar form factor from the axial current matrix element is defined by

\begin{equation}
\begin{aligned}
\label{N_axial}
    &\langle N'|\bar{\psi}\gamma_\mu\gamma^5\psi|N\rangle=\bar{u}_{s'}(p')\left(\gamma_\mu\gamma^5G_A(Q^2)+\gamma^5\frac{q_\mu}{2m_N}G_P(Q^2)\right)u_s(p)
\end{aligned}
\end{equation}

By ABJ anomaly, the pseudoscalar form factor can be related to the axial form factor in

\begin{equation}
    G_A(Q^2)-\frac{Q^2}{2m^2_N}G_P(Q^2)=N_f \tilde{G}_N(Q^2)+\tilde{G}_P(Q^2)
\end{equation}

At zero momentum transfer, this relation connects the relation between the intrinsic quark spin and axial anomaly, as shown in Eq.~\eqref{eq:ABJ} where the instrinsic quark spin $\Sigma_q=G_A(0)$, anomalous gluon helicity $\Delta\tilde{g}=\tilde{G}_N(0)$, and pseudoscalar current density $\Delta\tilde{q}=\tilde{G}_P(0)$.

At non-zero momentum transfer, the nucleon gluonic pseudoscalar form factor is approximated by \cite{Kacir:1996qn},

\begin{equation}
\begin{aligned}
    \tilde{G}_N(Q^2)\approx&\Delta\tilde{g}\frac{\chi_t(Q^2)}{\chi_t}\frac{\left[1+\mathcal{O}(\rho^2Q^2)\right]}{(1+Q^2/m^2_{\eta'})^2}
\end{aligned}
\end{equation}
where the nucleon gluonic pseudoscalar form factors $\tilde{G}_N$ is dominated by pseudoscalar gluball exchange with mixing from $\eta'$ meson channel. The gluonic pseudoscalar correlation function is parameterized by a monopole form with the pseudoscalar glueball mass $m_{0^{-+}}=2.5$ GeV (see Table~\ref{tab:mass_ILM})

\begin{equation}
    \chi_t(Q^2)=\frac{\chi_t}{1+Q^2/m_{0^{-+}}^2}
\end{equation} 

For the pseudoscalar quark form factor $\tilde{G}_P(Q^2)$, we parameterize it by isoscalar pseudoscalar meson dominance ($\eta'$) 

\begin{equation}
    \tilde{G}_P(Q^2)=\frac{\Delta \tilde{q}}{(1+Q^2/m^2_{\eta'})^2}
\end{equation}
with $\eta'$ meson mass $m_{\eta'}=682$ MeV estimated by ILM in Table~\ref{tab:mes0_spec}. 

\section{Hadron mass structure in ILM}

Although the trace anomaly in Eq.~\eqref{3X} explains the strong-interaction nature of the hadronic mass origin in, it does not provide a detailed decomposition of this mass in terms of the underlying hadronic constituents.

In a strongly interacting theory, the concept of constituents is resolution dependent. Fortunately, the QCD instanton vacuum emerging from cooled lattice
simulations, allows for a quantitative description, all within the well-defined framework of semi-classics. In this spirit, a  physically  motivated proposal to budget the mass, was 
put forth by Ji in~\cite{Ji:1994av,Ji:1995sv},  and  since revisited by many~\cite{Lorce:2017xzd,Roberts:2021xnz,Metz:2020vxd} (and references therein). The ensuing mass composition
involves the sum of partonic contributions, some of which may be measurable using deep inelastic scattering (DIS) experiments.
The proposal relies on an invariant decomposition of the energy momentum tensor (\ref{EMT_decop}).

The tracefull and traceless part of the
energy momentum tensor  (\ref{EMT_decop}) correspond to the spin-2 and spin-0 representations of the Lorentz group, and do not mix  under renormalization by symmetry, as we noted earlier. 
Their renormalization  at  the  instanton size scale $\rho\approx 0.3~{\rm fm}$ is  achieved by cooling through gradient flow, under the constaint of a fixed topological susceptibility.   Note that this renormalization scale is softer than the one  used in currently  fine lattices with $1/\mu\approx 0.1\,{\rm fm}$ ($\overline{\rm MS}$ scheme)~\cite{Yang:2018nqn}.

 With this in mind, the  corresponding Hamiltonian in Minkowski signature, follows from the 00-component of (\ref{EMT_decop})
\begin{align}
\label{T5X}
H_g&=\int d^3x\,\frac 12(E^2+B^2)\nonumber\\
H_q&=\int d^3x\, \overline\psi \gamma\cdot  i\overleftrightarrow D\psi\nonumber\\
H_a&=\int d^3x\,\frac 14 \left(\frac{\beta(g^2)}{4g^2}F^2\right)\nonumber\\
H_\sigma&=\int d^3x\, m\overline\psi\psi
\end{align}

We rearranged the quark mass term so that the hadron mass distribution is then defined by

\begin{equation}
\begin{aligned}
\label{T6}
m_h=&\frac{\langle h|H_g+H_q+H_a+H_\sigma|h\rangle}{\langle h|h\rangle}\\
=&m^g_h+m^q_h+m^a_h+m^\sigma_h
\end{aligned}
\end{equation}
with the identification
\begin{equation}
m^{(\rm inv)}_h=m^g_h+m^q_h+m^a_h
\end{equation}

\subsection{Mass sum rule}

With the scalar and gravitational form factors determined, we can now calculate the mass sum rule for a pion and a nucleon.
\begin{equation}
\begin{aligned}
\frac{m^q_h}{m_h} &= \frac{3}{4} \left(A^q_h(0)-\frac{\sigma_{\pi h}}{m_h}\right), 
&\quad \frac{m^g_h}{m_h} &= \frac{3}{4} A^g_h(0), \\
\frac{m^a_h}{m_h} &= \frac{1}{4}\left(1-\frac{\sigma_{\pi h}}{m_h}\right), 
&\quad \frac{m^\sigma_h}{m_h} &= \frac{\sigma_{\pi h}}{m_h}.
\end{aligned}
\end{equation}

\subsubsection{Pion}
\begin{figure}
    \centering
\subfloat[\label{fig:PIEVAC1}]{\includegraphics[width=.5\linewidth]{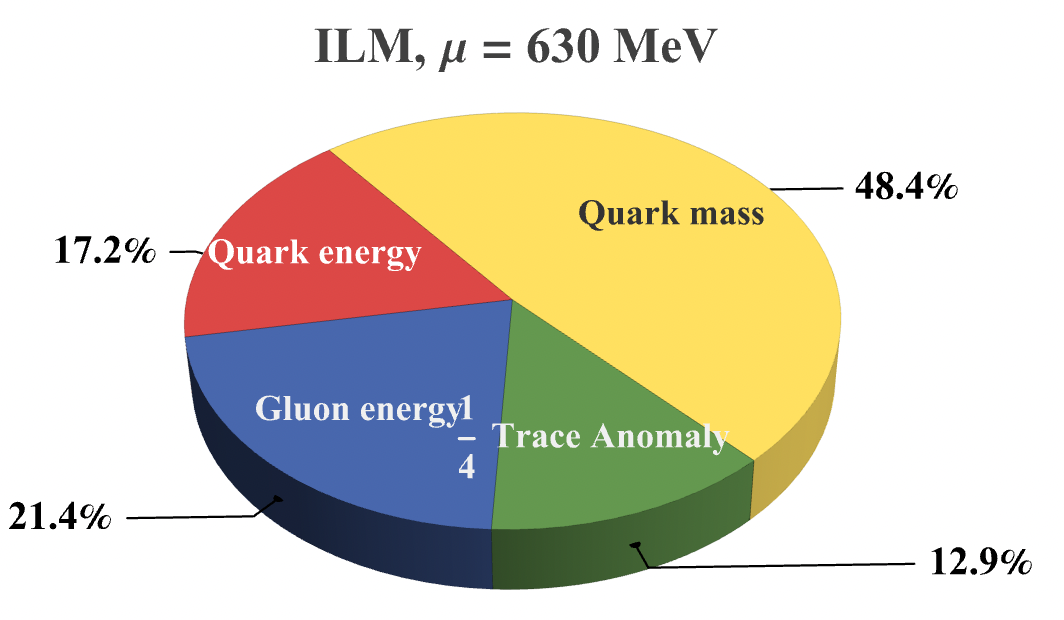}}
\hfill
\subfloat[\label{fig:PIEVAC2}]{\includegraphics[width=.5\linewidth]{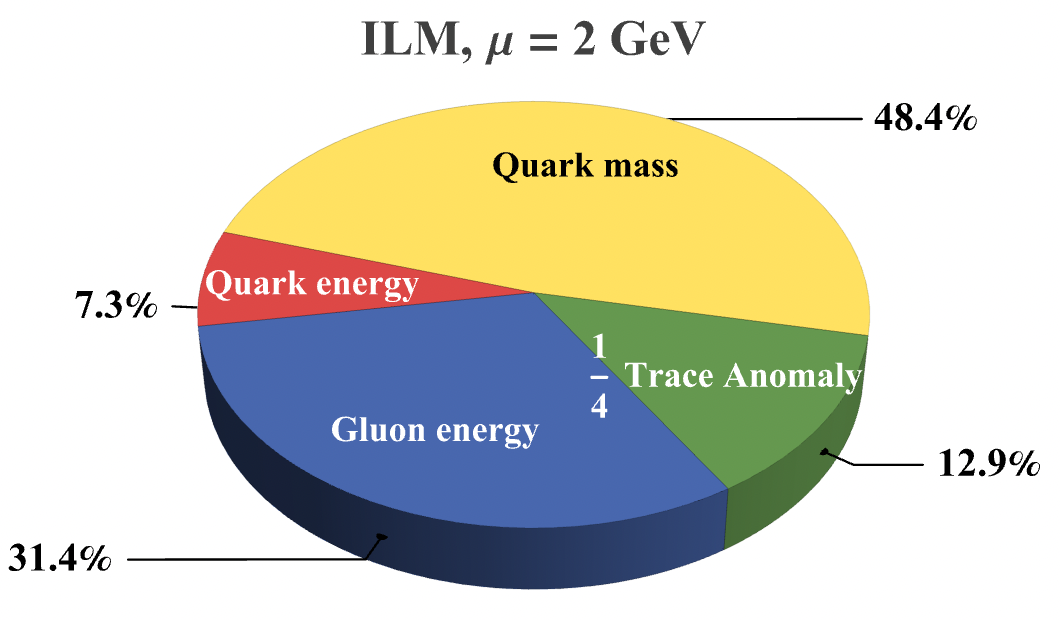}}
    \caption{Pion mass decomposition using Ji's mass sum rule, in the QCD instanton vacuum at the resolution $\mu\sim1/\rho$ (a), and after DGLAP evolution at a resolution $\mu=2~\mathrm{GeV}$ (b)}
    \label{fig:PIEVAC}
\end{figure}

At the resolution of the order of the inverse instanton size, we have

\begin{equation}
\begin{aligned}
\label{T8}
\frac{m_\pi^q}{m_\pi}&\approx   17.25\%, \quad \frac{m_\pi^g}{m_\pi}&\approx  21.43\%, \quad
\frac{m_\pi^a}{m_\pi}&\approx 12.89\%, \quad
\frac{m_\pi^\sigma}{m_\pi}&\approx 48.42\%
\end{aligned}
\end{equation}
using the empirical pion sigma term~\cite{Hoferichter:2016ocj,Alarcon:2021dlz}.
\eqref{T8} shows that in the QCD instanton vacuum, about  17\%  of the pion mass stems from the valence quarks hopping around zero modes, 13\% from the gluon condensate (displaced vacuum instanton field), 
and 21\% from emerging perturbative gluons. This is illustrated in Fig.~\ref{fig:PIEVAC1}.

The budgeting of the pion mass evolves as the energy scale varies. At $\mu=2$ GeV, the valence quark and gluon energy contribution redistributes as illustrated in Fig.~\ref{fig:PIEVAC2}. Under Dokshitzer–Gribov–Lipatov–Altarelli–Parisi (DGLAP) evolution, the gluons carry a larger energy fraction at the expense of the quarks. 

\begin{equation}
\frac{m_\pi^q}{m_\pi}\approx   7.29\%, \qquad \frac{m_\pi^g}{m_\pi}\approx   31.39\%
\end{equation}

\subsubsection{Nucleon}
\begin{figure}
    \centering
\subfloat[\label{fig:PIEVACN1}]{\includegraphics[width=.5\linewidth]{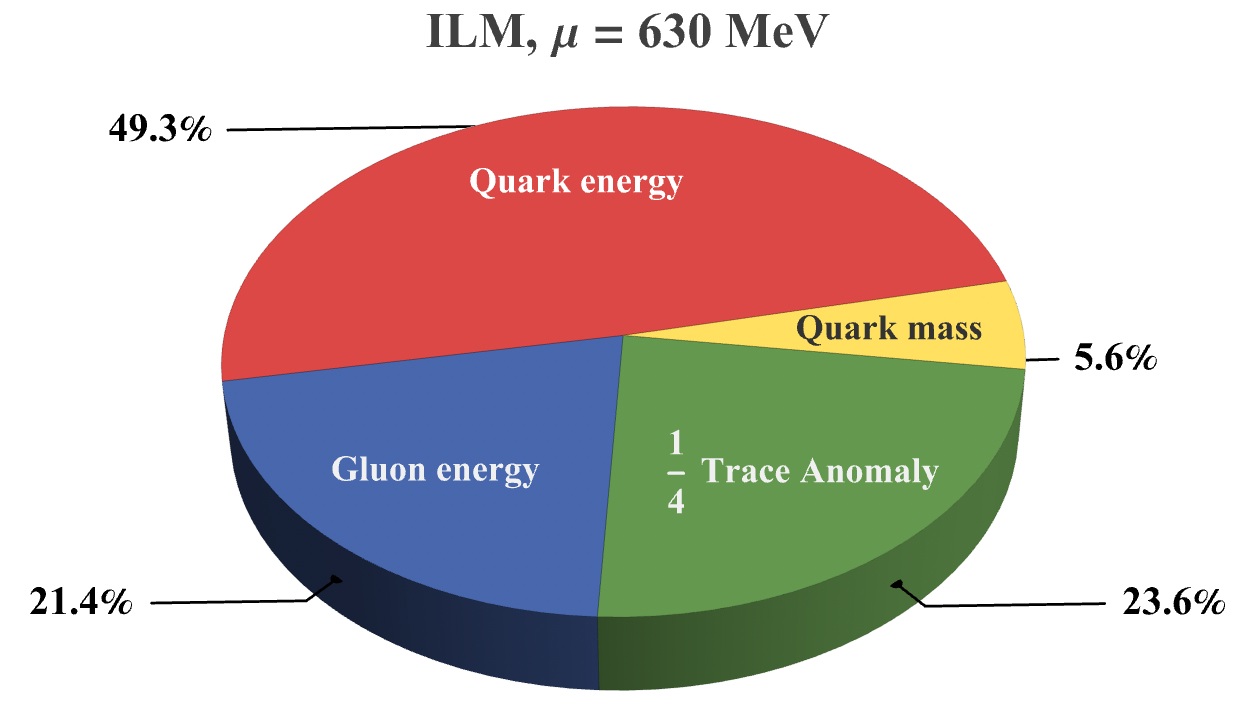}}
\hfill
\subfloat[\label{fig:PIEVACN2}]{\includegraphics[width=.5\linewidth]{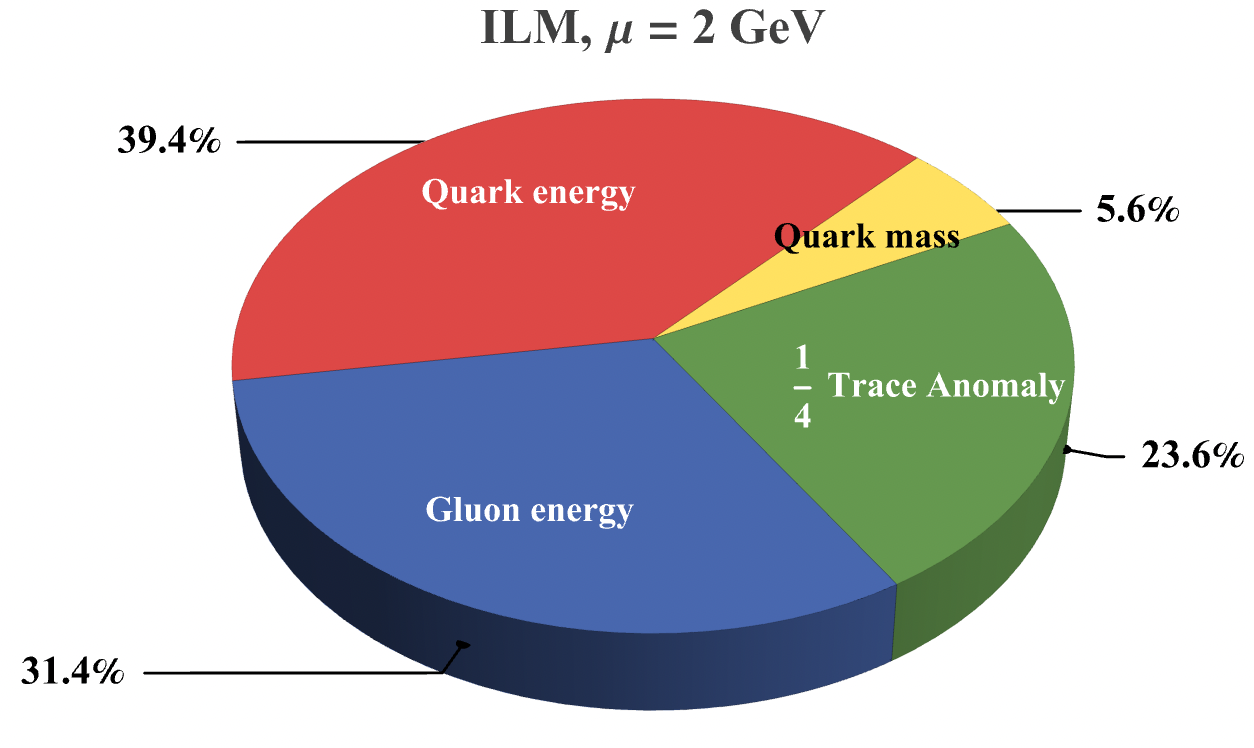}}
\hfill
\subfloat[\label{PIELAT}]{\includegraphics[width=.5\linewidth]{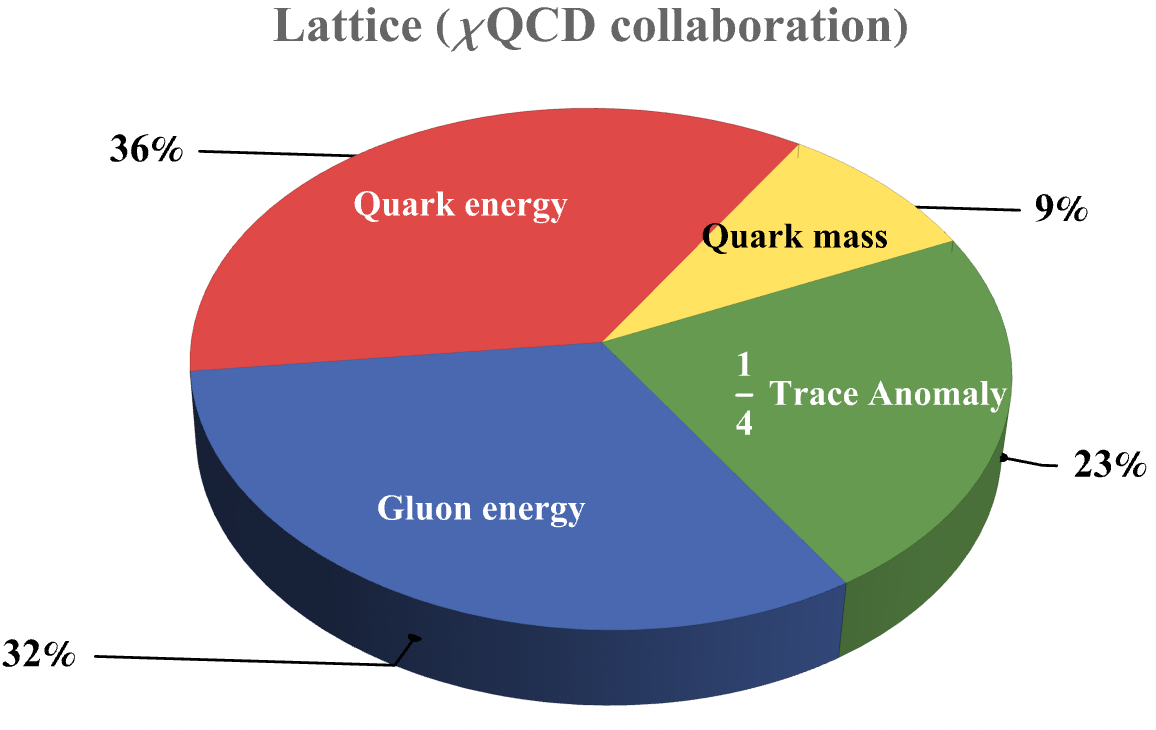}}
\caption{Nucleon mass decomposition using Ji's mass sum rule in the QCD instanton vacuum (a) at the resolution $\mu\sim1/\rho$, and (b) after DGLAP evolution at a resolution $\mu=2~\mathrm{GeV}$ ; (c) Ji's mass decomposition obtained from the lattice collaboration~\cite{Yang:2018nqn} at a resolution of $\mu=2$ GeV.}
    \label{fig:PIEVACN}
\end{figure}

At the resolution of the order of the inverse instanton size, we have
\begin{equation}
\label{T8N}
\begin{aligned}
\frac{m_N^q}{m_N} &\approx 49.33\%, \quad 
\frac{m_N^g}{m_N} &\approx 21.44\%, \quad 
\frac{m_N^a}{m_N} &\approx 23.59\%, \quad
\frac{m_N^\sigma}{m_N} &\approx 5.65\%
\end{aligned}
\end{equation}
using the empirical pion-nucleon sigma term~\cite{Hoferichter:2016ocj,Alarcon:2021dlz}.
\eqref{T8N} shows that in the QCD instanton vacuum, about 49\%  of the nucleon mass stems from the valence quarks (hopping zero modes),  24\% from the gluon condensate (displaced vacuum instanton field), 
and 21\% from emerging  valence gluons. This is illustrated in Fig.~\ref{fig:PIEVACN}. The nucleon is composed mostly of quark constituents hopping and scooping the vacuum gluon fields.  This result is consistent with the one observed in~\cite{Zahed:2021fxk}.

The budgeting of the nucleon mass evolves as the energy scale varies. At $\mu=2$ GeV, the valence quark and gluon energy contribution redistributes as illustrated in~\ref{fig:PIEVACN2}. The  budgeting of the nucleon mass in
(\ref{T9}) from the QCD instanton vacuum, is consistent with the one reported on the lattice in~\cite{Yang:2018nqn} from $\chi$QCD as illustrated in Fig.~\ref{PIELAT}. Under DGLAP evolution,   the gluons carry a larger energy fraction at the expense of the quarks. 

\begin{equation}
\label{T9}
\frac{m_N^q}{m_N}\approx   39.37\%\nonumber, \qquad \frac{m_N^g}{m_N}\approx   31.39\%\nonumber\\
\end{equation}

\section{Nucleon spin structure in ILM}
\label{sec:N_spin}
The spin structure of the nucleon has been addressed both empirically and theoretically by many, and we refer to the review in~\cite{Deur:2018roz} for an exhaustive account and references. Here, we  will address it in the QCD instanton vacuum following  recent estimates by one of us~\cite{Zahed:2021fxk}, using Ji's spin decomposition~\cite{Ji:1996ek},
\begin{equation}
    S_N=\frac{1}{2}\Sigma_{q}+L_{q}+J_{g}
\end{equation}
where $S_N=1/2$ is the nucleon spin. 

\subsubsection{Quark intrinsic spin}

\begin{table}[]
    \centering
    \begin{tabular}{|c|c|c|c|}
    \hline
       $\Sigma_q$  & $N_f\Delta \tilde g$ & $\Delta \tilde q$   \\
    \hline
         0.25 & 0.0138 & 0.236 \\ 
         0.30 & 0.0138 & 0.286 \\ 
         0.35 & 0.0138 & 0.336 \\ 
    \hline
    \end{tabular}
    \caption{ILM ($N_f=3$) estimate of the pseudoscalar charge in Eq.~\eqref{eq:ABJ} as a function of the intrinsic quark spin, with fixed $n_{I+A}=1~\mathrm{fm}^{-4}$, quark condensate $\langle \bar\psi \psi \rangle = (272~\mathrm{MeV})^3$~\cite{FlavourLatticeAveragingGroup:2019iem}, quark mass $m=3.387~\mathrm{MeV}$~\cite{FlavourLatticeAveragingGroupFLAG:2024oxs}, and quark mass screened topological susceptibility $\chi_t$ given in Eq.~\eqref{SUS}.}
    \label{tab:quark_spin}
\end{table}

The quark intrinsic spin $\Sigma_q$, given by
\begin{align}
\label{spin}
\Sigma_q=&\int d^3\vec{x} \bar{\psi}\vec{\gamma}\gamma^5\psi
\end{align}
captures the isoscalar axial charge inside the hadron, which is best described by the hadronic matrix element of the flavor-singlet axial current,
\begin{equation}
    \langle N|\bar{\psi}\gamma_\mu\gamma^5\psi|N\rangle=2m_N\Sigma_qS_\mu
\end{equation}
Here $S_\mu$ is the spin vector of the nucleon with the normalization $S^2=-1$ and $p\cdot S=0$.

The quark intrinsic spin is tied to the pseudoscalar glueball operator by the ABJ anomaly~\cite{Adler:1969gk}, as defined in Eq.~\eqref{eq:ABJ_clean}.

\begin{equation}
\label{chiral_anomaly}
\partial_\mu \bar{\psi}\gamma_\mu\gamma^5\psi = \frac{N_f}{16\pi^2} g^2F_{\mu\nu}\tilde{F}_{\mu\nu}+2m\bar{\psi}i\gamma^5\psi
\end{equation}

The chiral anomaly, or the ABJ anomaly reflects on the general fact that all intrinsic quark spin inside hadron are tied to the quantum breaking of $U(1)$ chiral symmetry, and should be enforced by any non-perturbative QCD description. The  quark intrinsic spin consists of the anomalous gluonic contribution $\Delta\tilde{g}$ proportional to flavor number $N_f$ plus the explicit breaking of the $U(1)$ axial symmetry by the quark current mass $m$ which is associated to the pseudoscalar current density $\Delta\tilde{q}$.
Thus the intrinsic quark spin can be decomposed into \cite{Zahed:2022wae} (and references therein).

\begin{equation}
\label{eq:ABJ}
    \Sigma_q=N_f\Delta\tilde{g}+\Delta\tilde{q}
\end{equation}
where the instrinsic quark spin $\Sigma_q=G_A(0)$, anomalous gluon helicity $\Delta\tilde{g}=\tilde{G}_N(0)$, and pseudoscalar current density $\Delta\tilde{q}=\tilde{G}_P(0)$.

The intrinsic quark spin has been the subject of extensive experimental and theoretical investigation for decades. It is therefore important for us to provide at least an estimate. While a direct calculation of the intrinsic quark spin would require the full nucleon wave function in ILM, one may still obtain an estimate of the order of magnitude from the anomaly equation. In Table~\ref{tab:quark_spin}, we present estimates of $\Delta \tilde{g}$ and $\Delta \tilde{q}$ for different choices of the intrinsic quark spin, taken within the range $\Sigma_q = 0.25$ -- $0.35$ \cite{Nocera:2014gqa}. In light of this, we neglect the pseudoscalar current density and restrict our discussion to the quenched vacuum first, where sea quarks are absent. In this limit, the intrinsic quark spin is fully determined by quenched $\Delta \tilde{g}^{(\rm quench)}$, which originates purely from topological fluctuations. Using $M = 395~\mathrm{MeV}$, $m_N = 938~\mathrm{MeV}$, and $n_{I+A} = 1~\mathrm{fm}^{-4}$, together with $(\chi_t/V)^{1/4}$ in the range $153$--$180~\mathrm{MeV}$ for the quenched vacuum as listed in Table~\ref{tab:chi}, we obtain the following estimate for $N_f = 3$ in the quenched limit:

\begin{equation}
\label{eq:qspin_1}
   \Sigma_q^{u+d+s}\approx N_f\frac{M}{m_N}\frac{\chi_t/V}{n_{I+A}}\approx 0.457 - 0.875
\end{equation}
The result is very sensitive to the details of the interaction among pseudoparticles, as discussed in Sec.~\ref{sec:ILM_QCD}. In contrast, the mechanism changes significantly in the unquenched vacuum. The topological susceptibility is strongly suppressed by quark-mass screening, with $(\chi_t/V)^{1/4} \simeq 73$--$76~\mathrm{MeV}$, reflecting the substantial reduction of the gluonic anomalous contribution, and hence lowering the quenched estimate. Using the previous estimate on $\Delta\tilde{g}$ \eqref{eq:unq} and $\Delta\tilde{q}$ \eqref{eq:dtq}, the intrinsic quark spin in unquenched ILM ($N_f=3$) is given by

\begin{equation}
\label{eq:qspin_2}
    \Sigma_q^{u+d+s}=0.09\pm0.04
\end{equation}

\begin{table*}[]
    \centering
    \begin{tabular}{|c|c|c|c|c|c|}
    \hline
         $N_f$ & \footnotesize MSU \cite{Lin:2018obj} & \footnotesize $\chi$QCD \cite{Liang:2018pis} & \footnotesize ETMC \cite{Alexandrou:2020sml}  & \footnotesize NNPDF\cite{Nocera:2014gqa} & \footnotesize DSSV\cite{deFlorian:2009vb} \\
    \hline
       $2$  & 0.339(30)(42) & 0.440(24)(40) & 0.438(23) & 0.35 $\pm$ 0.06 & $0.38 \pm 0.04$  \\
       $3$  & 0.286(31)(36) & 0.405(25)(37) & 0.392(24) & $0.25\pm0.10$ & $0.366^{+0.042}_{-0.062}$ \\
    \hline
     \end{tabular}
    \caption{$\Sigma_q$ obtained by various lattice group, compared to ILM prediction in Eqs.~\eqref{eq:qspin_1}amd \eqref{eq:qspin_2}.}
\end{table*}

\subsubsection{Quark OAM}

The quark orbital angular momentum (OAM) is given by

\begin{align}
L_q=&\int d^3\vec{x} \bar{\psi}\gamma^0\vec{x}\times i\vec{D}\psi
\end{align}

Combined with the intrinsic quark spin, we have the quark total angular momentum

\begin{equation}
    J_q=\frac{1}{2}\Sigma_q+L_q
\end{equation}

The quark total angular momentum $J_q$ is calculated by the forward hadronic matrix element of the traceless part of the quark $0j$-EMT, 
\begin{equation}
\label{SPINQ}
\begin{aligned}
J_q &= \frac{\langle PS|\int d^3\vec{x}\,\epsilon^{3ij}x^i T_q^{0j}|PS\rangle}{\langle PS|PS\rangle} \\
&= \epsilon^{3ij} i \partial_q^i
\frac{\langle P'S|T_q^{0j}|PS\rangle}{\langle PS|PS\rangle}
\end{aligned}
\end{equation}
where $\partial^i_q$ refers to the derivative with respect to the momentum transfer, followed by the zero momentum transfer limit. 

\subsubsection{Gluon angular momentum}

Similarly, the angular momentum carried by the gluons is

\begin{align}
J_g=&\int d^3\vec{x} \vec{x}\times(\vec{E}^a\times\vec{B}^a)
\end{align}
which translates to the forward hadronic matrix element.
\begin{equation}
\label{SPING}
\begin{aligned}
J_g &= \frac{\langle PS|\int d^3\vec{x}\,\epsilon^{3ij} x^i T_g^{0j}|PS\rangle}{\langle PS|PS\rangle} \\
&= \epsilon^{3ij} i \partial_q^i
\frac{\langle P'S|T_g^{0j}|PS\rangle}{\langle PS|PS\rangle}
\end{aligned}
\end{equation}

\subsection{Spin sum rule}

\begin{table}[]
    \centering
    \begin{tabular}{|c|c|c|}
    \hline
         & $J_q$ & $J_g$ \\
    \hline
        ILM ($\mu=2$ GeV) & $0.291$ & $0.209$\\
        QCDSF-UKQCD \cite{QCDSF-UKQCD:2007gdl} & $0.226(13)$ & -- \\
        LHPC  \cite{LHPC:2007blg} & $0.213(44)$ & -- \\
        MIT \cite{Hackett:2023rif} & 0.251(21) &  0.255(13) \\
    \hline
    \end{tabular}
    \caption{The quark and gluon angular momentum estimated in ILM is compared to various lattice result. We assume $B^q_N$ and $B^g_N$ are zero, as indicated in Sec.~\ref{sec:GFF}}
    \label{tab:J}
\end{table}

Now with the gravitational form factors, pseudoscalar form factor, and axial form factors are discussed and evaluated. Now we can calculate the spin sum rule. The following hadron spin distribution is then defined by
\begin{align}
\frac{\frac{1}{2}\Sigma_q}{S_N} &= \Delta \tilde{q}(0)+N_f \Delta\tilde{g} \\
\frac{L_q}{S_N} &= 2J_q(0)-\frac{\frac{1}{2}\Sigma_q}{S_N} \\
\frac{J_g}{S_N} &= 2J_g(0)
\end{align}

We begin by comparing the total angular momentum carried by quarks and gluons inside the nucleon. In this analysis, we set $B_N^q$ and $B_N^g$ to zero, as discussed in Sec.~\ref{sec:GFF}. The resulting ILM estimates, summarized in Table~\ref{tab:J}, are then compared with lattice QCD calculations. In particular, we refer to the results of the QCDSF Collaboration~\cite{QCDSF-UKQCD:2007gdl}, obtained using $N_f = 2$ clover-improved Wilson fermions with chiral extrapolation to physical pion mass, the LHPC~\cite{LHPC:2007blg}, based on $N_f = 2 + 1$ mixed-action simulations using chiral extrapolation with the pion mass as small as $350$ MeV and volumes as large as $(3.5~\mathrm{fm})^3$ for a lattice spacing of $0.124$ fm, and those of the \cite{Hackett:2023rif} at physical pion mass $170$ MeV.

In unquenched $N_f=2$ QCD in the instanton vacuum, the topological susceptibility screened. At the resolution of the order of the inverse instanton size, Ji's spin sum rule is given by

\begin{equation}
\label{S8}
\frac{\frac{1}{2}\Sigma_q}{S_N}\approx 35\%,\quad
\frac{L_q}{S_N}\approx 36.4\%,\quad
\frac{J_g}{S_N}\approx 28.6\%
\end{equation}
Here we assumed $J_N^q(0)=\frac12A_N^q(0)$ \cite{Mamo:2022eui}, hence $J_N^g(0)=\frac12A_N^g(0)$. This assumption is well-consistent with various lattice studies.

Eq.~\eqref{S8} shows that in the QCD instanton vacuum, about $35\%$  of the nucleon spin stems from the spin of the valence quarks as they hop and mix with the vacuum topological charge fluctuations, $36\%$ stems from their orbital angular motion (OAM), and $29\%$ from the emerging gluons as the topological charge fluctuates.

The budgeting of the nucleon spin evolves as the energy scale varies. For the QCD instanton vacuum with 2 flavors at $\mu=2$ GeV, the valence quark OAM and gluon angular momentum redistributes as
\begin{equation}
\label{S9}
\frac{\frac{1}{2}\Sigma_q}{S_N}\approx 35\%,\quad
\frac{L_q}{S_N}\approx 23.14\%,\quad
\frac{J_g}{S_N}\approx 41.86\%
\end{equation}

In the QCD instanton vacuum, the intrinsic spin does not evolve with energy scale as it captures the vacuum topological
susceptibility scooped by the nucleon. As a result, DGLAP
evolution enhances the gluon contribution at the sole expense of the quark orbital contribution, both of which are not topological in our analysis.

\begin{figure}
    \centering
\includegraphics[width=1\linewidth]{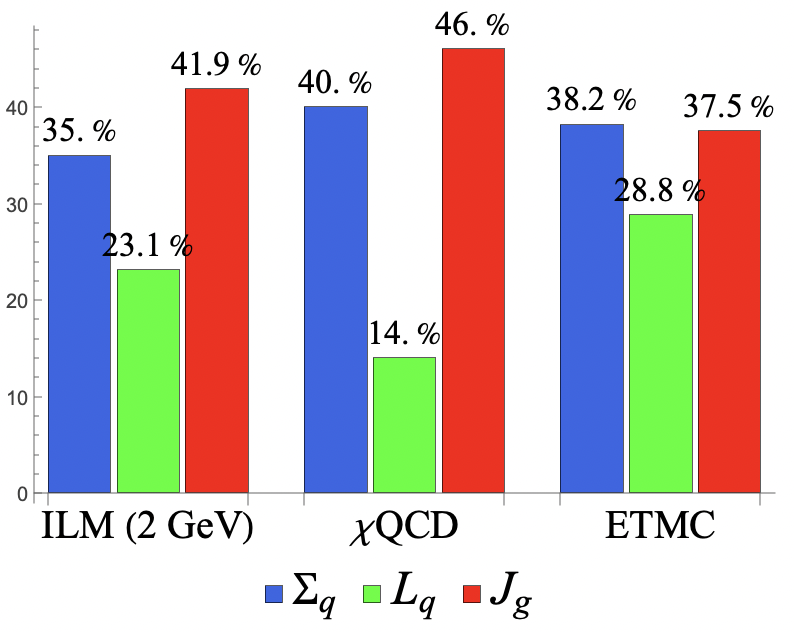}
    \caption{Spin decomposition using Ji's sum rule using $\Sigma_q=0.35$ as input at a resolution of $\mu=2\,{\rm GeV}$, compared to the lattice collaboration $\chi$QCD ~\cite{Wang:2021vqy} and ETMC~\cite{Alexandrou:2020sml}.}
    \label{fig:spin}
\end{figure}

The results for Ji's spin decomposition in the QCD instanton vacuum, are illustrated in Fig.~\ref{fig:spin} (unquenched)  at $\mu=0.63$ GeV and at $\mu=2$ GeV after one-loop DGLAP evolution. 
They are compared to the reported 
results from the $\chi$QCD collaboration~\cite{Wang:2021vqy} in Fig.~\ref{fig:spin},
and from the ETMC collaboration~\cite{Alexandrou:2020sml} 
in Fig.~\ref{fig:spin}.

While the gluonic contributions are comparable to the one reported by both lattic collaborations, there is a difference in the way the quarks 
are carrying the spin. In the lattice, the intrinsic spin to OAM ratio is about 3:1 ($\chi$QCD) and 2:1 (ETMC), which is to be compared to 
1:1 in the QCD instanton vacuum with two flavors. The origin is the substantial depletion of the intrinsic spin at low resolution, owing to the strong screening of the large volume topological susceptibility. 

\chapter{Wilson lines in ILM}
\label{ch:wilson}
Similar to the discussion in Ch.~\ref{ch:ILM} for the local QCD operators with gluonic fields, the framework can now be extended to non-local gluonic operators such as Wilson lines and Wilson loop.

For a generic Wilson loop, the nonperturbative contribution  at low resolution (following from say gradient cooling) can be estimated using 
an ensemble of topologically active gauge configurations made chiefly of instantons and anti-instantons. In the dilute regime, the full instanton contribution to the Wilson loop is given by the exponent of the all-order single instanton result~\cite{Shuryak:2000df,Shuryak:2021hng,Dorokhov:2002qf}
\begin{equation}
    W(C)=\exp\left[\frac{1}{N_cV}\sum_I\int d^4z ~\mathrm{Tr}_c\left(W_I(\rho,z)-1\right)\right]
\end{equation}
with the single instanton insertion
\begin{equation}
W_I(\rho,z)=e^{i\tau^a\phi_I^a(\rho,z)}\,,
\end{equation}
where the single-instanton profile in the soft Wilson loop is defined as 
\begin{equation}
\label{single_inst_wilson}
        \phi_I^a(\rho,z)=\int_{C} dx_\mu  \frac{\bar\eta^a_{\mu\nu}(x-z)_\nu\rho^2}{(x-z)^2[(x-z)^2+\rho^2]}\,.
\end{equation}
$\eta^a_{\mu\nu}$ is the 't Hooft symbol defined as \cite{Liu:2024rdm,Vainshtein:1981wh} (see also Appendix~\ref{App:singular}).
The exponentiation
arises due to taking into account the many-instanton configurations effect.
Note that the path ordering operator $\mathcal{P}$ in Wilson loop \eqref{eq:loop} is not needed, since the instanton field is a hedgehog in color space, with the color-space orientations locked.

\section{YM vacuum}

The investigation of the infrared behavior of
QCD is the formulation of the non-Abelian gauge theory on loop space that becomes possible by virtue of the close correspondence between gauge and chiral fields \cite{Polyakov:1980ca}. One of the advantages
of this method is that it allows one to consider both the perturbative and nonperturbative
contributions. The basic object of study in such an approach is the gauge invariant vacuum
average of the Wilson loop operator 

\begin{equation}
\label{eq:loop}
    W=\mathrm{Tr}~\mathcal{P} \exp\left(ig\int_C dx_\mu A_\mu\right)
\end{equation}



\begin{figure}
    \centering
\subfloat[\label{fig:quenched_vac1}]{\includegraphics[width=.45\linewidth]{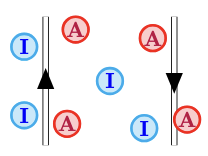}}
\hfill
\subfloat[\label{fig:quenched_vac2}]{\includegraphics[width=.45\linewidth]{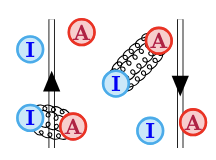}}
    \caption{Wilson loop in quenched vacuum (YM)}
    \label{fig:quenched_vac}
\end{figure}

In the quenched vacuum, the nonperturbative contribution to a generic Wilson loop at low resolution (following from say gradient cooling) can be estimated using an ensemble of topologically active gauge configurations made chiefly of instantons and anti-instantons. In this dilute regime, the vacuum can be treated in a mean-field sense as a random instanton medium. Accordingly, for a dilute configuration one may neglect non-commutativity between different pseudoparticles, $[A_I, A_J] \simeq 0$, which leads to~\cite{Callan:1977gz}

\begin{equation}
    W\approx\prod_{I}W_I
\end{equation}
where $W_I$ is the same Wilson line evaluated in single instanton (anti-instanton) background given by

\begin{equation}
W_I(\rho,z)=e^{i\tau^a\phi_I^a(\rho,z)}\,,
\end{equation}
where the single-instanton profile in the soft Wilson loop is defined as 
\begin{equation}
        \phi_I^a(\rho,z)=\int_{C} dx_\mu  \frac{\bar\eta^a_{\mu\nu}(x-z)_\nu\rho^2}{(x-z)^2[(x-z)^2+\rho^2]}\,.
\end{equation}
and $\eta^a_{\mu\nu}$ is the 't Hooft symbol defined as \cite{Liu:2024rdm,Vainshtein:1981wh} (see also Appendix~\ref{App:singular}).

Thus the deep cooled Wilson line can be evaluated by \cite{Shuryak:2000df,Dorokhov:2002qf,Shuryak:2021hng}

\begin{equation}
\label{loop_inst}
    W(C)\approx\exp\left[\frac{1}{N_cV}\sum_I\int d^4z ~\mathrm{Tr}_c\left(W_I(\rho,z)-1\right)\right]
\end{equation}

Beyond mean field, the instantons and anti-instantons have non-trivial attraction induced by exchanging color force (soft gluons). The resulting instanton and anti-instanton pairs introduce extra contribution to the Wilson line as argued by \cite{Shuryak:2021fsu,Shuryak:2021hng}. With this extra contribution in mind, the Wilson line will become

\begin{equation}
\begin{aligned}
\label{wilson_mf}
        \langle W\rangle\approx\sum_{N_{\rm mol}} \binom{N_+}{N_{\rm mol}}\binom{N_-}{N_{\rm mol}}\exp\bigg[&\frac{N_++N_--2N_{\rm mol}}{N_cV}\int d^4z \mathrm{tr}_c(W_I-1)\\
    &+\frac{N_{\rm mol}}{N_cV}\int d^4z\int \frac{d^4Rdu}{V}~\mathrm{tr}_c\left(W_{IA}e^{-S_{\rm int}}-1\right)\bigg]
    \end{aligned}
\end{equation}
where $W_{IA}$ is the same Wilson line evaluated in an instanton-anti-instanton pair background and $S_{\rm int}$ is the semi-classical gauge interaction between instanton and anti-instanton defined by \cite{Zakharov:1992bx,Liu:2025ldh,Shuryak:1981ff} (see also Sec.~\ref{sec:semi-gauge-IA}).

Here the instanton numbers and molecular numbers are controlled by deep cooled instanton density $n_{I+A}$ and molecular density $n_{IA}$.

\begin{equation}
    N_+ + N_- = V \int d\rho\, n(\rho)= V(n_{I+A} +n_{IA})
\end{equation}

The extra color integral over the relative color orientation $u$ between the pair will give extra $1/N_c$. Therefore, mean field approximation is strictly followed $1/N_c$ counting.

\section{Unquenched QCD vacuum}

\begin{figure}
    \centering
\subfloat[\label{fig:unquenched_vac1}]{\includegraphics[width=.45\linewidth]{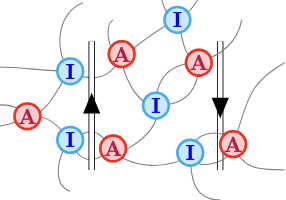}}
\hfill
\subfloat[\label{fig:unquenched_vac2}]{\includegraphics[width=.45\linewidth]{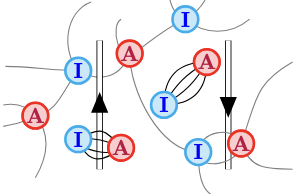}}
    \caption{Wilson loop in chiral broken QCD vacuum}
    \label{fig:unquenched_vac}
\end{figure}

In the unquenched vacuum, we begin with a mean-field background in which instantons and anti-instantons are fully connected through quark zero modes, as illustrated in Fig.~\ref{fig:unquenched_vac}. This replaces the quark mass disordering by an effective determinantal mass $m^*$. Within this framework, the mean-field approximation in \eqref{wilson_mf} remains applicable. The key difference relative to the quenched case lies in the weighting of the instanton ensemble: the averaged total number of instantons and anti-instantons in the deeply cooled vacuum is modified by the fermion determinant, leading to
\begin{equation}
N_+ + N_- = V\int d\rho\, n(\rho) \;\longrightarrow\; V\int d\rho\, n(\rho)\,(\rho m^*)^{N_f}.
\end{equation}

Beyond mean field, the instantons and anti-instantons have non-trivial attraction induced by exchanging light quarks. This attraction is much stronger than the semi-classical gauge interaction in quenched vacuum such that the vacuum mean field is disordered. The resulting instanton and anti-instanton pairs introduce extra contribution to the Wilson line as argued by \cite{Shuryak:2021fsu,Shuryak:2021hng}. With this extra contribution in mind, the Wilson line will become

\begin{equation}
\begin{aligned}
    \langle W\rangle\approx&\sum_{N_{\rm mol}} \binom{N_+}{N_{\rm mol}}\binom{N_-}{N_{\rm mol}}\exp\bigg[\frac{N-2N_{\rm mol}}{N_cV}\int d^4z \mathrm{tr}_c(W_I-1)\\
    &+\frac{N_{\rm mol}}{N_cV}\int d^4z\int \frac{d^4Rdu}{V}~\mathrm{tr}_c\left(W_{IA}\sum_{N_f=1}\left|\frac{T_{IA}}{m^*}\right|^{2N_f}-1\right)\bigg]
    \end{aligned}
\end{equation}

Similarly, the extra color integral over the relative color orientation $u$ between the pair gives extra $1/N_c^{N_f}$. Therefore, the molecular correction starts at subleading order in $1/N_c$.

\chapter{Meson Photoproduction Near Threshold}
\label{ch:near_th}

\begin{figure}
    \centering
    \includegraphics[width=.6\linewidth]{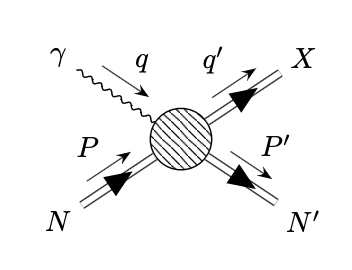}
    \caption{Kinematics for the $\gamma N\rightarrow XN'$ process.}
    \label{fig:scattering}
\end{figure}

Recent experiments carried at JLAB~\cite{Hafidi:2017bsg,GlueX:2019mkq,Meziani:2020oks,Anderle:2021wcy} may have started to
reveal some aspects of the primordial glue in the QCD vacuum. Near threshold diffractive photo- or electro-production of heavy mesons, such as charmonium or bottomonium, are exclusively sensitive to the nonperturbative gluon content of the nucleon as scooped from the QCD vacuum.  The recently reported data by JLAB have spurred a renewed interest by many~\cite{Hatta:2018ina,Mamo:2019mka,Kharzeev:2021qkd,Ji:2021mtz,Hatta:2021can,Guo:2021ibg,Sun:2021gmi,Wang:2022vhr}.

The fact that gluons dominate the diffractive vector meson production at large center of mass energy $\sqrt s$ is
not surprising. What is a bit surprising, is that they may still dominate the threshold production of heavy quarkonia. Indeed, at large $\sqrt s$ diffractive $pp$  and $p\bar p$ is dominated by Pomeron exchange a tower a $C$-even soft gluons with positive signature, with a small Odderon admixture~\cite{Bartels:1980pe,Kwiecinski:1980wb,Braun:1998fs},a tower of $C$-odd soft gluons with negative signature~\cite{TOTEM:2020zzr}.  Negative signature  Reggeons add in the $pp$ channel, and subtract in the $p\bar p$ channel. 

Threshold $J/\Psi$ photo-production at JLAB has allowed for a measurement of the gluonic gravitational form factors of the proton~\cite{GlueX:2019mkq,Meziani:2020oks}. This is supported by dual gravity~\cite{Hatta:2018ina,Mamo:2019mka,Mamo:2021krl,Mamo:2022eui} and leading twist pQCD calculations at large skewness~\cite{Boussarie:2020vmu,Guo:2021aik,Hatta:2021can,Guo:2023pqw},
although threshold coupled channels were also suggested~\cite{Du:2020bqj}. More specifically,
the tensor $A$ and scalar $D$ form-factors, with the latter providing for a potential map of the gluonic shear and pressure content of the proton. The $A$-form factor is dominated by the 
$2^{++}$ glueball exchange, and the $D$-form factor is a balance between the $2^{++}$ and $0^{++}$ glueball exchanges, in dual gravity~\cite{Mamo:2021krl,Mamo:2022eui}. In dual gravity, the Reggeized $2^{++}$ trajectory gives rise to the Pomeron, a $C$-even 
gluon exchange away from threshold~\cite{Polchinski:2002jw}.

Threshold $\eta_c$ photoproduction on a proton, if detectable at JLAB or other facilities, probes $C$-odd gluon exchanges. The latters Reggeize to the
Odderon at higher center of mass energy. The aim of this chapter is to analyze the near threshold production of the $\eta_{c,b}$, by factorizing out the
leading gluonic contribution to the production process, in the large skewness limit. This $C$-odd
and twist-3 contribution in the proton, is then evaluated in the QCD instanton vacuum. For completeness, we note a recent
analysis of the process in dual gravity~\cite{Hechenberger:2024abg}.

\section{Kinematics}
The kinematics of the threshold production is captured by the Mandelstam invariants $s,t$, with  $s=(P+q)^2$ related to the center of mass energy $W=\sqrt{s}$, and $t=\Delta^2$  related to the momentum transfer $\Delta^\mu=(P'-P)^\mu$. The  $Q^2$ in photoproduction is exactly set to be $0$ although similar analysis can be easily extended for the large-$Q^2$ leptoproduction.
Without loss of generality, we can work in the center of mass frame. In Fig.~\ref{fig:scattering}, The four-momenta of the incoming photon, incoming proton, outgoing proton and outgoing meson $X$ are denoted by $q$, $P$, $P'$, and $q'$ respectively. Each external state is given by the on-shell conditions defined as
\begin{align*}
P^2=P'^2&=M_N^2\ , & q^2=&0\ , & q'^2=M_X^2
\end{align*} 
With the on-shell conditions, the four-momenta in the center of mass frame, can be written as
\begin{align}
\label{eq:kinematics}
q=&\left(\frac{s-M^2_N}{2\sqrt{s}},\ 0,\ -\frac{s-M^2_N}{2\sqrt{s}}\right)  & \\[5pt] \nonumber 
q'=&\left(\frac{s+M_X^2-M^2_N}{2\sqrt{s}},\ -|\vec{P}'_{\mathrm{cm}}|\sin\theta,\ -|\vec{P}'_{\mathrm{cm}}|\cos\theta\right) \\[5pt] \nonumber
P=&\left(\frac{s+M^2_N}{2\sqrt{s}},\ 0,\ \frac{s-M^2_N}{2\sqrt{s}}\right)  &  \\[5pt] \nonumber
P'=&\left(\frac{s-M_X^2+M^2_N}{2\sqrt{s}},|\vec{P}'_{\mathrm{cm}}|\sin\theta,\ |\vec{P}'_{\mathrm{cm}}|\cos\theta \right)
\end{align}
where $M_N$ is the nucleon mass, $M_X$ is the produced meson mass, and $\theta$ is the scattering angle in the center of mass frame. 

The magnitude of the outgoing three-momentum reads
\begin{equation}
|\vec{P}'_{\mathrm{cm}}|=\left(\frac{[s-(M_X+M_N)^2][s-(M_X-M_N)^2]}{4s}\right)^{1/2}    
\end{equation}   

\begin{figure*}
\subfloat[\label{fig:t-W_c}]{%
\includegraphics[height=5.2cm,width=.49\linewidth]{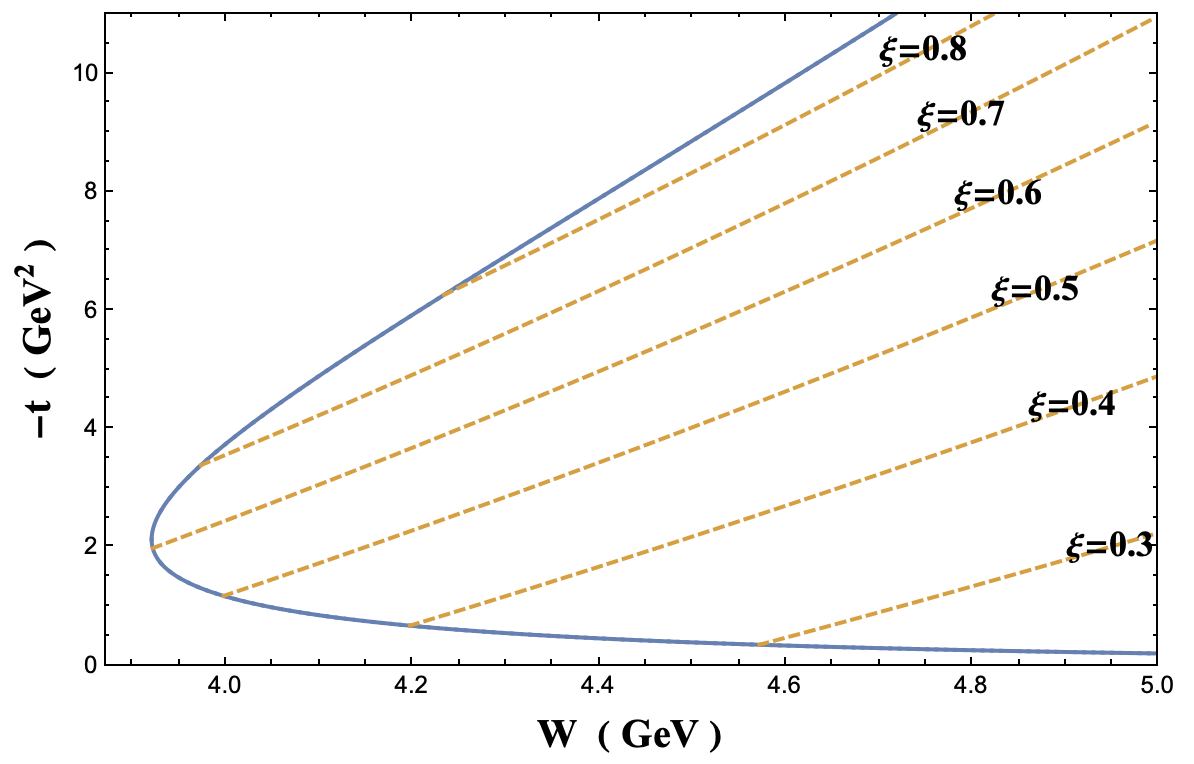}%
}\hfill
\subfloat[\label{fig:t-W_b}]{%
\includegraphics[height=5.25cm,width=.49\linewidth]{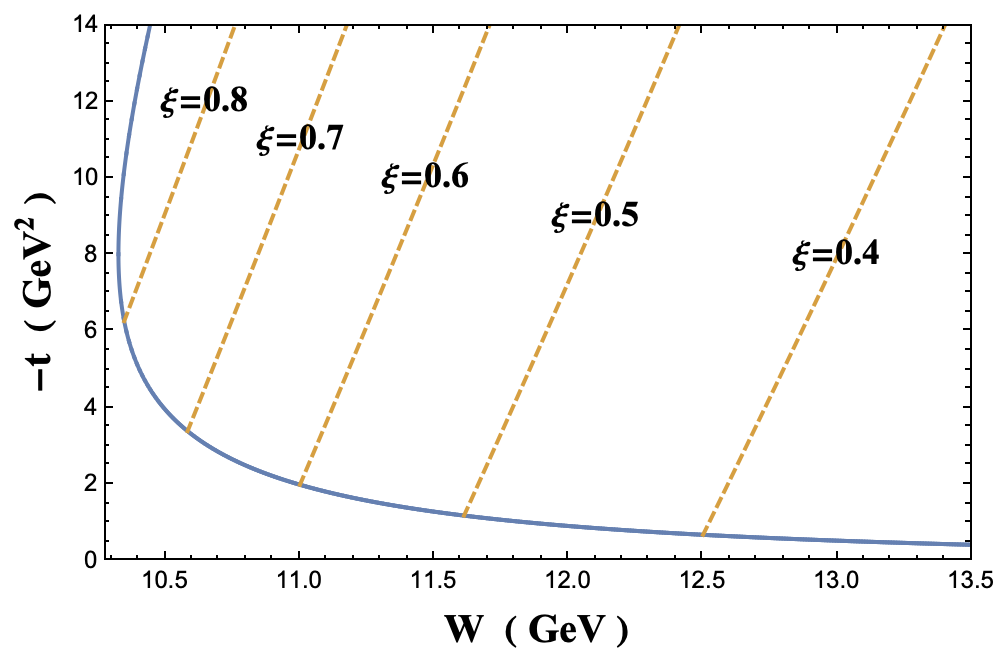}%
}
\caption{ (a) Skewness $\xi$ in  the $(W,-t)$ plane in the kinematically allowed region with $M_{\eta_c}=2.982$ GeV. The kinematically allowed region for charmonium $J/\psi$ is similar due to the similar mass  $M_{J/\psi}=3.097$ GeV. (b) Skewness $\xi$ in the $(W,-t)$ plane in the kinematically allowed region with $M_{\eta_b}=9.388$ GeV. The kinematically allowed region for bottomonium $\Upsilon$ is similar due to the similar mass $M_{\Upsilon}=9.46$ GeV.}
\end{figure*}

The scattering angle is fixed by the invariant $t$,
\begin{equation}
    \cos\theta=\frac{2st+(s-M^2_N)^2-M^2_X(s+M_N^2)}{2\sqrt{s}|\vec{P}'_{\mathrm{cm}}|(s-M_N^2)}
\end{equation}
Also, the skewness $\xi$ is
\begin{equation}
    \xi=-\frac{\Delta\cdot q}{2\bar{P}\cdot q}
\end{equation}
where $\bar{P}^\mu=(P+P')^\mu/2$.
In the threshold limit $\sqrt{s}\rightarrow M_N+M_X$, the momentum transfer $t$ is constrained in the neighborhood of $t_{th}$
\begin{equation}
\label{eq:t_th}
    t_{th}=-\frac{M_NM_X^2}{M_N+M_X}
\end{equation}
The kinematically allowed regions are shown on the $(W,-t)$ plane in Fig.~\ref{fig:t-W_c} for charmonium and in Fig.~\ref{fig:t-W_b} for bottomonium.

In the near threshold region $s\gtrsim (M_N+M_X)^2$, the factorization on the proton side only works when the outgoing meson is heavy enough, such that the proton target moves fast enough to be factorized using the parton picture. In the heavy meson limit, the incoming and outgoing nucleon velocity is of order 1 up to some correction proportional to the mass ratio $M_N^2/M_X^2$. Therefore, the factorization method  using the parton description  still holds near the photoproduction threshold. On the other hand, near the threshold region, there is not much energy left to move the heavy meson. The outgoing meson velocity becomes non-relativistic. Therefore, the meson part can be analysed using  non-relativistic QCD (NRQCD). Close to threshold, the skewness $\xi$ is close to one. Similar arguments for the photoproduction near threshold can also be found in \cite{Guo:2021ibg,Sun:2021pyw} for $J/\psi$ and in \cite{Ma:2003py} for $\eta_c$.

\subsection{$J/\psi$ meson photoproduction}
For a fast moving hadron, the constituent becomes collinear. In the heavy limit, the dominant process of the scattering in Fig.~\ref{fig:scattering} can be factorized into a hard scattering with outgoing partons carrying momentum $k^+_i$ and the parton correlation in the nucleon. Thus, the factorization amplitude of the photoproduction near threshold can be written as

\begin{equation}
\begin{aligned}
    i\mathcal{M}(\gamma N\rightarrow J/\psi\ N')
    =&\int_{-P^+}^{\infty} dk^+_1\int_{-P^+}^{\infty}dk^+_2W^{ab}_{\mu\nu}(k_1^+,k_2^+)\frac{i}{k_1^++i0^+}\frac{i}{k_2^++i0^+}\nonumber\\
    &\times\int \frac{dz_1^-}{2\pi}\frac{dz_2^-}{2\pi}e^{-ik^+_1z_1^--ik^+_2z_2^-}\langle N'|F^{a\mu+}\left(z_2^-\right) F^{b\nu+}\left(z_1^-\right)|N\rangle
\end{aligned}
\end{equation}

We use the light-cone gauge,  where the unphysical longitudinally polarized collinear gluons vanish. The lower bound of the parton momentum-integral assures that the spectators in the nucleon remains physical, when the parton comes out from the nucleon.

\begin{figure}
\subfloat[\label{fig_J1}]{%
  \includegraphics[width=.5\linewidth]{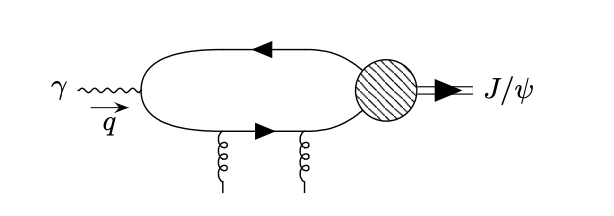}%
}\hfill
\subfloat[\label{fig_J2}]{%
  \includegraphics[width=.5\linewidth]{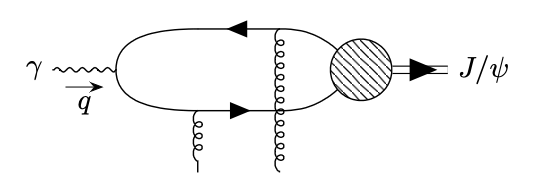}%
}
\caption{Feynman diagram for the hard kernel in $J/\psi$ photoproduction.}
\label{fig_Jpsi}
\end{figure}
The hard kernel does not depend on $\Delta^+$ due to translational symmetry in the theory. The relative momentum $k^+$ dependence can be computed using the Feynman diagrams in Fig.~\ref{fig_Jpsi}. The transversely polarized outgoing quarkonium dominates the amplitude in the heavy meson limit \cite{Sun:2021pyw}. Thus, the hard kernel reads
\begin{equation}
    W^{ab}_{\mu\nu}=\frac{g^2}{2}\frac{\delta^{ab}g_{\perp\mu\nu}}{\sqrt{N_c}}\frac{\psi^*_{J/\psi}(0)}{m_c}(8\epsilon_\gamma\cdot\epsilon_V^*)
\end{equation}
where $\psi_{J/\psi}$ is the non-relativistic wave function defined in Sec.~\ref{sec:NRWF}. 
In the heavy meson photoproduction process, the mesonic part in the hard kernels are treated by non-relativistic QCD. The relative momentum between the quark-antiquark pair is of order $\mathcal{O}(\alpha_sM_X)$, and the heavy meson mass is assumed to be $M_X=2m_Q$. 

To simplify the amplitude, we use the  symmetric parameterization
\begin{align*}
z_c^-=&\frac{z_1^-+z_2^-}{2}  & \Delta^+=&k^+_1+k_2^+=-2\xi\bar{P}^+ \\
z^-=&z_1^--z_2^- & k^+=&\frac{k^+_1-k_2^+}{2}=x\bar{P}^+
\end{align*}
Translational symmetry implies no dependence on $z_c^-$ in the parton correlation. With this in mind, the photoproduction amplitude reads

\begin{equation}
\begin{aligned}
    &i\mathcal{M}(\gamma N\rightarrow J/\psi\ N')= \frac{g^2}{\sqrt{N_c}}\frac{\psi^*_{J/\psi}(0)}{m_c}(4\epsilon_\gamma\cdot\epsilon_V^*)\mathcal{W}_{2g}(t,\xi)
\end{aligned}
\end{equation}
Here $\mathcal{W}_{2g}(t,\xi)$  depends implicitly on the polarization of the initial and final proton, through
\begin{equation}
    \mathcal{W}_{2g}(t,\xi)=\int_{-1}^{1} dx\frac{1}{x-\xi+i0^+}\frac{1}{x+\xi-i0^+}f_{2g}(x,t,\xi)
\end{equation}
where the two-gluon distribution is defined as
\begin{equation}
    f_{2g}(x,\xi,t)=\int \frac{dz^-}{2\pi}e^{-ix\bar{P}^+z^-}\frac{1}{\bar{P}^+}\langle N'|F^{a+i}\left(-z^-/2\right) F^{a+}{}_i\left(z^-/2\right)|N\rangle
\end{equation}

\subsection{$\eta_c$ meson photoproduction}
The threshold photoproduction analysis of heavy pseudoscalars
follow a similar reasoning. The dominant process in the heavy meson limit, is factorized into a hard scattering with three outgoing gluons carrying momenta $k^+_1, k_2^+, k^+_3$, times the three-gluon light-front correlation in the nucleon,

\begin{align}
&i\mathcal{M}(\gamma N\rightarrow \eta_c\ N')=\\
&\int^{\infty}_{-P^+} dk^+_1\int^{\infty}_{-P^+}dk^+_2\int^{\infty}_{-P^+}dk^+_3W^{abc}_{\mu\nu\sigma}(k_1^+,k_2^+,k_3^+)\frac{i}{k_1^++i0^+}\frac{i}{k_2^++i0^+}\frac{i}{k_3^++i0^+}\nonumber\\
    &\times\int\frac{dz_1^-}{2\pi}\frac{dz_2^-}{2\pi}\frac{dz_3^-}{2\pi}e^{-ik^+_1z_1^--ik^+_2z_2^--ik^+_3z_3^-}\langle N'|F^{a\mu+}\left(z_3^-\right) F^{b\nu+}\left(z_2^-\right)F^{c\sigma+}\left(z_1^-\right)|N\rangle
\end{align}

The  light-cone gauge is used again to remove the unphysical longitudinally polarized collinear gluons. The lower bound of the parton momentum-integral assures the spectators in the nucleon remains physical when the parton comes out from the nucleon.

\begin{figure}
\subfloat[\label{fig_ec1}]{%
  \includegraphics[width=.5\linewidth]{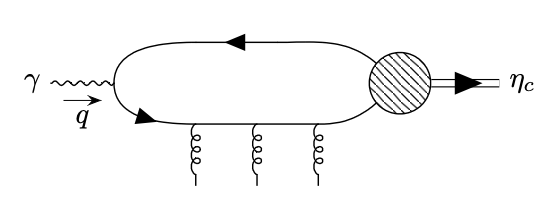}%
}\hfill
\subfloat[\label{fig_ec2}]{%
  \includegraphics[width=.5\linewidth]{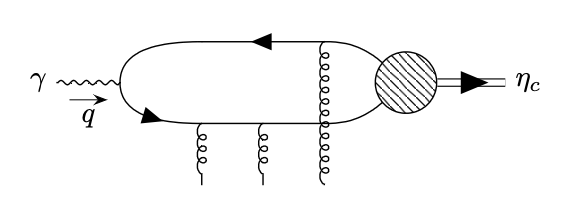}%
}
\caption{Feynman diagram for the hard kernel for $\eta_c$ photoproduction.}
\label{fig_etac}
\end{figure}

Again, the hard kernel $W^{abc}_{\mu\nu\sigma}$ does not depend on $\Delta^+$ due to translational symmetry. The relative momenta $k_\lambda^+$ and $k^+_\rho$ dependence can be computed using the Feynman diagrams in Fig.~\ref{fig_etac}. In the heavy meson limit, the hard kernel reads

\begin{equation}
\begin{aligned}
W^{abc}_{\mu\nu\sigma}=&\frac{ig^3}{2}d^{abc}\epsilon_{\mu-+\alpha}\epsilon^\alpha_{\gamma}(q)\frac{g_{\perp\nu\sigma}}{\sqrt{N_c}}\frac{\psi^*_{\eta_c}(0)}{m_c}\\
&\times\left[4\sqrt{2}\left(\frac{2i}{k_1^++k_2^++i0^+}+\frac{i}{k_1^++k_3^++i0^+}+\frac{i}{k_2^++k_3^++i0^+}\right)\right]    
\end{aligned}
\end{equation}
where $\psi_{\eta_c}$ is the non-relativistic wave function of the $\eta_c$ meson defined in Sec.~\ref{sec:NRWF}. 
To further simplify the amplitude, we use the symmetric parameterization
\begin{align*}
z_c^- &= \frac{z_1^-+z_2^-+z_3^-}{3}, &
\Delta^+ &= k^+_1+k_2^++k_3^+=-2\xi\bar{P}^+ \\
\lambda^- &= z_1^- - z_2^-, &
k_\lambda^+ &= \frac{k^+_1-k_2^+}{2}=x_\lambda\bar{P}^+ \\
\rho^- &= \frac{1}{2}\left(z_1^-+z_2^- - 2z^-_3\right), &
k_\rho^+ &= \frac{k^+_1+k_2^+-2k_3^+}{3}=x_\rho\bar{P}^+
\end{align*}
The dependence on $z_c^-$ drops out from the parton correlation because of translational symmetry, hence
\begin{equation}
\begin{aligned}
    &i\mathcal{M}(\gamma N\rightarrow \eta_c\ N')=-\frac{ig^3}{\sqrt{N_c}}\frac{\psi^*_{\eta_c}(0)}{m_c}2\sqrt{2}\epsilon_{\perp ij}\epsilon_{\gamma}^i(q)\mathcal{W}^j_{3g}(t,\xi)
\end{aligned}
\end{equation}
The Levi-Civita tensor in the transverse space is defined as $\epsilon_{\perp ij}=\epsilon_{0ij3}$ with the convention $\epsilon_{0123}=1$. The transverse vector function $\mathcal{W}^i_{3g}(t,\xi)$ depends implicitly on the polarization of the initial and final proton, and is defined as
\begin{equation}
\begin{aligned}
    &\mathcal{W}^i_{3g}(t,\xi)=\int_{-2-\xi/3}^{1+\xi/3} dx_\rho\int_{-1-\xi/3-x_\rho/2}^{1+\xi/3+x_\rho/2} dx_\lambda\\
    &\times\left(\frac{2}{\frac{4}{3}\xi-x_\lambda-i0^+}+\frac{1}{\frac{4}{3}\xi-x_\rho+\frac{1}{2}x_\lambda-i0^+}+\frac{1}{\frac{4}{3}\xi+x_\rho+\frac{1}{2}x_\lambda-i0^+}\right)\\
    &\times\bigg(\frac{1}{\frac{2}{3}\xi-x_\lambda-\frac{1}{2}x_\rho-i0^+}\frac{1}{\frac{2}{3}\xi+x_\lambda-\frac{1}{2}x_\rho-i0^+}\frac{1}{\frac{2}{3}\xi+x_\rho-i0^+}f^i_{3g}(x_\rho,x_\lambda,\xi,t)\bigg)
\end{aligned}
\end{equation}
The three-gluon distribution is defined as
\begin{equation}
\begin{aligned}
    &f^i_{3g}(x_\rho,x_\lambda,\xi,t)=\int \frac{d\lambda^-}{2\pi}\int \frac{d\rho^-}{2\pi}e^{-ix_\lambda\bar{P}^+\lambda^--ix_\rho\bar{P}^+\rho^-}\frac{d^{abc}}{(\bar{P}^+)^2}\\
    &\times\langle N'|F^{a+i}\left(-\frac{2}{3}\rho^-\right) F^{b+j}\left(\frac{1}{2}\lambda^-+\frac{1}{3}\rho^-\right)F^{c+}{}_j\left(\frac{1}{2}\lambda^-+\frac{1}{3}\rho^-\right)|N\rangle
    \end{aligned}
\end{equation}

\section{Non-relativistic wave function for quarkonia}
\label{sec:NRWF}
Generally, the non-relativistic wave function (NRWF) for mesons are defined as

\begin{equation}
    |X(P)\rangle=\int\frac{d^3\vec{k}}{(2\pi)^3}\Psi_X(k,s_1,s_2)b^\dagger_{s_1}(\vec{k}_1)c^\dagger_{s_2}(\vec{k}_2)|0\rangle
\end{equation}
where $b_s(k)$ and $c_s(k)$ are normalized in non-relativistic conventions and the constituent momenta are defined as $\vec{k}_1=\vec{k}$ and $\vec{k}_2=\vec{P}-\vec{k}$.
For vector quarkonium, the NRWF is defined as \cite{Liu:2023fpj}
\begin{equation}
    \Psi_X(k,s_1,s_2)=\frac{1}{\sqrt{N_c}}\phi_{X}(\vec{k})\bar{u}_{s_1}(\vec{k}_1)\gamma\cdot\epsilon_{V} v_{s_2}(\vec{k}_2)
\end{equation}
and for pseudoscalar quarkonium, the light front wave function is defined as \cite{Liu:2023yuj}
\begin{equation}
    \Psi_X(k,s_1,s_2)=\frac{1}{\sqrt{N_c}}\phi_{X}(\vec{k})\bar{u}_{s_1}(\vec{k}_1)i\gamma^5 v_{s_2}(\vec{k}_2)
\end{equation}
where $v^\mu$ is the four velocity of the quarkonium and $\phi_X(\vec{k})$ is the spin-independent wave function in momentum space. In heavy limit, the relative momentum between quark and antiquark is of order $\mathcal{O}(\alpha_sM_X)$. Thus at the leading order of $\alpha_s$, the wave function can be further simplied by

\begin{equation}
    \phi_{X}(\vec{k})\simeq (2\pi)^3\delta^3(\vec{k})\psi_X(0)
\end{equation}
where we already assume $M_X=2m_Q$ for the quarkonium at the leading order of $\alpha_s$ and $\psi_X$ is the quarkonium wave function in coordinate space.
$$
\psi_X(0)=\int\frac{d^3\vec{k}}{(2\pi)^3}\phi_X(\vec{k})
$$

Thus, in the heavy limit, the wave functions for $J/\psi$ and $\eta_c$ read

\begin{equation}
\begin{aligned}
    |J/\psi\rangle=\frac{1}{\sqrt{N_c}}\psi_{J/\psi}(0)\bar{u}_{s_1}(m_cv)\gamma\cdot\epsilon_{V} v_{s_2}(m_cv)b^\dagger_{s_1}(m_cv)c^\dagger_{s_2}(m_cv)|0\rangle
\end{aligned}
\end{equation}

\begin{equation}
\begin{aligned}
    |\eta_c\rangle=\frac{1}{\sqrt{N_c}}\psi_{\eta_c}(0)\bar{u}_{s_1}(m_cv)i\gamma^5v_{s_2}(m_cv)b^\dagger_{s_1}(m_cv)c^\dagger_{s_2}(m_cv)|0\rangle
\end{aligned}
\end{equation}

The heavy wave function can be estimated by the decay rate with the NLO QCD radiative correction included. For vector quarkonia \cite{Bodwin:2006yd}

\begin{equation}
    \Gamma(J/\psi\rightarrow e^+e^-)=\frac{16\pi\alpha_{em}^2Q_c^2}{3M_{J/\psi}}N_c|\psi_{J/\psi}(0)|^2\left(1-\frac{16}{3}\frac{\alpha_s}{\pi}\right)
\end{equation}
For pseudoscalar quarkonia \cite{Lansberg:2008cq,Fabiano:2002se}:
\begin{equation}
    \Gamma(\eta_c\rightarrow \gamma\gamma)=\frac{16\pi\alpha_{em}^2Q_c^4}{M_{\eta_c}}N_c|\psi_{\eta_c}(0)|^2\left(1-\frac{20-\pi^2}{3}\frac{\alpha_s}{\pi}\right)
\end{equation}
Note that our non-relativistic wavefunctions maps  onto 
those in~\cite{Guo:2021aik} through $M_X|\psi_X|^2\rightarrow |\psi_X|^2$. The experimental decay rates for each heavy meson are shown in Table~\ref{tab:decay_rate}. Each value is obtained by PDG group in \cite{ParticleDataGroup:2018ovx}.
With the QED coupling $\alpha_{em}$ at charmonium mass scale fixed to be $1/134$ and at bottomonia mass scale $1/132$ \cite{Erler:1998sy}, thus, we have
\begin{equation}
\begin{aligned}
|\psi_{J/\psi}(0)|^2 &= 7.237 \times 10^{-3},\quad
|\psi_{\eta_c}(0)|^2 = 3.361 \times 10^{-3},\\
|\psi_{\Upsilon}(0)|^2 &= 1.463 \times 10^{-2},\quad
|\psi_{\eta_b}(0)|^2 = 1.463 \times 10^{-2}
\end{aligned}
\end{equation}

\begin{table}
    \centering
    \begin{tabular}{|c|c|c|c|c|}
    \hline
         Decay modes 
         & $\Gamma(J/\psi\rightarrow e^+e^-)$ 
         & $\Gamma(\eta_c\rightarrow \gamma\gamma)$ 
         & $\Gamma(\Upsilon\rightarrow e^+e^-)$ 
         & $\Gamma(\eta_b\rightarrow \gamma\gamma)$ \\
    \hline
    (keV) & 5.55 & 5.0 & 1.29 & 0.52 \\
    \hline
    \end{tabular}
    \caption{The decay rates of the heavy vector meson decay to electron-positron pair and the heavy pseudoscalar meson decay to two photons are listed. $\eta_b$ decay is not available from the experiment. The value is estimated by assuming $|\psi_{\Upsilon}(0)|^2=|\psi_{\eta_b}(0)|^2$ based on the heavy quark symmetry.}
    \label{tab:decay_rate}
\end{table}

\section{Large skewness expansion}
In the threshold region, the exchanged gluons in the factorized proton matrix elements can be organized using a twist expansion (see Sec.~\ref{sec:OPE}). Each of the local operators in the OPE,  sources a colorless gluonic exchange in the form of a glueball. The corresponding glueball exchanges, can be organized using charge conjugation and parity. However, parity changes the direction of spatial momentum and flip the helicity. Parity symmetry on the light front has to be combined with time reversal, so that the light front direction is  unchanged. We adopt the $\Lambda$-parity in \cite{Ji:2003yj} by making an additional $180^{\circ}$ rotation around the $y$-axis to restore the longitudinal momentum, i.e. $ Y= e^{-i\pi J_y}P$.
As a result, the plus component is transformed in the same way as the time component under the usual parity symmetry.

\subsection{Glueball excitations in conformal moment expansion}
\label{sec:OPE}
The OPE of the multi-gluon distributions is presented here. The local operators probe glueball states lying on the Regge trajectory associated with multi-gluon exchanges within hadronic bound states.

In the process of two gluon exchange, the twist expansion corresponds to the Regge trajectory with the lowest $2^{++}$ glueball. 
\begin{equation}
\begin{aligned}
F^{a+i}\left(-\frac{z^-}2\right) F^{a+}{}_i\left(\frac{z^-}2\right)=&\sum_{n_1,n_2=0}^\infty\left(i\frac{z^-}{2}\right)^{n_1}\left(-i\frac{z^-}{2}\right)^{n_1}\frac{(iD^+)^{n_1}}{n_1!}F^{a+i}\frac{(iD^+)^{n_2}}{n_2!}F^{a+}{}_i
\end{aligned}
\end{equation}

In the process of two gluon exchange, the twist expansion corresponds to the Regge trajectory with the lowest $1^{+-}$ glueball.
 
\begin{equation}
\begin{aligned}
d^{abc}\,F^{a+\mu}\!\left(-\tfrac{2}{3}\rho^-\right)
F^{b+i}\!\left(-\tfrac{1}{2}\lambda^-+\tfrac{1}{3}\rho^-\right)
F^{c+}{}_i\!\left(\tfrac{1}{2}\lambda^-+\tfrac{1}{3}\rho^-\right)
\\
=\, d^{abc}\sum_{n_1,n_2,n_3=0}^{\infty}
\left(-\tfrac{2}{3}\rho^-\right)^{n_1}
\left(-\tfrac{1}{2}\lambda^-+\tfrac{1}{3}\rho^-\right)^{n_2}
\left(\tfrac{1}{2}\lambda^-+\tfrac{1}{3}\rho^-\right)^{n_3}
\\
\quad\times
\frac{(iD^+)^{n_1}}{n_1!}F^{a+\mu}(0)\,
\frac{(iD^+)^{n_2}}{n_2!}F^{b+i}(0)\,
\frac{(iD^+)^{n_3}}{n_3!}F^{c+}{}_i(0)\,.
\end{aligned}
\end{equation}
where we adopt the light-cone coordinate defined by $x^\pm=(x^0\pm x^3)/\sqrt{2}$.

Among these operators with the same twist, operators with higher dimension corresponds to higher glueball excitations in the Regge trajectory. Since there is not much energy to excite these gluebalsl near threshold, only the operator with lowest conformal spin and the same charge conjugation and parity symmetry, which represents the ground state of the glueball in the corresponding channel, dominates the non-perturbative nucleon matrix element. This shows that the OPE for $f_{2g}(x,\xi,t)$ is dominated by a $2^{++}$ glueball operator, and the OPE of $f^i_{3g}(x_\rho,x_\lambda,\xi,t)$ by a  $1^{+-}$ glueball operator. Therefore, the non-local gluonic operators can be approximately treated as a local operator

\begin{equation}
\begin{aligned}
F^{a+i}\left(-\lambda^-/2\right) F^{a+}{}_i\left(\lambda^-/2\right)\simeq F^{a+i}F^{a+}{}_i
\end{aligned}
\end{equation}
\begin{equation}
\begin{aligned}
    d^{abc}F^{a+\mu}\left(-\frac{2}{3}\rho^-\right) F^{b+i}\left(\frac{1}{3}\rho^--\frac{1}{2}\lambda^-\right)F^{c+}{}_i\left(\frac{1}{3}\rho^-+\frac{1}{2}\lambda^-\right)
    \simeq d^{abc}F^{a+\mu} F^{b+i}F^{c+}{}_i
\end{aligned}
\end{equation}

In the local approximation and the large skewness $\xi$ limit, the functions $\mathcal{W}_{2g}$ and $\mathcal{W}^\mu_{3g}$ in the leading moment are
\begin{equation}
   \mathcal{W}_{2g}(t,\xi\rightarrow1)=\frac{1}{\xi^2(\bar{P}^+)^2}\langle N'|F^{a+i} F^{a+}{}_i|N\rangle
\end{equation}
\begin{equation}
   \mathcal{W}^\mu_{3g}(t,\xi\rightarrow1)=\left(\frac{3}{2}\right)^3\frac{3}{\xi^4(\bar{P}^+)^4} d^{abc}\langle N'|F^{a+\mu}F^{b+i}F^{c+}{}_i|N\rangle
\end{equation}


\subsection{Twist-2 GPD and two-gluon distribution}
The threshold photoproduction of $J/\Psi$ is dominated by the $C$-even $2^{++}$ glueball  exchange,  parameterized by the gluon generalized parton distribution (GPD)

\begin{equation}
\begin{aligned}
    f_{2g}(x,\xi,t)
    =\frac{1}{\bar{P}^+}\bar{u}_{s'}(P')\left[xH_g(x,\xi,t)\gamma^++xE_g(x,\xi,t)\frac{i\sigma^{+\alpha}\Delta_\alpha}{2M_N}\right]u_{s}(P)
\end{aligned}
\end{equation}
with  $H_g(x,\xi,t), E_g(x,\xi,t)$ the gluon GPDs defined in \cite{Ji:1998pc}. The process near threshold is dominated by the zeroth moment of the gluon distribution $f_{2g}(x,\xi,t)$, which is related to the nucleon matrix element of the energy momentum tensor, the matrix element can be parameterized by 

\begin{equation}
\begin{aligned}
\label{eq:f2g}
    \int_{-1}^1dx f_{2g}(x,\xi,t)=&\frac{1}{(\bar{P}^+)^2}\langle P'|F^{a+ i} F^{a+}{}_i|P\rangle\\
    =&\frac{1}{\bar{P}^+}\bar{u}_{s'}(P')\left[H_{2g}(t,\xi)\gamma^++E_{2g}(t,\xi)\frac{i\sigma^{+\alpha}\Delta_\alpha}{2M_N}\right]u_s(P)
\end{aligned}
\end{equation}

These twist-$2$ functions $H_{2g}(t,\xi)$ and $E_{2g}(t,\xi)$ are related to the second Mellin moment of the gluon GPD. Hermicity and time reversal symmetry imply that the twist-$2$ functions are even functions of the skewness $\xi$, with 

\begin{equation}
H_{2g}(t,-\xi)=H_{2g}(t,\xi),\quad
E_{2g}(t,-\xi)=E_{2g}(t,\xi)
\end{equation}

\subsection{Twist-3 GPD and three-gluon distribution}
The threshold photoproduction of heavy pseudoscalars is dominated by $C$-odd $1^{+-}$ glueball exchanges.
Following~\cite{Ma:2003py,Guo:2021aik}, the three-gluon distribution can be parameterized by the four twist-3 gluon GPDs $G_{g1}(x_\rho,x_\lambda,t,\xi)$, $G_{g2}(x_\rho,x_\lambda,t,\xi)$, $G_{g3}(x_\rho,x_\lambda,t,\xi)$ and $G_{g4}(x_\rho,x_\lambda,t,\xi)$~\cite{Chen:2023hvu}. The parameterization is similar to the generalized helicity flip quark distribution in \cite{Diehl:2001pm}

\begin{equation}
\begin{aligned}
    &f^i_{3g}(x_\rho,x_\lambda,\xi,t)=\\
    &\frac{M_N}{(\bar{P}^+)^2}\bar{u}_{s'}(P')\left[G_{g1}\sigma^{+i}+G_{g2}\frac{i\bar{P}^{[+}\Delta^{i]}}{M_N^2}+G_{g3}\frac{i\gamma^{[+}\Delta^{i]}}{M_N}+G_{g4}\frac{\bar{P}^{[+}\sigma^{i]\alpha}\Delta_\alpha}{2M_N^2}\right]u_s(P)
\end{aligned}
\end{equation}
where the variables in the twist-3 gluon GPDs are dropped for simplicity. These twist-3 distributions are especially unique as they are not related to any other types of twist-3 gluon
distributions \cite{Koike:2019zxc}.  The process near threshold is dominated by the zeroth moment of this specific gluon distribution $f^i_{3g}(x,\xi,t)$, which is related to the nucleon matrix element of the $C$-odd three gluon operator, the matrix element can be parameterized by 

\begin{equation}
\begin{aligned}
\label{eq:odd_gluball1}
    &\int_{-2-\xi/3}^{1+\xi/3} dx_\rho\int_{-1-\xi/3-x_\rho/2}^{1+\xi/3+x_\rho/2} dx_\lambda f^i_{3g}(x_\rho,x_\lambda,\xi,t)=\frac{1}{(\bar{P}^+)^4} d^{abc}\langle P'|F^{a+i}F^{b+j}F^{c+}{}_j|P\rangle\\
    =&\frac{M_N}{(\bar{P}^+)^2}\bar{u}_{s'}(P')\left(f_1\sigma^{+i}+f_2\frac{i\bar{P}^{[+}\Delta^{i]}}{M_N^2}+f_3\frac{i\gamma^{[+}\Delta^{i]}}{M_N}+f_4\frac{\bar{P}^{[+}\sigma^{i]\alpha}\Delta_\alpha}{2M_N^2}\right)u_s(P)
\end{aligned}
\end{equation}

These twist-$3$ generalizaed form factors $f_{i}(t,\xi)$ are related to the zeroth moment of the $C$-odd dynamical twist-3 gluon GPDs $G_{gi}(x_\rho x_\lambda,\xi,t)$
and all proportional to $M_N/\bar{P}^+$ since they correspond to twist-3 operators. The hermicity of the matrix element implies
\begin{equation}
f_{1,2,3}(t,-\xi)=f_{1,2,3}(t,\xi),\quad
f_4(t,-\xi)=-f_4(t,\xi)
\end{equation}

In our evaluations to follow, these twist-$3$ functions vanish in forward limit $\xi\rightarrow0$. They are related to the magnetic moment form factor $f_g(t)$. Higher order terms corresponds to other form factors that are induced by different resonance in the glueball exchange channels. 

In the limit $\Delta^i_\perp\rightarrow0$ near threshold, $f_2$ and $f_3$ vanish while $f_1$ and $f_4$ do not. Since the initial photon has the helicity $\pm1$, and $\eta_c$ has the helicity $0$, the proton has to flip its total angular momentum in the $z$-direction by $\pm1$ in order to satisfy  angular momentum conservation.
This change can be fulfilled by the change of the orbital angular momentum, whose effects are captured by $\Delta^i_\perp$ in \eqref{eq:odd_gluball1}, or made by changing the helicity of the proton, which is denoted by the $z$-component of the proton spin. It is clear that in the limit $\Delta^i_\perp\rightarrow0$, the orbital contributions parameterized by $f_2$ and $f_3$ become zero. $f_1$ corresponds to the chiral flipping process, and $f_4$ can be mapped to the chiral conserving process by $$\bar{u}_s(P')\frac{\sigma^{i\alpha}\Delta_\alpha}{2M_N}u_s(P)\bigg|_{\Delta^i_\perp=0}=-\bar{u}_s(P')i\gamma^i_\perp u_s(P)$$ 
Therefore, the contribution comes from the helicity flipping, which corresponds to the chirality flipping term $f_1$ and chirality conserving $f_4$. In the massless limit of the proton, the chirality is exactly equal to helicity. The nonzero contribution of $f_4$ in the forward limit is due to the finite mass correction between the helicity and chirality.

\section{Gluonic form factors}
\label{sec:glu_FF}
At low momentum transfer, the $C$-even and $C$-odd gluonic matrix elements in the proton state can be evaluated in the QCD instanton vacuum. The derivation can be found in Ch.~\ref{ch:FF}, with most of the details presented in~\cite{Liu:2024rdm}.

\subsection{$C$-even gluonic operator}
\begin{figure}
\centering
\subfloat[]{\includegraphics[width=0.45\linewidth]{figures/v3.png}}
\hfill
\subfloat[]{\includegraphics[width=0.45\linewidth]{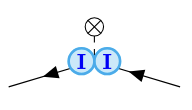}}
\caption{One-body operator arising from (a) instanton--anti-instanton molecular configurations in gluon EMT and (b) $II$ and $AA$ pair in $C$-odd three gluon operator (see text)}
\label{fig:ILM_mole_v}
\end{figure}

The dominant $C$-even gluonic local operator in the threshold photoproduction of $J/\Psi$ is $F^{a+i} F^{a+}{}_i$. It is related to the gluonic EMT \eqref{eq:EMTg} in light front signature and reads 
\begin{equation}
\begin{aligned}
   &F^{a+i}F^{a+}{}_{i}=-\frac{1}{2}\left(\vec{E}_\perp^a\cdot \vec{E}_\perp^a+2\hat{z}\cdot(\vec{E}_\perp^a\times \vec{B}_\perp^a)+\vec{B}_\perp^a\cdot \vec{B}_\perp^a\right)
\end{aligned}
\end{equation}
Since isolated pseudoparticles are self-dual tunneling configurations with $\vec{E}^a=\mp i\vec{B}^a$, the leading contribution in the pseudoparticle density vanishes. The non-vanishing contributions solely come from the molecules illustrated in Fig.~\ref{fig:ILM_mole_v}~\cite{Zahed:2021fxk}. 

Without loss of generality, we focus on the component of the $C$-even two-gluon operator with the indices $\mu$, $\nu$ taken to be symmetric and traceless, since this is the component that is relevant to twist-2.

\begin{equation}
\begin{aligned}
    &\langle N'|F^{a\{\mu}{}_{\alpha}F^{a\nu\}\alpha}|N\rangle=\\
    &\bar{u}_{s'}(p')\bigg(A_{g}(t)\frac{\gamma^{\{\mu} \bar{p}^{\nu\}}}{M_N}+B_{g}(t)\frac{i\bar{p}^{\{\mu}\sigma^{\nu\}\alpha}q_\alpha}{2M_N}+D_g(t)\frac{q^{\{\mu}q^{\nu\}}}{4M_N}\bigg)u_{s}(p)
    \end{aligned}
\end{equation}

Compared to Eq.~\eqref{eq:f2g}, the second Mellin moment of twist-$2$ GPD $H_{2g}$ and $E_{2g}$ are related to the gluonic gravitational form factors $A_g,B_g,D_g$ by
\begin{equation}
H_{2g}(t,\xi)=A_g(t)+\xi^2D_g(t),\quad
E_{2g}(t,\xi)=B_g(t)-\xi^2D_g(t)
\end{equation}
The lattice simulations~\cite{Shanahan:2018pib} and dual gravity arguments~\cite{Mamo:2021krl,Hatta:2021can},
suggest that $B_g$ is about zero. This will be assumed throughout this chapter. 

To proceed the evaluation in $J/\psi$ production, we briefly summarize the conclusion about EMT form factor in Sec.~\ref{sec:GFF}. Using the result of Eq.~\eqref{eq:EMTFF}, together with the traceless quark EMT form factors of the nucleon parameterized in Eq.~\eqref{T_traceless}~\cite{Ji:2000id,Polyakov:2019lbq}, the gluonic contribution can be constrained by Eq.~\eqref{eq:EMTFF} in ILM. For $A$-form factor, in forward limit, one obtains $A_g(0)=0.419$ with the QCD instanton vacuum parameters $n_{I+A}=1~\mathrm{fm}^{-4}$ and $\rho=0.313~\mathrm{fm}$, momentum sum rule $A_g(0)+A_q(0)=1$, together with DGLAP evolution to $\mu=2~\mathrm{GeV}$. This value is close to the gluon momentum fraction $\langle x\rangle_g=0.414$ obtained from the CTEQ global analysis~\cite{Hou:2019efy,Guo:2023pqw}, and is also comparable to the lattice result $A_g(0)=0.501$~\cite{Hackett:2023rif}. 

The charge $D_{g}(0)$ is evaluated using the QCD instanton vacuum parameters listed in Table~\ref{tab:parameters_ILM_FF} and evolved to the resolution $\mu=2~\mathrm{GeV}$ using DGLAP evolution. The result is (see Sec.~\ref{sec:GFF}) $D_g(0) = -1.299$. Here $B_g(0)$ is assumed to vanish, as suggested by lattice simulations~\cite{Shanahan:2018pib} and supported by dual gravity arguments~\cite{Mamo:2021krl,Hatta:2021can}.

\subsection{$C$-odd gluon operator}
Similarly to the $C$-even gluonic operator for threshold photoproduction of $J/\Psi$, the $C$-odd gluonic operator $d^{abc}F^{a+i}F^{b+j}F^{c+}{}_j$ is dominant in the photoproduction of heavy pseudoscalar mesons
such as $\eta_{c,b}$~\cite{Ma:2003py}. 



We now proceed to evaluate the matrix element in the QCD 
instanton vacuum with the ensemble average presented in Chs.~\ref{ch:ILM} and \ref{ch:FF}, through the substitution of the sum ansatz 
(see Sec.~\ref{sec:ILMEFT}). The symmetric $SU(3)$ structure constant $d^{abc}$ has no support in the $SU(2)$ subgroup. Hence, individual $SU(2)$ pseudoparticles cannot contribute in leading order, even if embedded in $SU(N)$ since it is always true $\mathrm{tr}(\tau^a\{\tau^b,\tau^c\})=0$. Alternatively, the hadronic matrix element receives contributions from quark hoping around the nearby instantons with the same duality.

Without loss of generality, we focus on the component of the $C$-odd three-gluon operator with the indices $\nu,\rho,\lambda$ taken to be symmetric and traceless
$
d^{abc}\,F^{a\mu\{\nu}F^{b\rho}{}_{\alpha}F^{c\lambda\}\alpha}
$, since this is the component that contributes to twist-3. For such off-forward nucleon matrix element, the most general Lorentz-covariant decomposition is characterized by 
$7$ independent generalized form factors. 
This counting follows from a Lorentz group $SL(2,\mathbb{C})$ representation argument. With this in mind, the $C$-odd gluonic nucleon form factor can be parameterized in Lorentz covariant form by, we fully symmetrize and traceless the indices $\nu\rho\lambda$

\begin{equation}
\begin{aligned}
&\langle N' | d^{abc}\,F^{a\mu\{\nu}F^{b\rho}{}_{\alpha}F^{c\lambda\}\alpha}| N \rangle
=\\
&M_N\bar u_{s'}(p') \, \Bigg\{
 \sigma^{\mu\{\nu}
\left[
A^g_{T1}(t)\bar{P}^{\rho} \bar{P}^{\lambda\}}
+
A^g_{T2}(t)\Delta^{\rho} \Delta^{\lambda\}}
\right]\\
&+ \frac{i\gamma^{[\mu}\Delta^{\{\nu]}}{2M_N}
\left[
B^g_{T1}(t)\bar{P}^{\rho} \bar{P}^{\lambda\}}
+
B^g_{T2}(t)\Delta^{\rho} \Delta^{\lambda\}}
\right]\\
&+ \frac{i\bar{P}^{[\mu}\Delta^{\{\nu]}}{M_N^2}
\,\left[\tilde{A}^g_{T1}(t)\bar{P}^{\rho} \bar{P}^{\lambda\}}+\tilde{A}^g_{T2}(t)\Delta^{\rho} \Delta^{\lambda\}}\right]\\
&+ \frac{i\gamma^{[\mu}\bar{P}^{\{\nu]}}{M_N}
\,\tilde{B}^g_{T1}(t)\bar{P}^{\rho} \Delta^{\lambda\}}
\Bigg\}\,
u_s(p).
\end{aligned}
\end{equation}

Compared to Eq.~\eqref{eq:odd_gluball1}, the moment of twist-$3$ gluon distribution are related to those twist-3 form factors by

\begin{align}
\label{eq:f1234}
f_1(t,\xi) &= A^g_{T1}+4\xi^2A^g_{T2}
&\qquad
f_2(t,\xi) &= \tilde{A}^g_{T1}+4\xi^2\tilde{A}^g_{T2} \nonumber\\
f_3(t,\xi) &= B^g_{T1}+4\xi^2B^g_{T2}
&\qquad
f_4(t,\xi) &= -2\xi \tilde{B}^g_{T}
\end{align}


In ILM, the gluonic contribution to the $C$-odd non-forward hadronic matrix element is given by

\begin{equation}
\begin{aligned}
\label{eq:had_3g_2}
    &\langle N'|g^3d^{abc}F^{a+i}F^{b+j}F^{c+}{}_{j}|N\rangle=\frac{N_c-2}{8N_c^2(N_c^2-1)(N_c+2)}n_{II}\gamma_{II}m(\Delta^+)^2\\
    &\times \frac{2\pi^2\rho^2}{225}\beta^{(II)}_{dGGG}(\rho \sqrt{-t})\langle N'|\bar{\psi}\left(\sigma^{i\alpha}\Delta_\alpha\Delta^+-\sigma^{+\alpha}\Delta_\alpha\Delta^i-\frac{1}{2}\Delta^2\sigma^{+i}\right)\psi|N\rangle\\
\end{aligned}
\end{equation}
where
\begin{equation}
    \beta^{(II)}_{dGGG}(\rho q)=\frac{450}{q}\int_0^\infty dx\frac{512x^4}{(1+x^2)^6}\frac{J_5(qx)}{q^4x^4}
\end{equation}
which is seen to vanish in the forward direction. For details on $n_{II}$ and $\gamma_{II}$, see Appendix~\ref{App:pair}. At non-zero momentum transfer, Eq.~\eqref{eq:had_3g_2} shows that the $C$-odd three-gluon operator in a hadron encodes the tensor charge and tensor magnetic moment, which are related to the nucleon transiversity GPD (see Sec.~\ref{sec:GPD}). 


Using the result in Eq.~\eqref{eq:had_3g_2}, the nucleon form factor is given by

\begin{equation}
\begin{aligned}
\label{eq:gluon_tw3}
A^g_{T2}(t) &= \frac{(N_c-2)n_{II}\gamma_{II}mM_N}{8N_c^2(N_c^2-1)(N_c+2)}\frac{2\pi^2\rho^2}{225}\beta^{(II)}_{dGGG}(\rho \sqrt{-t})\frac{-t}{2M_N^2}G^q_T(t), \\
\tilde{A}^g_{T2}(t) &= -\frac{(N_c-2)n_{II}\gamma_{II}mM_N}{8N_c^2(N_c^2-1)(N_c+2)}\frac{8\pi^2\rho^2}{225}\beta^{(II)}_{dGGG}(\rho \sqrt{-t})\left[G^q_T(t)-\frac{3t}{8M_N^2}\tilde{H}_T^q(t)\right], \\
B^g_{T2}(t) &= \frac{(N_c-2)n_{II}\gamma_{II}mM_N}{8N_c^2(N_c^2-1)(N_c+2)}\frac{16\pi^2\rho^2}{225}\beta^{(II)}_{dGGG}(\rho \sqrt{-t})\left[G^q_T(t)+\frac{3t}{16M_N^2}E_T^q(t)\right]
\end{aligned}
\end{equation}
The remaining form factors \(A^g_{T1}\), \(B^g_{T1}\), \(\tilde{A}^g_{T1}\), and \(\tilde{B}^g_{T}\) vanish at the current order of the ILM calculation. Note that in the forward limit $\Delta^\mu\rightarrow0$, all 3 form factors vanishes. However, near the threshold region in the heavy meson limit, the allowed kinematic region is restricted to $t\rightarrow t_{th}$.

The values of the $C$-odd gluonic contributions $A^g_{T2}(0)$ and $2\tilde{A}^g_{T2}(0)+B^g_{T2}(0)$ are tied to the nucleon tensor charge 
$\delta q$ and anomalous tensor magnetic moment $\kappa_T^q$. The tensor charge $\delta q$ can be extracted from the experimental data obtained in the back-to-back di-hadron productions in $e^+e^-$ annihilation, with the result $\delta q(0)=0.185$ at $\mu=1.55$ GeV~\cite{YE201791,Kang:2015msa} and the tensor magnetic moment $\kappa_T^q$ can also be extracted from the transversity GPD $\kappa_T^u=3.0$ and $\kappa_T^d=1.9$ at $\mu=2$ GeV~\cite{QCDSF:2006tkx}, related to Boer-Mulders function in TMD factorization (see Ch.~\ref{ch:tmd}). Although the tensor charge and tenrsor magnetic moment evolves with the renormalization scale \(\mu\) due to its anomalous dimension~\cite{artru1990transversely,Gamberg:2001xc}, we can still obtain an approximate estimate of the gluonic charge by using its value at $\mu=1$--$2\,\mathrm{GeV}$. This allows us to estimate the gluonic contribution at low resolution \(\mu \sim 0.6\text{--}1~\mathrm{GeV}\), which is given by
\begin{equation}
\label{F00}
A_{T2}^g(t\rightarrow0)=3.78\times10^{-5}\frac{-t}{2M_N^2}
\end{equation}
and
\begin{equation}
\label{F002}
[B_{T2}^g+2\tilde{A}_{T2}^g](t\rightarrow0)=3.00\times10^{-3}\frac{t}{2M_N^2}
\end{equation}

This charge characterizes a $C$-odd gluonic twist-3 operator, which is strongly suppressed in the QCD instanton vacuum as the natural coupling to the pseudoparticles is through their charges $F\tilde F=\pm FF$. The reason that Eqs.~(\ref{F00}) and (\ref{F002}) are substantially smaller than $A_g(0)$ can be also seen by a direct comparison 
\begin{equation}
\label{RATIO23}
\frac{\rm twist\text{-}3}{\rm twist\text{-}2}\approx 
\frac{N_c-2}{16N_c(N_c+2)} \frac{3}{225}\rho^2mM_N\left(\frac{n_{II}\gamma_{II}}{n_{IA}\gamma_{IA}}\right)\frac{\delta q}{A_q}\approx 10^{-4}
\end{equation}
for $(n_{II}\gamma_{II})/(n_{IA}\gamma_{IA})\approx 2.1$. In the QCD instanton vacuum, the $C$-odd twist-3 contribution is substantially small compared to the $C$-even twist-2 contribution, and the ratio vanishes in the
large $N_c$ limit. The extra suppression in $1/N_c$ stems
from the non-Abelian crossing and color orientation average in moduli space.

A full determination of the form factors using the nucleon light front wavefunctions
derived in the QCD instanton vacuum in~\cite{Shuryak:2022wtk} will be discussed in the future. Alternatively, in this chapter, we adopt a tripole parametrization for each gluonic form factor
\begin{equation}
\label{ACF}
A_g(t)=\frac{A_g(0)}{\left(1-\frac{t}{m_A^2}\right)^3},\quad
D_g(t)=\frac{D_g(0)}{\left(1-\frac{t}{m_D^2}\right)^3},\quad
A_{T2}^g(t)=\frac{-t}{2M_N^2}\frac{f_g(0)}{\left(1-\frac{t}{m_{3g}^2}\right)^3}
\end{equation}
with the charges given respectively by
\begin{equation}
A_g(0)=0.419,\qquad D_g(0)=-1.299,\qquad f_g(0)=3.78\times10^{-5}
\end{equation}
and the tripole masses $m_A, m_{D}$ are fitted to the recent lattice data~\cite{Shanahan:2018pib}, with the results
\begin{equation}
m_A=2.02~\mathrm{GeV},\quad
m_D=1.226~\mathrm{GeV},\quad
m_{3g}=1.49~\mathrm{GeV}
\end{equation}
The tripole mass $m_A$ reflects on the tensor  $2^{++}$ glueball mass in dual gravity, and the tripole mass $m_D$ on the mixing between the tensor $2^{++}$ and scalar  $0^{++}$ glueball masses. Similarly, the tripole mass $m_{3g}$ is fixed by dual gravity~\cite{Hechenberger:2024abg}, and reflects on the $C$-odd $1^{+-}$ glueball mass. Similar parameterization for the form factors were used in dual gravity~\cite{Mamo:2021krl,Hatta:2021can}
and the QCD factorization method~\cite{Guo:2021ibg,Guo:2023pqw}, in the analysis of coherent $J/\Psi$ production.

\begin{figure*}
\subfloat[\label{fig:dJPsi}]{%
\includegraphics[height=5cm,width=.49\linewidth]{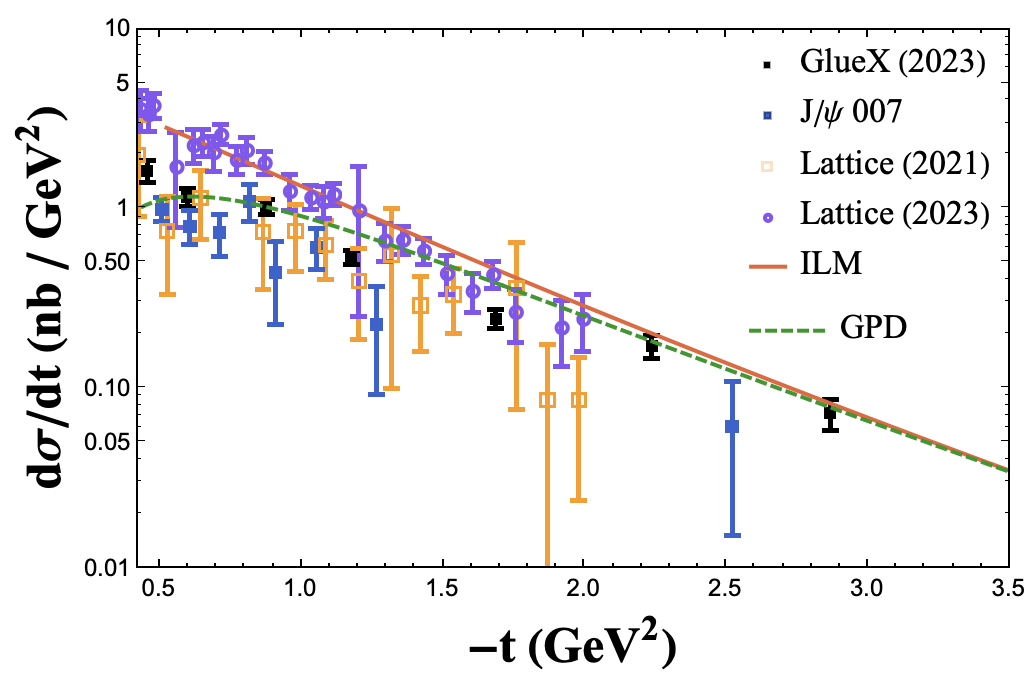}%
}\hfill
\subfloat[\label{fig:JPsi}]{%
\includegraphics[height=5cm,width=.49\linewidth]{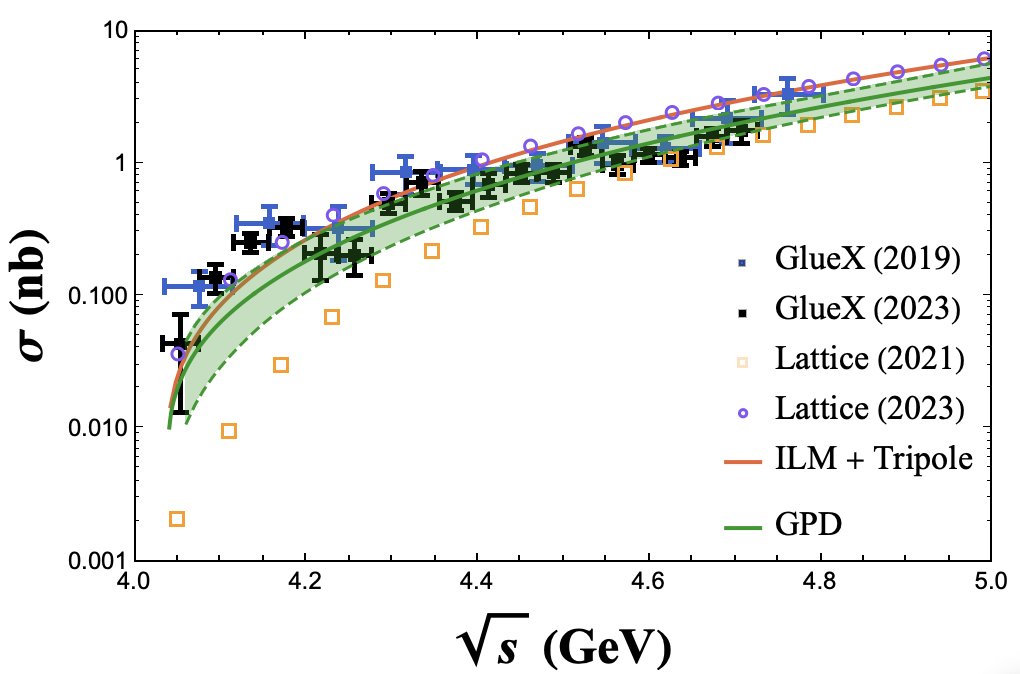}%
}\hfill
\subfloat[\label{fig:dUpsilon}]{%
\includegraphics[height=5cm,width=.49\linewidth]{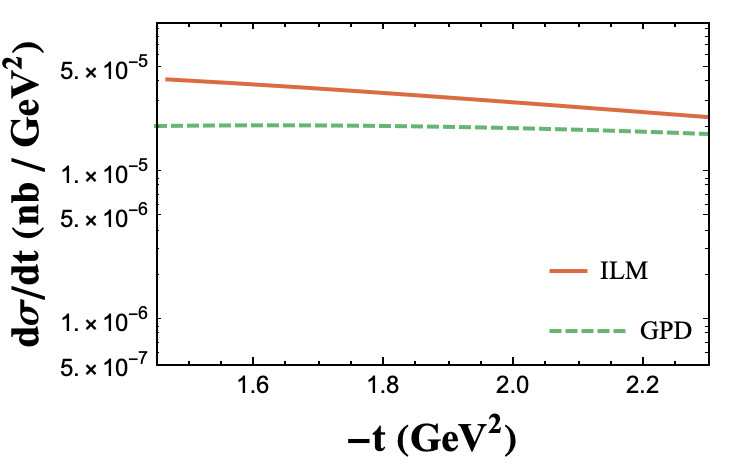}%
}\hfill
\subfloat[\label{fig:Upsilon}]{%
\includegraphics[height=5cm,width=.49\linewidth]{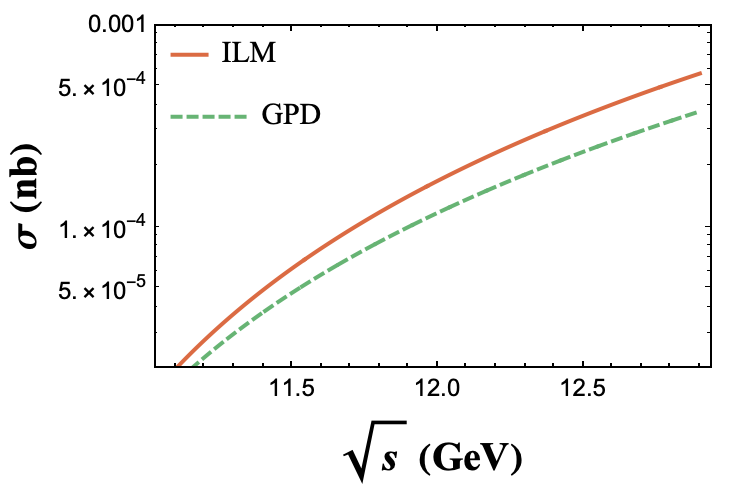}%
}
\caption{ (a) The ILM prediction with the meson dominance exchange on the $t$-dependence \cite{Mamo:2022eui} using tripole approximation is compared to the GlueX experiment in 2023 \cite{GlueX:2023pev} and $J/\psi$ 007 \cite{Duran:2022xag}. The lattice calculations \cite{Pefkou:2021fni} and \cite{Hackett:2023rif} are also compared. We also compare with the GPD prediction \cite{Guo:2023pqw};
(b) The instanton estimation with the holographic prediction on the $t$-dependence \cite{Mamo:2022eui} using tripole approximation is compared to the GlueX experiment in 2019 \cite{GlueX:2019mkq} and in 2023 \cite{GlueX:2023pev}. The lattice calculations \cite{Pefkou:2021fni} and \cite{Hackett:2023rif} are also compared. We also compare with the GPD prediction \cite{Guo:2023pqw};
(c) The instanton estimation on the $\Upsilon$ production differential cross section; 
(d) The cross section  with the holographic prediction on the $t$-dependence \cite{Mamo:2022eui} is compared to the GPD prediction \cite{Guo:2021ibg}.}
\end{figure*}

\section{Photoproduction cross sections}
\label{SEC4}
The differential cross sections for both the coherent production of heavy vector and pseudoscalar mesons off a nucleon, can now be evaluated in parallel for comparison,
and consistency with previous calculations for the vector mesons. 

\subsection{$J/\Psi, \Upsilon$ photo-production}
In the case of $J/\psi$ photoproduction, we average over the proton initial spin and sum over the final spin. The cross section reads

\begin{equation}
\label{DIFFPSI}
\begin{aligned}
  \frac{d\sigma}{dt}
=& 4\pi\alpha_{em}Q_c^2\frac{16\pi\alpha^2_s}{(s-M_N^2)^2}\frac{4}{N_cM^2_{J/\psi}}\left|\psi_{J/\psi}(0)\right|^2\frac{1}{4}\sum_{\lambda_\gamma \lambda_Xss'}(\epsilon_\gamma\cdot\epsilon_V^*)^2\left|\mathcal{W}_{2g}(t,\xi)\right|^2\\
=&4\pi \alpha_{em}Q_c^2\frac{16\pi\alpha^2_s}{(s-M_N^2)^2}\frac{4}{N_cM^2_{J/\psi}}\left|\psi_{J/\psi}(0)\right|^2\\
&\times\frac{4}{\xi^4}\left[(H_{2g}+E_{2g})^2(1-\xi^2)-2(H_{2g}+E_{2g})E_{2g}+\left(1-\frac{t}{4M_N^2}\right)E_{2g}^2\right]    
\end{aligned}
\end{equation}

The dominant amplitude for the photo-production of $J/\Psi$ stems solely from
their transverse polarizations~\cite{Sun:2021pyw}, where we used
$$\sum_{\lambda_\gamma \lambda_X}(\epsilon_\gamma\cdot\epsilon_V^*)^2=2$$
and in overall agreement with~\cite{Guo:2021ibg,Guo:2023pqw}.

The heavy vector meson wave function is fixed by the decay constant. For  $J/\psi$ ($\eta_c$ below), the strong coupling constant is fixed by the charmonium mass scale $\alpha_s(\mu=2m_c)=0.308$, with   
$M_{J/\psi}=3.097~\mathrm{GeV}$ ~\cite{ParticleDataGroup:2018ovx}.
Similarly, for the heavier mesons  $\alpha_s(\mu=2m_b)=0.207$, with $M_{\Upsilon}=9.460~\mathrm{GeV}$~\cite{ParticleDataGroup:2004fcd,ParticleDataGroup:2014cgo}.

In Fig.~\ref{fig:dJPsi} we show our result using the tripole 
form factors (\ref{ACF}) for the differential cross section for photo-production of $J/\Psi$  (\ref{DIFFPSI}) (orange-solid line),  in comparison to the  recently reported measurement by GlueX collaboration (black-data points) \cite{GlueX:2023pev} and by the $J/\Psi~007$ collaboration  (blue-data points) \cite{Duran:2022xag}. Also for comparison, we show the same  differential cross section using the lattice data for the form factors (orange-open circles) and 
from~\cite{Pefkou:2021fni} and (purple-open circles) from~\cite{Hackett:2023rif}, and from the  GPD analysis in~\cite{Guo:2023pqw}. The results for the total cross section for the same process and using the same labeling are shown in  Fig.~\ref{fig:JPsi}.

In Fig.~\ref{fig:dUpsilon} we show our prediction for the differential cross section for the coherent photoproduction of $\Upsilon$ (orange-solid line) 
in comparison to the GPD result (green-dashed line)~\cite{Guo:2023pqw}. In Fig.~\ref{fig:Upsilon} our prediction for the cross section for the same process (orange-solid line) also in comparison to the GPD result (green-dashed line)~\cite{Guo:2023pqw}.  The
results are totally compatible, since our approach follows their construction. In light of this, we now
proceed to the analysis of the coherent production of heavy pseudoscalars.

\begin{figure*}
\includegraphics[width=\linewidth]{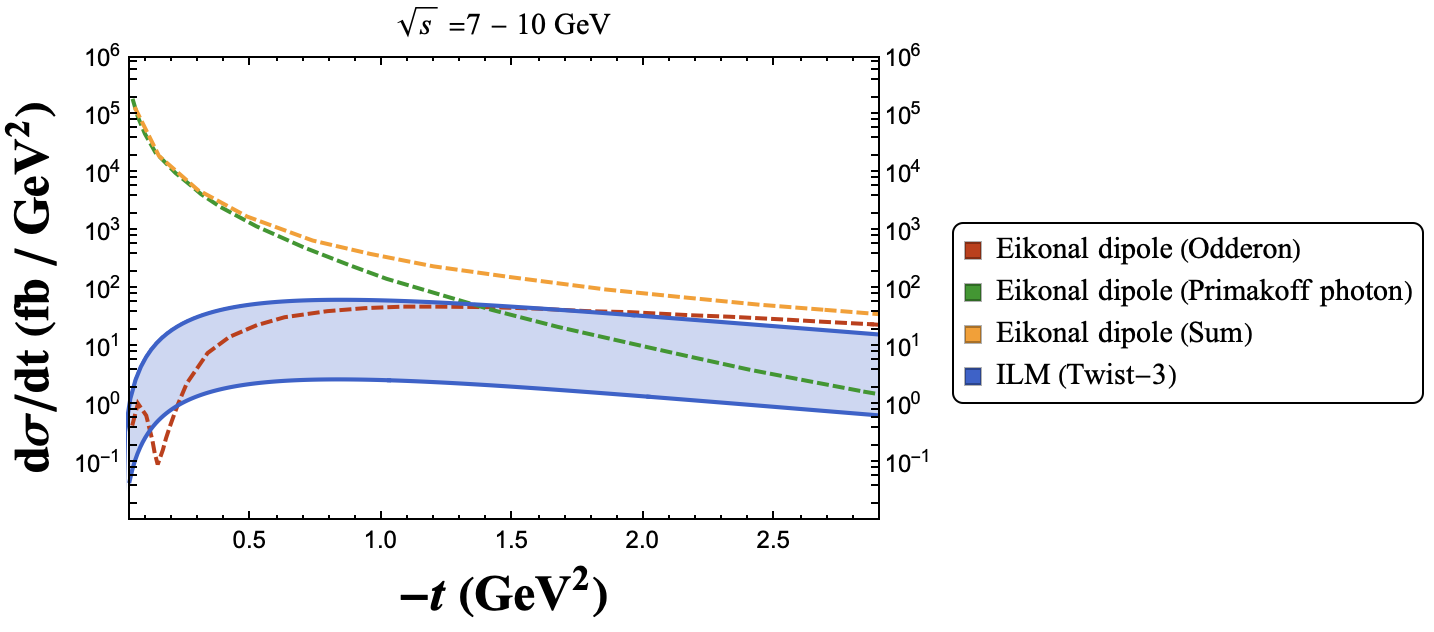}
\caption{ 
The instanton estimation on the $\eta_c$ production differential cross section with the holographic prediction on the $t$-dependence \cite{Hechenberger:2024abg} using fitted mass $m_{3g}=1.49~\mathrm{GeV}$ is compared to the model calculation using eikonal dipole approximation \cite{Dumitru:2019qec}.}
\label{fig:eta_c_2}
\end{figure*}

\begin{figure*}
\subfloat[\label{fig:eta_c}]{%
\includegraphics[height=4.5cm,width=.49\linewidth]{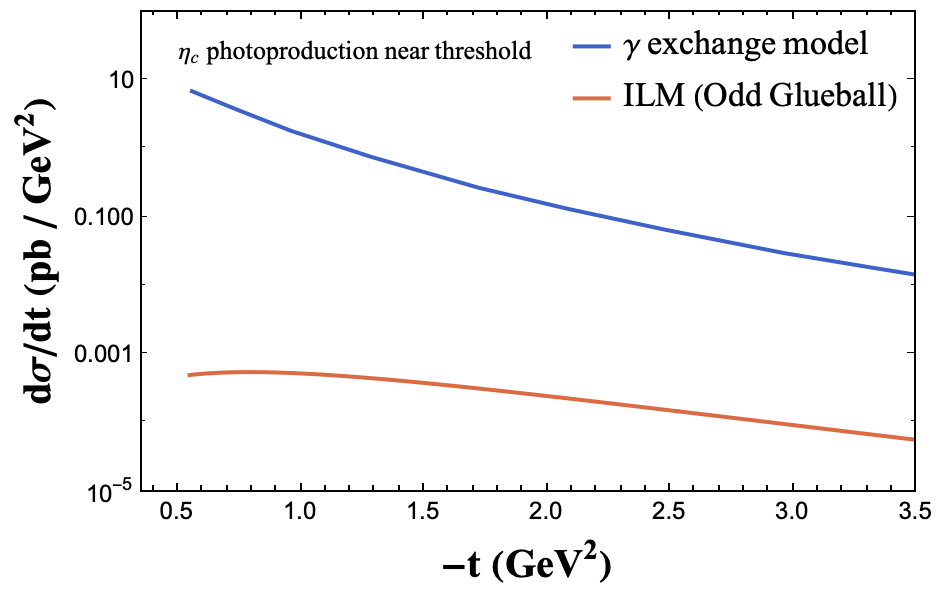}%
}\hfill
\subfloat[\label{fig:eta_b}]{%
\includegraphics[height=4.5cm,width=.49\linewidth]{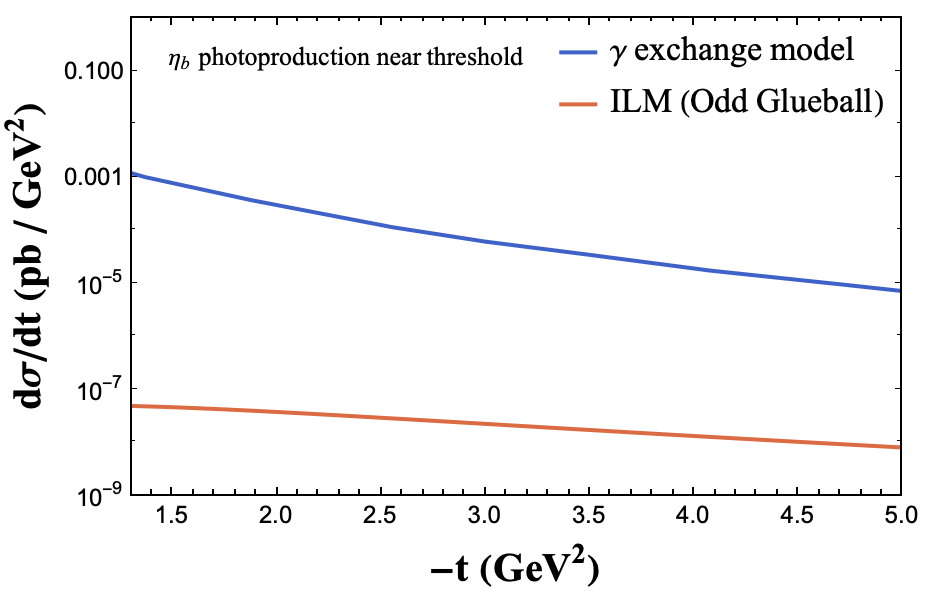}%
}
\caption{ILM estimation on the differential cross section of the $\eta_c$ production (a) and $\eta_b$ production (b) using twist-3 three gluon operator predicted by QCD factorization  with fitted tripole mass $m_{3g}=1.49~\mathrm{GeV}$ \cite{Hechenberger:2024abg} is compared to the photon exchange process predicted in the perturbative QCD calculation \cite{Jia:2022oyl}.}
\end{figure*}

\subsection{$\eta_{c,b}$ photo-production}
The coherent photoproduction of heavy pseudoscalars  $\eta_{c,b}$ in the near threshold region follows 
a similar reasoning. The corresponding differential cross section for $\eta_c$ production is given by

\begin{equation}
\label{DTETAC}
\begin{aligned}
    &\frac{d\sigma}{dt}=4\pi\alpha_{em}Q_c^2\frac{32\pi^2\alpha_s^3}{(s-M_N^2)^2}\frac{4}{N_cM^2_{\eta_c}}|\psi_{\eta_c}(0)|^2\frac{1}{4}\sum_{\lambda_\gamma ss'} \left|\epsilon_{\perp ij}\epsilon^i_{\gamma}\mathcal{W}^j_{3g}(t,\xi)\right|^2\\
    &=4\pi \alpha_{em}Q_c^2\frac{32\pi^2\alpha_s^3}{(s-M_N^2)^2}\frac{4}{N_cM^2_{\eta_c}}|\psi_{\eta_c}(0)|^2\left(\frac{3}{2}\right)^6\frac{36}{\xi^8}\left(\frac{M_N}{\bar{P}^+}\right)^2\\
    &\times\left[f_1^2(1-\xi^2)-2\xi f_1f_4-f_4^2\frac{t}{4M_N^2}+\frac{\Delta_\perp^2}{2M^2_N}L_1-\frac{\bar{P}_\perp\cdot\Delta_\perp}{M^2_N}L_2-\frac{\bar{P}_\perp^2}{2M^2_N}L_3\right]
\end{aligned}
\end{equation}
The orbital form factor contributions are defined as
\begin{equation}
    L_1(t,\xi)=f_2^2\left(1-\frac{t}{4M^2_N}\right)+f^2_3(1-\xi^2)-f_1f_2+2f_2f_3-\xi f_3f_4-\frac{1}{2}f_4^2
\end{equation}
\begin{equation}
    L_2(t,\xi)=\xi f_1f_2+\frac{1}{2}f_1f_4+\frac{t}{4M^2_N}f_2f_4
\end{equation}
\begin{equation}
    L_3(t,\xi)=2\xi f_1f_4+\frac{t}{4M^2_N}f_4^2
\end{equation}

Using Eqs.~\eqref{eq:odd_gluball1}, \eqref{eq:f1234}, and \eqref{eq:gluon_tw3}, we see that the twist-3 GPD moments $f_{1,2,3,4}(t,\xi)$ are all tied to our $C$-odd quark tensor charge and magnetic form factors \eqref{eq:tensor_mag}. Although the cross section vanishes in the forward limit $\Delta^\mu\rightarrow0$, near the threshold region in the heavy quark limit, the allowed kinematic region is restricted to $\xi\rightarrow1$ and $t\rightarrow t_{th}$. The leftover of the center-of-mass energy $\sqrt{s}$ is not enough to excite a large orbital motion (large $\Delta_\perp$ and $\bar{P}_\perp$) during the scattering. Therefore, the momentum transfer $t$ would be constrained in a small range around $t_{th}\sim M_NM_X$. As the orbital motion inside the hadron bound state does not have significant effect in this regime, the remaining contribution to the differential cross section comes from the intrinsic property of the constituent quark which can be estimated in the ILM. The differential cross sections near the threshold in the heavy meson limit read

\begin{equation}
\begin{aligned}
    \frac{d\sigma}{dt}\bigg|_{\eta_c}\simeq&4\pi \alpha_{em}Q_c^2\frac{32\pi^2\alpha_s^3}{(s-M_N^2)^2}\frac{4|\psi_{\eta_c}(0)|^2}{N_cM^2_{\eta_c}}\left(\frac{3}{2}\right)^6\frac{36}{\xi^8}\left(\frac{M_N}{\bar{P}^+}\right)^2\\
    &\times\left[f_1^2(1-\xi^2)-2\xi f_1f_4-f_4^2\frac{t}{4M_N^2}\right]
\end{aligned}
\end{equation}

Here the heavy pseudoscalar meson wave function is fixed by the decay constant. For $\eta_c$, the strong coupling constant is fixed by the charmonium mass scale $\alpha_s(\mu=2m_c)=0.308$, with   
$ M_{\eta_c}=2.984~\mathrm{GeV}$~\cite{ParticleDataGroup:2018ovx}, and bottomium mass scale $\alpha_s(\mu=2m_b)=0.207$, with $M_{\eta_b}=9.398~\mathrm{GeV}$~\cite{ParticleDataGroup:2004fcd,ParticleDataGroup:2014cgo}.

In Fig.~\ref{fig:eta_c_2} we show our result (\ref{DTETAC}) for the differential cross section for the coherent photo-production of $\eta_c$ near threshold (orange-solid line), to 
eikonalized odderon exchange (blue-solid curve), eikonalized photon exchange (green-dashed line) and their sum (black-dashed line)~\cite{Dumitru:2019qec}. Our 
result is substantially smaller than the one reported using the eikonal dipole approximation~\cite{Dumitru:2019qec}.
In Fig.~\ref{fig:eta_c} we compare our result for the $\eta_c$ production (orange-solid line), to the photon exchange model (blue-solid line)~\cite{Jia:2022oyl}. The same comparison is
carried in Fig.~\ref{fig:eta_b} for the $\eta_b$ production.  The photon rate dwarfs our estimate for the  $C$-odd gluon exchange production of heavy $\eta_{c,b}$ near threshold.

\begin{figure}
    \centering
    \includegraphics[width=\linewidth]{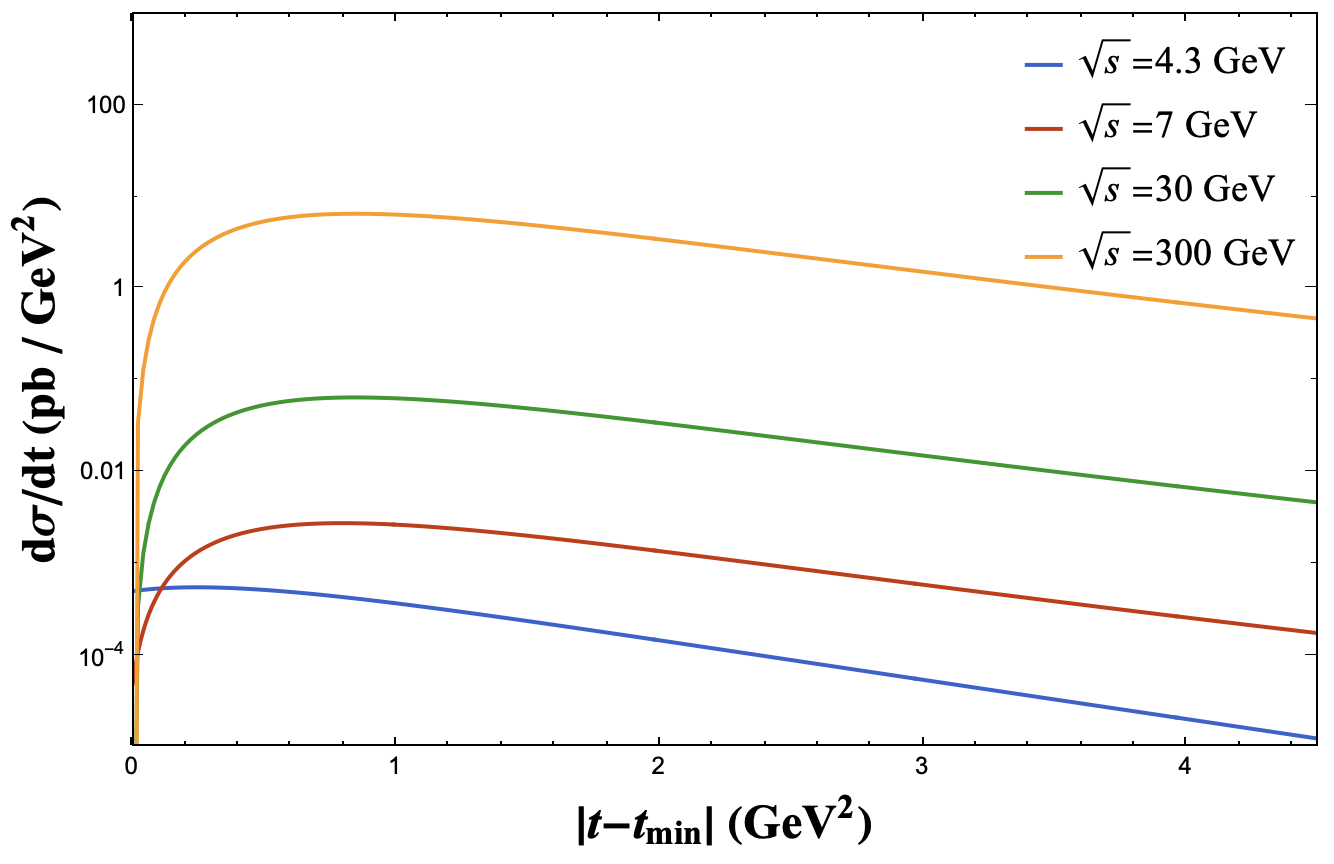}
    \caption{The $t$ dependence for the threshold photo-production of $\eta_c$ using near-threshold approximated ILM formula ($f_1$ dominated crossection) with different the center of mass energy $\sqrt{s}$.}
    \label{fig:s_dep}
\end{figure}

\begin{figure}
    \centering
    \includegraphics[width=\linewidth]{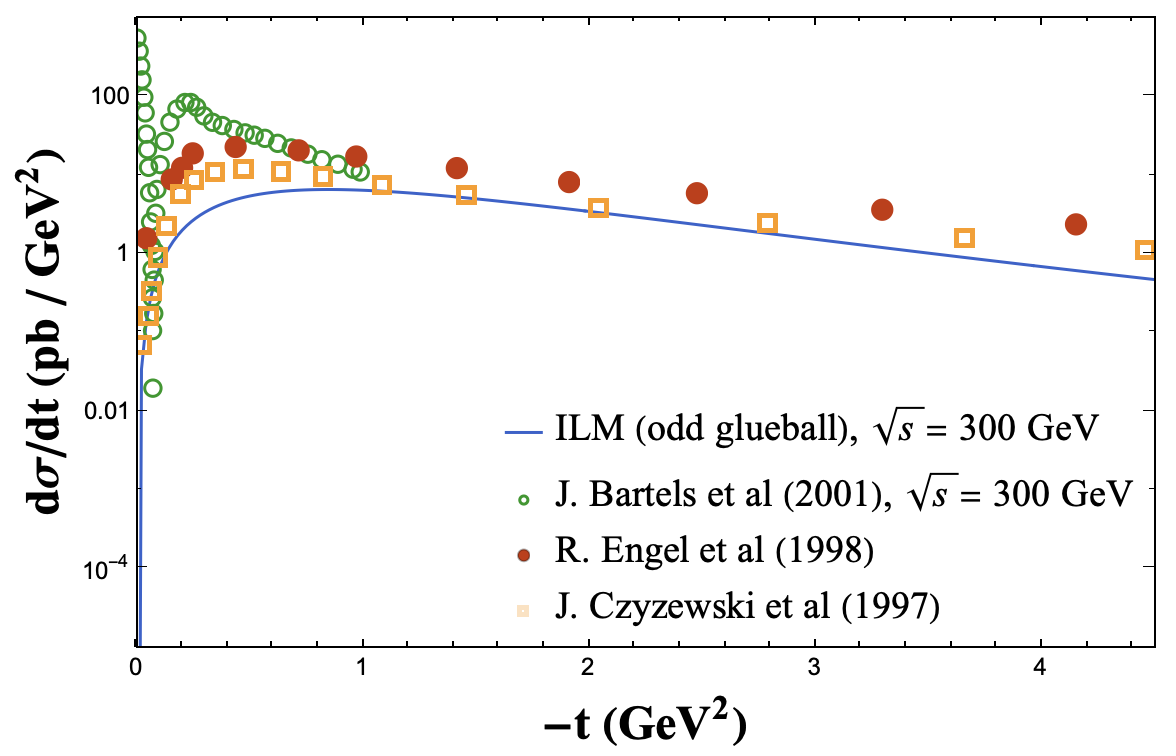}
    \caption{ILM differential cross section for $\eta_c$ photoproduction using near-threshold approximated formula ($f_1$ dominated) at $\sqrt{s}=300\rm GeV$, using the GPD arguments with the $C$-odd holographic form factor~\cite{Hechenberger:2024abg} 
    (fitted mass $m_{3g}=1.49~\mathrm{GeV}$)
    (blue solid line). The comparison is to the perturbative QCD results  \cite{Czyzewski:1996bv} (orange box), \cite{Engel:1997cga} (red dot), and \cite{Bartels:2001hw} (green circle).}
    \label{fig:odderon}
\end{figure}

In the threshold region and in the heavy meson limit ($\xi\rightarrow1$, $t\rightarrow t_{th}$), the ratios of the vector to pseudoscalar differential cross sections for the charmed and bottom mesons respectively, are
\begin{equation}
\frac{d\sigma/dt(\eta_c)}{d\sigma/dt(J/\psi)}\Bigg|_{\substack{\sqrt{s}=4.05~\mathrm{GeV}\\ -t=1.8~\mathrm{GeV}^2}}=4.424\times10^{-6},\quad
\frac{d\sigma/dt(\eta_b)}{d\sigma/dt(\Upsilon)}\Bigg|_{\substack{\sqrt{s}=10.4~\mathrm{GeV}\\ -t=7.2~\mathrm{GeV}^2}}=1.577\times10^{-5}
\end{equation}
This illustrates the smallness of the gluonic mechanism for pseudoscalar production predicted by QCD factorisation.  These ratios depend sensitively on the $C$-even twist-2 gluonic gravitational charges $A_g$, and the $C$-odd twist-3 gluonic charge, as shown in (\ref{RATIO23}). 

Our analysis of the photo-production of $\eta_c$, is limited to the threshold region. It relies on the skewness expansion of the $C$-odd twist-3 GPD, much like the $C$-even twist-2 GPD for $J/\Psi$ threshold photo production~\cite{Guo:2023pqw}, both of which are not Reggeized. This notwithstanding, 
the dependence of the ILM results on the center of mass energy are  shown in Fig.~\ref{fig:s_dep}. The
differential cross section is seen to rise
with increasing $\sqrt{s}$. In Fig.~\ref{fig:odderon} we compare the ILM results
at $\sqrt{s}=300\,\rm GeV$ (solid-blue line)
to the perturbative QCD results in \cite{Czyzewski:1996bv,Engel:1997cga,Bartels:2001hw} in . The ILM results are in the  pb/GeV$^2$ range, while the recent calculation in \cite{Benic:2023ybl}  is substantially higher in the fb/GeV$^2$ range.

Finally, we note that in dual gravity the photo-production of heavy pseudoscalars in the threshold region, was recently found to be substantially distinct from the QCD factorization argument~\cite{Hechenberger:2024abg}. This is due to the fact that the boundary dual of the bulk Kalb-Ramond field is a twist-5 operator $d^{abc}F^{a}_{\mu\nu}F^{b}_{\rho\lambda}F^{c}_{\rho\lambda}$, in contrast to the QCD factorization process, which is driven by the twist-3 operator. In ILM, the twist-3 parametrically suppressed near threshold $\rho q\approx 1$ and the ratio between twist-3 and twist-5 again vanishingly small in the large $N_c$ limit.

\chapter{Twist-3 color Lorentz force}
\label{ch:twist3}
In DIS, the twist-3 contribution describes the average Lorentz force acting on a quark in the nucleon. This contribution is accessible in DIS process on a polarized nucleon target.
In contrast to the  structure function $g_1(x, Q^2)$ which is leading twist-2 and polarization independent, the structure function $g_2(x, Q^2)$ is a twist-3 and polarization dependent~\cite{Shuryak:1981kj,Shuryak:1981pi}. We now proceed to its quantitative description in the QCD instanton vacuum, using the general framework developed in~\cite{Liu:2024rdm} that includes 
individual pseudoparticles (instantons, anti-instantons) and their molecular configurations (instanton-anti-instanton). We will focus on this average force acting on the emerging quark, the pion and the nucleon. 

DIS process on a polarized target, splits into longitudinal and transverse contributions
\begin{equation}
\begin{aligned}
    &g_1(x)s_L=\int_{-\infty}^{\infty} \frac{d\xi^-}{4\pi} e^{i \xi^-p^+ x}
\langle ps | \bar{\psi}(0) \gamma^+ \gamma^5 W(0,\xi^-) \psi(\xi^-) | ps \rangle
\end{aligned}
\end{equation}
and
\begin{equation}
\begin{aligned}
    &\frac{m_N}{p^+} g_T(x)s_\perp=\int_{-\infty}^{\infty} \frac{d\xi^-}{4\pi}e^{i\xi^-p^+x} 
\langle ps | \bar{\psi}(0) \gamma^\perp \gamma^5 W(0,\xi^-)\psi(\xi^-) | ps \rangle
\end{aligned}
\end{equation}
where the transverse distribution is the sum of $g_2$ and $g_1$ 
\begin{equation}
    g_T(x)=g_1(x)+g_2(x)
\end{equation}
as a measure of the transverse spin. We now recall that $g_2$ satisfies the 
Burkhardt–Cottingham (BC) sum rule  \cite{Bhattacharya:2021boh}
\begin{equation}
    \int dx\, g_2(x,Q^2) = 0 
\end{equation}
connecting the twist-2 and twist-3 parton distributions. 
Since the structure functions $g_1, g_2$ can be separated kinematically in experiments, this observation  allows for the description of higher twist effects initially developed in~\cite{Shuryak:1981pi,Shuryak:1981dg,Jaffe:1981td,Jaffe:1982pm}.

For a transversely polarized nucleon  $g_1(x, Q^2)$ vanishes and the remaining DIS amplitude is purely $g_2$ of twist-3. By analogy with the twist-2 PDFs $f_1(x), g_1(x), h_1(x)$, the twist-3 PDFs are referred to as 
$e(x), g_T(x), h_L(x)$. In particular,  the twist-3 PDF $g_T(x)$ can be expressed as a sum of the Wandzura-Wilczek (WW) term, a piece that is determined entirely in terms of twist-2 helicity PDF $g_1(x)$, and an interaction dependent dynamical twist-3 term $\bar g_2(x)$, which involves quark-gluon correlations. \cite{Aslan:2019jis}
\begin{equation}
    g_T(x, Q^2)=g_T^{WW}(x, Q^2)+\bar{g}_2(x, Q^2)
\end{equation}
The first contribution is the twist-2  Wandzura-Wilczek fixed by~\cite{Wandzura:1977qf}
\begin{equation}
g_T^{WW}(x, Q^2) \equiv \int_x^1 dy \frac{g_1(y, Q^2)}{y}
\end{equation}
%
After subtraction of this twist-2 contribution,
the second Mellin moment of $\bar g_2(x)$  is related to the Lorentz color force matrix element ~\cite{Aslan:2019jis,Vladimirov:2025qrh}
\begin{equation}
\frac {d_2}3=\int_0^1dx\,x^2 \bar g_2(x)
\end{equation}
where the $d_2$ moment can be related directly to the matrix element 
\begin{equation}
\label{FY}
\begin{aligned}
d_2 \epsilon_{\perp ij} S^j
=
-\frac{\langle ps|\bar \psi(0)\gamma^+ g F^{+i}(0) \psi(0)|ps\rangle}
{2m_N (p^+)^2 }
\end{aligned}
\end{equation}
This result can be seen to follow from the identity~\cite{Aslan:2019jis}
\begin{equation}
\begin{aligned}
&2S^xm_N(P^+)^2\int_0^1dx\,x^2 g_T(x) =-\langle ps | \bar{\psi}(0) \gamma^x \gamma^5 (\overleftrightarrow{D}^+)^2 \psi(0) | ps\rangle\nonumber\\
&= 2 S^x m_N (P^+)^2 \int_{-1}^{1} \mathrm{d}x \, x^2 g_T^{WW}(x) - \frac{1}{3} \langle ps| \bar{\psi}(0) \gamma^+ g F^{+y}(0) \psi(0) | ps \rangle
\end{aligned}
\end{equation}

For a polarized nucleon along the $x$-direction  with light-cone momentum $P^+$ along the $z$-direction and struck by a transverse probe $q_\perp$, the average color Lorentz force density on a transverse plane is \cite{Crawford:2024wzx}
\begin{equation}
\begin{aligned}
    &F_{q/h}^i(b)=i\int \frac{d^2q_\perp}{(2\pi)^2}e^{-iq_\perp\cdot b}\frac {\langle h(p')|\bar \psi \gamma^+igF^{+i}\psi|h(p)\rangle}{\sqrt{2}\bar{p}^+ }
\end{aligned}
\end{equation}
The presence of $\bar \psi\gamma^+ \psi$ in the operator means that this force is proportional to the quark density~\cite{Aslan:2020eqo}
\begin{equation}
    \rho_{q/h}(b)=\int \frac{d^2q_\perp}{(2\pi)^2}e^{-iq\cdot b_\perp}\frac{\langle h(p')|\bar \psi \gamma^+\psi|h(p)\rangle}{\sqrt{2}\bar{p}^+}
\end{equation}
which is a measure of the electromagnetic form factor in the transverse plane.


We note that the force involves the quark current
 with a single $\gamma_\mu$ which is chiral
even ($LL+RR$). Since the instanton-induced operators are
chiral-odd ($LR+RL$), their contribution to the force is a priori suppressed,
except for their non-zero mode contributions which are chiral even. This
brings about the role and contribution of the molecular instanton-anti-instanton  configurations to the force as we discussed earlier, although formally suppressed by an extra power of the packing fraction.

\begin{figure}
\centering
\subfloat[\label{fig:ONEv1}]{\includegraphics[width=.35\linewidth]{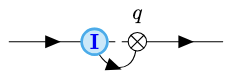}}
\hfill
\subfloat[\label{fig:ONEv2}]
{\includegraphics[width=.35\linewidth]{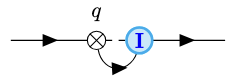}}
\hfill
\subfloat[\label{fig:ONEv3}]{\includegraphics[width=.28\linewidth]{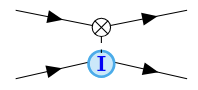}}
    \caption{The single instanton/anti-instanton  vertices with the insertion of the color Lorentz operator (crossed-circle): (a), (b) along a fermion line and (c) with a pair of fermion lines. The latter (two-body operator) is surppessed by $1/N_c$ compared to the former (one-body operator).}
    \label{fig:ONEvertices}
\end{figure}

\begin{figure}
\centering
\subfloat[\label{fig:v1}]{\includegraphics[width=.35\linewidth]{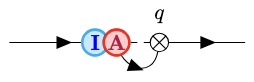}}
\hfill
\subfloat[\label{fig:v2}]{\includegraphics[width=.35\linewidth]{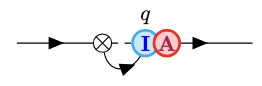}}
\hfill
\subfloat[\label{fig:v3}]{\includegraphics[width=.28\linewidth]{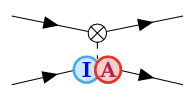}}
    \caption{A molecular pair of instanton-anti-instanton with the insertion of the color Lorentz operator (crossed-circle): (a), (b) along a fermion line and (c) with a pair of fermion lines. The latter (two-body operator) is suppressed by $1/N_c$ compared to the former (one-body operator).}
    \label{fig:vertices}
\end{figure}

\section{Emergent form factors and the Color Lorentz Force  }
\label{SEC-FF}
The color Lorentz force contributions from
individual instantons or anti-instantons are illustrated in Fig.~\ref{fig:ONEvertices}, and the molecular contributions illustrated in Fig.~\ref{fig:vertices}. The details of their analytic forms are given in \cite{Liu:2025ypg}. More specifically, the  non-forward amplitude of the twist-3 operator in Fig.~\ref{fig:ONEvertices} can be deduced from the 4-fermion contribution given by

\begin{equation}
\begin{aligned}
\label{eq:force_op_0}
    \langle h'|ig\bar \psi F_{\mu\nu}\gamma_\sigma\psi|h\rangle=&\frac{1}{4(N_c^2-1)}\frac{n_{I+A}}2\left(\frac{4\pi^2\rho^2}{m^*}\right)i\int \frac{d^4k}{(2\pi)^4}8\rho^2G_I(\rho k)\left(\frac{k_\lambda k_\nu}{k^2}-\frac14g_{\lambda\nu}\right)\\
    &~\times\int d^4xe^{-i(q+k)x}\langle h'|\bar\psi(x)\gamma_{\sigma}\lambda^A\psi(x)\bar\psi(0) \sigma_{\mu\lambda}\lambda^A\psi(0)|h\rangle\\
    &-\frac{n_{mol}}{4(N_c^2-1)^2}\gamma_{IA}\, i\,t_{\mu\nu\rho\lambda\alpha\beta}\int \frac{d^4k}{(2\pi)^4}\rho^2\frac{k_{\rho}k_{\lambda}}{k^2}G_{IA}(\rho k)\\
    &~\times\int d^4xe^{-i(q+k)x}\langle h'|\bar\psi(x)\gamma_\sigma\lambda^A\psi(x)\bar\psi(0)i\gamma_{(\alpha}\gamma^5\overleftrightarrow{\partial}_{\beta)}\lambda^A\psi(0)|h\rangle
\end{aligned}
\end{equation}

Compared with Ref.~\cite{Polyakov:2018exb}, we have additional molecular contribution. Here each emergent quark carries a non-local form factor defined  in \eqref{ZMform}, reflecting on its origin as a quark zero mode. The Fourier transform of the instanton field strength profile gives 
\begin{equation}
\begin{aligned}
    &G_I(k)
    =\frac{4\pi^2}{k^2}\left(\frac{8}{k^{2}} -\frac{k^{2}}2 \, K_{2}(k) - k \, K_{3}(k) \right)
\end{aligned}
\end{equation}
and
\begin{equation}
\begin{aligned}
    &G_{IA}(k)
    =
    \frac{4\pi^2}{k^2}\left( 1-\frac{16}{k^{2}} +\frac{k^{2}}2 \, K_{2}(k) + 2 k \, K_{3}(k) \right)
\end{aligned}
\end{equation}
with $K_{2,3}$ are  modified Bessel functions of the second kind.
The emerging instanton-anti-instanton coupling $\gamma_{IA}$ is estimated as (see also Eq.~\eqref{eq:gammaIA}) 

\begin{equation}
    \gamma_{IA}=2\pi^2n_{I+A}\int_0^{\bar{R}} dRR^3\, \sum_{n=1}^{N_f}\left|\frac{T_{IA}}{m^*}\right|^{2n}
   \left(\frac{4\pi^2\rho^2}{|T_{IA}|}\right)^{2}\left[\frac{-1}4R\frac{dT(R)}{dR}\right]
\end{equation}
where the value is 311 fm$^4$ for $m^*=67.55$ MeV after summing over all the 3-flavors in the molecular pairing, as detailed in \cite{Liu:2025ypg}.

For the quark-gluon operator inside a hadron, the resulting effective quark operators are usually related to a four point correlation (two quark current and two hadronic source).
Since the instanton profile is highly localized, the separation between the two-quark source in \eqref{eq:force_op_0} is controlled by $|x|\lesssim\rho\ll \sqrt[4]{1/n_{I+A}}$. Thus, we can further approximate the hopping quark propagator in Fig.~\ref{fig:ONEv1} and \ref{fig:ONEv2} for a single instanton, and \ref{fig:v1} and \ref{fig:v2} for an instanton-anti-instanton pair by 

\begin{equation}
\begin{aligned}
&\sqrt{\mathcal{F}(i\rho\partial)}S(x)\equiv\langle\psi(x)\sqrt{\mathcal{F}(i\rho\partial)}\bar\psi(0)\rangle\\
\simeq&\frac{i\slashed{x}}{2\pi^2x^4}K_D(x/\rho)+\frac{m}{4\pi^2x^2}K_m(x/\rho)+\mathcal{O}(m^2)
\end{aligned}
\end{equation}

The vertices in Fig.~\ref{fig:ONEv3} and \ref{fig:v3} will be neglected as they contribute to higher power of $1/N_c$ (two-body operators).




With this in mind, the contribution to the (non-forward) matrix element of Fig.~\ref{fig:ONEvertices}  is

\begin{equation}
\begin{aligned}
\label{eq:CFOp2}
    &\langle h'|ig\bar\psi\gamma_\sigma F_{\mu\nu}\psi|h\rangle_{I+A}=-\left(\frac{n_{I+A}}2\right)\frac{1}{2N_c}\left(\frac{4\pi^2\rho^2}{m^*}\right)\beta^{(+)}_{\bar{q}Gq,1}(\rho q)\left(g_{\mu\sigma}q_\nu-g_{\nu\sigma}q_\mu\right)\langle h'|\bar\psi\psi|h\rangle\\
    &+\left(\frac{n_{I+A}}2\right)\frac{1}{2N_c}\left(\frac{4\pi^2\rho^2}{m^*}\right)\beta^{(+)}_{\bar{q}Gq,1}(\rho q)\epsilon_{\mu\nu\sigma\alpha}q_\alpha\langle h'|\bar\psi i\gamma^5\psi|h\rangle\\
    &-\left(\frac{n_{I+A}}2\right)\frac{1}{2N_c}\left(\frac{4\pi^2\rho^2}{m^*}\right)\rho^2\beta^{(+)}_{\bar{q}Gq,2}(\rho q)2i\epsilon_{[\mu\lambda\sigma\rho}\left(q_\lambda q_{\nu]}-\frac 1{4}g_{\lambda\nu]}q^2\right)m^*\langle h'|\bar\psi\gamma_\rho\gamma^5\psi|h\rangle\\
\end{aligned}
\end{equation}
where two new instanton form factors are defined as 
\begin{align}
    \beta^{(+)}_{\bar q Gq,1}(q)=&\frac1q\int_0^\infty dx \frac{16}{(x^2+1)^2}\frac{J_2(qx)}{qx}K_D(x) \\
    \beta^{(+)}_{\bar q Gq,2}(q)=&\frac1q\int_0^\infty dx \frac{16x^2}{(x^2+1)^2}\frac{J_3(qx)}{q^2x^2}K_m(x)
\end{align}
with the quark zero mode induced modification during the quark propagation
\begin{equation}
    K_D(x)=
    \frac{x^3}{(1 + x^2)^{3/2}}
\end{equation}
and
\begin{equation}
    K_m(x)=\int_0^\infty dk x J_1(kx)\sqrt{\mathcal{F}(k)}
\end{equation}
At zero momentum transfer, the values of those two instanton form factors are
$\beta^{(+)}_{\bar q Gq,1}(0)=\frac25$ and
$\beta^{(+)}_{\bar q Gq,2}(q\rightarrow0)=-\frac13\ln q$. 

The ``molecular'' contribution from Fig.~\ref{fig:vertices} is

\begin{equation}
\begin{aligned}
\label{eq:CFOp3}
    &\langle h'|\bar\psi igF_{\mu\nu}\gamma_\sigma\psi|h\rangle_{IA}=
    \\
    &-\frac{n_{mol}\gamma_{IA}}{2N_c(N_c^2-1)}\beta^{(IA)}_{\bar{q}Gq,1}(\rho q)\left(g_{\mu\sigma}g_{\nu\alpha}-g_{\nu\sigma}g_{\mu\alpha}\right)q_\beta\langle h'|\bar{\psi}\left(\gamma_{(\alpha} i\overleftrightarrow{\partial}_{\beta)}-\frac{1}{4}g_{\alpha\beta}i\overleftrightarrow{\slashed{\partial}}\right)\psi|h\rangle\\
    &-\frac{n_{mol}\gamma_{IA}}{2N_c(N_c^2-1)}\beta^{(IA)}_{\bar{q}Gq,2}(\rho q)\left(g_{\mu\alpha}q_\nu-g_{\nu\alpha}q_\mu\right)\langle h'|\bar{\psi}\left(\gamma_{(\alpha} i\overleftrightarrow{\partial}_{\sigma)}-\frac{1}{4}g_{\alpha\sigma}i\overleftrightarrow{\slashed{\partial}}\right)\psi|h\rangle\\
    &+\frac{n_{mol}\gamma_{IA}}{2N_c(N_c^2-1)}\frac1{\rho^2}\left[\frac14\rho^2Q^2\beta^{(IA)}_{\bar{q}Gq,2}(\rho q)-\beta^{(IA)}_{\bar{q}Gq,3}(\rho q)\right]i\epsilon_{\mu\nu\sigma\rho}\langle h'|\bar\psi\gamma_\rho\gamma^5\psi|h\rangle\\
    &-\frac{n_{mol}\gamma_{IA}}{2N_c(N_c^2-1)}\left[\frac14\beta^{(IA)}_{\bar{q}Gq,2}(\rho q)-\beta^{(IA)}_{\bar{q}Gq,4}(\rho q)\right]i\epsilon_{\mu\nu\sigma\lambda}q_\lambda q_\rho\langle h'|\bar\psi\gamma_\rho\gamma^5\psi|h\rangle\\
    &-\frac{n_{mol}\gamma_{IA}}{2N_c(N_c^2-1)}\left[\frac34\beta^{(IA)}_{\bar{q}Gq,2}(\rho q)-\beta^{(IA)}_{\bar{q}Gq,4}(\rho q)\right]i\epsilon_{\mu\nu\lambda\rho}q_\lambda q_\sigma\langle h'|\bar\psi\gamma_\rho\gamma^5\psi|h\rangle\\
   &-\frac{n_{mol}\gamma_{IA}}{2N_c(N_c^2-1)}\beta^{(IA)}_{\bar{q}Gq,5}(\rho q)m\epsilon_{\mu\nu\sigma\lambda}q_\lambda \langle h'|\bar\psi i\gamma^5\psi|h\rangle\\
    &-\frac{n_{mol}\gamma_{IA}}{2N_c(N_c^2-1)}\rho^2\beta^{(IA)}_{\bar{q}Gq,6}(\rho q)im\epsilon_{\mu\nu\alpha\lambda}q_\lambda q_\beta\langle h'|\bar\psi\sigma_{\sigma(\alpha}\gamma^5\overleftrightarrow{\partial}_{\beta)}\psi|h\rangle\\
   &+\frac{n_{mol}\gamma_{IA}}{2N_c(N_c^2-1)}\beta^{(IA)}_{\bar{q}Gq,7}(\rho q)im\epsilon_{\mu\nu\alpha\lambda}\langle h'|\bar\psi\sigma_{\sigma\alpha}\gamma^5\overleftrightarrow{\partial}_{\lambda}\psi|h\rangle
\end{aligned}
\end{equation}
where the new {\it molecular form factors} $\beta^{(IA)}$ entering \eqref{eq:CFOp2} are

\begin{align}
    \beta^{(IA)}_{\bar q Gq,1}(q)=&\frac{1}{q}\int_0^\infty dx\frac{16}{(x^2+1)^2x^2}\left(\frac{3J_3(qx)}{q^2x^2}-\frac{J_4(qx)}{qx}\right)K_D(x)\\
    \beta^{(IA)}_{\bar q Gq,2}(q)=&\frac{1}{q}\int_0^\infty dx\frac{16}{(x^2+1)^2x^2}\frac{J_3(qx)}{q^2x^2}K_D(x)
\end{align}

\begin{align}
    \beta^{(IA)}_{\bar q Gq,3}(q)=&\frac1q\int_0^\infty dx \Bigg[\frac{128}{x^4(x^2+1)^2}\left(K_D(x)-\frac14xK'_D(x)\right)\frac{J_2(qx)}{qx}\nonumber\\
    &\qquad\qquad-\frac{16}{(x^2+1)^2x^4}K_D(x)J_3(qx)\Bigg]\\
    \beta^{(IA)}_{\bar q Gq,4}(q)=&\frac1q\int_0^\infty dx \Bigg[\frac{64}{x^2(x^2+1)^2}\left(K_D(x)-\frac14xK'_D(x)\right)\frac{J_3(qx)}{q^2x^2}\nonumber\\
    &\qquad\qquad-\frac{8}{(x^2+1)^2x^2}K_D(x)\left(\frac{J_4(qx)}{qx}-\frac{2J_3(qx)}{q^2x^2}\right)\Bigg]\\
\beta^{(IA)}_{\bar q Gq,5}(q)=&\frac1q\int_0^\infty \frac{dx}{x^2} \Bigg[\frac{16}{(x^2+1)^2}\left(K_m(x)-\frac x2K'_m(x)\right)\frac{J_2(qx)}{qx}-\frac{4x^2K_m(x)}{(x^2+1)^2}J_3(qx)\Bigg]\\
\beta^{(IA)}_{\bar q Gq,6}(q)=&\frac1q\int_0^\infty dx \frac{16}{(x^2+1)^2}\frac{J_3(qx)}{q^2x^2}K_m(x)\\
\beta^{(IA)}_{\bar q Gq,7}(q)=&\frac1q\int_0^\infty dx \Bigg[\frac{16}{x^2(x^2+1)^2}\left(K_m(x)-\frac12xK'_m(x)\right)\frac{J_2(qx)}{qx}\Bigg]
\end{align}

At zero momentum transfer, their values are
\begin{equation}
\begin{aligned}
\beta^{(+)}_{\bar q Gq,1}(0) &= \frac{2}{5}, 
& \beta^{(+)}_{\bar q Gq,2}(q \to 0) &= -\frac{1}{3}\ln q, 
& \beta^{(IA)}_{\bar q Gq,1}(0) &= \frac{2}{15}, \\
\beta^{(IA)}_{\bar q Gq,2}(0) &= 0.0444, 
& \beta^{(IA)}_{\bar q Gq,3}(0) &= 3.048, 
& \beta^{(IA)}_{\bar q Gq,4}(0) &= 0.1460, \\
\beta^{(IA)}_{\bar q Gq,5}(0) &= 0.2331, 
& \beta^{(IA)}_{\bar q Gq,6}(0) &= 0.0935, 
& \beta^{(IA)}_{\bar q Gq,7}(0) &= 0.2262.
\end{aligned}
\end{equation}

\begin{figure}
    \centering
    \includegraphics[width=.8\linewidth]{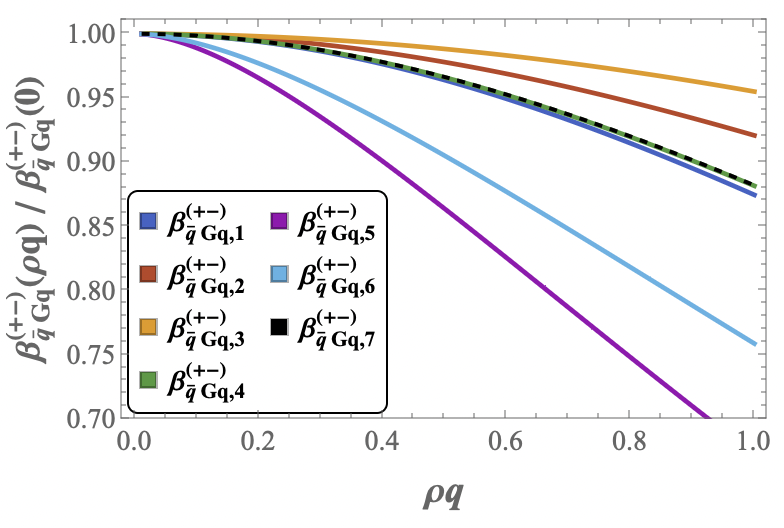}
    \caption{Emergent non-local form factors $\beta^{(IA)}_{\bar{q}Gq}$ associated to the color Lorentz operator in a molecular (instanton-anti-instanton)  pair.}
    \label{fig:beta}
\end{figure}

We note that the derived relations in Eqs.~\eqref{eq:CFOp2} and ~\eqref{eq:CFOp3} relies on a local approximation for the quark-gluon operator. This approximation is justified by the strong localization of the instanton profile and the restriction on the quark separation, $|x|\lesssim \rho$, as follows from the structure of the zero modes and the instanton-induced interaction. However, this simplification neglects nonlocal contributions inherent in the full expression \eqref{eq:force_op_0}, and therefore contains an additional source of systematic uncertainty in the evaluation of the color-force operator. The details of the derivation can be found in Appendix~\ref{app:mole}.

\section{Color force on a pion}
\label{SEC-PION}

To illustrate these ideas, let us start with a simpler case, that of 
deep inelastic scattering on a pion, as its description in the instanton vacuum is well established.


The non-forward amplitude for the twist-3 operator in a pion target,
is constrained by intrinsic parity and hermiticity. Its Lorentz covariant form
is characterized by two invariant form factors,

\begin{equation}
\label{TWIST3FFPION}
\begin{aligned}
    \langle \pi(p') |\bar \psi igF^{\mu\nu} \gamma^\sigma \psi| \pi(p)\rangle=&2\Big(\bar{p}^\mu q^{\nu} - \bar{p}^\nu q^\mu \Big)\bar{p}^\sigma \Phi^q_{\pi,1}(Q^2)\\
&+2m_\pi^2\Big(q^\mu g^{\nu\sigma} - q^\nu g^{\mu\sigma} \Big)\Phi^q_{\pi,2}(Q^2)
\end{aligned}
\end{equation}
where

\begin{align}
    \Phi^q_{\pi,1}(Q^2)=&-\frac{n_{mol}}{2N_c(N_c^2-1)}\gamma_{IA}\beta^{(IA)}_{\bar{q}Gq,2}(\rho q)A^q_\pi(Q^2)\\
    \Phi^q_{\pi,2}(Q^2)=&\left(\frac{n_{I+A}}2\right)\frac{1}{2N_c}\left(\frac{4\pi^2\rho^2}{m^*}\right)\beta^{(+)}_{\bar{q}Gq,1}(\rho q)\frac{\sigma^q_{\pi}(Q^2)}{m}\nonumber\\
    &-\frac{n_{mol}}{8N_c(N_c^2-1)}\gamma_{IA}\left[\beta^{(IA)}_{\bar{q}Gq,1}(\rho q)+\beta^{(IA)}_{\bar{q}Gq,2}(\rho q)\right]\left(1+\frac{Q^2}{4m_\pi^2}\right)A^q_\pi(Q^2)\nonumber\\
    &-\frac{n_{mol}}{8N_c(N_c^2-1)}\gamma_{IA}\left[\beta^{(IA)}_{\bar{q}Gq,1}(\rho q)-\frac13\beta^{(IA)}_{\bar{q}Gq,2}(\rho q)\right]\frac{3Q^2}{4m_\pi^2}D^q_\pi(Q^2)
\end{align}

Here $\sigma_\pi^q$ is the quark scalar form factor for each flavor $q$ in the pion, and $A_\pi^q$ and $D_\pi^q$ are the quark gravitational form factors for each flavor $q$ in the pion. Their detailed definitions are given in Eqs.~\eqref{pi_scalar} and \eqref{THETA123}.

At zero momentum transfer, the forward matrix element of the color Lorentz force vanishes 
\begin{equation}
\begin{aligned}
    &\langle \pi(p) |\bar \psi igF^{\mu\nu} \gamma^\sigma \psi| \pi(p)\rangle=0 
\end{aligned}
\end{equation}
This is expected, since the pion does not carry spin.
However, the off-forward matrix element -- the form factor of the color Lorenttz force -- is non-vanishing. Indeed, 
in the Drell-Yan frames with $q^+=0$,  a collection of frames related by light-front boosts, we specialize to the Breit frame with $p_\perp=0$, where the pion momenta have light-front components
\begin{equation}
p^\mu=\left(p^+,-\tfrac{1}{2}q_\perp,\frac{m_\pi^2+\tfrac{1}{4}q_\perp^2}{2p^+}\right),\quad
p'^\mu=\left(p^+,\tfrac{1}{2}q_\perp,\frac{m_\pi^2+\tfrac{1}{4}q_\perp^2}{2p^+}\right)
\end{equation}
The color force form factor in momentum space, is then
\begin{equation}
\begin{aligned}
\label{eq:pi_f}
    &\langle \pi (p')|\bar \psi \gamma^+igF^{+i}\psi|\pi(p)\rangle=2(p^+)^2q^i\mathcal{F}^q_{\pi,1}(Q^2)
\end{aligned}
\end{equation}
where
\begin{equation}
    \mathcal{F}^q_{\pi,1}(Q^2)=\Phi^q_{\pi,1}(Q^2)
\end{equation}
In transverse coordinate, it is given by
\eqref{eq:pi_f}
\begin{equation}
\begin{aligned}
    F_{q/\pi}^i(b)
    =&\frac{2b^i p^+}{2\pi b}\int_0^\infty dQ Q^2 \mathcal{F}^q_{\pi,1}(Q^2)J_1(Q b)
\end{aligned}
\end{equation}

\begin{figure}
    \centering
    \includegraphics[width=0.9\linewidth]{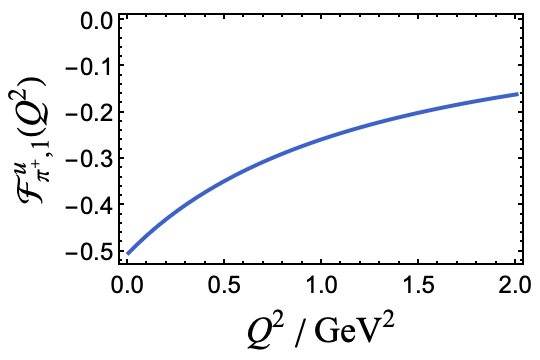}
    \caption{Emergent form factor  $\mathcal{F}^q_{\pi,1}(Q^2)$ induced by a color Lorentz operator in the pion in the ILM enhanced by ``molecular'' $IA$ pairs. }
    \label{fig:pi_force}
\end{figure}

\begin{figure}
    \centering
    \includegraphics[width=\linewidth]{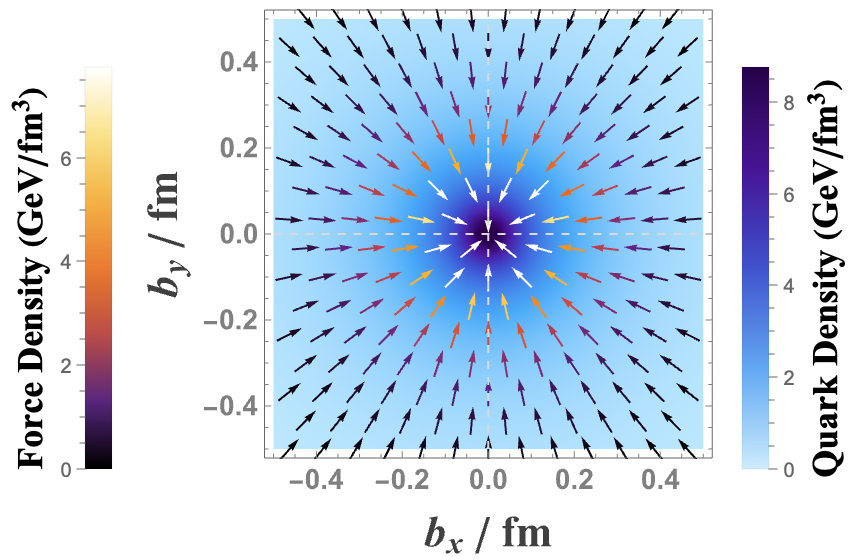}
    \caption{Transverse field distribution of the color Lorentz force
in an unpolarized up quark (arrows), along with the  up quark density distribution (heat map), in impact parameter space for a pion in the ILM enhanced by molecular pairs.}
    \label{fig:pi_den}
\end{figure}

The behavior of the color force form factor in the pion \eqref{eq:pi_f} is shown in Fig.~\ref{fig:pi_force} for a range of $Q^2$. 
We used the ILM  parameters $n_{\rm mol}=7.248$ \rm fm$^{-4}$ and the physical pion mass $m_\pi=140$ MeV. The final result is evolved from $\mu=1/\rho\approx650$ MeV to 2 GeV for possible comparison to future lattice simulations.
Note that the vanishing at $Q=0$ of (\ref{eq:pi_f}) is caused by the extra factor of $Q^i$.
In Fig.~\ref{fig:pi_den} we show the distribution of the color-Lorentz force
acting on an unpolarized up quark in the transverse plane (indicated by the vector field), superimposed on the up quark density distribution in
impact parameter space in a  pion. The quark density we get from the pion electromagnetic form factor, using a  vector meson dominance model with the vector meson mass $m_\rho=791$ MeV \cite{Liu:2023yuj,Liu:2023fpj}.

\begin{table*}
    \centering
\resizebox{0.98\textwidth}{!}{%
\begin{tabular}{|c|c|c|c|c|c|} \hline & ILM & Göckeler et. al \cite{Gockeler:2005vw} & QCDSF \cite{Crawford:2024wzx} & RQCD \cite{Burger:2021knd} & E143 \cite{E143:1998hbs} \\ \hline $d_{2}^u$ & 0.0255 & $0.010(12)$ & 0.079(18) & 0.025(4)(12) & 0.02(4)\\ \hline $d_{2}^d$ & $-0.00496$ & $-0.0056(50)$ & $-0.007(6)$ & $-0.0081(25)(138)$ & 0.02(4) \\ \hline $d_2^p$ & 0.0108& 0.004(5) & 0.046(7)(16) & 0.0105(68) & 0.0122(106) \\ \hline $d_2^n$ & 0.000636& $-0.001(3)$ & 0.023(5)(8) & $-0.0009(70)$ & 0.0106(443) \\ \hline 
\end{tabular}
}
    \caption{ILM calculation with $n_{mol}=7.248$ fm$^{-4}$ and a physical pion mass $m_\pi=140$ MeV, evolved to 2 GeV by \eqref{eq:d2_evol}. Our results are compared to the lattice calculation in~\cite{Gockeler:2005vw} at $\mu=2.24$ GeV, the recent lattice calculation from the QCDSF collaboration in~\cite{Crawford:2024wzx} at a renormalization scale of 2 GeV and a heavy pion mass =450 MeV, and the RQCD lattice collaboration in \cite{Burger:2021knd} at a renormalization scale of 2.24 GeV. Our  results are also compared to the experimental fit from the E143 collaboration \cite{E143:1998hbs} at $\mu=2.24$ GeV.}
    \label{tab:d_2}
\end{table*}

\section{Color force on a nucleon}
\label{SEC-NUCLEON}
The color Lorentz force on a quark in a nucleon can be carried out in the same spirit. Parity invariance, time reversal symmetry and Lorentz covariance, imply that the  matrix element of the Lorentz force in the nucleon is characterized by 8 form factors,

\begin{equation}
\begin{aligned}
    &\langle N(p',s') |\bar \psi igF^{\mu\nu} \gamma^\sigma \psi| N(p,s)\rangle = \bar{u}_{s'}(p') \Bigg\{\left(\bar{p}^{\mu} q^{\nu}-\bar{p}^{\nu} q^{\mu}\right)\frac{\bar{p}^{\sigma}}{m_N}\Phi^q_{N,1}(Q^2)\\
    &+m_N\left(q^{\mu}g^{\sigma\nu}-q^{\nu}g^{\sigma\mu}\right)\Phi^q_{N,2}(Q^2)+m_N i\sigma^{\mu\nu}\bar{p}^\sigma \Phi^q_{N,3}(Q^2) \\
    &+m^2_Ni\epsilon^{\mu\nu\sigma\lambda}\gamma_\lambda\gamma^5\Phi^q_{N,4}(Q^2)+m_Ni \epsilon^{\mu\nu\sigma\lambda}q_\lambda\gamma^5\Phi^q_{N,5}(Q^2)+\frac{i\sigma^{\mu\alpha}\bar{p}^{\nu} -i\sigma^{\nu\alpha}\bar{p}^{\mu}}{2m_N}  q_\alpha q^{\sigma}\Phi^q_{N,6}(Q^2)\\
    &+ \left(\bar{p}^{\mu} q^{\nu}-\bar{p}^{\nu} q^{\mu}\right)\frac{i\sigma^{\sigma\lambda}q_\lambda}{2m_N}\Phi^q_{N,7}(Q^2)+ \frac{i\sigma^{\mu\alpha} q^{\nu}-i\sigma^{\nu\alpha}q^{\mu}}{2m_N}q_\alpha \bar{p}^{\sigma}\Phi^q_{N,8}(Q^2)\Bigg\} u_s(p)
\end{aligned}
\end{equation}
With the help of the short distance approximation from \eqref{eq:CFOp2} and \eqref{eq:CFOp3}, each form factor can be expressed in terms of more standard nucleon form factors defined in Eqs. \eqref{N_scalar}, \eqref{T_traceless}, \eqref{N_pseuoscalar}, and \eqref{N_axial}.  
\begin{align}
    \Phi^q_{N,1}(Q^2)=&
    -\frac{n_{mol}\gamma_{IA}}{2N_c(N_c^2-1)}\beta^{(IA)}_{\bar{q}Gq,2}(\rho q)A^q_N(Q^2)\nonumber\\
    &+\frac{n_{mol}\gamma_{IA}}{2N_c(N_c^2-1)}\beta^{(IA)}_{\bar{q}Gq,7}(\rho q) \frac{m}{m_N} \left[\tilde{A}^q_{T}(Q^2)+\frac12B^q_{T}(Q^2)\right]
\end{align}

\begin{align}
    \Phi^q_{N,2}(Q^2)=&\left(\frac{n_{I+A}}2\right)\frac{1}{2N_c}\left(\frac{4\pi^2\rho^2}{m^*}\right)\beta^{(I)}_{\bar{q}Gq,1}(\rho q)\frac{\sigma^q_N(Q^2)}{m}\nonumber\\
    &-\frac{n_{mol}\gamma_{IA}}{8N_c(N_c^2-1)}\left[\beta^{(IA)}_{\bar{q}Gq,1}(\rho q)+\beta^{(IA)}_{\bar{q}Gq,2}(\rho q)\right]\left(1+\frac{Q^2}{4m_N^2}\right)A^q_N(Q^2)\nonumber\\
    &+\frac{n_{mol}\gamma_{IA}}{8N_c(N_c^2-1)}\left[\beta^{(IA)}_{\bar{q}Gq,1}(\rho q)+\beta^{(IA)}_{\bar{q}Gq,2}(\rho q)\right]\frac{Q^2}{2m_N^2}J^q_N(Q^2)\nonumber\\
    &-\frac{n_{mol}\gamma_{IA}}{8N_c(N_c^2-1)}\left[\beta^{(IA)}_{\bar{q}Gq,1}(\rho q)-\frac{1}3\beta^{(IA)}_{\bar{q}Gq,2}(\rho q)\right]\frac{3Q^2}{4m_N^2}D_N^q(Q^2)\nonumber\\
    &+\frac{n_{mol}\gamma_{IA}}{2N_c(N_c^2-1)}\beta^{(IA)}_{\bar{q}Gq,7}(\rho q) \frac{m}{m_N} \left[\frac12A^q_{T}(Q^2)+\frac12B^q_{T}(Q^2)\right]\\
    &+\frac{n_{mol}\gamma_{IA}}{2N_c(N_c^2-1)}\beta^{(IA)}_{\bar{q}Gq,7}(\rho q) \frac{m}{m_N} \left[\left(1+\frac{Q^2}{4m_N^2}\right)\tilde{A}^q_{T}(Q^2)+2C^q_T(Q^2)\right]\\[7pt]
    \Phi^q_{N,3}(Q^2)=&\frac{n_{mol}\gamma_{IA}}{2N_c(N_c^2-1)}\frac{m}{m_N}\left[\beta^{(IA)}_{\bar{q}Gq,7}(\rho q)-\frac12\rho^2Q^2\beta^{(IA)}_{\bar{q}Gq,6}(\rho q)\right]A^q_{T}(Q^2)\\
    \Phi^q_{N,4}(Q^2)=&-\left(\frac{n_{I+A}}2\right)\frac{1}{2N_c}\left(\frac{4\pi^2\rho^2}{m^*}\right)\frac{\rho^2Q^2}{2}\beta^{(I)}_{\bar{q}Gq,2}(\rho q)\frac{m}{m_N^2}G^q_A(Q^2)\nonumber\\
    &-\frac{n_{mol}\gamma_{IA}}{2N_c(N_c^2-1)}\frac{1}{\rho^2m^2_N}\left[\beta^{(IA)}_{\bar{q}Gq,3}(\rho q)-\frac14\rho^2Q^2\beta^{(IA)}_{\bar{q}Gq,2}(\rho q)\right]G^q_{A}(Q^2)
\end{align}

\begin{align}
    \Phi^q_{N,5}(Q^2)=&\left(\frac{n_{I+A}}2\right)\frac{1}{2N_c}\left(\frac{4\pi^2\rho^2}{m^*}\right)\left[\beta^{(I)}_{\bar{q}Gq,1}\frac{\tilde{G}^q_P(Q^2)}{m}-\frac14\rho^2Q^2\beta^{(I)}_{\bar{q}Gq,2}\frac{m}{m_N^2}G^q_P(Q^2)\right]\nonumber\\
    &+\frac{n_{mol}\gamma_{IA}}{N_c(N_c^2-1)}\left[\beta^{(IA)}_{\bar{q}Gq,4}-\frac14\beta^{(IA)}_{\bar{q}Gq,2}\right]G^q_{A}(Q^2)\nonumber\\
    &+\frac{n_{mol}\gamma_{IA}}{2N_c(N_c^2-1)}\frac{1}{2\rho^2m^2_N}\left[\frac12\rho^2Q^2\beta^{(IA)}_{\bar{q}Gq,2}-\beta^{(IA)}_{\bar{q}Gq,3}-\rho^2Q^2\beta^{(IA)}_{\bar{q}Gq,4}\right]G^q_{P}(Q^2)\nonumber\\
    &-\frac{n_{mol}\gamma_{IA}}{2N_c(N_c^2-1)}\beta^{(IA)}_{\bar{q}Gq,5}\tilde{G}^q_P(Q^2)\\
    &+\frac{n_{mol}\gamma_{IA}}{2N_c(N_c^2-1)}\beta^{(IA)}_{\bar{q}Gq,6}\frac{m}{m_N}\left[\rho^2m_N^2D_T^q(Q^2)-\frac12\rho^2Q^2\tilde{D}_T^q(Q^2)\right]\\
    \Phi^q_{N,6}(Q^2)=
    &-\frac{n_{mol}\gamma_{IA}}{N_c(N_c^2-1)}\left[\frac34\beta^{(IA)}_{\bar{q}Gq,2}(\rho q)-\beta^{(IA)}_{\bar{q}Gq,4}(\rho q)\right]G^q_A(Q^2)\nonumber\\
    &+\frac{n_{mol}\gamma_{IA}}{2N_c(N_c^2-1)}\left[\beta^{(IA)}_{\bar{q}Gq,7}(\rho q)+\frac12\rho^2Q^2\beta^{(IA)}_{\bar{q}Gq,6}(\rho q)\right]\frac{m}{m_N} \tilde{B}^q_{T}(Q^2)\\
    &-\frac{n_{mol}\gamma_{IA}}{2N_c(N_c^2-1)}\beta^{(IA)}_{\bar{q}Gq,6}(\rho q)\rho^2m m_N D^q_{T}(Q^2)
\end{align}

\begin{align}
    \Phi^q_{N,7}(Q^2)=&-\left(\frac{n_{I+A}}2\right)\frac{1}{N_c}\left(\frac{4\pi^2\rho^2}{m^*}\right)\beta^{(I)}_{\bar{q}Gq,2}(\rho q)\rho^2mG^q_A(Q^2)\nonumber\\
    &-\frac{n_{mol}\gamma_{IA}}{2N_c(N_c^2-1)}\beta^{(IA)}_{\bar{q}Gq,2}(\rho q)J^q_{N}(Q^2)\\
    \Phi^q_{N,8}(Q^2)=&\left(\frac{n_{I+A}}2\right)\frac{1}{N_c}\left(\frac{4\pi^2\rho^2}{m^*}\right)\beta^{(I)}_{\bar{q}Gq,2}(\rho q)\rho^2mG^q_A(Q^2)\nonumber\\
    &-\frac{n_{mol}\gamma_{IA}}{2N_c(N_c^2-1)}\beta^{(IA)}_{\bar{q}Gq,2}(\rho q)J^q_{N}(Q^2)\nonumber\\
    &+\frac{n_{mol}\gamma_{IA}}{2N_c(N_c^2-1)}\beta^{(IA)}_{\bar{q}Gq,7}(\rho q) \frac{m}{2m_N} B^q_{T}(Q^2)
\end{align}

The nucleon $\Phi^q_{N,1}$ and $\Phi^q_{N,2}$ form factors are related to the quark scalar form factor $\sigma_N^q$  and the EMT form factors $A_N^q$, $J_N^q$, and $D_N^q$, both in the isoscalar and isovector channels, where the angular momentum form factor is defined by $J_N^q=\frac12\left(A^q_N+B^q_N\right)$. Those form factors are also related to the generalized form factors derived from the unpolarized nucleon GPD $H^{q/N}$ and $E^{q/N}$ with skewness $2\xi=-q^+/\bar p^+ $ \cite{Diehl:2005jf} (see also Secs.~\ref{sec:GFF}, \ref{sec:glu_FF}, and \ref{sec:GPD}).

The nucleon $\Phi^q_{N,4-8}$ form factors are related the quark axial form factor $G_A^q$, and the pseudoscalar form factors $\tilde{G}_P^q$ and $G_P^q$ both in isoscalar and isovector channels. However, these form factors $G^q_{A,P}$ and $\tilde{G}^q_{P}$ do not contribute to the twist-3 combination. The remaining color force form factors $\Phi^q_{N,1-3}$, $\Phi^q_{N,5}$, $\Phi^q_{N,6}$, and $\Phi^q_{N,8}$ are further related to the 7 tensor quark gravitational form factors $A^q_{T}$, $\tilde{A}^q_{T}$, $B^q_{T}$, $\tilde{B}^q_{T}$, $C^q_{T}$, $D^q_{T}$, and $\tilde{D}^q_{T}$ \cite{Bhoonah:2017olu} parameterized by (see also Sec.~\ref{sec:GPD})


\begin{equation}
\begin{aligned}
    &\langle N'|\bar\psi\sigma_{\mu\nu}\overleftrightarrow{\partial}_{\rho}\psi|N\rangle=\bar{u}_{s'}(p') \Bigg\{-i\sigma^{\mu\nu}\bar{p}^\rho  A_{T}^q-\frac{\bar{p}^\mu q^\nu-\bar{p}^\nu q^\mu}{m_N^2}\bar{p}^{\rho}\tilde{A}_{T}^q-\frac{\gamma^\mu q^\nu-\gamma^\nu q^\mu}{2m_N}\bar{p}^\rho B_{T}^q\\
    &-\frac{\gamma^\mu \bar{p}^\nu-\gamma^\nu \bar{p}^\mu}{m_N}q^{\rho}\tilde{B}_{T}^q-(g^{\mu\rho}q^{\nu}-g^{\nu\rho}q^{\mu})C_T^q-im_N\epsilon^{\mu\nu\rho\lambda}\gamma_\lambda\gamma^5D_T^q-i\epsilon^{\mu\nu\rho\lambda}q_\lambda\gamma^5\tilde{D}_T^q\Bigg\} u_s(p)
\end{aligned}
\end{equation}
However, only the first 4 contribute to the nucleon twist-3 transversity gravitational form factors with their relations to the second moment of the nucleon transversity GPD defined in Sec.~\ref{sec:GPD}. (see also  \cite{Kim:2025mol,Bhoonah:2017olu,Alexandrou:2024awx})

Using equation of motion \cite{Bhoonah:2017olu,Freese:2019bhb},

\begin{equation}
\bar{\psi}\, i\sigma^{\lambda\mu}\gamma^5\, i\overleftrightarrow{\partial}_\mu \psi
= 2M\,\bar{\psi}\gamma^\lambda\gamma^5\psi
+ i\partial^\lambda(\bar{\psi}\gamma^5\psi)
\end{equation}
and
\begin{equation}
\epsilon_{\lambda\mu\nu\alpha}\bar{\psi}\, i\sigma^{\lambda\mu}\gamma^5\, i\overleftrightarrow{\partial}^{\nu} \psi
=2\,
\partial_\alpha(\bar{\psi}\psi).
\end{equation}
those tensor quark gravitational form factors can be further constrained by the relations

\begin{equation}
\begin{aligned}
    \frac{2M}{m_N}G_A^q=&A_T^q-\frac{Q^2}{m_N^2}\tilde{A}_T^q-\frac{Q^2}{4m_N^2}B_T^q+\frac{Q^2}{2m_N^2}\tilde{B}_T^q-2C_T^q-3D_T^q
\end{aligned}
\end{equation}

\begin{equation}
\begin{aligned}
    \frac{2M}{m_N}G_P^q-\frac{2m_N}m\tilde{G}^q_P=&-A_T^q+\frac{Q^2}{m_N^2}\tilde{A}_T^q-B_T^q+2\tilde{B}_T^q+2C_T^q-6\tilde{D}_T^q
\end{aligned}
\end{equation}

\begin{equation}
\begin{aligned}
    -\frac{2m_N}{m}\sigma_N^q=&A_T^q+2\tilde{A}_T^q-\frac{Q^2}{m_N^2}\tilde{A}_T^q+B_T^q+4C_T^q
\end{aligned}
\end{equation}

As we are only interested in the twist-3 contribution in the Breit frame $q^+=0$, we will only focus on the $\Phi^q_{N,1}$, $\Phi^q_{N,3}$, $\Phi^q_{N,7}$, and $\Phi^q_{N,8}$ form factors in this work.

\subsection{$d_2$ moment of the nucleon}
At zero momentum transfer, the single instanton contribution is suppressed. The zero momentum transfer receives contributions from molecular configurations. $\Phi^q_{3,N}(0)$ is equal to $d^q_{2,N}$, 
\begin{equation}
\begin{aligned}
    d^q_{2,N}=&\frac{n_{mol}\gamma_{IA}}{2N_c(N_c^2-1)}\beta^{(IA)}_{\bar{q}Gq,7}(0)\frac{m}{m_N}A^{q}_{T}(0)
\end{aligned}
\end{equation}

If we assume the evolution does not mix flavors at one-loop, the one-loop evolution equation is \cite{Burger:2021knd} 
\begin{equation}
\label{eq:d2_evol}
    d_{2}(\mu) = 
\left( \frac{\alpha_{s}(\mu')}{\alpha_{s}(\mu)} \right)^{-\frac{\gamma_{d_2}}{\beta_0}}
\, d_{2}(\mu')
\end{equation}
where $\beta_0=\frac{11}{3}N_c - \frac{2}{3}N_f$ and the one-loop anomalous dimension is defined by
\begin{equation}
  \gamma_{d_2}= 3N_c - \frac{1}{6}\left( N_c - \frac{1}{N_c}\right) 
\end{equation}
The total proton and neutron $d_2$ can be constructed from $d^q_{2,N}$ using~\cite{Burger:2021knd}
\begin{align}
\label{eq:d2N}
d_{2}^{p} &= \left( \tfrac{2}{3} \right)^{2} d_{2,N}^{u} 
             + \left( -\tfrac{1}{3} \right)^{2} d_{2,N}^{d} \\[6pt]
d_{2}^{n} &= \left( -\tfrac{1}{3} \right)^{2} d_{2,N}^{u} 
             + \left( \tfrac{2}{3} \right)^{2} d_{2,N}^{d}
\end{align}

To compare our results with the lattice and global analyses, we evolve our ILM result from $\mu=1/\rho\approx650$ MeV to $\mu=2$ GeV. The one-loop perturbative correction gives around $$d_2(\mu=2~\mathrm{GeV})\approx0.48\,d_2(\mu=1/\rho)$$ 
The final result is shown in Table~\ref{tab:d_2} with our result using $n_{\rm mol}=7.248$ fm$^{-4}$ and the physical pion mass $m_\pi=140$ MeV, corresponding to current quark mass $m(\mu=1/\rho)=10$ MeV, which has been suggested in many low energy vacuum-based models \cite{Liu:2025ldh,Liu:2023yuj,Schafer:1996wv,Hatsuda:1994pi,Liu:2025kuc,Diakonov:2002fq}. 
Note that the result of QCDSF collaboration in Table~\ref{tab:d_2}, there is a mismatch between the $d_2^p$ and $d_2^n$ obtained from $d_2^u$ and $d_2^d$ by \eqref{eq:d2N} presented in their plot, and the direct value of $d_2^p$ and $d_2^n$  they present in \cite{Crawford:2024wzx}.

We note that the one-loop evolution from the low renormalization point $\mu \sim 1/\rho \approx 650$ MeV to higher scales introduces sizable uncertainties and scheme dependence, particularly below 1 GeV where perturbation theory is marginal. Therefore, the evolution presented here should be regarded as qualitative and illustrative. A quantitatively reliable matching to higher scales would require nonperturbative renormalization or lattice input as outlined in~\cite{Shuryak:2026pqt}, which is still under development and lies beyond the scope of the present work.

\subsection{Twist-3 color Lorentz force FFs of the nucleon}
To evaluate the color force form factors (FFs) in the nucleon, we specialize 
again to the Breit frames with $q^+=0$ and $p_\perp=0$ for the nucleon. 
The form factors defined on the light front are

\begin{equation}
\begin{aligned}
    &\langle N(p's)|\bar \psi \gamma^+igF^{+i}\psi|N(ps)\rangle=\\
    &\bar u_s(p')\Bigg[\gamma^+q^i\mathcal{F}^q_{N,1}(Q^2)+m_Ni\sigma^{+i}\mathcal{F}^q_{N,2}(Q^2)-\frac{i\sigma^{+j}q_jq^i}{m_N}\mathcal{F}^q_{N,3}(Q^2)\Bigg]p^+u_s(p)
\end{aligned}
\end{equation}
with 
\begin{equation}
\mathcal{F}^q_{N,1}(Q^2)=\Phi^q_{N,1}(Q^2),\quad
\mathcal{F}^q_{N,2}(Q^2)=\Phi^q_{N,3}(Q^2)
\end{equation} 
and
\begin{equation}
\mathcal{F}^q_{N,3}(Q^2)
=
-\frac{1}{2}\left[\Phi^q_{N,7}(Q^2)+\Phi^q_{N,8}(Q^2)-\Phi^q_{N,1}(Q^2)\right]
\end{equation}
Note that all  $\mathcal{F}^q_{N,1,2,3}(Q^2)$ are proportional to $n_{\rm mol}$. The single instanton contribution in $\Phi_{N,7}$ and $\Phi_{N,8}$ cancel each other. Thus, even though in general the off-foward matrix element of $\bar\psi\gamma_\sigma F_{\mu\nu}\psi$ receives contribution from single instantons, the leading twist-3 contribution only receives contribution from $IA$ molecules.

In order to facilitate comparison with lattice results obtained at heavier pion masses ($m_\pi = 450$ MeV), we adjust the quark mass $m$ in the ILM using the chiral relation $m_\pi \propto \sqrt{m}$. Under the assumption that the other ILM parameters ($n_{mol}$, $\gamma_{IA}$) and the nucleon mass $m_N$ are only weakly dependent on the quark mass $m$, this procedure provides an effective matching between the model parameters and the lattice setup. For completeness, we also present results over a range of the quark mass $m=10$ MeV ($m_\pi=140$ MeV) and $m=58$ MeV ($m_\pi=338$ MeV), which allows for a comparison of the quark-mass sensitivity. The plots suggest strong quark mass dependence in the second color force fomr factor $\mathcal{F}^q_{N,2}(Q^2)$, which is opposite to the lattice observation in \cite{Burger:2021knd}.

\begin{figure}
    \centering
\subfloat[\label{fig:nuc_force1}]{\includegraphics[width=.5\linewidth]{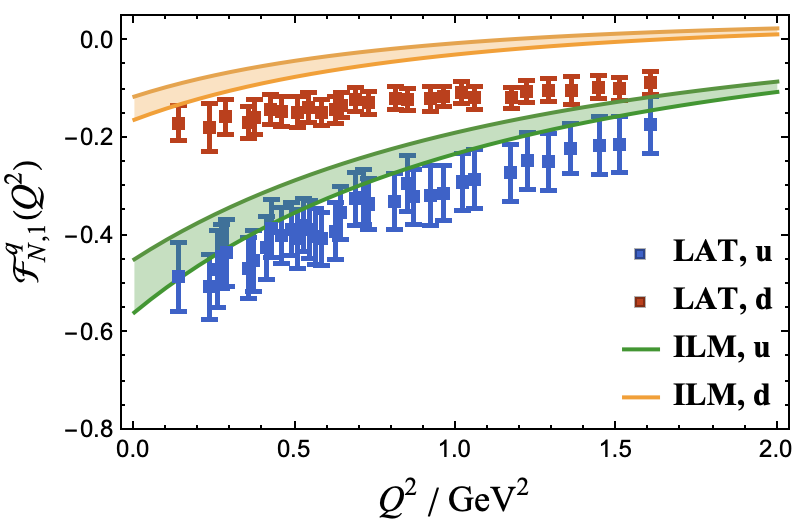}}
\hfill
\subfloat[\label{fig:nuc_force2}]{\includegraphics[width=.5\linewidth]{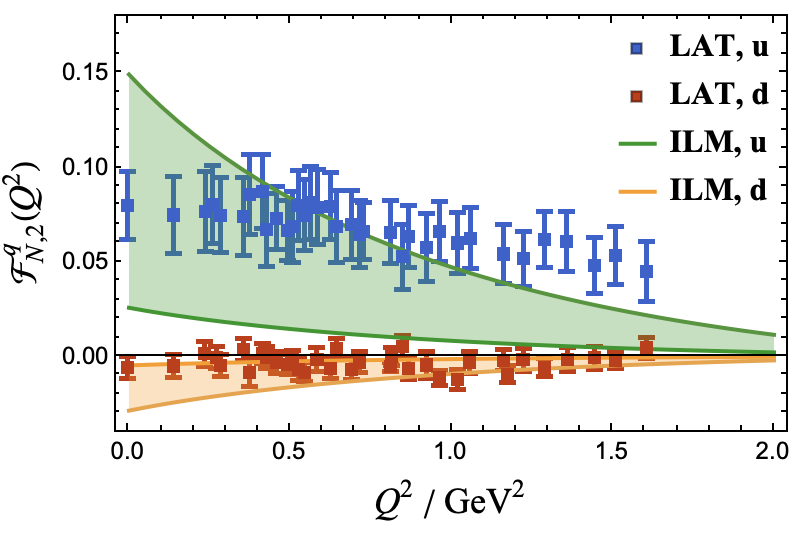}}
\hfill
\subfloat[\label{fig:nuc_force3}]{\includegraphics[width=.6\linewidth]{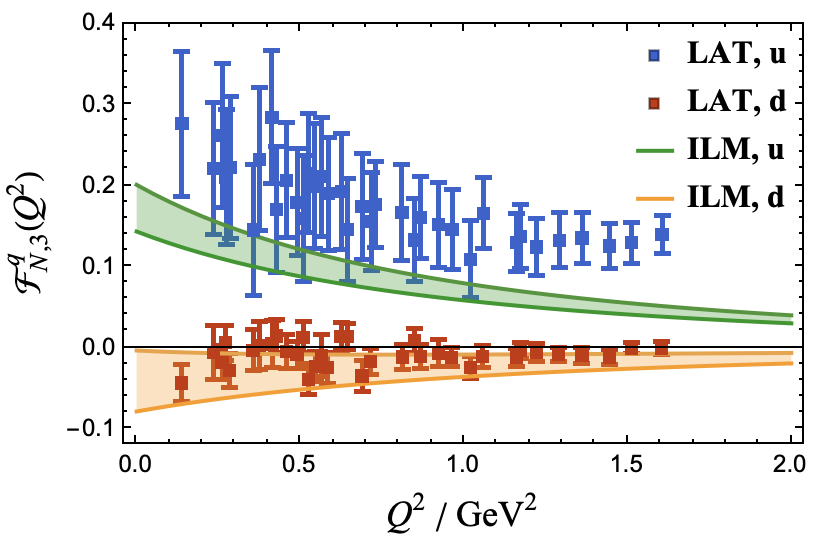}}
\caption{Color force form factors (a) $\mathcal{F}^q_{N,1}(Q^2)$, (b) $\mathcal{F}^q_{N,2}(Q^2)$, and (c) $\mathcal{F}^q_{N,3}(Q^2)$ from ILM with parameters $n_{mol}=7.248$ fm$^{-4}$ and current quark mass $m=10-58$ MeV ($m_\pi=140-338$ MeV) showing with bands, evolved to 2 GeV compared to the lattice calculation with pion mass $450$ MeV in \cite{Crawford:2024wzx}.}
\label{fig:nuc_force}
\end{figure}

The color force form factor $\mathcal{F}^q_{N,1}(Q^2)$ is related to the {\it unpolarized nucleon gravitational form factor} $A^q_{N}$ and {\it transversity gravitational form factor} $2\tilde{A}^q_{T}+B^q_{T}$. Its behavior is shown in Fig.~\ref{fig:nuc_force1}, 
using the ILM with molecule density $n_{mol}=7.248$ fm$^{-4}$. The results are evolved  to 2 GeV, and compared to the lattice calculation with pion mass $450$ MeV in~\cite{Crawford:2024wzx}. Both the $u$- and $d$-flavor components of the Lorentz force are in good agreement with the reported lattice results, over a relatively broad range of $Q^2$.

The color force form factor $\mathcal{F}^q_{N,2}(Q^2)$ is fully related to the {\it transversity gravitational form factor}  $A^q_{T}$. Its behavior is shown in Fig.~\ref{fig:nuc_force2} using the ILM parameters. The comparison to the recently reported lattice results in~\cite{Crawford:2024wzx} is fair. Note that  the smallness of this form factor originates from its  suppression by the quark mass $m$ and in the ILM enhanced by molecules.  

The color force form factor $\mathcal{F}^q_{N,3}(Q^2)$ is a combination of  {\it gravitational form factor} $B^q_N$ and {\it transversity gravitational form factor} $\tilde{A}^q_{T}$. This form factor is displayed in Fig.~\ref{fig:nuc_force3} using the ILM enhanced by molecules, and compared 
to the recent lattice results~\cite{Crawford:2024wzx}. Again the comparison is satisfactory.  Since $B^u_N+B^d_N\approx0$, the asymmetry between $u$ and $d$ arises from the non-vanishing isoscalar transversity GPD moment $\tilde{A}^{u+d}_{T}(0)$.

\begin{figure}
    \centering
\subfloat[]{\includegraphics[width=.8\linewidth]{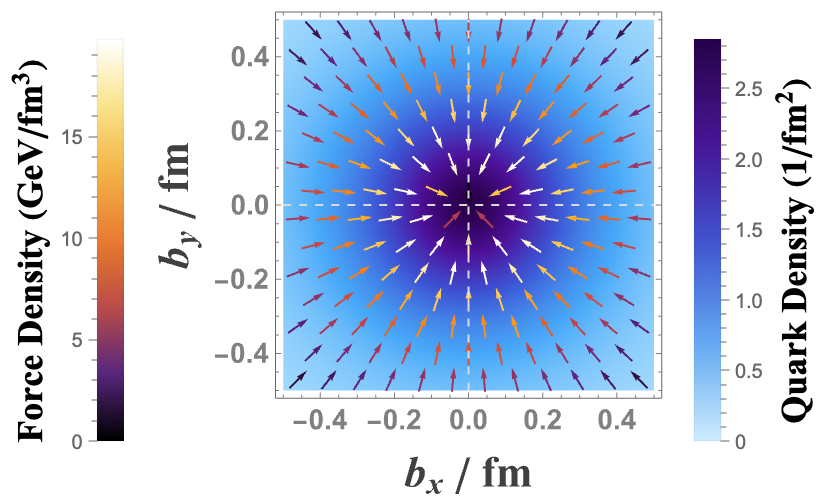}}
\hfill
\subfloat[]{\includegraphics[width=.8\linewidth]{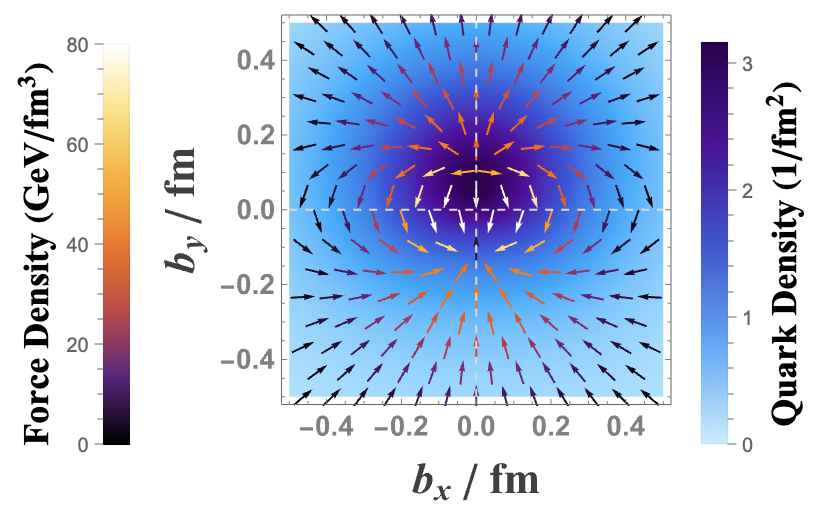}}
    \caption{(a) Transverse field distribution of the color Lorentz force in an unpolarized up quark (arrows), along with the up quark density distribution (heat map), for an unpolarized proton in the enhanced ILM. The up quark density follows from the proton Dirac and Pauli form factors from the phenomenological fit in \cite{Alberico:2008sz}. The results are compared to the lattice results in~\cite{Crawford:2024wzx}. (b) Same as (a), but for a proton transversely polarized in the $\hat{x}$ direction.}
    \label{fig:nuc_den}
\end{figure}

In Fig.~\ref{fig:nuc_den}, we plot the color force density on a transverse plane at physical pion mass ($m_\pi=140$ MeV). The result is also compared to the lattice calculation in \cite{Crawford:2024wzx} at a heavy pion mass ($m_\pi=450$ MeV).

We have developed a comprehensive and unified treatment of the twist-3 color Lorentz force in pions and nucleons, within the framework of the instanton liquid model and its {\it molecular} $IA$ extension. 

The color Lorentz force arises naturally from the non-perturbative QCD vacuum fluctuations of the gauge field, including  correlated instanton–anti-instanton  pairs.
By combining semiclassical strength and spatial structure of these fields  with the operator structure of higher-twist quark–gluon correlators, we have shown  that   molecular correlations provide an important mechanism behind the large higher-twist effects observed in polarized DIS, and confirmed in recent lattice simulations.
This agrees with our previous derivation of the confining potential
in quarkonia, from the same setting.

One new result of this study is the explicit construction and analysis of the gauge field structure of the molecular ensemble,
and its role in the quark-colored force. By employing the ratio ansatz \eqref{eq:Ratio} for the correlated instanton–anti-instanton pair, we derived explicit expressions for the color-electric and color-magnetic field components and computed their spatial profiles. The overlap of the quark zero modes across the molecular pair was analyzed in detail, revealing a strong delocalization effect that enhances the color force.  Using Monte Carlo sampling over molecular orientations and separations, we obtained a robust estimate of the average color Lorentz force, $F\approx2$–$3\,\mathrm{GeV/fm}$, acting on a single quark. This result indicates that the nonperturbative color forces are comparable to, or can even exceed the magnitude of the confining string tension.

Our  analysis shows a  direct connection between these microscopic nonperturbative fields and the experimentally accessible twist-3 observables. The color Lorentz force operator $\bar{\psi}gF^{+i}\gamma^+\psi$--whose expectation value defines the twist-3 matrix element $d_2$--is shown to acquire contributions from the molecular component of the instanton liquid. The induced form factors  encode the nonlocal structure of this coupling and provide a unified description of the quark–gluon interaction over a broad range of momentum transfers. In particular, the molecular contributions dominate at low $Q^2$, leading to enhanced color forces consistent with phenomenological extractions from polarized structure functions.  These findings confirm that the twist-3 sector is a direct manifestation of the underlying topological fields in the QCD vacuum.
Remarkably, the emergent color Lorentz force form factors are shown to be intimately related to the hadronic gravitational and transversity form factors, offering additional insights to the nature of the mass and force distributions in hadrons. However, one has to keep in mind that the relations shown in \eqref{eq:CFOp2} and \eqref{eq:CFOp3} are built on the local approximation of the original relation in \eqref{eq:force_op_0}, which involves a more complicated nonlocal two-current structure.

From a phenomenological perspective, the implications of our results are as follows. The pion, as a spinless bound state, exhibits a vanishing $d^\pi_2$, consistent with the absence of a net transverse Lorentz force. This clear dichotomy between spin-1/2 and spin-0 systems demonstrates that the twist-3 color force is intrinsically spin-dependent and directly tied to the internal color-magnetic structure of the hadron. Moreover, the ability of the instanton–molecule framework to simultaneously describe both systems underscores its universality as a source of nonperturbative QCD dynamics.

Conceptually, this study shows how topological excitations of the QCD vacuum—once considered peripheral to the hadron structure—directly govern measurable high-energy observables. The emergence of strong, localized color forces from correlated instanton configurations provides a compelling microscopic picture that complements and extends conventional models based on confinement and gluon exchange. The instanton–molecule framework offers a natural bridge between the semiclassical field picture and the partonic description of hadrons, thereby unifying aspects of nonperturbative vacuum structure and experimental phenomenology.

The results also suggest new directions for future research. Extending the present formalism to generalized parton distributions (GPDs) and transverse-momentum-dependent distributions (TMDs) would enable mapping of the spatial and dynamical structure of color forces in both position and momentum space. Lattice simulations that isolate instanton and molecular contributions could provide stringent tests of the predicted force magnitudes and their dependence on quark flavor and spin.

The present analysis is subject to several limitations inherent to the ILM framework. First, confinement is not explicitly incorporated, and the results depend on phenomenological input for the instanton ensemble, including the typical instanton size and density. Second, the semiclassical treatment itself is not systematically controlled by a small parameter although the many-body expansion in our current framework is, and therefore the quantitative predictions should be interpreted at the level of order-of-magnitude estimates. Third, the use of a local approximation for the quark-gluon operator neglects nonlocal contributions that may become relevant at larger distances or lower momenta. Finally, the perturbative evolution from low scales introduces additional uncertainties. 
These notwithstanding, the framework captures essential nonperturbative features of the QCD vacuum and provides a coherent and physically transparent mechanism for the emergence of sizable color Lorentz forces, in qualitative agreement with recent lattice observations.

In summary, the instanton–anti-instanton molecular component of the QCD vacuum is an important  source of the nonperturbative color Lorentz forces responsible for twist-3 phenomena in light hadrons. By  relating the semiclassical field theory to measurable quantities, we have provided both qualitative understanding and quantitative predictions that connect the topology of the QCD vacuum to hadronic observables. The framework presented here not only enhances our comprehension of the QCD vacuum but also lays the foundation for a broader, unified picture of nonperturbative dynamics in strong interaction physics, ranging from hadronic spectroscopy  to 
partonic physics on the light front.

\chapter{Strong CP and Nucleon EDM}
\label{ch:CP}
In the QCD vacuum, topologically active pseudoparticles, namely instantons and anti-instantons, are CP-conjugated pairs, which naturally serve as sources of local CP violation. At a finite $\theta$-angle, CP symmetry is significantly violated in QCD.

This parameter is actually a physical parameter of the gauge theory and cannot be removed by gauge transformations. 
It can be introduced explicitly by adding the term
\begin{equation}
\mathcal{L}_\theta 
= \frac{\theta g^2}{32\pi^2} F^a_{\mu\nu} \tilde{F}^{a\mu\nu}
\end{equation}
to the QCD Lagrangian. This term violates CP symmetry. Therefore, a nonzero value of $\theta$ implies CP violation in strong interactions. 
Experimentally, $\theta$ is constrained to be extremely small,
\begin{equation}
\theta \lesssim 10^{-11},
\end{equation}
leading to the so-called \emph{strong CP problem}.

At finite vacuum angle $\theta$, the QCD ground vacuum is CP violating.
In the ILM, this is seen by noting that the added $\theta$ term acts as a chemical potential for the topological charge $\Delta$, therefore enhancing instantons and depleting anti-instantons. This affects most observables, and in particular the  electromagnetic (EM) form factors of hadrons.

\begin{figure}
    \centering
    \includegraphics[scale=0.8]{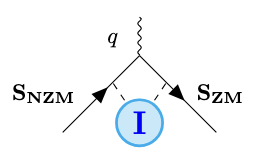}
    \caption{Leading pseudoparticle (single instanton) contribution to the quark EM current in the ILM. The dashed lines refer to the exchange induced by a classical instanton and a propagating quark. The details can be found in Sec.~\ref{sec:Inst_QCD} and Appendix \ref{App:NZM}}
    \label{fig:VERTEX}
\end{figure}

In general, the nucleon-photon coupling can be described by Dirac $F_1$, Pauli $F_2$, the electric dipole moment form factor $F_3$, and the axial tensor form factor $F_A$

\begin{equation}
\label{EM_form}
\begin{aligned}
\langle N'|J^{\mu}_{\rm EM}|N\rangle
&= \bar{u}_{s'}(p')\bigg[
\gamma^\mu F_1(Q^2)
+ \frac{i\sigma^{\mu\nu}q_\nu}{2M_N}\left(F_2(Q^2)-i\gamma^5 F_3(Q^2)\right) \\
&\qquad
+ \frac{1}{M_N^2}\left(\slashed{q}q^\mu - q^2\gamma^\mu\right)\gamma^5 F_A(Q^2)
\bigg]u_s(p)
\end{aligned}
\end{equation}
where $J^{\mu}_{\rm EM}=\sum_f Q_f\,\bar{\psi}_f\gamma^\mu\psi_f$
with 
$F_3$ and $F_A$ vanishing for $\theta=0$. 
The ensuing magnetic and electric dipole moments are,
\begin{equation}
\mu_N = \frac{e}{2m_N}\bigl(F_1(0)+F_2(0)\bigr), \qquad
d_N = \frac{e}{2m_N}F_3(0)
\end{equation}
In Fig.~\ref{fig:VERTEX} we show how a single pseudoparticle (instanton or anti-instanton)  affects the electromagnetic vertex in the ILM. A light quark propagating in the instanton liquid background undergoes scattering into a zero mode (ZM) or a non-zero-mode (NZM) states, the eigenmodes of a Dirac operator. The ZMs are left handed in an instanton and right-handed in an anti-instanton~\cite{Liu:2024rdm,Liu:2025ldh,Zahed:2022wae}. This chirality selection rule implies that the EM vector-like vertex can only couple light quarks undergoing a ZM-NZM transition or vice-versa. This interaction induces a non-perturbative helicity-flip magnetic type coupling~\cite{Kochelev:2003cp}.

The explicit forms of the ZM and NZM  propagators are well-known~\cite{Schafer:1996wv} (see also Eqs.~\eqref{eq:SIA_prop_q} and \eqref{eq:ZM_prop}). With this in mind and using the  
on-shell reduction scheme developed in~\cite{Liu:2021evw},
the instanton plus anti-instanton contributions presented in Fig.~\ref{fig:VERTEX}
modify the vector current operator for a single flavor

\begin{equation}
\begin{aligned}
\label{VPM}
  &\frac1V\int d^4x e^{-iq\cdot x}\bar\psi_f(x)\gamma^\mu\psi_f(x)\\ &\rightarrow 
  \int\frac{d^4k}{(2\pi)^4}\frac{d^4k'}{(2\pi)^4}
  \left\langle\bar{\psi}_f(k')\left[\frac{N_+}{V}V^\mu_+(k',k)+\frac{N_-}{V}V^\mu_-(k',k)\right]\psi_f(k)\right\rangle
\end{aligned}
\end{equation}
with the vertex for the Pauli contribution
\begin{equation}
\begin{aligned}
\label{eq:V_inst}
    V^\mu_\pm(k',k)
    &\simeq~ (2\pi)^4\delta(k-k'+q)\frac{8\pi^2\rho^4}{N_c}
\frac{1\mp\gamma^5}{2}\frac{i\sigma^{\mu\nu}q_\nu}{2m^*}\nonumber\\
&\times\int_0^1dt\left[tK_0(u\sqrt{1-t})-\frac{1}{8}\frac{\sqrt{1-t}}{u}\frac{\partial}{\partial u}\left(uK_1(u\sqrt{1-t})\right)\right]\bigg|_{u=\rho Q}
\end{aligned}
\end{equation}
where $K_n(x)$ is a  modified Bessel function of the second kind. 

At zero vacuum angle, 
the Pauli form factor $F_2$ 
in~\eqref{EM_form} receives a large contribution from pseudoparticles in the QCD instanton vacuum~\cite{Kochelev:2003cp}. This observation carries to the  CP-odd contribution $F_3$  at finite vacuum angle. For that, we need to average the EM form factor over the numbers of pseudoparticles  $N_\pm$ in the ILM. This procedure is by now standard~\cite{Diakonov:1995qy,Schafer:1996wv,Zahed:2021fxk},
with a more detailed review given in~\cite{Liu:2024rdm,Liu:2025ldh}. Some key steps can also be found in Sec.~\ref{sec:had_grand} and Appendix \ref{App:NZM} 


\textcolor{black}{For simplicity, we model the nucleon by a simple description of a quark–scalar diquark compound, e.g., the proton with a spin up is $(uud)_\uparrow\approx u_\uparrow[ud]_0$. This schematic description captures certain part of the correlations in the nucleon, but not the full set~\cite{Schafer:1996wv}. Hence, the quark contribution to the proton spin structure is mostly carried through the unpaired $u$-quark.}

With this in mind, the Pauli form factor of a quark of flavor $f$
in a nucleon ($u$ for proton and $d$ for neutron), can be identified from \eqref{VPM} and \eqref{eq:V_inst} by comparison and after inserting (\ref{SUS}), the result gives

\begin{equation}
\label{EDM5}
\langle q_f(k')|J^{\mu}_{\rm EM}|q_f(k)\rangle_{\rm Pauli}
\simeq
Q_f e\,\bar{u}_s(k')\,F_I(\rho Q)\left(1 -
\frac{\langle\Delta^2\rangle}{\bar N}\,\theta\,
i\gamma^5\right)\frac{i\sigma_{\mu\nu}q_\nu}{2m^*}\,u_s(k)
\end{equation}

The pseudoparticle induced form factor $F_I(\rho Q)$ is related to \eqref{eq:V_inst} and given by

\begin{equation}
\begin{aligned}
\label{EDM6}
    F_I(q)= \frac{8\pi^2\rho^4n_{I+A}}{N_c} \frac{1}{q^2}\left[\bigg(2-2qK_1(q)\bigg)-\left(\frac7{q^2}-\frac{7}{4}qK_1(q)-\frac72K_2(q)\right)\right]
\end{aligned}
\end{equation}


\begin{figure}
    \centering
\subfloat[]{\includegraphics[width=.5\linewidth]{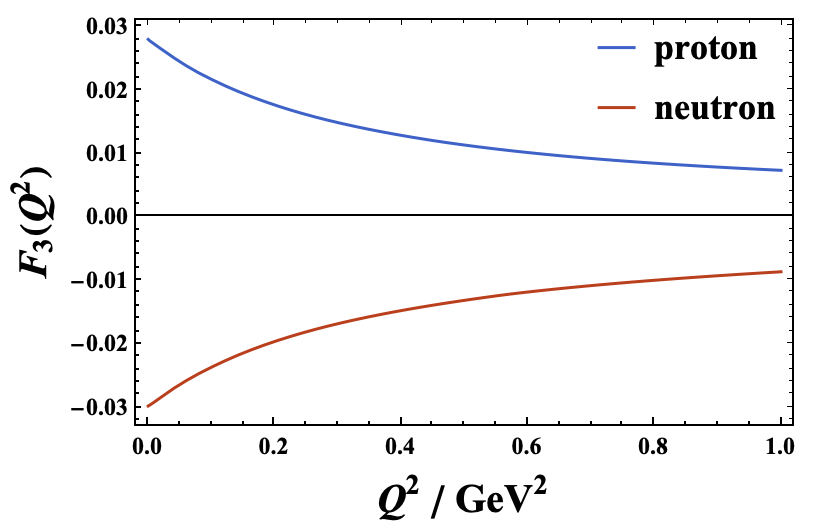}}
\hfill
\subfloat[]{\includegraphics[width=.5\linewidth]{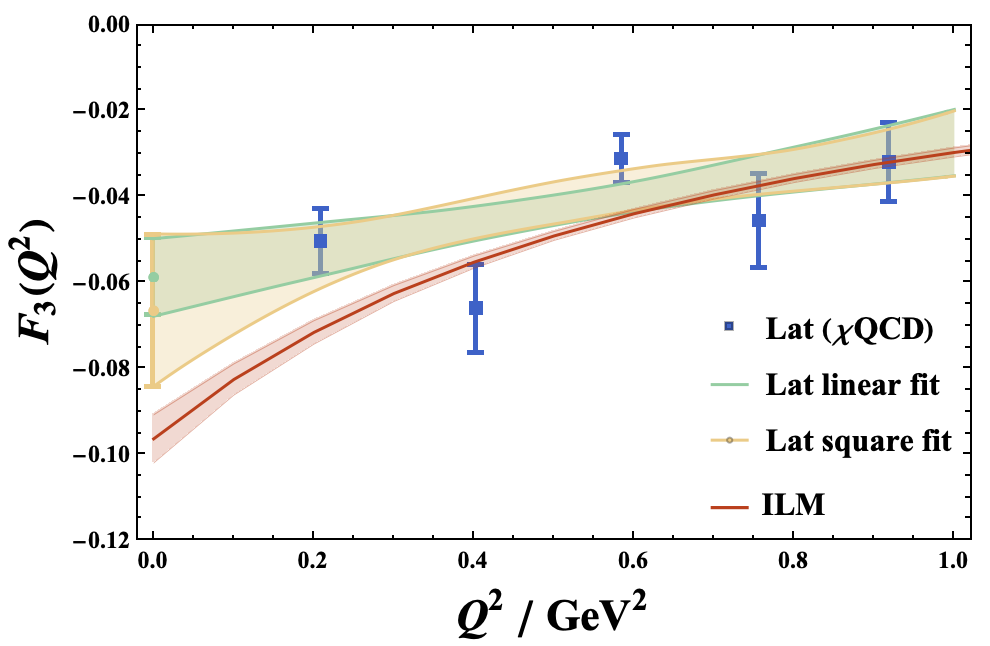}}
    \caption{The CP-odd electric dipole form factor $F_3(Q^2)$ from the ILM for proton (blue) and neutron (red) is obtained using \eqref{F32F} with ILM parameters given in Table~\ref{tab:parameters_ILM} and combining the parameters in \eqref{F32F} with the measurement \cite{Perdrisat:2006hj} for $F_2(Q^2)$ of proton and neutron fitted by dispersion analysis \cite{Belushkin:2006qa}. The neutron electric dipole form factor predicted using the ILM relation \eqref{F32F} combined with $F_2$ lattice results obtained by CSSM and QCDSF/UKQCD Collaborations \cite{CSSM:2014knt} in a 4-volume $32^3 \times64$ ensemble with pion mass $m_\pi=310$ MeV (red band) and lattice spacing $a=0.074(2)$ fm. 
    The $Q^2$ dependence is compared to the lattice results from $\chi$QCD collaboration~\cite{Liang:2023jfj} with $m_\pi=339$ MeV (blue data). The green band is a lattice 
    linear fit and the yellow band is a lattice 
    square fit with additional $Q^4$ terms.}
    \label{fig:F2}
\end{figure}

A comparison of (\ref{EDM5}) to (\ref{EM_form}) yields the pseudoparticle contribution to the $F_{2,3}$ $f$-quark form factors in leading order in the pseudoparticle density.

\begin{equation}
F^f_2(Q^2)\rightarrow Q_f e\,F_I(\rho Q),\quad
F^f_3(Q^2)\rightarrow Q_f e\,F_I(\rho Q)\frac{\langle\Delta^2\rangle}{\bar N}\theta
\end{equation}
  
This connects the $f$-quark Pauli form factor to its electric dipole moment form factor
\begin{equation}
    \begin{aligned}
    \label{F32F}
    \frac{F^f_3(Q^2)}{ \theta}=&\frac{\chi_t}{n_{I+A}}F^f_2(Q^2)
    \end{aligned}
\end{equation}
for a small vacuum angle $\theta$ where the topological susceptibility $\chi_t$ is defined in \eqref{SUS}.
Eq.~(\ref{F32F}) provides for a
relationship between the CP odd and even EM form factors. 

\begin{table*}
\begin{center}
\begin{tabular}{|l|l|l|c|}
    \hline
    & Neutron ($10^{-3}~e\theta$$\cdot\rm fm$) & Proton ($10^{-3}~e\theta$$\cdot\rm fm$) & Ratio $|d_n/d_p|$ \\
    \hline
    ILM  & $d_n=-3.414$ & $d_p=3.200$ & $1.06693$  \\
    ILM \cite{Faccioli:2004jz}  & $|d_n|=6\sim14$ & $-$ & $-$  \\
    ChPT \cite{Mereghetti:2010kp} & $|d_n|=2.10$ & $|d_p|=2.38$ & $0.882$ \\
    $\chi$QCD \cite{Liang:2023jfj} & $d_n=-1.48^{+0.14}_{-0.31}$ & $d_p=3.8^{+1.1}_{-0.8}$ & $0.39^{+0.12}_{-0.12}$  \\
    lattice \cite{Bhattacharya:2021lol} & $d_n=-3^{+7}_{-20}$ & $d_p=24^{+10}_{-30}$ & $0.13^{+0.80}_{-0.30}$ \\
    lattice \cite{Dragos:2019oxn} & $d_n=-1.52\pm 0.71$ & $d_p=1.1 \pm 1.0$ & $1.4 \pm 1.4$ \\
    ETMC \cite{Alexandrou:2020mds} & $|d_n|=0.9\pm2.4$ & $-$ & $-$  \\
    \hline
\end{tabular}
\caption{proton and neutron EDM}
\label{TAB_1}
\end{center}
\end{table*}

Finally, using again (\ref{F32F}) at zero momentum transfer, we arrive at
a direct relationship between the nucleon dipole moments and the nucleon magnetic
moments

\begin{equation}
\label{DNDP}
\begin{aligned}
\frac{d_p(0)}{\theta} &= \frac{\chi_t}{n_{I+A}}\left(\mu_p - Q_p\frac{e}{2M_N}\right) \\
\frac{d_n(0)}{\theta} &= \frac{\chi_t}{n_{I+A}}\left(\mu_n - Q_n\frac{e}{2M_N}\right)
\end{aligned}
\end{equation}
where $Q_p=1$ and $Q_n=0$ is the charge of protons and neutrons.
In this section, (\ref{EDM5}), (\ref{F32F}) and (\ref{DNDP}) are the new and main results of this work.

\begin{table}
    \centering
    \begin{tabular}{|c|c|c|c|}
    \hline
     $\rho$ & $n_{I+A}$ & $m$ & $m^*$   \\[5pt] 
    \hline
   $0.313$ fm & 1 fm$^{-4}$ & $6$ MeV & $103.6$ MeV \\
   \hline
\end{tabular}
    \caption{ILM parameters used in this chapter.}
    \label{tab:parameters_ILM_EDM}
\end{table}

\begin{table}
    \centering
    \begin{tabular}{|c|c|c|c|}
    \hline
     & $m$ &  $|\langle\bar{\psi}\psi\rangle|$ & $m|\langle\bar{\psi}\psi\rangle|$   \\[5pt] 
    \hline
    ILM & $6$ MeV & ($223.5$ MeV)$^3$ & $6.69\times10^{-5}$ \\
     \hline
    FLAG & $3.381(40)$ MeV & ($272(5)$ MeV)$^3$ & $6.80\times10^{-5}$ \\
   \hline
\end{tabular}
    \caption{Chiral condensate in the ILM using the parameters listed in Table~\ref{tab:parameters_ILM_EDM} at the resolution
    $\mu=1/\rho\approx600$ MeV. We compare the ILM results to the Flavour Lattice Averaging Group (FLAG) \cite{FlavourLatticeAveragingGroupFLAG:2021npn}
    at the resolution $\mu=$2 GeV. The last RG invariant combination is in GeV$^4$.}
    \label{tab:parameters_ILM_EDM2}
\end{table}

In Fig.~\ref{fig:F2}, we present the electric dipole form factor $F_3(Q^2)$ predicted by the ILM at low $Q^2$, by using \eqref{F32F} with  $F_2(Q^2)$ borrowed from the dispersion analysis in~\cite{Perdrisat:2006hj} of the proton and neutron form factor measurements in~\cite{Belushkin:2006qa}.

In Fig.~\ref{fig:F2} (red band), we use \eqref{F32F} to derive the CP odd $F_3$, from the CP even lattice $F_2$ in~\cite{CSSM:2014knt} in a 4-volume $32^3 \times64$, with a pion mass $m_\pi=310$ MeV and lattice spacing $a=0.074(2)$ fm. The results are compared to the recently reported 
lattice results from the  $\chi$QCD collaboration (blue squares). We choose the ILM parameters $m$ and $\langle\bar\psi\psi\rangle$ in Table~\ref{tab:parameters_ILM_EDM2} for Fig.~\ref{fig:F2} by extending the physical point $(m=6~\mathrm{MeV},m_\pi=139~\mathrm{MeV})$ to the lattice choice of unphysical mass $(m=28.3~\mathrm{MeV},m_\pi=310~\mathrm{MeV})$ MeV with GOR relation, assuming other low energy parameters such as $n_{I+A}$, $\rho$, and pion decay constant are insensitive to quark mass. The new determinantal mass $m^*$ is $142.4$ MeV ($\langle\bar\psi\psi\rangle=(200~\mathrm{MeV})^3$), determined by Eq.~\eqref{eq:det_gap} or \eqref{eq:det_gap_2}. (see Sec.~\ref{sec:det}) The green band is a lattice linear fit, and the yellow band is a lattice square fit with additional $Q^4$ terms. Our ILM prediction ($m_\pi=310$ MeV) shows consistency with the recent lattice result in $\chi$QCD \cite{Liang:2023jfj} ($m_\pi=339$ MeV).

Using the empirical  values $\mu_p=2.793\, e/2M_N\simeq0.2937$ e$\cdot$fm and $\mu_n=-1.913\, e/2M_N\simeq-0.2012$ e$\cdot$fm~\cite{ParticleDataGroup:2016lqr},
 the ILM predicted electric dipole moments are listed in Table~\ref{TAB_1},
for the ILM parameters $N_f=2$, $m=6.0~\mathrm{MeV}$ and  $\langle\bar\psi\psi\rangle=\langle\bar uu\rangle+\langle\bar dd\rangle= -(223.5~\mathrm{MeV})^3$. One should keep in mind that $d_{p,n}$ in \eqref{DNDP} can be directly determined by the values of topological susceptibility and instanton density $n_{I+A}$ measured by experiments or lattice computations as well. The results are compared to the recently reported lattice results and chiral perturbation calculation (ChPT). Overall, our results appear to be in the range of the proton and
neutron EDM reported by some lattice collaborations. For completeness, we note an earlier  estimate of the neutron dipole moment $|d_n(0)|=(6-14)\times10^{-3}\,(e\theta\cdot\rm fm)$ using  a numerical ensemble of pseudoparticles to describe the ILM, with a moment approximation to extract the EDM~\cite{Faccioli:2004jz}.

In the current analysis of the ILM, the parameters entering our estimation of the nucleon EDM are fixed in Table~\ref{tab:parameters_ILM_EDM} and \ref{tab:parameters_ILM_EDM2}. Corrections to the present results are expected to arise from: 
\begin{enumerate}
    \item omitted instanton-anti-instanton molecules; in the deep cooling regime where most of the UV configurations are smeared and the corresponding renormalization scale is a few hundred MeV, their contribution compared to the single-instanton one (current work) is about 10\% \cite{Liu:2025ldh}
    \item deviations from the quark-diquark approximation used here. 
\end{enumerate}
To address 1. and 2. thoroughly goes beyond the scope of this work.

In QCD, the breaking of conformal symmetry puts stringent constraints on the bulk hadronic correlations in the form of low energy theorems~\cite{Novikov:1981xi}. These constraints are enforced in the QCD instanton vacuum in the form of weaker than Poisson fluctuations in the number of pseudoparticles, with a vacuum compressibility 
indicative of a quantum liquid. The fluctuations in the difference of the pseudoparticles are peaked around a neutral topological charge, with the variance fixed by the topological susceptibility. The latter 
is large in gluodynamics, but substantially screened in QCD with the presence of light quarks. These are essential features of the QCD vacuum as captured by the ILM.

In the ILM, the Pauli  form factor of  a constituent quark, receives a large contribution as noted initially in~\cite{Kochelev:2003cp}. In the presence of a  CP violation by a small vacuum angle $\theta$, we have shown that a $CP$-odd contribution develops in the Pauli form factor, driven mostly by the vacuum topological susceptibility. We have used this result to derive a nontrivial relationship between the $CP$-even $F_2$ and 
$CP$-odd $F_3$ EM form factors, and to estimate the electric dipole moment of
both the  proton and neutron in terms of their respective magnetic moments. Our results are in the range of recently reported lattice results extrapolated at the physical pion mass~\cite{Liang:2023jfj}. The new relationship between $F_2$ and $F_3$ we derived, should prove useful for more accurate  lattice estimates of the nucleon dipole form factor.

\chapter{Bridges from hadrons to partons}
\label{ch:had-to-par}

Since the 1970s, it has been well established that hard processes at large momentum transfer, $Q^2 \gg 1~\mathrm{GeV}^2$, admit a description in terms of weakly interacting partonic degrees of freedom—quarks, antiquarks, and gluons. In this regime, the internal structure of hadrons is encoded in parton distribution functions (PDFs), which characterize the probability densities for finding a parton carrying a longitudinal momentum fraction $x$ at a given resolution scale $Q^2$, along with a variety of related partonic distributions.

At sufficiently high resolution, these pointlike constituents undergo perturbative radiation, emitting and absorbing one another according to splitting functions derived directly from the perturbative QCD. As a result, these parton observables defined at different scales are not independent, but are systematically related through the perturbative evolution governed by the Dokshitzer–Gribov–Lipatov–Altarelli–Parisi (DGLAP) equations.

The combination of perturbative QCD, factorization, and global analyses of hard-scattering data has evolved into a highly successful and quantitatively precise framework for describing hard scattering processes. For a comprehensive overview of modern partonic distribution determinations and global fitting strategies, we refer to recent reviews by the CTEQ Collaboration~\cite{Hou:2019efy}.

While the partonic description provides a robust framework in high energy limit, it relies on several key approximations. In particular, the underlying assumption of decoherence in DGLAP evolution renders the equations effectively probabilistic, making them analogous to kinetic equations. As emphasized in \cite{Kharzeev:2017qzs}, this framework naturally introduces an associated entanglement entropy. However, hadrons are fundamentally pure quantum states, and a fully consistent description should ultimately be formulated in terms of their complete light-front wave functions, without invoking such entropy.

Evolving parton observables toward lower $Q^2$ reveals the emergence of a characteristic scale below which gluon radiation effectively ceases. Empirically, the lightest glueball masses, $M_{\text{glueball}} \sim 2~\text{GeV}$, suggest an effective gluon mass of order $\sim 1~\text{GeV}$. This, in turn, indicates a lower bound at $Q^2 \sim 1~\text{GeV}^2$ for perturbative evolution to be valid.

Below this scale, the relevant degrees of freedom are no longer quarks and gluons, but rather collective phenomena associated with spontaneous chiral symmetry breaking. In this regime, the dynamics are governed by Nambu--Goldstone pions and the scalar condensate $\sigma$. Correspondingly, the QCD Lagrangian is replaced by an effective chiral theory with an UV cutoff $\Lambda_\chi \sim 1~\text{GeV}$ (see Ch.~\ref{ch:low_QCD}). This mechanism was later understood to arise from instanton-induced interactions~\cite{Nowak:1988bh,Vainshtein:1981wh,Diakonov:1985eg}, with the renormalization scale naturally set by the typical instanton size, $\rho \sim 1/3~\text{fm}$.

These observations suggest a natural strategy: the perturbative and chiral regimes of the theory may be smoothly connected around a common scale of order $1~\text{GeV}$. In practice, however, many observables require a more refined treatment than a simple sharp transition between effective descriptions.

\begin{figure}
    \centering
    \includegraphics[width=0.6\linewidth]{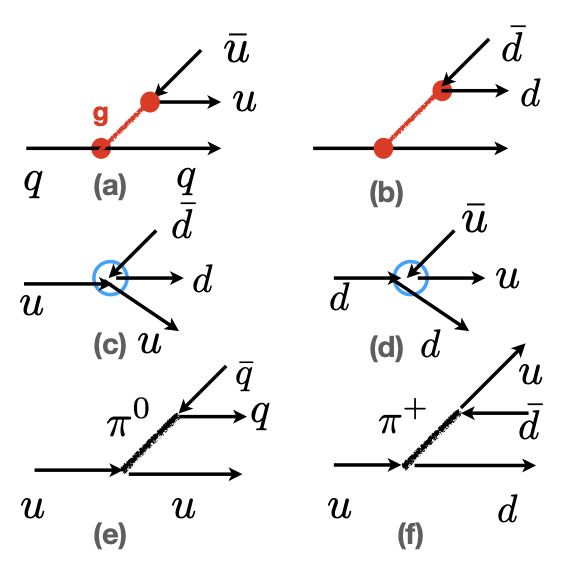}
    \caption{(a,b): pQCD gluon-mediated quark
pair production; (c,d):
instanton-induced ’t Hooft four-fermion interaction;
 (e,f): pion-mediated quark pair
production.}
    \label{fig:dglap3}
\end{figure}

As an illustrative example, we consider the antiquark sea PDF. The mechanisms that generate additional $\bar{q}q$ pairs in the nucleon include three possible scenarios as shown in Fig.~\ref{fig:dglap3}. First, the perturbative gluon splitting produces $\bar{u}u$ and $\bar{d}d$ pairs symmetrically, reflecting the flavor-blind nature of gluons. In contrast, the instanton-induced 't Hooft interaction generates sea quarks in a maximally flavor-asymmetric pattern, dictated by the Pauli exclusion principle acting on fermionic zero modes. Specifically, the dominant processes are $u \rightarrow u d \bar{d}$ and $d \rightarrow d u \bar{u}$, while transitions such as $u \rightarrow u u \bar{u}$ are suppressed. Given that the proton contains two valence $u$ quarks and one valence $d$ quark, a simple leading-order estimate based on the instanton vertex only yields the ratio
\begin{equation}
\frac{\bar{d}}{\bar{u}} \rightarrow 2 \, .
\label{eq:sea_ratio}
\end{equation}
while for the $\Delta^{++}$, which has valence content uuu, the sea should consist solely of $\bar d$ quarks, with no $\bar u$
component at all. Last but not least, one may think of virtual process $N \rightarrow N + \pi$ creating a ``pion cloud''
around a nucleon. 

\begin{figure}
    \centering
    \includegraphics[width=0.8\linewidth]{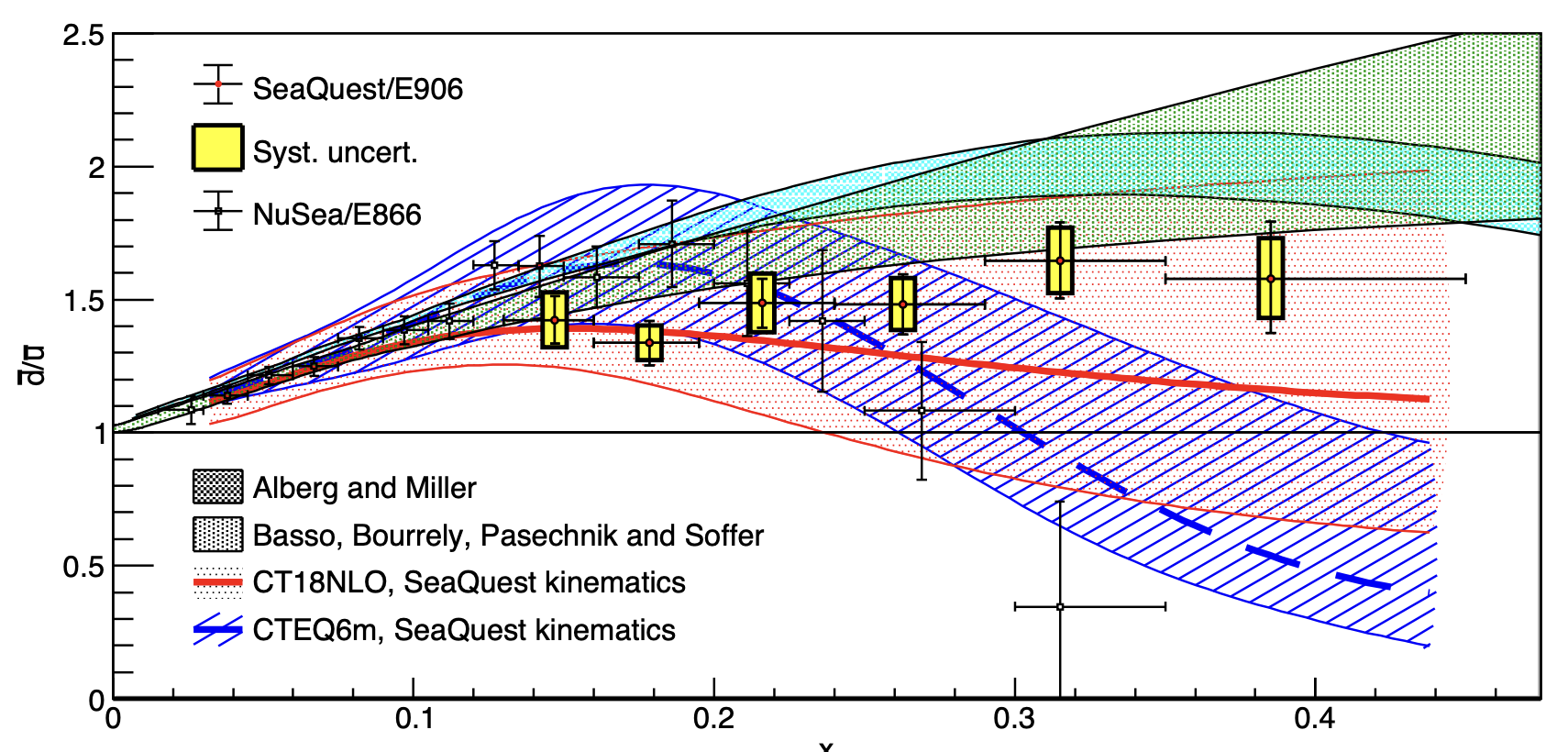}
    \caption{Flavor asymmetry of the antiquark sea as the ratio $\bar d/\bar u$ versus momentum fraction $x$.}
    \label{fig:anti-sea}
\end{figure}

The experimental data for the ratio $\bar d/\bar u$ PDFs are shown in Fig~\ref{fig:anti-sea} from \cite{SeaQuest:2021zxb}.
The data indicate that this ratio is approximately $3/2$, lying between the values of one and two, as expected from instanton-induced interactions. The observed flavor asymmetry of the sea therefore provides a sensitive probe for disentangling the relative contributions of chiral dynamics and perturbative QCD to the nucleon light-front wave function.

\section{Light-front criticality and Wilsonian resolution}


Building on the previous discussion, the ILM provides a well-defined nonperturbative framework for describing the QCD vacuum and its excitations (hadrons) at a fixed resolution $\mu_0 \sim 1/\rho \simeq 0.6\text{--}1~\mathrm{GeV}$ associated with the gauge configurations with typical size and density $\rho \simeq \frac{1}{3}\,\mathrm{fm}$ and $n \simeq 1~\mathrm{fm}^{-4}$, which can be realized in gradient-flow RG scheme. The gradient flow smooths gauge fields over a radius $\sqrt{8t}$, thereby removing UV fluctuations above the scale $\mu_0$. Within this setup, light-front wave functions (LFWFs) are constructed by taking the infinite-momentum limit at fixed resolution. Unlike the full QCD, this limit is well defined, as the underlying cooled gauge configurations are free of ultraviolet fluctuations. The resulting LFWFs should therefore be understood in the Wilsonian sense, corresponding to degrees of freedom resolved at a finite scale. 

In QCD, the definition of renormalized collinear parton distribution functions
\begin{equation}
q(x,\mu) = \int \frac{dz}{2\pi} e^{ixP^+ z^-} \langle P | \bar{\psi}_{\alpha}(0)\gamma^+ W(0,z) \psi(z) | P \rangle,
\end{equation}
and
\begin{equation}
g(x,\mu) = \int \frac{dz}{2\pi} e^{ixP^+ z^-} \langle P | F^{a+\alpha}(0)W^{ab}(0,z)F^{b}{}_{\alpha}{}^{+}(z) | P \rangle,
\end{equation}
require that the UV cutoff $\Lambda$ (or inverse lattice spacing $1/a$) be removed before taking the infinite-momentum limit in the equal-time definitions. 
\begin{equation}
\lim_{P_z\to\infty} \lim_{\Lambda\to\infty} q(x,\Lambda,P_z)=q(x,\mu)
\end{equation}
The order of the two operations is essential to prevent the UV regulator from mixing with the boost, thus ensures regulator independence, boost invariance, and the validity of factorization. In other word, QCD itself is critical in the $P_z \to \infty$ limit, where the corresponding correlation functions become singular unless UV renormalization is performed first.

In ILM, however, one first fixes a finite smoothing scale via gradient flow and then takes $P_z \to \infty$
\begin{equation}
\lim_{P_z\to\infty} q(x,\mu_0,P_z;t)= q(x,\mu)
\tag{14.43}
\end{equation}
Thus the parton distributions correspond to a resolution $\mu_0$ associated with effective hadronic scale. Because gradient flow removes UV modes, the resulting effective theory has no critical behavior and allows a finite, well-defined $P_z \to \infty$ limit. Thus, the light-front formulation and its resulting light-front wave function are meaningful at the finite resolution $\mu_0$, distinct from full QCD. These wavefunctions encode nonperturbative physics at the scale $\mu_0$ and are used to construct various partonic observables such as PDFs, transverse momentum distributions (TMDs), generalized parton distributions (GPDs), and distribution amplitudes (DAs).


One important conclusion from the discussion above is that the light-front (LF) formulation from an IR effective QCD essentially recovers the intrinsically IR nature of LF QCD. 

In full QCD, direct LF quantization effectively places the theory at a critical point. The LF vacuum appears trivial, while long-distance dynamics are encoded in constrained fields and zero modes. As a result, LF QCD is inherently plagued by severe IR singularities. These singularities are manifested from several aspect of LF QCD. For example, constrained fermion fields involve inverse longitudinal derivatives, such as $\psi_- \sim \frac{1}{iD^-}\psi_+$, and instantaneous gluon interactions generate terms proportional to $1/(k^+)^2$. Both of them diverge in the limit $k^+ \to 0$. These divergences are not merely technical artifacts. They reflect the absence of an intrinsic IR scale in the naive LF formulation, leading to a critical theory.

A variety of LF approaches attempt to regulate these singularities through explicit cutoffs, zero-mode subtractions, or basis truncations. While these methods are often practically effective, they tend to obscure the distinction between genuine nonperturbative physics and artifacts introduced by the critical nature of the formulation.

In contrast, the LF formulation of ILM framework adopts an explicit Wilsonian perspective. Instead of constructing the LF formulation for the full theory, we construct LF QCD at a finite resolution scale. At this scale, the relevant gauge fields are smooth and effective LF degrees of freedom are introduced by gauge-covariant gradient flow, which systematically suppresses short-distance fluctuations and generates a natural separation of scales. Consequently, both the LF Hamiltonian and the corresponding wave functions become well-defined and free from spurious IR divergences.

The connection to perturbative QCD is then established through a matching procedure. Gradient-flow–renormalized operators and LF wave functions are matched to the $\bar{MS}$ scheme at a fixed scale (see Sec.~\ref{sec:vac_GF}), after which the perturbative evolution can be applied. This perspective enables a controlled and systematic bridge between hadronic structure at low energies and partonic observables at high energies.

We emphasize that the Wilsonian perspective advocated here clarifies how such constructions should be interpreted. In particular, infrared-sensitive modes which are associated with longitudinal zero modes and light-cone criticality must be treated nonperturbatively and systematically incorporated into effective interactions.



\section{Comparison to LaMET approach}

\begin{figure}
    \centering
    \subfloat[]{\includegraphics[height=4.2cm, width=0.33\linewidth]{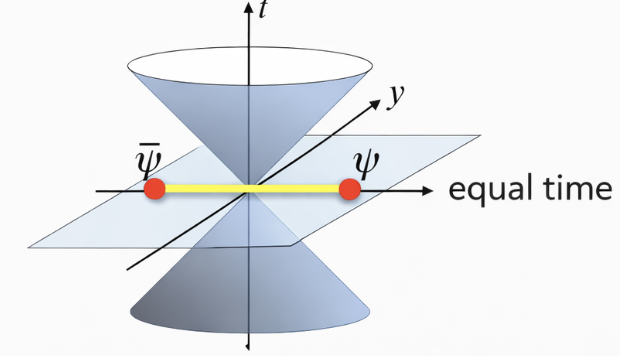}}
    \hfill
    \subfloat[]{\includegraphics[height=4.2cm, width=0.33\linewidth]{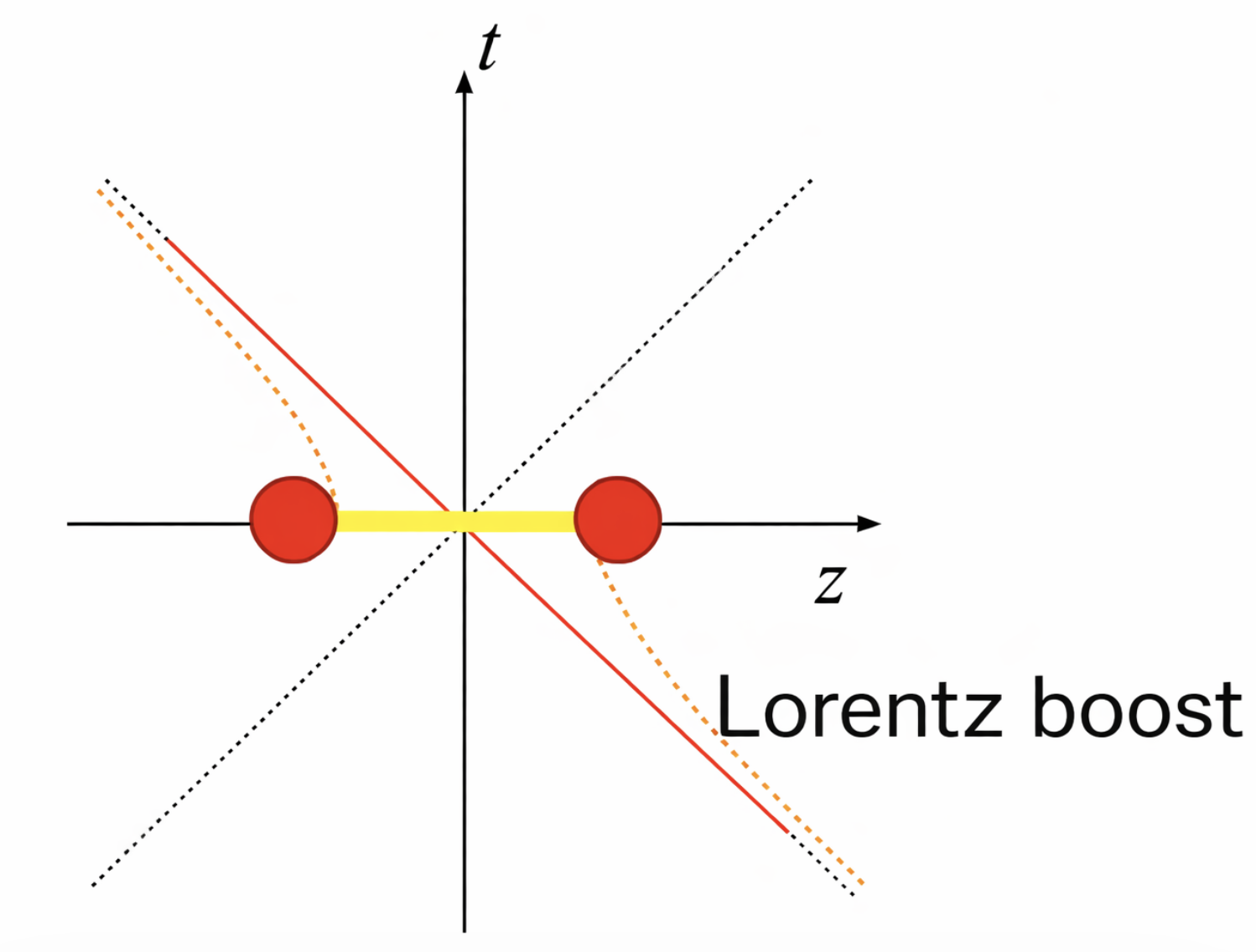}}
    \hfill
    \subfloat[]{\includegraphics[height=4.2cm, width=0.33\linewidth]{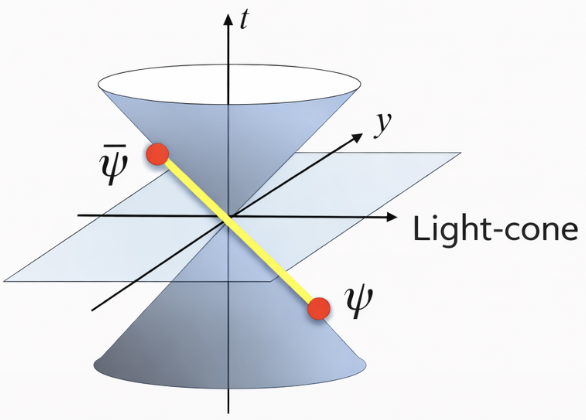}}
    \caption{The picture of LaMET framework. The equal-time correlation is related to light-cone correlation though Lorentz boost.}
    \label{fig:LaMET}
\end{figure}

It is instructive to draw a parallel between the present strategy and the Large Momentum Effective Theory (LaMET) approach. In LaMET, partonic structure is accessed through equal-time correlation functions evaluated at large but finite hadron momentum as shown in Fig.~\ref{fig:LaMET}. Crucially, this formulation avoids the LF criticality by the finite momentum $P_z$ as an intrinsic IR regulator, preventing the appearance of singular structures such as $k^+\rightarrow0$ in LF quantization. The connection to light-cone correlations is established only after taking the large-momentum limit, implemented through systematic factorization and perturbative matching.

From this perspective, LaMET and the LF ILM framework share a common conceptual foundation. In both cases, one avoids formulating QCD directly at the LF critical point, and works instead with an effective theory defined by a finite control scale that measures the distance away from the critical point, that is, finite hadron momentum $P_z$ in LaMET, and a finite gradient-flow scale in ILM. In each framework, the connection to light-cone physics is recovered only after perturbative matching and RG evolution. This shared Wilsonian logic explains why both approaches yield well-defined nonperturbative inputs for parton physics while bypassing the intrinsic IR criticality of LF QCD.

In LaMET, equal-time correlators are computed at large but finite $P_z$, where the theory is IR safe and free from zero-mode singularities. The limit $P_z\rightarrow\infty$ is never taken at the nonperturbative level. Alternatively, quasi distributions are factorized and matched onto light-cone PDFs in the 
$\bar MS$ scheme through an expansion in powers of $1/P_z$ \cite{Ji:2013dva,Izubuchi:2018srq,Ji:2020byp,Zhao:2025oto} 

\begin{equation}
    \mathcal{O}_{\rm LaMET}^R(x,P_z)=\int \frac{dy}yC\left(\frac xy,\frac{\mu}{P_z}\right)\otimes \mathcal{O}_{\rm LF}^R(y,\mu) +\mathcal{O}\left(\frac{\Lambda^2_{\rm QCD}}{P_z^2}\right)
\end{equation}

In this sense, $P_z$ plays the role of a Wilsonian scale that keeps the theory away from the LF critical surface. One should keep in mind that as in any Wilsonian factorization construction, power-suppressed corrections in $1/P_z$ encode genuine long-distance physics and may exhibit ambiguities related to renormalons, which need to be systematically absorbed into higher-twist matrix elements.

In ILM, the analogous control parameter is the gradient-flow scale $\sqrt{t}$, or equivalently the matching scale $\mu_0\sim1/\sqrt{8t}$, as discussed in Sec.~\ref{sec:vac_GF}. For more details, see \cite{Brambilla:2023vwm,Shuryak:2026pqt,Hieda:2016lly,Mereghetti:2021nkt}. 

\begin{equation}
    \mathcal{O}_{\rm ILM}(z,\mu,t)=C\left(t,\mu\right)e^{\delta mz} \mathcal{O}_{\rm LF}(z,\mu) +\mathcal{O}\left(t\Lambda^2_{\rm QCD}\right)
\end{equation}
where $\delta m$ denotes the linear renormalon contribution from the Wilson lines~\cite{Liu:2020rqi,Braun:2024snf,Zhang:2025mer,Braun:2018brg,}.
The LF Hamiltonian and wave functions are constructed at finite resolution using flowed gauge fields, for which constrained LF quark fields are smooth and IR safe. The LF critical point is not imposed as a starting point, but is approached only indirectly through perturbative matching and RG evolution.

This comparison highlights a unifying perspective: both LaMET and ILM realize a controllable Wilsonian interpolation between nonperturbative hadronic physics and partonic observables, while systematically avoiding the singularities associated with LF criticality.

\begin{align}
\text{LaMET:} \quad & P_z < \infty \;\Longrightarrow\; \text{matching} \;\Longrightarrow\; P_z \to \infty, \\
\text{ILM} \quad & \sqrt{t} > 0 \;\Longrightarrow\; \text{matching} \;\Longrightarrow\; \sqrt{t} \to 0.
\end{align}

In both frameworks, it is essential that the formal limits $P_z \to \infty$ and $\sqrt{t} \to 0$ are \emph{never} taken at the nonperturbative level. Instead, the light-cone theory emerges only after a controlled perturbative matching procedure. 

\section{PDF}

At leading twist (twist-2), the partonic structure of a hadron in collinear factorization is encoded in six parton distribution functions (PDFs) defined through gauge-invariant light-front correlators. For quarks, the unpolarized, helicity, and transversity distributions, $f_1^{q/h}(x)$, $g_1^{q/h}(x)$, and $h_1^{q/h}(x)$, arise from the operators $\bar\psi \gamma^+ \psi$, $\bar\psi \gamma^+\gamma^5 \psi$, and $\bar\psi i\sigma^{+i} \psi$, and describe the momentum density, longitudinal spin alignment, and transverse spin structure, respectively. These quark PDFs are given by
\begin{equation}
\begin{aligned}
\label{eq:qPDF}
&q_h(x)\equiv f_1^{q/h}(x) = \frac{1}{2} \int \frac{dz^-}{2\pi} \, e^{ixp^+ z^-}
\left\langle h \left| \bar{\psi}\left(0\right)\gamma^+W[0,z^-] \psi\left(z^-\right) \right| h \right\rangle  \\
&\Delta q_h(x)\equiv g_1^{q/h}(x)  = \frac{1}{2} \int \frac{dz^-}{2\pi} \, e^{ixp^+ z^-}
\left\langle h \left| \bar{\psi}\left(0\right)\gamma^+\gamma^5W[0,z^-] \psi\left(z^-\right) \right| h \right\rangle \\
&\delta q_h(x)\equiv h_1^{q/h}(x) =\frac{1}{2}\int \frac{dz^-}{2\pi}\, e^{ixp^+ z^-}
\left\langle h \left| \bar{q}\left(0\right)
i\sigma^{+i}W[0,z^-] q\left(z^-\right) \right| h \right\rangle
\end{aligned}
\end{equation}

Similarly for gluons, the corresponding distributions $f_1^{g/h}(x)$, $g_1^{g/h}(x)$, and $h_1^{g/h}(x)$ are associated with the correlators of $F^{+\alpha}F_{\alpha}^{\ +}$, $F^{+\alpha}\tilde F_{\alpha}^{\ +}$, and the symmetric traceless combination $F^{+i}F^{+j}$, encoding momentum, helicity, and tensor polarization of gluons, given by

\begin{equation}
\begin{aligned}
&g_h(x)\equiv f_1^{g/h} = \frac{1}{2xp^+} \int \frac{dz^-}{2\pi} \, e^{ixp^+ z^-}
\left\langle h \left| F^{+\alpha}\left(0\right)
W[0,z^-]F_{\alpha}^{\ +}\left(z^-\right) \right| h \right\rangle 
 \\
&\Delta g_h(x)\equiv g_1^{g/h} = -\frac{i}{2xp^+} \int \frac{dz^-}{2\pi} \, e^{ixp^+ z^-}
\langle h | F^{+\alpha}\left(0\right)
W[0,z^-]\tilde{F}_{\alpha}^{\ +}\left(z^-\right) | h \rangle 
\\
&\delta g_h(x)\equiv h_1^{g/h}=\frac{1}{2xp^+}\int \frac{dz^-}{2\pi}\, e^{ixp^+ z^-}
\left\langle h \left| F^{+\{i}\left(0\right)
W[0,z^-]F^{j\}+}\left(z^-\right) \right| h \right\rangle
\end{aligned}
\end{equation}
where $a^{\{i}b^{j\}}=\frac12(a^ib^j+a^jb^i)-\frac12g^{ij}_\perp a\cdot b$ denotes symmetric traceless in transverse plane.  

For spin-$0$ hadrons, such as the pion and kaon, only the unpolarized quark and gluon distributions are nonvanishing at leading twist. For spin-$\tfrac{1}{2}$ hadrons, such as the nucleon, five leading-twist PDFs are nonzero, while the gluon transversity distribution vanishes. This follows from the fact that gluon transversity carries two units of helicity, which cannot be supported by a spin-$\tfrac{1}{2}$ target, though it can arise for higher-spin hadrons such as the $\rho$ meson. In general, hadron PDFs satisfy $q_{\bar{h}}(x)=\bar{q}_{h}(x)=-q_{h}(-x)$ due to charge conjugation symmetry. Flavor symmetry further relates different parton species, for example $u_{K^+}(x)=\bar{s}_{K^+}(1-x)=s_{K^-}(1-x)$ and $u_{p}(x)=d_{n}(x)$.

To illustrate this, we use the LFWFs derived in Eqs.~\ref{WFX} and \ref{WFX2} and substitute them into Eq.~\eqref{eq:qPDF} to compute the corresponding unpolarized quark PDF for mesons. The resulting pion and kaon PDF read

\begin{equation}
\begin{aligned}
\label{eq:pionPDF_RIV}
    u_{\pi^+}(x)=&\int d^2k_\perp\frac{2N_cZ_{\pi}}{(2\pi)^3}\frac{k^2_\perp+M^2}{(x\bar{x}m_{\pi}^2-k_\perp^2-M^2)^2}\mathcal{F}^2\left(\frac{k_\perp}{\lambda_\pi\sqrt{x\bar x}}\right)
\end{aligned}
\end{equation}

and

\begin{equation}
\begin{aligned}
\label{eq:kaonPDF_RIV}
    u_{K^+}(x)
    =&\int d^2k_\perp\frac{2N_cZ_K}{(2\pi)^3}\frac{k_\perp^2+x^2M_s^2+2x\bar{x}M_sM_u+\bar{x}^2M_u^2}{[x\bar{x}m^2_K-(k_\perp^2+xM_s^2+\bar{x}M_u^2)]^2}\mathcal{F}^2\left(\frac{k_\perp}{\lambda_K\sqrt{x\bar x}}\right)
\end{aligned}
\end{equation}

\begin{figure}
    \centering
    \includegraphics[width=\linewidth]{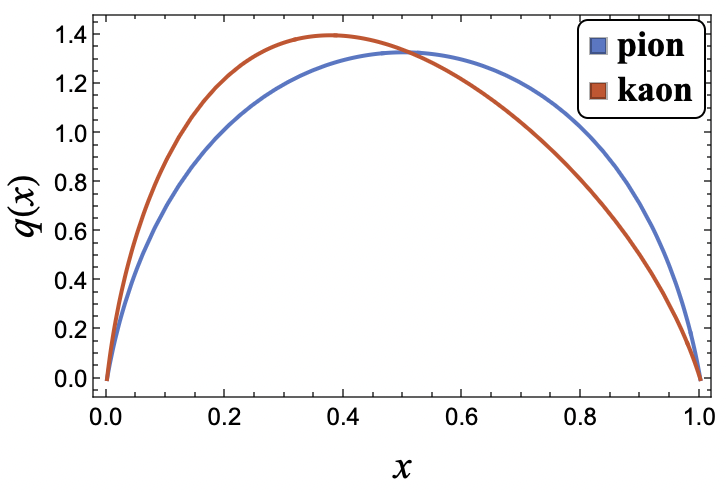}
    \caption{Pion PDF $u_{\pi^+}$ and kaon PDF $u_{K^+}$ obtained by ILM at low resolution $\mu_0\sim630$ MeV}
    \label{fig:pi_K}
\end{figure}

In Fig.~\ref{fig:pi_K}, we present the pion and kaon LFWFs with an instanton size $\rho = 0.313~\mathrm{fm}$. For the pion, we use $\lambda_\pi = 2.46$, $m_\pi = 135~\mathrm{MeV}$, $M = 398~\mathrm{MeV}$, and $\sqrt{Z_\pi} = 4.27$. For the kaon, we take $\lambda_K = 2.98$, $m_K = 445~\mathrm{MeV}$, $M_u = 366~\mathrm{MeV}$, $M_s = 606~\mathrm{MeV}$, and $\sqrt{Z_K} = 4.01$. Due to the mass asymmetry between the $u$ and $s$ quarks, the heavier $s$ quark carries a larger fraction of the kaon momentum. As a result, the $u$-quark distribution in the positively charged kaon, $u_{K^+}(x)$, is shifted toward smaller $x$.

The wave-function normalizations $Z_{\pi,K}$ are related to the pion and kaon decay constants through the Goldberger–Treiman relation, 
\begin{equation}
\sqrt{Z_\pi}\approx\frac{M}{f_\pi}, \qquad
\sqrt{Z_K}\approx\frac{M_u+M_s}{2f_K},
\end{equation}
with $f_\pi = 93.3~\mathrm{MeV}$ and $f_K = 121.2~\mathrm{MeV}$, indicating that low-energy chiral relations are preserved even within the light-front formulation of ILM.

\begin{figure}
    \centering
    \subfloat[\label{fig:pdf_pi}]{\includegraphics[width=0.86\linewidth]{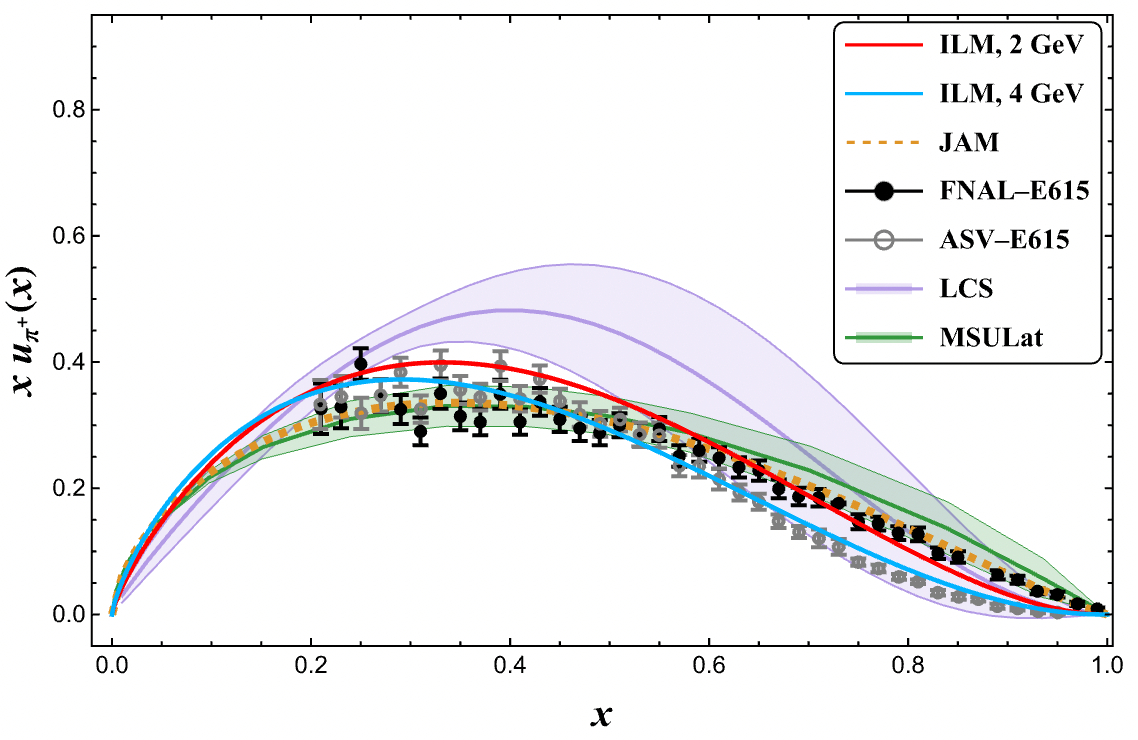}}
    \hfill
    \subfloat[\label{fig:pdf_K}]{\includegraphics[width=0.88\linewidth]{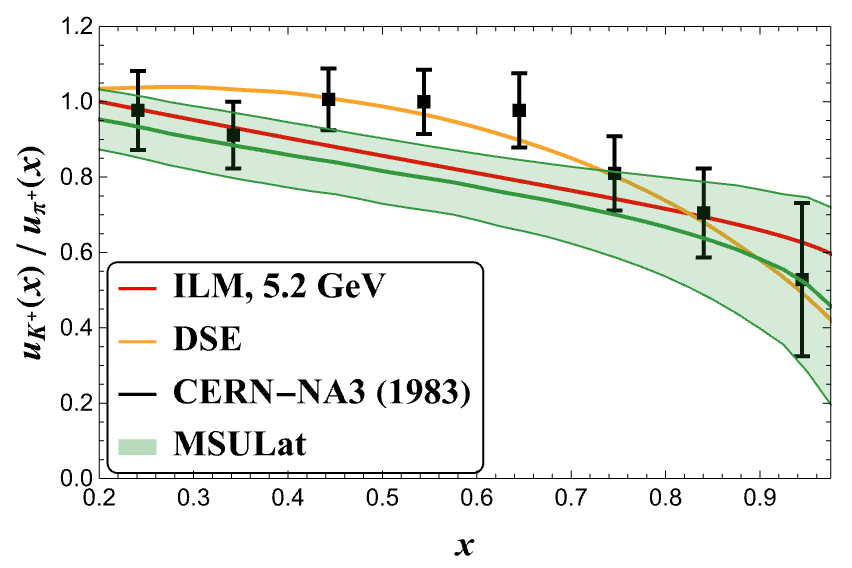}}
    \caption{Pion and kaon PDF evolved to a few GeV indicated in the plot are compared with the experimental data from E615 and NA3 and several lattice calculations (see text).}
    \label{fig:pdf}
\end{figure}

In Fig.~\ref{fig:pdf_pi}, the resulting pion LFWF is evolved to $2$ and $4~\mathrm{GeV}$ using one-loop DGLAP with $\Lambda_{\mathrm{QCD}} = 226~\mathrm{MeV}$ ($N_f=3$). The evolved distributions are compared with the original LO/NLO extraction from the Fermilab E615 experiment \cite{Conway:1989fs} (black circle), obtained from $\mu^+\mu^-$ production in the Drell–Yan process; a reanalysis of the same data improved by threshold soft-gluon resummation at NLL accuracy \cite{Aicher:2010cb} (gray open circle); the global QCD fit by the Jefferson Lab Angular Momentum (JAM) Collaboration \cite{Barry:2023qqh} (orange dashed curve); and lattice-QCD calculations, including the lattice cross-section (LCS) approach \cite{Sufian:2020vzb} (purple band) and results from the Michigan State University (MSU) lattice group \cite{Lin:2020ssv} (green band).

In Fig.~\ref{fig:pdf_K}, the ratio of the $u$-quark PDFs in the positively charged kaon and pion, $u_{K^+}(x)/u_{\pi^+}(x)$, obtained from Eqs.~\eqref{eq:pionPDF_RIV} and \eqref{eq:kaonPDF_RIV}, is evolved to $5.2~\mathrm{GeV}$ ($\mu^2=27~\mathrm{GeV}^2$). The result is compared with the experimental extraction (red circles) from muon-pair production in the NA3 experiment \cite{Saclay-CERN-CollegedeFrance-EcolePoly-Orsay:1980fhh}, where an invariant mass cut of $4.1$–$8.5~\mathrm{GeV}$ is imposed to suppress resonance contributions. We further compare with results obtained from the Dyson--Schwinger equation (DSE) \cite{Nguyen:2011jy}, as well as recent lattice-QCD calculations from the MSU lattice group \cite{Lin:2020ssv}.

\section{GPD}
\label{sec:GPD}

The modern development of GPDs began in 1996, when it was realized that hard exclusive processes, such as deeply virtual Compton scattering (DVCS) and exclusive meson production, probe non-forward (off-diagonal) parton correlations~\cite{Ji:1996ek,Radyushkin:1996nd,Ji:1996nm,Radyushkin:1996ru}. First, the proof of QCD factorization for meson electroproduction in both diffractive and non-diffractive regimes was established in~\cite{Collins:1996fb}, demonstrating that the amplitude separates into a perturbatively calculable hard kernel and universal nonperturbative functions, namely GPDs and meson DAs. In~\cite{Ji:1996ek}, it was further shown that GPDs encode essential information about the spin structure of the nucleon. In particular, their moments obey a sum rule (Sec.~\ref{sec:N_spin}) that provides access to the total angular momentum carried by quarks and gluons, including both intrinsic spin and orbital contributions.

More generally, GPDs interpolate between PDFs and elastic form factors. They reduce to ordinary PDFs in the forward limit and to form factors upon integration over $x$. They thus provide a unified description of parton momentum and spatial structure (tomography). 

\begin{figure}
    \centering
    \includegraphics[width=1\linewidth]{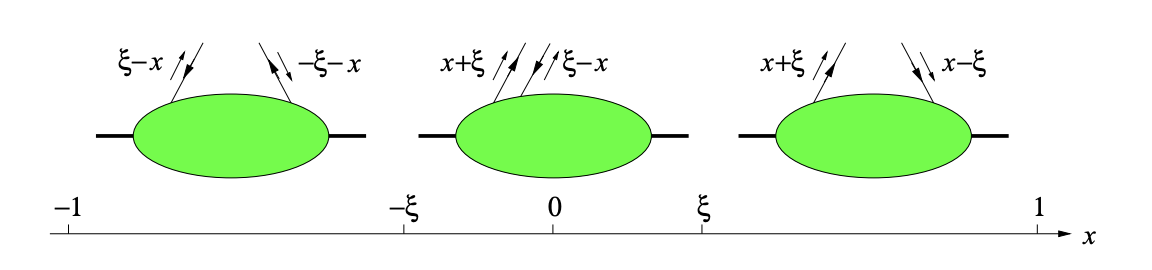}
    \caption{The physically allowed kinematic region in GPD}
    \label{fig:gpd}
\end{figure}

As illustrated in Fig.~\ref{fig:gpd}, the distributions have support in the interval $x \in [-1,1]$ with the skewness parameter $\xi=-q^+/(2\bar{p}^+)$, which can be divided into three regions:

\begin{enumerate}
\item For $x \in [\xi,1]$, both momentum fractions $x+\xi$ and $x-\xi$ are positive; the distribution describes the emission and reabsorption of a quark.

\item For $x \in [-\xi,\xi]$, one has $x+\xi \ge 0$ but $x-\xi \le 0$. The second momentum fraction is then interpreted as belonging to an antiquark with momentum fraction $\xi - x$ emitted from the initial proton.

\item For $x \in [-1,-\xi]$, both $x+\xi$ and $x-\xi$ are negative; this corresponds to the emission and reabsorption of antiquarks with respective momentum fractions $\xi - x$ and $-\xi - x$.
\end{enumerate}

The first and third regions are commonly referred to as the DGLAP regions, while the second is known as the Efremov--Radyushkin--Brodsky--Lepage (ERBL) region, reflecting their respective evolution patterns with the factorization scale.

Generally, GPDs can be classified into two categories based on the properties of physical processes. The first involves quark and gluon operators with zero net helicity transfer, which, in the DGLAP region, preserve the helicity of the emitted and reabsorbed parton. The second consists of distributions that induce helicity flip. In the DGLAP region, these correspond to processes where the parton helicity changes, while in both the DGLAP and ERBL regions they carry a net helicity transfer of one unit for quarks and two units for gluons. These helicity-flipping distributions are off-diagonal in the parton helicity basis and become diagonal in transversity basis defined by eigenstates of transversity, therefore referred to as transversity distributions.

In the following, we adopt the definition in \cite{Ji:1998pc} for chiral even distribution and \cite{Diehl:2003ny} for chiral odd, similar definition with silghtly different convention can be found in \cite{Goeke:2001tz,Hoodbhoy:1998vm}. For chiral-even quark GPD, we have
\begin{equation}
\begin{aligned}
F^{q/N} &= \frac{1}{2} \int \frac{dz^-}{2\pi} \, e^{ix\bar{p}^+ z^-}
\left\langle p' \left| \bar{\psi}\left(-\tfrac{1}{2}z\right)\gamma^+W[-\tfrac{1}{2}z,\tfrac{1}{2}z] \psi\left(\tfrac{1}{2}z\right) \right| p \right\rangle \Big|_{z^+=0,\, z_\perp=0} \\
&= \frac{1}{2\bar{p}^+} \bar{u}_{s'}(p')\left[
H^{q/N}(x,\xi,t)\,\gamma^+
+ E^{q/N}(x,\xi,t)\, \frac{i\sigma^{+\alpha}q_\alpha}{2m_N}
\right]u_s(p), \\
\tilde{F}^{q/N} &= \frac{1}{2} \int \frac{dz^-}{2\pi} \, e^{ix\bar{p}^+ z^-}
\left\langle p' \left| \bar{\psi}\left(-\tfrac{1}{2}z\right)\gamma^+\gamma^5W[-\tfrac{1}{2}z,\tfrac{1}{2}z] \psi\left(\tfrac{1}{2}z\right) \right| p \right\rangle \Big|_{z^+=0,\, z_\perp=0} \\
&= \frac{1}{2\bar{p}^+} \bar{u}_{s'}(p')\left[
\tilde{H}^{q/N}(x,\xi,t)\,\gamma^+\gamma^5
+ \tilde{E}^{q/N}(x,\xi,t)\, \frac{\gamma^5 q^+}{2m_N} 
\right]u_s(p).
\end{aligned}
\end{equation}
For chiral-even gluon GPD, we have
\begin{equation}
\begin{aligned}
F^{g/N} &= \frac{1}{2x\bar{p}^+} \int \frac{dz^-}{2\pi} \, e^{ix\bar{p}^+ z^-}
\left\langle p' \left| F^{+\alpha}\left(-\tfrac{1}{2}z\right)
W[-\tfrac{1}{2}z,\tfrac{1}{2}z]F_{\alpha}^{\ +}\left(\tfrac{1}{2}z\right) \right| p \right\rangle 
\Big|_{z^+=0,\, z_\perp=0} \\
&= \frac{1}{2\bar{p}^+} \bar{u}_{s'}(p')\left[
H^{g/N}(x,\xi,t)\,\gamma^+ 
+ E^{g/N}(x,\xi,t)\,\frac{i\sigma^{+\alpha}q_\alpha}{2m_N} 
\right]u_s(p), \\
\tilde{F}^{g/N} &= -\frac{i}{2x\bar{p}^+} \int \frac{dz^-}{2\pi} \, e^{ix\bar{p}^+ z^-}
\left\langle p' \left| F^{+\alpha}\left(-\tfrac{1}{2}z\right)
W[-\tfrac{1}{2}z,\tfrac{1}{2}z]\tilde{F}_{\alpha}^{\ +}\left(\tfrac{1}{2}z\right) \right| p \right\rangle 
\Big|_{z^+=0,\, z_\perp=0} \\
&= \frac{1}{2\bar{p}^+} \bar{u}_{s'}(p')\left[
\tilde{H}^{g/N}(x,\xi,t)\,\gamma^+\gamma^5 
+ \tilde{E}^{g/N}(x,\xi,t)\,\frac{\gamma^5 q^+}{2m_N}
\right]u_s(p).
\end{aligned}
\end{equation}
where $F^{g/N}(x,\xi) = -\,F^{g/N}(-x,\xi)$ due to the Bose symmetry. One should keep in mind that chiral-even GPDs preserve quark helicity but may still induce nucleon helicity flip. Specifically, $H$ and $\tilde{H}$ correspond to helicity-conserving amplitudes, while $E$ and $\tilde E$ describe nucleon helicity-flip contributions.

Similarly, the quark (one unit helicity transfer) and gluon (two unit helicity transfer) chiral-odd GPD can be defined by \cite{Diehl:2003ny,Kim:2025mol}

\begin{equation}
\begin{aligned}
F^{q/N}_T=&\frac{1}{2}\int \frac{dz^-}{2\pi}\, e^{ix\bar{p}^+ z^-}
\left\langle p' \left| \bar{q}\left(-\tfrac{1}{2}z\right)
i\sigma^{+i}W[-\tfrac{1}{2}z,\tfrac{1}{2}z] q\left(\tfrac{1}{2}z\right) \right| p \right\rangle
\Big|_{z^+=0,\, z_\perp=0}
\\
=& \frac{1}{2\bar{p}^+}\,\bar{u}_{s'}(p') \Bigg[
H_T^{q/N}\, i\sigma^{+i}
+ \tilde{H}_T^{q/N} \frac{\bar{p}^+ q^i - q^+ \bar{p}^i}{m_N^2}
\\
&+ E_T^{q/N} \frac{\gamma^+ q^i - q^+ \gamma^i}{2m_N}
+ \tilde{E}_T^{q/N} \frac{\gamma^+ \bar{p}^i - \bar{p}^+ \gamma^i}{m_N}
\Bigg] u_{s}(p),
\\[1em]
F^{g/N}_T=&\frac{1}{2x\bar{p}^+}\int \frac{dz^-}{2\pi}\, e^{ix\bar{p}^+ z^-}
\left\langle p' \left| F^{+\{i}\left(-\tfrac{1}{2}z\right)
W[-\tfrac{1}{2}z,\tfrac{1}{2}z]F^{j\}+}\left(\tfrac{1}{2}z\right) \right| p \right\rangle
\Big|_{z^+=0,\, z_\perp=0}
\\
=& \frac{1}{2\bar{p}^+}
\,\bar{u}_{s'}(p') \Bigg[
H_T^{g/N}\, i\sigma^{+\{i}
+ \tilde{H}_T^{g/N} \frac{\bar{p}^+ q^{\{i} - q^+ \bar{p}^{\{i}}{m_N^2}
\\
&+ E_T^{g/N} \frac{\gamma^+ q^{\{i} - q^+ \gamma^{\{i}}{2m_N}
+ \tilde{E}_T^{g/N} \frac{\gamma^+ \bar{p}^{\{i} - \bar{p}^+ \gamma^{\{i}}{m_N}
\Bigg]\frac{q^{j\}}\bar{p}^+ - \bar{p}^{j\}}q^+ }{2m_N\,\bar{p}^+} u_{s}(p).
\end{aligned}
\end{equation}

Further symmetry properties follow from time-reversal invariance. Inserting antiunitary operator implementing time reversal in Hilbert space in the matrix elements defining the GPDs, one obtains
\begin{equation}
H(x,-\xi,t) = H(x,\xi,t),
\end{equation}
As under time reversal, interchanging initial and final states of the matrix elements interchanges the momenta $p$ and $p'$, time reversal thus changes the sign of $\xi$. Taking the complex conjugate of the defining matrix elements (Hermicity) gives
\begin{equation}
\bigl[ H(x,-\xi,t) \bigr]^* = H(x,\xi,t).
\end{equation}
Similar relations also appear for $E$, $\tilde{H}$, $\tilde{E}$, $H_T$, $\tilde{H}_T$, $E_T$ both for quark and gluon distributions.

\subsection{Form factors of first Mellin moment}

The first Mellin moment of twist-2 nucleon GPDs corresponds to nucleon form factors associated with local operators of the lowest conformal spin $j=1$. In this limit, the nonlocal light-cone operators reduce to local currents, and the corresponding matrix elements probe the exchange of spin-$1$ quantum numbers in the $t$-channel. The corresponding off-forward hadronic matrix elements are given by

\begin{equation}
\langle N'| \bar \psi\, \gamma^{\mu} \psi |N\rangle
=
\bar u_{s'}(p')
\left[
\gamma^{\mu}\, F_1^q(t)
+ \frac{i\sigma^{\mu\nu}q_{\nu}}{2m_N}\, F_2^q(t)
\right]
u_s(p),
\end{equation}

\begin{equation}
\langle N'| \bar \psi\,\gamma^{\mu}\gamma^5 \psi |N\rangle
=
\bar u_{s'}(p')
\left[
\gamma^{\mu}\gamma^5\, G_A^q(t)
+ \frac{q^{\mu}\gamma^5}{2m_N}\, G_P^q(t)
\right]
u_s(p),
\end{equation}

\begin{equation}
\label{eq:tensor_mag}
\langle N'| \bar \psi\, i\sigma^{\mu\nu} \psi |N\rangle
=
\bar u_{s'}(p')
\left[
i\sigma^{\mu\nu}\, G_T^q(t)
+ \frac{\gamma^{[\mu}q^{\nu]}}{2m_N}\, E_T^q(t)
+ \frac{\bar p^{[\mu}q^{\nu]}}{m_N^2}\, \widetilde H_T^q(t)
\right]
u_s(p),
\end{equation}
with antisymmetrization $a^{[\mu} b^{\nu]}=\frac12(a^{\mu} b^{\nu}-a^{\nu} b^{\mu})$.

At zero momentum transfer, they correspond to electric charge $F^q_1(0)=Q_q$, magnetic moment $F^q_2(0)=1$, axial charge $G^q_A(0)=g^q_A\equiv\Delta q$, pseudoscalar charge $G^q_P(0)=g^q_P$, tensor charge $G^q_T(0)=g_T^q\equiv\delta q$, and anomalous tensor magnetic moment $\bar{E}^q_T(0)\equiv E^q_T(0)+2\tilde{H}^q_T=\kappa_T^q$.

For the chiral-even distributions, this reduction reproduces the matrix elements of the conserved vector and axial-vector currents. Consequently, the first moments of the GPDs $H$ and $E$ yield the electromagnetic Dirac and Pauli form factors, $F_1(t)$ and $F_2(t)$, respectively, while the moments of $\widetilde H$ and $\widetilde E$ give the axial and induced pseudoscalar form factors, $G_A(t)$ and $G_P(t)$. These relations follow from Lorentz covariance and current conservation, and are independent of the skewness, reflecting the polynomiality property of GPDs at lowest order.

\begin{equation}
\begin{aligned}
\int_{-1}^1 dx\, H^{q/N}(x,\xi,t) &= F^q_1(t), \qquad
\int_{-1}^1 dx\, E^{q/N}(x,\xi,t) = F^q_2(t), \\
\int_{-1}^1 dx\, \tilde{H}^{q/N}(x,\xi,t) &= G^q_A(t), \qquad
\int_{-1}^1 dx\, \tilde{E}^{q/N}(x,\xi,t) = G^q_P(t).
\end{aligned}
\end{equation}

For chiral odd distributions,
\begin{equation}
\begin{aligned}
\int_{-1}^1 dx\, H_T^{q/N}(x,\xi,t) &= G_T^q(t), 
\qquad
\int_{-1}^1 dx\, \tilde{E}_T^{q/N}(x,\xi,t) = 0, \\
\multicolumn{2}{c}{$
\int_{-1}^1 dx\, \Big[ E_T^{q/N}(x,\xi,t) + 2\tilde{H}_T^{q/N}(x,\xi,t) \Big] 
= \bar{E}_T^q(t)$}
\end{aligned}
\end{equation}

\subsection{Form factors of second Mellin moment}

The second Mellin moment of twist-2 nucleon GPDs corresponds to nucleon form factors associated with local operators of the next-to-lowest conformal spin $j$. In this case, the nonlocal light-cone operators reduce to components of the EMT, and the corresponding matrix elements probe the exchange of spin-2 quantum numbers in the $t$-channel. The corresponding off-forward hadronic matrix elements for quarks are given by

\begin{equation}
\begin{aligned}
\langle N' |\bar{\psi}\gamma^{\{\mu} i\overleftrightarrow{D}^{\nu\}}\psi| N \rangle
=&\bar{u}_{s'}(p')\bigg[A_N^{q}(t)\gamma^{\{\mu} \bar{p}^{\nu\}}+B_N^{q}(t)\frac{i\bar{p}^{\{\mu}\sigma^{\nu\}\alpha}q_\alpha}{2m_N}\\
    &+D_N^{q}(t)\frac{1}{4m_N}\left(q^{\mu}q^{\nu}-\frac14g^{\mu\nu}q^2\right)\bigg]u_{s}(p)
\end{aligned}
\end{equation}

\begin{equation}
\begin{aligned}
&\langle N' |\bar{\psi}\gamma^{\{\mu} i\overleftrightarrow{D}^{\nu\}}\gamma^5\psi| N \rangle
=\bar{u}_{s'}(p')
\left[
\widetilde{A}^q_N(t)\, \gamma^{\{\mu}\, \bar{p}^{\nu\}}\gamma^5
+
\widetilde{B}^q_N(t)\, \frac{q^{\{\mu} \bar{p}^{\nu\}}}{2m_N}\gamma^5\, 
\right]
u_s(p)
\end{aligned}
\end{equation}

\begin{equation}
\begin{aligned}
    \langle N'|\bar\psi i\sigma^{[\mu\{\nu]}i\overleftrightarrow{D}^{\rho\}}\psi|N\rangle=&\bar{u}_{s'}(p') \bigg[i\sigma^{[\mu\{\nu]}\bar{p}^{\rho\}}  A_{T}^q(t)+\frac{2\bar{p}^{[\mu} q^{\{\nu]}\bar{p}^{\rho\}}}{m_N^2}\tilde{A}_{T}^q(t)\\
    &+\frac{\gamma^{[\mu} q^{\{\nu]}\bar{p}^{\rho\}}}{m_N} B_{T}^q(t)
    +\frac{2\gamma^{[\mu} \bar{p}^{\{\nu]}q^{\rho\}}}{m_N}\tilde{B}_{T}^q(t)\bigg] u_s(p)
\end{aligned}
\end{equation}
where the last operator symmetrization follows 

$$
\sigma^{[\mu\{\nu]}\overleftrightarrow{D}^{\rho\}}=\frac{1}{2}\left(
\sigma^{\mu\nu} \overleftrightarrow{D}^{\rho}
- \sigma^{\rho[\mu} \overleftrightarrow{D}^{\nu]}
+ \frac{1}{2}\, g^{\rho[\mu}\, \sigma^{\nu]\lambda} \overleftrightarrow{D}_{\lambda}
\right)
$$

Similarly, the corresponding off-forward hadronic matrix elements for gluons are given by corresponding gluonic operators $F^{\{\mu\alpha}F_{\alpha}{}^{\nu\}}$ and $F^{\{\mu\alpha}\tilde{F}_{\alpha}{}^{\nu\}}$ for gravitational current and axial gravitational current respectively with the same Lorentz structure. For the tensor gravitational current, the gluonic form factors are differently defined by

\begin{equation}
\begin{aligned}
    &\langle N'|F^{\{\mu\{\alpha}F^{\beta\}\nu\}}|N\rangle=\bar{u}_{s'}(p') \bigg[-i\sigma^{\mu\nu}\bar{p}^\rho  A_{T}^q(t)-\frac{\bar{p}^\mu q^\nu-\bar{p}^\nu q^\mu}{m_N^2}\bar{p}^{\rho}\tilde{A}_{T}^q(t)\\
    &-\frac{\gamma^\mu q^\nu-\gamma^\nu q^\mu}{2m_N}\bar{p}^\rho B_{T}^q(t)
    -\frac{\gamma^\mu \bar{p}^\nu-\gamma^\nu \bar{p}^\mu}{m_N}q^{\rho}\tilde{B}_{T}^q(t)\bigg]\frac{q^{\beta\}}\bar{p}^\nu - \bar{p}^{\beta\}}q^\nu }{2m_N} u_s(p)
\end{aligned}
\end{equation}

For chiral-even distributions, the second moments of $H$ and $E$ are related to the gravitational form factors $A(t)$, $B(t)$, and $D(t)$, which parameterize the nucleon matrix element of the symmetric, conserved EMT. Similarly, the second moments of $\widetilde H$ and $\widetilde E$ define the axial gravitational form factors, associated with the matrix elements of the axial-vector twist-2 operators with one derivative. These relations follow from Lorentz covariance and the polynomiality property of GPDs.

\begin{align}
\int_{-1}^1 dx\, x\, H^{q,g/N}(x,\xi,t) &= A_N^{q,g}(t) + \xi^2 D^{q,g}(t), \\
\int_{-1}^1 dx\, x\, E^{q,g/N}(x,\xi,t) &= B_N^{q,g}(t) - \xi^2 D^{q,g}(t), \\
\int_{-1}^1 dx\, x\, \widetilde{H}^{q,g/N}(x,\xi,t) &= \widetilde{A}_N^{q,g}(t), \\
\int_{-1}^1 dx\, x\, \widetilde{E}^{q,g/N}(x,\xi,t) &= \widetilde{B}_N^{q,g}(t).
\end{align}

For chiral-odd distributions, the second Mellin moments give rise to tensor gravitational form factors, corresponding to the nucleon matrix elements of twist-2 transversity operators with one covariant derivative, and probing spin-2 exchange in the $t$-channel.
\begin{equation}
\begin{aligned}
\int_{-1}^1 dx\, x\, H_T^{q/N}(x,\xi,t) &= A_T^q(t), \qquad
\int_{-1}^1 dx\, x\, E_T^{q/N}(x,\xi,t) = B_T^q(t), \\
\int_{-1}^1 dx\, x\, \widetilde{H}_T^{q/N}(x,\xi,t) &= \widetilde{A}_T^q(t), \qquad
\int_{-1}^1 dx\, x\, \widetilde{E}_T^{q/N}(x,\xi,t) = -2\xi\, \widetilde{B}_T^q(t).
\end{aligned}
\end{equation}

\subsection{Polynomiality}

The general polynomiality relation for the Mellin moments can be expressed as

\begin{align}
\int_{-1}^{1} dx\, x^{n-1} H(x,\xi,t)
&= \sum_{\substack{k=0 \\ k\ \mathrm{even}}}^{n-1}
(-2\xi)^k\, A_{n,k}(t)
+ \delta_{n\,\mathrm{even}}\,(-2\xi)^n C_n(t), \\
\int_{-1}^{1} dx\, x^{n-1} E(x,\xi,t)
&= \sum_{\substack{k=0 \\ k\ \mathrm{even}}}^{n-1}
(-2\xi)^k\, B_{n,k}(t)
- \delta_{n\,\mathrm{even}}\,(-2\xi)^n C_n(t).
\end{align}
where $C_2(t)=D(t)/4$.

\begin{align}
\int_{-1}^{1} dx\, x^{n-1}\, \widetilde{H}(x,\xi,t)
&= \sum_{\substack{k=0 \\ k\ \mathrm{even}}}^{n-1} (-2\xi)^k\, \widetilde{A}_{n,k}(t), \\
\int_{-1}^{1} dx\, x^{n-1}\, \widetilde{E}(x,\xi,t)
&= \sum_{\substack{k=0 \\ k\ \mathrm{even}}}^{n-1} (-2\xi)^k\, \widetilde{B}_{n,k}(t).
\end{align}

\begin{align}
\int_{-1}^{1} dx\, x^{n-1} H_T(x,\xi,t)
&= \sum_{\substack{k=0 \\ k\ \mathrm{even}}}^{n-1}
(-2\xi)^k\, A_{T\,n,k}(t), \\
\int_{-1}^{1} dx\, x^{n-1} E_T(x,\xi,t)
&= \sum_{\substack{k=0 \\ k\ \mathrm{even}}}^{n-1}
(-2\xi)^k\, B_{T\,n,k}(t), \\
\int_{-1}^{1} dx\, x^{n-1} \widetilde{H}_T(x,\xi,t)
&= \sum_{\substack{k=0 \\ k\ \mathrm{even}}}^{n-1}
(-2\xi)^k\, \widetilde{A}_{T\,n,k}(t), \\
\int_{-1}^{1} dx\, x^{n-1} \widetilde{E}_T(x,\xi,t)
&= \sum_{\substack{k=1 \\ k\ \mathrm{odd}}}^{n-1}
(-2\xi)^k\, \widetilde{B}_{T\,n,k}(t).
\end{align}

\chapter{TMD factorization}
\label{ch:tmd}
Fast moving hadrons carry an increasing number of sub-constituents quarks and gluons in QCD, deemed partons. The parton distribution functions (PDFs) capture their longitudinal momentum distributions, the simplest of all partonic distributions in a hadron~\cite{Blumlein:2012bf}. PDFs play a central role in the description of inclusive and semi-inclusive
processes in high energy scattering, thanks to factorization. Transverse momentum distributions (TMDs)
allows for a spatial description of the partons in a fast moving hadron, by recording both the longitudinal momentum and transverse location of a given parton~\cite{Boussarie:2023izj}.

Transverse momentum parton distribution functions (TMDPDFs)  play a central role in the analyses of a wide range of high energy data both at electron and hadron facilities worldwide. Their understanding is part of a large experimental effort at COMPASS at CERN, JLAB in
Virginia, LHC at CERN and the future EIC at BNL. Currently, TMDPDFs are empirically extracted from 
Drell-Yan processes and semi-inclusive deep inelastic scattering (SIDIS), with small final hadron momentum transfer~\cite{Rogers:2015sqa}. 

TMDPDFs are defined as correlation functions of bilocal quarks stapled with light-like Wilson loops, which are inherently non-perturbative. Their unnderstanding from first principles has been challenging. Recent lattice developments using the large momentum effective theory (LaMET) have proven useful for their possible Euclidean extraction and matching using quasi-TMDs, where the gluonic staple (soft function) can be related to a mesonic form factor~\cite{LatticeParton:2020uhz}.

The purpose of this chapter is to develop an understanding of the TMDPDFs for the pion and kaon at low resolution, using the ILM.  

\section{Semi-inclusive hadronic scattering}

After factoring out the parton distribution and fragmentation functions, the remaining contribution to the cross section corresponds to the short-distance hard scattering of partons. This hard part is perturbatively calculable within the framework of OPE. In this section, we focus on processes involving two hadrons, which serve as typical examples where such factorization theorems are expected to hold.

\begin{itemize}
    \item Drell-Yan process
\begin{equation}
    h_A(p)+h_B(p')\rightarrow l+l'+X
\end{equation}
    \item Semi-inclusive deep inelastic scattering (SIDIS)
\begin{equation}
    h_A(p)+l\rightarrow h_B(p_h)+l'+X
\end{equation}
    \item lepton pair annihilation for hadron pair production ($e^+e^-$)
\begin{equation}
    l+l'\rightarrow h_A(p_h)+h_B(p_h')+X
\end{equation}\\[1pt]
\end{itemize}

In the leptonic part, $l$ and $l'$ denotes the momenta of the leptons involved. In the hadronic part, $p$ and $p'$ denotes the momenta for the initial hadrons and $p_h$ and $p_h'$ denotes for the final hadrons. With this in mind, the differential cross section for each process can be defined as

\begin{align}
\label{cross_section}
    d\sigma_{hh'\rightarrow ll'X}=&\frac{1}{2s}\left(\frac{d^3\vec{l}}{(2\pi)^32E_l}\right)\left(\frac{d^3\vec{l}\,'}{(2\pi)^32E_l'}\right)\frac{e^4}{q^4}(2\pi)^4L_{\mu\nu} W^{\mu\nu}_{hh'\rightarrow\gamma^*X} \nonumber\\
    d\sigma_{lh\rightarrow l'h'X}=&\frac{1}{2s}\left(\frac{d^3\vec{l}\,'}{(2\pi)^32E_l'}\right)\left(\frac{d^3\vec{p}_h}{(2\pi)^32E_h}\right)\frac{e^4}{q^4}(2\pi)^4L_{\mu\nu} W^{\mu\nu}_{\gamma^*h\rightarrow h'X} \nonumber\\
    d\sigma_{ll'\rightarrow  hh'X}=&\frac{1}{2s}\left(\frac{d^3\vec{p}_h}{(2\pi)^32E_h}\right)\left(\frac{d^3\vec{p}\,'_h}{(2\pi)^32E_h'}\right)\frac{e^4}{q^4}(2\pi)^4L_{\mu\nu} W^{\mu\nu}_{\gamma^*\rightarrow hh'X}
\end{align}
where the hadronic tensors with electromagnetic current $J^\mu$ are defined as

\begin{align}
\label{cross_sectionW}
    W^{\mu\nu}_{hh'\rightarrow\gamma^*X}(p,p',q)=& \frac1{(2\pi)^4}\int d^4xe^{iq\cdot x}\langle hh'|J^\mu(x)J^\nu(0)|hh'\rangle \nonumber\\
    W^{\mu\nu}_{\gamma^*h\rightarrow h'X}(p,p_h,q)=&\frac1{(2\pi)^4}\sum_X\int d\Pi_X\int d^4xe^{iq\cdot x}\langle h|J^\mu(x)|h'X\rangle\langle h'X|J^\nu(0)|h\rangle \nonumber\\
    W^{\mu\nu}_{\gamma^*\rightarrow hh'X}(p_h,p'_h,q)=&\frac1{(2\pi)^4}\sum_X\int d\Pi_X\int d^4xe^{iq\cdot x}\langle0|J^\mu(x)|hh'X\rangle\langle hh'X|J^\nu(0)|0\rangle 
\end{align}

The hadronic part is described by the hadronic tensors which consist of structure functions as a function of dimensionless kinematics in each type of processes (details will be shown later). The definition of the cross sections includes the exclusive final state phase space integral, one-photon exchange characterized by $e^4/q^4$, leptonic tensor $L_{\mu\nu}$, and hadronic part with inclusive states summed. The leptonic tensor is universally given by 
\begin{equation}
\label{lep}
L^{\mu\nu}=2\left(l^\mu l'^\nu+l'^\mu l^\nu-l\cdot l'g^{\mu\nu} -2s i\epsilon^{\mu\nu\alpha\beta}q_\alpha s_{l\beta}\right)
\end{equation}
where $s_l$ is the spin polarization vector normalized to $s^2_l=-m_l^2$ in the lepton system. $s$ corresponds to the leptonic helicity.
For unpolarized scattering, the leptonic tensor is averaged over the lepton spin. $$L^{\mu\nu}\rightarrow\frac12\sum_sL^{\mu\nu}.$$ The definition in \eqref{cross_section} can be extended to include the weak interaction by introducing the quark weak current and weak charges. 

\section{TMD factorization}

When the transverse momentum is sufficiently large ($q_{\perp}^2 \gtrsim Q^2$), the cross section can be systematically described within collinear factorization, where all intrinsic transverse momenta are power suppressed and absorbed into universal parton distribution and fragmentation functions. If the transverse momentum is integrated over, the same collinear factorization framework remains valid.
However, in the regime of small transverse momentum ($q_{\perp}^2 \ll Q^2$), the separation of scales becomes nontrivial: the partonic cross section develops logarithmic enhancements of the form $\ln(Q^2/q_{\perp}^2)$, and the transverse motion of partons can no longer be neglected. In this limit, the cross section admits an approximate description in terms of TMD factorization, where the dynamics is encoded in TMD parton distributions and fragmentation functions supplemented by soft factors, and the large logarithms are resummed through evolution equations such as the Collins--Soper (CS) and RG equations. In this sense, the partonic cross section in TMD factorization only depends on $Q^2$ since the initial and the final phase space are both factorized into the TMD functions.

\subsection{Drell-Yan}

In Drell-Yan, two hadrons collide in the center of mass frame along the $z$ axis with momentum $p$ and $p'$. The kinematics includes the center of mass energy of the two hadrons $s=(p+p')^2$ and the invariant mass of the leptonic pair (virtuality of the photon) $q^2=(l+l')^2=Q^2$. The common kinematics in Drell-Yan $x'$ and $x_b'$ can be written in the form of Bjorken variables.

\begin{align}
     x_b=&\frac{Q^2}{2p\cdot q} & 
     x_b'=&\frac{Q^2}{2p'\cdot q}
\end{align}

By changing of the variables, the exclusive final lepton-pair phase space can be expressed by the 4-momenta of photon $q$, which can be further defined by kinematic $(x_b,~x_b',~d^2q_\perp)$. Thus, the cross section in \eqref{cross_section} can be written as
\begin{equation}
    d\sigma_{hh'\rightarrow ll'X}=\frac{4\pi \alpha_e^2}{3Q^2}dx_bdx_b'd^2q_\perp\left(-g_{\mu\nu}\right)W^{\mu\nu}_{hh'\rightarrow\gamma^*X}
\end{equation}

In the parton model, one introduces the parton variables $\xi$ and $\xi'$ defined similarly to the hadronic variables $x_b$ and $x_b'$ by defining the four-momenta of the incident partons. The ratio thus defines $x=x_b/\xi$ and $x'=x_b'/\xi'$ are the usual light-cone momentum fractions.

\begin{figure}
    \centering
    \includegraphics[width=.85\linewidth]{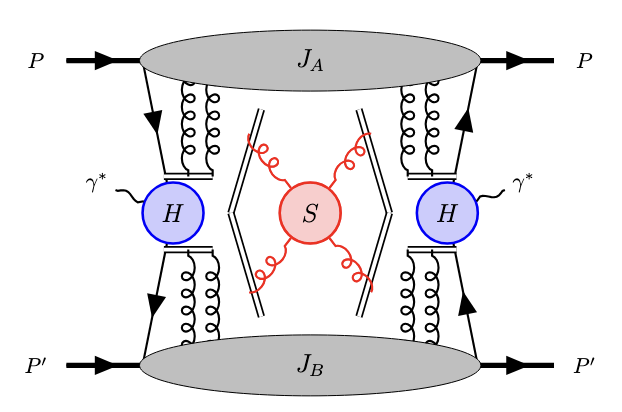}
    \caption{Feynman diagram of TMD factorization for Drell-Yan with hard processes \( H \), soft interaction \( S \), and two incoming hadrons \( A \) and \( B \) }
    \label{fig:DY}
\end{figure}

As illustrated in Fig.~\ref{fig:DY}, the Drell-Yan differential cross section within this regime can be factorized as

\begin{equation}
\begin{aligned}
    \frac{d\sigma_{hh'\rightarrow ll'X}}{dx_bdx_b'd^2q_\perp}=&\hat{\sigma}_{q\bar q\rightarrow ll'}\int \frac{d^2b_\perp}{(2\pi)^2} e^{-iq_\perp\cdot b_\perp}\tilde{f}^{q/h}(x_b,b_\perp)\tilde{f}^{\bar{q}/h'}(x_b',b_\perp)S^{--}(b_\perp)\\
    &+\mathcal{O}(q_\perp/Q)
\end{aligned}
\end{equation}
where $\tilde{f}^{q/h}(x,b_\perp)$ is the TMD parton distribution function in transverse coordinate space defined by

\begin{equation}
\label{eq:TMDPDF}
    \tilde{f}^{q/h}(x,b_\perp)=\int d^2k_\perp e^{ik_\perp\cdot b_\perp}f^{q/h}(x,k_\perp)
\end{equation}
and $S^{(--)}$ is the soft factor defined in \eqref{tmd_soft-1}.

\subsection{SIDIS}

In SIDIS $l+h \rightarrow l'+ h'+ X$, the photon produced by the lepton  collides with the hadron along the $z$ axis with momenta $q$ and $p$ respectively. The virtuality of the photon is described by the negative squared four-momentum transfer $Q^2$. The usual SIDIS kinematics are given by

\begin{align}
    x_b=&\frac{Q^2}{2p\cdot q} & y=&\frac{p\cdot q}{p\cdot l} & z_h=&\frac{p\cdot p_h}{p\cdot q}
\end{align}
Here $x_b$ is the Bjorken variable, which is crucial to the Bjorken scaling. $y$ denotes the fractional energy loss of the lepton in the target frame. This is the kinematics describes the final state lepton phase space.  $z_h$ represents the fraction of the energy of the virtual photon carried by the detected hadron. This helps to understand the hadronization process and how energy is distributed among the final-state hadrons. By an appropriate change of variables, the exclusive phase space can be parametrized by the kinematic set $(x_b, y, z_h, p_{h\perp})$, where $p_{h\perp}$ is defined relative to the photon direction, and the two transverse momenta are related by $q_\perp = p_{h\perp}/z_h$.
Thus, the cross section in \eqref{cross_section} can be written as
\begin{equation}
\begin{aligned}
&d\sigma_{lh\rightarrow l'h'X}=dx_bdydz_hd^2p_{h\perp}\frac{\pi y\alpha_e^2}{2z_hQ^4}L_{\mu\nu} W^{\mu\nu}_{\gamma^*q\rightarrow q'X}  
\end{aligned}
\end{equation}

\begin{figure}
    \centering
    \includegraphics[width=.85\linewidth]{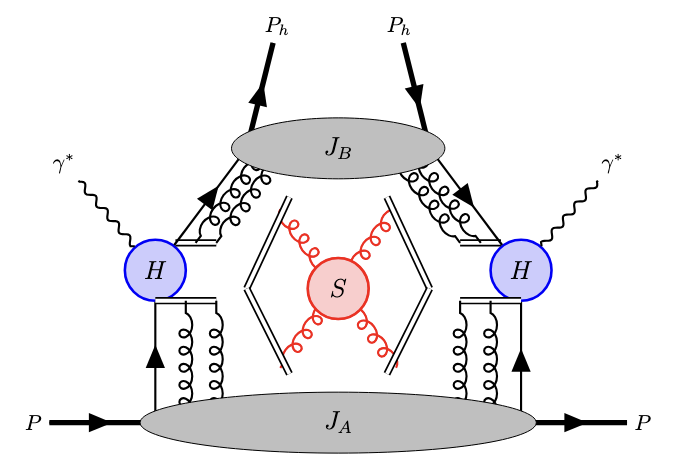}
    \caption{Feynman diagram of TMD factorization for SIDIS with hard processes \( H \), soft interaction \( S \), incoming hadron \( A \) and outgoing hadron \( B \) }
    \label{fig:SIDIS}
\end{figure}

As illustrated in Fig.~\ref{fig:SIDIS}, the same factorization can be applied when the transverse momentum is small ($p^2_{h\perp}\ll Q^2$) \cite{Ji:2004wu,Tangerman:1995hw},

\begin{equation}
\begin{aligned}
    \frac{d\sigma_{lh\rightarrow l'h'X}}{dx_bdydz_hd^2p_{h\perp}}=&\frac{1}{z^2_h}\frac{d\hat{\sigma}_{lq\rightarrow l'q'}}{dy}\int \frac{d^2b_\perp}{(2\pi)^2} e^{-i\frac{p_{h\perp}\cdot b_\perp}{z_h}}\tilde{f}^{q/h_A}(x_b,b_\perp)S^{-+}(b_\perp)\tilde{D}^{h_B/q'}(z_h,b_\perp)\\
    &+\mathcal{O}\left(\frac{p_{h\perp}}{Q}\right)
\end{aligned}
\end{equation}
where $\tilde{f}^{q/h}(x,b_\perp)$ is the TMD parton distribution function in transverse coordinate space defined in \eqref{eq:TMDPDF} and $\tilde{D}^{h'/q}(z,b_\perp)$ is the TMD fragmentation function in transverse coordinate space defined by

\begin{equation}
    \tilde{D}^{h/q}(z,b_\perp)=\int d^2p_\perp e^{ip_\perp\cdot b_\perp/z}D^{h/q}(z,p_\perp)
\end{equation}
and the soft factor  $S^{(-+)}$ is defined in \eqref{tmd_soft-2}. 

\subsection{$e^+e^-$ annihilation}

In electron-positron annihilation $e^+e^- \rightarrow h+ h'+ X$, the photon produced by the lepton pair (electron-positron pair) along the $z$ axis with momenta $q$ and $p$ respectively. The virtuality of the photon is described by the positive center of mass energy $Q^2=s$. The commonly used kinematics in electron-positron are given by

\begin{align}
    z_h=&\frac{2p_h\cdot q}{q^2} & z_h'=&\frac{2p_h'\cdot q}{q^2}
\end{align}
that describes the final state hadron phase space by representing the fraction of the energy of the virtual photon carried by the detected hadron. This helps to understand the hadronization process and how energy is distributed among the final-state hadrons. For the interest of the theories, we choose $z$-axis along the lepton-pair in the photon rest frame for convenience. By changing of the variables, the exclusive final state phase space can be expressed by the kinematics $(z_h,~p_{h\perp},~z'_h,~p'_{h\perp})$ where $p_{h\perp}$ and $p'_{h\perp}$ is the momenta of produced hadrons perpendicular to the lepton axis. Thus, the cross section in \eqref{cross_section} can be written as

\begin{equation}
    \frac{d\sigma_{ll'\rightarrow hh'X}}{dz_hdz'_hd^2p_{h\perp} d^2p'_{h\perp}}=\frac{1}{z_hz_h'}\frac{\alpha_e^2}{2Q^6}L_{\mu\nu} W^{\mu\nu}_{\gamma^*\rightarrow hh'X}
\end{equation}

In the parton model, one introduces the parton variables $\zeta$ and $\zeta'$ defined parton analog of the hadronic variables $z_h$ and $z_h'$ by defining the momenta of the fragmenting partons. The ratio thus defines $z=z_h/\zeta$ and $z'=z_h'/\zeta'$, the light-cone momentum fractions of final hadrons $h$ and $h'$.

\begin{figure}
    \centering
    \includegraphics[width=.85\linewidth]{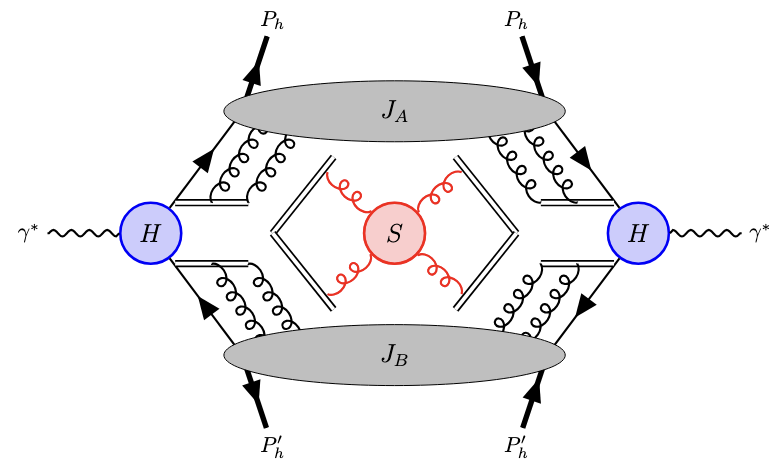}
    \caption{Feynman diagram of TMD factorization for $e^+e^-$ production with hard processes \( H \), soft interaction \( S \), and two outgoing hadron states \( A \) and \( B \) }
    \label{fig:ee}
\end{figure}

As illustrated in Fig.~\ref{fig:ee}, the differential cross section within the regime, where transverse momentum is small ($p^{\prime2}_{h\perp} \ll Q^2$), can be factorized as
 
\begin{equation}
\begin{aligned}
&\frac{d\sigma_{ll'\rightarrow hh'X}}{dz_h\,dz_h'\,d^2p_{h\perp}\, d^2p'_{h\perp}}
=
\frac{Q^2}{4z_h^2 z_h'^2}
\frac{d\hat{\sigma}_{ll'\rightarrow q\bar{q}}}{d\Omega}
\\
&\times\int \frac{d^2 b_\perp}{(2\pi)^2}\,
e^{-i\left(\frac{p_{h\perp}}{z_h}+\frac{p'_{h\perp}}{z_h'}\right)\cdot b_\perp}
S^{++}(b_\perp)\,
\tilde{D}^{h/q}(z_h,b_\perp)\,
\tilde{D}^{h'/\bar{q}}(z_h',b_\perp)+\mathcal{O}\left(\frac{q_{\perp}}{Q}\right)
\end{aligned}
\end{equation}
where $S^{(++)}$ is the soft factor defined in \eqref{tmd_soft-3}.

\section{TMD parton distribution functions}
The definitions of various quark TMD parton distributions read

\begin{align}
\label{tmd}
    &q_h(x,k_\perp)=\int\frac{dz^-d^2b_\perp}{(2\pi)^3}e^{ix p^+z^--ik_\perp\cdot b_\perp}\langle h|\bar{\psi}(0) \gamma^+W^{(\pm)}[0,z]\psi(z)|h\rangle \Big|_{z^+=0,\, z_\perp=b_\perp} \nonumber\\
    &\Delta q_h(x,k_\perp)=\int\frac{dz^-d^2b_\perp}{(2\pi)^3}e^{ix p^+z^--ik_\perp\cdot b_\perp}\langle h|\bar{\psi}(0) \gamma^+\gamma^5W^{(\pm)}[0,z]\psi(z)|h\rangle\Big|_{z^+=0,\, z_\perp=b_\perp} \nonumber\\
    &\delta q_h(x,k_\perp)=\int\frac{dz^-d^2b_\perp}{(2\pi)^3}e^{ix p^+z^--ik_\perp\cdot b_\perp}\langle h|\bar{\psi}(0)i\sigma^{\perp+}\gamma^5 W^{(\pm)}[0,z]\psi(z)|h\rangle\Big|_{z^+=0,\, z_\perp=b_\perp}
\end{align}

The Wilson line in TMD functions in \eqref{tmd} is different in SIDIS process (space-like) and in Drell-Yan process (time-like) \cite{Kumano:2020ijt}. Resummations of the SIDIS processes with intermediate collinear gluons from the final state result in the Wilson line defined as $W^{(+)}$ while from the initial state collinear gluon resummations of the Drell-Yan processes lead to the Wilson line defined as $W^{(-)}$ \cite{Angeles-Martinez:2015sea,GrossePerdekamp:2015xdx,Aidala:2012mv,Barone:2010zz,DAlesio:2007bjf}. 

Generally, a straight Wilson line connecting two spacetime points $x$ and $y$ is defined by
\begin{equation}
W[x;y]=\mathcal{P}\exp\left[ig\int_{x}^{y} dz_\mu A^\mu(z)\right],
\end{equation}

For convenience, one can also define a half-infinite Wilson line with the initial point $x$ extending along a direction $v_\mu$, which is given by
\begin{equation}
W_v(x)=W[v\infty + x; x ]=\mathcal{P} \exp \left[ig \int_0^\infty dz\, v_\mu A^\mu(x+vz)\right],
\end{equation}
and a half-infinite Wilson line pointing in the opposite direction is given by 
\begin{equation}
W_{-v}(x)=W[-v\infty + x; x ]=\mathcal{P} \exp \left[ig \int_0^{-\infty} dz\, v_\mu A^\mu(x+vz)\right].
\end{equation}
where $W_v(x)$ denotes the resummation of the eikonalized gluons emitted by the final quark state moving in the direction $v_\mu$ while $W^\dagger_{-v}(x)$ represents the resummation of the eikonalized gluons emitted by the initial quark state moving in the direction $v_\mu$.

With these definitions, the Wilson line structure entering the TMD correlator is
\begin{equation}
\label{TMD_wilson}
W^{(\pm)}[0,0_\perp; z^-,b_\perp]
=
W^\dagger_{\pm n}(0)\,
W[\pm n\infty ; \pm n\infty+b_\perp]\,
W_{\pm n}(nz^-+b_\perp).
\end{equation}
This object has a staple-like geometry, consisting of two segments along the light-cone direction $n^\mu$ and one transverse segment. It systematically resums soft gluon contributions and encodes the process-dependent gauge links characteristic of TMD factorization~\cite{Collins:2011zzd}.

\begin{figure}
    \centering
\subfloat[\label{fig:wilson_1}]{\includegraphics[height=2cm,width=0.5\linewidth]{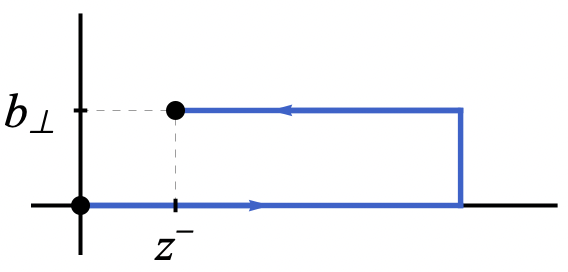}}
\subfloat[\label{fig:wilson_2}]{\includegraphics[height=2cm,width=0.5\linewidth]{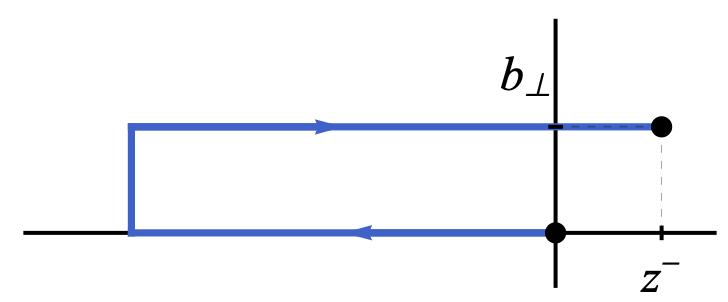}} 
    \caption{Wilson lines in Eq.~\eqref{tmd} for (a) SIDIS process with the space-like correlation function and (b) Drell-Yan process with the timelike correlation.}
\end{figure}


\section{Wilson loops and  soft functions}
\label{SECII}
Wilson lines are important ingredients in QCD factorization, as they enforce gauge invariance of bi-local operators, in the twist expansion of pertinent correlators on the light cone~\cite{Collins:2011zzd,Balitsky:2001gj}. 
They resum the gluon emission between the bi-local sources, and 
also compose the building parts of the above mentioned Wilson loops
in Minkowski signature.  Wilson loops  made of Wilson lines capture important aspects of non-perturbative physics in QCD. In Euclidean signature, they are central for our understanding of IR physics in QCD, such as confinement in the vacuum~\cite{Diakonov:2009jq,Simonov:1996ati}, deconfinement in matter~\cite{DeMartini:2021xkg,Zhitnitsky:2006sr}, and the spin forces between heavy quarks ~\cite{Shuryak:2021hng,Shuryak:2021fsu}. A full formulation of gauge theories in loop space
was also pursued in~\cite{Makeenko:1980vm,Makeenko:1979pb,Polyakov:1987ez}.
In Minkowski signature, Wilson loops can be used to define the forward scattering amplitudes of selected hadronic processes in the eikonal approximation~\cite{Korchemskaya:1994qp}, and appear in 
a number of inclusive and semi-inclusive processes in high energy scattering using factorization~\cite{Cherednikov:2020mtu,Cherednikov:2017qbt}. 

In general,  a Wilson loop along a smooth and simple contour accumulate divergences (self-energies) that can be eliminated consistently by multiplicative renormalization of the strong coupling $g$ and gauge field $A_\mu$. However, for the contour including cusps where the derivatives of points are not smooth, the Wilson loop yields new cusp divergences~\cite{Polyakov:1980ca,Dotsenko:1979wb,Brandt:1982gz,Korchemskaya:1992je}. These divergences are sensitive to the cusp angle
 (rapidity angle). A standard procedure for the removal of these divergences has been discussed in~\cite{Brandt:1981kf}. These geometrical singularities, along with their corresponding anomalous dimensions, are important in defining the asymptotic states of partonic hard processes \cite{Korchemsky:1993hr}. The multiplicative renormalizability to any loop integral with a finite number of cusps has been proven in~\cite{Brandt:1981kf,Korchemskaya:1992je}.

\begin{figure}
    \centering
    \subfloat[\label{wl_DY}]{\includegraphics[width=1\linewidth]{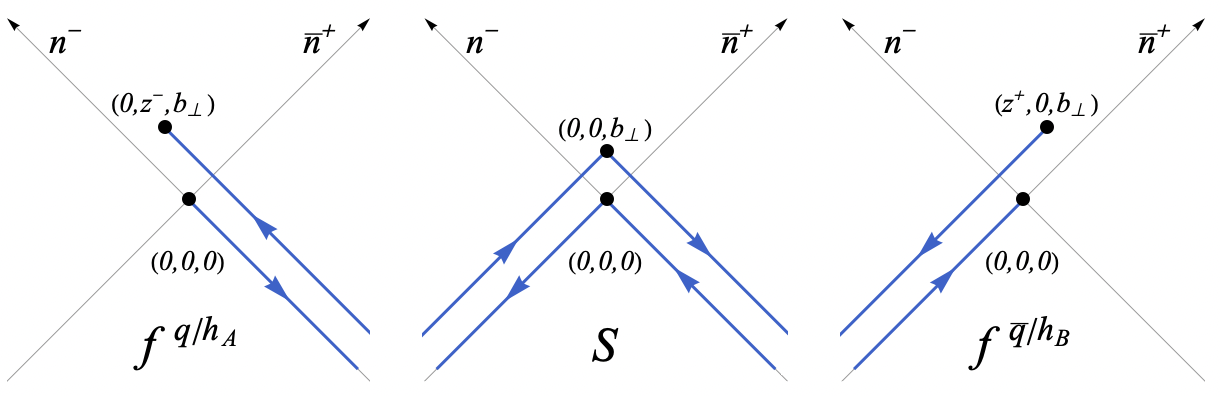}}
    \hfill
    \subfloat[\label{wl_ee}]{\includegraphics[width=1\linewidth]{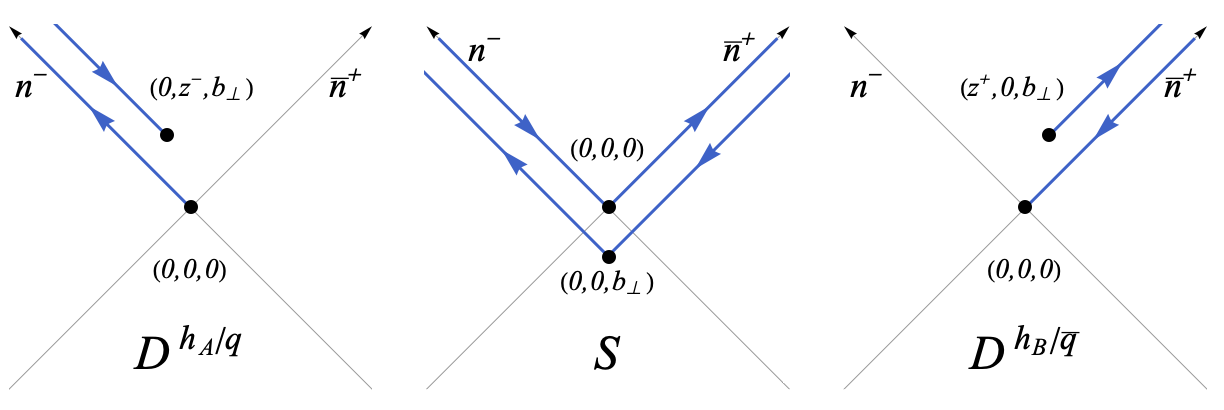}}
     \hfill
    \subfloat[\label{wl_DIS}]{\includegraphics[width=1\linewidth]{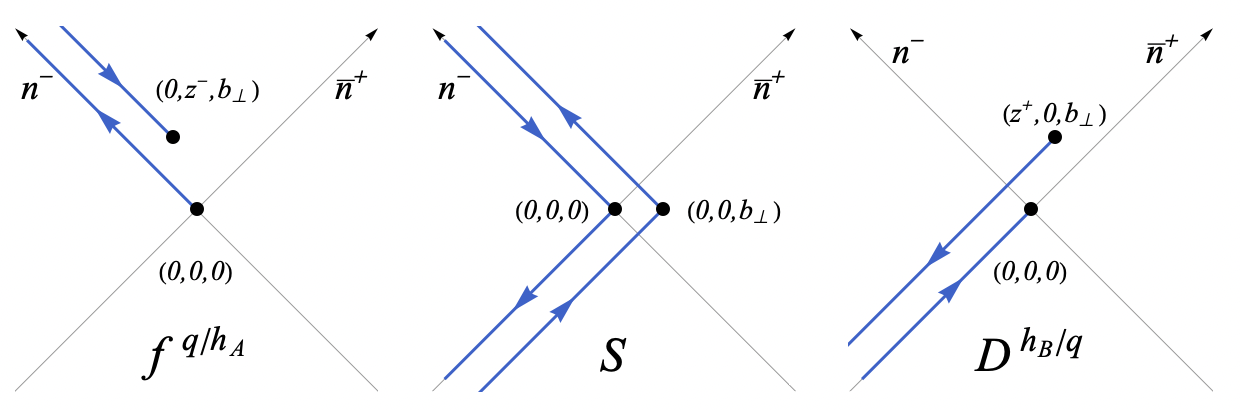}}
    \caption{The contours of the Wilson lines for the soft function of the cross section in (a) Drell-Yan,  (b) electron-positron annihilation,
    and (c) SIDIS. Hadron $A$ is moving near the light-cone direction $n$,
    and hadron $B$ is moving near the  light-cone direction $\bar{n}$~\cite{Vladimirov:2014hla}.}
    \label{fig:wl}
\end{figure}


The Wilson loop for soft functions are process dependent \cite{Collins:2011zzd,Ji:2004wu}. Resummation of the intermediate soft gluons from the initial state hadrons leads to the Wilson lines defined on the past light-cone and final state resummation of the soft gluon gives the Wilson lines defined on the future light-cone.

The Wilson loop defining the soft function for the Drell-Yan process is illustrated in Fig.~\ref{wl_DY}. The corresponding soft-function for the Drell-Yan kinematics 
with support on the past light-cone is
\begin{equation}
\begin{aligned}
\label{tmd_soft-1}
&S^{(--)}(b_\perp)=\frac1{N_c}\mathrm{Tr}\left\langle W_{-\bar n}(0)W^\dagger_{-n}(0)W_{-n}(b_\perp)W^\dagger_{-\bar n}(b_\perp)\right\rangle\,,
\end{aligned}
\end{equation}
where $\bar{n}$ and $n$ denotes the two light-cone directions of the moving hadrons involving in the process, given by the unit vectors $n=(1,0,0_\perp)$ and $\bar n=(0,1,0_\perp)$ in light-cone signature. The time ordering is not needed, since the initial points of the Wilson lines are space-like separated. 

The Wilson loop defining the soft function for the $e^+e^-$ annihilation  process is illustrated in Fig.~\ref{wl_ee}. The soft factor in the annihilation process has support in the forward light cone, with 
\begin{equation}
\begin{aligned}
\label{tmd_soft-3}
    &S^{(++)}(b_\perp)=\frac1{N_c}\mathrm{Tr}\langle W_{\bar n}(0)W^\dagger_{n}(0)W_{n}(b_\perp)W^\dagger_{\bar n}(b_\perp)\rangle\,,
\end{aligned}
\end{equation}
which again does not need time ordering. 

The Wilson loop defining the soft function for the SIDIS annihilation  process is illustrated in Fig.~\ref{wl_DIS}. The soft factor is composed of Wilson lines on the past-light-cone and future light-cone,
\begin{equation}
\begin{aligned}
\label{tmd_soft-2}
    &S^{(-+)}(b_\perp)=\frac1{N_c}\mathrm{Tr}\left\langle{\cal P}\bigg( W_{-\bar n}(0)W^\dagger_{n}(0)W_{n}(b_\perp)W^\dagger_{-\bar n}(b_\perp)\bigg)\right\rangle
\end{aligned}
\end{equation}
In this case, not all distances are space-like, so the  time-ordering cannot be eliminated.
The Wilson line along $\bar n$ resums the soft gluons emitted from the initial hadron moving in $\bar n$, and the Wilson line along $n$ resums the gluons from the final hadron.
Note that all the soft functions depend only on the transverse coordinate $b_\perp$.

Generally the soft functions defined in \eqref{tmd_soft-1}, \eqref{tmd_soft-3}, and \eqref{tmd_soft-2} have light-cone divergences that can be regulated by tilting $n$ and $\bar n$ slightly off-lightcone~\cite{Collins:2011zzd}. Thus the soft functions form a Wilson loop with two cusp angles on each side, with the rapidity angle defined as $$\cosh\chi=n\cdot\bar n\,.$$
With this in mind, the renormalized soft function can be parameterized as~\cite{Vladimirov:2017ksc}
\begin{equation}
\label{eq:sot_func}
S(\mu,b_\perp,\chi)=\exp\left[K(\mu,b_\perp,\chi)+P(\mu,b_\perp)\right]\,,
\end{equation}
where $K(\mu,b_\perp,\chi)$ is the cusp-angle kernel and $P(\mu,b_\perp)$ is the scheme-dependent non-cusp term that solely depends on $b_\perp$. In a scheme-independent low energy model such as the ILM, we can consider $P(\mu,b_\perp)=0$ for simplicity. Such consideration does not affect the CS kernel extraction since only the rapidity dependence is required to determine the CS kernel. 

As we noted earlier, the soft function includes UV divergences that cannot be eliminated by multiplicative renormalization of the strong coupling and gauge field. Around the cusps where the derivatives of points are not smooth, the light-cone divergences further introduce a dependence on the rapidity angle $ \chi $.

With this in mind, the RG equation of the soft function, with a Minkowski cusp angle $\chi$, reads
\begin{equation}
    \frac{dK(\mu,b_\perp,\chi)}{d \ln\mu^2}= -\gamma_{K}(\chi,\alpha_s)\,,
\end{equation}
where
\begin{equation}
    \frac{d}{d\ln\mu}=\left(\mu\frac{\partial}{\partial \mu}+\beta(g)\frac{\partial}{\partial g}\right)\,.
\end{equation}
The angle dependence can be factorized out under different limits~\cite{Grozin:2015kna,Korchemsky:1987wg}.
When the angle approaches zero, we have
\begin{equation}
\label{KBX}
    K(\mu,b_\perp,\chi)\xrightarrow{\chi\rightarrow0}K_{B}(\mu,b_\perp)\chi^2\,.
\end{equation}
The corresponding anomalous dimension is called bremsstrahlung function~\cite{Korchemsky:1991zp,Mitev:2015oty}
\begin{equation}
\label{KCSX}
    \gamma_{K}(\chi,\alpha_s) \xrightarrow{\chi\rightarrow0} B(\alpha_s)\chi^2\,.
\end{equation}
On the other hand, when the angle approaches infinity, we have
\begin{equation}
    K(\mu,b_\perp,\chi)\xrightarrow{\chi\rightarrow\infty}K_{\rm CS}(\mu,b_\perp)\chi\,,
\end{equation} 
where $K_{\rm CS}$ is the CS kernel~\cite{Collins:1981uk,Collins:1981va}, or rapidity anomalous dimension (RAD), defining the scaling properties of
transverse momentum dependent distributions. It is a nonperturbative function, which represents the interaction of light-quarks in the
QCD vacuum~\cite{Vladimirov:2020umg}.

Finally, the cusp anomalous dimension in the light-cone limit is defined as
\begin{equation}
    \gamma_{K}(\chi,\alpha_s) \xrightarrow{\chi\rightarrow\infty} \Gamma_{\rm cusp}(\alpha_s)\chi\,.
\end{equation}
The perturbative contributions to the light-cone cusp anomalous dimension are given in \cite{Liu:2024sqj,Liu:2025mbl}.

\begin{figure}
    \centering
    \includegraphics[width=0.6\linewidth]{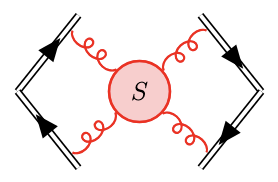}
    \caption{Soft function}
    \label{fig:soft_wilson}
\end{figure}

\subsection{Soft function in ILM}
\label{SECIII}

In the ILM, the soft function associated to Fig.~\ref{fig:soft_wilson} 
is composed of four contributions,

\begin{equation}
\begin{aligned}
\label{soft_wilson}
        &\phi^a(z,b_\perp,\rho)=\bar{\eta}^a_{\mu\nu}v_\mu z_\nu \int_0^\infty ds \varphi_v(s,z,\rho)+\bar{\eta}^a_{\mu\nu}\bar{v}_\mu z_\nu\int_{-\infty}^0 ds \varphi_{\bar v}(s,z,\rho)\\
        &-\bar{\eta}^a_{\mu\nu}v_\mu (z-b_\perp)_\nu \int_0^\infty ds \varphi_v(s,z-b_\perp,\rho)-\bar{\eta}^a_{\mu\nu}\bar{v}_\mu (z-b_\perp)_\nu \int_{-\infty}^0 ds \varphi_{\bar v}(s,z-b_\perp,\rho)\,,
\end{aligned}
\end{equation}
where the instanton profiling function along the Wilson line is defined as
\begin{equation}
\begin{aligned}
    \varphi_{v}(s,z,\rho)=\frac{\rho^2}{(s^2-2v\cdot zs+z^2)(s^2-2v\cdot zs+z^2+\rho^2)}\,.
\end{aligned}
\end{equation}

As we noted for the generic Wilson loop \eqref{loop_inst}, there is no need for the path ordering. 
Each term in \eqref{soft_wilson} represents a half-infinite segment in the Wilson loop in Fig.~\ref{fig:soft_wilson}.
In terms of Eq.~(\ref{soft_wilson}), the  Wilson loop related to the soft function can be written as
\begin{equation}
\begin{aligned}\
\label{eq:W}
    &W(\rho,b_\perp,\theta)=\exp\left[\frac{2n_{I+A}}{N_c}\int d^4z \left(\cos\phi(z,\rho,b_\perp,\theta)-1\right)\right]\,,
\end{aligned}
\end{equation}
where the Euclidean angle is identified as $v\cdot \bar{v}=\cos\theta$, which can be analytically continued to the Minkowskian rapidity through
$\theta\rightarrow i\chi$. 
The norm  $\phi=\sqrt{(\phi^a)^2}$ follows from Eq.~\eqref{soft_wilson}. 

When $\theta=0$, only the Wilson line self-energy and the Coulomb-type interaction between the parallel Wilson lines contribute. Following translational symmetry in  Euclidean space, this integral will reduce from the four dimensional $z$-integral  to three dimensional integral (one longitudinal and two transverse directions) with linear dependence on $z_4$,
\begin{equation}
\begin{aligned}
\ln W(\rho,b_\perp,0)=&-\int dz_4 V_{\rm Coul}(b_\perp/\rho)\,,
\end{aligned}
\end{equation}
where the Coulomb potential is defined as

\begin{equation}
\label{V_inst}
V_{\rm Coul}(b_\perp/\rho)=\frac{2n_{I+A}}{N_c}\int dz_3d^2z_\perp \left[1-\cos\phi(z,\rho,b_\perp,0)\right]\,.
\end{equation}

This linear $z_4$ divergence reflects on the self-energy in the Wilson line. After subtracting the linear divergent and angle-independent term from the exponent in \eqref{eq:W}, the remaining contribution comes from the cusp-angle,

\begin{equation}
\begin{aligned}
\label{K}
K(\rho,b_\perp,\theta)=\ln \frac{W(b_\perp/\rho,\theta)}{W(b_\perp/\rho,0)}=&\frac{2n_{I+A}}{N_c}\int d^4z \left[\cos\phi(z,\rho,b_\perp,\theta)-\cos\phi(z,\rho,b_\perp,0)\right]\,.
\end{aligned}
\end{equation}

The subtraction ensures $K=0$ for $\theta\rightarrow0$ as we only need the angle dependence to extract the CS kernel.


For simplicity, we adopt the symmetric parameterization for the angle
\begin{equation}
v_\mu=\left(\cos\frac{\theta}{2},\sin\frac{\theta}{2},0_\perp\right),\quad
\bar v_\mu=\left(\cos\frac{\theta}{2},-\sin\frac{\theta}{2},0_\perp\right)
\end{equation}
The zero angle direction is thus along the $z_4$-axis, i.e. $v\cdot z|_{\theta=0}=\bar v\cdot z|_{\theta=0}=z_4$.  


\begin{figure}
    \centering
\subfloat[\label{fig:KCSX_a}]{\includegraphics[width=.46\linewidth]{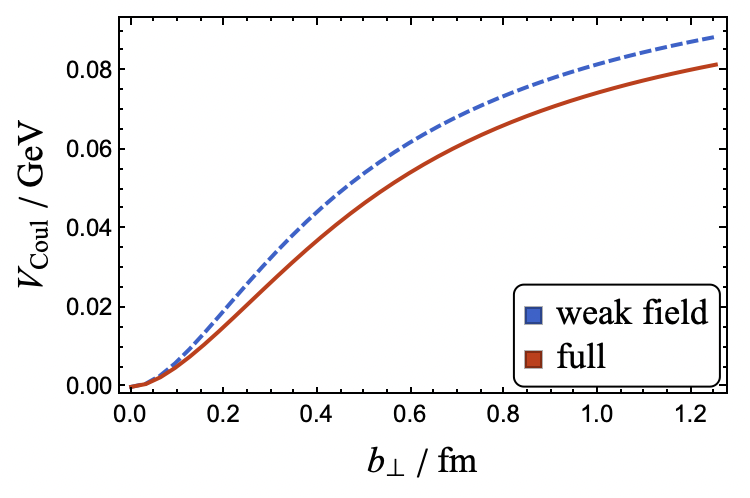}}
\hfill
\subfloat[\label{fig:KCSX_b}]{\includegraphics[width=.5\linewidth]{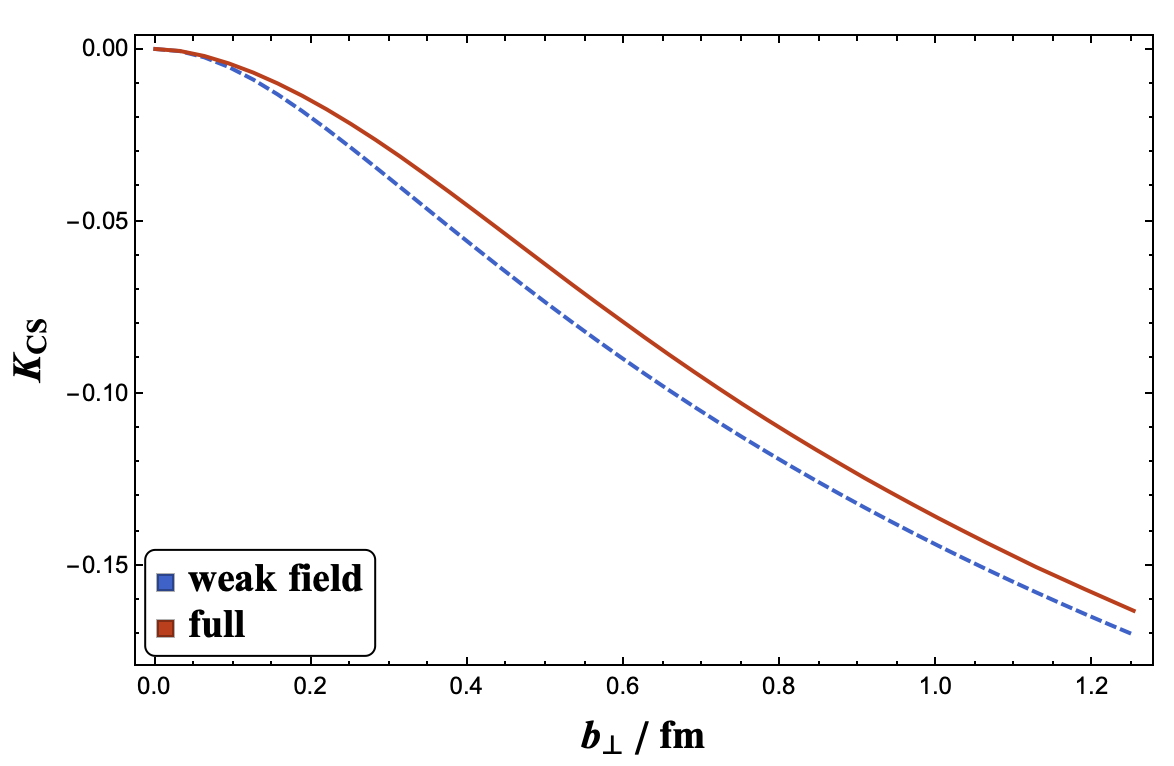}}
    \caption{Instanton liquid estimation on (a) Coulomb potential (b) CS kernel with the full calculation \eqref{K} and weak field approximation \eqref{CS_weak}.}
    \label{fig:KCSX}
\end{figure}

\subsubsection{Weak field}
A full analytical analysis of the results is too involved and beyond the scope of this work. Below, we will rely instead on a numerical evaluation. However, in the weak field limit, the analytical form can be worked out for an initial estimate. Indeed,  in the weak field limit ($|\phi|\ll1$), we have

\begin{equation}
    \ln W(\rho,b_\perp,\theta)\simeq\ln W^{(1)}(\rho,b_\perp,\theta)\equiv-\frac{n_{I+A}}{N_c}\int d^4z \phi^2(z,\rho,b_\perp,\theta)\,,
\end{equation}

In the zero angle limit, the two parallel Wilson lines give rise to the Coulomb potential,
\begin{equation}
\begin{aligned}
\label{V_weak}
    &V^{(1)}_{\rm Coul}=\frac{4n_{I+A}\pi^2\rho^4}{N_c}\int_0^\infty dk\mathcal{F}^2_g(\rho k)\left(1-\frac{\sin kb_\perp}{kb_\perp}\right)\,,
\end{aligned}
\end{equation}
with the UV divergence cutoff by the instanton size $\rho$. 
In Fig.~\ref{fig:KCSX_a} we show the result of the  Coulomb self-energy for the weak field limit \eqref{V_weak} 
in dashed-blue line, and the full result in solid-green line. The
weak field approximation is in good agreement with the full result over a large range of separations $b_\perp$.  The accumulated self-energy at large separation is about $80\,\rm MeV$ in agreement with earlier estimates~\cite{Shuryak:2021fsu}.

In the weak field limit, the analytical result exhibit a universal angular dependence. 
\begin{equation}
\label{csX}
    K^{(1)}(\rho,b_\perp,\theta)=K_{\rm CS}^{(1)}(b_\perp/\rho)h(\theta)\,,
\end{equation}
with the cusp factor
$$h(\theta)=\theta \cot\theta-1$$
and the CS kernel in weak field 
\begin{equation}
\begin{aligned}
\label{CS_weak}
    &K_{\rm CS}^{(1)}=\frac{4n_{I+A}\pi^2\rho^4}{N_c}\int_0^\infty \frac{dk}k\mathcal{F}^2_g(\rho k) \left(J_0(kb_\perp)-1\right)\,.
\end{aligned}
\end{equation}
This separation is exact in the weak field limit. In the large $b_\perp$ limit, the asymptotic form of \eqref{CS_weak} in weak field reduces to a logarithmic function in $b_\perp$,
\begin{equation}
    K^{(1)}_{\rm CS}\xrightarrow{b_\perp\rightarrow\infty} -\frac{4n_{I+A}\pi^2\rho^4}{N_c}\left[\ln (b_\perp/\rho) -0.072\right]
\end{equation}

\begin{figure}
    \centering
    \includegraphics[width=\linewidth]{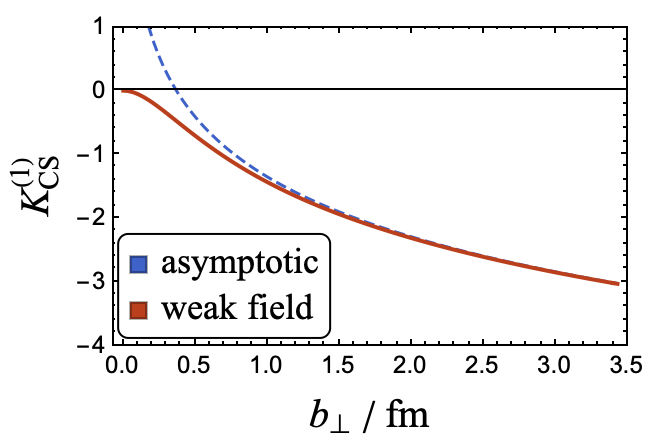}
    \caption{Asymptotic logarithmic curve compared to CS kernel in weak field approximation}
    \label{fig:asym}
\end{figure}

In Fig.~\ref{fig:asym}, we show the asymptotic logarithmic behavior of the CS kernel is expected to be observed for $b_\perp\gtrsim1.5$ fm. 

\subsubsection{Full result}

\begin{figure}
    \centering
\includegraphics[width=.7\linewidth]{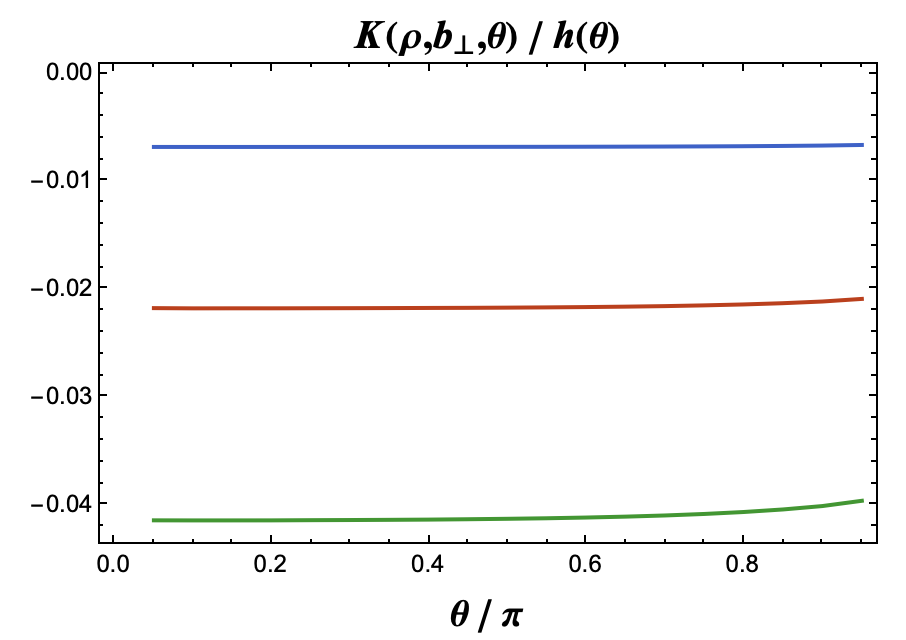}
    \caption{The angular dependence of $K$ in \eqref{K}. }
    \label{fig:theta_b}
\end{figure}

In the strong field limit, the evaluation of the soft function (\ref{eq:W}) requires full consideration of the instanton ensemble. This can only be carried out numerically. In Fig.~\ref{fig:KCSX_b} we show the ILM results for the CS kernel at low resolution, from
the full or strong field limit \eqref{K} red-solid curve, in comparison to the weak field limit \eqref{CS_weak}, versus the separation $b_\perp$. 

The numerical dependence on the cusp factor for the full result follows closely  $h(\theta)$, as illustrated in~Fig.~\ref{fig:theta_b}.  We  conclude that the weak field separation (\ref{csX}) holds in full result as well, 
\begin{equation}
\label{K_2}
K(\rho,b_\perp,\theta)\simeq K_{\rm CS}(b_\perp/\rho)h(\theta)\,.
\end{equation}

For future reference, we list the numerical values for the full $K_{\rm CS}$ \eqref{K} in Table~\ref{tab:CS_numbers}.

\begin{table}
\centering
\begin{minipage}{0.3\linewidth}
\centering
\begin{tabular}{@{}cc@{}}
\toprule
$b_\perp$ (GeV$^{-1}$) & $K_{\rm CS}$ \\ \midrule
0.000  &  0.000 \\
0.174  & -0.006 \\
0.348  & -0.022 \\
0.522  & -0.045 \\
0.695  & -0.074 \\
0.869  & -0.108 \\
1.043  & -0.146 \\
1.217  & -0.188 \\
1.565  & -0.284 \\
1.912  & -0.390 \\
2.260  & -0.503 \\
\bottomrule
\end{tabular}
\end{minipage}
\hfill
\begin{minipage}{0.3\linewidth}
\centering
\begin{tabular}{@{}cc@{}}
\toprule
$b_\perp$ (GeV$^{-1}$) & $K_{\rm CS}$ \\ \midrule
2.608  & -0.618 \\
2.955  & -0.733 \\
3.303  & -0.846 \\
3.651  & -0.954 \\
3.998  & -1.058 \\
4.346  & -1.157 \\
4.694  & -1.251 \\
5.042  & -1.340 \\
5.389  & -1.424 \\
5.737  & -1.504 \\
6.085  & -1.580 \\
\bottomrule
\end{tabular}
\end{minipage}
\hfill
\begin{minipage}{0.3\linewidth}
\centering
\begin{tabular}{@{}cc@{}}
\toprule
$b_\perp$ (GeV$^{-1}$) & $K_{\rm CS}$ \\ \midrule
6.432  & -1.653 \\
6.780  & -1.722 \\
7.128  & -1.787 \\
7.475  & -1.850 \\
7.823  & -1.910 \\
8.171  & -1.969 \\
8.519  & -2.025 \\
8.866  & -2.078 \\
9.214  & -2.129 \\
9.562  & -2.179 \\
9.909  & -2.226 \\
\bottomrule
\end{tabular}
\end{minipage}

\caption{Table of the full CS kernel $K_{\rm CS}(b_\perp)$ defined in \eqref{K} and \eqref{K_2} as a function of $b_\perp$ with $n_{I+A}=7.459$ fm$^{-4}$ at renormalization scale $\mu=2$ GeV.}
\label{tab:CS_numbers}
\end{table}

The ILM result \eqref{K} can be analytically continued to Minkowski signature by $\theta\rightarrow i\chi$, 
\begin{equation}
    v\cdot \bar v=\cos\theta\rightarrow\cosh\chi\simeq1+\frac{Q^2}{2m_h^2}\,,
\end{equation}
with the universal cusp factor
$h(\chi)\simeq\ln Q^2/m_h^2$.
Here $m_h$ is the hadron mass, and $Q$ is the momentum transfer in the hadron, assuming  $v_\mu=p_\mu /m_h$ for simplicity. 
A comparison of \eqref{K} to (\ref{KBX}) shows that in the ILM
\begin{equation}
    K_B\simeq-\frac13 K_{\rm CS}\,.
\end{equation}

The factorization of angular dependence in \eqref{K_2} over the full range of $\theta$ ensures that analytic continuation of $\theta$ does not alter the $b_\perp$-dependence of the CS kernel. 
This property is generally absent in perturbative calculations, leading to different cusp anomalous dimension beyond one-loop in Euclidean and Minkowskian spaces \cite{Ji:2019sxk}. Our prediction suggests that in lattice QCD, the Euclidean soft function at large \(b_\perp\) can be computed with real $\theta$, and the CS kernel can be extracted by factoring out $h(\theta)$.

Furthermore, understanding the asymptotic form at large $b_\perp$ is of importance for the phenomenological extractions, as various groups employ different models for $b_\perp$. From Fig.~\ref{fig:KCSX_b} we observe that the full CS kernel agrees with the weak field limit both for $b_\perp/\rho\ll 1$ (kinematics) and $b_\perp/\rho\gg 1$ (asymptotics). The deviation from the weak field limit is only about $10\%$ for $b_\perp/\rho\sim 1$. 
Thus, this implies the large-$b_\perp$ asymptotic behavior of the full CS kernel is also logarithmic with the form given by
\begin{equation}
    K_{\rm CS}\xrightarrow{b_\perp\rightarrow\infty}-\frac{4n_{I+A}\pi^2\rho^4}{N_c}\left[\ln (b_\perp/\rho) -0.114\right],
\end{equation}
in agreement with an early phenomenological paramterization used in~\cite{Collins:1985xx}. Thus, our ILM calculations provide strong constraints on the models for phenomenological fitting.

\begin{figure}
    \centering
\subfloat[\label{CS1}]{\includegraphics[width=.8\linewidth]{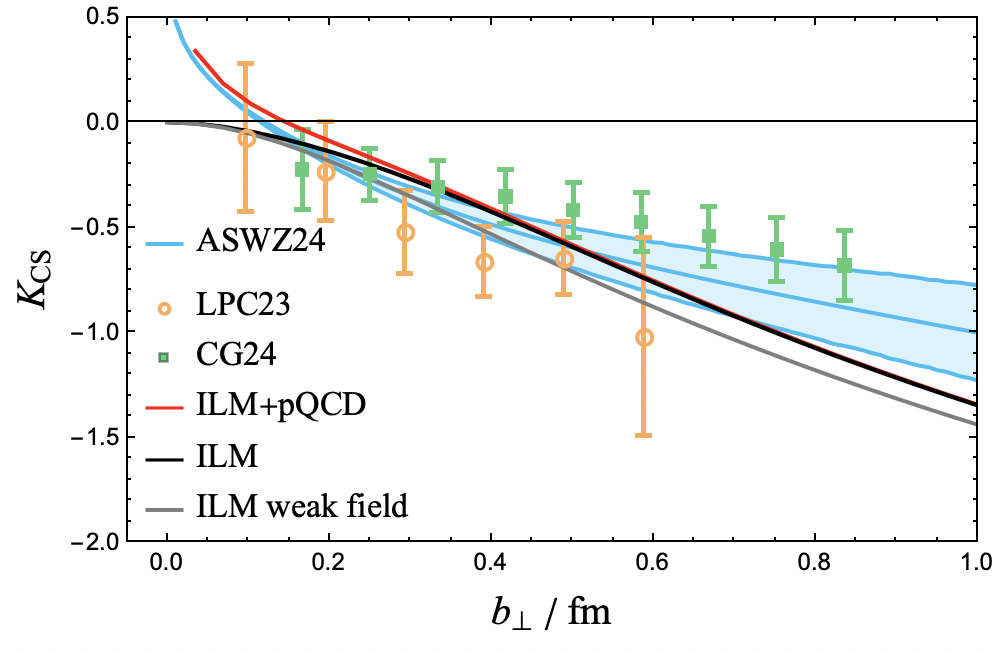}}
\hfill
\subfloat[\label{CS2}]{\includegraphics[width=.8\linewidth]{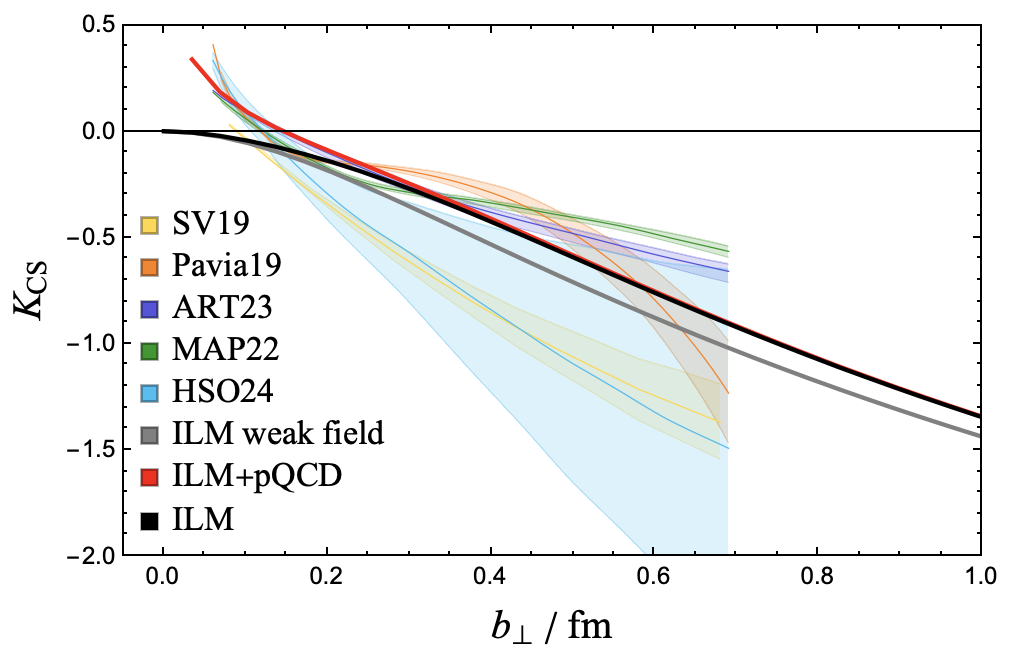}}
    \caption{(a) The black-solid curve is the  ILM result for the full CS kernel ($\mu=2$ GeV) with $\rho=0.343$ fm, $n_{I+A}=7.46$ fm$^{-4}$, compared to the weak field approximation in \eqref{CS_weak} gray-solid curve. The optimized between large distance and short distance in (\ref{KINTER}-\ref{b_star}) are shown in red solid curve. The short-distance CS kernel is modified by perturbative contribution. The results are compared to the recent lattice calculation by ASWZ24~\cite{Avkhadiev:2023poz},  a new lattice calculation using Coulomb-gauge TMD correlators~\cite{Zhao:2023ptv} by CG24~\cite{Bollweg:2024zet} and results from Lattice Parton Collaboration LPC23~\cite{LatticePartonLPC:2023pdv}. (b) shows the comparison of our results with the phenomenologically extracted CS kernel  by SV19~\cite{Scimemi:2019cmh}, Pavia19~\cite{Bacchetta:2019sam}, MAP22~\cite{Bacchetta:2022awv}, ART23~\cite{Moos:2023yfa}, and HSO24~\cite{Gonzalez-Hernandez:2022ifv}.}
    \label{fig:CS}
\end{figure}

\subsection{Collins-Soper kernel and RG evolution}
\label{SECIV}
In the perturbative calculation, the Wilson lines in TMD functions introduces rapidity divergence which requires different regularization other than the UV divergences~\cite{Collins:1981uk,Collins:2011zzd,Becher:2010tm,Echevarria:2011epo,Chiu:2012ir,Collins:2008ht,Collins:1992tv,Chiu:2011qc,Manohar:2006nz,Li:2016axz,Ebert:2018gsn}. As a regularization, the path of Wilson lines can be taken off the lightcone~\cite{Collins:2011zzd}, specified by the variable $\zeta$ related to finite rapidity. 
The variable $\zeta$
is the energy of the hadron \cite{Boer:2014tka,Idilbi:2004vb}, 
\begin{equation}
    \zeta^2=m_h^2x^2e^{2(y-y_v)}
    =\frac{(2xp\cdot v)^2}{v^2}=2(xp^+)^2\frac{v^-}{v^+}\,,
\end{equation}
where $y=\frac12\ln \frac{p^+}{p^-}$ is the hadron rapidity with mass $m_h$, and $y_v=\frac12\ln \frac{v^+}{v^-}$  the off lightcone rapidity velocity $v$. In general, $v$ can be different from the momenta of both involving hadrons $h$ and $h'$ in the TMD scattering process. For the simplest case, we can assume $v_\mu=p_\mu/m_h$ ($y_v=y$), $v_\mu=p'_\mu/m_{h'}$ ($y_v=y'$), or  which simply assume $\zeta^2=Q^2$.

The variation in $\zeta$ is determined by the CS kernel $K_{\rm CS}$ \cite{Collins:1981uk,Collins:1981va}. The integrability of the TMD renormalization group (RG) equations guarantees that
\begin{equation}
\label{RG_K}
    \frac{dK_{\rm CS}(b_\perp,\mu) }{d\ln \mu^2}=-\Gamma_{\mathrm{cusp}}(\alpha_s(\mu))\,.
\end{equation}

The ILM emerges from the QCD vacuum under gradient-flow cooling~\cite{Leinweber:1999cw,Athenodorou:2018jwu}. As discussed in Sec.~\ref{sec:vac_GF} and Ch.~\ref{ch:had-to-par}, the CS kernel is likewise sensitive to the resolution scale of the vacuum and therefore requires RG evolution across different scales.

With this in mind, we evaluate \eqref{K} using the larger instanton density 
to account for the non-perturbative part of the CS kernel at high resolution and  large separation $b_\perp$. This is in line with the
'dense' ILM suggested in~\cite{Shuryak:2021fsu}.

At small separations $b_\perp\ll b_{\rm max}$, most of the instanton annihilates with anti-instanton, and perturbative effects are dominant, hence the interpolating solution for the evolution equation \eqref{RG_K} is \cite{Collins:2011zzd} 
\begin{equation}
\begin{aligned}
\label{KINTER}
K_{\rm CS}&=K^{(\rm np)}_{\rm CS}(b_\perp,\mu) + K^{(\rm pert)}_{\rm CS}(b_*, \mu_b) -2\int_{\mu_b}^\mu \frac{d\mu'}{\mu'}\Gamma_{\rm cusp}(\alpha_s(\mu'))\,.
\end{aligned}
\end{equation}
The non-perturbative part $K^{(\rm np)}_{\rm CS}$ is fixed from the ILM evolved to higher resolution,
\begin{equation}
    K^{(\rm np)}_{\rm CS}(b_\perp,\mu)=K^{(\rm inst)}_{\rm CS}(b_\perp/\rho,\mu) - K^{(\rm inst)}_{\rm CS}(b_*/\rho,\mu)\,.
\end{equation}
The perturbative contribution  $K^{(\rm pert)}_{\rm CS}(b_*, \mu_b)$ (integration constant obtained by solving the RG equation of $K_{\rm CS}$) and cusp anomalous dimension $\Gamma_{\rm cusp}$are given in \cite{Liu:2024sqj,Liu:2025mbl}. The transverse scale is parameterized by
\begin{equation}
    \mu_b=\frac{2e^{-\gamma_E}}{b^*(b_\perp)}\xrightarrow{b_\perp\rightarrow\infty}\frac{2e^{-\gamma_E}}{b_{\rm max}}
\end{equation}
with the asymptotic saturation following from the conventional $b^*$-parameterization~\cite{Collins:2011zzd},
\begin{equation}
\label{b_star}
    b^*(b_\perp)=\frac{b_\perp}{\sqrt{1+b_\perp^2/b_{\rm max}^2}}\,,
\end{equation}
which reduces to $b_\perp$ in the limit $b_\perp\to0$.
Here $\gamma_E$ is  Euler constant, and the optimal non-perturbative distance is  $b_{\rm max}=0.56$ fm, which is chosen to optimize the interpolation between non-perturbative and perturbative contributions in the lattice results of ASWZ24~\cite{Avkhadiev:2023poz}. 
The lattice results are extracted from the quasi-TMD and matched to the $\overline{\rm MS}$ scheme. Thus our result of the CS kernel depends on the model parameterization \eqref{b_star} and renormalization scheme. Since $b_\perp$ near zero is not available in lattice, the parameterization in \eqref{b_star} is still applicable.

In Fig.~\ref{fig:CS} we show our results for the CS kernel.
The solid-black curve is the full non-perturbative contribution to
the CS from the ILM for $\rho=0.343$ fm and  $n_{I+A}=7.46$ fm$^{-4}$,
compared to the gray-solid curve in the weak-field limit, as a function of $b_\perp$. The deviation between the strong and weak field limits is small. 
The amended result for small $b_\perp$ using the interpolating solution 
(\ref{KINTER}-\ref{b_star})  for the CS kernel with $b_{\rm max}$ parameter, is shown as the solid-red curve. In Fig.~\ref{CS1} our results are compared 
to the recent lattice calculation by ASWZ24~\cite{Avkhadiev:2023poz}, a new lattice calculation using Coulomb-gauge TMD correlators~\cite{Zhao:2023ptv} by CG24~\cite{Bollweg:2024zet} and the results from the Lattice Parton Collaboration LPC23~\cite{LatticePartonLPC:2023pdv}. The ILM are in overall agreement with the lattice results, especially with those reported 
by ASWZ24~\cite{Avkhadiev:2023poz}. In Fig.~\ref{CS2}
 our results are compared with the  phenomenologically  extracted CS kernel by SV19~\cite{Scimemi:2019cmh}, Pavia19~\cite{Bacchetta:2019sam}, MAP22~\cite{Bacchetta:2022awv}, ART23~\cite{Moos:2023yfa}, and HSO24~\cite{Gonzalez-Hernandez:2022ifv}. At short distances, the phenomenological fittings converge and align with the predictions of the ILM, while at large distances the latter remains within the substantial error margins among different phenomenological fits.


\section{Pion and Kaon TMDs}

The computation of TMD quark distribution in pions can be simply addressed by the light front wave functions (LFWFs). The contribution of the staple-shaped Wilson line $W^{(\pm)}$ is addressed in Eq.~\eqref{TMD_wilson}. With this in mind, the pion TMD reads

\begin{equation}
    q_\pi(x,k_\perp)=\frac{1}{(2\pi)^3}\sum_{s_1,s_2}\Phi^\dagger_\pi(x,k_\perp,s_1,s_2)\Phi_\pi(x,k_\perp,s_1,s_2)
\end{equation}

\begin{equation}
    \Delta q_\pi(x,k_\perp)=\frac{1}{(2\pi)^3}\sum_{s_1,s_1',s_2}\Phi^\dagger_\pi(x,k_\perp,s_1,s_2)\sigma^3_{s_1s_1'}\Phi_\pi(x,k_\perp,s_1',s_2)
\end{equation}
and
\begin{equation}
    \delta q_{\pi}(x,k_\perp)=\frac{1}{(2\pi)^3}\sum_{s_1,s_1',s_2}\Phi^\dagger_\pi(x,k_\perp,s_1,s_2)\sigma^{\perp}_{s_1s_1'}\Phi_\pi(x,k_\perp,s_1',s_2)
\end{equation}
where $\sigma^{3,\perp}_{ss'}$ are the Pauli matrices and the pseudoscalar meson wave functions are defined in Sec.~\ref{sec:LFWF}.
with $u_{s_1}(k_1)$ denotes the light front quark spinor with spin $s_1$ carrying internal momentum $k^+_1=xp^+$, $k_{1\perp}=k_\perp$ and $v_{s_2}(k_2)$ denotes the light front anti-quark spinor with spin $s_2$ carrying internal momentum $k^+_2=\bar{x}p^+$, $k_{2\perp}=-k_\perp$. The instanton size induces the non-local form factor $\mathcal{F}(k)$ profiling the wave function with $\rho=0.313$ fm. The pion $\lambda_\pi$ is determined to make sure the wave function is properly normalized with given pion wave function renormalization $Z_\pi$. 
Also see \cite{Liu:2023yuj,Liu:2023fpj}. With given $m_\pi=139$ MeV, $M=398.2$ MeV, and $f_\pi=94$ MeV, the value of $\lambda_\pi$ is chosen to be $ \lambda_\pi=2.550$

After a few algebraic steps, we have unpolarized TMD with only constituent quark contribution considered

\begin{equation}
\begin{aligned}
\label{eq:pion_tmd}
    &f^{q/\pi}_{1}(x,k_\perp)=\frac{N_cZ_{\pi}}{(2\pi)^3}\frac{2(k_\perp^2+M^2)}{\left(x\bar{x}m^2_\pi-k^2_\perp-M^2\right)^2}\mathcal{F}^2\left(\frac{k_\perp}{\lambda_\pi\sqrt{x\bar x}}\right)
\end{aligned}
\end{equation}
and with the RPA on Wilson line, Boer-Mulders function is zero~\cite{Lorce:2016ugb}.
\begin{equation}
    h^{\perp,q/\pi}_{1}(x,k_\perp)=0
\end{equation}

For the Boer-Mulders function, the gauge link must be taken into consideration~\cite{Noguera:2015iia}. Dynamically, $T$-odd PDFs emerge from the gauge link structure of the multiparton quark and/or gluon
correlation functions which describe initial and final-state interactions of the active parton via
soft gluon exchanges with the target remnant \cite{Kou:2023ady}. For simplicity, we will only focus on $f^{q/\pi}_1$ TMD PDF.

Similar calculation can also apply to kaon. The kaon wave functions in Sec~\ref{sec:mes_3}
with quark-$1$ assigned to $u$ quark with constituent mass $M_u$ and quark-$2$ assigned to $s$ quark with constituent mass $M_s$. 
where the kaon $\lambda_K$ is determined to make sure the kaon wave function is properly normalized by the kaon wave function normalization $Z_K$. With given $m_K=458$ MeV, $M_u=394.4$ MeV, $M_s=556.5$ MeV, and $f_K=112$ MeV, the value is chosen to be $\lambda_K=3.069$

After a few algebraic steps, we can also have unpolarized TMD

\begin{equation}
\begin{aligned}
\label{eq:kaon_tmd}
    f^{u/K^+}_{1}(x,k_\perp)=\frac{N_cZ_{K}}{(2\pi)^3}\frac{2(k^2_\perp+\bar{x}^2M_u^2+x^2M_s^2+2x\bar xM_uM_s)}{(x\bar{x}m_{K}^2-k_\perp^2-\bar xM_u^2- xM_s^2)^2}\mathcal{F}^2\left(\frac{k_\perp}{\lambda_K\sqrt{x\bar x}}\right)
\end{aligned}
\end{equation}

\begin{figure}
\centering
\subfloat[\label{fig:pion_contour}]{\includegraphics[width=0.45\linewidth]{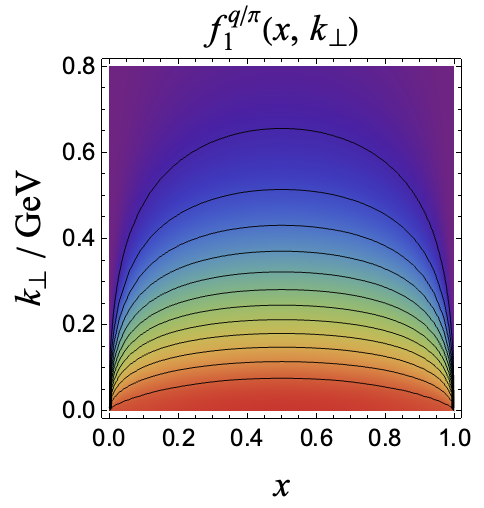}}
\hfill
\subfloat[\label{fig:pion_contourk}]{\includegraphics[width=0.45\linewidth]{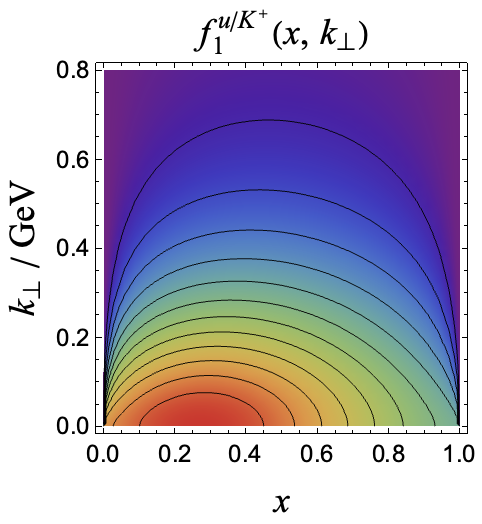}}
\hfill
\subfloat[\label{fig:pion_3d}]{\includegraphics[width=0.45\linewidth]{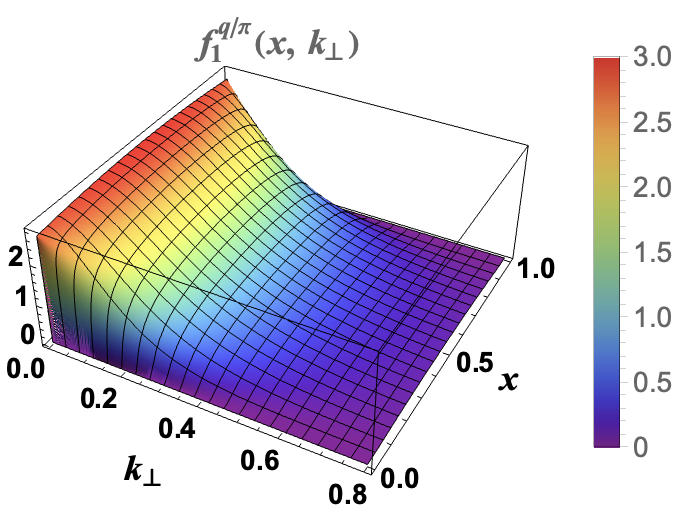}}
\hfill
\subfloat[\label{fig:pion_3dk}]{\includegraphics[width=0.45\linewidth]{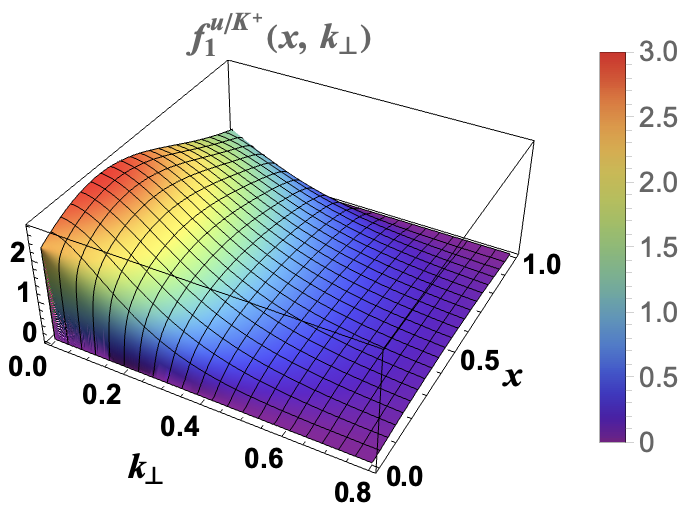}}
\caption{Pion (left) \eqref{eq:pion_tmd}  and kaon (right) 
\eqref{eq:kaon_tmd} TMDPDFs at low resolution $\mu=1/\rho$
with $\rho=0.313$ fm: (a,b) are the density plots, (c,d) the 3D plots,
(e,f) the transverse momentum dependent plots for fixed $x$, and (g,h) the longitudinal momentum dependence for fixed $k_\perp$. The pion parameters are
$\sqrt{N_cZ_\pi}=7.240$, $m_\pi=139.0$ MeV, $M=398.17$ MeV.
The kaon parameters are $\sqrt{N_cZ_K}=6.60$, $m_K=458.0$ MeV, $M_u=394.4$ MeV, $M_s=556.5$ MeV.}
\label{fig:pion_TMD}
\end{figure}

\begin{figure}
\centering
\subfloat[\label{fig:pion_k}]{\includegraphics[width=0.45\linewidth]{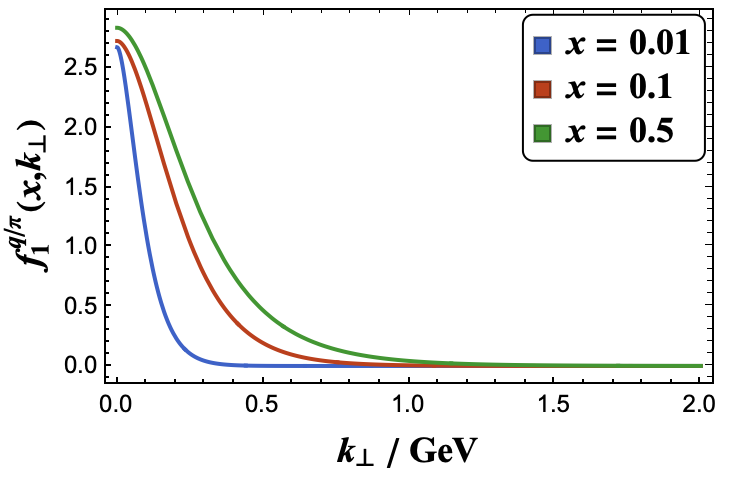}}
\hfill
\subfloat[\label{fig:pion_kk}]{\includegraphics[width=0.45\linewidth]{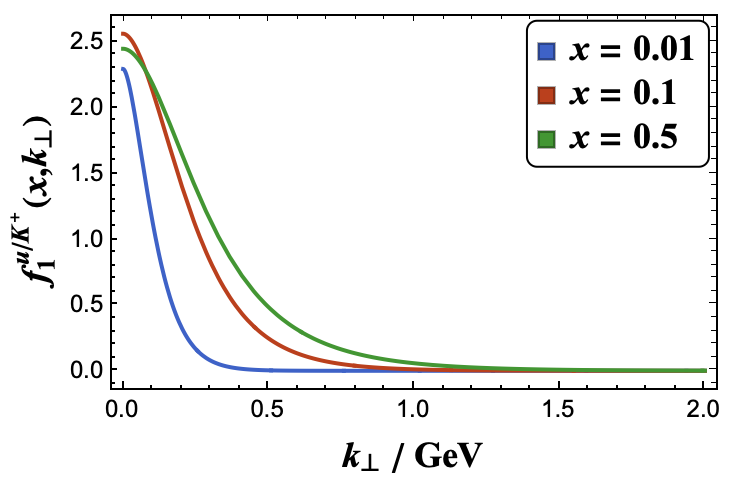}}
\hfill
\subfloat[\label{fig:pion_x}]{\includegraphics[width=0.45\linewidth]{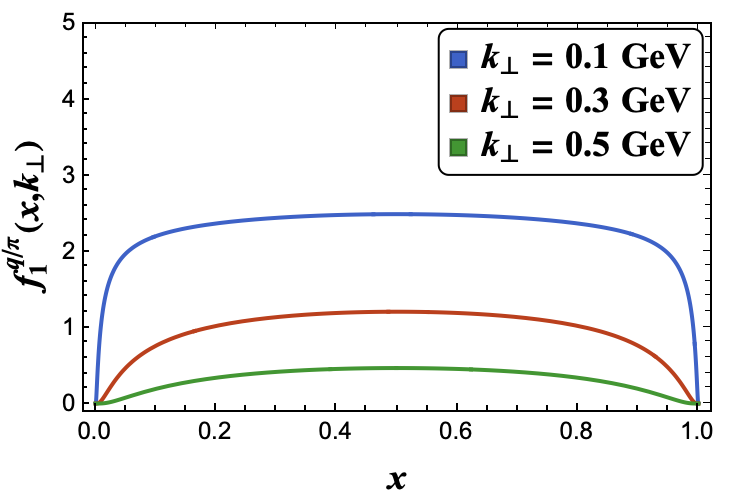}}
\hfill
\subfloat[\label{fig:pion_xk}]{\includegraphics[width=0.45\linewidth]{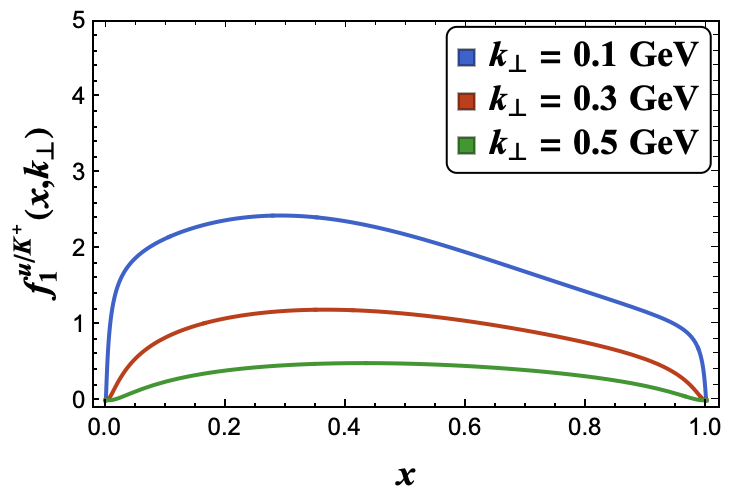}}
\caption{Pion (left) \eqref{eq:pion_tmd}  and kaon (right) 
\eqref{eq:kaon_tmd} TMDPDFs at low resolution $\mu=1/\rho$
with $\rho=0.313$ fm: (a,b) are the density plots, (c,d) the 3D plots,
(e,f) the transverse momentum dependent plots for fixed $x$, and (g,h) the longitudinal momentum dependence for fixed $k_\perp$. The pion parameters are
$\sqrt{N_cC_\pi}=7.240$, $m_\pi=139.0$ MeV, $M=398.17$ MeV.
The kaon parameters are $\sqrt{N_cZ_K}=6.60$, $m_K=458.0$ MeV, $M_u=394.4$ MeV, $M_s=556.5$ MeV.}
\end{figure}

\section{TMD Evolution}
\label{TMD}
The rapidity scale plays an important role in TMD evolution, as it distangles the soft exchanges between the collinear hadrons in phase space, in the context of TMD factorization. 
The rapidity divergences in \eqref{tmd} are controlled by Wilson line rapidity $y_n\rightarrow-\infty$.
With this in mind, the TMD beam functions are renormalized by subtracting the rapidity divergence with soft function $S$. The renormalized (soft-subtracted) TMD functions read~\cite{Ji:2004wu}.
\begin{equation}
    \begin{aligned}
        &\tilde{F}^{q/\pi}_1(x,b_\perp;\mu,\zeta)=\lim_{y_n\rightarrow-\infty}\frac{\tilde{f}^{q/\pi}_1(x,b_\perp,\mu,y_\pi-y_n)}{S(b_\perp,\mu,y-y_n)}
    \end{aligned}
\end{equation}
where $S$ is the soft factor discussed in Sec.~\ref{SECIII}.
Here rapidity scale \cite{Collins:2011zzd}
$$\zeta=2(k^+)^2e^{-2y}
$$ 
represents the energy of a parton in the hadron, and  $y$ an arbitrary rapidity mark to 
account for the rapidity subtraction under evolution.

\begin{figure}
    \centering
\subfloat[\label{tmd_b_a}]{\includegraphics[width=0.4\linewidth]{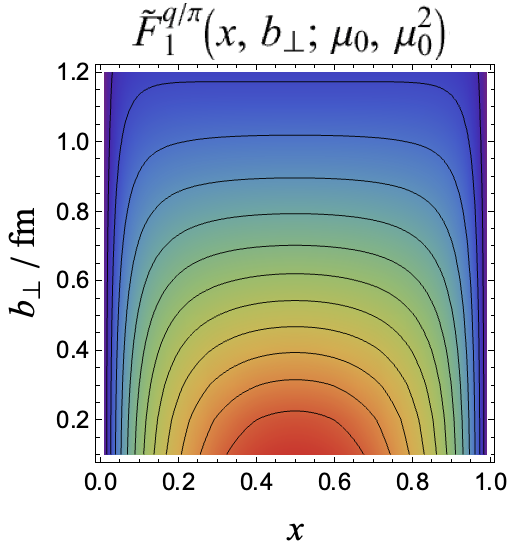}}
\hfill
\subfloat[]{\includegraphics[width=0.4\linewidth]{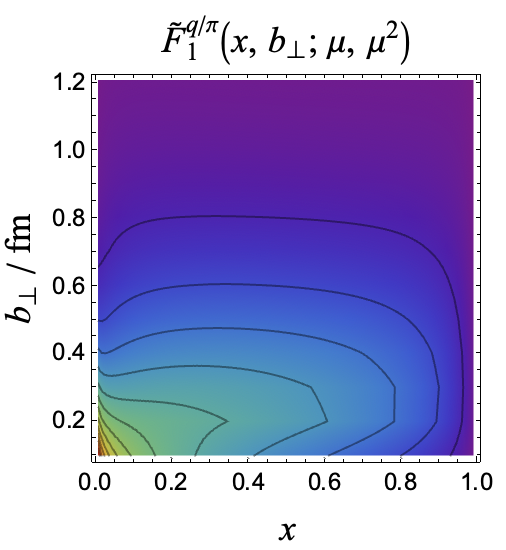}}
\hfill
\subfloat[]{\includegraphics[width=0.4\linewidth]{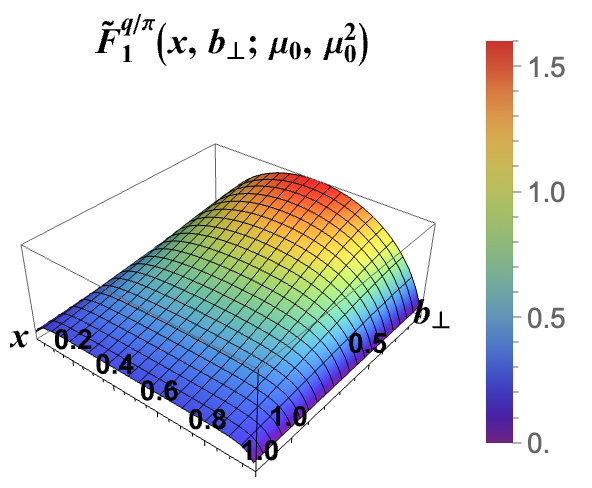}}
\hfill
\subfloat[]{\includegraphics[width=0.4\linewidth]{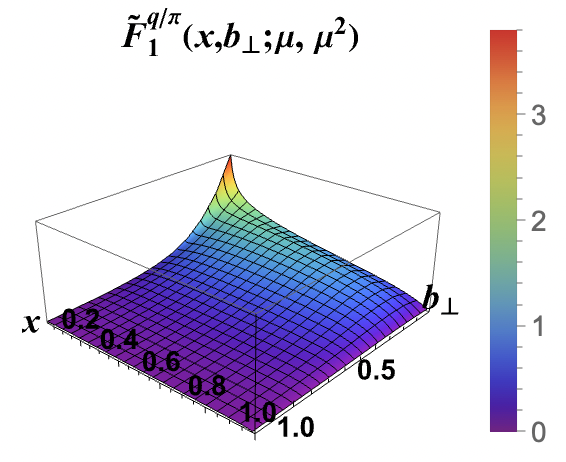}}
    \caption{Upper panel is the contour plot of the pion TMD parton distribution $\tilde{F}_1^{q/\pi}(x,b_\perp;\mu,\zeta)$: (a) soft subtracted quark TMD at intrinsic rapidity scale $\zeta=\mu_0^2$ and $\mu_0=1/\rho=0.63$ GeV 
    and (b) soft subtracted TMD evolved to $\mu=2$ GeV with chosen scheme $\zeta=\mu^2$. The lower panel is the 3D plot of the pion TMD parton distribution $\tilde{F}_1^{q/\pi}$: (c) quark TMD with the soft subtraction at intrinsic rapidity scale $\zeta=\mu_0^2$ and $\mu_0=1/\rho=0.63$ GeV
    and (d) subtracted TMD evolved to $\mu=2$ GeV. Here $b_\perp$ is in fm}
    \label{fig:tmd_b}
\end{figure}

\begin{figure}
    \centering
\subfloat[]{\includegraphics[width=0.4\linewidth]{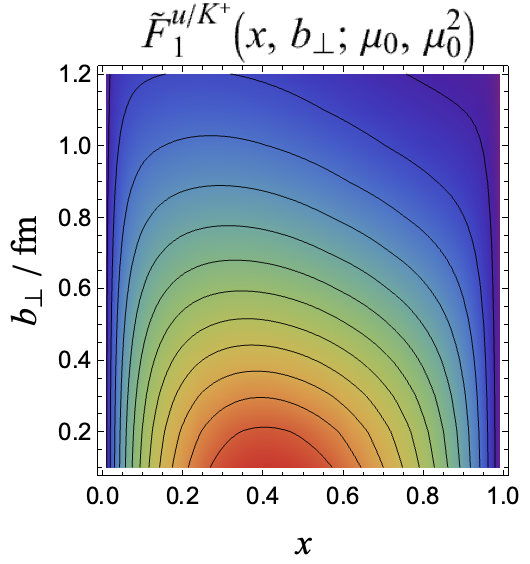}}
\hfill
\subfloat[]{\includegraphics[width=0.4\linewidth]{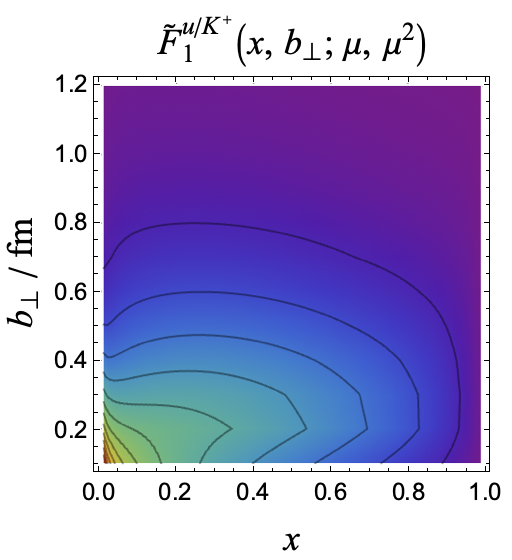}}
\hfill
\subfloat[]{\includegraphics[width=0.44\linewidth]{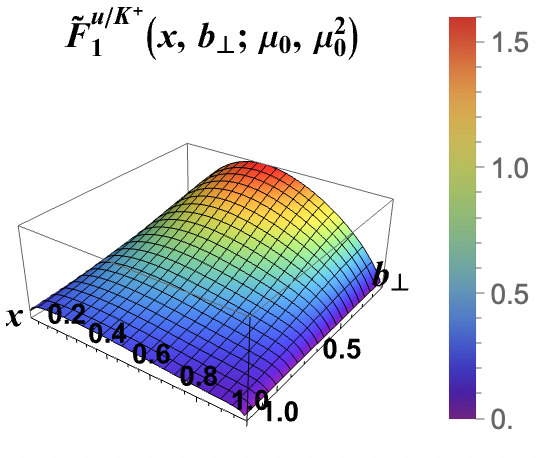}}
\hfill
\subfloat[]{\includegraphics[width=0.43\linewidth]{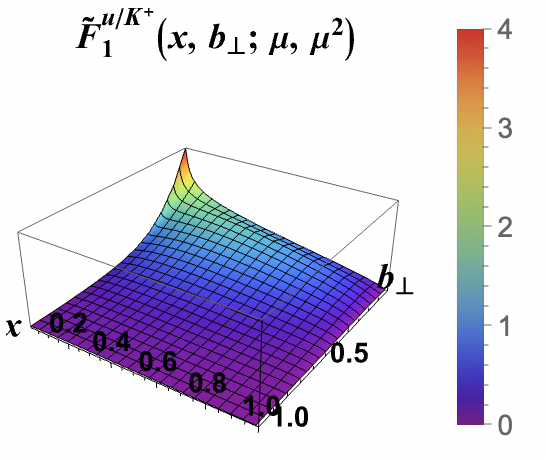}}
    \caption{Upper panel is the contour plot of the kaon TMD parton distribution $\tilde{F}_1^{u/K^+}$: (a) quark TMD with the soft subtraction at minimal rapidity scale $\zeta=\mu_0^2$ and $\mu_0=1/\rho=0.63$ GeV
    and (b) subtracted TMD  evolved to $\mu=2$ GeV. The lower panel is the 3D plot of the kaon TMD parton distribution $\tilde{F}_1^{u/K^+}$: (c) quark TMD with the soft subtraction at minimal rapidity scale $\zeta=\mu_0^2$ and $\mu_0=1/\rho=0.63$ GeV
    and (d) subtracted TMD evolved to $\mu=2$ GeV. Here $b_\perp$ is in fm}
    \label{fig:tmd_b_K}
\end{figure}

The amount of subtraction is controlled by the  rapidity dependence $\zeta$. 
At low resolution $\mu_0$, the soft-subtracted TMD $\tilde{F}^{q/\pi}_1(x,b_\perp;\mu_0,\zeta_0)$ in the ILM is defined as
\begin{equation}
\begin{aligned}
\label{pi_tmd}
    \tilde{F}&_1^{q/\pi}(x,b_\perp;\mu_0,\zeta_0)=\tilde{f}_{1}^{q/\pi}(x,b_\perp)e^{K^{(\rm inst)}_{\rm CS}(b_\perp/\rho)\ln\sqrt{\frac{\zeta_0}{1/\rho^2}}}
\end{aligned}
\end{equation}
where 
$\tilde{f}^{q/\pi}_{1}(x,b_\perp)$ is evaluated in \eqref{eq:pion_tmd} and \eqref{eq:kaon_tmd} for kaons.
%
%
The variation in the rapidity $y$ determines the $\zeta$ evolution of TMDs  \cite{Boer:2014tka,Idilbi:2004vb,Aybat:2011zv,Collins:2003fm}.
Generally, this  is implemented in the $b_\perp$ space Fourier-conjugate to $k_\perp$, allowing for a clearer separation of perturbative and non-perturbative effects~\cite{Collins:2011zzd}.
For any leading-twist TMD  with soft subtraction $\tilde{F}$, the CSS renormalization group equations reads \cite{Boer:2015ala,Collins:2014loa}
\begin{align}
\label{tmd_evol_rg}
    \frac{d}{d\ln\sqrt{\zeta}} \ln \tilde{F}(x, b_\perp; \mu,\zeta)=&K_{\rm CS}(b_\perp,\mu)\nonumber\\
    \frac{d}{d\ln\mu} \ln\tilde{F}(x, b_\perp; \mu,\zeta)=& \Gamma_F (\mu,\zeta)
\end{align}
where 
\begin{equation}
    \Gamma_F(\mu,\zeta)=\gamma_F (\alpha_s(\mu))\nonumber- \Gamma_{\mathrm{cusp}}(\alpha_s(\mu)) \ln\left(\frac{\zeta}{\mu^2}\right)
\end{equation}
The CS kernel $K_{\rm CS}$ drives
 the rapidity $\zeta$ scale evolution. The explicit forms of the anomalous dimensions $\gamma_F$,  depend on the quark operator insertion $\gamma^+,\gamma^+\gamma^5, i\sigma^{\alpha+}\gamma^5$. Their perturbative expansions are given in \cite{Liu:2024sqj,Liu:2025mbl}. The evolved  TMDs involve one intrinsic transverse scale $b_\perp$ and two evolution scales $\mu$, $\zeta$, both driven
 by perturbative and nonperturbative physics. 

A consistent matching between the TMD evolution in the non-perturbative part of the transverse space,  to the standard and perturbative collinear factorization part, is  notoriously subtle. For completeness, we note the recent development on the $k_\perp$ momentum space TMD approach, related to the non-perturbative TMD evolution in~\cite{Aslan:2024nqg,Gonzalez-Hernandez:2023iso,Gonzalez-Hernandez:2022ifv}. For a more thorough review, see \cite{Rogers:2024cci}. Alternatively, we use a simple  procedure in $b_\perp$ space, that consists in minimal  matching of the large and small $b_\perp$ regions.

 \begin{figure}
    \centering
    \includegraphics[width=0.8\linewidth]{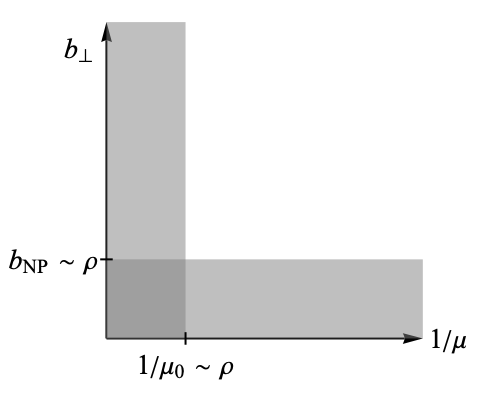}
    \caption{Perturbative (shaded) and non-perturbative (unshaded) regions
    for TMD evolution in transverse $b_\perp$ and resolution $\mu$.}
    \label{fig:scale}
\end{figure}
 
 For simplicity, we can choose a simple renormalization scheme where the rapidity evolution follows $\zeta=\mu^2$. In Fig.~\ref{fig:scale} we illustrate the regime of scales pertinent to perturbative
 evolution in shade. 
In the perturbative regime, the perturbative gluons are dominant, while in the opposite regime the primordial gluon epoxy (hard glue) \cite{Shuryak:2021fsu} is dominant. More specifically, in the perturbative regime ($2e^{-\gamma_E}/\mu\lesssim b_\perp\ll b_{\rm NP}$), we have~\cite{Bacchetta:2017vzh,Bacchetta:2017gcc,Barry:2023qqh,Collins:2011zzd}

\begin{equation}
\begin{aligned}
\label{convo}
    &\tilde{F}_1^{q/\pi}(x, b_\perp; \mu,\mu^2)\big|_{b_\perp\ll b_{\rm NP}}
    \rightarrow\sum_{i=q, \bar q, g}\int_x^1 \frac{dx'}{x'}C_{q/i}(x/x',b_\perp,\mu_b)f_1^{i/\pi} (x',\mu_b) e^{-S_{\rm sud}(\mu_b,\mu)}
\end{aligned}
\end{equation}

where the transverse scale $\mu_b$ is defined as 
\begin{equation}
\label{mu_b}
\mu_b = 2 e^{-\gamma_E}/b_\perp
\end{equation}
with $\gamma_E$ the Euler–Mascheroni constant. This formulation follows the CSS formalism in~\cite{Collins:2017oxh},
and is perturbatively meaningful only at small values of $b_\perp$, with a scale $\mu_b$ that is sufficiently larger than the Landau pole $\Lambda_{\rm QCD}$.
The convolution is the operator product expansion (OPE) which describes the small-$b_\perp$ behavior of the TMDs in terms of the collinear PDFs $f_1^{q/\pi}$ convoluted with the perturbative Wilson coefficients $C_{i/j}$ \cite{Scimemi:2017etj} defined in Appendix~\ref{App:pQCD}. The PDFs with the evolution are shown in Fig.~\ref{fig:pdf}. The Sudakov factor contains the perturbative effects of soft gluon radiation, driving the evolution of the TMDs from $\mu_b$ to the UV renormalization scale $\mu$ perturbatively. The  Sudakov kernel $S_{\rm sud}$ is defined as

\begin{equation}
\begin{aligned}
\label{sudakov}
    S_{\rm sud}(\mu_b, \mu)\approx& \int_{\mu_b}^{\mu} \frac{d\mu'}{\mu'} \left[ \Gamma_{\rm cusp}(\alpha_s(\mu'^2)) \ln \left( \frac{\mu^2}{\mu'^2} \right) - \gamma_F(\alpha_s(\mu'^2)) \right]\\
    &-K^{(\rm pert)}_{\rm CS}(b_{\perp},\mu_b)\ln\left(\frac{\mu}{\mu_b}\right)
\end{aligned}
\end{equation}

We note that when $b_\perp\rightarrow 0$, the perturbative formulation \eqref{convo} is not valid, and should be matched to the fixed-order collinear calculation at $\mu$ following from the OPE,
\begin{equation}
\begin{aligned}
    &\tilde{F}_1^{q/\pi}(x, b_\perp; \mu,\mu^2)\big|_{b_\perp\rightarrow0}
    \rightarrow f_1^{q/\pi} (x,\mu)
\end{aligned}
\end{equation}
This can be improved by modifying the transverse scale \eqref{mu_b} in the Sudakov factor $S_{\rm sud}$ through a smooth substitution $\mu_b \rightarrow \mu$ as $b_\perp \rightarrow 0$,  so that the lower limit of the integral in \eqref{sudakov} never exceeds the upper limit. This is similar to the transverse scale introduced in \cite{Bacchetta:2017gcc,Bacchetta:2022awv,Rogers:2024cci,Boer:2014tka,Collins:2016hqq}. In a way, this modification corresponds to a re-summation of logarithms,  such that the Sudakov exponent $S_{\rm sud}$ interpolates to $0$ at $b_\perp = 0$, as it should~\cite{Parisi:1979se,Altarelli:1984pt}.

In the non-perturbative region ($b_\perp\gg b_{\rm NP}$) as illustrated by the unshaded region
in Fig.~\ref{fig:scale}, the glue contribution in the ILM starts to modify the perturbative prediction in \eqref{convo}. As the transverse distance grows, the soft (transverse) scale $\mu_b$ in the Sudakov factor saturates to $\mu_0$. A similar $b_*$ prescription with the transverse scale $\mu_b$ saturated at large $b_\perp$, has been discussed in~\cite{Bozzi:2010xn,Boer:2014tka}.  The perturbative gluon radiation at large transverse distance is suppressed, and the nonperturbative component of the TMD evolution compensates the gluon radiation. A phenomenological illustration of this mechanism is discussed in~\cite{Collins:2011zzd,Collins:2014jpa}, where the transverse scale is smoothly saturated at large distance $\mu_b(b_\perp\rightarrow\infty)\rightarrow\mu_0$, to ensure that the perturbative evolution remains within the perturbative regime. With this in mind, the soft-subtracted TMD in large distance can be defined as

\begin{equation}
\begin{aligned}
\label{F_np}
    &\tilde{F}_1^{q/\pi}(x,b_\perp;\mu,\mu^2)\big|_{b_\perp\gg b_{\rm NP}}\simeq~\tilde{F}^{q/\pi}_1(x,b_\perp;\mu_0,\mu_0^2) e^{K^{(\rm inst)}_{\rm CS}(b_\perp/\rho)\ln\sqrt{\frac{\mu^2}{\mu^2_0}}}\\
    &\times \exp\left[\int_{\mu_0}^{\mu} \frac{d\mu'}{\mu'}\left(\gamma_F(\alpha_s(\mu'^2))-\Gamma_{\rm cusp}(\alpha_s(\mu'^2)) \ln \left( \frac{\mu_0^2}{\mu'^2} \right) \right)\right]
\end{aligned}
\end{equation}

The perturbative Sudakov form factor $S_{\rm sud}$ evolves from $\mu_0$ to $\mu$.
At the initial scale $\mu=\mu_0\sim1/\rho$, the large distance TMD can be fully estimated by the instanton vacuum. The amount of the soft subtraction is controlled by the rapidity dependence $\zeta$. 
The full TMD function at high resolution $\mu$ interpolates between the large $b_\perp$ and small $b_\perp$ by optimal matching with the smooth cut-off $b_{\rm NP}$ in linear combination.

\begin{equation}
\begin{aligned}
\label{INTER}
\tilde{F}_1^{q/h}(x, b_\perp; \mu,\mu^2)=&c_{\mathrm{s}}(b_\perp,b_{\rm NP})\tilde{F}_1^{q/h}(x, b_\perp\ll b_{\rm NP}; \mu,\mu^2)
\\
&+c_{\mathrm{l}}(b_\perp,b_{\rm NP})\tilde{F}_1^{q/h}(x, b_\perp\gg b_{\rm NP}; \mu,\mu^2)
\end{aligned}
\end{equation}

Here  $b_{\rm NP}$ is an optimized cross-over with a value of about $\rho$, that interpolates 
the perturbative and non-perturbative contributions. The matching coefficient functions $(c_{\mathrm{s}}, c_{\mathrm{l}})$ can be parameterized by any smooth functions with the boundary condition where at short distance ($b_\perp\ll b_{\rm NP}$),  $(c_{\mathrm{s}}, c_{\mathrm{l}})\rightarrow (1, 0)$ and at long distance ($b_\perp\gg b_{\rm NP}$),  $(c_{\mathrm{s}}, c_{\mathrm{l}})\rightarrow (0, 1)$ and a cross-over occurs at $b_{\rm NP}$.



\begin{figure}
    \centering
\subfloat[]{\includegraphics[width=.85\linewidth]{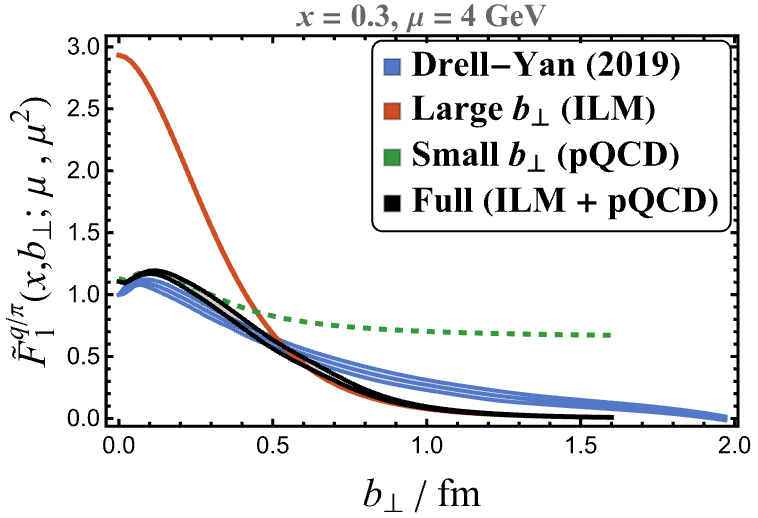}}
\hfill
\subfloat[]{\includegraphics[width=.85\linewidth]{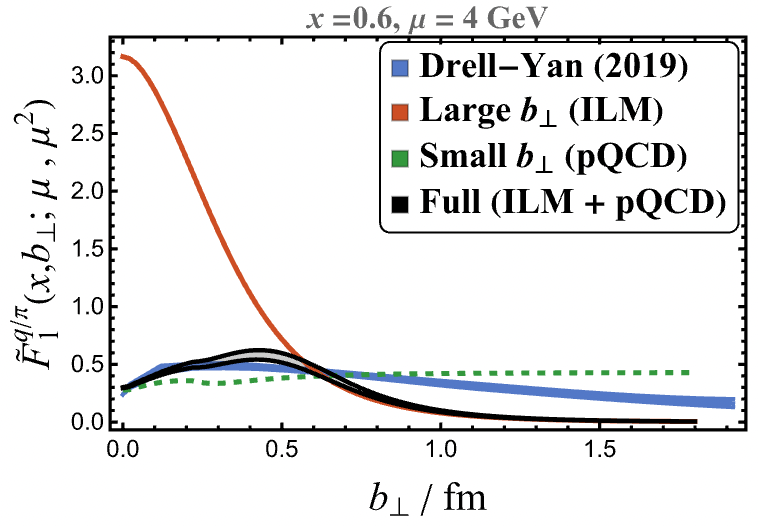}}
\caption{The evolved  pion TMD versus  $b_\perp$ with (a) $x=0.3$ and (b) $x=0.6$, compared to the Drell-Yan experimental data extraction \cite{Vladimirov:2019bfa} from three available measurements of the transverse momentum cross section for pion-induced Drell-Yan process performed by NA3 \cite{NA3:1982ntq}, E537 \cite{Anassontzis:1987hk} and E615 \cite{Conway:1989fs}.  The black band represents the optimized result with the propagated uncertainty from the original data.}
    \label{fig:f_b}
\end{figure}

\chapter{Conclusion}
\label{ch:conclusion}
This dissertation has developed a quantitative framework for understanding nonperturbative QCD in the continuum limit, grounded in the topological structure of the infrared vacuum and constrained by lattice QCD and experimental observations. By modeling the infrared gluon fields as a correlated ensemble of instantons and anti-instantons, the framework provides a unified description of key nonperturbative phenomena, including the emergence of trace and axial anomalies, spontaneous chiral symmetry breaking, and confinement at intermediate distance scales. The equivalence between the statistical ensemble formulation and the effective field theory description further clarifies how quark–gluon dynamics can be systematically encoded in terms of nonlocal interactions.

A central outcome of this work is the explicit connection established between Euclidean vacuum structure and light-front partonic dynamics. Through the construction of light-front wave functions and the derivation of distribution amplitudes, parton distribution functions, and transverse-momentum-dependent distributions, the framework demonstrates how hadronic structure at the partonic level originates from underlying topological fluctuations. In particular, the embedding into TMD factorization and the computation of soft functions provide a vacuum-based interpretation of rapidity evolution, linking nonperturbative QCD dynamics to observable high-energy processes.

The application of this framework to hadronic form factors, higher-twist color forces, and multigluon correlations further illustrates its predictive power across a broad range of observables, including mass and spin decomposition and near-threshold quarkonium production. These results highlight the essential role of topological vacuum configurations in shaping hadronic properties over a wide range of scales.

Looking forward, this approach opens several avenues for further investigation, including systematic improvements in matching to perturbative QCD at higher resolutions, extensions to polarized and multi-hadron processes, and closer integration with future lattice and experimental inputs. Overall, this work establishes that the topological QCD vacuum is not only a conceptual foundation but also a quantitatively robust framework for bridging nonperturbative dynamics with partonic phenomenology.

\bibliographystyle{unsrtnat}
\renewcommand{\baselinestretch}{1}
\normalsize

\clearpage
\newpage
\phantomsection%
\addcontentsline{toc}{chapter}{\numberline{}{Bibliography}}%
\bibliography{ref,cs,tmd,eta}

\begin{thebibliography}{466}
\providecommand{\natexlab}[1]{#1}
\providecommand{\url}[1]{\texttt{#1}}
\expandafter\ifx\csname urlstyle\endcsname\relax
  \providecommand{\doi}[1]{doi: #1}\else
  \providecommand{\doi}{doi: \begingroup \urlstyle{rm}\Url}\fi

\bibitem['t~Hooft(1974)]{tHooft:1974kcl}
Gerard 't~Hooft.
\newblock {Magnetic Monopoles in Unified Gauge Theories}.
\newblock \emph{Nucl. Phys. B}, 79:\penalty0 276--284, 1974.
\newblock \doi{10.1016/0550-3213(74)90486-6}.

\bibitem[Polyakov(1974)]{Polyakov:1974ek}
Alexander~M. Polyakov.
\newblock {Particle Spectrum in Quantum Field Theory}.
\newblock \emph{JETP Lett.}, 20:\penalty0 194--195, 1974.

\bibitem[Belavin et~al.(1975)Belavin, Polyakov, Schwartz, and
  Tyupkin]{Belavin:1975fg}
A.~A. Belavin, Alexander~M. Polyakov, A.~S. Schwartz, and Yu.~S. Tyupkin.
\newblock {Pseudoparticle Solutions of the Yang-Mills Equations}.
\newblock \emph{Phys. Lett. B}, 59:\penalty0 85--87, 1975.
\newblock \doi{10.1016/0370-2693(75)90163-X}.

\bibitem[Kraan and van Baal(1998{\natexlab{a}})]{Kraan:1998sn}
Thomas~C. Kraan and Pierre van Baal.
\newblock {Monopole constituents inside SU(n) calorons}.
\newblock \emph{Phys. Lett. B}, 435:\penalty0 389--395, 1998{\natexlab{a}}.
\newblock \doi{10.1016/S0370-2693(98)00799-0}.

\bibitem[Diakonov and Petrov(1984)]{Diakonov:1983hh}
Dmitri Diakonov and V.~Yu. Petrov.
\newblock {Instanton Based Vacuum from Feynman Variational Principle}.
\newblock \emph{Nucl. Phys. B}, 245:\penalty0 259--292, 1984.
\newblock \doi{10.1016/0550-3213(84)90432-2}.

\bibitem[Callan et~al.(1976)Callan, Dashen, and Gross]{Callan:1976je}
Curtis~G. Callan, Jr., R.~F. Dashen, and David~J. Gross.
\newblock {The Structure of the Gauge Theory Vacuum}.
\newblock \emph{Phys. Lett. B}, 63:\penalty0 334--340, 1976.
\newblock \doi{10.1016/0370-2693(76)90277-X}.

\bibitem[Callan et~al.(1978)Callan, Dashen, and Gross]{Callan:1977gz}
Curtis~G. Callan, Jr., Roger~F. Dashen, and David~J. Gross.
\newblock {Toward a Theory of the Strong Interactions}.
\newblock \emph{Phys. Rev. D}, 17:\penalty0 2717, 1978.
\newblock \doi{10.1103/PhysRevD.17.2717}.

\bibitem[Shuryak(1982)]{Shuryak:1981ff}
Edward~V. Shuryak.
\newblock {The Role of Instantons in Quantum Chromodynamics. 1. Physical
  Vacuum}.
\newblock \emph{Nucl. Phys. B}, 203:\penalty0 93, 1982.
\newblock \doi{10.1016/0550-3213(82)90478-3}.

\bibitem[Diakonov and Petrov(1986)]{Diakonov:1985eg}
Dmitri Diakonov and V.~Yu. Petrov.
\newblock {A Theory of Light Quarks in the Instanton Vacuum}.
\newblock \emph{Nucl. Phys. B}, 272:\penalty0 457--489, 1986.
\newblock \doi{10.1016/0550-3213(86)90011-8}.

\bibitem[Vainshtein et~al.(1982)Vainshtein, Zakharov, Novikov, and
  Shifman]{Vainshtein:1981wh}
A.~I. Vainshtein, Valentin~I. Zakharov, V.~A. Novikov, and Mikhail~A. Shifman.
\newblock {ABC's of Instantons}.
\newblock \emph{Sov. Phys. Usp.}, 25:\penalty0 195, 1982.
\newblock \doi{10.1070/PU1982v025n04ABEH004533}.

\bibitem[Diakonov(1996)]{Diakonov:1995ea}
Dmitri Diakonov.
\newblock {Chiral symmetry breaking by instantons}.
\newblock \emph{Proc. Int. Sch. Phys. Fermi}, 130:\penalty0 397--432, 1996.
\newblock \doi{10.3254/978-1-61499-215-8-397}.

\bibitem[Miesch et~al.(2023)Miesch, Shuryak, and Zahed]{Miesch:2023hvl}
Nicholas Miesch, Edward Shuryak, and Ismail Zahed.
\newblock {Hadronic structure on the light-front. IX. Orbital-spin-isospin wave
  functions of baryons}.
\newblock \emph{Phys. Rev. D}, 108\penalty0 (9):\penalty0 094033, 2023.
\newblock \doi{10.1103/PhysRevD.108.094033}.

\bibitem[Shuryak and Zahed(2023{\natexlab{a}})]{Shuryak:2022wtk}
Edward Shuryak and Ismail Zahed.
\newblock {Hadronic structure on the light front. V. Diquarks, nucleons, and
  multiquark Fock components}.
\newblock \emph{Phys. Rev. D}, 107\penalty0 (3):\penalty0 034027,
  2023{\natexlab{a}}.
\newblock \doi{10.1103/PhysRevD.107.034027}.

\bibitem[Shuryak and Zahed(2023{\natexlab{b}})]{Shuryak:2022thi}
Edward Shuryak and Ismail Zahed.
\newblock {Hadronic structure on the light front. IV. Heavy and light baryons}.
\newblock \emph{Phys. Rev. D}, 107\penalty0 (3):\penalty0 034026,
  2023{\natexlab{b}}.
\newblock \doi{10.1103/PhysRevD.107.034026}.

\bibitem['t~Hooft(1986)]{tHooft:1986ooh}
Gerard 't~Hooft.
\newblock {How Instantons Solve the U(1) Problem}.
\newblock \emph{Phys. Rept.}, 142:\penalty0 357--387, 1986.
\newblock \doi{10.1016/0370-1573(86)90117-1}.

\bibitem[Shuryak and Zahed(2026)]{Shuryak:2026pqt}
Edward Shuryak and Ismail Zahed.
\newblock {The Hadron-Parton Bridge, From the QCD Vacuum to Partons}.
\newblock 1 2026.

\bibitem[Shuryak(2021)]{Shuryak:2018fjr}
Edward Shuryak.
\newblock \emph{{Nonperturbative Topological Phenomena in QCD and Related
  Theories}}.
\newblock Springer Cham, March 2021.
\newblock \doi{10.1007/978-3-030-62990-8}.

\bibitem[Sch\"afer and Shuryak(1998)]{Schafer:1996wv}
Thomas Sch\"afer and Edward~V. Shuryak.
\newblock {Instantons in QCD}.
\newblock \emph{Rev. Mod. Phys.}, 70:\penalty0 323--426, 1998.
\newblock \doi{10.1103/RevModPhys.70.323}.

\bibitem[Michael and Spencer(1995{\natexlab{a}})]{Michael:1994uu}
Christopher Michael and P.~S. Spencer.
\newblock {Instanton size distributions from calibrated cooling}.
\newblock \emph{Nucl. Phys. B Proc. Suppl.}, 42:\penalty0 261--263,
  1995{\natexlab{a}}.
\newblock \doi{10.1016/0920-5632(95)00220-4}.

\bibitem[Michael and Spencer(1995{\natexlab{b}})]{Michael:1995br}
Christopher Michael and P.~S. Spencer.
\newblock {Cooling and the SU(2) instanton vacuum}.
\newblock \emph{Phys. Rev. D}, 52:\penalty0 4691--4699, 1995{\natexlab{b}}.
\newblock \doi{10.1103/PhysRevD.52.4691}.

\bibitem[Leinweber(1999)]{Leinweber:1999cw}
Derek~B. Leinweber.
\newblock {Visualizations of the QCD vacuum}.
\newblock In \emph{{Workshop on Light-Cone QCD and Nonperturbative Hadron
  Physics}}, pages 138--143, 12 1999.

\bibitem[Bakas(2011)]{Bakas:2010by}
Ioannis Bakas.
\newblock {Gradient flows and instantons at a Lifshitz point}.
\newblock \emph{J. Phys. Conf. Ser.}, 283:\penalty0 012004, 2011.
\newblock \doi{10.1088/1742-6596/283/1/012004}.

\bibitem[Biddle et~al.(2018)Biddle, Kamleh, and Leinweber]{Biddle:2018bst}
James~C. Biddle, Waseem Kamleh, and Derek~B. Leinweber.
\newblock {Visualizations of Centre Vortex Structure in Lattice Simulations}.
\newblock \emph{PoS}, LATTICE2018:\penalty0 256, 2018.
\newblock \doi{10.22323/1.334.0256}.

\bibitem[Hasenfratz and Witzel(2020)]{Hasenfratz:2019hpg}
Anna Hasenfratz and Oliver Witzel.
\newblock {Continuous renormalization group $\beta$ function from lattice
  simulations}.
\newblock \emph{Phys. Rev. D}, 101\penalty0 (3):\penalty0 034514, 2020.
\newblock \doi{10.1103/PhysRevD.101.034514}.

\bibitem[Athenodorou et~al.(2018)Athenodorou, Boucaud, De~Soto,
  Rodr{\'\i}guez-Quintero, and Zafeiropoulos]{Athenodorou:2018jwu}
A.~Athenodorou, Ph. Boucaud, F.~De~Soto, J.~Rodr{\'\i}guez-Quintero, and
  S.~Zafeiropoulos.
\newblock {Instanton liquid properties from lattice QCD}.
\newblock \emph{JHEP}, 02:\penalty0 140, 2018.
\newblock \doi{10.1007/JHEP02(2018)140}.

\bibitem[Biddle et~al.(2020{\natexlab{a}})Biddle, Kamleh, and
  Leinweber]{Biddle:2020eec}
James~C. Biddle, Waseem Kamleh, and Derek~B. Leinweber.
\newblock {Visualisations of Centre Vortices}.
\newblock \emph{EPJ Web Conf.}, 245:\penalty0 06010, 2020{\natexlab{a}}.
\newblock \doi{10.1051/epjconf/202024506010}.

\bibitem[Zimmermann(2024)]{Zimmermann:2024mar}
Falk Zimmermann.
\newblock {Extracting Instantons from the Lattice}.
\newblock \emph{PoS}, LATTICE2023:\penalty0 376, 2024.
\newblock \doi{10.22323/1.453.0376}.

\bibitem[Ringwald and Schrempp(1999)]{Ringwald:1999ze}
A.~Ringwald and F.~Schrempp.
\newblock {Confronting instanton perturbation theory with QCD lattice results}.
\newblock \emph{Phys. Lett. B}, 459:\penalty0 249--258, 1999.
\newblock \doi{10.1016/S0370-2693(99)00682-6}.

\bibitem[Faccioli and DeGrand(2003)]{Faccioli:2003qz}
Pietro Faccioli and Thomas~A. DeGrand.
\newblock {Evidence for instanton induced dynamics, from lattice QCD}.
\newblock \emph{Phys. Rev. Lett.}, 91:\penalty0 182001, 2003.
\newblock \doi{10.1103/PhysRevLett.91.182001}.

\bibitem[Liu et~al.(2024{\natexlab{a}})Liu, Shuryak, and Zahed]{Liu:2024rdm}
Wei-Yang Liu, Edward Shuryak, and Ismail Zahed.
\newblock {Glue in hadrons at medium resolution and the QCD instanton vacuum}.
\newblock \emph{Phys. Rev. D}, 110\penalty0 (5):\penalty0 054005,
  2024{\natexlab{a}}.
\newblock \doi{10.1103/PhysRevD.110.054005}.

\bibitem[Liu(2025)]{Liu:2025ldh}
Wei-Yang Liu.
\newblock {Generic framework for non-perturbative QCD in light hadrons}.
\newblock 1 2025.

\bibitem[Liu et~al.(2023)Liu, Shuryak, and Zahed]{Liu:2023yuj}
Wei-Yang Liu, Edward Shuryak, and Ismail Zahed.
\newblock {Hadronic structure on the light-front. VII. Pions and kaons and
  their partonic distributions}.
\newblock \emph{Phys. Rev. D}, 107\penalty0 (9):\penalty0 094024, 2023.
\newblock \doi{10.1103/PhysRevD.107.094024}.

\bibitem[Liu et~al.(2024{\natexlab{b}})Liu, Shuryak, and Zahed]{Liu:2023fpj}
Wei-Yang Liu, Edward Shuryak, and Ismail Zahed.
\newblock {Hadronic structure on the light front. VIII. Light scalar and vector
  mesons}.
\newblock \emph{Phys. Rev. D}, 109\penalty0 (7):\penalty0 074029,
  2024{\natexlab{b}}.
\newblock \doi{10.1103/PhysRevD.109.074029}.

\bibitem[Liu et~al.(2024{\natexlab{c}})Liu, Shuryak, Weiss, and
  Zahed]{Liu:2024jno}
Wei-Yang Liu, Edward Shuryak, Christian Weiss, and Ismail Zahed.
\newblock {Pion gravitational form factors in the QCD instanton vacuum. I}.
\newblock \emph{Phys. Rev. D}, 110\penalty0 (5):\penalty0 054021,
  2024{\natexlab{c}}.
\newblock \doi{10.1103/PhysRevD.110.054021}.

\bibitem[Liu et~al.(2024{\natexlab{d}})Liu, Shuryak, and Zahed]{Liu:2024vkj}
Wei-Yang Liu, Edward Shuryak, and Ismail Zahed.
\newblock {Pion gravitational form factors in the QCD instanton vacuum. II}.
\newblock \emph{Phys. Rev. D}, 110\penalty0 (5):\penalty0 054022,
  2024{\natexlab{d}}.
\newblock \doi{10.1103/PhysRevD.110.054022}.

\bibitem[Liu et~al.(2024{\natexlab{e}})Liu, Zahed, and Zhao]{Liu:2024sqj}
Wei-Yang Liu, Ismail Zahed, and Yong Zhao.
\newblock {Collins-Soper Kernel in the QCD Instanton Vacuum}.
\newblock 12 2024{\natexlab{e}}.

\bibitem[Liu and Zahed(2024)]{Liu:2024yqa}
Wei-Yang Liu and Ismail Zahed.
\newblock {Photoproduction of {\ensuremath{\eta}}c,b near threshold}.
\newblock \emph{Phys. Rev. D}, 110\penalty0 (5):\penalty0 054025, 2024.
\newblock \doi{10.1103/PhysRevD.110.054025}.

\bibitem[Liu et~al.(2025)Liu, Shuryak, and Zahed]{Liu:2025ypg}
Wei-Yang Liu, Edward Shuryak, and Ismail Zahed.
\newblock {The color force acting on a quark in the pion and nucleon}.
\newblock 11 2025.

\bibitem[Liu and Zahed(2025{\natexlab{a}})]{Liu:2025kuc}
Wei-Yang Liu and Ismail Zahed.
\newblock {Nucleon electric dipole form factor in QCD vacuum}.
\newblock 1 2025{\natexlab{a}}.

\bibitem[Liu and Zahed(2025{\natexlab{b}})]{Liu:2025mbl}
Wei-Yang Liu and Ismail Zahed.
\newblock {Tomography of pions and kaons in the QCD vacuum: Transverse momentum
  dependent parton distribution functions}.
\newblock \emph{Phys. Rev. D}, 112\penalty0 (3):\penalty0 034039,
  2025{\natexlab{b}}.
\newblock \doi{10.1103/9t4h-zpwm}.

\bibitem[Liu and Zahed(2025{\natexlab{c}})]{Liu:2025fuf}
Wei-Yang Liu and Ismail Zahed.
\newblock {Tomography of the rho meson in the QCD instanton vacuum: Transverse
  momentum dependent parton distribution functions}.
\newblock \emph{Phys. Rev. D}, 112\penalty0 (3):\penalty0 034028,
  2025{\natexlab{c}}.
\newblock \doi{10.1103/6ffp-qs8p}.

\bibitem[Luscher(2010)]{Luscher:2009eq}
Martin Luscher.
\newblock {Trivializing maps, the Wilson flow and the HMC algorithm}.
\newblock \emph{Commun. Math. Phys.}, 293:\penalty0 899--919, 2010.
\newblock \doi{10.1007/s00220-009-0953-7}.

\bibitem[Luscher and Weisz(2011)]{Luscher:2011bx}
Martin Luscher and Peter Weisz.
\newblock {Perturbative analysis of the gradient flow in non-abelian gauge
  theories}.
\newblock \emph{JHEP}, 02:\penalty0 051, 2011.
\newblock \doi{10.1007/JHEP02(2011)051}.

\bibitem[L{\"u}scher(2014)]{Luscher:2013vga}
Martin L{\"u}scher.
\newblock {Future applications of the Yang-Mills gradient flow in lattice QCD}.
\newblock \emph{PoS}, LATTICE2013:\penalty0 016, 2014.
\newblock \doi{10.22323/1.187.0016}.

\bibitem[Makino et~al.(2018)Makino, Morikawa, and Suzuki]{Makino:2018rys}
Hiroki Makino, Okuto Morikawa, and Hiroshi Suzuki.
\newblock {Gradient flow and the Wilsonian renormalization group flow}.
\newblock \emph{PTEP}, 2018\penalty0 (5):\penalty0 053B02, 2018.
\newblock \doi{10.1093/ptep/pty050}.

\bibitem[Narayanan and Neuberger(2006)]{Narayanan:2006rf}
R.~Narayanan and H.~Neuberger.
\newblock {Infinite N phase transitions in continuum Wilson loop operators}.
\newblock \emph{JHEP}, 03:\penalty0 064, 2006.
\newblock \doi{10.1088/1126-6708/2006/03/064}.

\bibitem[Moran and Leinweber(2008)]{Moran:2008xq}
P.~J. Moran and D.~B. Leinweber.
\newblock {Buried treasure in the sand of the QCD vacuum}.
\newblock In \emph{{QCD Downunder II}}, 5 2008.

\bibitem[Musakhanov(2023)]{Musakhanov:2023dsn}
Mirzayusuf Musakhanov.
\newblock {Gluons, light and heavy quarks and their interactions in the
  instanton vacuum}.
\newblock 3 2023.

\bibitem[Hutter(2001)]{hutter2001instantonsqcdtheoryapplication}
Marcus Hutter.
\newblock Instantons in qcd: Theory and application of the instanton liquid
  model, 2001.
\newblock URL \url{https://arxiv.org/abs/hep-ph/0107098}.

\bibitem[Rapp et~al.(1998)Rapp, Sch{\"a}fer, Shuryak, and
  Velkovsky]{Rapp:1997zu}
R.~Rapp, Thomas Sch{\"a}fer, Edward~V. Shuryak, and M.~Velkovsky.
\newblock {Diquark Bose condensates in high density matter and instantons}.
\newblock \emph{Phys. Rev. Lett.}, 81:\penalty0 53--56, 1998.
\newblock \doi{10.1103/PhysRevLett.81.53}.

\bibitem[Rapp et~al.(2000)Rapp, Sch\"afer, Shuryak, and Velkovsky]{Rapp:1999qa}
R.~Rapp, Thomas Sch\"afer, Edward~V. Shuryak, and M.~Velkovsky.
\newblock {High density QCD and instantons}.
\newblock \emph{Annals Phys.}, 280:\penalty0 35--99, 2000.
\newblock \doi{10.1006/aphy.1999.5991}.

\bibitem[Ilgenfritz and Shuryak(1989)]{Ilgenfritz:1988dh}
Ernst-Michael Ilgenfritz and Edward~V. Shuryak.
\newblock {Chiral Symmetry Restoration at Finite Temperature in the Instanton
  Liquid}.
\newblock \emph{Nucl. Phys. B}, 319:\penalty0 511--520, 1989.
\newblock \doi{10.1016/0550-3213(89)90617-2}.

\bibitem[Shuryak and Zahed(2023{\natexlab{c}})]{Shuryak:2021fsu}
Edward Shuryak and Ismail Zahed.
\newblock {Hadronic structure on the light front. I. Instanton effects and
  quark-antiquark effective potentials}.
\newblock \emph{Phys. Rev. D}, 107\penalty0 (3):\penalty0 034023,
  2023{\natexlab{c}}.
\newblock \doi{10.1103/PhysRevD.107.034023}.

\bibitem[Shuryak(1988{\natexlab{a}})]{Shuryak:1987tr}
Edward~V. Shuryak.
\newblock {Toward the Quantitative Theory of the 'Instanton Liquid' 4.
  Tunneling in the Double Well Potential}.
\newblock \emph{Nucl. Phys. B}, 302:\penalty0 621--644, 1988{\natexlab{a}}.
\newblock \doi{10.1016/0550-3213(88)90191-5}.

\bibitem[Balitsky and Yung(1986)]{Balitsky:1986qn}
I.~I. Balitsky and A.~V. Yung.
\newblock {Collective - Coordinate Method for Quasizero Modes}.
\newblock \emph{Phys. Lett. B}, 168:\penalty0 113--119, 1986.
\newblock \doi{10.1016/0370-2693(86)91471-1}.

\bibitem[Yung(1988)]{Yung:1987zp}
A.~V. Yung.
\newblock {Instanton Vacuum in Supersymmetric QCD}.
\newblock \emph{Nucl. Phys. B}, 297:\penalty0 47, 1988.
\newblock \doi{10.1016/0550-3213(88)90199-X}.

\bibitem[Verbaarschot(1991)]{Verbaarschot:1991sq}
J.~J.~M. Verbaarschot.
\newblock {Streamlines and conformal invariance in Yang-Mills theories}.
\newblock \emph{Nucl. Phys. B}, 362:\penalty0 33--53, 1991.
\newblock \doi{10.1016/0550-3213(91)90554-B}.
\newblock [Erratum: Nucl.Phys.B 386, 236--236 (1992)].

\bibitem[Ostrovsky et~al.(2002)Ostrovsky, Carter, and
  Shuryak]{Ostrovsky:2002cg}
D.~M. Ostrovsky, G.~W. Carter, and E.~V. Shuryak.
\newblock {Forced tunneling and turning state explosion in pure Yang-Mills
  theory}.
\newblock \emph{Phys. Rev.}, D66:\penalty0 036004, 2002.
\newblock \doi{10.1103/PhysRevD.66.036004}.

\bibitem[Klinkhamer and Manton(1984)]{Klinkhamer:1984di}
Frans~R. Klinkhamer and N.~S. Manton.
\newblock {A Saddle Point Solution in the Weinberg-Salam Theory}.
\newblock \emph{Phys. Rev.}, D30:\penalty0 2212, 1984.
\newblock \doi{10.1103/PhysRevD.30.2212}.

\bibitem[Shuryak and Zahed(2021{\natexlab{a}})]{Shuryak:2021iqu}
Edward Shuryak and Ismail Zahed.
\newblock {How to observe the QCD instanton/sphaleron processes at hadron
  colliders?}
\newblock 1 2021{\natexlab{a}}.

\bibitem[Atiyah and Singer(1963)]{Atiyah:1963index}
Michael~F. Atiyah and Isadore~M. Singer.
\newblock The index of elliptic operators on compact manifolds.
\newblock \emph{Bulletin of the American Mathematical Society}, 69:\penalty0
  322--433, 1963.
\newblock \doi{10.1090/S0002-9904-1963-10957-X}.

\bibitem[Greensite(2017)]{Greensite:2016pfc}
Jeff Greensite.
\newblock {Confinement from Center Vortices: A review of old and new results}.
\newblock \emph{EPJ Web Conf.}, 137:\penalty0 01009, 2017.
\newblock \doi{10.1051/epjconf/201713701009}.

\bibitem[Diakonov(1999)]{Diakonov:1999gg}
Dmitri Diakonov.
\newblock {Potential energy of Yang-Mills vortices in three-dimensions and
  four-dimensions}.
\newblock \emph{Mod. Phys. Lett. A}, 14:\penalty0 1725--1732, 1999.
\newblock \doi{10.1142/S0217732399001826}.

\bibitem[Diakonov and Maul(2002)]{Diakonov:2002bx}
Dmitri Diakonov and Martin Maul.
\newblock {Center vortex solutions of the Yang-Mills effective action in three
  and four dimensions}.
\newblock \emph{Phys. Rev. D}, 66:\penalty0 096004, 2002.
\newblock \doi{10.1103/PhysRevD.66.096004}.

\bibitem[Maas(2006)]{Maas:2005qt}
Axel Maas.
\newblock {On the spectrum of the Faddeev-Popov operator in topological
  background fields}.
\newblock \emph{Eur. Phys. J. C}, 48:\penalty0 179--192, 2006.
\newblock \doi{10.1140/epjc/s10052-006-0003-y}.

\bibitem[Biddle et~al.(2020{\natexlab{b}})Biddle, Kamleh, and
  Leinweber]{Biddle:2019gke}
James~C. Biddle, Waseem Kamleh, and Derek~B. Leinweber.
\newblock {Visualization of center vortex structure}.
\newblock \emph{Phys. Rev. D}, 102\penalty0 (3):\penalty0 034504,
  2020{\natexlab{b}}.
\newblock \doi{10.1103/PhysRevD.102.034504}.

\bibitem[Langfeld et~al.(1999)Langfeld, Tennert, Engelhardt, and
  Reinhardt]{Langfeld:1998cz}
K.~Langfeld, O.~Tennert, M.~Engelhardt, and H.~Reinhardt.
\newblock {Center vortices of Yang-Mills theory at finite temperatures}.
\newblock \emph{Phys. Lett. B}, 452:\penalty0 301, 1999.
\newblock \doi{10.1016/S0370-2693(99)00252-X}.

\bibitem[Kamleh et~al.(2024)Kamleh, Leinweber, and Virgili]{Kamleh:2023gho}
Waseem Kamleh, Derek~B. Leinweber, and Adam Virgili.
\newblock {Numerical indication that center vortices drive dynamical mass
  generation in QCD}.
\newblock \emph{Phys. Rev. D}, 110\penalty0 (5):\penalty0 L051502, 2024.
\newblock \doi{10.1103/PhysRevD.110.L051502}.

\bibitem[Nguyen et~al.(2025)Nguyen, Sulejmanpasic, and
  {\"U}nsal]{Nguyen:2024ikq}
Mendel Nguyen, Tin Sulejmanpasic, and Mithat {\"U}nsal.
\newblock {Phases of Theories with ZN 1-Form Symmetry, and the Roles of Center
  Vortices and Magnetic Monopoles}.
\newblock \emph{Phys. Rev. Lett.}, 134\penalty0 (14):\penalty0 141902, 2025.
\newblock \doi{10.1103/PhysRevLett.134.141902}.

\bibitem[Nguyen and {\"U}nsal(2025)]{Nguyen:2025voy}
Mendel Nguyen and Mithat {\"U}nsal.
\newblock {Self-dual monopole loops, instantons and confinement}.
\newblock 9 2025.

\bibitem[Hayashi and Tanizaki(2024)]{Hayashi:2024yjc}
Yui Hayashi and Yuya Tanizaki.
\newblock {Unifying Monopole and Center Vortex as the Semiclassical Confinement
  Mechanism}.
\newblock \emph{Phys. Rev. Lett.}, 133\penalty0 (17):\penalty0 171902, 2024.
\newblock \doi{10.1103/PhysRevLett.133.171902}.

\bibitem[G{\"u}vendik et~al.(2024)G{\"u}vendik, Schaefer, and
  {\"U}nsal]{Guvendik:2024umd}
Canberk G{\"u}vendik, Thomas Schaefer, and Mithat {\"U}nsal.
\newblock {The metamorphosis of semi-classical mechanisms of confinement: from
  monopoles on {\ensuremath{\mathbb{R}}}$^{3}$ {\texttimes} S$^{1}$ to
  center-vortices on {\ensuremath{\mathbb{R}}}$^{2}$ {\texttimes} T$^{2}$}.
\newblock \emph{JHEP}, 11:\penalty0 163, 2024.
\newblock \doi{10.1007/JHEP11(2024)163}.

\bibitem[Kraan and van Baal(1998{\natexlab{b}})]{Kraan:1998pm}
Thomas~C. Kraan and Pierre van Baal.
\newblock {Periodic instantons with nontrivial holonomy}.
\newblock \emph{Nucl. Phys. B}, 533:\penalty0 627--659, 1998{\natexlab{b}}.
\newblock \doi{10.1016/S0550-3213(98)00590-2}.

\bibitem[Lee and Lu(1998)]{Lee:1998bb}
Ki-Myeong Lee and Chang-hai Lu.
\newblock {SU(2) calorons and magnetic monopoles}.
\newblock \emph{Phys. Rev. D}, 58:\penalty0 025011, 1998.
\newblock \doi{10.1103/PhysRevD.58.025011}.

\bibitem[Kraan and van Baal(1998{\natexlab{c}})]{Kraan:1998kp}
Thomas~C. Kraan and Pierre van Baal.
\newblock {Exact T duality between calorons and Taub - NUT spaces}.
\newblock \emph{Phys. Lett. B}, 428:\penalty0 268--276, 1998{\natexlab{c}}.
\newblock \doi{10.1016/S0370-2693(98)00411-0}.

\bibitem[Zhitnitsky(2006)]{Zhitnitsky:2006sr}
Ariel~R. Zhitnitsky.
\newblock {Confinement- deconfinement phase transition and fractional instanton
  quarks in dense matter}.
\newblock In \emph{{Light-Cone QCD and Nonperturbative Hadron Physics}}, pages
  207--213, 1 2006.
\newblock \doi{10.1142/9789812708267_0023}.

\bibitem[Unsal and Yaffe(2008)]{Unsal:2008ch}
Mithat Unsal and Laurence~G. Yaffe.
\newblock {Center-stabilized Yang-Mills theory: Confinement and large N volume
  independence}.
\newblock \emph{Phys. Rev. D}, 78:\penalty0 065035, 2008.
\newblock \doi{10.1103/PhysRevD.78.065035}.

\bibitem[Liu et~al.(2015{\natexlab{a}})Liu, Shuryak, and Zahed]{Liu:2015jsa}
Yizhuang Liu, Edward Shuryak, and Ismail Zahed.
\newblock {Light quarks in the screened dyon-antidyon Coulomb liquid model.
  II.}
\newblock \emph{Phys. Rev. D}, 92\penalty0 (8):\penalty0 085007,
  2015{\natexlab{a}}.
\newblock \doi{10.1103/PhysRevD.92.085007}.

\bibitem[Liu et~al.(2015{\natexlab{b}})Liu, Shuryak, and Zahed]{Liu:2015ufa}
Yizhuang Liu, Edward Shuryak, and Ismail Zahed.
\newblock {Confining dyon-antidyon Coulomb liquid model. I.}
\newblock \emph{Phys. Rev. D}, 92\penalty0 (8):\penalty0 085006,
  2015{\natexlab{b}}.
\newblock \doi{10.1103/PhysRevD.92.085006}.

\bibitem[Diakonov(2009)]{Diakonov:2009jq}
Dmitri Diakonov.
\newblock {Topology and confinement}.
\newblock \emph{Nucl. Phys. B Proc. Suppl.}, 195:\penalty0 5--45, 2009.
\newblock \doi{10.1016/j.nuclphysbps.2009.10.010}.

\bibitem[Larsen and Shuryak(2015)]{Larsen:2015vaa}
Rasmus Larsen and Edward Shuryak.
\newblock {Interacting ensemble of the instanton-dyons and the deconfinement
  phase transition in the SU(2) gauge theory}.
\newblock \emph{Phys. Rev. D}, 92\penalty0 (9):\penalty0 094022, 2015.
\newblock \doi{10.1103/PhysRevD.92.094022}.

\bibitem[Larsen and Shuryak(2016)]{Larsen:2015tso}
Rasmus Larsen and Edward Shuryak.
\newblock {Instanton-dyon Ensemble with two Dynamical Quarks: the Chiral
  Symmetry Breaking}.
\newblock \emph{Phys. Rev. D}, 93\penalty0 (5):\penalty0 054029, 2016.
\newblock \doi{10.1103/PhysRevD.93.054029}.

\bibitem[Shuryak(2018)]{Shuryak:2017kct}
E.~Shuryak.
\newblock {Instanton-dyon ensembles reproduce deconfinement and chiral
  restoration phase transitions}.
\newblock \emph{EPJ Web Conf.}, 175:\penalty0 12001, 2018.
\newblock \doi{10.1051/epjconf/201817512001}.

\bibitem[Unsal(2009)]{Unsal:2007jx}
Mithat Unsal.
\newblock {Magnetic bion condensation: A New mechanism of confinement and mass
  gap in four dimensions}.
\newblock \emph{Phys. Rev. D}, 80:\penalty0 065001, 2009.
\newblock \doi{10.1103/PhysRevD.80.065001}.

\bibitem[Poppitz et~al.(2012)Poppitz, Sch\"afer, and Unsal]{Poppitz:2012sw}
Erich Poppitz, Thomas Sch\"afer, and Mithat Unsal.
\newblock {Continuity, Deconfinement, and (Super) Yang-Mills Theory}.
\newblock \emph{JHEP}, 10:\penalty0 115, 2012.
\newblock \doi{10.1007/JHEP10(2012)115}.

\bibitem[Dorey and Parnachev(2001)]{Dorey:2000qc}
Nick Dorey and Andrei Parnachev.
\newblock {Instantons, compactification and S duality in N=4 SUSY Yang-Mills
  theory. 2.}
\newblock \emph{JHEP}, 08:\penalty0 059, 2001.
\newblock \doi{10.1088/1126-6708/2001/08/059}.

\bibitem[Poppitz and Unsal(2011)]{Poppitz:2011wy}
Erich Poppitz and Mithat Unsal.
\newblock {Seiberg-Witten and 'Polyakov-like' magnetic bion confinements are
  continuously connected}.
\newblock \emph{JHEP}, 07:\penalty0 082, 2011.
\newblock \doi{10.1007/JHEP07(2011)082}.

\bibitem[Ramamurti et~al.(2018)Ramamurti, Shuryak, and
  Zahed]{Ramamurti:2018evz}
Adith Ramamurti, Edward Shuryak, and Ismail Zahed.
\newblock {Are there monopoles in the quark-gluon plasma?}
\newblock \emph{Phys. Rev. D}, 97\penalty0 (11):\penalty0 114028, 2018.
\newblock \doi{10.1103/PhysRevD.97.114028}.

\bibitem['t~Hooft(1976)]{tHooft:1976snw}
Gerard 't~Hooft.
\newblock {Computation of the Quantum Effects Due to a Four-Dimensional
  Pseudoparticle}.
\newblock \emph{Phys. Rev. D}, 14:\penalty0 3432--3450, 1976.
\newblock \doi{10.1103/PhysRevD.14.3432}.
\newblock [Erratum: Phys.Rev.D 18, 2199 (1978)].

\bibitem[Shuryak and Verbaarschot(1990)]{Shuryak:1989cx}
Edward~V. Shuryak and J.~J.~M. Verbaarschot.
\newblock {Chiral Symmetry Breaking and Correlations in the Instanton Liquid}.
\newblock \emph{Nucl. Phys. B}, 341:\penalty0 1--26, 1990.
\newblock \doi{10.1016/0550-3213(90)90260-K}.

\bibitem[Shuryak(1988{\natexlab{b}})]{Shuryak:1987iz}
Edward~V. Shuryak.
\newblock {Toward the Quantitative Theory of the Topological Effects in Gauge
  Field Theories. 2. The SU(2) Gluodynamics}.
\newblock \emph{Nucl. Phys. B}, 302:\penalty0 574--598, 1988{\natexlab{b}}.
\newblock \doi{10.1016/0550-3213(88)90189-7}.

\bibitem[Shuryak(1999)]{Shuryak:1999fe}
Edward~V. Shuryak.
\newblock {Probing the boundary of the nonperturbative QCD by small size
  instantons}.
\newblock 9 1999.

\bibitem[Dorokhov and Cherednikov(2002)]{Dorokhov:2002qf}
Alexander~E. Dorokhov and Igor~O. Cherednikov.
\newblock {Instanton contributions to the quark form-factor}.
\newblock \emph{Phys. Rev. D}, 66:\penalty0 074009, 2002.
\newblock \doi{10.1103/PhysRevD.66.074009}.

\bibitem[Shuryak(1995)]{Shuryak:1995pv}
Edward~V. Shuryak.
\newblock {Instanton size distribution: Repulsion or the infrared fixed point?}
\newblock \emph{Phys. Rev. D}, 52:\penalty0 5370--5373, 1995.
\newblock \doi{10.1103/PhysRevD.52.5370}.

\bibitem[Diakonov et~al.(1996)Diakonov, Polyakov, and Weiss]{Diakonov:1995qy}
Dmitri Diakonov, Maxim~V. Polyakov, and C.~Weiss.
\newblock {Hadronic matrix elements of gluon operators in the instanton
  vacuum}.
\newblock \emph{Nucl. Phys. B}, 461:\penalty0 539--580, 1996.
\newblock \doi{10.1016/0550-3213(95)00675-3}.

\bibitem[Millo and Faccioli(2011)]{Millo:2011zn}
Raffaele Millo and Pietro Faccioli.
\newblock {Computing the Effective Hamiltonian of Low-Energy Vacuum Gauge
  Fields}.
\newblock \emph{Phys. Rev. D}, 84:\penalty0 034504, 2011.
\newblock \doi{10.1103/PhysRevD.84.034504}.

\bibitem[Hasenfratz and Nieter(1998)]{Hasenfratz:1998qk}
Anna Hasenfratz and Chet Nieter.
\newblock {Instanton content of the SU(3) vacuum}.
\newblock \emph{Phys. Lett. B}, 439:\penalty0 366--372, 1998.
\newblock \doi{10.1016/S0370-2693(98)01058-2}.

\bibitem[Smith and Teper(1998)]{Smith:1998wt}
Douglas~A. Smith and Michael~J. Teper.
\newblock {Topological structure of the SU(3) vacuum}.
\newblock \emph{Phys. Rev. D}, 58:\penalty0 014505, 1998.
\newblock \doi{10.1103/PhysRevD.58.014505}.

\bibitem[Negele(1999)]{Negele:1998ev}
John~W. Negele.
\newblock {Instantons, the QCD vacuum, and hadronic physics}.
\newblock \emph{Nucl. Phys. B Proc. Suppl.}, 73:\penalty0 92--104, 1999.
\newblock \doi{10.1016/S0920-5632(99)85010-5}.

\bibitem[Sch\"afer et~al.(1995)Sch\"afer, Shuryak, and
  Verbaarschot]{Schafer:1994nv}
Thomas Sch\"afer, Edward~V. Shuryak, and J.~J.~M. Verbaarschot.
\newblock {The Chiral phase transition and instanton - anti-instanton
  molecules}.
\newblock \emph{Phys. Rev. D}, 51:\penalty0 1267--1281, 1995.
\newblock \doi{10.1103/PhysRevD.51.1267}.

\bibitem[Khoze and Ringwald(1991)]{Khoze:1991sa}
Valentin~V. Khoze and A.~Ringwald.
\newblock {Valley trajectories in gauge theories}.
\newblock 5 1991.

\bibitem[Shuryak and Verbaarschot(1992)]{Shuryak:1991pn}
Edward~V. Shuryak and J.~J.~M. Verbaarschot.
\newblock {On baryon number violation and nonperturbative weak processes at SSC
  energies}.
\newblock \emph{Phys. Rev. Lett.}, 68:\penalty0 2576--2579, 1992.
\newblock \doi{10.1103/PhysRevLett.68.2576}.

\bibitem[Forster(1977)]{Forster:1977jv}
D.~Forster.
\newblock {On the Structure of Instanton Plasma in the Two-Dimensional O(3)
  Nonlinear Sigma Model}.
\newblock \emph{Nucl. Phys. B}, 130:\penalty0 38--60, 1977.
\newblock \doi{10.1016/0550-3213(77)90391-1}.

\bibitem[Zakharov(1992)]{Zakharov:1992bx}
Valentin~I. Zakharov.
\newblock {QCD perturbative expansions in large orders}.
\newblock \emph{Nucl. Phys. B}, 385:\penalty0 452--480, 1992.
\newblock \doi{10.1016/0550-3213(92)90054-F}.

\bibitem[Zahed(2021)]{Zahed:2021fxk}
Ismail Zahed.
\newblock {Mass sum rule of hadrons in the QCD instanton vacuum}.
\newblock \emph{Phys. Rev. D}, 104\penalty0 (5):\penalty0 054031, 2021.
\newblock \doi{10.1103/PhysRevD.104.054031}.

\bibitem[Zahed(2022)]{Zahed:2022wae}
Ismail Zahed.
\newblock {Spin Sum Rule of the Nucleon in the QCD Instanton Vacuum}.
\newblock \emph{Symmetry}, 14\penalty0 (5):\penalty0 932, 2022.
\newblock \doi{10.3390/sym14050932}.

\bibitem[Kacir et~al.(1999)Kacir, Prakash, and Zahed]{Kacir:1996qn}
M.~Kacir, M.~Prakash, and I.~Zahed.
\newblock {Hadrons and QCD instantons: A Bosonized view}.
\newblock \emph{Acta Phys. Polon. B}, 30:\penalty0 287--348, 1999.

\bibitem[Nowak et~al.(1996{\natexlab{a}})Nowak, Rho, and Zahed]{Nowak:1996aj}
Maciej~A. Nowak, Mannque Rho, and I.~Zahed.
\newblock \emph{{Chiral nuclear dynamics}}.
\newblock World Scientific Publishing Company, 1996{\natexlab{a}}.

\bibitem[Witten(1979)]{Witten:1979vv}
Edward Witten.
\newblock {Current Algebra Theorems for the U(1) Goldstone Boson}.
\newblock \emph{Nucl. Phys. B}, 156:\penalty0 269--283, 1979.
\newblock \doi{10.1016/0550-3213(79)90031-2}.

\bibitem[Veneziano(1979)]{Veneziano:1979ec}
G.~Veneziano.
\newblock {U(1) Without Instantons}.
\newblock \emph{Nucl. Phys. B}, 159:\penalty0 213--224, 1979.
\newblock \doi{10.1016/0550-3213(79)90332-8}.

\bibitem[Novikov et~al.(1981)Novikov, Shifman, Vainshtein, and
  Zakharov]{Novikov:1981xi}
V.~A. Novikov, Mikhail~A. Shifman, A.~I. Vainshtein, and Valentin~I. Zakharov.
\newblock {Are All Hadrons Alike?~}.
\newblock \emph{Nucl. Phys. B}, 191:\penalty0 301--369, 1981.
\newblock \doi{10.1016/0550-3213(81)90303-5}.

\bibitem[{Verbaarschot}(1991)]{1991NuPhB.362...33V}
J.~J.~M. {Verbaarschot}.
\newblock {Streamlines and conformal invariance in Yang-Mills theories}.
\newblock \emph{Nuclear Physics B}, 362\penalty0 (1):\penalty0 33--53,
  September 1991.
\newblock \doi{10.1016/0550-3213(91)90554-B}.

\bibitem[Wantz(2010{\natexlab{a}})]{Wantz:2009qk}
Olivier Wantz.
\newblock {The Topological susceptibility from grand canonical simulations in
  the interacting instanton liquid model: Zero temperature calibrations and
  numerical framework}.
\newblock \emph{Nucl. Phys. B}, 829:\penalty0 48--90, 2010{\natexlab{a}}.
\newblock \doi{10.1016/j.nuclphysb.2009.12.007}.

\bibitem[Dunne et~al.(2005)Dunne, Hur, Lee, and Min]{Dunne:2005te}
Gerald~V. Dunne, Jin Hur, Choonkyu Lee, and Hyunsoo Min.
\newblock {Calculation of QCD instanton determinant with arbitrary mass}.
\newblock \emph{Phys. Rev. D}, 71:\penalty0 085019, 2005.
\newblock \doi{10.1103/PhysRevD.71.085019}.

\bibitem[Glozman et~al.(2012)Glozman, Lang, and Schrock]{Glozman:2012fj}
L.~Ya. Glozman, C.~B. Lang, and M.~Schrock.
\newblock {Symmetries of hadrons after unbreaking the chiral symmetry}.
\newblock \emph{Phys. Rev. D}, 86:\penalty0 014507, 2012.
\newblock \doi{10.1103/PhysRevD.86.014507}.

\bibitem[Pobylitsa(1989{\natexlab{a}})]{Pobylitsa1989TheQP}
P.~V. Pobylitsa.
\newblock The quark propagator and correlation functions in the instanton
  vacuum.
\newblock \emph{Physics Letters B}, 226:\penalty0 387--392, 1989{\natexlab{a}}.
\newblock URL \url{https://api.semanticscholar.org/CorpusID:120858371}.

\bibitem[Kock et~al.(2020)Kock, Liu, and Zahed]{Kock:2020frx}
Arthur Kock, Yizhuang Liu, and Ismail Zahed.
\newblock {Pion and kaon parton distributions in the QCD instanton vacuum}.
\newblock \emph{Phys. Rev. D}, 102\penalty0 (1):\penalty0 014039, 2020.
\newblock \doi{10.1103/PhysRevD.102.014039}.

\bibitem[Kock and Zahed(2021)]{Kock:2021spt}
Arthur Kock and Ismail Zahed.
\newblock {Pion and kaon distribution amplitudes up to twist-3 in the QCD
  instanton vacuum}.
\newblock \emph{Phys. Rev. D}, 104\penalty0 (11):\penalty0 116028, 2021.
\newblock \doi{10.1103/PhysRevD.104.116028}.

\bibitem[Oliveira et~al.(2019)Oliveira, Silva, Skullerud, and
  Sternbeck]{Oliveira:2018lln}
Orlando Oliveira, Paulo~J. Silva, Jon-Ivar Skullerud, and Andre Sternbeck.
\newblock {Quark propagator with two flavors of O(a)-improved Wilson fermions}.
\newblock \emph{Phys. Rev. D}, 99\penalty0 (9):\penalty0 094506, 2019.
\newblock \doi{10.1103/PhysRevD.99.094506}.

\bibitem[Bowman et~al.(2004)Bowman, Heller, Leinweber, Williams, and
  Zhang]{Bowman:2004xi}
Patrick~O. Bowman, Urs~M. Heller, Derek~B. Leinweber, Anthony~G. Williams, and
  Jian-bo Zhang.
\newblock {Infrared and ultraviolet properties of the Landau gauge quark
  propagator}.
\newblock \emph{Nucl. Phys. B Proc. Suppl.}, 128:\penalty0 23--29, 2004.
\newblock \doi{10.1016/S0920-5632(03)02454-X}.

\bibitem[Chernyshev et~al.(1996)Chernyshev, Nowak, and
  Zahed]{Chernyshev:1995gj}
S.~Chernyshev, Maciej~A. Nowak, and I.~Zahed.
\newblock {Heavy hadrons and QCD instantons}.
\newblock \emph{Phys. Rev. D}, 53:\penalty0 5176--5184, 1996.
\newblock \doi{10.1103/PhysRevD.53.5176}.

\bibitem[Shuryak and Zahed(2000)]{Shuryak:2000df}
Edward~V. Shuryak and Ismail Zahed.
\newblock {Instanton induced effects in QCD high-energy scattering}.
\newblock \emph{Phys. Rev. D}, 62:\penalty0 085014, 2000.
\newblock \doi{10.1103/PhysRevD.62.085014}.

\bibitem[Shuryak and Zahed(2023{\natexlab{d}})]{Shuryak:2021hng}
Edward Shuryak and Ismail Zahed.
\newblock {Hadronic structure on the light front. II. QCD strings, Wilson
  lines, and potentials}.
\newblock \emph{Phys. Rev. D}, 107\penalty0 (3):\penalty0 034024,
  2023{\natexlab{d}}.
\newblock \doi{10.1103/PhysRevD.107.034024}.

\bibitem[Nowak et~al.(1996{\natexlab{b}})Nowak, Rho, and
  Zahed]{doi:10.1142/1681}
Maciej~A Nowak, Mannque Rho, and Ismail Zahed.
\newblock \emph{Chiral Nuclear Dynamics}.
\newblock WORLD SCIENTIFIC, 1996{\natexlab{b}}.
\newblock \doi{10.1142/1681}.
\newblock URL \url{https://www.worldscientific.com/doi/abs/10.1142/1681}.

\bibitem[Leutwyler and Smilga(1992)]{Leutwyler:1992yt}
H.~Leutwyler and Andrei~V. Smilga.
\newblock {Spectrum of Dirac operator and role of winding number in QCD}.
\newblock \emph{Phys. Rev. D}, 46:\penalty0 5607--5632, 1992.
\newblock \doi{10.1103/PhysRevD.46.5607}.

\bibitem[Navas et~al.(2024)]{ParticleDataGroup:2024cfk}
S.~Navas et~al.
\newblock {Review of particle physics}.
\newblock \emph{Phys. Rev. D}, 110\penalty0 (3):\penalty0 030001, 2024.
\newblock \doi{10.1103/PhysRevD.110.030001}.

\bibitem[Aoki et~al.(2022)]{FlavourLatticeAveragingGroupFLAG:2021npn}
Y.~Aoki et~al.
\newblock {FLAG Review 2021}.
\newblock \emph{Eur. Phys. J. C}, 82\penalty0 (10):\penalty0 869, 2022.
\newblock \doi{10.1140/epjc/s10052-022-10536-1}.

\bibitem[Harnett et~al.(2021)Harnett, Ho, and Steele]{Harnett:2021zug}
D.~Harnett, J.~Ho, and T.~G. Steele.
\newblock {Correlations Between the Strange Quark Condensate, Strange Quark
  Mass, and Kaon PCAC Relation}.
\newblock \emph{Phys. Rev. D}, 103\penalty0 (11):\penalty0 114005, 2021.
\newblock \doi{10.1103/PhysRevD.103.114005}.

\bibitem[Liang et~al.(2023)Liang, Alexandru, Draper, Liu, Wang, Wang, and
  Yang]{Liang:2023jfj}
Jian Liang, Andrei Alexandru, Terrence Draper, Keh-Fei Liu, Bigeng Wang, Gen
  Wang, and Yi-Bo Yang.
\newblock {Nucleon electric dipole moment from the \ensuremath{\theta} term
  with lattice chiral fermions}.
\newblock \emph{Phys. Rev. D}, 108\penalty0 (9):\penalty0 094512, 2023.
\newblock \doi{10.1103/PhysRevD.108.094512}.

\bibitem[Bhattacharya et~al.(2021)Bhattacharya, Cirigliano, Gupta, Mereghetti,
  and Yoon]{Bhattacharya:2021lol}
Tanmoy Bhattacharya, Vincenzo Cirigliano, Rajan Gupta, Emanuele Mereghetti, and
  Boram Yoon.
\newblock {Contribution of the QCD $\Theta$-term to the nucleon electric dipole
  moment}.
\newblock \emph{Phys. Rev. D}, 103\penalty0 (11):\penalty0 114507, 2021.
\newblock \doi{10.1103/PhysRevD.103.114507}.

\bibitem[Alexandrou et~al.(2021)Alexandrou, Athenodorou, Hadjiyiannakou, and
  Todaro]{Alexandrou:2020mds}
C.~Alexandrou, A.~Athenodorou, K.~Hadjiyiannakou, and A.~Todaro.
\newblock {Neutron electric dipole moment using lattice QCD simulations at the
  physical point}.
\newblock \emph{Phys. Rev. D}, 103\penalty0 (5):\penalty0 054501, 2021.
\newblock \doi{10.1103/PhysRevD.103.054501}.

\bibitem[Pobylitsa(1989{\natexlab{b}})]{Pobylitsa:1989uq}
P.~V. Pobylitsa.
\newblock {The Quark Propagator and Correlation Functions in the Instanton
  Vacuum}.
\newblock \emph{Phys. Lett. B}, 226:\penalty0 387--392, 1989{\natexlab{b}}.
\newblock \doi{10.1016/0370-2693(89)91216-1}.

\bibitem[Faccioli and Shuryak(2001)]{Faccioli:2001ug}
P.~Faccioli and Edward~V. Shuryak.
\newblock {Systematic study of the single instanton approximation in QCD}.
\newblock \emph{Phys. Rev. D}, 64:\penalty0 114020, 2001.
\newblock \doi{10.1103/PhysRevD.64.114020}.

\bibitem[Shuryak(1988{\natexlab{c}})]{Shuryak:1987ja}
Edward~V. Shuryak.
\newblock {Toward the Quantitative Theory of the Topological Phenomena in Gauge
  Theories. 3. Instantons and Light Fermions}.
\newblock \emph{Nucl. Phys. B}, 302:\penalty0 599--620, 1988{\natexlab{c}}.
\newblock \doi{10.1016/0550-3213(88)90190-3}.

\bibitem[Weiss(2021)]{Weiss:2021kpt}
C.~Weiss.
\newblock {Nucleon matrix element of Weinberg's CP-odd gluon operator from the
  instanton vacuum}.
\newblock \emph{Phys. Lett. B}, 819:\penalty0 136447, 2021.
\newblock \doi{10.1016/j.physletb.2021.136447}.

\bibitem[Shuryak(1988{\natexlab{d}})]{Shuryak:1988ff}
Edward~V. Shuryak.
\newblock {THE 'INSTANTON LIQUID.'}.
\newblock \emph{Z. Phys. C}, 38:\penalty0 165--172, 1988{\natexlab{d}}.
\newblock \doi{10.1007/BF01574532}.

\bibitem[Nowak et~al.(1989)Nowak, Verbaarschot, and Zahed]{Nowak:1988bh}
Maciej~A. Nowak, J.~J.~M. Verbaarschot, and I.~Zahed.
\newblock {Flavor Mixing in the Instanton Vacuum}.
\newblock \emph{Nucl. Phys. B}, 324:\penalty0 1--33, 1989.
\newblock \doi{10.1016/0550-3213(89)90178-8}.

\bibitem[Shuryak(1984)]{Shuryak:1983ni}
Edward~V. Shuryak.
\newblock {Theory and phenomenology of the QCD vacuum}.
\newblock \emph{Phys. Rept.}, 115:\penalty0 151, 1984.
\newblock \doi{10.1016/0370-1573(84)90037-1}.

\bibitem[Wantz(2010{\natexlab{b}})]{Wantz:2009kh}
Olivier Wantz.
\newblock {The Topological susceptibility from grand canonical simulations in
  the interacting instanton liquid model: strongly associating fluids and
  biased Monte Carlo}.
\newblock \emph{Nucl. Phys. B}, 829:\penalty0 91--109, 2010{\natexlab{b}}.
\newblock \doi{10.1016/j.nuclphysb.2009.12.006}.

\bibitem[Wantz and Shellard(2010)]{Wantz:2009mi}
Olivier Wantz and E.~P.~S. Shellard.
\newblock {The Topological susceptibility from grand canonical simulations in
  the interacting instanton liquid model: Chiral phase transition and axion
  mass}.
\newblock \emph{Nucl. Phys. B}, 829:\penalty0 110--160, 2010.
\newblock \doi{10.1016/j.nuclphysb.2009.12.005}.

\bibitem[Woodbury(1950)]{woodbury1950inverting}
Max~A Woodbury.
\newblock \emph{Inverting modified matrices}.
\newblock Department of Statistics, Princeton University, 1950.

\bibitem[Sherman and Morrison(1950)]{sherman1950adjustment}
Jack Sherman and Winifred~J Morrison.
\newblock Adjustment of an inverse matrix corresponding to a change in one
  element of a given matrix.
\newblock \emph{The Annals of Mathematical Statistics}, 21\penalty0
  (1):\penalty0 124--127, 1950.

\bibitem[Sch{\"a}fer and Shuryak(1996)]{Schafer:1995pz}
Thomas Sch{\"a}fer and Edward~V. Shuryak.
\newblock {The Interacting instanton liquid in QCD at zero and finite
  temperature}.
\newblock \emph{Phys. Rev. D}, 53:\penalty0 6522--6542, 1996.
\newblock \doi{10.1103/PhysRevD.53.6522}.

\bibitem[Faccioli and Shuryak(2002)]{Faccioli:2001qg}
P.~Faccioli and Edward~V. Shuryak.
\newblock {Proton electromagnetic form-factors in the instanton liquid model}.
\newblock \emph{Phys. Rev. D}, 65:\penalty0 076002, 2002.
\newblock \doi{10.1103/PhysRevD.65.076002}.

\bibitem[Kirkwood(1935)]{kirkwood1935statistical}
John~G Kirkwood.
\newblock Statistical mechanics of fluid mixtures.
\newblock \emph{The Journal of chemical physics}, 3\penalty0 (5):\penalty0
  300--313, 1935.

\bibitem[Balla et~al.(1998{\natexlab{a}})Balla, Polyakov, and
  Weiss]{Balla:1998rt}
J.~Balla, Maxim~V. Polyakov, and C.~Weiss.
\newblock {Nucleon matrix elements of twist - three and twist -4 operators from
  the instanton vacuum}.
\newblock In \emph{{8th International Conference on the Structure of Baryons}},
  pages 310--315, 9 1998{\natexlab{a}}.

\bibitem[Balla et~al.(1998{\natexlab{b}})Balla, Polyakov, and
  Weiss]{Balla:1997hf}
J.~Balla, Maxim~V. Polyakov, and C.~Weiss.
\newblock {Nucleon matrix elements of higher twist operators from the instanton
  vacuum}.
\newblock \emph{Nucl. Phys. B}, 510:\penalty0 327--364, 1998{\natexlab{b}}.
\newblock \doi{10.1016/S0550-3213(98)00638-5}.

\bibitem[Kochelev(1998)]{Kochelev:1996pv}
N.~I. Kochelev.
\newblock {Anomalous quark chromomagnetic moment induced by instantons}.
\newblock \emph{Phys. Lett. B}, 426:\penalty0 149--153, 1998.
\newblock \doi{10.1016/S0370-2693(98)00262-7}.

\bibitem[Qian and Zahed(2016)]{Qian:2015wyq}
Yachao Qian and Ismail Zahed.
\newblock {Spin Physics through QCD Instantons}.
\newblock \emph{Annals Phys.}, 374:\penalty0 314--337, 2016.
\newblock \doi{10.1016/j.aop.2016.09.002}.

\bibitem[Diakonov(2003)]{Diakonov:2002fq}
Dmitri Diakonov.
\newblock {Instantons at work}.
\newblock \emph{Prog. Part. Nucl. Phys.}, 51:\penalty0 173--222, 2003.
\newblock \doi{10.1016/S0146-6410(03)90014-7}.

\bibitem[Kochelev et~al.(2016)Kochelev, Lee, Zhang, and
  Zhang]{Kochelev:2015pqd}
Nikolai Kochelev, Hee-Jung Lee, Baiyang Zhang, and Pengming Zhang.
\newblock {Gluonic Structure of the Constituent Quark}.
\newblock \emph{Phys. Lett. B}, 757:\penalty0 420--425, 2016.
\newblock \doi{10.1016/j.physletb.2016.04.027}.

\bibitem[Kochelev and Korchagin(2014)]{Kochelev:2013csa}
Nikolai Kochelev and Nikolai Korchagin.
\newblock {Anomalous Quark Chromomagnetic Moment and Dynamics of Elastic
  Scattering}.
\newblock \emph{Phys. Rev. D}, 89\penalty0 (3):\penalty0 034028, 2014.
\newblock \doi{10.1103/PhysRevD.89.034028}.

\bibitem[Zhang et~al.(2017)Zhang, Radzhabov, Kochelev, and
  Zhang]{Zhang:2017zpi}
Baiyang Zhang, Andrey Radzhabov, Nikolai Kochelev, and Pengming Zhang.
\newblock {Pauli form factor of quark and nontrivial topological structure of
  the QCD}.
\newblock \emph{Phys. Rev. D}, 96\penalty0 (5):\penalty0 054030, 2017.
\newblock \doi{10.1103/PhysRevD.96.054030}.

\bibitem[Cherednikov et~al.(2006)Cherednikov, D'Alesio, Kochelev, and
  Murgia]{Cherednikov:2006zn}
I.~O. Cherednikov, U.~D'Alesio, N.~I. Kochelev, and F.~Murgia.
\newblock {Instanton contribution to the Sivers function}.
\newblock \emph{Phys. Lett. B}, 642:\penalty0 39--47, 2006.
\newblock \doi{10.1016/j.physletb.2006.09.019}.

\bibitem[Kochelev(2008)]{Kochelev:2008fy}
N.~I. Kochelev.
\newblock {Instantons and Spin-Flavor effects in Hadron Physics}.
\newblock In \emph{{15th Annual Seminar Nonlinear Phenomena in Complex Systems:
  Chaos, Fractals, Phase Transitions, Self-organization}}, 9 2008.

\bibitem[Kim and Weiss(2024)]{Kim:2023pll}
June-Young Kim and Christian Weiss.
\newblock {Instanton effects in twist-3 generalized parton distributions}.
\newblock \emph{Phys. Lett. B}, 848:\penalty0 138387, 2024.
\newblock \doi{10.1016/j.physletb.2023.138387}.

\bibitem[Shuryak(1993)]{Shuryak:1993kg}
Edward~V. Shuryak.
\newblock {Correlation functions in the QCD vacuum}.
\newblock \emph{Rev. Mod. Phys.}, 65:\penalty0 1--46, 1993.
\newblock \doi{10.1103/RevModPhys.65.1}.

\bibitem[Shuryak and Zahed(2023{\natexlab{e}})]{Shuryak:2021yif}
Edward Shuryak and Ismail Zahed.
\newblock {Hadronic structure on the light front. III. The Hamiltonian, heavy
  quarkonia, spin, and orbit mixing}.
\newblock \emph{Phys. Rev. D}, 107\penalty0 (3):\penalty0 034025,
  2023{\natexlab{e}}.
\newblock \doi{10.1103/PhysRevD.107.034025}.

\bibitem[Osipov and Hiller(2004)]{Osipov:2003xu}
Alexander~A. Osipov and Brigitte Hiller.
\newblock {Path integral bosonization of the 't Hooft determinant:
  Quasiclassical corrections}.
\newblock \emph{Eur. Phys. J. C}, 35:\penalty0 223--241, 2004.
\newblock \doi{10.1140/epjc/s2004-01779-3}.

\bibitem[Diakonov(1997)]{Diakonov:1997sj}
Dmitri Diakonov.
\newblock {Chiral quark - soliton model}.
\newblock In \emph{{Advanced Summer School on Nonperturbative Quantum Field
  Physics}}, pages 1--55, 6 1997.

\bibitem[Weinberg(1991)]{Weinberg:1991um}
Steven Weinberg.
\newblock {Effective chiral Lagrangians for nucleon - pion interactions and
  nuclear forces}.
\newblock \emph{Nucl. Phys. B}, 363:\penalty0 3--18, 1991.
\newblock \doi{10.1016/0550-3213(91)90231-L}.

\bibitem[Liu and Zahed(2021)]{Liu:2021evw}
Yizhuang Liu and Ismail Zahed.
\newblock {Small size instanton contributions to the quark quasi-PDF and
  matching kernel}.
\newblock 2 2021.

\bibitem[Gasser and Leutwyler(1984)]{Gasser:1983yg}
J.~Gasser and H.~Leutwyler.
\newblock {Chiral Perturbation Theory to One Loop}.
\newblock \emph{Annals Phys.}, 158:\penalty0 142, 1984.
\newblock \doi{10.1016/0003-4916(84)90242-2}.

\bibitem[Scherer(2003)]{Scherer:2002tk}
Stefan Scherer.
\newblock {Introduction to chiral perturbation theory}.
\newblock \emph{Adv. Nucl. Phys.}, 27:\penalty0 277, 2003.

\bibitem[Lee et~al.(2021)Lee, Ohmori, and Tachikawa]{Lee:2020ojw}
Yasunori Lee, Kantaro Ohmori, and Yuji Tachikawa.
\newblock {Revisiting Wess-Zumino-Witten terms}.
\newblock \emph{SciPost Phys.}, 10\penalty0 (3):\penalty0 061, 2021.
\newblock \doi{10.21468/SciPostPhys.10.3.061}.

\bibitem[Ishii et~al.(1995)Ishii, Bentz, and Yazaki]{Ishii:1995bu}
N.~Ishii, W.~Bentz, and K.~Yazaki.
\newblock {Baryons in the NJL model as solutions of the relativistic Faddeev
  equation}.
\newblock \emph{Nucl. Phys. A}, 587:\penalty0 617--656, 1995.
\newblock \doi{10.1016/0375-9474(95)00032-V}.

\bibitem[Oertel et~al.(2000)Oertel, Buballa, and Wambach]{Oertel:2000sr}
M.~Oertel, M.~Buballa, and J.~Wambach.
\newblock {Meson properties in the 1/N(c) corrected NJL model}.
\newblock \emph{Nucl. Phys. A}, 676:\penalty0 247--272, 2000.
\newblock \doi{10.1016/S0375-9474(00)00198-6}.

\bibitem[Shuryak and Verbaarschot(1993{\natexlab{a}})]{Shuryak:1992jz}
Edward~V. Shuryak and J.~J.~M. Verbaarschot.
\newblock {Quark propagation in the random instanton vacuum}.
\newblock \emph{Nucl. Phys. B}, 410:\penalty0 37--54, 1993{\natexlab{a}}.
\newblock \doi{10.1016/0550-3213(93)90572-7}.

\bibitem[Shuryak and Verbaarschot(1993{\natexlab{b}})]{Shuryak:1992ke}
Edward~V. Shuryak and J.~J.~M. Verbaarschot.
\newblock {Mesonic correlation functions in the random instanton vacuum}.
\newblock \emph{Nucl. Phys. B}, 410:\penalty0 55--89, 1993{\natexlab{b}}.
\newblock \doi{10.1016/0550-3213(93)90573-8}.

\bibitem[Olive et~al.(2014)]{ParticleDataGroup:2014cgo}
K.~A. Olive et~al.
\newblock {Review of Particle Physics}.
\newblock \emph{Chin. Phys. C}, 38:\penalty0 090001, 2014.
\newblock \doi{10.1088/1674-1137/38/9/090001}.

\bibitem[Patrignani et~al.(2016)]{ParticleDataGroup:2016lqr}
C.~Patrignani et~al.
\newblock {Review of Particle Physics}.
\newblock \emph{Chin. Phys. C}, 40\penalty0 (10):\penalty0 100001, 2016.
\newblock \doi{10.1088/1674-1137/40/10/100001}.

\bibitem[Tanabashi et~al.(2018)]{ParticleDataGroup:2018ovx}
M.~Tanabashi et~al.
\newblock {Review of Particle Physics}.
\newblock \emph{Phys. Rev. D}, 98\penalty0 (3):\penalty0 030001, 2018.
\newblock \doi{10.1103/PhysRevD.98.030001}.

\bibitem[Aoki et~al.(2026)]{FlavourLatticeAveragingGroupFLAG:2024oxs}
Y.~Aoki et~al.
\newblock {FLAG review 2024}.
\newblock \emph{Phys. Rev. D}, 113\penalty0 (1):\penalty0 014508, 2026.
\newblock \doi{10.1103/nfzp-p5dn}.

\bibitem[Ji et~al.(2004)Ji, Ma, and Yuan]{Ji:2003yj}
Xiang-dong Ji, Jian-Ping Ma, and Feng Yuan.
\newblock {Classification and asymptotic scaling of hadrons' light cone wave
  function amplitudes}.
\newblock \emph{Eur. Phys. J. C}, 33:\penalty0 75--90, 2004.
\newblock \doi{10.1140/epjc/s2003-01563-y}.

\bibitem[McNeile et~al.(2013)McNeile, Bazavov, Davies, Dowdall, Hornbostel,
  Lepage, and Trottier]{McNeile:2012xh}
C.~McNeile, A.~Bazavov, C.~T.~H. Davies, R.~J. Dowdall, K.~Hornbostel, G.~P.
  Lepage, and H.~D. Trottier.
\newblock {Direct determination of the strange and light quark condensates from
  full lattice QCD}.
\newblock \emph{Phys. Rev. D}, 87\penalty0 (3):\penalty0 034503, 2013.
\newblock \doi{10.1103/PhysRevD.87.034503}.

\bibitem[Oller(2003)]{Oller:2003vf}
Jose~A. Oller.
\newblock {The Mixing angle of the lightest scalar nonet}.
\newblock \emph{Nucl. Phys. A}, 727:\penalty0 353--369, 2003.
\newblock \doi{10.1016/j.nuclphysa.2003.08.002}.

\bibitem[Agaev et~al.(2018)Agaev, Azizi, and Sundu]{Agaev:2017cfz}
S.~S. Agaev, K.~Azizi, and H.~Sundu.
\newblock {The structure, mixing angle, mass and couplings of the light scalar
  $f_0(500)$ and $f_0(980)$ mesons}.
\newblock \emph{Phys. Lett. B}, 781:\penalty0 279--282, 2018.
\newblock \doi{10.1016/j.physletb.2018.03.085}.

\bibitem[Napsuciale(1998)]{Napsuciale:1998ip}
M.~Napsuciale.
\newblock {Scalar meson masses and mixing angle in a U(3) x U(3) linear sigma
  model}.
\newblock 3 1998.

\bibitem[Bramon et~al.(1999)Bramon, Escribano, and Scadron]{Bramon:1997va}
A.~Bramon, R.~Escribano, and M.~D. Scadron.
\newblock {The eta - eta-prime mixing angle revisited}.
\newblock \emph{Eur. Phys. J. C}, 7:\penalty0 271--278, 1999.
\newblock \doi{10.1007/s100529801009}.

\bibitem[Bramon et~al.(2001)Bramon, Escribano, and Scadron]{Bramon:2000fr}
A.~Bramon, R.~Escribano, and M.~D. Scadron.
\newblock {Radiative V P gamma transitions and eta - eta-prime mixing}.
\newblock \emph{Phys. Lett. B}, 503:\penalty0 271--276, 2001.
\newblock \doi{10.1016/S0370-2693(01)00161-7}.

\bibitem[Di~Donato et~al.(2012)Di~Donato, Ricciardi, and Bigi]{DiDonato:2011kr}
Camilla Di~Donato, Giulia Ricciardi, and Ikaros Bigi.
\newblock {$\eta - \eta'$ Mixing - From electromagnetic transitions to weak
  decays of charm and beauty hadrons}.
\newblock \emph{Phys. Rev. D}, 85:\penalty0 013016, 2012.
\newblock \doi{10.1103/PhysRevD.85.013016}.

\bibitem[Kucukarslan and Meissner(2006)]{Kucukarslan:2006wk}
Ayse Kucukarslan and Ulf-G. Meissner.
\newblock {Omega-phi mixing in chiral perturbation theory}.
\newblock \emph{Mod. Phys. Lett. A}, 21:\penalty0 1423--1430, 2006.
\newblock \doi{10.1142/S0217732306020743}.

\bibitem[Sch\"afer et~al.(1994)Sch\"afer, Shuryak, and
  Verbaarschot]{Schafer:1993ra}
Thomas Sch\"afer, Edward~V. Shuryak, and J.~J.~M. Verbaarschot.
\newblock {Baryonic correlators in the random instanton vacuum}.
\newblock \emph{Nucl. Phys. B}, 412:\penalty0 143--168, 1994.
\newblock \doi{10.1016/0550-3213(94)90497-9}.

\bibitem[Roberts et~al.(2011)Roberts, Chang, Cloet, and
  Roberts]{Roberts:2011cf}
Hannes L.~L. Roberts, Lei Chang, Ian~C. Cloet, and Craig~D. Roberts.
\newblock {Masses of ground and excited-state hadrons}.
\newblock \emph{Few Body Syst.}, 51:\penalty0 1--25, 2011.
\newblock \doi{10.1007/s00601-011-0225-x}.

\bibitem[Bi et~al.(2016)Bi, Cai, Chen, Gong, Liu, Qiao, and Yang]{Bi:2015ifa}
Yujiang Bi, Hao Cai, Ying Chen, Ming Gong, Zhaofeng Liu, Hao-Xue Qiao, and
  Yi-Bo Yang.
\newblock {Diquark mass differences from unquenched lattice QCD}.
\newblock \emph{Chin. Phys. C}, 40\penalty0 (7):\penalty0 073106, 2016.
\newblock \doi{10.1088/1674-1137/40/7/073106}.

\bibitem[Shuryak(2005)]{Shuryak:2005pk}
Edward~V. Shuryak.
\newblock {Toward dynamical understanding of the diquarks, pentaquarks and
  dibaryons}.
\newblock \emph{J. Phys. Conf. Ser.}, 9:\penalty0 213--217, 2005.
\newblock \doi{10.1088/1742-6596/9/1/039}.

\bibitem[Cloet et~al.(2005)Cloet, Bentz, and Thomas]{Cloet:2005pp}
I.~C. Cloet, Wolfgang Bentz, and Anthony~William Thomas.
\newblock {Nucleon quark distributions in a covariant quark-diquark model}.
\newblock \emph{Phys. Lett. B}, 621:\penalty0 246--252, 2005.
\newblock \doi{10.1016/j.physletb.2005.06.065}.

\bibitem[Oettel et~al.(1998)Oettel, Hellstern, Alkofer, and
  Reinhardt]{Oettel:1998bk}
M.~Oettel, G.~Hellstern, Reinhard Alkofer, and H.~Reinhardt.
\newblock {Octet and decuplet baryons in a covariant and confining diquark -
  quark model}.
\newblock \emph{Phys. Rev. C}, 58:\penalty0 2459--2477, 1998.
\newblock \doi{10.1103/PhysRevC.58.2459}.

\bibitem[Buck et~al.(1992)Buck, Alkofer, and Reinhardt]{Buck:1992wz}
A.~Buck, Reinhard Alkofer, and H.~Reinhardt.
\newblock {Baryons as bound states of diquarks and quarks in the
  Nambu-Jona-Lasinio model}.
\newblock \emph{Phys. Lett. B}, 286:\penalty0 29--35, 1992.
\newblock \doi{10.1016/0370-2693(92)90154-V}.

\bibitem[Clo\"et et~al.(2014)Clo\"et, Bentz, and Thomas]{Cloet:2014rja}
Ian~C. Clo\"et, Wolfgang Bentz, and Anthony~W. Thomas.
\newblock {Role of diquark correlations and the pion cloud in nucleon elastic
  form factors}.
\newblock \emph{Phys. Rev. C}, 90:\penalty0 045202, 2014.
\newblock \doi{10.1103/PhysRevC.90.045202}.

\bibitem[Rezaeian et~al.(2004)Rezaeian, Walet, and Birse]{Rezaeian:2003kq}
Amir~H. Rezaeian, Niels~R. Walet, and Michael~C. Birse.
\newblock {Relativistic Faddeev approach to a nonlocal NJL model}.
\newblock \emph{AIP Conf. Proc.}, 717\penalty0 (1):\penalty0 690--694, 2004.
\newblock \doi{10.1063/1.1799781}.

\bibitem[Oettel et~al.(2000)Oettel, Alkofer, and von Smekal]{Oettel:2000jj}
M.~Oettel, Reinhard Alkofer, and L.~von Smekal.
\newblock {Nucleon properties in the covariant quark diquark model}.
\newblock \emph{Eur. Phys. J. A}, 8:\penalty0 553--566, 2000.
\newblock \doi{10.1007/s100500070078}.

\bibitem[Hellstern et~al.(1997)Hellstern, Alkofer, Oettel, and
  Reinhardt]{Hellstern:1997pg}
G.~Hellstern, Reinhard Alkofer, M.~Oettel, and H.~Reinhardt.
\newblock {Nucleon form-factors in a covariant diquark - quark model}.
\newblock \emph{Nucl. Phys. A}, 627:\penalty0 679--709, 1997.
\newblock \doi{10.1016/S0375-9474(97)00514-9}.

\bibitem[Mineo et~al.(2002)Mineo, Bentz, Ishii, and Yazaki]{Mineo:2002bg}
H.~Mineo, W.~Bentz, N.~Ishii, and K.~Yazaki.
\newblock {Axial vector diquark correlations in the nucleon: Structure
  functions and static properties}.
\newblock \emph{Nucl. Phys. A}, 703:\penalty0 785--820, 2002.
\newblock \doi{10.1016/S0375-9474(02)00656-5}.

\bibitem[Mineo et~al.(1999)Mineo, Bentz, and Yazaki]{Mineo:1999eq}
H.~Mineo, W.~Bentz, and K.~Yazaki.
\newblock {Quark distributions in the nucleon based on a relativistic
  three-body approach to the NJL model}.
\newblock \emph{Phys. Rev. C}, 60:\penalty0 065201, 1999.
\newblock \doi{10.1103/PhysRevC.60.065201}.

\bibitem[Bloch et~al.(1999)Bloch, Roberts, Schmidt, Bender, and
  Frank]{Bloch:1999ke}
Jacques C.~R. Bloch, Craig~D. Roberts, S.~M. Schmidt, A.~Bender, and M.~R.
  Frank.
\newblock {Nucleon form-factors and a nonpointlike diquark}.
\newblock \emph{Phys. Rev. C}, 60:\penalty0 062201, 1999.
\newblock \doi{10.1103/PhysRevC.60.062201}.

\bibitem[Anselmino et~al.(1993)Anselmino, Predazzi, Ekelin, Fredriksson, and
  Lichtenberg]{Anselmino:1992vg}
Mauro Anselmino, Enrico Predazzi, Svante Ekelin, Sverker Fredriksson, and D.~B.
  Lichtenberg.
\newblock {Diquarks}.
\newblock \emph{Rev. Mod. Phys.}, 65:\penalty0 1199--1234, 1993.
\newblock \doi{10.1103/RevModPhys.65.1199}.

\bibitem[Braaten and Tse(1987)]{Braaten:1987yy}
Eric Braaten and Sze-Man Tse.
\newblock {Perturbative {QCD} Correction to the Hard Scattering Amplitude for
  the Meson Form-factor}.
\newblock \emph{Phys. Rev. D}, 35:\penalty0 2255, 1987.
\newblock \doi{10.1103/PhysRevD.35.2255}.

\bibitem[Sterman and Stoler(1997)]{Sterman:1997sx}
George~F. Sterman and Paul Stoler.
\newblock {Hadronic form-factors and perturbative QCD}.
\newblock \emph{Ann. Rev. Nucl. Part. Sci.}, 47:\penalty0 193--233, 1997.
\newblock \doi{10.1146/annurev.nucl.47.1.193}.

\bibitem[Dittes and Radyushkin(1984)]{Dittes:1983dy}
F.~M. Dittes and A.~V. Radyushkin.
\newblock {TWO LOOP CONTRIBUTION TO THE EVOLUTION OF THE PION WAVE FUNCTION}.
\newblock \emph{Phys. Lett. B}, 134:\penalty0 359--362, 1984.
\newblock \doi{10.1016/0370-2693(84)90016-9}.

\bibitem[Shuryak and Zahed(2021{\natexlab{b}})]{Shuryak:2020ktq}
Edward Shuryak and Ismail Zahed.
\newblock {Nonperturbative quark-antiquark interactions in mesonic form
  factors}.
\newblock \emph{Phys. Rev. D}, 103\penalty0 (5):\penalty0 054028,
  2021{\natexlab{b}}.
\newblock \doi{10.1103/PhysRevD.103.054028}.

\bibitem[Brodsky and Farrar(1973)]{Brodsky:1973kr}
Stanley~J. Brodsky and Glennys~R. Farrar.
\newblock {Scaling Laws at Large Transverse Momentum}.
\newblock \emph{Phys. Rev. Lett.}, 31:\penalty0 1153--1156, 1973.
\newblock \doi{10.1103/PhysRevLett.31.1153}.

\bibitem[Huber et~al.(2008)]{JeffersonLab:2008jve}
G.~M. Huber et~al.
\newblock {Charged pion form-factor between Q**2 = 0.60-GeV**2 and 2.45-GeV**2.
  II. Determination of, and results for, the pion form-factor}.
\newblock \emph{Phys. Rev. C}, 78:\penalty0 045203, 2008.
\newblock \doi{10.1103/PhysRevC.78.045203}.

\bibitem[Br\"ommel et~al.(2007)]{QCDSFUKQCD:2006gmg}
D.~Br\"ommel et~al.
\newblock {The Pion form-factor from lattice QCD with two dynamical flavours}.
\newblock \emph{Eur. Phys. J. C}, 51:\penalty0 335--345, 2007.
\newblock \doi{10.1140/epjc/s10052-007-0295-6}.

\bibitem[Bakulev et~al.(2004)Bakulev, Passek-Kumericki, Schroers, and
  Stefanis]{Bakulev:2004cu}
A.~P. Bakulev, K.~Passek-Kumericki, W.~Schroers, and N.~G. Stefanis.
\newblock {Pion form-factor in QCD: From nonlocal condensates to NLO analytic
  perturbation theory}.
\newblock \emph{Phys. Rev. D}, 70:\penalty0 033014, 2004.
\newblock \doi{10.1103/PhysRevD.70.033014}.
\newblock [Erratum: Phys.Rev.D 70, 079906 (2004)].

\bibitem[Gronberg et~al.(1998)]{CLEO:1997fho}
J.~Gronberg et~al.
\newblock {Measurements of the meson - photon transition form-factors of light
  pseudoscalar mesons at large momentum transfer}.
\newblock \emph{Phys. Rev. D}, 57:\penalty0 33--54, 1998.
\newblock \doi{10.1103/PhysRevD.57.33}.

\bibitem[Behrend et~al.(1991)]{CELLO:1990klc}
H.~J. Behrend et~al.
\newblock {A Measurement of the pi0, eta and eta-prime electromagnetic
  form-factors}.
\newblock \emph{Z. Phys. C}, 49:\penalty0 401--410, 1991.
\newblock \doi{10.1007/BF01549692}.

\bibitem[Geshkenbein(2000)]{Geshkenbein:1998gu}
B.~V. Geshkenbein.
\newblock {Pion electromagnetic form-factor in the space - like region and P
  phase delta(1) in one-dimension (s) of pi pi scattering from the value of the
  modulus of form-factor in the time - like region.}
\newblock \emph{Phys. Rev. D}, 61:\penalty0 033009, 2000.
\newblock \doi{10.1103/PhysRevD.61.033009}.

\bibitem[Tarrach(1982)]{Tarrach:1981bi}
R.~Tarrach.
\newblock {The renormalization of FF}.
\newblock \emph{Nucl. Phys. B}, 196:\penalty0 45--61, 1982.
\newblock \doi{10.1016/0550-3213(82)90301-7}.

\bibitem[Nielsen(1977)]{Nielsen:1977sy}
N.~K. Nielsen.
\newblock {The Energy Momentum Tensor in a Nonabelian Quark Gluon Theory}.
\newblock \emph{Nucl. Phys. B}, 120:\penalty0 212--220, 1977.
\newblock \doi{10.1016/0550-3213(77)90040-2}.

\bibitem[Collins et~al.(1977)Collins, Duncan, and Joglekar]{Collins:1976yq}
John~C. Collins, Anthony Duncan, and Satish~D. Joglekar.
\newblock {Trace and Dilatation Anomalies in Gauge Theories}.
\newblock \emph{Phys. Rev. D}, 16:\penalty0 438--449, 1977.
\newblock \doi{10.1103/PhysRevD.16.438}.

\bibitem[Sch\"afer and Shuryak(1995)]{Schafer:1994fd}
Thomas Sch\"afer and Edward~V. Shuryak.
\newblock {Glueballs and instantons}.
\newblock \emph{Phys. Rev. Lett.}, 75:\penalty0 1707--1710, 1995.
\newblock \doi{10.1103/PhysRevLett.75.1707}.

\bibitem[Chen et~al.(2006)]{Chen:2005mg}
Y.~Chen et~al.
\newblock {Glueball spectrum and matrix elements on anisotropic lattices}.
\newblock \emph{Phys. Rev. D}, 73:\penalty0 014516, 2006.
\newblock \doi{10.1103/PhysRevD.73.014516}.

\bibitem[Sun et~al.(2018)Sun, Gui, Chen, Gong, Liu, Liu, Liu, Ma, and
  Zhang]{Sun:2017ipk}
Wei Sun, Long-Cheng Gui, Ying Chen, Ming Gong, Chuan Liu, Yu-Bin Liu, Zhaofeng
  Liu, Jian-Ping Ma, and Jian-Bo Zhang.
\newblock {Glueball spectrum from $N_f=2$ lattice QCD study on anisotropic
  lattices}.
\newblock \emph{Chin. Phys. C}, 42\penalty0 (9):\penalty0 093103, 2018.
\newblock \doi{10.1088/1674-1137/42/9/093103}.

\bibitem[Bijnens et~al.(1998)Bijnens, Colangelo, and Talavera]{Bijnens:1998fm}
J.~Bijnens, G.~Colangelo, and P.~Talavera.
\newblock {The Vector and scalar form-factors of the pion to two loops}.
\newblock \emph{JHEP}, 05:\penalty0 014, 1998.
\newblock \doi{10.1088/1126-6708/1998/05/014}.

\bibitem[Bijnens et~al.(1997)Bijnens, Colangelo, Ecker, Gasser, and
  Sainio]{Bijnens:1997vq}
J.~Bijnens, G.~Colangelo, G.~Ecker, J.~Gasser, and M.~E. Sainio.
\newblock {Pion-pion scattering at low energy}.
\newblock \emph{Nucl. Phys. B}, 508:\penalty0 263--310, 1997.
\newblock \doi{10.1016/S0550-3213(97)00621-4}.
\newblock [Erratum: Nucl.Phys.B 517, 639--639 (1998)].

\bibitem[Alexandrou et~al.(2022)Alexandrou, Bacchio, Cloet, Constantinou,
  Delmar, Hadjiyiannakou, Koutsou, Lauer, and Vaquero]{Alexandrou:2021ztx}
Constantia Alexandrou, Simone Bacchio, Ian Cloet, Martha Constantinou, Joseph
  Delmar, Kyriakos Hadjiyiannakou, Giannis Koutsou, Colin Lauer, and Alejandro
  Vaquero.
\newblock {Scalar, vector, and tensor form factors for the pion and kaon from
  lattice QCD}.
\newblock \emph{Phys. Rev. D}, 105\penalty0 (5):\penalty0 054502, 2022.
\newblock \doi{10.1103/PhysRevD.105.054502}.

\bibitem[Wang et~al.(2024{\natexlab{a}})Wang, He, Wang, Draper, Liang, Liu, and
  Yang]{wang2024trace}
Bigeng Wang, Fangcheng He, Gen Wang, Terrence Draper, Jian Liang, Keh-Fei Liu,
  and Yi-Bo Yang.
\newblock Trace anomaly form factors from lattice qcd, 2024{\natexlab{a}}.

\bibitem[Hackett et~al.(2023)Hackett, Oare, Pefkou, and
  Shanahan]{Hackett:2023nkr}
Daniel~C. Hackett, Patrick~R. Oare, Dimitra~A. Pefkou, and Phiala~E. Shanahan.
\newblock {Gravitational form factors of the pion from lattice QCD}.
\newblock 7 2023.

\bibitem[Wang et~al.(2024{\natexlab{b}})Wang, He, Wang, Draper, Liang, Liu, and
  Yang]{Wang:2024lrm}
Bigeng Wang, Fangcheng He, Gen Wang, Terrence Draper, Jian Liang, Keh-Fei Liu,
  and Yi-Bo Yang.
\newblock {Trace anomaly form factors from lattice QCD}.
\newblock \emph{Phys. Rev. D}, 109\penalty0 (9):\penalty0 094504,
  2024{\natexlab{b}}.
\newblock \doi{10.1103/PhysRevD.109.094504}.

\bibitem[Nakamura et~al.(2010)]{ParticleDataGroup:2010dbb}
K.~Nakamura et~al.
\newblock {Review of particle physics}.
\newblock \emph{J. Phys. G}, 37:\penalty0 075021, 2010.
\newblock \doi{10.1088/0954-3899/37/7A/075021}.

\bibitem[Hackett et~al.(2024)Hackett, Pefkou, and Shanahan]{Hackett:2023rif}
Daniel~C. Hackett, Dimitra~A. Pefkou, and Phiala~E. Shanahan.
\newblock {Gravitational Form Factors of the Proton from Lattice QCD}.
\newblock \emph{Phys. Rev. Lett.}, 132\penalty0 (25):\penalty0 251904, 2024.
\newblock \doi{10.1103/PhysRevLett.132.251904}.

\bibitem[Schweitzer(2004)]{Schweitzer:2003sb}
P.~Schweitzer.
\newblock {The Sigma term form-factor of the nucleon in the large N(C) limit}.
\newblock \emph{Phys. Rev. D}, 69:\penalty0 034003, 2004.
\newblock \doi{10.1103/PhysRevD.69.034003}.

\bibitem[Hoferichter et~al.(2023)Hoferichter, de~Elvira, Kubis, and
  Mei{\ss}ner]{Hoferichter:2023ptl}
Martin Hoferichter, Jacobo~Ruiz de~Elvira, Bastian Kubis, and Ulf-G.
  Mei{\ss}ner.
\newblock {On the role of isospin violation in the pion{\textendash}nucleon
  {\ensuremath{\sigma}}-term}.
\newblock \emph{Phys. Lett. B}, 843:\penalty0 138001, 2023.
\newblock \doi{10.1016/j.physletb.2023.138001}.

\bibitem[Alarc{\'o}n and Weiss(2017)]{Alarcon:2017ivh}
J.~M. Alarc{\'o}n and C.~Weiss.
\newblock {Nucleon form factors in dispersively improved chiral effective field
  theory: Scalar form factor}.
\newblock \emph{Phys. Rev. C}, 96\penalty0 (5):\penalty0 055206, 2017.
\newblock \doi{10.1103/PhysRevC.96.055206}.

\bibitem[Dong et~al.(1996)Dong, Lagae, and Liu]{Dong:1995ec}
S.~J. Dong, J.~F. Lagae, and K.~F. Liu.
\newblock {Pi N sigma term, anti-s s in nucleon, and scalar form-factor: A
  Lattice study}.
\newblock \emph{Phys. Rev. D}, 54:\penalty0 5496--5500, 1996.
\newblock \doi{10.1103/PhysRevD.54.5496}.

\bibitem[Freese and Clo{\"e}t(2019)]{Freese:2019bhb}
Adam Freese and Ian~C. Clo{\"e}t.
\newblock {Gravitational form factors of light mesons}.
\newblock \emph{Phys. Rev. C}, 100\penalty0 (1):\penalty0 015201, 2019.
\newblock \doi{10.1103/PhysRevC.100.015201}.
\newblock [Erratum: Phys.Rev.C 105, 059901 (2022)].

\bibitem[Polyakov and Son(2018)]{Polyakov:2018exb}
Maxim~V. Polyakov and Hyeon-Dong Son.
\newblock {Nucleon gravitational form factors from instantons: forces between
  quark and gluon subsystems}.
\newblock \emph{JHEP}, 09:\penalty0 156, 2018.
\newblock \doi{10.1007/JHEP09(2018)156}.

\bibitem[Ji(2021)]{Ji:2021mtz}
Xiangdong Ji.
\newblock {Proton mass decomposition: naturalness and interpretations}.
\newblock \emph{Front. Phys. (Beijing)}, 16\penalty0 (6):\penalty0 64601, 2021.
\newblock \doi{10.1007/s11467-021-1065-x}.

\bibitem[Ji(1995{\natexlab{a}})]{Ji:1995sv}
Xiang-Dong Ji.
\newblock {Breakup of hadron masses and energy - momentum tensor of QCD}.
\newblock \emph{Phys. Rev. D}, 52:\penalty0 271--281, 1995{\natexlab{a}}.
\newblock \doi{10.1103/PhysRevD.52.271}.

\bibitem[Hatta et~al.(2018)Hatta, Rajan, and Tanaka]{Hatta:2018sqd}
Yoshitaka Hatta, Abha Rajan, and Kazuhiro Tanaka.
\newblock {Quark and gluon contributions to the QCD trace anomaly}.
\newblock \emph{JHEP}, 12:\penalty0 008, 2018.
\newblock \doi{10.1007/JHEP12(2018)008}.

\bibitem[Chen and Savage(1998)]{Chen:1997zza}
Jiunn-Wei Chen and Martin~J. Savage.
\newblock {Hadronic and electromagnetic interactions of quarkonia}.
\newblock \emph{Phys. Rev. D}, 57:\penalty0 2837--2846, 1998.
\newblock \doi{10.1103/PhysRevD.57.2837}.

\bibitem[Novikov and Shifman(1981)]{Novikov:1980fa}
V.~A. Novikov and Mikhail~A. Shifman.
\newblock {Comment on the psi-prime ---\ensuremath{>} J/psi pi pi Decay}.
\newblock \emph{Z. Phys. C}, 8:\penalty0 43, 1981.
\newblock \doi{10.1007/BF01429829}.

\bibitem[Shanahan and Detmold(2019)]{Shanahan:2018pib}
P.~E. Shanahan and W.~Detmold.
\newblock {Gluon gravitational form factors of the nucleon and the pion from
  lattice QCD}.
\newblock \emph{Phys. Rev. D}, 99\penalty0 (1):\penalty0 014511, 2019.
\newblock \doi{10.1103/PhysRevD.99.014511}.

\bibitem[Pefkou et~al.(2022)Pefkou, Hackett, and Shanahan]{Pefkou:2021fni}
Dimitra~A. Pefkou, Daniel~C. Hackett, and Phiala~E. Shanahan.
\newblock {Gluon gravitational structure of hadrons of different spin}.
\newblock \emph{Phys. Rev. D}, 105\penalty0 (5):\penalty0 054509, 2022.
\newblock \doi{10.1103/PhysRevD.105.054509}.

\bibitem[Alexandrou et~al.(2020)Alexandrou, Bacchio, Constantinou, Finkenrath,
  Hadjiyiannakou, Jansen, Koutsou, Panagopoulos, and
  Spanoudes]{Alexandrou:2020sml}
C.~Alexandrou, S.~Bacchio, M.~Constantinou, J.~Finkenrath, K.~Hadjiyiannakou,
  K.~Jansen, G.~Koutsou, H.~Panagopoulos, and G.~Spanoudes.
\newblock {Complete flavor decomposition of the spin and momentum fraction of
  the proton using lattice QCD simulations at physical pion mass}.
\newblock \emph{Phys. Rev. D}, 101\penalty0 (9):\penalty0 094513, 2020.
\newblock \doi{10.1103/PhysRevD.101.094513}.

\bibitem[Gross and Wilczek(1974)]{Gross:1974cs}
D.~J. Gross and Frank Wilczek.
\newblock {ASYMPTOTICALLY FREE GAUGE THEORIES. 2.}
\newblock \emph{Phys. Rev. D}, 9:\penalty0 980--993, 1974.
\newblock \doi{10.1103/PhysRevD.9.980}.

\bibitem[Politzer(1974)]{Politzer:1974sm}
H.~David Politzer.
\newblock {Setting the scale for predictions of asymptotic freedom}.
\newblock \emph{Phys. Rev. D}, 9:\penalty0 2174--2175, 1974.
\newblock \doi{10.1103/PhysRevD.9.2174}.

\bibitem[Hou et~al.(2021)]{Hou:2019efy}
Tie-Jiun Hou et~al.
\newblock {New CTEQ global analysis of quantum chromodynamics with
  high-precision data from the LHC}.
\newblock \emph{Phys. Rev. D}, 103\penalty0 (1):\penalty0 014013, 2021.
\newblock \doi{10.1103/PhysRevD.103.014013}.

\bibitem[Burkert et~al.(2018)Burkert, Elouadrhiri, and Girod]{Burkert:2018bqq}
V.~D. Burkert, L.~Elouadrhiri, and F.~X. Girod.
\newblock {The pressure distribution inside the proton}.
\newblock \emph{Nature}, 557\penalty0 (7705):\penalty0 396--399, 2018.
\newblock \doi{10.1038/s41586-018-0060-z}.

\bibitem[Jung et~al.(2014)Jung, Yakhshiev, and Kim]{Jung:2013bya}
Ju-Hyun Jung, Ulugbek Yakhshiev, and Hyun-Chul Kim.
\newblock {Energy\textendash{}momentum tensor form factors of the nucleon
  within a
  \ensuremath{\pi}\textendash{}\ensuremath{\rho}\textendash{}\ensuremath{\omega}
  soliton model}.
\newblock \emph{J. Phys. G}, 41:\penalty0 055107, 2014.
\newblock \doi{10.1088/0954-3899/41/5/055107}.

\bibitem[Cebulla et~al.(2007)Cebulla, Goeke, Ossmann, and
  Schweitzer]{Cebulla:2007ei}
C.~Cebulla, K.~Goeke, J.~Ossmann, and P.~Schweitzer.
\newblock {The Nucleon form-factors of the energy momentum tensor in the Skyrme
  model}.
\newblock \emph{Nucl. Phys. A}, 794:\penalty0 87--114, 2007.
\newblock \doi{10.1016/j.nuclphysa.2007.08.004}.

\bibitem[Wakamatsu(2007)]{Wakamatsu:2007uc}
M.~Wakamatsu.
\newblock {On the D-term of the nucleon generalized parton distributions}.
\newblock \emph{Phys. Lett. B}, 648:\penalty0 181--185, 2007.
\newblock \doi{10.1016/j.physletb.2007.03.013}.

\bibitem[Singh et~al.(2011)Singh, Lee, and Wang]{Singh:2010wd}
Janardan~P. Singh, Frank~X. Lee, and Lai Wang.
\newblock {Eta-nucleon coupling constant in QCD with SU(3) symmetry breaking}.
\newblock \emph{Int. J. Mod. Phys. A}, 26:\penalty0 947--963, 2011.
\newblock \doi{10.1142/S0217751X11051561}.

\bibitem[Nasrallah(2007)]{Nasrallah:2005hn}
N.~F. Nasrallah.
\newblock {Couplings of the eta and eta-prime mesons to the nucleon}.
\newblock \emph{Phys. Lett. B}, 645:\penalty0 335--338, 2007.
\newblock \doi{10.1016/j.physletb.2006.12.048}.

\bibitem[Feldmann(2000)]{Feldmann:1999uf}
Thorsten Feldmann.
\newblock {Quark structure of pseudoscalar mesons}.
\newblock \emph{Int. J. Mod. Phys. A}, 15:\penalty0 159--207, 2000.
\newblock \doi{10.1142/S0217751X00000082}.

\bibitem[Beneke and Neubert(2003)]{Beneke:2002jn}
Martin Beneke and Matthias Neubert.
\newblock {Flavor singlet B decay amplitudes in QCD factorization}.
\newblock \emph{Nucl. Phys. B}, 651:\penalty0 225--248, 2003.
\newblock \doi{10.1016/S0550-3213(02)01091-X}.

\bibitem[Ji(1995{\natexlab{b}})]{Ji:1994av}
Xiang-Dong Ji.
\newblock {A QCD analysis of the mass structure of the nucleon}.
\newblock \emph{Phys. Rev. Lett.}, 74:\penalty0 1071--1074, 1995{\natexlab{b}}.
\newblock \doi{10.1103/PhysRevLett.74.1071}.

\bibitem[Lorc\'e(2018)]{Lorce:2017xzd}
C\'edric Lorc\'e.
\newblock {On the hadron mass decomposition}.
\newblock \emph{Eur. Phys. J. C}, 78\penalty0 (2):\penalty0 120, 2018.
\newblock \doi{10.1140/epjc/s10052-018-5561-2}.

\bibitem[Roberts(2021)]{Roberts:2021xnz}
Craig~D. Roberts.
\newblock {On Mass and Matter}.
\newblock \emph{AAPPS Bull.}, 31:\penalty0 6, 2021.
\newblock \doi{10.1007/s43673-021-00005-4}.

\bibitem[Metz et~al.(2021)Metz, Pasquini, and Rodini]{Metz:2020vxd}
Andreas Metz, Barbara Pasquini, and Simone Rodini.
\newblock {Revisiting the proton mass decomposition}.
\newblock \emph{Phys. Rev. D}, 102\penalty0 (11):\penalty0 114042, 2021.
\newblock \doi{10.1103/PhysRevD.102.114042}.

\bibitem[Yang et~al.(2018)Yang, Liang, Bi, Chen, Draper, Liu, and
  Liu]{Yang:2018nqn}
Yi-Bo Yang, Jian Liang, Yu-Jiang Bi, Ying Chen, Terrence Draper, Keh-Fei Liu,
  and Zhaofeng Liu.
\newblock {Proton Mass Decomposition from the QCD Energy Momentum Tensor}.
\newblock \emph{Phys. Rev. Lett.}, 121\penalty0 (21):\penalty0 212001, 2018.
\newblock \doi{10.1103/PhysRevLett.121.212001}.

\bibitem[Hoferichter et~al.(2016)Hoferichter, Ruiz~de Elvira, Kubis, and
  Mei\ss{}ner]{Hoferichter:2016ocj}
Martin Hoferichter, Jacobo Ruiz~de Elvira, Bastian Kubis, and Ulf-G.
  Mei\ss{}ner.
\newblock {Remarks on the pion\textendash{}nucleon \ensuremath{\sigma}-term}.
\newblock \emph{Phys. Lett. B}, 760:\penalty0 74--78, 2016.
\newblock \doi{10.1016/j.physletb.2016.06.038}.

\bibitem[Alarc\'on(2021)]{Alarcon:2021dlz}
J.~M. Alarc\'on.
\newblock {Brief history of the pion\textendash{}nucleon sigma term}.
\newblock \emph{Eur. Phys. J. ST}, 230\penalty0 (6):\penalty0 1609--1622, 2021.
\newblock \doi{10.1140/epjs/s11734-021-00145-6}.

\bibitem[Deur et~al.(2018)Deur, Brodsky, and De~T\'eramond]{Deur:2018roz}
Alexandre Deur, Stanley~J. Brodsky, and Guy~F. De~T\'eramond.
\newblock {The Spin Structure of the Nucleon}.
\newblock 7 2018.
\newblock \doi{10.1088/1361-6633/ab0b8f}.

\bibitem[Ji(1997{\natexlab{a}})]{Ji:1996ek}
Xiang-Dong Ji.
\newblock {Gauge-Invariant Decomposition of Nucleon Spin}.
\newblock \emph{Phys. Rev. Lett.}, 78:\penalty0 610--613, 1997{\natexlab{a}}.
\newblock \doi{10.1103/PhysRevLett.78.610}.

\bibitem[Aoki et~al.(2020)]{FlavourLatticeAveragingGroup:2019iem}
S.~Aoki et~al.
\newblock {FLAG Review 2019: Flavour Lattice Averaging Group (FLAG)}.
\newblock \emph{Eur. Phys. J. C}, 80\penalty0 (2):\penalty0 113, 2020.
\newblock \doi{10.1140/epjc/s10052-019-7354-7}.

\bibitem[Adler(1969)]{Adler:1969gk}
Stephen~L. Adler.
\newblock {Axial vector vertex in spinor electrodynamics}.
\newblock \emph{Phys. Rev.}, 177:\penalty0 2426--2438, 1969.
\newblock \doi{10.1103/PhysRev.177.2426}.

\bibitem[Nocera et~al.(2014)Nocera, Ball, Forte, Ridolfi, and
  Rojo]{Nocera:2014gqa}
Emanuele~R. Nocera, Richard~D. Ball, Stefano Forte, Giovanni Ridolfi, and Juan
  Rojo.
\newblock {A first unbiased global determination of polarized PDFs and their
  uncertainties}.
\newblock \emph{Nucl. Phys. B}, 887:\penalty0 276--308, 2014.
\newblock \doi{10.1016/j.nuclphysb.2014.08.008}.

\bibitem[Lin et~al.(2018)Lin, Gupta, Yoon, Jang, and Bhattacharya]{Lin:2018obj}
Huey-Wen Lin, Rajan Gupta, Boram Yoon, Yong-Chull Jang, and Tanmoy
  Bhattacharya.
\newblock {Quark contribution to the proton spin from 2+1+1-flavor lattice
  QCD}.
\newblock \emph{Phys. Rev. D}, 98\penalty0 (9):\penalty0 094512, 2018.
\newblock \doi{10.1103/PhysRevD.98.094512}.

\bibitem[Liang et~al.(2018)Liang, Yang, Draper, Gong, and Liu]{Liang:2018pis}
Jian Liang, Yi-Bo Yang, Terrence Draper, Ming Gong, and Keh-Fei Liu.
\newblock {Quark spins and Anomalous Ward Identity}.
\newblock \emph{Phys. Rev. D}, 98\penalty0 (7):\penalty0 074505, 2018.
\newblock \doi{10.1103/PhysRevD.98.074505}.

\bibitem[de~Florian et~al.(2009)de~Florian, Sassot, Stratmann, and
  Vogelsang]{deFlorian:2009vb}
Daniel de~Florian, Rodolfo Sassot, Marco Stratmann, and Werner Vogelsang.
\newblock {Extraction of Spin-Dependent Parton Densities and Their
  Uncertainties}.
\newblock \emph{Phys. Rev. D}, 80:\penalty0 034030, 2009.
\newblock \doi{10.1103/PhysRevD.80.034030}.

\bibitem[Brommel et~al.(2007)]{QCDSF-UKQCD:2007gdl}
Dirk Brommel et~al.
\newblock {Moments of generalized parton distributions and quark angular
  momentum of the nucleon}.
\newblock \emph{PoS}, LATTICE2007:\penalty0 158, 2007.
\newblock \doi{10.22323/1.042.0158}.

\bibitem[Hagler et~al.(2008)]{LHPC:2007blg}
Ph. Hagler et~al.
\newblock {Nucleon Generalized Parton Distributions from Full Lattice QCD}.
\newblock \emph{Phys. Rev. D}, 77:\penalty0 094502, 2008.
\newblock \doi{10.1103/PhysRevD.77.094502}.

\bibitem[Mamo and Zahed(2022)]{Mamo:2022eui}
Kiminad~A. Mamo and Ismail Zahed.
\newblock {J/\ensuremath{\psi} near threshold in holographic QCD: A and D
  gravitational form factors}.
\newblock \emph{Phys. Rev. D}, 106\penalty0 (8):\penalty0 086004, 2022.
\newblock \doi{10.1103/PhysRevD.106.086004}.

\bibitem[Wang et~al.(2022{\natexlab{a}})Wang, Yang, Liang, Draper, and
  Liu]{Wang:2021vqy}
Gen Wang, Yi-Bo Yang, Jian Liang, Terrence Draper, and Keh-Fei Liu.
\newblock {Proton momentum and angular momentum decompositions with overlap
  fermions}.
\newblock \emph{Phys. Rev. D}, 106\penalty0 (1):\penalty0 014512,
  2022{\natexlab{a}}.
\newblock \doi{10.1103/PhysRevD.106.014512}.

\bibitem[Polyakov(1980)]{Polyakov:1980ca}
Alexander~M. Polyakov.
\newblock {Gauge Fields as Rings of Glue}.
\newblock \emph{Nucl. Phys. B}, 164:\penalty0 171--188, 1980.
\newblock \doi{10.1016/0550-3213(80)90507-6}.

\bibitem[Hafidi et~al.(2017)Hafidi, Joosten, Meziani, and Qiu]{Hafidi:2017bsg}
K.~Hafidi, S.~Joosten, Z.~E. Meziani, and J.~W. Qiu.
\newblock {Production of Charmonium at Threshold in Hall A and C at Jefferson
  Lab}.
\newblock \emph{Few Body Syst.}, 58\penalty0 (4):\penalty0 141, 2017.
\newblock \doi{10.1007/s00601-017-1305-3}.

\bibitem[Ali et~al.(2019)]{GlueX:2019mkq}
A.~Ali et~al.
\newblock {First Measurement of Near-Threshold J/\ensuremath{\psi} Exclusive
  Photoproduction off the Proton}.
\newblock \emph{Phys. Rev. Lett.}, 123\penalty0 (7):\penalty0 072001, 2019.
\newblock \doi{10.1103/PhysRevLett.123.072001}.

\bibitem[Meziani and Joosten(2020)]{Meziani:2020oks}
Zein-Eddine Meziani and Sylvester Joosten.
\newblock {Origin of the Proton Mass? Heavy Quarkonium Production at Threshold
  from Jefferson Lab to an Electron Ion Collider}.
\newblock In \emph{{Probing Nucleons and Nuclei in High Energy Collisions}:
  {Dedicated to the Physics of the Electron Ion Collider}}, pages 234--237,
  2020.
\newblock \doi{10.1142/9789811214950_0048}.

\bibitem[Anderle et~al.(2021)]{Anderle:2021wcy}
Daniele~P. Anderle et~al.
\newblock {Electron-ion collider in China}.
\newblock \emph{Front. Phys. (Beijing)}, 16\penalty0 (6):\penalty0 64701, 2021.
\newblock \doi{10.1007/s11467-021-1062-0}.

\bibitem[Hatta and Yang(2018)]{Hatta:2018ina}
Yoshitaka Hatta and Di-Lun Yang.
\newblock {Holographic $J/\psi$ production near threshold and the proton mass
  problem}.
\newblock \emph{Phys. Rev. D}, 98\penalty0 (7):\penalty0 074003, 2018.
\newblock \doi{10.1103/PhysRevD.98.074003}.

\bibitem[Mamo and Zahed(2020)]{Mamo:2019mka}
Kiminad~A. Mamo and Ismail Zahed.
\newblock {Diffractive photoproduction of $J/\psi$ and $\Upsilon$ using
  holographic QCD: gravitational form factors and GPD of gluons in the proton}.
\newblock \emph{Phys. Rev. D}, 101\penalty0 (8):\penalty0 086003, 2020.
\newblock \doi{10.1103/PhysRevD.101.086003}.

\bibitem[Kharzeev(2021)]{Kharzeev:2021qkd}
Dmitri~E. Kharzeev.
\newblock {Mass radius of the proton}.
\newblock \emph{Phys. Rev. D}, 104\penalty0 (5):\penalty0 054015, 2021.
\newblock \doi{10.1103/PhysRevD.104.054015}.

\bibitem[Hatta and Strikman(2021)]{Hatta:2021can}
Yoshitaka Hatta and Mark Strikman.
\newblock {$\phi$-meson lepto-production near threshold and the strangeness
  $D$-term}.
\newblock \emph{Phys. Lett. B}, 817:\penalty0 136295, 2021.
\newblock \doi{10.1016/j.physletb.2021.136295}.

\bibitem[Guo et~al.(2021{\natexlab{a}})Guo, Ji, and Liu]{Guo:2021ibg}
Yuxun Guo, Xiangdong Ji, and Yizhuang Liu.
\newblock {QCD Analysis of Near-Threshold Photon-Proton Production of Heavy
  Quarkonium}.
\newblock \emph{Phys. Rev. D}, 103\penalty0 (9):\penalty0 096010,
  2021{\natexlab{a}}.
\newblock \doi{10.1103/PhysRevD.103.096010}.

\bibitem[Sun et~al.(2021)Sun, Tong, and Yuan]{Sun:2021gmi}
Peng Sun, Xuan-Bo Tong, and Feng Yuan.
\newblock {Perturbative QCD analysis of near threshold heavy quarkonium
  photoproduction at large momentum transfer}.
\newblock \emph{Phys. Lett. B}, 822:\penalty0 136655, 2021.
\newblock \doi{10.1016/j.physletb.2021.136655}.

\bibitem[Wang et~al.(2022{\natexlab{b}})Wang, Zeng, and Wang]{Wang:2022vhr}
Xiao-Yun Wang, Fancong Zeng, and Quanjin Wang.
\newblock {Systematic analysis of the proton mass radius based on
  photoproduction of vector charmoniums}.
\newblock \emph{Phys. Rev. D}, 105\penalty0 (9):\penalty0 096033,
  2022{\natexlab{b}}.
\newblock \doi{10.1103/PhysRevD.105.096033}.

\bibitem[Bartels(1980)]{Bartels:1980pe}
Jochen Bartels.
\newblock {High-Energy Behavior in a Nonabelian Gauge Theory (II)}: {First
  Corrections to $T_{n\to m}$ Beyond the Leading $\ln s$ Approximation}.
\newblock \emph{Nucl. Phys. B}, 175:\penalty0 365--401, 1980.
\newblock \doi{10.1016/0550-3213(80)90019-X}.

\bibitem[Kwiecinski and Praszalowicz(1980)]{Kwiecinski:1980wb}
J.~Kwiecinski and M.~Praszalowicz.
\newblock {Three Gluon Integral Equation and Odd c Singlet Regge Singularities
  in QCD}.
\newblock \emph{Phys. Lett. B}, 94:\penalty0 413--416, 1980.
\newblock \doi{10.1016/0370-2693(80)90909-0}.

\bibitem[Braun(1998)]{Braun:1998fs}
M.~A. Braun.
\newblock {Odderon and QCD}.
\newblock 5 1998.

\bibitem[Abazov et~al.(2021)]{TOTEM:2020zzr}
V.~M. Abazov et~al.
\newblock {Odderon Exchange from Elastic Scattering Differences between $pp$
  and $p \bar{p}$ Data at 1.96~TeV and from pp Forward Scattering
  Measurements}.
\newblock \emph{Phys. Rev. Lett.}, 127\penalty0 (6):\penalty0 062003, 2021.
\newblock \doi{10.1103/PhysRevLett.127.062003}.

\bibitem[Mamo and Zahed(2021)]{Mamo:2021krl}
Kiminad~A. Mamo and Ismail Zahed.
\newblock {Nucleon mass radii and distribution: Holographic QCD, Lattice QCD
  and GlueX data}.
\newblock \emph{Phys. Rev. D}, 103\penalty0 (9):\penalty0 094010, 2021.
\newblock \doi{10.1103/PhysRevD.103.094010}.

\bibitem[Boussarie and Hatta(2020)]{Boussarie:2020vmu}
Renaud Boussarie and Yoshitaka Hatta.
\newblock {QCD analysis of near-threshold quarkonium leptoproduction at large
  photon virtualities}.
\newblock \emph{Phys. Rev. D}, 101\penalty0 (11):\penalty0 114004, 2020.
\newblock \doi{10.1103/PhysRevD.101.114004}.

\bibitem[Guo et~al.(2021{\natexlab{b}})Guo, Ji, and Shiells]{Guo:2021aik}
Yuxun Guo, Xiangdong Ji, and Kyle Shiells.
\newblock {Novel twist-three transverse-spin sum rule for the proton and
  related generalized parton distributions}.
\newblock \emph{Nucl. Phys. B}, 969:\penalty0 115440, 2021{\natexlab{b}}.
\newblock \doi{10.1016/j.nuclphysb.2021.115440}.

\bibitem[Guo et~al.(2023)Guo, Ji, Liu, and Yang]{Guo:2023pqw}
Yuxun Guo, Xiangdong Ji, Yizhuang Liu, and Jinghong Yang.
\newblock {Updated analysis of near-threshold heavy quarkonium production for
  probe of proton\textquoteright{}s gluonic gravitational form factors}.
\newblock \emph{Phys. Rev. D}, 108\penalty0 (3):\penalty0 034003, 2023.
\newblock \doi{10.1103/PhysRevD.108.034003}.

\bibitem[Du et~al.(2020)Du, Baru, Guo, Hanhart, Mei\ss{}ner, Nefediev, and
  Strakovsky]{Du:2020bqj}
Meng-Lin Du, Vadim Baru, Feng-Kun Guo, Christoph Hanhart, Ulf-G Mei\ss{}ner,
  Alexey Nefediev, and Igor Strakovsky.
\newblock {Deciphering the mechanism of near-threshold $J/\psi$
  photoproduction}.
\newblock \emph{Eur. Phys. J. C}, 80\penalty0 (11):\penalty0 1053, 2020.
\newblock \doi{10.1140/epjc/s10052-020-08620-5}.

\bibitem[Polchinski and Strassler(2003)]{Polchinski:2002jw}
Joseph Polchinski and Matthew~J. Strassler.
\newblock {Deep inelastic scattering and gauge / string duality}.
\newblock \emph{JHEP}, 05:\penalty0 012, 2003.
\newblock \doi{10.1088/1126-6708/2003/05/012}.

\bibitem[Hechenberger et~al.(2024)Hechenberger, Mamo, and
  Zahed]{Hechenberger:2024abg}
Florian Hechenberger, Kiminad~A. Mamo, and Ismail Zahed.
\newblock {Threshold production of $\eta_{c,b}$ using holographic QCD}.
\newblock 1 2024.

\bibitem[Sun et~al.(2022)Sun, Tong, and Yuan]{Sun:2021pyw}
Peng Sun, Xuan-Bo Tong, and Feng Yuan.
\newblock {Near threshold heavy quarkonium photoproduction at large momentum
  transfer}.
\newblock \emph{Phys. Rev. D}, 105\penalty0 (5):\penalty0 054032, 2022.
\newblock \doi{10.1103/PhysRevD.105.054032}.

\bibitem[Ma(2003)]{Ma:2003py}
J.~P. Ma.
\newblock {Diffractive photoproduction of eta(c)}.
\newblock \emph{Nucl. Phys. A}, 727:\penalty0 333--352, 2003.
\newblock \doi{10.1016/j.nuclphysa.2003.08.016}.

\bibitem[Bodwin et~al.(2006)Bodwin, Braaten, Lee, and Yu]{Bodwin:2006yd}
Geoffrey~T. Bodwin, Eric Braaten, Jungil Lee, and Chaehyun Yu.
\newblock {Exclusive two-vector-meson production from e+ e- annihilation}.
\newblock \emph{Phys. Rev. D}, 74:\penalty0 074014, 2006.
\newblock \doi{10.1103/PhysRevD.74.074014}.

\bibitem[Lansberg and Pham(2008)]{Lansberg:2008cq}
J.~P. Lansberg and T.~N. Pham.
\newblock {Two-photon decay of pseudoscalar quarkonia}.
\newblock \emph{AIP Conf. Proc.}, 1038\penalty0 (1):\penalty0 259, 2008.
\newblock \doi{10.1063/1.2987179}.

\bibitem[Fabiano and Pancheri(2003)]{Fabiano:2002se}
Nicola Fabiano and Giulia Pancheri.
\newblock {Two photons width of heavy pseudoscalar mesons}.
\newblock \emph{Frascati Phys. Ser.}, 31:\penalty0 417--419, 2003.

\bibitem[Erler(1999)]{Erler:1998sy}
Jens Erler.
\newblock {Calculation of the QED coupling alpha (M(Z)) in the modified minimal
  subtraction scheme}.
\newblock \emph{Phys. Rev. D}, 59:\penalty0 054008, 1999.
\newblock \doi{10.1103/PhysRevD.59.054008}.

\bibitem[Ji(1998)]{Ji:1998pc}
Xiang-Dong Ji.
\newblock {Off forward parton distributions}.
\newblock \emph{J. Phys. G}, 24:\penalty0 1181--1205, 1998.
\newblock \doi{10.1088/0954-3899/24/7/002}.

\bibitem[Chen et~al.(2023)Chen, Xing, and Yoshida]{Chen:2023hvu}
Longjie Chen, Hongxi Xing, and Shinsuke Yoshida.
\newblock {The twist-3 gluon contribution to Sivers asymmetry in $J/\psi$
  production in semi-inclusive deep inelastic scattering}.
\newblock \emph{Phys. Rev. D}, 108\penalty0 (5):\penalty0 054021, 2023.
\newblock \doi{10.1103/PhysRevD.108.054021}.

\bibitem[Diehl(2001)]{Diehl:2001pm}
M.~Diehl.
\newblock {Generalized parton distributions with helicity flip}.
\newblock \emph{Eur. Phys. J. C}, 19:\penalty0 485--492, 2001.
\newblock \doi{10.1007/s100520100635}.

\bibitem[Koike et~al.(2020)Koike, Yabe, and Yoshida]{Koike:2019zxc}
Yuji Koike, Kenta Yabe, and Shinsuke Yoshida.
\newblock {Exact Relations for Twist-3 Gluon Distribution and Fragmentation
  Functions from Operator Identities}.
\newblock \emph{Phys. Rev. D}, 101\penalty0 (5):\penalty0 054017, 2020.
\newblock \doi{10.1103/PhysRevD.101.054017}.

\bibitem[Ji and Lebed(2001)]{Ji:2000id}
Xiang-Dong Ji and Richard~F. Lebed.
\newblock {Counting form-factors of twist-two operators}.
\newblock \emph{Phys. Rev. D}, 63:\penalty0 076005, 2001.
\newblock \doi{10.1103/PhysRevD.63.076005}.

\bibitem[Polyakov and Sun(2019)]{Polyakov:2019lbq}
Maxim~V. Polyakov and Bao-Dong Sun.
\newblock {Gravitational form factors of a spin one particle}.
\newblock \emph{Phys. Rev. D}, 100\penalty0 (3):\penalty0 036003, 2019.
\newblock \doi{10.1103/PhysRevD.100.036003}.

\bibitem[Ye et~al.(2017)Ye, Sato, Allada, Liu, Chen, Gao, Kang, Prokudin, Sun,
  and Yuan]{YE201791}
Zhihong Ye, Nobuo Sato, Kalyan Allada, Tianbo Liu, Jian-Ping Chen, Haiyan Gao,
  Zhong-Bo Kang, Alexei Prokudin, Peng Sun, and Feng Yuan.
\newblock Unveiling the nucleon tensor charge at jefferson lab: A study of the
  solid case.
\newblock \emph{Physics Letters B}, 767:\penalty0 91--98, 2017.
\newblock ISSN 0370-2693.
\newblock \doi{https://doi.org/10.1016/j.physletb.2017.01.046}.
\newblock URL
  \url{https://www.sciencedirect.com/science/article/pii/S0370269317300643}.

\bibitem[Kang et~al.(2016)Kang, Prokudin, Sun, and Yuan]{Kang:2015msa}
Zhong-Bo Kang, Alexei Prokudin, Peng Sun, and Feng Yuan.
\newblock {Extraction of Quark Transversity Distribution and Collins
  Fragmentation Functions with QCD Evolution}.
\newblock \emph{Phys. Rev. D}, 93\penalty0 (1):\penalty0 014009, 2016.
\newblock \doi{10.1103/PhysRevD.93.014009}.

\bibitem[G{\"o}ckeler et~al.(2007)G{\"o}ckeler, H{\"a}gler, Horsley, Nakamura,
  Pleiter, Rakow, Sch{\"a}fer, Schierholz, St{\"u}ben, and
  Zanotti]{QCDSF:2006tkx}
M.~G{\"o}ckeler, Ph. H{\"a}gler, R.~Horsley, Y.~Nakamura, D.~Pleiter, P.~E.~L.
  Rakow, A.~Sch{\"a}fer, G.~Schierholz, H.~St{\"u}ben, and J.~M. Zanotti.
\newblock {Transverse spin structure of the nucleon from lattice QCD
  simulations}.
\newblock \emph{Phys. Rev. Lett.}, 98:\penalty0 222001, 2007.
\newblock \doi{10.1103/PhysRevLett.98.222001}.

\bibitem[Artru and Mekhfi(1990)]{artru1990transversely}
Xavier Artru and Mustapha Mekhfi.
\newblock Transversely polarized parton densities, their evolution and their
  measurement.
\newblock \emph{Zeitschrift f{\"u}r Physik C Particles and Fields}, 45\penalty0
  (4):\penalty0 669--676, 1990.

\bibitem[Gamberg and Goldstein(2001)]{Gamberg:2001xc}
Leonard~P. Gamberg and Gary~R. Goldstein.
\newblock {Estimates of the nucleon tensor charge}.
\newblock In \emph{{20th International Symposium on Lepton and Photon
  Interactions at High Energies (LP 01)}}, 6 2001.

\bibitem[Adhikari et~al.(2023)]{GlueX:2023pev}
S.~Adhikari et~al.
\newblock {Measurement of the J/$\psi $ photoproduction cross section over the
  full near-threshold kinematic region}.
\newblock \emph{Phys. Rev. C}, 108\penalty0 (2):\penalty0 025201, 2023.
\newblock \doi{10.1103/PhysRevC.108.025201}.

\bibitem[Duran et~al.(2023)]{Duran:2022xag}
B.~Duran et~al.
\newblock {Determining the gluonic gravitational form factors of the proton}.
\newblock \emph{Nature}, 615\penalty0 (7954):\penalty0 813--816, 2023.
\newblock \doi{10.1038/s41586-023-05730-4}.

\bibitem[Eidelman et~al.(2004)]{ParticleDataGroup:2004fcd}
S.~Eidelman et~al.
\newblock {Review of particle physics. Particle Data Group}.
\newblock \emph{Phys. Lett. B}, 592\penalty0 (1-4):\penalty0 1, 2004.
\newblock \doi{10.1016/j.physletb.2004.06.001}.

\bibitem[Dumitru and Stebel(2019)]{Dumitru:2019qec}
Adrian Dumitru and Tomasz Stebel.
\newblock {Multiquark matrix elements in the proton and three gluon exchange
  for exclusive $\eta_c$ production in photon-proton diffractive scattering}.
\newblock \emph{Phys. Rev. D}, 99\penalty0 (9):\penalty0 094038, 2019.
\newblock \doi{10.1103/PhysRevD.99.094038}.

\bibitem[Jia et~al.(2023)Jia, Mo, Pan, and Zhang]{Jia:2022oyl}
Yu~Jia, Zhewen Mo, Jichen Pan, and Jia-Yue Zhang.
\newblock {Photoproduction of C-even quarkonia at the EIC and EicC}.
\newblock \emph{Phys. Rev. D}, 108\penalty0 (1):\penalty0 016015, 2023.
\newblock \doi{10.1103/PhysRevD.108.016015}.

\bibitem[Czyzewski et~al.(1997)Czyzewski, Kwiecinski, Motyka, and
  Sadzikowski]{Czyzewski:1996bv}
J.~Czyzewski, J.~Kwiecinski, L.~Motyka, and M.~Sadzikowski.
\newblock {Exclusive eta(c) photoproduction and electroproduction at HERA as a
  possible probe of the odderon singularity in QCD}.
\newblock \emph{Phys. Lett. B}, 398:\penalty0 400--406, 1997.
\newblock \doi{10.1016/S0370-2693(97)00249-9}.
\newblock [Erratum: Phys.Lett.B 411, 402 (1997)].

\bibitem[Engel et~al.(1998)Engel, Ivanov, Kirschner, and
  Szymanowski]{Engel:1997cga}
R.~Engel, D.~Yu. Ivanov, R.~Kirschner, and L.~Szymanowski.
\newblock {Diffractive meson production from virtual photons with odd charge -
  parity exchange}.
\newblock \emph{Eur. Phys. J. C}, 4:\penalty0 93--99, 1998.
\newblock \doi{10.1007/s100520050188}.

\bibitem[Bartels et~al.(2001)Bartels, Braun, Colferai, and
  Vacca]{Bartels:2001hw}
Jochen Bartels, M.~A. Braun, D.~Colferai, and G.~P. Vacca.
\newblock {Diffractive eta(c) photoproduction and electroproduction with the
  perturbative QCD odderon}.
\newblock \emph{Eur. Phys. J. C}, 20:\penalty0 323--331, 2001.
\newblock \doi{10.1007/s100520100676}.

\bibitem[Beni\'c et~al.(2023)Beni\'c, Horvati\'c, Kaushik, and
  Vivoda]{Benic:2023ybl}
Sanjin Beni\'c, Davor Horvati\'c, Abhiram Kaushik, and Eric~Andreas Vivoda.
\newblock {Exclusive \ensuremath{\eta}c production from small-x evolved Odderon
  at an electron-ion collider}.
\newblock \emph{Phys. Rev. D}, 108\penalty0 (7):\penalty0 074005, 2023.
\newblock \doi{10.1103/PhysRevD.108.074005}.

\bibitem[Shuryak and Vainshtein(1982{\natexlab{a}})]{Shuryak:1981kj}
Edward~V. Shuryak and A.~I. Vainshtein.
\newblock {Theory of Power Corrections to Deep Inelastic Scattering in Quantum
  Chromodynamics. 1. Q**2 Effects}.
\newblock \emph{Nucl. Phys. B}, 199:\penalty0 451--481, 1982{\natexlab{a}}.
\newblock \doi{10.1016/0550-3213(82)90355-8}.

\bibitem[Shuryak and Vainshtein(1982{\natexlab{b}})]{Shuryak:1981pi}
Edward~V. Shuryak and A.~I. Vainshtein.
\newblock {Theory of Power Corrections to Deep Inelastic Scattering in Quantum
  Chromodynamics. 2. Q**4 Effects: Polarized Target}.
\newblock \emph{Nucl. Phys. B}, 201:\penalty0 141, 1982{\natexlab{b}}.
\newblock \doi{10.1016/0550-3213(82)90377-7}.

\bibitem[Bhattacharya and Metz(2022)]{Bhattacharya:2021boh}
Shohini Bhattacharya and Andreas Metz.
\newblock {Burkhardt-Cottingham-type sum rules for light-cone and quasi-PDFs}.
\newblock \emph{Phys. Rev. D}, 105\penalty0 (5):\penalty0 054027, 2022.
\newblock \doi{10.1103/PhysRevD.105.054027}.

\bibitem[Shuryak and Vainshtein(1981)]{Shuryak:1981dg}
Edward~V. Shuryak and A.~I. Vainshtein.
\newblock {QCD POWER CORRECTIONS TO DEEP INELASTIC SCATTERING}.
\newblock \emph{Phys. Lett. B}, 105:\penalty0 65--67, 1981.
\newblock \doi{10.1016/0370-2693(81)90042-3}.

\bibitem[Jaffe and Soldate(1981)]{Jaffe:1981td}
R.~L. Jaffe and M.~Soldate.
\newblock {Twist Four in the QCD Analysis of Leptoproduction}.
\newblock \emph{Phys. Lett. B}, 105:\penalty0 467--472, 1981.
\newblock \doi{10.1016/0370-2693(81)91206-5}.

\bibitem[Jaffe and Soldate(1982)]{Jaffe:1982pm}
R.~L. Jaffe and M.~Soldate.
\newblock {Twist Four in Electroproduction: Canonical Operators and Coefficient
  Functions}.
\newblock \emph{Phys. Rev. D}, 26:\penalty0 49--68, 1982.
\newblock \doi{10.1103/PhysRevD.26.49}.

\bibitem[Aslan et~al.(2019)Aslan, Burkardt, and Schlegel]{Aslan:2019jis}
Fatma~P. Aslan, Matthias Burkardt, and Marc Schlegel.
\newblock {Transverse Force Tomography}.
\newblock \emph{Phys. Rev. D}, 100\penalty0 (9):\penalty0 096021, 2019.
\newblock \doi{10.1103/PhysRevD.100.096021}.

\bibitem[Wandzura and Wilczek(1977)]{Wandzura:1977qf}
S.~Wandzura and Frank Wilczek.
\newblock {Sum Rules for Spin Dependent Electroproduction: Test of Relativistic
  Constituent Quarks}.
\newblock \emph{Phys. Lett. B}, 72:\penalty0 195--198, 1977.
\newblock \doi{10.1016/0370-2693(77)90700-6}.

\bibitem[Vladimirov et~al.(2025)Vladimirov, Portela, and
  Rodini]{Vladimirov:2025qrh}
Alexey Vladimirov, Guillermo Portela, and Simone Rodini.
\newblock {Determination of quark-gluon-quark interference within the proton}.
\newblock 11 2025.

\bibitem[Crawford et~al.(2024)Crawford, Can, Horsley, Rakow, Schierholz,
  St\"uben, Young, and Zanotti]{Crawford:2024wzx}
J.~A. Crawford, K.~U. Can, R.~Horsley, P.~E.~L. Rakow, G.~Schierholz,
  H.~St\"uben, R.~D. Young, and J.~M. Zanotti.
\newblock {Transverse force distributions in the proton from lattice QCD}.
\newblock 8 2024.

\bibitem[Aslan et~al.(2020)Aslan, Burkardt, and Schlegel]{Aslan:2020eqo}
Fatma~P. Aslan, Matthias Burkardt, and Marc Schlegel.
\newblock {Transverse Force Tomography}.
\newblock In \emph{{Probing Nucleons and Nuclei in High Energy Collisions}:
  {Dedicated to the Physics of the Electron Ion Collider}}, pages 186--189,
  2020.
\newblock \doi{10.1142/9789811214950_0038}.

\bibitem[Gockeler et~al.(2005)Gockeler, Horsley, Pleiter, Rakow, Schafer,
  Schierholz, Stuben, and Zanotti]{Gockeler:2005vw}
M.~Gockeler, R.~Horsley, D.~Pleiter, Paul E.~L. Rakow, A.~Schafer,
  G.~Schierholz, H.~Stuben, and J.~M. Zanotti.
\newblock {Investigation of the second moment of the nucleon's g(1) and g(2)
  structure functions in two-flavor lattice QCD}.
\newblock \emph{Phys. Rev. D}, 72:\penalty0 054507, 2005.
\newblock \doi{10.1103/PhysRevD.72.054507}.

\bibitem[B{\"u}rger et~al.(2022)B{\"u}rger, Wurm, L{\"o}ffler, G{\"o}ckeler,
  Bali, Collins, Sch{\"a}fer, and Sternbeck]{Burger:2021knd}
S.~B{\"u}rger, T.~Wurm, M.~L{\"o}ffler, M.~G{\"o}ckeler, G.~Bali, S.~Collins,
  A.~Sch{\"a}fer, and A.~Sternbeck.
\newblock {Lattice results for the longitudinal spin structure and color forces
  on quarks in a nucleon}.
\newblock \emph{Phys. Rev. D}, 105\penalty0 (5):\penalty0 054504, 2022.
\newblock \doi{10.1103/PhysRevD.105.054504}.

\bibitem[Abe et~al.(1998)]{E143:1998hbs}
K.~Abe et~al.
\newblock {Measurements of the proton and deuteron spin structure functions
  g(1) and g(2)}.
\newblock \emph{Phys. Rev. D}, 58:\penalty0 112003, 1998.
\newblock \doi{10.1103/PhysRevD.58.112003}.

\bibitem[Diehl and Hagler(2005)]{Diehl:2005jf}
M.~Diehl and Ph. Hagler.
\newblock {Spin densities in the transverse plane and generalized transversity
  distributions}.
\newblock \emph{Eur. Phys. J. C}, 44:\penalty0 87--101, 2005.
\newblock \doi{10.1140/epjc/s2005-02342-6}.

\bibitem[Bhoonah and Lorc{\'e}(2017)]{Bhoonah:2017olu}
Amit Bhoonah and C{\'e}dric Lorc{\'e}.
\newblock {Quark transverse spin{\textendash}orbit correlations}.
\newblock \emph{Phys. Lett. B}, 774:\penalty0 435--440, 2017.
\newblock \doi{10.1016/j.physletb.2017.10.003}.

\bibitem[Kim(2025)]{Kim:2025mol}
June-Young Kim.
\newblock {Chiral-odd generalized parton distributions in the large-$N_{c}$
  limit of QCD: Next-to-leading-order contributions}.
\newblock 6 2025.

\bibitem[Alexandrou(2024)]{Alexandrou:2024awx}
Constantia Alexandrou.
\newblock {Nucleon Transversity from lattice QCD}.
\newblock \emph{PoS}, Transversity2024:\penalty0 002, 2024.
\newblock \doi{10.22323/1.477.0002}.

\bibitem[Hatsuda and Kunihiro(1994)]{Hatsuda:1994pi}
Tetsuo Hatsuda and Teiji Kunihiro.
\newblock {QCD phenomenology based on a chiral effective Lagrangian}.
\newblock \emph{Phys. Rept.}, 247:\penalty0 221--367, 1994.
\newblock \doi{10.1016/0370-1573(94)90022-1}.

\bibitem[Alberico et~al.(2009)Alberico, Bilenky, Giunti, and
  Graczyk]{Alberico:2008sz}
W.~M. Alberico, S.~M. Bilenky, C.~Giunti, and K.~M. Graczyk.
\newblock {Electromagnetic form factors of the nucleon: New Fit and analysis of
  uncertainties}.
\newblock \emph{Phys. Rev. C}, 79:\penalty0 065204, 2009.
\newblock \doi{10.1103/PhysRevC.79.065204}.

\bibitem[Kochelev(2003)]{Kochelev:2003cp}
N.~I. Kochelev.
\newblock {The Pauli form-factor of the quark induced by instantons}.
\newblock \emph{Phys. Lett. B}, 565:\penalty0 131--136, 2003.
\newblock \doi{10.1016/S0370-2693(03)00727-5}.

\bibitem[Perdrisat et~al.(2007)Perdrisat, Punjabi, and
  Vanderhaeghen]{Perdrisat:2006hj}
C.~F. Perdrisat, V.~Punjabi, and M.~Vanderhaeghen.
\newblock {Nucleon Electromagnetic Form Factors}.
\newblock \emph{Prog. Part. Nucl. Phys.}, 59:\penalty0 694--764, 2007.
\newblock \doi{10.1016/j.ppnp.2007.05.001}.

\bibitem[Belushkin et~al.(2007)Belushkin, Hammer, and
  Meissner]{Belushkin:2006qa}
M.~A. Belushkin, H.~W. Hammer, and U.~G. Meissner.
\newblock {Dispersion analysis of the nucleon form-factors including meson
  continua}.
\newblock \emph{Phys. Rev. C}, 75:\penalty0 035202, 2007.
\newblock \doi{10.1103/PhysRevC.75.035202}.

\bibitem[Shanahan et~al.(2014)Shanahan, Thomas, Young, Zanotti, Horsley,
  Nakamura, Pleiter, Rakow, Schierholz, and St\"uben]{CSSM:2014knt}
P.~E. Shanahan, A.~W. Thomas, R.~D. Young, J.~M. Zanotti, R.~Horsley,
  Y.~Nakamura, D.~Pleiter, P.~E.~L. Rakow, G.~Schierholz, and H.~St\"uben.
\newblock {Magnetic form factors of the octet baryons from lattice QCD and
  chiral extrapolation}.
\newblock \emph{Phys. Rev. D}, 89:\penalty0 074511, 2014.
\newblock \doi{10.1103/PhysRevD.89.074511}.

\bibitem[Faccioli et~al.(2004)Faccioli, Guadagnoli, and
  Simula]{Faccioli:2004jz}
P.~Faccioli, D.~Guadagnoli, and S.~Simula.
\newblock {The Neutron electric dipole moment in the instanton vacuum: Quenched
  versus unquenched simulations}.
\newblock \emph{Phys. Rev. D}, 70:\penalty0 074017, 2004.
\newblock \doi{10.1103/PhysRevD.70.074017}.

\bibitem[Mereghetti et~al.(2011)Mereghetti, de~Vries, Hockings, Maekawa, and
  van Kolck]{Mereghetti:2010kp}
E.~Mereghetti, J.~de~Vries, W.~H. Hockings, C.~M. Maekawa, and U.~van Kolck.
\newblock {The Electric Dipole Form Factor of the Nucleon in Chiral
  Perturbation Theory to Sub-leading Order}.
\newblock \emph{Phys. Lett. B}, 696:\penalty0 97--102, 2011.
\newblock \doi{10.1016/j.physletb.2010.12.018}.

\bibitem[Dragos et~al.(2021)Dragos, Luu, Shindler, de~Vries, and
  Yousif]{Dragos:2019oxn}
Jack Dragos, Thomas Luu, Andrea Shindler, Jordy de~Vries, and Ahmed Yousif.
\newblock {Confirming the Existence of the strong CP Problem in Lattice QCD
  with the Gradient Flow}.
\newblock \emph{Phys. Rev. C}, 103\penalty0 (1):\penalty0 015202, 2021.
\newblock \doi{10.1103/PhysRevC.103.015202}.

\bibitem[Kharzeev and Levin(2017)]{Kharzeev:2017qzs}
Dmitri~E. Kharzeev and Eugene~M. Levin.
\newblock {Deep inelastic scattering as a probe of entanglement}.
\newblock \emph{Phys. Rev.}, D95\penalty0 (11):\penalty0 114008, 2017.
\newblock \doi{10.1103/PhysRevD.95.114008}.

\bibitem[Dove et~al.(2021)]{SeaQuest:2021zxb}
J.~Dove et~al.
\newblock {The asymmetry of antimatter in the proton}.
\newblock \emph{Nature}, 590\penalty0 (7847):\penalty0 561--565, 2021.
\newblock \doi{10.1038/s41586-022-04707-z}.
\newblock [Erratum: Nature 604, E26 (2022)].

\bibitem[Ji(2013)]{Ji:2013dva}
Xiangdong Ji.
\newblock {Parton Physics on a Euclidean Lattice}.
\newblock \emph{Phys. Rev. Lett.}, 110:\penalty0 262002, 2013.
\newblock \doi{10.1103/PhysRevLett.110.262002}.

\bibitem[Izubuchi et~al.(2018)Izubuchi, Ji, Jin, Stewart, and
  Zhao]{Izubuchi:2018srq}
Taku Izubuchi, Xiangdong Ji, Luchang Jin, Iain~W. Stewart, and Yong Zhao.
\newblock {Factorization Theorem Relating Euclidean and Light-Cone Parton
  Distributions}.
\newblock \emph{Phys. Rev. D}, 98\penalty0 (5):\penalty0 056004, 2018.
\newblock \doi{10.1103/PhysRevD.98.056004}.

\bibitem[Ji(2020)]{Ji:2020byp}
Xiangdong Ji.
\newblock {Why is LaMET an effective field theory for partonic structure?}
\newblock 7 2020.

\bibitem[Zhao(2026)]{Zhao:2025oto}
Yong Zhao.
\newblock {Improving the Precision of First-Principles Calculation of Parton
  Physics from Lattice Quantum Chromodynamics}.
\newblock \emph{Research}, 9:\penalty0 1181, 2026.
\newblock \doi{10.34133/research.1181}.

\bibitem[Brambilla and Wang(2024)]{Brambilla:2023vwm}
Nora Brambilla and Xiang-Peng Wang.
\newblock {Off-lightcone Wilson-line operators in gradient flow}.
\newblock \emph{JHEP}, 06:\penalty0 210, 2024.
\newblock \doi{10.1007/JHEP06(2024)210}.

\bibitem[Hieda and Suzuki(2016)]{Hieda:2016lly}
Kenji Hieda and Hiroshi Suzuki.
\newblock {Small flow-time representation of fermion bilinear operators}.
\newblock \emph{Mod. Phys. Lett. A}, 31\penalty0 (38):\penalty0 1650214, 2016.
\newblock \doi{10.1142/S021773231650214X}.

\bibitem[Mereghetti et~al.(2022)Mereghetti, Monahan, Rizik, Shindler, and
  Stoffer]{Mereghetti:2021nkt}
Emanuele Mereghetti, Christopher~J. Monahan, Matthew~D. Rizik, Andrea Shindler,
  and Peter Stoffer.
\newblock {One-loop matching for quark dipole operators in a gradient-flow
  scheme}.
\newblock \emph{JHEP}, 04:\penalty0 050, 2022.
\newblock \doi{10.1007/JHEP04(2022)050}.
\newblock [Erratum: JHEP 03, 101 (2025)].

\bibitem[Liu and Chen(2021)]{Liu:2020rqi}
Wei-Yang Liu and Jiunn-Wei Chen.
\newblock {Renormalon effects in quasiparton distributions}.
\newblock \emph{Phys. Rev. D}, 104\penalty0 (9):\penalty0 094501, 2021.
\newblock \doi{10.1103/PhysRevD.104.094501}.

\bibitem[Braun et~al.(2024)Braun, Koller, and Schoenleber]{Braun:2024snf}
Vladimir~M. Braun, Maria Koller, and Jakob Schoenleber.
\newblock {Renormalons and power corrections in pseudo- and quasi-GPDs}.
\newblock \emph{Phys. Rev. D}, 109\penalty0 (7):\penalty0 074510, 2024.
\newblock \doi{10.1103/PhysRevD.109.074510}.

\bibitem[Zhang(2025)]{Zhang:2025mer}
Jia-Lu Zhang.
\newblock {Renormalon effects on quasi-PDFs in the gradient flow formalism}.
\newblock \emph{Phys. Rev. D}, 112\penalty0 (9):\penalty0 094504, 2025.
\newblock \doi{10.1103/h5xf-pmww}.

\bibitem[Braun et~al.(2019)Braun, Vladimirov, and Zhang]{Braun:2018brg}
Vladimir~M. Braun, Alexey Vladimirov, and Jian-Hui Zhang.
\newblock {Power corrections and renormalons in parton quasidistributions}.
\newblock \emph{Phys. Rev. D}, 99\penalty0 (1):\penalty0 014013, 2019.
\newblock \doi{10.1103/PhysRevD.99.014013}.

\bibitem[()]{}


\bibitem[Conway et~al.(1989)]{Conway:1989fs}
J.~S. Conway et~al.
\newblock {Experimental Study of Muon Pairs Produced by 252-GeV Pions on
  Tungsten}.
\newblock \emph{Phys. Rev. D}, 39:\penalty0 92--122, 1989.
\newblock \doi{10.1103/PhysRevD.39.92}.

\bibitem[Aicher et~al.(2010)Aicher, Schafer, and Vogelsang]{Aicher:2010cb}
Matthias Aicher, Andreas Schafer, and Werner Vogelsang.
\newblock {Soft-gluon resummation and the valence parton distribution function
  of the pion}.
\newblock \emph{Phys. Rev. Lett.}, 105:\penalty0 252003, 2010.
\newblock \doi{10.1103/PhysRevLett.105.252003}.

\bibitem[Barry et~al.(2023)Barry, Gamberg, Melnitchouk, Moffat, Pitonyak,
  Prokudin, and Sato]{Barry:2023qqh}
P.~C. Barry, L.~Gamberg, W.~Melnitchouk, E.~Moffat, D.~Pitonyak, A.~Prokudin,
  and N.~Sato.
\newblock {Tomography of pions and protons via transverse momentum dependent
  distributions}.
\newblock \emph{Phys. Rev. D}, 108\penalty0 (9):\penalty0 L091504, 2023.
\newblock \doi{10.1103/PhysRevD.108.L091504}.

\bibitem[Sufian et~al.(2020)Sufian, Egerer, Karpie, Edwards, Jo{\'o}, Ma,
  Orginos, Qiu, and Richards]{Sufian:2020vzb}
Raza~Sabbir Sufian, Colin Egerer, Joseph Karpie, Robert~G. Edwards, B{\'a}lint
  Jo{\'o}, Yan-Qing Ma, Kostas Orginos, Jian-Wei Qiu, and David~G. Richards.
\newblock {Pion Valence Quark Distribution from Current-Current Correlation in
  Lattice QCD}.
\newblock \emph{Phys. Rev. D}, 102\penalty0 (5):\penalty0 054508, 2020.
\newblock \doi{10.1103/PhysRevD.102.054508}.

\bibitem[Lin et~al.(2021)Lin, Chen, Fan, Zhang, and Zhang]{Lin:2020ssv}
Huey-Wen Lin, Jiunn-Wei Chen, Zhouyou Fan, Jian-Hui Zhang, and Rui Zhang.
\newblock {Valence-Quark Distribution of the Kaon and Pion from Lattice QCD}.
\newblock \emph{Phys. Rev. D}, 103\penalty0 (1):\penalty0 014516, 2021.
\newblock \doi{10.1103/PhysRevD.103.014516}.

\bibitem[Badier
  et~al.(1980)]{Saclay-CERN-CollegedeFrance-EcolePoly-Orsay:1980fhh}
J.~Badier et~al.
\newblock {Measurement of the $K^- / \pi^-$ Structure Function Ratio Using the
  {Drell-Yan} Process}.
\newblock \emph{Phys. Lett. B}, 93:\penalty0 354--356, 1980.
\newblock \doi{10.1016/0370-2693(80)90530-4}.

\bibitem[Nguyen et~al.(2011)Nguyen, Bashir, Roberts, and Tandy]{Nguyen:2011jy}
Trang Nguyen, Adnan Bashir, Craig~D. Roberts, and Peter~C. Tandy.
\newblock {Pion and kaon valence-quark parton distribution functions}.
\newblock \emph{Phys. Rev. C}, 83:\penalty0 062201, 2011.
\newblock \doi{10.1103/PhysRevC.83.062201}.

\bibitem[Radyushkin(1996{\natexlab{a}})]{Radyushkin:1996nd}
A.~V. Radyushkin.
\newblock {Scaling limit of deeply virtual Compton scattering}.
\newblock \emph{Phys. Lett. B}, 380:\penalty0 417--425, 1996{\natexlab{a}}.
\newblock \doi{10.1016/0370-2693(96)00528-X}.

\bibitem[Ji(1997{\natexlab{b}})]{Ji:1996nm}
Xiang-Dong Ji.
\newblock {Deeply virtual Compton scattering}.
\newblock \emph{Phys. Rev. D}, 55:\penalty0 7114--7125, 1997{\natexlab{b}}.
\newblock \doi{10.1103/PhysRevD.55.7114}.

\bibitem[Radyushkin(1996{\natexlab{b}})]{Radyushkin:1996ru}
A.~V. Radyushkin.
\newblock {Asymmetric gluon distributions and hard diffractive
  electroproduction}.
\newblock \emph{Phys. Lett. B}, 385:\penalty0 333--342, 1996{\natexlab{b}}.
\newblock \doi{10.1016/0370-2693(96)00844-1}.

\bibitem[Collins et~al.(1997)Collins, Frankfurt, and Strikman]{Collins:1996fb}
John~C. Collins, Leonid Frankfurt, and Mark Strikman.
\newblock {Factorization for hard exclusive electroproduction of mesons in
  QCD}.
\newblock \emph{Phys. Rev. D}, 56:\penalty0 2982--3006, 1997.
\newblock \doi{10.1103/PhysRevD.56.2982}.

\bibitem[Diehl(2003)]{Diehl:2003ny}
M.~Diehl.
\newblock {Generalized parton distributions}.
\newblock \emph{Phys. Rept.}, 388:\penalty0 41--277, 2003.
\newblock \doi{10.1016/j.physrep.2003.08.002}.

\bibitem[Goeke et~al.(2001)Goeke, Polyakov, and Vanderhaeghen]{Goeke:2001tz}
K.~Goeke, Maxim~V. Polyakov, and M.~Vanderhaeghen.
\newblock {Hard exclusive reactions and the structure of hadrons}.
\newblock \emph{Prog. Part. Nucl. Phys.}, 47:\penalty0 401--515, 2001.
\newblock \doi{10.1016/S0146-6410(01)00158-2}.

\bibitem[Hoodbhoy and Ji(1998)]{Hoodbhoy:1998vm}
Pervez Hoodbhoy and Xiang-Dong Ji.
\newblock {Helicity flip off forward parton distributions of the nucleon}.
\newblock \emph{Phys. Rev. D}, 58:\penalty0 054006, 1998.
\newblock \doi{10.1103/PhysRevD.58.054006}.

\bibitem[Blumlein(2013)]{Blumlein:2012bf}
Johannes Blumlein.
\newblock {The Theory of Deeply Inelastic Scattering}.
\newblock \emph{Prog. Part. Nucl. Phys.}, 69:\penalty0 28--84, 2013.
\newblock \doi{10.1016/j.ppnp.2012.09.006}.

\bibitem[Boussarie et~al.(2023)]{Boussarie:2023izj}
Renaud Boussarie et~al.
\newblock {TMD Handbook}.
\newblock 4 2023.

\bibitem[Rogers(2016)]{Rogers:2015sqa}
Ted~C. Rogers.
\newblock {An overview of transverse-momentum\textendash{}dependent
  factorization and evolution}.
\newblock \emph{Eur. Phys. J. A}, 52\penalty0 (6):\penalty0 153, 2016.
\newblock \doi{10.1140/epja/i2016-16153-7}.

\bibitem[Zhang et~al.(2020)]{LatticeParton:2020uhz}
Qi-An Zhang et~al.
\newblock {Lattice-QCD Calculations of TMD Soft Function Through Large-Momentum
  Effective Theory}.
\newblock \emph{Phys. Rev. Lett.}, 125\penalty0 (19):\penalty0 192001, 2020.
\newblock \doi{10.22323/1.396.0477}.

\bibitem[Ji et~al.(2005)Ji, Ma, and Yuan]{Ji:2004wu}
Xiang-dong Ji, Jian-ping Ma, and Feng Yuan.
\newblock {QCD factorization for semi-inclusive deep-inelastic scattering at
  low transverse momentum}.
\newblock \emph{Phys. Rev. D}, 71:\penalty0 034005, 2005.
\newblock \doi{10.1103/PhysRevD.71.034005}.

\bibitem[Tangerman and Mulders(1995)]{Tangerman:1995hw}
R.~D. Tangerman and P.~J. Mulders.
\newblock {Probing transverse quark polarization in deep inelastic
  leptoproduction}.
\newblock \emph{Phys. Lett. B}, 352:\penalty0 129--133, 1995.
\newblock \doi{10.1016/0370-2693(95)00485-4}.

\bibitem[Kumano and Song(2021)]{Kumano:2020ijt}
S.~Kumano and Qin-Tao Song.
\newblock {Transverse-momentum-dependent parton distribution functions up to
  twist 4 for spin-1 hadrons}.
\newblock \emph{Phys. Rev. D}, 103\penalty0 (1):\penalty0 014025, 2021.
\newblock \doi{10.1103/PhysRevD.103.014025}.

\bibitem[Angeles-Martinez et~al.(2015)]{Angeles-Martinez:2015sea}
R.~Angeles-Martinez et~al.
\newblock {Transverse Momentum Dependent (TMD) parton distribution functions:
  status and prospects}.
\newblock \emph{Acta Phys. Polon. B}, 46\penalty0 (12):\penalty0 2501--2534,
  2015.
\newblock \doi{10.5506/APhysPolB.46.2501}.

\bibitem[Grosse~Perdekamp and Yuan(2015)]{GrossePerdekamp:2015xdx}
Matthias Grosse~Perdekamp and Feng Yuan.
\newblock {Transverse Spin Structure of the Nucleon}.
\newblock \emph{Ann. Rev. Nucl. Part. Sci.}, 65:\penalty0 429--456, 2015.
\newblock \doi{10.1146/annurev-nucl-102014-021948}.

\bibitem[Aidala et~al.(2013)Aidala, Bass, Hasch, and Mallot]{Aidala:2012mv}
Christine~A. Aidala, Steven~D. Bass, Delia Hasch, and Gerhard~K. Mallot.
\newblock {The Spin Structure of the Nucleon}.
\newblock \emph{Rev. Mod. Phys.}, 85:\penalty0 655--691, 2013.
\newblock \doi{10.1103/RevModPhys.85.655}.

\bibitem[Barone et~al.(2010)Barone, Bradamante, and Martin]{Barone:2010zz}
Vincenzo Barone, Franco Bradamante, and Anna Martin.
\newblock {Transverse-spin and transverse-momentum effects in high-energy
  processes}.
\newblock \emph{Prog. Part. Nucl. Phys.}, 65:\penalty0 267--333, 2010.
\newblock \doi{10.1016/j.ppnp.2010.07.003}.

\bibitem[D'Alesio and Murgia(2008)]{DAlesio:2007bjf}
U.~D'Alesio and F.~Murgia.
\newblock {Azimuthal and Single Spin Asymmetries in Hard Scattering Processes}.
\newblock \emph{Prog. Part. Nucl. Phys.}, 61:\penalty0 394--454, 2008.
\newblock \doi{10.1016/j.ppnp.2008.01.001}.

\bibitem[Collins(2011)]{Collins:2011zzd}
John Collins.
\newblock \emph{{Foundations of Perturbative QCD}}, volume~32.
\newblock Cambridge University Press, 2011.
\newblock ISBN 978-1-009-40184-5, 978-1-009-40183-8, 978-1-009-40182-1.
\newblock \doi{10.1017/9781009401845}.

\bibitem[Balitsky(2001)]{Balitsky:2001gj}
I.~Balitsky.
\newblock {High-energy QCD and Wilson lines}.
\newblock 1 2001.
\newblock \doi{10.1142/9789812810458_0030}.

\bibitem[Simonov(1996)]{Simonov:1996ati}
Yu.~A. Simonov.
\newblock {Confinement}.
\newblock \emph{Phys. Usp.}, 39:\penalty0 313--336, 1996.
\newblock \doi{10.1070/PU1996v039n04ABEH000140}.

\bibitem[DeMartini and Shuryak(2021)]{DeMartini:2021xkg}
Dallas DeMartini and Edward Shuryak.
\newblock {Chiral symmetry breaking and confinement from an interacting
  ensemble of instanton dyons in two-flavor massless QCD}.
\newblock \emph{Phys. Rev. D}, 104\penalty0 (9):\penalty0 094031, 2021.
\newblock \doi{10.1103/PhysRevD.104.094031}.

\bibitem[Makeenko and Migdal(1980)]{Makeenko:1980vm}
Yu. Makeenko and Alexander~A. Migdal.
\newblock {Quantum Chromodynamics as Dynamics of Loops}.
\newblock \emph{Sov. J. Nucl. Phys.}, 32:\penalty0 431, 1980.
\newblock \doi{10.1016/0550-3213(81)90258-3}.

\bibitem[Makeenko and Migdal(1979)]{Makeenko:1979pb}
Yu.~M. Makeenko and Alexander~A. Migdal.
\newblock {Exact Equation for the Loop Average in Multicolor QCD}.
\newblock \emph{Phys. Lett. B}, 88:\penalty0 135, 1979.
\newblock \doi{10.1016/0370-2693(79)90131-X}.
\newblock [Erratum: Phys.Lett.B 89, 437 (1980)].

\bibitem[Polyakov(1987)]{Polyakov:1987ez}
Alexander~M. Polyakov.
\newblock \emph{{Gauge Fields and Strings}}, volume~3.
\newblock 1987.

\bibitem[Korchemskaya and Korchemsky(1995)]{Korchemskaya:1994qp}
I.~A. Korchemskaya and G.~P. Korchemsky.
\newblock {High-energy scattering in QCD and cross singularities of Wilson
  loops}.
\newblock \emph{Nucl. Phys. B}, 437:\penalty0 127--162, 1995.
\newblock \doi{10.1016/0550-3213(94)00553-Q}.

\bibitem[Cherednikov et~al.(2020)Cherednikov, Mertens, and Van~der
  Veken]{Cherednikov:2020mtu}
Igor~Olegovich Cherednikov, Tom Mertens, and Frederik Van~der Veken.
\newblock \emph{{Wilson Lines in Quantum Field Theory}}, volume~24 of \emph{De
  Gruyter Studies in Mathematical Physics}.
\newblock De Gruyter, 1 2020.
\newblock ISBN 978-3-11-065169-0, 978-3-11-065092-1.
\newblock \doi{10.1515/9783110651690}.

\bibitem[Cherednikov and Van~der Veken(2017)]{Cherednikov:2017qbt}
Igor~Olegovich Cherednikov and Frederik~F. Van~der Veken.
\newblock \emph{{Parton Densities in QCD}}, volume~37 of \emph{De Gruyter
  Studies in Mathematical Physics}.
\newblock De Gruyter, 1 2017.
\newblock ISBN 978-3-11-043060-8, 978-3-11-043939-7.
\newblock \doi{10.1515/9783110430608}.

\bibitem[Dotsenko and Vergeles(1980)]{Dotsenko:1979wb}
V.~S. Dotsenko and S.~N. Vergeles.
\newblock {Renormalizability of Phase Factors in the Nonabelian Gauge Theory}.
\newblock \emph{Nucl. Phys. B}, 169:\penalty0 527--546, 1980.
\newblock \doi{10.1016/0550-3213(80)90103-0}.

\bibitem[Brandt et~al.(1982)Brandt, Gocksch, Sato, and Neri]{Brandt:1982gz}
Richard~A. Brandt, A.~Gocksch, M.~A. Sato, and F.~Neri.
\newblock {LOOP SPACE}.
\newblock \emph{Phys. Rev. D}, 26:\penalty0 3611, 1982.
\newblock \doi{10.1103/PhysRevD.26.3611}.

\bibitem[Korchemskaya and Korchemsky(1992)]{Korchemskaya:1992je}
I.~A. Korchemskaya and G.~P. Korchemsky.
\newblock {On lightlike Wilson loops}.
\newblock \emph{Phys. Lett. B}, 287:\penalty0 169--175, 1992.
\newblock \doi{10.1016/0370-2693(92)91895-G}.

\bibitem[Brandt et~al.(1981)Brandt, Neri, and Sato]{Brandt:1981kf}
Richard~A. Brandt, Filippo Neri, and Masa-aki Sato.
\newblock {Renormalization of Loop Functions for All Loops}.
\newblock \emph{Phys. Rev. D}, 24:\penalty0 879, 1981.
\newblock \doi{10.1103/PhysRevD.24.879}.

\bibitem[Korchemsky(1994)]{Korchemsky:1993hr}
Gregory~P. Korchemsky.
\newblock {On Near forward high-energy scattering in QCD}.
\newblock \emph{Phys. Lett. B}, 325:\penalty0 459--466, 1994.
\newblock \doi{10.1016/0370-2693(94)90040-X}.

\bibitem[Vladimirov and Stefanis(2014)]{Vladimirov:2014hla}
A.~A. Vladimirov and N.~G. Stefanis.
\newblock {Key features of the TMD soft-factor structure}.
\newblock \emph{Few Body Syst.}, 55:\penalty0 297--302, 2014.
\newblock \doi{10.1007/s00601-014-0851-1}.

\bibitem[Vladimirov(2018)]{Vladimirov:2017ksc}
Alexey Vladimirov.
\newblock {Structure of rapidity divergences in multi-parton scattering soft
  factors}.
\newblock \emph{JHEP}, 04:\penalty0 045, 2018.
\newblock \doi{10.1007/JHEP04(2018)045}.

\bibitem[Grozin et~al.(2016)Grozin, Henn, Korchemsky, and
  Marquard]{Grozin:2015kna}
Andrey Grozin, Johannes~M. Henn, Gregory~P. Korchemsky, and Peter Marquard.
\newblock {The three-loop cusp anomalous dimension in QCD and its
  supersymmetric extensions}.
\newblock \emph{JHEP}, 01:\penalty0 140, 2016.
\newblock \doi{10.1007/JHEP01(2016)140}.

\bibitem[Korchemsky and Radyushkin(1987)]{Korchemsky:1987wg}
G.~P. Korchemsky and A.~V. Radyushkin.
\newblock {Renormalization of the Wilson Loops Beyond the Leading Order}.
\newblock \emph{Nucl. Phys. B}, 283:\penalty0 342--364, 1987.
\newblock \doi{10.1016/0550-3213(87)90277-X}.

\bibitem[Korchemsky and Radyushkin(1992)]{Korchemsky:1991zp}
G.~P. Korchemsky and A.~V. Radyushkin.
\newblock {Infrared factorization, Wilson lines and the heavy quark limit}.
\newblock \emph{Phys. Lett. B}, 279:\penalty0 359--366, 1992.
\newblock \doi{10.1016/0370-2693(92)90405-S}.

\bibitem[Mitev and Pomoni(2016)]{Mitev:2015oty}
Vladimir Mitev and Elli Pomoni.
\newblock {Exact Bremsstrahlung and Effective Couplings}.
\newblock \emph{JHEP}, 06:\penalty0 078, 2016.
\newblock \doi{10.1007/JHEP06(2016)078}.

\bibitem[Collins and Soper(1981)]{Collins:1981uk}
John~C. Collins and Davison~E. Soper.
\newblock {Back-To-Back Jets in QCD}.
\newblock \emph{Nucl. Phys. B}, 193:\penalty0 381, 1981.
\newblock \doi{10.1016/0550-3213(81)90339-4}.
\newblock [Erratum: Nucl.Phys.B 213, 545 (1983)].

\bibitem[Collins and Soper(1982)]{Collins:1981va}
John~C. Collins and Davison~E. Soper.
\newblock {Back-To-Back Jets: Fourier Transform from B to K-Transverse}.
\newblock \emph{Nucl. Phys. B}, 197:\penalty0 446--476, 1982.
\newblock \doi{10.1016/0550-3213(82)90453-9}.

\bibitem[Vladimirov(2020)]{Vladimirov:2020umg}
Alexey~A. Vladimirov.
\newblock {Self-contained definition of the Collins-Soper kernel}.
\newblock \emph{Phys. Rev. Lett.}, 125\penalty0 (19):\penalty0 192002, 2020.
\newblock \doi{10.1103/PhysRevLett.125.192002}.

\bibitem[Ji et~al.(2020)Ji, Liu, and Liu]{Ji:2019sxk}
Xiangdong Ji, Yizhuang Liu, and Yu-Sheng Liu.
\newblock {TMD soft function from large-momentum effective theory}.
\newblock \emph{Nucl. Phys. B}, 955:\penalty0 115054, 2020.
\newblock \doi{10.1016/j.nuclphysb.2020.115054}.

\bibitem[Collins and Soper(1987)]{Collins:1985xx}
John~C. Collins and Davison~E. Soper.
\newblock {The Two Particle Inclusive Cross-section in $e^+ e^-$ Annihilation
  at {PETRA}, {PEP} and {LEP} Energies}.
\newblock \emph{Nucl. Phys. B}, 284:\penalty0 253--270, 1987.
\newblock \doi{10.1016/0550-3213(87)90035-6}.

\bibitem[Avkhadiev et~al.(2023)Avkhadiev, Shanahan, Wagman, and
  Zhao]{Avkhadiev:2023poz}
Artur Avkhadiev, Phiala~E. Shanahan, Michael~L. Wagman, and Yong Zhao.
\newblock {Collins-Soper kernel from lattice QCD at the physical pion mass}.
\newblock \emph{Phys. Rev. D}, 108\penalty0 (11):\penalty0 114505, 2023.
\newblock \doi{10.1103/PhysRevD.108.114505}.

\bibitem[Zhao(2024)]{Zhao:2023ptv}
Yong Zhao.
\newblock {Transverse Momentum Distributions from Lattice QCD without Wilson
  Lines}.
\newblock \emph{Phys. Rev. Lett.}, 133\penalty0 (24):\penalty0 241904, 2024.
\newblock \doi{10.1103/PhysRevLett.133.241904}.

\bibitem[Bollweg et~al.(2024)Bollweg, Gao, Mukherjee, and
  Zhao]{Bollweg:2024zet}
Dennis Bollweg, Xiang Gao, Swagato Mukherjee, and Yong Zhao.
\newblock {Nonperturbative Collins-Soper kernel from chiral quarks with
  physical masses}.
\newblock \emph{Phys. Lett. B}, 852:\penalty0 138617, 2024.
\newblock \doi{10.1016/j.physletb.2024.138617}.

\bibitem[Chu et~al.(2023)]{LatticePartonLPC:2023pdv}
Min-Huan Chu et~al.
\newblock {Lattice calculation of the intrinsic soft function and the
  Collins-Soper kernel}.
\newblock \emph{JHEP}, 08:\penalty0 172, 2023.
\newblock \doi{10.1007/JHEP08(2023)172}.

\bibitem[Scimemi and Vladimirov(2020)]{Scimemi:2019cmh}
Ignazio Scimemi and Alexey Vladimirov.
\newblock {Non-perturbative structure of semi-inclusive deep-inelastic and
  Drell-Yan scattering at small transverse momentum}.
\newblock \emph{JHEP}, 06:\penalty0 137, 2020.
\newblock \doi{10.1007/JHEP06(2020)137}.

\bibitem[Bacchetta et~al.(2020)Bacchetta, Bertone, Bissolotti, Bozzi, Delcarro,
  Piacenza, and Radici]{Bacchetta:2019sam}
Alessandro Bacchetta, Valerio Bertone, Chiara Bissolotti, Giuseppe Bozzi,
  Filippo Delcarro, Fulvio Piacenza, and Marco Radici.
\newblock {Transverse-momentum-dependent parton distributions up to N$^{3}$LL
  from Drell-Yan data}.
\newblock \emph{JHEP}, 07:\penalty0 117, 2020.
\newblock \doi{10.1007/JHEP07(2020)117}.

\bibitem[Bacchetta et~al.(2022)Bacchetta, Bertone, Bissolotti, Bozzi, Cerutti,
  Piacenza, Radici, and Signori]{Bacchetta:2022awv}
Alessandro Bacchetta, Valerio Bertone, Chiara Bissolotti, Giuseppe Bozzi,
  Matteo Cerutti, Fulvio Piacenza, Marco Radici, and Andrea Signori.
\newblock {Unpolarized transverse momentum distributions from a global fit of
  Drell-Yan and semi-inclusive deep-inelastic scattering data}.
\newblock \emph{JHEP}, 10:\penalty0 127, 2022.
\newblock \doi{10.1007/JHEP10(2022)127}.

\bibitem[Moos et~al.(2024)Moos, Scimemi, Vladimirov, and Zurita]{Moos:2023yfa}
Valentin Moos, Ignazio Scimemi, Alexey Vladimirov, and Pia Zurita.
\newblock {Extraction of unpolarized transverse momentum distributions from the
  fit of Drell-Yan data at N$^{4}$LL}.
\newblock \emph{JHEP}, 05:\penalty0 036, 2024.
\newblock \doi{10.1007/JHEP05(2024)036}.

\bibitem[Gonzalez-Hernandez et~al.(2022)Gonzalez-Hernandez, Rogers, and
  Sato]{Gonzalez-Hernandez:2022ifv}
J.~O. Gonzalez-Hernandez, T.~C. Rogers, and N.~Sato.
\newblock {Combining nonperturbative transverse momentum dependence with TMD
  evolution}.
\newblock \emph{Phys. Rev. D}, 106\penalty0 (3):\penalty0 034002, 2022.
\newblock \doi{10.1103/PhysRevD.106.034002}.

\bibitem[Becher and Neubert(2011)]{Becher:2010tm}
Thomas Becher and Matthias Neubert.
\newblock {Drell-Yan Production at Small $q_T$, Transverse Parton Distributions
  and the Collinear Anomaly}.
\newblock \emph{Eur. Phys. J. C}, 71:\penalty0 1665, 2011.
\newblock \doi{10.1140/epjc/s10052-011-1665-7}.

\bibitem[Echevarria et~al.(2012)Echevarria, Idilbi, and
  Scimemi]{Echevarria:2011epo}
Miguel~G. Echevarria, Ahmad Idilbi, and Ignazio Scimemi.
\newblock {Factorization Theorem For Drell-Yan At Low $q_T$ And Transverse
  Momentum Distributions On-The-Light-Cone}.
\newblock \emph{JHEP}, 07:\penalty0 002, 2012.
\newblock \doi{10.1007/JHEP07(2012)002}.

\bibitem[Chiu et~al.(2012{\natexlab{a}})Chiu, Jain, Neill, and
  Rothstein]{Chiu:2012ir}
Jui-Yu Chiu, Ambar Jain, Duff Neill, and Ira~Z. Rothstein.
\newblock {A Formalism for the Systematic Treatment of Rapidity Logarithms in
  Quantum Field Theory}.
\newblock \emph{JHEP}, 05:\penalty0 084, 2012{\natexlab{a}}.
\newblock \doi{10.1007/JHEP05(2012)084}.

\bibitem[Collins(2008)]{Collins:2008ht}
John Collins.
\newblock {Rapidity divergences and valid definitions of parton densities}.
\newblock \emph{PoS}, LC2008:\penalty0 028, 2008.
\newblock \doi{10.22323/1.061.0028}.

\bibitem[Collins and Tkachov(1992)]{Collins:1992tv}
John~C. Collins and F.~V. Tkachov.
\newblock {Breakdown of dimensional regularization in the Sudakov problem}.
\newblock \emph{Phys. Lett. B}, 294:\penalty0 403--411, 1992.
\newblock \doi{10.1016/0370-2693(92)91541-G}.

\bibitem[Chiu et~al.(2012{\natexlab{b}})Chiu, Jain, Neill, and
  Rothstein]{Chiu:2011qc}
Jui-yu Chiu, Ambar Jain, Duff Neill, and Ira~Z. Rothstein.
\newblock {The Rapidity Renormalization Group}.
\newblock \emph{Phys. Rev. Lett.}, 108:\penalty0 151601, 2012{\natexlab{b}}.
\newblock \doi{10.1103/PhysRevLett.108.151601}.

\bibitem[Manohar and Stewart(2007)]{Manohar:2006nz}
Aneesh~V. Manohar and Iain~W. Stewart.
\newblock {The Zero-Bin and Mode Factorization in Quantum Field Theory}.
\newblock \emph{Phys. Rev. D}, 76:\penalty0 074002, 2007.
\newblock \doi{10.1103/PhysRevD.76.074002}.

\bibitem[Li et~al.(2020)Li, Neill, and Zhu]{Li:2016axz}
Ye~Li, Duff Neill, and Hua~Xing Zhu.
\newblock {An exponential regulator for rapidity divergences}.
\newblock \emph{Nucl. Phys. B}, 960:\penalty0 115193, 2020.
\newblock \doi{10.1016/j.nuclphysb.2020.115193}.

\bibitem[Ebert et~al.(2019)Ebert, Moult, Stewart, Tackmann, Vita, and
  Zhu]{Ebert:2018gsn}
Markus~A. Ebert, Ian Moult, Iain~W. Stewart, Frank~J. Tackmann, Gherardo Vita,
  and Hua~Xing Zhu.
\newblock {Subleading power rapidity divergences and power corrections for
  q$_{T}$}.
\newblock \emph{JHEP}, 04:\penalty0 123, 2019.
\newblock \doi{10.1007/JHEP04(2019)123}.

\bibitem[Boer and den Dunnen(2014)]{Boer:2014tka}
Dani\"el Boer and Wilco~J. den Dunnen.
\newblock {TMD evolution and the Higgs transverse momentum distribution}.
\newblock \emph{Nucl. Phys. B}, 886:\penalty0 421--435, 2014.
\newblock \doi{10.1016/j.nuclphysb.2014.07.006}.

\bibitem[Idilbi et~al.(2004)Idilbi, Ji, Ma, and Yuan]{Idilbi:2004vb}
Ahmad Idilbi, Xiang-dong Ji, Jian-Ping Ma, and Feng Yuan.
\newblock {Collins-Soper equation for the energy evolution of
  transverse-momentum and spin dependent parton distributions}.
\newblock \emph{Phys. Rev. D}, 70:\penalty0 074021, 2004.
\newblock \doi{10.1103/PhysRevD.70.074021}.

\bibitem[Lorc\'e et~al.(2016)Lorc\'e, Pasquini, and Schweitzer]{Lorce:2016ugb}
C.~Lorc\'e, B.~Pasquini, and P.~Schweitzer.
\newblock {Transverse pion structure beyond leading twist in constituent
  models}.
\newblock \emph{Eur. Phys. J. C}, 76\penalty0 (7):\penalty0 415, 2016.
\newblock \doi{10.1140/epjc/s10052-016-4257-8}.

\bibitem[Noguera and Scopetta(2015)]{Noguera:2015iia}
Santiago Noguera and Sergio Scopetta.
\newblock {Pion transverse momentum dependent parton distributions in the Nambu
  and Jona-Lasinio model}.
\newblock \emph{JHEP}, 11:\penalty0 102, 2015.
\newblock \doi{10.1007/JHEP11(2015)102}.

\bibitem[Kou et~al.(2023)Kou, Shi, Chen, and Jia]{Kou:2023ady}
Wei Kou, Chao Shi, Xurong Chen, and Wenbao Jia.
\newblock {Transverse momentum dependent parton distributions of pion at
  leading twist}.
\newblock \emph{Phys. Rev. D}, 108\penalty0 (3):\penalty0 036021, 2023.
\newblock \doi{10.1103/PhysRevD.108.036021}.

\bibitem[Aybat and Rogers(2011)]{Aybat:2011zv}
S.~Mert Aybat and Ted~C. Rogers.
\newblock {TMD Parton Distribution and Fragmentation Functions with QCD
  Evolution}.
\newblock \emph{Phys. Rev. D}, 83:\penalty0 114042, 2011.
\newblock \doi{10.1103/PhysRevD.83.114042}.

\bibitem[Collins(2003)]{Collins:2003fm}
John~C. Collins.
\newblock {What exactly is a parton density?}
\newblock \emph{Acta Phys. Polon. B}, 34:\penalty0 3103, 2003.

\bibitem[Boer(2016)]{Boer:2015ala}
Daniel Boer.
\newblock {Overview of TMD evolution}.
\newblock \emph{Int. J. Mod. Phys. Conf. Ser.}, 40:\penalty0 1660014, 2016.
\newblock \doi{10.1142/S2010194516600144}.

\bibitem[Collins(2015)]{Collins:2014loa}
John Collins.
\newblock {Different approaches to TMD Evolution with scale}.
\newblock \emph{EPJ Web Conf.}, 85:\penalty0 01002, 2015.
\newblock \doi{10.1051/epjconf/20158501002}.

\bibitem[Aslan et~al.(2024)Aslan, Boglione, Gonzalez-Hernandez, Rainaldi,
  Rogers, and Simonelli]{Aslan:2024nqg}
F.~Aslan, M.~Boglione, J.~O. Gonzalez-Hernandez, T.~Rainaldi, T.~C. Rogers, and
  A.~Simonelli.
\newblock {Phenomenology of TMD parton distributions in Drell-Yan and Z0 boson
  production in a hadron structure oriented approach}.
\newblock \emph{Phys. Rev. D}, 110\penalty0 (7):\penalty0 074016, 2024.
\newblock \doi{10.1103/PhysRevD.110.074016}.

\bibitem[Gonzalez-Hernandez et~al.(2023)Gonzalez-Hernandez, Rainaldi, and
  Rogers]{Gonzalez-Hernandez:2023iso}
J.~O. Gonzalez-Hernandez, T.~Rainaldi, and T.~C. Rogers.
\newblock {Resolution to the problem of consistent large transverse momentum in
  TMDs}.
\newblock \emph{Phys. Rev. D}, 107\penalty0 (9):\penalty0 094029, 2023.
\newblock \doi{10.1103/PhysRevD.107.094029}.

\bibitem[Rogers et~al.(2024)Rogers, Aslan, Boglione, Rainaldi, Simonelli, and
  Gonzalez-Hernandez]{Rogers:2024cci}
Ted Rogers, Fatma Aslan, Mariaelena Boglione, Tommaso Rainaldi, Andrea
  Simonelli, and J.~Osvaldo Gonzalez-Hernandez.
\newblock {TMD phenomenology and nonperturbative structures}.
\newblock \emph{PoS}, Transversity2024:\penalty0 030, 2024.
\newblock \doi{10.22323/1.477.0030}.

\bibitem[Bacchetta et~al.(2017{\natexlab{a}})Bacchetta, Cotogno, and
  Pasquini]{Bacchetta:2017vzh}
Alessandro Bacchetta, Sabrina Cotogno, and Barbara Pasquini.
\newblock {The transverse structure of the pion in momentum space inspired by
  the AdS/QCD correspondence}.
\newblock \emph{Phys. Lett. B}, 771:\penalty0 546--552, 2017{\natexlab{a}}.
\newblock \doi{10.1016/j.physletb.2017.05.072}.

\bibitem[Bacchetta et~al.(2017{\natexlab{b}})Bacchetta, Delcarro, Pisano,
  Radici, and Signori]{Bacchetta:2017gcc}
Alessandro Bacchetta, Filippo Delcarro, Cristian Pisano, Marco Radici, and
  Andrea Signori.
\newblock {Extraction of partonic transverse momentum distributions from
  semi-inclusive deep-inelastic scattering, Drell-Yan and Z-boson production}.
\newblock \emph{JHEP}, 06:\penalty0 081, 2017{\natexlab{b}}.
\newblock \doi{10.1007/JHEP06(2017)081}.
\newblock [Erratum: JHEP 06, 051 (2019)].

\bibitem[Collins and Rogers(2017)]{Collins:2017oxh}
John Collins and Ted~C. Rogers.
\newblock {Connecting Different TMD Factorization Formalisms in QCD}.
\newblock \emph{Phys. Rev. D}, 96\penalty0 (5):\penalty0 054011, 2017.
\newblock \doi{10.1103/PhysRevD.96.054011}.

\bibitem[Scimemi and Vladimirov(2018)]{Scimemi:2017etj}
Ignazio Scimemi and Alexey Vladimirov.
\newblock {Analysis of vector boson production within TMD factorization}.
\newblock \emph{Eur. Phys. J. C}, 78\penalty0 (2):\penalty0 89, 2018.
\newblock \doi{10.1140/epjc/s10052-018-5557-y}.

\bibitem[Collins et~al.(2016)Collins, Gamberg, Prokudin, Rogers, Sato, and
  Wang]{Collins:2016hqq}
J.~Collins, L.~Gamberg, A.~Prokudin, T.~C. Rogers, N.~Sato, and B.~Wang.
\newblock {Relating Transverse Momentum Dependent and Collinear Factorization
  Theorems in a Generalized Formalism}.
\newblock \emph{Phys. Rev. D}, 94\penalty0 (3):\penalty0 034014, 2016.
\newblock \doi{10.1103/PhysRevD.94.034014}.

\bibitem[Parisi and Petronzio(1979)]{Parisi:1979se}
G.~Parisi and R.~Petronzio.
\newblock {Small Transverse Momentum Distributions in Hard Processes}.
\newblock \emph{Nucl. Phys. B}, 154:\penalty0 427--440, 1979.
\newblock \doi{10.1016/0550-3213(79)90040-3}.

\bibitem[Altarelli et~al.(1984)Altarelli, Ellis, Greco, and
  Martinelli]{Altarelli:1984pt}
Guido Altarelli, R.~Keith Ellis, Mario Greco, and G.~Martinelli.
\newblock {Vector Boson Production at Colliders: A Theoretical Reappraisal}.
\newblock \emph{Nucl. Phys. B}, 246:\penalty0 12--44, 1984.
\newblock \doi{10.1016/0550-3213(84)90112-3}.

\bibitem[Bozzi et~al.(2011)Bozzi, Catani, Ferrera, de~Florian, and
  Grazzini]{Bozzi:2010xn}
Giuseppe Bozzi, Stefano Catani, Giancarlo Ferrera, Daniel de~Florian, and
  Massimiliano Grazzini.
\newblock {Production of Drell-Yan lepton pairs in hadron collisions:
  Transverse-momentum resummation at next-to-next-to-leading logarithmic
  accuracy}.
\newblock \emph{Phys. Lett. B}, 696:\penalty0 207--213, 2011.
\newblock \doi{10.1016/j.physletb.2010.12.024}.

\bibitem[Collins and Rogers(2015)]{Collins:2014jpa}
John Collins and Ted Rogers.
\newblock {Understanding the large-distance behavior of
  transverse-momentum-dependent parton densities and the Collins-Soper
  evolution kernel}.
\newblock \emph{Phys. Rev. D}, 91\penalty0 (7):\penalty0 074020, 2015.
\newblock \doi{10.1103/PhysRevD.91.074020}.

\bibitem[Vladimirov(2019)]{Vladimirov:2019bfa}
Alexey Vladimirov.
\newblock {Pion-induced Drell-Yan processes within TMD factorization}.
\newblock \emph{JHEP}, 10:\penalty0 090, 2019.
\newblock \doi{10.1007/JHEP10(2019)090}.

\bibitem[Badier et~al.(1982)]{NA3:1982ntq}
J.~Badier et~al.
\newblock {MEASUREMENT OF THE TRANSVERSE MOMENTUM OF DIMUONS PRODUCED BY
  HADRONIC INTERACTIONS AT 150-GEV/C, 200-GEV/C AND 280-GEV/C}.
\newblock \emph{Phys. Lett. B}, 117:\penalty0 372--376, 1982.
\newblock \doi{10.1016/0370-2693(82)90738-9}.

\bibitem[Anassontzis et~al.(1988)]{Anassontzis:1987hk}
E.~Anassontzis et~al.
\newblock {High mass dimuon production in $\bar{p} n$ and $\pi^- n$
  interactions at 125-GeV/c}.
\newblock \emph{Phys. Rev. D}, 38:\penalty0 1377, 1988.
\newblock \doi{10.1103/PhysRevD.38.1377}.

\bibitem[Brown and Creamer(1978)]{Brown:1978yj}
Lowell~S. Brown and Dennis~B. Creamer.
\newblock {VACUUM POLARIZATION ABOUT INSTANTONS}.
\newblock \emph{Phys. Rev. D}, 18:\penalty0 3695, 1978.
\newblock \doi{10.1103/PhysRevD.18.3695}.

\bibitem[Brown et~al.(1978)Brown, Carlitz, Creamer, and Lee]{PhysRevD.17.1583}
Lowell~S. Brown, Robert~D. Carlitz, Dennis~B. Creamer, and Choonkyu Lee.
\newblock Propagation functions in pseudoparticle fields.
\newblock \emph{Phys. Rev. D}, 17:\penalty0 1583--1597, Mar 1978.
\newblock \doi{10.1103/PhysRevD.17.1583}.
\newblock URL \url{https://link.aps.org/doi/10.1103/PhysRevD.17.1583}.

\bibitem[Zubkov et~al.(1999)Zubkov, Dubasov, and Kerbikov]{Zubkov:1997fn}
A.~G. Zubkov, O.~V. Dubasov, and B.~O. Kerbikov.
\newblock {Instanton - anti-instanton molecule with nonzero modes of quarks
  included}.
\newblock \emph{Int. J. Mod. Phys. A}, 14:\penalty0 241--252, 1999.
\newblock \doi{10.1142/S0217751X99000129}.

\bibitem[Creutz(1978)]{creutz1978invariant}
Michael Creutz.
\newblock On invariant integration over su (n).
\newblock \emph{Journal of Mathematical Physics}, 19\penalty0 (10):\penalty0
  2043--2046, 1978.

\bibitem[Miesch et~al.(2024)Miesch, Shuryak, and Zahed]{Miesch:2023hjt}
Nicholas Miesch, Edward Shuryak, and Ismail Zahed.
\newblock {Baryons and tetraquarks using instanton-induced interactions}.
\newblock \emph{Phys. Rev. D}, 109\penalty0 (1):\penalty0 014022, 2024.
\newblock \doi{10.1103/PhysRevD.109.014022}.

\bibitem[Diakonov et~al.()Diakonov, Jaenicke, and Polyakov]{Diakonov:1991}
D.~Diakonov, J.~Jaenicke, and M.~Polyakov.
\newblock Gluon exchange corrections to the nucleon mass in the chiral theory.
\newblock Preprint LNPI-1738 (1991), unpublished.

\bibitem[Jaffe et~al.(1997)Jaffe, Meyer, and Piller]{Jaffe1997}
R.~L. Jaffe, H.~Meyer, and G.~Piller.
\newblock \emph{Spin, twist and hadron structure in deep inelastic processes},
  pages 178--249.
\newblock Springer Berlin Heidelberg, Berlin, Heidelberg, 1997.
\newblock ISBN 978-3-540-46967-4.
\newblock \doi{10.1007/BFb0105860}.
\newblock URL \url{https://doi.org/10.1007/BFb0105860}.

\bibitem[Jaffe and Manohar(1989)]{Jaffe:1988up}
R.~L. Jaffe and Aneesh Manohar.
\newblock {Deep Inelastic Scattering from Arbitrary Spin Targets}.
\newblock \emph{Nucl. Phys. B}, 321:\penalty0 343, 1989.
\newblock \doi{10.1016/0550-3213(89)90347-7}.

\bibitem[Leader and Lorc{\'e}(2014)]{Leader:2013jra}
E.~Leader and C.~Lorc{\'e}.
\newblock {The angular momentum controversy: What{\textquoteright}s it all
  about and does it matter?}
\newblock \emph{Phys. Rept.}, 541\penalty0 (3):\penalty0 163--248, 2014.
\newblock \doi{10.1016/j.physrep.2014.02.010}.

\bibitem[Moch et~al.(2004)Moch, Vermaseren, and Vogt]{Moch:2004pa}
S.~Moch, J.~A.~M. Vermaseren, and A.~Vogt.
\newblock {The Three loop splitting functions in QCD: The Nonsinglet case}.
\newblock \emph{Nucl. Phys. B}, 688:\penalty0 101--134, 2004.
\newblock \doi{10.1016/j.nuclphysb.2004.03.030}.

\bibitem[Henn et~al.(2020)Henn, Korchemsky, and Mistlberger]{Henn:2019swt}
Johannes~M. Henn, Gregory~P. Korchemsky, and Bernhard Mistlberger.
\newblock {The full four-loop cusp anomalous dimension in $\mathcal{N}=4$ super
  Yang-Mills and QCD}.
\newblock \emph{JHEP}, 04:\penalty0 018, 2020.
\newblock \doi{10.1007/JHEP04(2020)018}.

\bibitem[von Manteuffel et~al.(2020)von Manteuffel, Panzer, and
  Schabinger]{vonManteuffel:2020vjv}
Andreas von Manteuffel, Erik Panzer, and Robert~M. Schabinger.
\newblock {Cusp and collinear anomalous dimensions in four-loop QCD from form
  factors}.
\newblock \emph{Phys. Rev. Lett.}, 124\penalty0 (16):\penalty0 162001, 2020.
\newblock \doi{10.1103/PhysRevLett.124.162001}.

\bibitem[Li and Zhu(2017)]{Li:2016ctv}
Ye~Li and Hua~Xing Zhu.
\newblock {Bootstrapping Rapidity Anomalous Dimensions for Transverse-Momentum
  Resummation}.
\newblock \emph{Phys. Rev. Lett.}, 118\penalty0 (2):\penalty0 022004, 2017.
\newblock \doi{10.1103/PhysRevLett.118.022004}.

\bibitem[Vladimirov(2017)]{Vladimirov:2016dll}
Alexey~A. Vladimirov.
\newblock {Correspondence between Soft and Rapidity Anomalous Dimensions}.
\newblock \emph{Phys. Rev. Lett.}, 118\penalty0 (6):\penalty0 062001, 2017.
\newblock \doi{10.1103/PhysRevLett.118.062001}.

\bibitem[Moult et~al.(2022)Moult, Zhu, and Zhu]{Moult:2022xzt}
Ian Moult, Hua~Xing Zhu, and Yu~Jiao Zhu.
\newblock {The four loop QCD rapidity anomalous dimension}.
\newblock \emph{JHEP}, 08:\penalty0 280, 2022.
\newblock \doi{10.1007/JHEP08(2022)280}.

\bibitem[Duhr et~al.(2022)Duhr, Mistlberger, and Vita]{Duhr:2022yyp}
Claude Duhr, Bernhard Mistlberger, and Gherardo Vita.
\newblock {Four-Loop Rapidity Anomalous Dimension and Event Shapes to Fourth
  Logarithmic Order}.
\newblock \emph{Phys. Rev. Lett.}, 129\penalty0 (16):\penalty0 162001, 2022.
\newblock \doi{10.1103/PhysRevLett.129.162001}.

\end{thebibliography}

\clearpage
\newpage

\appendix
\chapter{Conventions in Euclidean space}
\label{App:conv}
The covariantized Pauli matrices in Euclidean space are defined as

\begin{equation}
\sigma_\mu = (-i\vec{\sigma},\,1), \qquad \bar{\sigma}_\mu = (i\vec{\sigma},\,1)
\end{equation}
with 
\begin{equation}
\sigma_\mu\bar{\sigma}_\nu+\sigma_\nu\bar{\sigma}_\mu=2\delta_{\mu\nu}
\end{equation}

The corresponding gamma matrices  are defined as
\begin{align}
    \gamma^\mu&=\begin{pmatrix}
    0 & \sigma^\mu \\
    \bar{\sigma}^\mu & 0 
    \end{pmatrix} & \gamma^5&=\begin{pmatrix}
    -1 & 0 \\
    0 & 1 
    \end{pmatrix}
\end{align}

Similarly, in $SU(N_c)$ color space, $\vec{\tau}$ in instantons is an $N_c\times N_c$ valued matrix with the $2\times2$ Pauli matrices embedded in the upper left corner
\begin{align}
     \tau^+_\mu&=(\vec{\tau},-i)& \tau^-_\mu&=(\vec{\tau},i)
\end{align}

They satisfy the identities
\begin{align}
      \tau^-_\mu\tau^+_\nu-\tau^-_\nu\tau^+_\mu&=2i\bar{\eta}^a_{\mu\nu}\tau^a \\ \tau^+_\mu\tau^-_\nu-\tau^+_\nu\tau^-_\mu&=2i\eta^a_{\mu\nu}\tau^a
\end{align}
where the 't-Hooft symbol is defined in \cite{Liu:2024rdm,Vainshtein:1981wh} 
\begin{equation}
    \eta^{a}_{\mu\nu}=\begin{cases}
        \epsilon^{a}{}_{\mu\nu} ,\ & \mu\neq4,\ \nu\neq4\\
        \delta^{a}_{\mu} ,\ & \mu\neq4,\ \nu=4 \\
        -\delta^{a}_\nu ,\ & \mu=4,\ \nu\neq4
    \end{cases}
\end{equation}
and its  conjugate,
\begin{equation}
    \bar{\eta}^{a}_{\mu\nu}=\begin{cases}
        \epsilon^{a}{}_{\mu\nu} ,\ & \mu\neq4,\ \nu\neq4\\
        -\delta^{a}_\mu ,\ & \mu\neq4,\ \nu=4 \\
        \delta^{a}_\nu ,\ & \mu=4,\ \nu\neq4
    \end{cases}
\end{equation}

The 't Hooft symbol $\eta^a_{\mu\nu}$ satisfies the following identities.

\begin{align*}
&\eta^{a}_{\mu \nu} \eta^{a}_{\rho \lambda} = \delta_{\mu \rho} \delta_{\nu \lambda} - \delta_{\mu \lambda} \delta_{\nu \rho} + \epsilon_{\mu \nu \rho \lambda} \\
&\epsilon_{\mu \nu \rho \lambda} \eta^{a}_{\sigma \lambda} = \delta_{\sigma \mu} \eta^{a}_{\nu \rho} - \delta_{\sigma \nu} \eta^{a}_{\mu \rho} + \delta_{\sigma \rho} \eta^{a}_{\mu \nu} \\
&\eta^{a}_{\mu \nu} \eta^{b}_{\mu \lambda} = \delta_{ab} \delta_{\nu \lambda} + \epsilon_{abc} \eta^{c}_{\nu \lambda} \\
&\epsilon_{abc} \eta^{b}_{\mu \nu} \eta^{c}_{\rho \lambda} = \delta_{\mu \rho} \eta^{a}_{\nu \lambda} - \delta_{\nu \rho} \eta^{a}_{\mu \lambda} - \delta_{\mu \lambda} \eta^{a}_{\nu \rho} + \delta_{\nu \lambda} \eta^{a}_{\mu \rho}
\end{align*}

Similar identities for $\bar\eta^a_{\mu\nu}$ can be obtained by replacing 4D Euclidean Levi-Civita symbol $\epsilon_{\mu\nu\rho\lambda}\rightarrow-\epsilon_{\mu\nu\rho\lambda}$

\chapter{Instanton in singular gauge}
\label{App:singular}
The BPST instanton in singular gauge is given by
\begin{equation}
\label{FSTX}
A^a_{\mu}(x;\Omega_I)=O^{ab}(U_I)A^b_\mu(x-z_I)
\end{equation}
which is parameterized by $\Omega_I=(z_I,\rho,U_I)$ including the color orientation $U_I$, instanton 
location $z_I$ and size $\rho$, and is seen to satisfy both fixed-point and covariant gauge. 

The color rotation  
$$O^{ab}(U_I)=\frac{1}{2}\mathrm{Tr}(\tau^aU_I\tau^bU_I^\dagger)$$
is defined with $\tau^a$ as an  $N_c\times N_c$ matrix with $2\times2$ Pauli matrices embedded in the upper left corner. For the anti-instanton field, we  substitute  $\bar{\eta}^a_{\mu\nu}$ by $\eta^a_{\mu\nu}$ and flip the sign in front of Levi-Cevita tensor, $\epsilon_{\mu\nu\rho\lambda}\rightarrow-\epsilon_{\mu\nu\rho\lambda}$.

The profile function is defined as
\begin{equation}
\begin{aligned}
   A^a_\mu(x)=&-\frac{1}{g}\bar\eta^{a}_{\mu\nu}\partial_\nu\ln\Pi(x)=\frac{1}{g}\frac{2\bar\eta^{a}_{\mu\nu}x_\nu\rho^2}{x^2(x^2+\rho^2)}
\end{aligned}
\end{equation}
which in momentum space reads
\begin{equation}
A_\mu^a(q)=i\frac{4\pi^2\rho^2}{g}\frac{\bar\eta^a_{\mu\nu}q_\nu}{q^2}\mathcal{F}_g(\rho q)
\end{equation}
with $\mathcal{F}_g(\rho q)$ defined in Eq. \eqref{eq:g_form}. 
and the singular gauge potential defined by
\begin{equation}
\label{INSINGULAR}
\Pi(x)=1+\frac{\rho^2}{x^2}
\end{equation}

The corresponding field strength profile is defined as

\begin{equation}
    F^a_{\mu\nu}(x)=
        \frac{1}{g}\frac{8\rho^2}{(x^2+\rho^2)^2}\left[\bar{\eta}^a_{\mu\rho}\left(\frac{x_\rho x_\nu}{x^2}-\frac{1}{4}\delta_{\rho\nu}\right)-\bar{\eta}^a_{\nu\rho}\left(\frac{x_\rho x_\mu}{x^2}-\frac{1}{4}\delta_{\rho\mu}\right)\right]
\end{equation}

\chapter{QCD instanton zero modes}
\label{App:ZM}
The quark zero mode is the eigen solution with zero eigenvalues in
\begin{equation}
i\slashed{D}\phi_I(x)=0
\end{equation}
with the instanton background where in fundamental representation, the covariant derivative is defined as
\begin{equation}
    D_\mu=\partial_\mu-i A^a_{\mu}\frac{\tau^a}2
\end{equation}
and $A^a_{\mu}$ here is the instanton field solution.
The solution is left-handed
\begin{equation}
    \phi_I(x)=\varphi(x)\slashed{x}\frac{1-\gamma^5}{2}U\chi
\end{equation}
where $\chi=\begin{pmatrix}
    \chi_L \\
    \chi_R
\end{pmatrix}$ is a 4-spinor composed of a $SU(2)$-spinor $\chi_{L,R}$ on both chirality carrying $SU(2)$-color with color-spin locked by $\chi^{i\alpha}_{L,R}=\epsilon^{i\alpha}$, $i$ for spin and $\alpha$ for color. Thus, the 4-spinor $\chi$ presents the following identity:
\begin{eqnarray}
    &&\chi_L\chi^\dagger_L=\frac{1}{8}\tau_\mu^-\tau_\nu^+\gamma_\mu\gamma_\nu\frac{1-\gamma^5}{2} \nonumber\\[5pt]
    &&\chi_R\chi^\dagger_R=\frac{1}{8}\tau_\mu^+\tau_\nu^-\gamma_\mu\gamma_\nu\frac{1+\gamma^5}{2} \nonumber\\[5pt]
    &&\chi_L\chi^\dagger_R=-\frac{i}{2}\tau_\mu^-\gamma_\mu\frac{1+\gamma^5}{2} \nonumber\\[5pt]
    &&\chi_R\chi^\dagger_L=\frac{i}{2}\tau_\mu^+\gamma_\mu\frac{1-\gamma^5}{2}
\end{eqnarray}
with normalization $(\chi^{i\alpha}_{L,R})^\dagger\chi^{i\alpha}_{L,R}=2$.

The zero mode profile in singular gauge is given by
\begin{equation}
    \varphi(x)=\frac{\rho}{\pi|x|(x^2+\rho^2)^{3/2}}
\end{equation}
which in momentum space reads
\begin{equation}
    \phi_I(k)=-\frac{i\slashed{k}}{k}\varphi^{\prime}(k)\frac{1-\gamma^5}{2}U\chi
\end{equation}
with
\begin{equation}
\label{VARPHIP}
    \varphi^{\prime}(k)=\pi\rho^2\frac{d}{dz}\bigg(I_0K_0(z)-I_1K_1(z)\bigg)\bigg|_{z=\rho k/2}
\end{equation}

The instanton zero mode is normalized by
\begin{equation}
    \int d^4x \phi^\dagger_{I}(x)\phi_{I}(x)=\int \frac{d^4k}{(2\pi)^4} \phi^\dagger_{I}(k)\phi_{I}(k) =1
\end{equation}
including sum over the spin and color indices.
For the anti-instanton, the zero mode is right handed and can be obtained through the substitution $\gamma^5\leftrightarrow-\gamma^5$.

\chapter{Quark propagator in single instanton background}
\label{App:NZM}
Here we quickly review the modification to the quark propagators in a single instanton background.

The effects of quark masses on the non-zero mode quark propagator in an instanton or anti-instanton background are not known in closed form \cite{Liu:2021evw}, but for small masses, the quark propagator can be expanded around the chiral limit~\cite{Brown:1978yj}.

\begin{equation}
\begin{aligned}
\label{eq:SIA_prop_q}
    S_I(x,y)=&S_{\mathrm{ZM}}(x,y)+S_0(x-y)\\
    &+\left[S_{\mathrm{NZM}}(x,y)-S_0(x-y)\right]-im\Delta_I(x,y)+\mathcal{O}(m^2)
\end{aligned}
\end{equation}
%
The quark zero mode propagator is defined as

\begin{equation}
\begin{aligned}
\label{eq:ZM_prop}
S_{\rm ZM}(x,y)=\frac{\phi_I(x)\phi_I^{\dagger}(y)}{im}
\end{aligned}
\end{equation}
and the quark propagator with zero modes substracted is defined as

\begin{equation}
\begin{aligned}
i\slashed{D}S_{\rm NZM}(x,y)=&\delta^4(x-y)-\phi_I(x)\phi_I^{\dagger}(y)\\
=&\frac{1\pm\gamma^5}2\delta^4(x-y) +i\overrightarrow{\slashed{D}}\Delta(x,y) i\overleftarrow{\slashed{D}}\frac{1\mp\gamma^5}2
\end{aligned}
\end{equation}
where the covariant derivative is defined in fundamental representation.

The subtraction of the quark zero mode can be expressed by the isospin-$1/2$ scalar propagator in the instanton background $\Delta(x,y)$. The massless scalar propagator in the single instanton background field is defined as \cite{PhysRevD.17.1583}

\begin{equation}
\begin{aligned}
    \Delta(x,y)=&\frac{1}{4\pi^2(x-y)^2}\left(1+\rho^2\frac{x_\mu y_\nu}{x^2y^2}U\tau_\mu^-\tau_\nu^+U^\dagger\right)\frac{1}{\Pi(x)^{1/2}\Pi(y)^{1/2}}\\
    =&\frac{1}{4\pi^2(x-y)^2}\left(1+\rho^2\frac{x\cdot y}{x^2y^2}+\rho^2\frac{i\bar{\eta}^b_{\mu\nu}x_\mu y_\nu}{x^2y^2}\tau^aO^{ab}(U)\right)\frac{1}{\Pi(x)^{1/2}\Pi(y)^{1/2}}
\end{aligned}
\end{equation}
where the singular gauge potential $\Pi(x)$ is defined in \eqref{INSINGULAR}.

The location of the instanton $z_I$ is set to be zero for simplicity and can be recovered by translational symmetry $x\rightarrow x-z_I$ and $y\rightarrow y-z_I$. Now the non-zero mode propagator for quarks in the chiral-split form reads \cite{PhysRevD.17.1583}
\begin{equation}
\begin{aligned}
\label{eq:NZM}
S_{\rm NZM}(x,y)&=i\overrightarrow{\slashed{D}}_x\Delta(x,y)\frac{1+\gamma^5}{2}+\Delta(x,y)i\overleftarrow{\slashed{D}}_{y}\frac{1-\gamma^5}{2}\\
&=S_{nz}(x,y)\frac{1+\gamma^5}{2}+\bar{S}_{nz}(x,y)\frac{1-\gamma^5}{2}
\end{aligned}
\end{equation}
where the overhead arrows are defined as $$\Delta(x,y)\overleftarrow{D}_\mu=-\frac{\partial}{\partial y_\mu}\Delta(x,y)-i\Delta(x,y)A_\mu(y)$$ and
$$\overrightarrow{D}_\mu\Delta(x,y)=\frac{\partial}{\partial x_\mu}\Delta(x,y)-iA_\mu(x)\Delta(x,y)$$

After a few steps of algebraic calculation, $S_{nz}$ and $\bar{S}_{nz}$ can be recast in the form~\cite{Schafer:1996wv,Zubkov:1997fn,Liu:2021evw}

\begin{equation}
\begin{aligned}
    S_{nz}(x,y)=&\frac{-i(\slashed{x}-\slashed{y})}{2\pi^2(x-y)^4}\left(1+\rho^2\frac{x_\mu y_\nu}{x^2y^2}U\tau_\mu^-\tau_\nu^+U^\dagger\right)\frac{1}{\Pi(x)^{1/2}\Pi(y)^{1/2}}\\
    &-\frac{\rho^2\gamma_\mu}{4\pi^2}\frac{x_\rho(x-y)_\nu y_\lambda}{(x^2+\rho^2)x^2(x-y)^2y^2}U\tau_\rho^-\tau^+_\mu\tau_\nu^-\tau_\lambda^+U^\dagger\frac{1}{\Pi(x)^{1/2}\Pi(y)^{1/2}}
\end{aligned}
\end{equation}
and
\begin{equation}
\begin{aligned}
    \bar{S}_{nz}(x,y)=&\frac{-i(\slashed{x}-\slashed{y})}{2\pi^2(x-y)^4}\left(1+\rho^2\frac{x_\mu y_\nu}{x^2y^2}U\tau_\mu^-\tau_\nu^+U^\dagger\right)\frac{1}{\Pi(x)^{1/2}\Pi(y)^{1/2}}\\
    &-\frac{\rho^2\gamma_\mu}{4\pi^2}\frac{x_\rho(x-y)_\nu y_\lambda}{(y^2+\rho^2)x^2(x-y)^2y^2}U\tau_\rho^-\tau^+_\nu\tau_\mu^-\tau_\lambda^+U^\dagger\frac{1}{\Pi(x)^{1/2}\Pi(y)^{1/2}}
\end{aligned}
\end{equation}

%
Note that the propagator in the anti-instanton background can be obtained via the substitutions  $\tau^-_\mu\leftrightarrow\tau^+_\mu$, and $\gamma^5\leftrightarrow-\gamma^5$. At short distances, as well as far away from the instanton, the propagator reduces to the free one. At intermediate distances, the propagator is modified due to gluon exchanges with the instanton field \cite{Schafer:1995pz,Kock:2020frx}
\begin{equation}
\begin{aligned}
    S_{\rm NZM}(x,y)|_{x\rightarrow y}&\simeq\frac{-i(\slashed{x}-\slashed{y})}{2\pi^2(x-y)^4}-\frac{i}{16\pi^2}\frac{(x-y)_\mu\gamma_\nu}{(x-y)^2}\gamma^5F_{\mu\nu}(x)
\end{aligned}
\end{equation}

This result is consistent with the OPE of the quark propagator in a general background field.

%
The on-shell reduction of the Euclidean and massless quark propagator in the instanton background is intricate. In principle, it can be achieved through LSZ reduction in the zero momentum limit. Here we show for massless quarks, the LSZ reduction in the zero momentum ($k^2\ll Q^2$) limit reads

\begin{equation}
    \int d^4ye^{-ik\cdot y}S_{nz}(x,y)\slashed{k}\psi_L\simeq\frac{e^{-ik\cdot x}}{(1+\rho^2/x^2)^{\frac12}}\left[1+(1-e^{ik\cdot x})\frac{\rho^2}{2x^2}\frac{x_\mu k_\nu}{k\cdot x}U\tau^-_\mu\tau^+_\nu U^\dagger\right]\psi_L
\end{equation}
\begin{equation}
    \int d^4ye^{ik\cdot y}\bar{\psi}_R\slashed{k}\bar{S}_{nz}(y,x)\simeq\frac{e^{ik\cdot x}}{(1+\rho^2/x^2)^{\frac12}}\left[1+(1-e^{-ik\cdot x})\frac{\rho^2}{2x^2}\frac{k_\mu x_\nu }{k\cdot x}U\tau^-_\mu\tau^+_\nu U^\dagger\right]\bar{\psi}_R
\end{equation}

In the asymptotic limit $x^2\gg \rho^2$, the reduction yields an on-shell free quark.

\chapter{Averaging over colors}
\label{App:average}

\section{Creutz formula}
One way to carry out the color averaging in \eqref{eq:tHooft_g} is by determinantal reduction~\cite{creutz1978invariant}
\begin{equation}
    \int dU\prod_{i=1}^{N_c}U_{a_ib_i}=\frac{1}{N_c!}\epsilon_{a_1\cdots a_{N_c}}\epsilon_{b_1\cdots b_{N_c}}
\end{equation}
and

\begin{equation}
\begin{aligned}
&U^\dagger_{ba}=\frac{1}{(N_c-1)!}\epsilon_{aa_1\cdots a_{N_c-1}}\epsilon_{bb_1\cdots b_{N_c-1}}U_{a_1b_1}\cdots U_{a_{N_c-1}b_{N_c-1}}    
\end{aligned}
\end{equation}
where $\epsilon_{a_1\cdots a_{N_c}}$ is the Levi-Civita tensor of rank-$N_c$ with $\epsilon_{12\cdots N_c}=1$. Now the color averagings of $(UU^\dagger)^p$ are

\begin{enumerate}
    \item $p=1$
      \begin{equation}
      \int dU U_{ab}U^{\dagger}_{cd}=\frac{1}{N_c}\delta_{ad}\delta_{cb}
      \end{equation}
    \item $p=2$
      \begin{equation}
      \begin{aligned}
      \int dU U_{a_1b_1}U^{\dagger}_{c_1d_1}U_{a_2b_2}&U^{\dagger}_{c_2d_2}=\frac{1}{N_c^2-1}\big(\delta_{a_1d_1}\delta_{a_2d_2}\delta_{c_1b_1}\delta_{c_2b_2}+\delta_{a_1d_2}\delta_{a_2d_1}\delta_{c_1b_2}\delta_{c_2b_1}\big)\\
      &-\frac{1}{N_c(N_c^2-1)}\big(\delta_{a_1d_1}\delta_{a_2d_2}\delta_{c_1b_2}\delta_{c_2b_1}+\delta_{a_1d_2}\delta_{a_2d_1}\delta_{c_1b_1}\delta_{c_2b_2}\big)
      \end{aligned}
      \end{equation}

    \item $p=3$
      \begin{equation}
      \begin{aligned}
      &\int dU U_{a_1b_1}U^{\dagger}_{c_1d_1}U_{a_2b_2}U^{\dagger}_{c_2d_2}U_{a_3b_3}U^{\dagger}_{c_3d_3}
      =\frac{N_c^2-2}{N_c(N^2_c-4)(N_c^2-1)}\\
      &\times(\delta_{a_1d_1}\delta_{a_2d_2}\delta_{a_3d_3}\delta_{c_1b_1}\delta_{c_2b_2}\delta_{c_3b_3}
      +\delta_{a_1d_2}\delta_{a_2d_1}\delta_{a_3d_3}\delta_{c_1b_2}\delta_{c_2b_1}\delta_{c_3b_3}\\
      &+\delta_{a_1d_3}\delta_{a_2d_2}\delta_{a_3d_1}\delta_{c_1b_3}\delta_{c_2b_2}\delta_{c_3b_1}
      +\delta_{a_1d_1}\delta_{a_3d_2}\delta_{a_2d_3}\delta_{c_1b_1}\delta_{c_3b_2}\delta_{c_2b_3}\\
      &+\delta_{a_1d_3}\delta_{a_3d_2}\delta_{a_2d_1}\delta_{c_1b_3}\delta_{c_3b_2}\delta_{c_2b_1}
      +\delta_{a_1d_2}\delta_{a_2d_3}\delta_{a_3d_1}\delta_{c_1b_2}\delta_{c_2b_3}\delta_{c_3b_1})\\
      &-\frac{1}{(N_c^2-4)(N_c^2-1)}\\
      &\times(\delta_{a_1d_1}\delta_{a_2d_2}\delta_{a_3d_3}\delta_{c_1b_2}\delta_{c_2b_1}\delta_{c_3b_3}
      +\delta_{a_1d_2}\delta_{a_2d_1}\delta_{a_3d_3}\delta_{c_1b_1}\delta_{c_2b_2}\delta_{c_3b_3}\\
      &+\delta_{a_1d_1}\delta_{a_2d_2}\delta_{a_3d_3}\delta_{c_1b_3}\delta_{c_2b_2}\delta_{c_3b_1}
      +\delta_{a_1d_3}\delta_{a_2d_2}\delta_{a_3d_1}\delta_{c_1b_1}\delta_{c_2b_2}\delta_{c_3b_3}\\
      &+\delta_{a_1d_1}\delta_{a_2d_2}\delta_{a_3d_3}\delta_{c_1b_1}\delta_{c_3b_2}\delta_{c_2b_3}
      +\delta_{a_1d_1}\delta_{a_3d_2}\delta_{a_2d_3}\delta_{c_1b_1}\delta_{c_2b_2}\delta_{c_3b_3}\\
      &+\delta_{a_1d_3}\delta_{a_3d_2}\delta_{a_2d_1}\delta_{c_1b_1}\delta_{c_3b_2}\delta_{c_2b_3}
      +\delta_{a_1d_3}\delta_{a_3d_2}\delta_{a_2d_1}\delta_{c_3b_1}\delta_{c_2b_2}\delta_{c_1b_3}\\
      &+\delta_{a_1d_3}\delta_{a_3d_2}\delta_{a_2d_1}\delta_{c_1b_2}\delta_{c_2b_1}\delta_{c_3b_3}
      +\delta_{a_1d_1}\delta_{a_3d_2}\delta_{a_2d_3}\delta_{c_1b_3}\delta_{c_3b_2}\delta_{c_2b_1}\\
      &+\delta_{a_1d_1}\delta_{a_3d_2}\delta_{a_2d_3}\delta_{c_1b_3}\delta_{c_3b_2}\delta_{c_2b_1}
      +\delta_{a_1d_1}\delta_{a_3d_2}\delta_{a_2d_3}\delta_{c_1b_3}\delta_{c_3b_2}\delta_{c_2b_1}\\
      &+\delta_{a_1d_2}\delta_{a_2d_3}\delta_{a_3d_1}\delta_{c_1b_1}\delta_{c_3b_2}\delta_{c_2b_3}
      +\delta_{a_1d_2}\delta_{a_2d_3}\delta_{a_3d_1}\delta_{c_3b_1}\delta_{c_2b_2}\delta_{c_1b_3}\\
      &+\delta_{a_1d_2}\delta_{a_2d_3}\delta_{a_3d_1}\delta_{c_1b_2}\delta_{c_2b_1}\delta_{c_3b_3}
      +\delta_{a_1d_1}\delta_{a_3d_2}\delta_{a_2d_3}\delta_{c_1b_2}\delta_{c_2b_3}\delta_{c_3b_1}\\
      &+\delta_{a_1d_1}\delta_{a_3d_2}\delta_{a_2d_3}\delta_{c_1b_2}\delta_{c_2b_3}\delta_{c_3b_1}
      +\delta_{a_1d_1}\delta_{a_3d_2}\delta_{a_2d_3}\delta_{c_1b_2}\delta_{c_2b_3}\delta_{c_3b_1})\\
      &+\frac{2}{N_c(N_c^2-4)(N_c^2-1)}\\
      &\times(\delta_{a_1d_2}\delta_{a_2d_3}\delta_{a_3d_1}\delta_{c_1b_1}\delta_{c_2b_2}\delta_{c_3b_3}
      +\delta_{a_1d_1}\delta_{a_2d_2}\delta_{a_3d_3}\delta_{c_1b_2}\delta_{c_2b_3}\delta_{c_3b_1}\\
      &+\delta_{a_1d_3}\delta_{a_3d_2}\delta_{a_2d_1}\delta_{c_1b_1}\delta_{c_2b_2}\delta_{c_3b_3}
      +\delta_{a_1d_1}\delta_{a_2d_2}\delta_{a_3d_3}\delta_{c_1b_3}\delta_{c_3b_2}\delta_{c_2b_1}\\
      &+\delta_{a_1d_2}\delta_{a_2d_3}\delta_{a_3d_1}\delta_{c_1b_3}\delta_{c_3b_2}\delta_{c_2b_1}
      +\delta_{a_1d_3}\delta_{a_3d_2}\delta_{a_2d_1}\delta_{c_1b_2}\delta_{c_2b_3}\delta_{c_3b_1}\\
      &+\delta_{a_1d_1}\delta_{a_3d_2}\delta_{a_2d_3}\delta_{c_3b_2}\delta_{c_2b_2}\delta_{c_2b_3}
      +\delta_{a_1d_1}\delta_{a_3d_2}\delta_{a_2d_3}\delta_{c_1b_2}\delta_{c_2b_1}\delta_{c_3b_3}\\
      &+\delta_{a_1d_3}\delta_{a_2d_2}\delta_{a_3d_1}\delta_{c_1b_2}\delta_{c_2b_1}\delta_{c_3b_3}
      +\delta_{a_1d_3}\delta_{a_2d_2}\delta_{a_3d_1}\delta_{c_1b_1}\delta_{c_3b_2}\delta_{c_2b_3}\\
      &+\delta_{a_1d_2}\delta_{a_2d_1}\delta_{a_3d_3}\delta_{c_1b_1}\delta_{c_3b_2}\delta_{c_2b_3}
      +\delta_{a_1d_2}\delta_{a_2d_1}\delta_{a_3d_3}\delta_{c_3b_2}\delta_{c_2b_2}\delta_{c_2b_3})
      \end{aligned}
      \end{equation}
  
\end{enumerate}

\section{CNZ formula}

However, For large values of $p$, this averaging method is tedious.  Since $N_c \otimes N_c = 1 \oplus (N_c^2-1)$, the group integral practically reduces to finding all projections of the product of adjoint representations onto the singlet for $SU(N_c)$. The result can be obtained by using the graphical color projection rules~\cite{Chernyshev:1995gj,Nowak:1988bh,Miesch:2023hjt}, 
with the following results

\begin{enumerate}
    \item $p=2$
      \begin{equation}
      \begin{aligned}
      \int dU U_{a_1b_1}U^{\dagger}_{c_1d_1}U_{a_2b_2}U^{\dagger}_{c_2d_2}=&\frac{1}{N_c^2}\delta_{a_1d_1}\delta_{a_2d_2}\delta_{c_1b_1}\delta_{c_2b_2}\\
      &+\frac{1}{4(N_c^2-1)}\lambda^\alpha_{a_1d_1}\lambda^\alpha_{a_2d_2}\lambda^\beta_{c_1b_1}\lambda^\beta_{c_2b_2}
      \end{aligned}
      \end{equation}

    \item $p=3$
      \begin{equation}
      \begin{aligned}
      &\int dU U_{a_1b_1}U^{\dagger}_{c_1d_1}U_{a_2b_2}U^{\dagger}_{c_2d_2}U_{a_3b_3}U^{\dagger}_{c_3d_3}=\frac{1}{N_c^3}\delta_{a_1d_1}\delta_{a_2d_2}\delta_{a_3d_3}\delta_{c_1b_1}\delta_{c_2b_2}\delta_{c_3b_3}\\
      &+\frac{1}{4N_c(N_c^2-1)}\lambda^\alpha_{a_1d_1}\lambda^\alpha_{a_2d_2}\delta_{a_3d_3}\lambda^\beta_{c_1b_2}\lambda^\beta_{c_2b_1}\delta_{c_3b_3}\\
      &+\frac{1}{4N_c(N_c^2-1)}\delta_{a_1d_1}\lambda^\alpha_{a_2d_2}\lambda^\alpha_{a_3d_3}\delta_{c_1b_1}\lambda^\beta_{c_2b_2}\lambda^\beta_{c_3b_3}\\
      &+\frac{1}{4N_c(N_c^2-1)}\lambda^\alpha_{a_1d_1}\delta_{a_2d_2}\lambda^\alpha_{a_3d_3}\lambda^\beta_{c_1b_1}\delta_{c_2b_2}\lambda^\beta_{c_3b_3}\\
       &+\frac{1}{4(N_c^2-1)}\left(\frac{N_c}{2(N_c^2-4)}d^{\alpha\beta\gamma}d^{\alpha'\beta'\gamma'}\lambda^\alpha_{a_1d_1}\lambda^\beta_{a_2d_2}\lambda^\gamma_{a_3d_3}\lambda^{\alpha'}_{c_1b_2}\lambda^{\beta'}_{c_2b_1}\lambda^{\gamma'}_{c_3b_3}\right)\\
       &+\frac{1}{4(N_c^2-1)}\left(\frac{1}{2N_c}f^{\alpha\beta\gamma}f^{\alpha'\beta'\gamma'}\lambda^\alpha_{a_1d_1}\lambda^\beta_{a_2d_2}\lambda^\gamma_{a_3d_3}\lambda^{\alpha'}_{c_1b_2}\lambda^{\beta'}_{c_2b_1}\lambda^{\gamma'}_{c_3b_3}\right)
      \end{aligned}
      \end{equation}
    \item $p=4$
      \begin{equation}
    \resizebox{0.95\textwidth}{!}{$
      \begin{aligned}
      &\int dU U_{a_1b_1}U^{\dagger}_{c_1d_1}U_{a_2b_2}U^{\dagger}_{c_2d_2}U_{a_3b_3}U^{\dagger}_{c_3d_3}U_{a_4b_4}U^{\dagger}_{c_4d_4}=\\
      &\frac{1}{N_c^4}\delta_{a_1d_1}\delta_{a_2d_2}\delta_{a_3d_3}\delta_{a_4d_4}\delta_{c_1b_1}\delta_{c_2b_2}\delta_{c_3b_3}\delta_{c_4b_4}\\
      &+\frac{1}{N_c}\delta_{a_4d_4}\delta_{c_4b_4}\left(\int dU U_{a_1b_1}U^{\dagger}_{c_1d_1}U_{a_2b_2}U^{\dagger}_{c_2d_2}U_{a_3b_3}U^{\dagger}_{c_3d_3}-\frac{1}{N_c^3}\delta_{a_1d_1}\delta_{a_2d_2}\delta_{a_3d_3}\delta_{c_1b_1}\delta_{c_2b_2}\delta_{c_3b_3}\right)\\
      &+\frac{1}{N_c}\delta_{a_3d_3}\delta_{c_3b_3}\left(\int dU U_{a_1b_1}U^{\dagger}_{c_1d_1}U_{a_2b_2}U^{\dagger}_{c_2d_2}U_{a_4b_4}U^{\dagger}_{c_4d_4}-\frac{1}{N_c^3}\delta_{a_1d_1}\delta_{a_2d_2}\delta_{a_4d_4}\delta_{c_1b_1}\delta_{c_2b_2}\delta_{c_4b_4}\right)\\
      &+\frac{1}{N_c}\delta_{a_2d_2}\delta_{c_2b_2}\left(\int dU U_{a_1b_1}U^{\dagger}_{c_1d_1}U_{a_3b_3}U^{\dagger}_{c_3d_3}U_{a_4b_4}U^{\dagger}_{c_4d_4}-\frac{1}{N_c^3}\delta_{a_1d_1}\delta_{a_3d_3}\delta_{a_4d_4}\delta_{c_1b_1}\delta_{c_3b_3}\delta_{c_4b_4}\right)\\
      &+\frac{1}{N_c}\delta_{a_1d_1}\delta_{c_1b_1}\left(\int dU U_{a_4b_4}U^{\dagger}_{c_4d_4}U_{a_2b_2}U^{\dagger}_{c_2d_2}U_{a_3b_3}U^{\dagger}_{c_3d_3}-\frac{1}{N_c^3}\delta_{a_4d_4}\delta_{a_2d_2}\delta_{a_3d_3}\delta_{c_4b_4}\delta_{c_2b_2}\delta_{c_3b_3}\right)\\
      &+\lambda^\alpha_{a_1d_1}\lambda^\beta_{a_2d_2}\lambda^\gamma_{a_3d_3}\lambda^\delta_{a_4d_4}\lambda^{\alpha'}_{c_1b_1}\lambda^{\beta'}_{c_2b_2}\lambda^{\gamma'}_{c_3b_3}\lambda^{\delta'}_{c_4b_4}\frac{1}{16(N_c^2-1)^2}\delta^{\alpha\beta}\delta^{\gamma\delta}\delta^{\alpha'\beta'}\delta^{\gamma'\delta'}\\
      &+\lambda^\alpha_{a_1d_1}\lambda^\beta_{a_2d_2}\lambda^\gamma_{a_3d_3}\lambda^\delta_{a_4d_4}\lambda^{\alpha'}_{c_1b_1}\lambda^{\beta'}_{c_2b_2}\lambda^{\gamma'}_{c_3b_3}\lambda^{\delta'}_{c_4b_4}\frac{1}{16(N_c^2-1)^2}\delta^{\alpha\gamma}\delta^{\beta\delta}\delta^{\alpha'\gamma'}\delta^{\beta'\delta'}\\
      &+\lambda^\alpha_{a_1d_1}\lambda^\beta_{a_2d_2}\lambda^\gamma_{a_3d_3}\lambda^\delta_{a_4d_4}\lambda^{\alpha'}_{c_1b_1}\lambda^{\beta'}_{c_2b_2}\lambda^{\gamma'}_{c_3b_3}\lambda^{\delta'}_{c_4b_4}\frac{1}{16(N_c^2-1)^2}\delta^{\alpha\delta}\delta^{\beta\gamma}\delta^{\alpha'\delta'}\delta^{\beta'\gamma'}\\
      &+\lambda^\alpha_{a_1d_1}\lambda^\beta_{a_2d_2}\lambda^\gamma_{a_3d_3}\lambda^\delta_{a_4d_4}\lambda^{\alpha'}_{c_1b_1}\lambda^{\beta'}_{c_2b_2}\lambda^{\gamma'}_{c_3b_3}\lambda^{\delta'}_{c_4b_4}\frac{1}{4(N_c^2-1)}\frac{N_c^2}{4(N_c^2-4)^2}d^{\alpha\beta\epsilon}d^{\gamma\delta\epsilon}d^{\alpha'\beta'\rho}d^{\gamma'\delta'\rho}\\
      &+\lambda^\alpha_{a_1d_1}\lambda^\beta_{a_2d_2}\lambda^\gamma_{a_3d_3}\lambda^\delta_{a_4d_4}\lambda^{\alpha'}_{c_1b_1}\lambda^{\beta'}_{c_2b_2}\lambda^{\gamma'}_{c_3b_3}\lambda^{\delta'}_{c_4b_4}\frac{1}{4(N_c^2-4)}d^{\alpha\beta\epsilon}f^{\gamma\delta\epsilon}d^{\alpha'\beta'\epsilon'}f^{\gamma'\delta'\epsilon'}\\
       &+\lambda^\alpha_{a_1d_1}\lambda^\beta_{a_2d_2}\lambda^\gamma_{a_3d_3}\lambda^\delta_{a_4d_4}\lambda^{\alpha'}_{c_1b_1}\lambda^{\beta'}_{c_2b_2}\lambda^{\gamma'}_{c_3b_3}\lambda^{\delta'}_{c_4b_4}\frac{1}{4(N_c^2-4)}f^{\alpha\beta\epsilon}d^{\gamma\delta\epsilon}f^{\alpha'\beta'\epsilon'}d^{\gamma'\delta'\epsilon'}\\
       &+\lambda^\alpha_{a_1d_1}\lambda^\beta_{a_2d_2}\lambda^\gamma_{a_3d_3}\lambda^\delta_{a_4d_4}\lambda^{\alpha'}_{c_1b_1}\lambda^{\beta'}_{c_2b_2}\lambda^{\gamma'}_{c_3b_3}\lambda^{\delta'}_{c_4b_4}\frac{1}{4N_c^2}f^{\alpha\beta\epsilon}f^{\gamma\delta\epsilon}f^{\alpha'\beta'\epsilon'}f^{\gamma'\delta'\epsilon}
      \end{aligned}
      $}
      \end{equation}
\end{enumerate}

\chapter{Pair configurations}
\label{App:pair}

\begin{figure}
    \centering
\subfloat[]{\includegraphics[height=2.5cm,width=.33\linewidth]{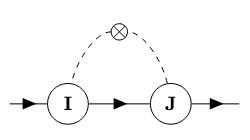}}
\hfill
\subfloat[]{\includegraphics[height=2.5cm,width=.33\linewidth]{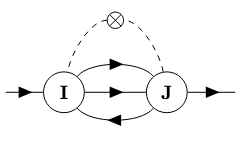}}
\hfill
\subfloat[]{\includegraphics[height=2.5cm,width=.33\linewidth]{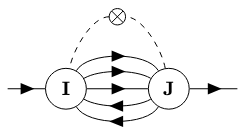}}
    \caption{Quark propagation between pair $IJ$ with presence of gluonic sources}
    \label{fig:LR}
\end{figure}

In this appendix, we will discuss the evaluation for the effective quark operator in \eqref{eq:dilute}, as illustrated in Fig.~\ref{fig:DILM_effOp} in ILM. Using $1/N_c$ counting, the leading contribution for a pair-induced operator comes from a one-body diagram illustrated in Fig.~\ref{fig:LR}. This effective operator for the quark passing through a pair of pseudoparticles $1$ and $2$ reads
\begin{equation}
\begin{aligned}
\label{eq:IIAA}
    \Theta^{(f)}_{12=II,\,AA}(z_1,z_2)=&\frac{(4\pi^2\rho^2)^2}{2N_c}
    \frac1{4}\mathrm{tr}_D\big(S(R)\big)\\
    &\times\bigg[ \mathrm{Tr}_c(u\mathds{1}_2)\mathrm{Tr}_c(\mathds{1}_2u^\dagger)\bar\psi(z_1)\frac{1\mp\gamma^5}2\psi(z_2)\\
    &\quad+\mathrm{Tr}_c(\mathds{1}_2u^\dagger)\mathrm{Tr}_c(u\mathds{1}_2)\bar\psi(z_2)\frac{1\mp\gamma^5}2\psi(z_1)\\
    &\quad-\frac i{4}\mathrm{Tr}_c(u\mathds{1}_2)\mathrm{Tr}_c(\tau^\mp_\mu\tau^\pm_\nu u^\dagger)\bar\psi(z_1)\sigma_{\mu\nu}\frac{1\mp\gamma^5}2\psi(z_2)\\
    &\quad-\frac i{4}\mathrm{Tr}_c(\mathds{1}_2u^\dagger)\mathrm{Tr}_c(u\tau^\mp_\mu\tau^\pm_\nu)\bar\psi(z_2)\sigma_{\mu\nu}\frac{1\mp\gamma^5}2\psi(z_1)\bigg]
\end{aligned}
\end{equation}
and
\begin{equation}
    \begin{aligned}
\label{eq:IA}
    \Theta^{(f)}_{12=IA}(z_1,z_2)=&\frac{(4\pi^2\rho^2)^2}{2N_c}
    \frac1{4}\mathrm{tr}_D\big(S(R)\gamma_\mu\big)\\
    &\times\bigg[\mathrm{Tr}_c(u\tau_\mu^+ )\mathrm{Tr}_c(\tau^-_\nu u^\dagger)\bar\psi(z_1)\gamma_\nu\frac{1+\gamma^5}{2}\psi(z_2)\\
    &\quad-\mathrm{Tr}_c(\tau_\mu^-u^\dagger)\mathrm{Tr}_c(u\tau^+_\nu)\bar\psi(z_2)\gamma_\nu\frac{1-\gamma^5}{2}\psi(z_1)\bigg]
\end{aligned}
\end{equation}
where $R=z_1-z_2$, $u=U^\dagger_1U_2$, and the modified Euclidean quark propagator \eqref{eq:Q_prop0} in RPA is defined as
\begin{equation}
    S(R)=\int \frac{d^4k}{(2\pi)^4}\frac{-\slashed{k}+iM}{k^2+M^2}\mathcal{F}(\rho k)e^{-ik\cdot R}=-i\frac{\slashed{R}}{R}\frac{dT(R)}{dR}+iMT(R)
\end{equation}
with
\begin{equation}
    T(R)=\int \frac{d^4k}{(2\pi)^4}\frac{1}{k^2+M^2}\mathcal{F}(\rho k)e^{-ik\cdot R}
\end{equation}

In general, the effective quark operators in \eqref{eq:IIAA} and \eqref{eq:IA} induced by long-distance pseudoparticle pairs are non-local, since the two pseudoparticles are separated by the
relative coordinate \(R=z_1-z_2\). To organize the discussion, we distinguish two relevant regimes. The first is the dilute instanton liquid, where the ensemble is probed at a coarse resolution scale and the resolved pseudoparticles are well separated, with typical separations $|z_1-z_2|\gtrsim \bar R$ where $\bar R$ is the resolution cut-off. In this regime, short-distance $IA$ correlations are not resolved as independent pairs and close $II$ and $AA$ configurations are further suppressed by the repulsive core. The dominant multi-instanton configuration is therefore associated with nearest-neighbor configurations within the dilute ensemble, but remains parametrically suppressed relative to the leading single-instanton contribution.

The second regime is the dense instanton liquid, where the resolution is increased enough to resolve short-distance correlated pairs. Since close same-charge pairs, $II$ and $AA$, are again strongly suppressed by the repulsive core, the relevant close pairs are dominated by instanton--anti-instanton $(IA)$ molecules. 

\section{Dilute regime and well-separated pairs}

In the dilute ensemble, as illustrated in Fig.~\ref{fig:DILM}, since the instanton fields are well-separated and highly localized, the only relevant contribution are those generated by the nearest neighboring pseudoparticles. Thus, at each order in the instanton-density expansion, the dominant contribution remains genuinely nonlocal, reflecting the finite separation between the relevant pseudoparticle centers. The analysis is therefore most naturally performed in momentum space instead of using $R$-expansion.

As presented in Fig.~\ref{fig:LR}, the ensuing pair coupling strength for one-body vertex are obtained by resumming all the quark looping diagrams with a finite separation $R$. That is, for $II$ and $AA$ pair 

\begin{equation}
\label{eq:gammaII_M}
\gamma^{(1)}_{II,AA}(R)=\sum_{n=0}^{N_f-1}\binom{N_f-1}{n}n!\left(\frac{2}{N_c}\left(\frac{4\pi^2\rho^2}{m^*}\right)^{2}\left(MT(R)\right)^{2}\right)^n\frac{(4\pi^2\rho^2)^2}{(m^*)^2}iMT(R)
\end{equation}
and for $IA$ pair 

\begin{equation}
\begin{aligned}
\label{eq:gammaIA_M}
\gamma^{(1)}_{IA}(R)=\sum_{n=0}^{N_f-1}\binom{N_f-1}{n}n!\left(\frac{2}{N_c}\left(\frac{4\pi^2\rho^2}{m^*}\right)^{2}\left(\frac{dT(R)}{dR}\right)^{2}\right)^n\frac{(4\pi^2\rho^2)^2}{(m^*)^2}
    \left(\frac{-1}{4}R\frac{dT(R)}{dR}\right)
\end{aligned}
\end{equation}
and the average over $R$ now is not separable in general.

\section{Dense regime and close pairs (molecules)}

In the dense instanton liquid ensemble, shown in Fig.~\ref{fig:DILM}, the increased resolution resolves short-distance $(IA)$ molecules, which yield stronger subleading correction compared to the ones in dilute regime. These molecular configurations provide the natural short-distance corrections that interpolate between the dilute instanton ensemble and the high-resolution quantum vacuum.


The hopping inside a close pair of pseudoparitcles with the opposite and the same duality are defined by $T_{IA}$ and $imD_{II,AA}$ respectively with their conjugation represents the reverse process.
\begin{equation}
\begin{aligned}
\label{eq:TIADII}
    &T_{IA}(u,R)
    =-4\pi^2\rho^2\mathrm{Tr}_c(\tau_\mu^+u)\frac{R_\mu}{R}\frac{dT(R)}{dR}\\[5pt]
    &imD_{II,AA}(u,R)
    =4\pi^2\rho^2\mathrm{Tr}_c(\mathds{1}_2u)imT(R)\\
\end{aligned}
\end{equation}
where the relative color orientation is $u=U^\dagger_IU_A$, the separation is defined by $R=z_I-z_A$, $\mathds{1}_2$ is the $N_c\times N_c$ diagonal matrix defined by $\mathrm{diag}(1,1,0,\cdots,0)$ and the hopping integral $T(R)$ is defined
\begin{equation}
\begin{aligned}
    T(R)=&\int\frac{d^4k}{(2\pi)^4}\frac{\mathcal{F}(\rho k)}{k^2}e^{-ik\cdot R}=\frac{1}{2\pi^2R}\int_0^\infty dk\mathcal{F}(\rho k)J_1(k R)
\end{aligned}
\end{equation}
with the quark zero mode non-local form factor $\mathcal{F}(k)$ defined in \eqref{ZMform} and $J_n$ Bessel functions of the first kind. 

The quark exchange between close pseudoparticles carrying the same topological charge ($II$ or $AA$) is strongly suppressed by the light quark mass. Consequently, fermionic exchanges do not provide additional attractions, and thus the interactions between $II$ and $AA$ remain dominated by the semiclassical Yang–Mills contribution, which is a repulsive hard core. As a result, correlated pairs of like-charge pseudoparticles are strongly suppressed and can be neglected \cite{Liu:2024rdm}. 

The $IA$ configurations do not contribute to chiral symmetry breaking due to their chiral preserving nature but their existence are crucial to the helicity preserving bound states such as vector mesons and chirality preserving process.


For each flavor passing through a close pair ($R<\bar{R}$) of instanton and anti-instanton, it generates 


\begin{equation}
\begin{aligned}
\label{eq:mol_vert}
    &\Theta^{(f)}_{IA}(z_I,z_A)=\frac 1{2N_c}\left((4\pi^2\rho^2)^2\frac{iR_\mu}{R}\frac{dT(R)}{dR}\right)\\
    &\times\bigg[\mathrm{Tr}_c(u\tau_\mu^+ )\mathrm{Tr}_c(\tau^-_\nu u^\dagger)\bar\psi(z_I)\gamma_\nu\frac{1+\gamma^5}{2}\psi(z_A)\\
    &\quad-\mathrm{Tr}_c(\tau_\mu^-u^\dagger)\mathrm{Tr}_c(u\tau^+_\nu)\bar\psi(z_A)\gamma_\nu\frac{1-\gamma^5}{2}\psi(z_I)\bigg]
\end{aligned}
\end{equation}
where the $1/N_c$ factor comes from the average over the overall color orientation. 

\begin{figure}
    \centering
\subfloat[]{\includegraphics[width=0.33\linewidth]{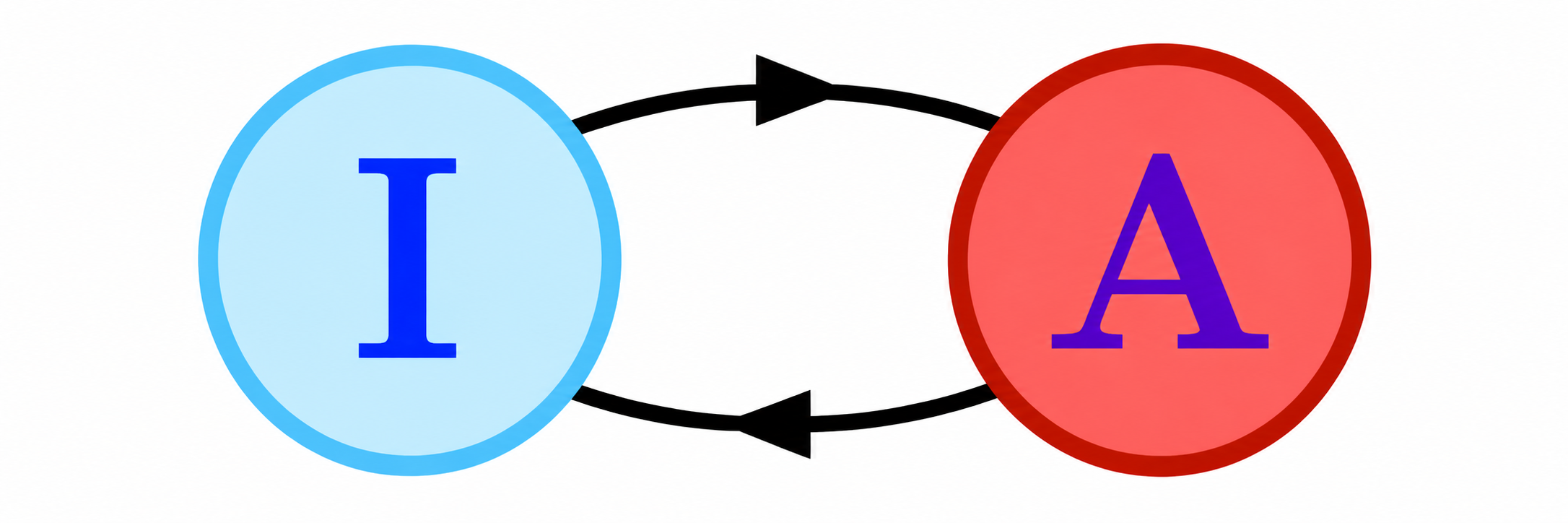}}
\hfill
\subfloat[]{\includegraphics[width=0.33\linewidth]{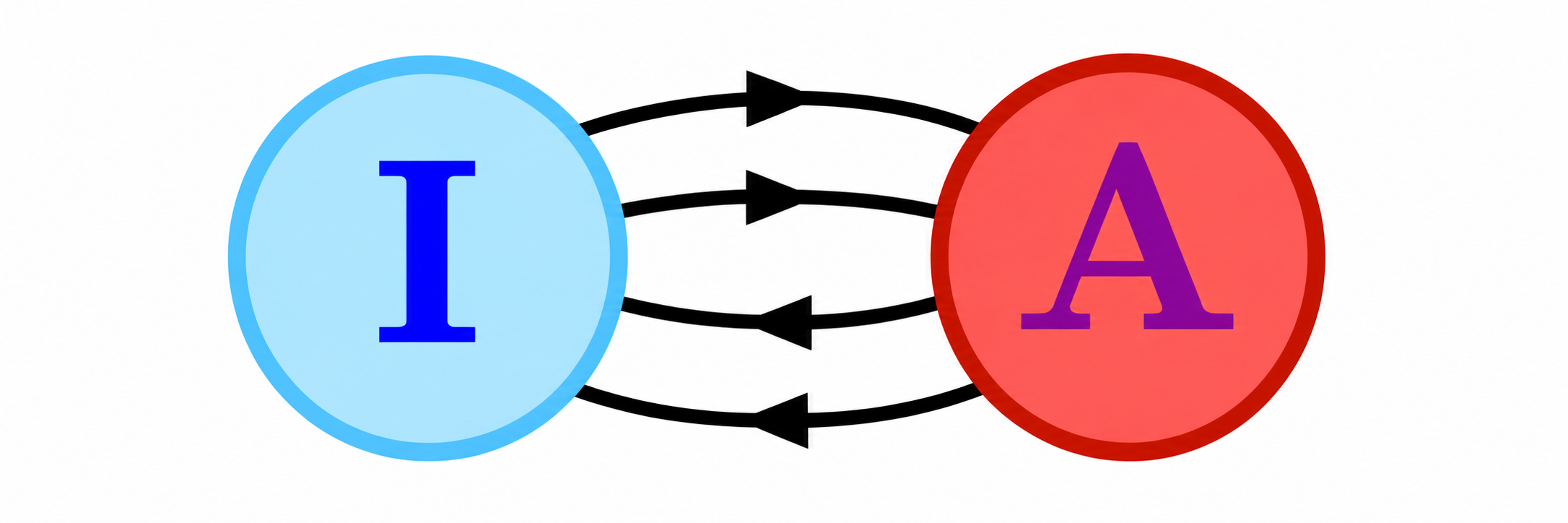}}
\hfill
\subfloat[]{\includegraphics[width=0.33\linewidth]{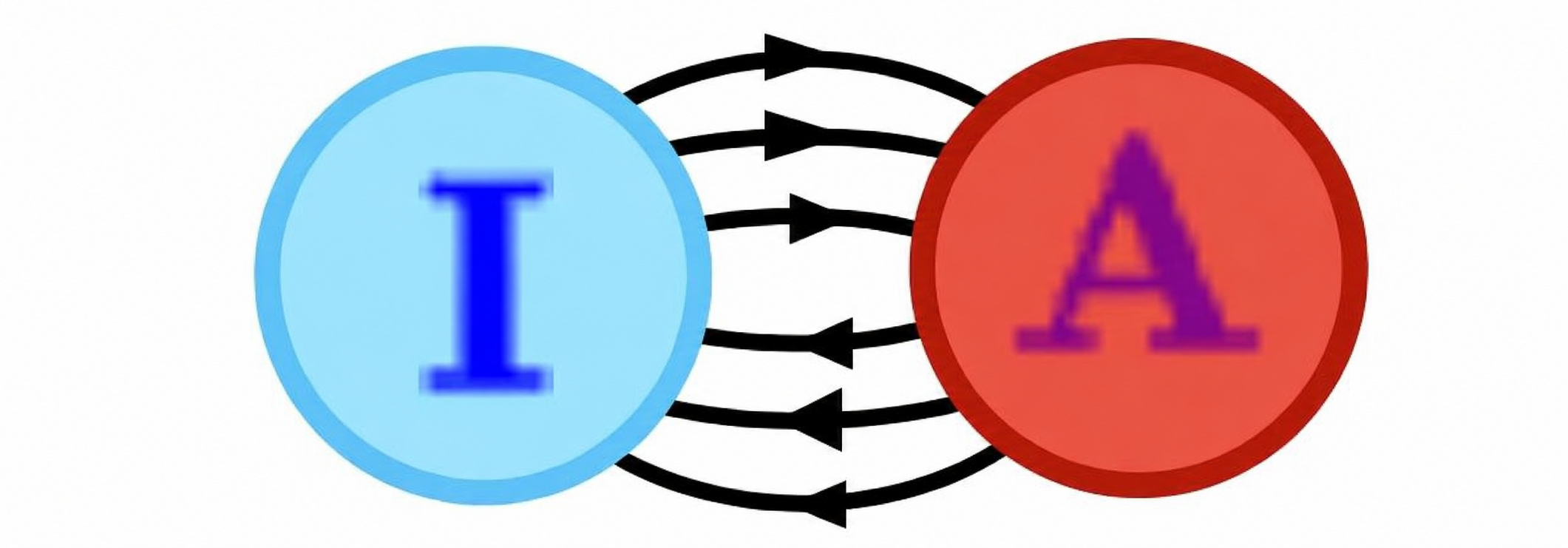}}
\hfill
\subfloat[]{\includegraphics[width=0.25\linewidth]{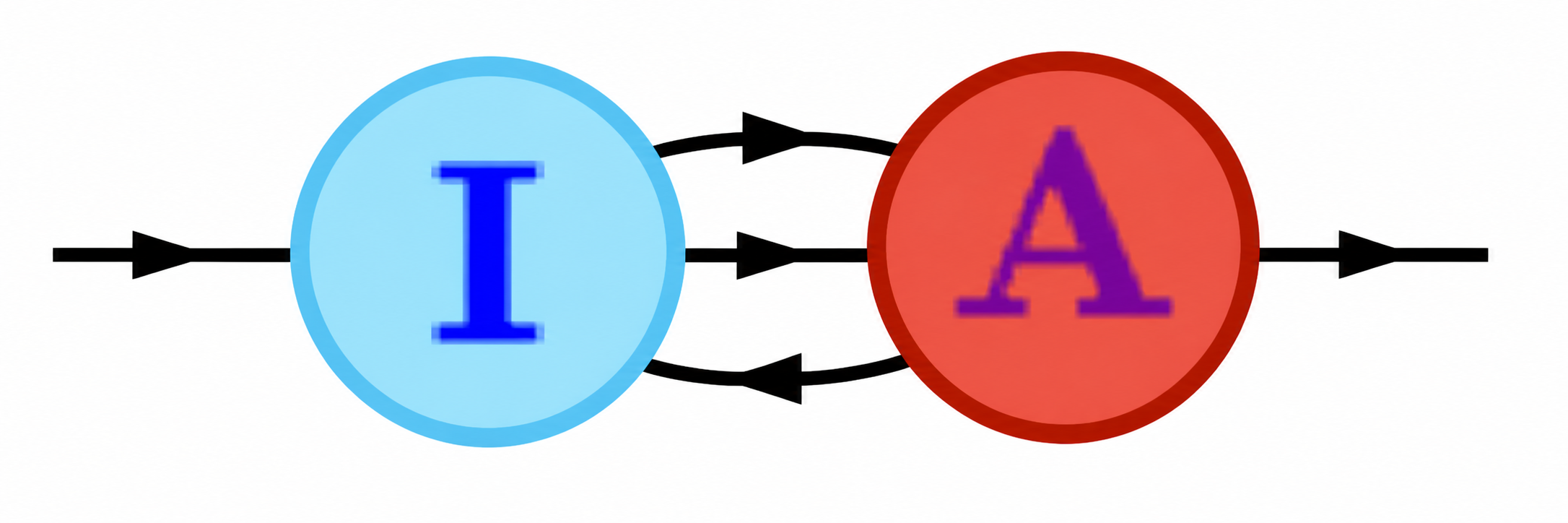}}
\hfill
\subfloat[]{\includegraphics[width=0.25\linewidth]{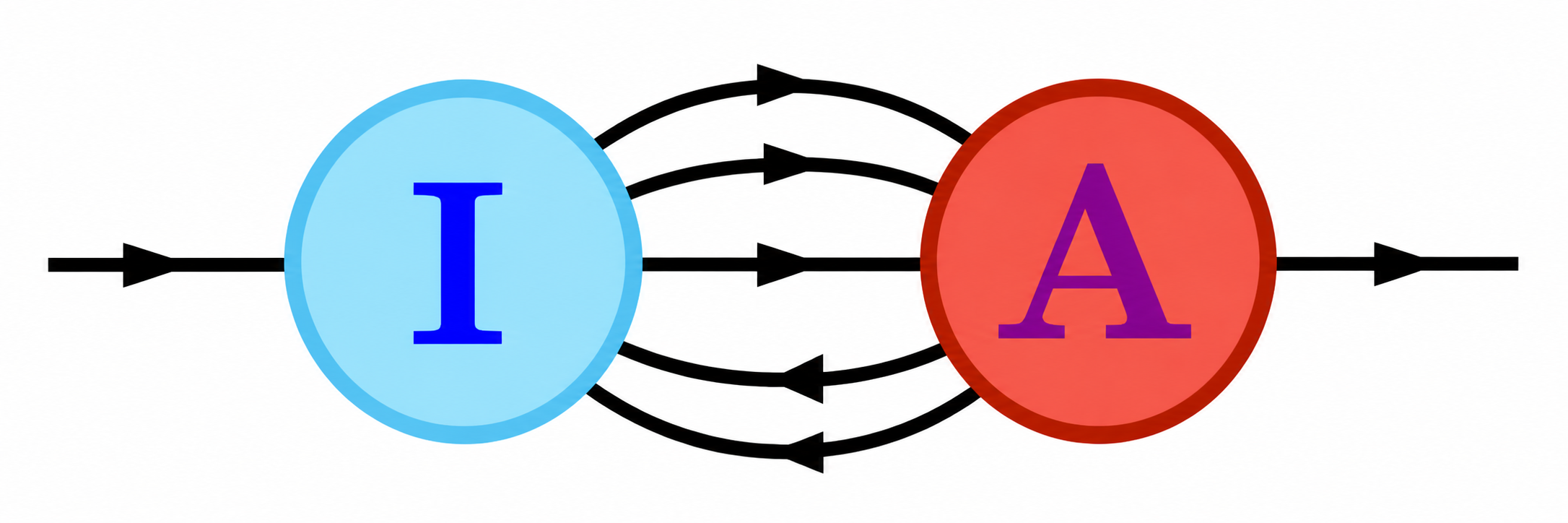}}
\hfill
\subfloat[]{\includegraphics[width=0.25\linewidth]{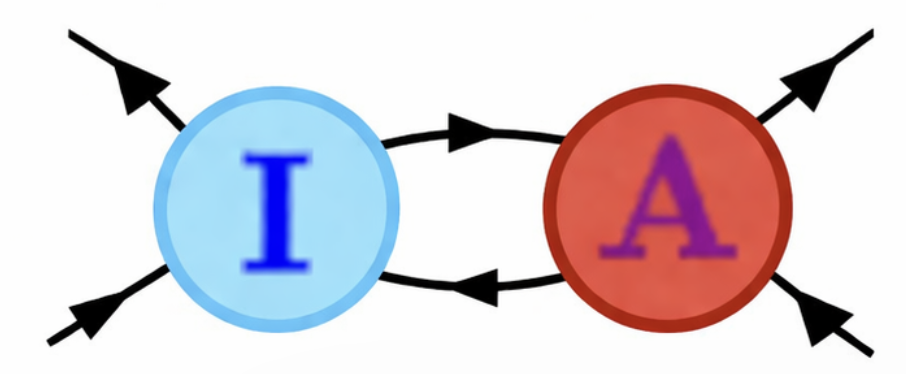}}
\hfill
\subfloat[]{\includegraphics[width=0.25\linewidth]{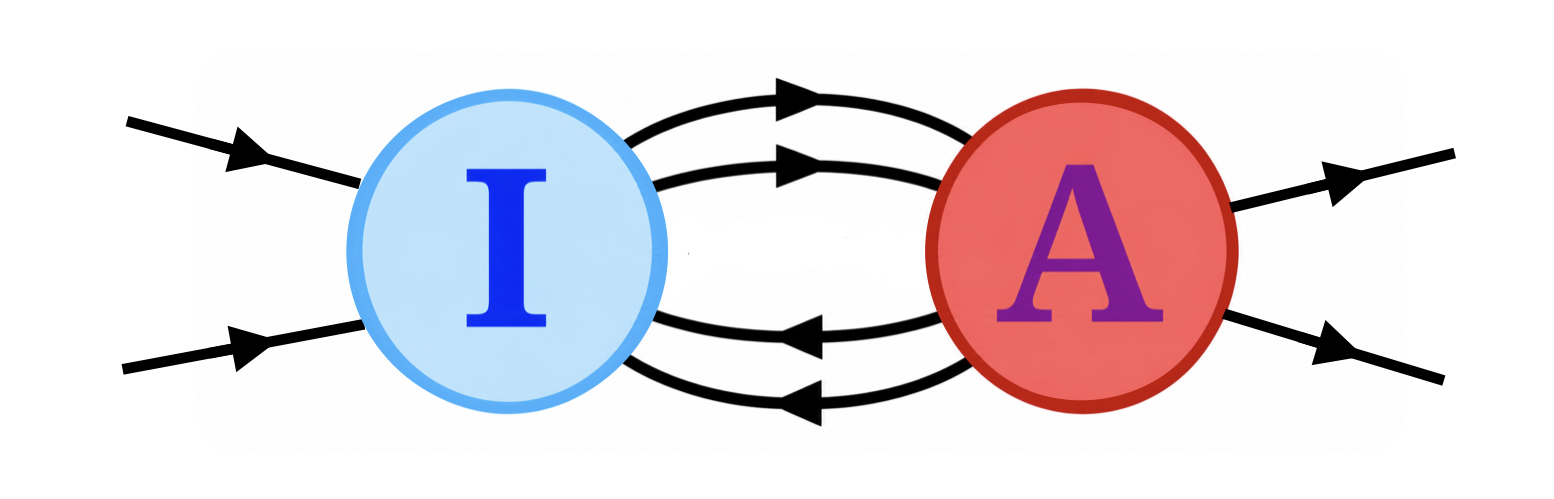}}
    \caption{Molecule diagrams in $N_f=3$ instanton vacuum.}
    \label{fig:onebody_IA}
\end{figure}

As presented in Fig.~\ref{fig:onebody_IA}, the ensuing molecular coupling strength for one-body vertex are obtained by resumming all the quark hopping diagrams inside the close pairs with one external quark leg (lower row (d), (e) in Fig.~\ref{fig:onebody_IA}), divided by the sum of diagrams (upper row (a-c) in Fig.~\ref{fig:onebody_IA}). That is, for $IA$ pair

\begin{equation}
\begin{aligned}
\label{eq:gammaIA}
\gamma^{(1)}_{IA}=&\left\langle\sum_{n=1}^{N_f-1}\binom{N_f-1}{n}\left(\frac{|T_{IA}|}{m^*}\right)^{2n}\frac{(4\pi^2\rho^2)^2}{(m^*)^2}\left(\frac{-1}4R\frac{dT(R)}{dR}\right)\right\rangle
\end{aligned}
\end{equation}
where the average is defined as $\langle\cdots\rangle=\int d^4R dU(\cdots)/\bar{R}^4|_{R<\bar{R}}$.

As shown in Table~\ref{tab:gammaIA}, the coupling  $\gamma_{IA}$ is dependent on the determinantal mass \(m^*\), which should be distinguished from
the constituent quark mass used in~\cite{Liu:2023yuj,Liu:2023fpj} (references
therein). The determinantal mass is associated with the fermion determinant in the zero-mode subspace (see~\cite{Shuryak:2021fsu} and \cite{Faccioli:2001ug}). 

\begin{table}[]
    \centering
    \begin{tabular}{|c|c|c|}
    \hline
       $m^*$ (MeV) & $\gamma^{(1)}_{IA}$ (fm$^4$)  \\
       \hline
       67.9 & $302.55$ \\
       77.1 & $152.99$ \\
       86.3 & $85.25$ \\
       95.5 & $51.41$ \\
       104.7 & $33.09$ \\
       113.9 & $22.50$ \\
       123.1 & $16.01$ \\
       132.3 & $11.84$ \\
    \hline
    \end{tabular}
    \caption{Molecular $\gamma_{IA}$ as a function of $m^*$ with resolution cut-off $\bar R\sim1$ fm.}
    \label{tab:gammaIA}
\end{table}

Similarly, the ensuing molecular coupling strength for two-body vertex can be obtained by resumming the diagrams with two external quark legs (lower row (f), (g) in Fig.~\ref{fig:onebody_IA}), divided by the sum of diagrams (upper row (a-c) in Fig.~\ref{fig:onebody_IA}).

\begin{equation}
\begin{aligned}
\label{eq:gammaIA2}
\gamma^{(2)}_{IA}=&\left\langle\sum_{n=1}^{N_f-1}\binom{N_f-1}{n}\left(\frac{|T_{IA}|}{m^*}\right)^{2n}\frac{(4\pi^2\rho^2)^2}{(m^*)^2}\left(\frac18\right)\right\rangle
\end{aligned}
\end{equation}
Compared to Eq.~\eqref{MOLX}, we have $G_{IA}=n_{IA}\gamma_{IA}^{(2)}$.

In general, the effective quark vertices \eqref{eq:mol_vert} induced by instanton--anti-instanton
molecules are non-local, since the two pseudoparticles are separated by the
relative coordinate \(R=z_I-z_A\). However, when the \(IA\) pair forms a compact
and strongly correlated molecule, its size is small compared with the relevant
hadronic scale, and the induced non-local operator can be expanded in powers of
\(R\).  Therefore, we can extract the leading contribution by the $R$-expansion (the local approximation) 
\begin{equation}
\label{eq:R-exp}
    \bar{\psi}(z_I)\psi(z_A)\simeq\bar{\psi}(z)\psi(z)-R_\mu\bar{\psi}(z)\overleftrightarrow{\partial_\mu}\psi(z)+\cdots
\end{equation}

This local approximation should be understood as a short-distance expansion, valid for tightly correlated \(IA\) pairs ($R\ll1/\Lambda_{QCD}$). It
does not describe the full long-distance instanton liquid, where the zero modes
delocalize over the ensemble and generate spontaneous chiral symmetry breaking. (see the discussion in Sec.~\ref{sec:DILM})

\chapter{Gluonic operators and quark-gluon operators in ILM }
\label{app:mole}

The gluonic operators for the multi-instanton configuration can be  constructed using  the gluonic field strength following from the sum ansatz \eqref{eq:sum}. Throughout, we will limit the discussion to instanton pairs. The consideration of higher clusters goes beyond the scope of this work. 

By the virtue of $R$-expansion (local approximation) \eqref{eq:R-exp} as originally suggested in~\cite{Liu:2024rdm}, we can approximate the induced non-local quark operators in ILM with local quark operators by localizing the instanton cluster profiles.

\section{Gluonic operators}

Here we present the detailed calculations using Eqs.~\eqref{eq:gluo_op} and \eqref{eq:op_eff} for scalar glueball $F^2$, pseudoscalar glueball $F\tilde{F}$, and tensor glueball $T^{\mu\nu}_g$ for example (see Ch.~\ref{ch:FF} for detailed analyses).

\subsection{Gluonic scalar $0^{++}$}

\begin{equation}
\begin{aligned}
    &g^2F^2(x)=\left(\frac{N_+}V\right)\int d^4z F^{(+)}_{GG}(x-z)+\left(\frac{N_-}V\right)\int d^4z F^{(-)}_{GG}(x-z)\\
    &-\left(\frac{N_+}V\right)\frac1{N_c}\left(\frac{4\pi^2\rho^2}{m^*}\right)\int d^4zF^{(+)}_{GG}(x-z)\bar{\psi}(z)\frac{1-\gamma^5}2\psi(z)\\
    &-\left(\frac{N_-}V\right)\frac{1}{N_c}\left(\frac{4\pi^2\rho^2}{m^*}\right)\int d^4zF^{(-)}_{GG}(x-z)\bar{\psi}(z)\frac{1+\gamma^5}2\psi(z)\\
    &+\left(\frac{N_+}V\right)\frac1{N_c}\left(\frac{4\pi^2\rho^2}{m^*}\right)\int d^4zF^{(+)}_{GG}(x-z)\left\langle\bar{\psi}(z)\frac{1-\gamma^5}2\psi(z)\right\rangle\left(\frac{1}{m^*}\right)^{N_f}\theta_+(x)\\
    &+\left(\frac{N_-}V\right)\frac{1}{N_c}\left(\frac{4\pi^2\rho^2}{m^*}\right)\int d^4zF^{(-)}_{GG}(x-z)\left\langle\bar{\psi}(z)\frac{1+\gamma^5}2\psi(z)\right\rangle\left(\frac{1}{m^*}\right)^{N_f}\theta_-(x)\\
    &-\left(\frac{N_+N_-}{V^2}\right)\frac{1}{2N_c(N_c^2-1)}\int d^4R\, \left(\frac{|T_{IA}|^2}{(m^*)^2}\right)^{N_f}\frac{(4\pi^2\rho^2)^2}{|T_{IA}|^2}\left[\frac{-1}4R\frac{dT(R)}{dR}\right]\\
    &\times\int d^4z F_{GG,\mu\nu}^{(+-)}(x-z)\bar{\psi}(z)\left(\gamma_{(\mu} i\overleftrightarrow{\partial}_{\nu)}-\frac14g_{\mu\nu}i\slashed{\partial}\right)\psi(z)
\end{aligned}
\end{equation}
where $\theta_\pm=\int dU \Theta_{I,A}$ and
\begin{align}
    F^{(+)}_{GG}(x)=&F^{(-)}_{GG}(x)=\frac{192\rho^4}{(x^2+\rho^2)^4}
\end{align}


\begin{align}
    F_{GG,\mu\nu}^{(+-)}(x)=&\frac{256\rho^6}{(x^2+\rho^2)^4x^2}\frac{x_\mu x_\nu}{x^2}
\end{align}

The terms in fourth and fifth lines proportional to $\theta_\pm$ come from the residual corrections of the inverse 't Hooft vertices $\Theta_{I}$, given by $$\frac1{(m^*)^{N_f}}\rightarrow\frac1{(m^*)^{N_f}}\left(1-\frac{\theta_\pm-\langle\theta_\pm\rangle}{\langle\theta_\pm\rangle}\right)$$ where $\langle\theta_\pm\rangle=(m^*)^{N_f}$. These correction is relevant only when computing connected averages of quark fields with effective operators which have a non-zero vacuum expectation value, such as the operator \(\bar\psi\psi\). See \cite{Diakonov:1995qy,Balla:1998rt,Balla:1997hf} for more details. As a result, the single instanton contributions cancel each other. The final result is simply

\begin{equation}
\begin{aligned}
    &g^2F^2(x)=\left(\frac{N_+}V\right)\int d^4z F^{(+)}_{GG}(x-z)+\left(\frac{N_-}V\right)\int d^4z F^{(-)}_{GG}(x-z)\\
    &-\left(\frac{N_+N_-}{V^2}\right)\frac{1}{2N_c(N_c^2-1)}\int d^4R\, \left(\frac{|T_{IA}|^2}{(m^*)^2}\right)^{N_f}\frac{(4\pi^2\rho^2)^2}{|T_{IA}|^2}\left[\frac{-1}4R\frac{dT(R)}{dR}\right]\\
    &\times\int d^4z F_{GG,\mu\nu}^{(+-)}(x-z)\bar{\psi}(z)\left(\gamma_{(\mu} i\overleftrightarrow{\partial}_{\nu)}-\frac14g_{\mu\nu}i\slashed{\partial}\right)\psi(z)
\end{aligned}
\end{equation}

\subsection{Gluonic pseudoscalar $0^{-+}$}

\begin{equation}
\begin{aligned}
    &g^2F\tilde{F}(x)=\left(\frac{N_+}V\right)\int d^4z F^{(+)}_{G\tilde{G}}(x-z)+\left(\frac{N_-}V\right)\int d^4z F^{(-)}_{G\tilde{G}}(x-z)\\
    &-\left(\frac{N_+}V\right)\frac1{N_c}\left(\frac{4\pi^2\rho^2}{m^*}\right)\int d^4zF^{(+)}_{G\tilde{G}}(x-z)\bar{\psi}(z)\frac{1-\gamma^5}2\psi(z)\\
    &-\left(\frac{N_-}V\right)\frac1{N_c}\left(\frac{4\pi^2\rho^2}{m^*}\right)\int d^4zF^{(-)}_{G\tilde{G}}(x-z)\bar{\psi}(z)\frac{1+\gamma^5}2\psi(z)\\
    &+\left(\frac{N_+}V\right)\frac1{N_c}\left(\frac{4\pi^2\rho^2}{m^*}\right)\int d^4zF^{(+)}_{G\tilde{G}}(x-z)\left\langle\bar{\psi}(z)\frac{1-\gamma^5}2\psi(z)\right\rangle\left(\frac{1}{m^*}\right)^{N_f}\theta_+(x)\\
    &+\left(\frac{N_-}V\right)\frac{1}{N_c}\left(\frac{4\pi^2\rho^2}{m^*}\right)\int d^4zF^{(-)}_{G\tilde{G}}(x-z)\left\langle\bar{\psi}(z)\frac{1+\gamma^5}2\psi(z)\right\rangle\left(\frac{1}{m^*}\right)^{N_f}\theta_-(x)\\
\end{aligned}
\end{equation}
where
\begin{align}
    F^{(+)}_{G\tilde{G}}(x)=&-F^{(-)}_{G\tilde{G}}(x)=\frac{192\rho^4}{(x^2+\rho^2)^4} \\
\end{align}

Similarly, the single instanton contribution cancels between the 'tHooft vertices and the residual corrections from inverse 'tHooft vertices. The molecule does not contribute in local approximation as the instanton contribution to the anti-instanton cancelled. Thus, the final result is simply governed by topological fluctuations.

\begin{equation}
\begin{aligned}
    &g^2F\tilde{F}(x)=\left(\frac{N_+}V\right)\int d^4z F^{(+)}_{G\tilde{G}}(x-z)+\left(\frac{N_-}V\right)\int d^4z F^{(-)}_{G\tilde{G}}(x-z)
\end{aligned}
\end{equation}

\subsection{Gluonic tensor $2^{++}$}

\begin{equation}
\begin{aligned}
    &g^2\left(\frac{1}{4}g_{\mu\nu}F^2(x)-F^{a}_{\mu\lambda}(x)F^{a}_{\nu\lambda}(x)\right)=\\
    &-\left(\frac{N_+N_-}{V^2}\right)\frac{1}{2N_c(N_c^2-1)}\int d^4R\, \left(\frac{|T_{IA}|^2}{(m^*)^2}\right)^{N_f}\frac{(4\pi^2\rho^2)^2}{|T_{IA}|^2}\left[\frac{-1}4R\frac{dT(R)}{dR}\right]\\
    &\times\int d^4z F_{T_g,\mu\nu\rho\lambda}^{(+-)}(x-z)\bar{\psi}(z)\left(\gamma_{(\rho} i\overleftrightarrow{\partial}_{\lambda)}-\frac14g_{\rho\lambda}i\slashed{\partial}\right)\psi(z)
\end{aligned}
\end{equation}

\section{Quark-gluon operator}

\begin{figure}
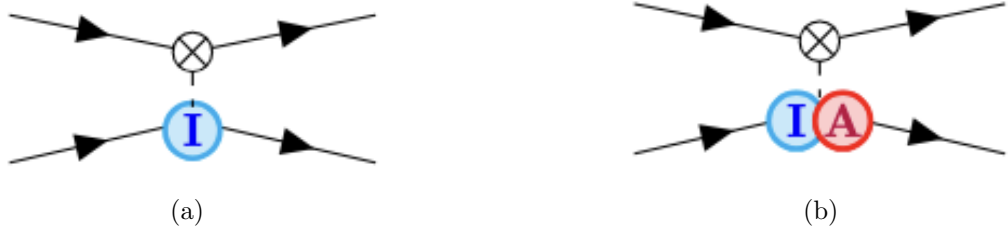

    \centering
    \subfloat[]{\includegraphics[width=.4\linewidth]{figures/I3.png}}
    \hfill
    \subfloat[]{\includegraphics[width=.4\linewidth]{figures/IA3.png}}
    \caption{Feynman diagrams for ILM expansion of quark-gluon operator denoted by the cross dot. The quark lines conncected to the cross dot at the top of each diagrams are located at $x$ while the quark lines at the bottom of each diagrams are surrounded around the center of the instanton $z$.}
    \label{fig:Oqg}
\end{figure}

For the gauge invariant  QCD operators with both quark and gluonic fields involved $\mathcal{O}[\psi,\bar\psi,A]$, the ensuing effective quark operators are usually related to a four point correlation (two quark current and two hadronic source). As presented in Fig.~\ref{fig:Oqg}, the original operator located at $x$ is presented by the cross dot. After $R$-expansion, the quark lines conncected to the cross dot at the top of each diagrams are a current source located at $x$ while the quark lines at the bottom of each diagrams are the current source around the center of the instanton $z$. As an illustration, we present the detailed calculations based on
Eqs.~\eqref{eq:gluo_op} and \eqref{eq:op_eff} for \(\bar\psi \slashed{A}\psi\) and
\(\bar\psi gF_{\mu\nu}\gamma_\sigma\psi\). A more complete analysis is
given in Ch.~\ref{ch:twist3}.

\subsubsection{Short distance propagation}
As we noted earlier, the pseudo-particle profile is highly localized. The separation between two quark sources is controlled by $|x|\lesssim\rho\ll \sqrt[4]{V_4/N}$, hence the approximation
\begin{equation}
\begin{aligned}
&\sqrt{\mathcal{F}(i\rho\partial)}S(x)\equiv\langle\psi(x)\sqrt{\mathcal{F}(i\rho\partial)}\bar\psi(0)\rangle\\
\simeq&\frac{i\slashed{x}}{2\pi^2x^4}K_D(x/\rho)+\frac{m}{4\pi^2x^2}K_m(x/\rho)+\mathcal{O}(m^2)
\end{aligned}
\end{equation}
where the distortions induced by the instanton zero mode profiles are given by
\begin{equation}
\begin{aligned}
    K_D(x)
    &= \frac{1}{2}\int_0^\infty dk\, k x^2 J_2(kx)\sqrt{\mathcal{F}(k)}
     = \frac{x^3}{(1+x^2)^{3/2}},\\
    K_m(x)
    &= \int_0^\infty dk\, x J_1(kx)\sqrt{\mathcal{F}(k)} .
\end{aligned}
\end{equation} 
and the quark propagator $S(x)$ in ILM is defined as 

\begin{equation}
S(x)
=
\frac{m^2}{4\pi^2}
\left[
\frac{i\slashed{x}}{x^2}
K_2\!\left(mx\right)
+
\frac{1}{x}
K_1\!\left(mx\right)
\right]\xrightarrow[]{|x|\ll 1/m}\frac{i\slashed{x}}{2\pi^2x^4}+\frac{m}{4\pi^2x^2}+\mathcal{O}(m^2)
\end{equation}
with $x=\sqrt{-x^2+i0}$ in Minkowski space.

\subsection{Quark EMT}
\begin{equation}
\begin{aligned}
\label{eq:CFOp2}
    &\bar{\psi}\gamma_{(\mu} i\overleftrightarrow{D}_{\nu)}\psi=\bar{\psi}\gamma_{(\mu} i\overleftrightarrow{\partial}_{\nu)}\psi\\
    &-\frac{1}{4(N_c^2-1)}\frac{N_+}V\left(\frac{4\pi^2\rho^2}{m^*}\right)\int d^4zF^{(+)}_{\bar{q}Aq,\alpha}(x-z)\bar\psi\gamma_{(\mu}\lambda^A\psi(x)\bar\psi \sigma_{\nu)\alpha}\frac{1-\gamma^5}2\lambda^A\psi(z)\\
    &-\frac{1}{4(N_c^2-1)}\frac{N_-}V\left(\frac{4\pi^2\rho^2}{m^*}\right)\int d^4zF^{(-)}_{\bar{q}Aq,\alpha}(x-z)\bar\psi\gamma_{(\mu}\lambda^A\psi(x)\bar\psi \sigma_{\nu)\alpha}\frac{1+\gamma^5}2\lambda^A\psi(z)\\
    &+\frac{1}{4(N_c^2-1)}\frac{N_+}V\left(\frac{4\pi^2\rho^2}{m^*}\right)\int d^4zF^{(+)}_{\bar{q}Aq,\alpha}(x-z)\left\langle\bar\psi\gamma_{(\mu}\lambda^A\psi(x)\bar\psi \sigma_{\nu)\alpha}\frac{1-\gamma^5}2\lambda^A\psi(z)\right\rangle\\
    &\qquad\times\left(\frac{1}{m^*}\right)^{N_f}\theta_+(x)\\
    &+\frac{1}{4(N_c^2-1)}\frac{N_-}V\left(\frac{4\pi^2\rho^2}{m^*}\right)\int d^4zF^{(-)}_{\bar{q}Aq,\alpha}(x-z)\left\langle\bar\psi\gamma_{(\mu}\lambda^A\psi(x)\bar\psi \sigma_{\nu)\alpha}\frac{1+\gamma^5}2\lambda^A\psi(z)\right\rangle\\
    &\qquad\times\left(\frac{1}{m^*}\right)^{N_f}\theta_-(x)\\
\end{aligned}
\end{equation}
where $\theta_\pm=\int dU \Theta_{I,A}$ and

\begin{equation}
    F^{(\pm)}_{\bar{q}Aq,\alpha}(x)=\frac{2\rho^2x_\alpha}{(x^2+\rho^2)x^2}
\end{equation}

The leading $1/N_c$ contribution comes from connecting one of the quarks from $x$ to $z$, yielding an effective one-body operator. Since instanton profile is highly localized, the quark propagation between $x$ and $z$ are mostly contributed from those high momentum modes ($k\gg M$). That is, we can further approximate the result using the short distance approximation


\begin{equation}
\begin{aligned}
    &F^{(\pm)}_{\bar{q}Aq,\alpha}(x)\bar\psi(x)\gamma_{\mu}\lambda^A\psi(x)\bar\psi(0) \sigma_{\nu\alpha}\frac{1\mp\gamma^5}2\lambda^A\psi(0)\\
    &\approx F^{(\pm)}_{\bar{q}Aq,\alpha}(x)\frac{-i}{2\pi^2x^4}K_D(x/\rho)\frac{N_c^2-1}{2N_c}\bar\psi(0)\left[\mathrm{tr}\left(\slashed{x}\gamma_{\mu}\sigma_{\nu\alpha}\right)-\mathrm{tr}\left(\sigma_{\nu\alpha}\gamma_{\mu}\slashed{x}\right)\right]\frac{1\mp\gamma^5}2\psi(0)\\
    &=\left[\frac{4(g_{\mu\nu}x_\alpha-g_{\alpha\mu}x_{\nu})}{\pi^2x^4}F^{(\pm)}_{\bar{q}Aq,\alpha}(x)K_D(x/\rho)\right]\frac{N_c^2-1}{2N_c}\bar\psi(0)\frac{1\mp\gamma^5}2\psi(0)
\end{aligned}
\end{equation}

Eventually by redefining the non-local form factors, we will have

\begin{equation}
\begin{aligned}
\label{eq:CFOp2}
    \bar{\psi}\gamma_{(\mu} i\overleftrightarrow{D}_{\nu)}\psi=&\bar{\psi}\gamma_{(\mu} i\overleftrightarrow{\partial}_{\nu)}\psi\\
    &-\left(\frac{N_+}V\right)\frac{1}{8N_c}\left(\frac{4\pi^2\rho^2}{m^*}\right)\int d^4zF^{(+)}_{T_q,\mu\nu}(x-z)\bar\psi(z)\frac{1-\gamma^5}2\psi(z)\\
    &-\left(\frac{N_-}V\right)\frac{1}{8N_c}\left(\frac{4\pi^2\rho^2}{m^*}\right)\int d^4zF^{(-)}_{T_q,\mu\nu}(x-z)\bar\psi(z)\frac{1+\gamma^5}2\psi(z)\\
    &+\left(\frac{N_+}V\right)\frac1{N_c}\left(\frac{4\pi^2\rho^2}{m^*}\right)\frac14g_{\mu\nu}\bar\psi(x)\frac{1-\gamma^5}2\psi(x)\\
    &+\left(\frac{N_-}V\right)\frac1{N_c}\left(\frac{4\pi^2\rho^2}{m^*}\right)\frac14g_{\mu\nu}\bar\psi(x)\frac{1+\gamma^5}2\psi(x)\\
\end{aligned}
\end{equation}
where

\begin{equation}
    F^{(\pm)}_{T_q,\mu\nu}(x)=\frac{8\rho^2(g_{\mu\nu}x^2-x_{\mu}x_{\nu})}{\pi^2(x^2+\rho^2)x^6}K_D(x/\rho)
\end{equation}

The details can be found in \cite{Balla:1997hf}. The last two terms come from the residual corrections of the inverse 't Hooft vertices $\Theta_{I}$, given by $$\frac1{(m^*)^{N_f}}\rightarrow\frac1{(m^*)^{N_f}}\left(1-\frac{\theta_\pm-\langle\theta_\pm\rangle}{\langle\theta_\pm\rangle}\right)$$ where $\langle\theta_\pm\rangle=(m^*)^{N_f}$. These correction is relevant only when computing connected averages of quark fields with effective operators which have a non-zero vacuum expectation value, such as the operator \(\bar\psi \slashed{A} \psi\). For effective operators with zero vacuum expectation value, which we shall mostly be concerned with, the residual vertices can be replaced simply by their vacuum expectation values, i.e. $1/(m^*)^{N_f}$.

\subsection{Color force}
\begin{equation}
\begin{aligned}
\label{eq:CFOpINS}
    &\bar \psi gF_{\mu\nu}\gamma_\sigma\psi(x)=\\
    &\frac{1}{2(N_c^2-1)}\frac{N_{+}}V\left(\frac{4\pi^2\rho^2}{m^*}\right)\int d^4zF^{(+)}_{\bar{q}Gq,\lambda[\mu}(x-z)\bar\psi(z) \sigma_{\nu]\lambda}\lambda^A\frac{1-\gamma^5}2\psi(z)\bar\psi(x)\gamma_{\sigma}\lambda^A\psi(x)\\
    &+\frac{1}{2(N_c^2-1)}\frac{N_{-}}V\left(\frac{4\pi^2\rho^2}{m^*}\right)\int d^4zF^{(-)}_{\bar{q}Gq,\lambda[\mu}(x-z)\bar\psi(z) \sigma_{\nu]\lambda}\lambda^A\frac{1+\gamma^5}2\psi(z)\bar\psi(x)\gamma_{\sigma}\lambda^A\psi(x)\\
    &+\frac{1}{4(N_c^2-1)^2}\frac{N_+N_{-}}{V^2} \int d^4R\,\left(\frac{|T_{IA}|^2}{(m^*)^2}\right)^{N_f}\left(\frac{4\pi^2\rho^2}{|T_{IA}|}\right)^{2}\left[\frac{-1}4R\frac{dT(R)}{dR}\right]\\
    &\times \int d^4zF_{\bar{q}Gq,\mu\nu\alpha\beta}^{(+-)}(x-z)\bar\psi(z) i\gamma_{(\alpha}\gamma^5\overleftrightarrow{\partial}_{\beta)}\lambda^A\psi(z)\bar\psi(x)\gamma_\sigma\lambda^A\psi(x)
\end{aligned}
\end{equation}
where the Gell-Mann matrices here are normalized to $\mathrm{tr}_c(\lambda^A\lambda^B)=2\delta^{AB}$ and
\begin{equation}
    F^{(\pm)}_{\bar{q}Gq,\lambda\nu}(x)=\frac{8\rho^2}{(x^2+\rho^2)^2}\left(\frac{x_\lambda x_\nu}{x^2}-\frac14 g_{\lambda\nu}\right)
\end{equation}

\begin{equation}
    F_{\bar{q}Gq,\mu\nu\alpha\beta}^{(+-)}(x)=t_{\mu\nu\rho\lambda\alpha\beta}\frac{\rho^4x_\rho x_\lambda}{(x^2+\rho^2)^2x^4}
\end{equation}
with the color factor defined as
\begin{equation}
t_{\mu\nu\rho\lambda\alpha\beta}=2i\left(\bar\eta^b_{\mu\rho}\eta^d_{\nu\lambda}-\bar\eta^b_{\nu\rho}\eta^d_{\mu\lambda}\right)
\mathrm{tr}(\tau^-_{\alpha}\tau^d\tau^+_{\beta}\tau^b)
\end{equation}
The last two indices are symmetric $t_{\mu\nu\rho\lambda(\alpha\beta)}=t_{\mu\nu\rho\lambda\alpha\beta}$, and the middle two indices are also symmetric but traceless.
The Lorentz operator takes the form of a product of two color–octet currents, analogous to a one–gluon exchange between quarks. The evaluation of its nucleon matrix element is therefore closely related to the calculation of gluon–exchange corrections to the nucleon mass \cite{Balla:1997hf,Diakonov:1991}

Similarly, in the short distance approximation and leading $1/N_c$, the effective Lorentz force operator can be further reduced to a sum of local fermionic operators induced by the zero modes in ILM. For single instanton contribution,

\begin{equation}
\resizebox{\textwidth}{!}{$
\begin{aligned}
    &\left[\frac{8\rho^2}{(x^2+\rho^2)^2}\left(\frac{x_\lambda x_\nu}{x^2}-\frac14 g_{\lambda\nu}\right)\right]\bar\psi(x)\gamma_{\sigma}\lambda^A\psi(x)\bar\psi(0) \sigma_{\mu\lambda}\frac{1\mp\gamma^5}2\lambda^A\psi(0)-(\mu\leftrightarrow\nu)\\
    =&\left[\frac{16\rho^2}{(x^2+\rho^2)^2}\frac{g_{\mu\sigma}x_\nu-g_{\nu\sigma}x_{\mu}\pm i\epsilon_{\mu\nu\sigma \alpha}x_\alpha}{2\pi^2x^4}K_D(x/\rho)\right]\frac{N_c^2-1}{2N_c}\bar\psi\frac12\left(e^{x\cdot\overrightarrow{\partial}}+e^{x\cdot\overleftarrow{\partial}}\right)\frac{1\mp\gamma^5}2\psi\\
    &\mp\left[\frac{16\rho^2}{(x^2+\rho^2)^2}\frac{g_{\mu\sigma}g_{\lambda\rho}-g_{\mu\rho} g_{\lambda\sigma}\mp i\epsilon_{\mu\lambda\sigma\rho}}{2\pi^2x^2}\left(\frac{x_\lambda x_{\nu}}{x^2}-\frac14 g_{\lambda\nu}\right)K_m(x/\rho)\right]\\
    &\quad\times\frac{N_c^2-1}{2N_c}im\bar\psi\gamma_\rho\gamma^5\frac12\left(e^{x\cdot\overrightarrow{\partial}}+e^{x\cdot\overleftarrow{\partial}}\right)\psi\\
    &-\left[\frac{16\rho^2}{(x^2+\rho^2)^2}\frac{g_{\mu\sigma}g_{\lambda\rho}-g_{\mu\rho} g_{\lambda\sigma}\mp i\epsilon_{\mu\lambda\sigma\rho}}{2\pi^2x^2}\left(\frac{x_\lambda x_{\nu}}{x^2}-\frac14 g_{\lambda\nu}\right)K_m(x/\rho)\right]\\
    &\quad\times\frac{N_c^2-1}{2N_c}im\bar\psi\gamma_\rho\frac12\left(e^{x\cdot\overrightarrow{\partial}}-e^{x\cdot\overleftarrow{\partial}}\right)\psi\\
    &-(\mu\leftrightarrow\nu)
\end{aligned}
$}
\end{equation}
and similar trick can apply to the molecular contribution,

\begin{equation}
\resizebox{\textwidth}{!}{$
\begin{aligned}
    &\frac{\rho^4}{(x^2+\rho^2)^2x^2}\frac{x_\rho x_\lambda}{x^2}\bar\psi(x)\gamma_\sigma\lambda^A\psi(x)\bar\psi(0) i\gamma_{\alpha}\gamma^5\overleftrightarrow{\partial}_{\beta}\lambda^A\psi(0)\\
    &\approx \frac{4\rho^4}{(x^2+\rho^2)^2x^2}\frac{x_\rho x_\lambda}{x^2}\frac{x_\mu}{x^4}K_D(x/\rho)\frac{1}{2\pi^2}i\epsilon_{\gamma\sigma\mu\alpha}\frac{N_c^2-1}{2N_c}\bar\psi\gamma_\gamma \frac12\left(e^{x\cdot\overleftarrow{\partial}}\overrightarrow{\partial}_{\beta}-\overleftarrow{\partial}_{\beta}e^{x\cdot\overrightarrow{\partial}}\right)\psi\\
    &-\frac{4\rho^4x_\rho x_\lambda x_\mu}{(x^2+\rho^2)^2x^8}K_D(x/\rho)\frac{g_{\sigma\mu}g_{\alpha\gamma}-g_{\sigma\alpha}g_{\mu\gamma}+g_{\sigma\gamma}g_{\mu\alpha}}{2\pi^2}\frac{N_c^2-1}{2N_c}\bar\psi\gamma_\gamma\gamma^5\frac12\left(e^{x\cdot\overleftarrow{\partial}}\overrightarrow{\partial}_{\beta}+\overleftarrow{\partial}_{\beta}e^{x\cdot\overrightarrow{\partial}}\right)\psi\\
    &+\frac{4\rho^4x_\rho x_\lambda}{(x^2+\rho^2)^2x^4}\partial_\beta\left(\frac{x_\mu}{x^4}K_D(x/\rho)\right)\frac{1}{2\pi^2}i\epsilon_{\gamma\sigma\mu\alpha}\frac{N_c^2-1}{2N_c}\bar\psi\gamma_\gamma\frac12\left(e^{x\cdot\overleftarrow{\partial}}-e^{x\cdot\overrightarrow{\partial}}\right)\psi\\
    &-\frac{4\rho^4x_\rho x_\lambda}{(x^2+\rho^2)^2x^4}\partial_\beta\left(\frac{x_\mu}{x^4}K_D(x/\rho)\right)\frac{g_{\sigma\mu}g_{\alpha\gamma}-g_{\sigma\alpha}g_{\mu\gamma}+g_{\sigma\gamma}g_{\mu\alpha}}{2\pi^2}\frac{N_c^2-1}{2N_c}\bar\psi\gamma_\gamma\gamma^5\frac12\left(e^{x\cdot\overleftarrow{\partial}}+e^{x\cdot\overrightarrow{\partial}}\right)\psi\\
    &+\frac{4\rho^4}{(x^2+\rho^2)^2x^2}\frac{x_\rho x_\lambda}{x^2}\frac{1}{4\pi^2x^2}K_m(x/\rho)\frac{N_c^2-1}{2N_c}m\bar\psi\sigma_{\sigma\alpha}\gamma^5\frac12\left(e^{x\cdot\overleftarrow{\partial}}\overrightarrow{\partial}_{\beta}-\overleftarrow{\partial}_{\beta}e^{x\cdot\overrightarrow{\partial}}\right)\psi\\
    &+\frac{4\rho^4}{(x^2+\rho^2)^2x^2}\frac{x_\rho x_\lambda}{x^2} \frac{1}{4\pi^2x^2}K_m(x/\rho)g_{\alpha\sigma}\frac{N_c^2-1}{2N_c}m\bar\psi i\gamma^5\frac12\left(e^{x\cdot\overleftarrow{\partial}}\overrightarrow{\partial}_{\beta}+\overleftarrow{\partial}_{\beta}e^{x\cdot\overrightarrow{\partial}}\right)\psi\\
    &+\frac{4\rho^4}{(x^2+\rho^2)^2x^2}\frac{x_\rho x_\lambda}{x^2}\frac{1}{4\pi^2}\partial_{\beta}\left(\frac{1}{x^2}K_m(x/\rho)\right)\frac{N_c^2-1}{2N_c}m\bar\psi\sigma_{\sigma\alpha}\gamma^5\frac12\left(e^{x\cdot\overleftarrow{\partial}}-e^{x\cdot\overrightarrow{\partial}}\right)\psi\\
        &+\frac{4\rho^4}{(x^2+\rho^2)^2x^2}\frac{x_\rho x_\lambda}{x^2}\frac{1}{4\pi^2}\partial_{\beta}\left(\frac{1}{x^2}K_m(x/\rho)\right)g_{\alpha\sigma}\frac{N_c^2-1}{2N_c}m\bar\psi i\gamma^5\frac12\left(e^{x\cdot\overleftarrow{\partial}}+e^{x\cdot\overrightarrow{\partial}}\right)\psi\\
\end{aligned}
$}
\end{equation}

Eventually by redefining the non-local form factors, we will have
\begin{equation}
\resizebox{\textwidth}{!}{$
\begin{aligned}
    &g\bar\psi\gamma_\sigma F_{\mu\nu}\psi(x)=-\left(\frac{N_+}V\right)\frac{1}{2N_c}\left(\frac{4\pi^2\rho^2}{m^*}\right)\\
    &\times\int d^4z\left[F^{(+)}_{\bar{q}Gq,1,\mu\nu\sigma}(x-z)\bar\psi(z)\frac{1-\gamma^5}2\psi(z)+F^{(+)}_{\bar{q}Gq,2,[\mu\nu]\sigma\rho}(x-z)m\bar\psi(z)\gamma_\rho\gamma^5\psi(z)\right]\\
    &-\left(\frac{N_-}V\right)\frac{1}{2N_c}\left(\frac{4\pi^2\rho^2}{m^*}\right)\\
    &\times\int d^4z\left[F^{(-)}_{\bar{q}Gq,1,\mu\nu\sigma}(x-z)\bar\psi(z)\frac{1+\gamma^5}2\psi(z)+F^{(-)}_{\bar{q}Gq,2,[\mu\nu]\sigma\rho}(x-z)m\bar\psi(z)\gamma_\rho\gamma^5\psi(z)\right]\\
    &+\left(\frac{N_+N_-}{V^2}\right)\frac{1}{2N_c(N_c^2-1)}\int d^4R\,\left(\frac{|T_{IA}|^2}{(m^*)^2}\right)^{N_f}\frac{(4\pi^2\rho^2)^2}{|T_{IA}|^2}\left[\frac{-1}4R\frac{dT(R)}{dR}\right]\\
    &\times\int d^4z\Bigg[F^{(+-)}_{\bar{q}Gq,1,\mu\nu\sigma\alpha\beta}(x-z)\bar\psi(z)\gamma_{\alpha}i\overleftrightarrow{\partial}_{\beta}\psi(z)+F^{(+-)}_{\bar{q}Gq,2,\mu\nu\sigma\rho}(x-z)\bar\psi(z)\gamma_\rho\gamma^5\psi(z)\\
    &\qquad\qquad+F^{(+-)}_{\bar{q}Gq,3,\mu\nu\alpha\beta}(x-z)m\bar\psi(z)\sigma_{\sigma\alpha}\gamma^5\overleftrightarrow{\partial}_{\beta}\psi(z)+F^{(+-)}_{\bar{q}Gq,4,\mu\nu\sigma}(x-z)m\bar\psi(z)i\gamma^5\psi(z)\Bigg]
\end{aligned}
$}
\end{equation}
with
\begin{equation}
    F^{(\pm)}_{\bar{q}Gq,1,\mu\nu\sigma}(x)=\frac{8\rho^2}{(x^2+\rho^2)^2}K_D(x/\rho)\frac{g_{\mu\sigma}x_\nu-g_{\nu\sigma}x_{\mu}\pm i\epsilon_{\mu\nu\sigma \alpha}x_\alpha}{2\pi^2x^4}
\end{equation}
\begin{equation}
\resizebox{0.94\textwidth}{!}{$
F^{(\pm)}_{\bar{q}Gq,2,\mu\nu\sigma\rho}(x)=\pm\frac{8\rho^2}{(x^2+\rho^2)^2}K_m(x/\rho)\frac{i\left(g_{\mu\rho} g_{\lambda\sigma}-g_{\lambda\rho} g_{\mu\sigma}\right)\pm\epsilon_{\mu\lambda\sigma\rho}}{2\pi^2x^2}\left(\frac{x_\lambda x_\nu}{x^2}-\frac14 g_{\lambda\nu}\right)
$}
\end{equation}
\begin{equation}
\begin{aligned}
    F^{(+-)}_{\bar{q}Gq,1,\mu\nu\sigma\alpha\beta}(x)=&\frac{\rho^4}{(x^2+\rho^2)^2x^2}K_D(x/\rho)\frac{x_\rho x_\lambda}{x^2}t_{\mu\nu\rho\lambda(\gamma\beta)}\frac{\epsilon_{\alpha\sigma\delta\gamma}x_\delta}{2\pi^2x^4}\\
\end{aligned}
\end{equation}

\begin{equation}
\resizebox{\textwidth}{!}{$
\begin{aligned}
&F^{(+-)}_{\bar{q}Gq,2,\mu\nu\sigma\rho}(x)=t_{\mu\nu\gamma\lambda\alpha\beta}\Bigg[\frac12\partial_{\beta}\left(\frac{\rho^4}{(x^2+\rho^2)^2x^2}\frac{x_{\gamma} x_\lambda}{x^2}\frac{x_\delta}{x^4}\right)-\frac{\rho^4}{(x^2+\rho^2)^2x^2}\frac{x_{\gamma} x_\lambda}{x^2}\partial_\beta\left(\frac{x_\delta}{x^4}\right)\\
&-\frac12\frac{\rho^4x_\rho x_\lambda x_\mu}{(x^2+\rho^2)^2x^8}\partial_\beta\ln K_D(x/\rho)\Bigg]K_D(x/\rho)\frac{g_{\sigma\delta}g_{\alpha\rho}-g_{\sigma\alpha}g_{\delta\rho}+g_{\sigma\rho}g_{\delta\alpha}}{2\pi^2}\\
=&\frac{1}{\pi^2}\left[-\frac{24\rho^4}{x^8\, (x^2 + \rho^2)^2}\left(K_D(x/\rho)-\frac16x\frac{d}{dx}K_D(x/\rho)\right)
+ \frac{16\rho^4}{x^6\, (x^2 + \rho^2)^3}K_D(x/\rho)\right]x_{[\rho}\epsilon_{\sigma]\mu \nu \lambda}x_\lambda
\end{aligned}
$}
\end{equation}

\begin{equation}
\begin{aligned}
    F^{(+-)}_{\bar{q}Gq,3,\mu\nu\alpha\beta}(x)=&t_{\mu\nu\rho\lambda\alpha\beta}\frac{\rho^4}{(x^2+\rho^2)^2x^2}\frac{x_\rho x_\lambda}{x^2}\frac{1}{4\pi^2x^2}K_m(x/\rho)\\
    &-t_{\mu\nu\rho\lambda\alpha\gamma}\frac{\rho^4}{(x^2+\rho^2)^2x^2}\frac{x_\rho x_\lambda x_\beta}{x^2}\frac{1}{4\pi^2}\partial_\gamma\left(\frac{K_m(x/\rho)}{x^2}\right)
\end{aligned}
\end{equation}
\begin{equation}
\begin{aligned}
   &F^{(+-)}_{\bar{q}Gq,4,\mu\nu\sigma}(x)=\frac{1}{2\pi^2}t_{\mu\nu\rho\lambda\sigma\beta}\Bigg[\frac12\partial_{\beta}\left(\frac{\rho^4}{(x^2+\rho^2)^2x^4}\frac{x_\rho x_\lambda}{x^2}\right)-\frac{\rho^4}{(x^2+\rho^2)^2x^2}\frac{x_\rho x_\lambda}{x^2}\partial_{\beta}\left(\frac{1}{x^2}\right)\\
   &-\frac12\frac{\rho^4}{(x^2+\rho^2)^2x^4}\frac{x_\rho x_\lambda}{x^2}\partial_\beta \ln K_m(x/\rho)\Bigg]K_m(x/\rho)\\
   =&\frac{1}{2\pi^2}\left[\frac{16\rho^4}
       {x^{6}\,(x^{2}+\rho^2)^2}\left(K_m(x/\rho)-\frac14x\frac{d}{dx}K_m(x/\rho)\right)-\frac{16\rho^4}
       {x^{4}\,(x^{2}+\rho^2)^3}K_m(x/\rho) \right]\epsilon_{\mu\nu\sigma\lambda}x_\lambda
\end{aligned}   
\end{equation}
with the help of the identity~\cite{Freese:2019bhb}, which still holds for the effective quark theory such as ILM.

\[
\,\bar{\psi}\,\gamma^{[\mu} i\overleftrightarrow{\partial}^{\nu]} \psi
= \frac{1}{4} \,\epsilon^{\mu\nu\rho\sigma} \,\partial_{\rho}
\left( \bar{\psi}\,\gamma_{\sigma}\,\gamma_{5}\,\psi \right)
\]

\chapter{Polarization and spin $4$-vectors}
\label{App:pol}
The kinematics of hadrons can be represented by the longitudinal light front unit vectors $n=(1,0,0_\perp)$ and $\bar n=(0,1,0_\perp)$ in light-cone signature and the transverse component. The hadronic momentum with energy $E_{\vec{p}}=\sqrt{m_h^2+\vec{p}^2}$ and $3$-momentum $\vec{p}$ can be decomposed into 

\begin{equation}
\begin{aligned}
    p^\mu=&p^+\bar n^\mu-\frac {m^2_h}{2p^+} n^\mu+p^\mu_\perp
\end{aligned}
\end{equation}
with hadron mass $m_h$. 

In rest frame, for spin-$j$ particles, the spin operator $\vec{\Sigma}$ is defined by the $(2j+1)\times(2j+1)$ matrices in the irreducible representation with the unit vector $\vec{s}$ represents the spin quantization axis, which can be chosen arbitrarily. 

The same argument can be generalized to Lorentz covariant formalism by boost transformation. By boosting the $4$-vector $(0,\vec{s})$ from the rest frame, the spin $4$-vector now is identified as

\begin{equation}
\label{spin-vec}
    S^\mu=\left(\frac{\vec{p}\cdot\vec s}{m_h}, \vec{s}+\frac{(\vec{p}\cdot\vec s) \vec p}{m_h(E_{\vec p}+m_h)}\right)
\end{equation}
with the condition $p\cdot S=0$ and $S^2=-1$ alongside. 

The spin operator $\vec\Sigma$ is generalized to so-called Pauli-Lubanski pseudovector $\Sigma^\mu$.

\begin{equation}
    \Sigma^\mu=\frac1{2m_h}\epsilon^{\mu\nu\rho\sigma}p_\nu J_{\rho\sigma}
\end{equation}
where $J_{\mu\nu}$ is the generators of Lorentz group.

For simplicity, one can choose the 3-momentum of the target along the $z$ direction ($p_\perp=0$) and denote the Cartesian components of the spin quantization
unit vector by $\vec s= (s_\perp,s_L)$, where $s_\perp = (s^1,s^2)$ is normal to the hadron target momentum and $s_L$ is the $z$ component. In the light-cone signature, the spin vector can be written as
\begin{equation}
    S^\mu=s_L\frac{p^+}{m_h}\bar n^\mu-s_L\frac {m_h}{p^+} n^\mu+s^\mu_\perp
\end{equation}
where $s_L$ can be viewed as the target averaged helicity and the transverse spin is defined by $s_\perp^2=1-s_L^2$.

Generally, for any given direction $\vec S$, the spin of the particle can be projected onto this direction. It is worth clarifying that longitudinal polarization means that $s_\perp$ = 0 and
$|s_L|=1$ and there are $2j+1$ spin projections for spin-$j$ particle along the direction of the momentum, whereas transverse polarization means that $s_L = 0$ and $|s_\perp| = 1$, and there
are $2j+1$ spin projections along the direction $s_\perp$ perpendicular to the momentum. 

\section{Spin-$1/2$ particles}

For spin-$1/2$ particles, the spin operator is a $2$ by $2$ Pauli matrix $\vec{\Sigma}=\vec\sigma/2$ defining the spin-$1/2$ eigenstates $\chi_s$ by 

\begin{equation}
\label{eq:spin}
    \frac12\vec{s}\cdot\vec{\sigma}~\chi_s=s~\chi_s
\end{equation}
with spin quantum number $s=\pm1/2$. If we set the spin quantization axis on $z$ direction, the spinor is simply $\chi_\uparrow=\begin{pmatrix}
    1 \\
    0
\end{pmatrix}$ and $\chi_\downarrow=\begin{pmatrix}
    0 \\
    1
\end{pmatrix}$.

By Lorentz boost, the equation \eqref{eq:spin} can be rewritten covariantly as
\begin{equation}
    \frac{1}{4m_h}\epsilon_{\mu\nu\rho\lambda}S^\mu p^\nu\sigma^{\rho\lambda}u_s(p)=su_s(p)
\end{equation}
with $4$-spinor generalization of the spin-$1/2$ eigenstate $\chi_s$. The $4$-spinor $u_s(p)$ of a spin-$1/2$ particle can be written as

\begin{equation}
    u_s(p)=
    \begin{pmatrix}
    \frac{E_{\vec p}+m_h-\vec\sigma\cdot \vec p}{\sqrt{2(E_{\vec p}+m_h)}}\chi_s \\[10pt]
    \frac{E_{\vec p}+m_h+\vec\sigma\cdot \vec p}{\sqrt{2(E_{\vec p}+m_h)}}\chi_s
    \end{pmatrix}
\end{equation}
with normalization $\bar{u}_s(p)u_s(p)=2m_h$. The following relation between the spinor and spin 4-vectors \eqref{spin-vec} is then easy to confirm

\begin{equation}
    \bar u_s(p)\frac{1}{2m_h}\gamma^\mu\gamma^5u_s(p)=2sS^\mu
\end{equation}

It is convenient to explore the covariant form of the spin-$1/2$ density matrix $\chi_s\chi_s^\dagger$ for pure spin states with some
fixed $s$, by the spin and momentum 4-vectors. The covariant spin density matrix with a fixed $s$ pure state can be expressed as
\begin{equation}
    u_s(p)\bar{u}_s(p)=(\slashed{p}+m_h)\frac12\left(1-2s\gamma^5\slashed{S}\right)
\end{equation}
\section{Spin-$1$ particles}

For the spin-$1$ particles, the spin operator is a $3$ by $3$ matrix with entries $(\Sigma^k)_{ij}=-i\epsilon_{kij}$, defining the spin-$1$ eigenstates $\vec\varepsilon_\lambda$ by

\begin{equation}
\label{eq:spin-1}
-is^k\epsilon_{kij}~\varepsilon^j_\lambda=\lambda~\varepsilon^i_\lambda
\end{equation}
with polarization $\lambda=0$, $\pm1$. If we set the spin quantization axis on $z$ direction, the polarization vector is simply $\vec\varepsilon_\pm=(1,\pm i,0)/\sqrt{2}$ and $\vec\varepsilon_0=(0,0,1)$.

The covariant generalization of the equation \eqref{eq:spin-1} is 
\begin{equation}
    \frac{i}{m_h}\epsilon_{\mu\nu\rho\sigma}S^\mu p^\nu \varepsilon^\sigma_\lambda(p)=\lambda\varepsilon^\rho_\lambda(p)
\end{equation}
with $4$-vector generalization of the spin-$1$ eigenstate $\vec\varepsilon_\lambda$. The polarization $4$-vectors of a spin-one particle are given by
\begin{equation}
\varepsilon^\mu_\lambda(p)=\left(\frac{\vec{p}\cdot\vec{\varepsilon_\lambda}}{{m_h}},\vec{\varepsilon}_\lambda+\frac{\vec{p}(\vec{p}\cdot\vec{\varepsilon}_\lambda)}{m_h(E_{\vec{p}} +m_h)}\right)
\end{equation}
with the Ward identity $p\cdot\varepsilon_\lambda(p)=0$ and normalization condition $\varepsilon^*_{\pm}\cdot\varepsilon_\pm=\varepsilon^*_{0}\cdot\varepsilon_0=-1$. The following relation between
the polarization and spin 4-vectors \eqref{spin-vec} is then easy to confirm

\begin{equation}
\label{spin_exp}
\frac{i}{m_h}\epsilon^{\mu\nu\rho\sigma}
p_\nu\varepsilon^*_{\lambda\rho}(p)\varepsilon_{\lambda\sigma}(p)=\lambda S^\mu
\end{equation}

It is convenient to explore the covariant form of the spin-$1$ density matrix $\varepsilon^i_\lambda\varepsilon^j_\lambda{}^*$ for pure spin states with some
fixed polarization $\lambda$, by the spin and momentum 4-vectors. The covariant spin density matrix with a fixed $\lambda$ pure state can be expressed as

\begin{equation}
\begin{aligned}
\varepsilon^{\mu*}_\lambda(p)\varepsilon^{\nu}_\lambda(p)=&\frac13\left(-g^{\mu\nu}+\frac{p^\mu p^\nu}{m^2_h}\right)-\frac{i\lambda}{2m_h}\epsilon^{\mu\nu\rho\sigma}p_\rho S_\sigma\\
&-\frac{3\lambda^2-2}{2}\left[S^\mu S^\nu-\frac13\left(-g^{\mu\nu}+\frac{p^\mu p^\nu}{m^2_h}\right)\right]
\end{aligned}
\end{equation}

The identity is achieved by decomposing the spin-$1$ density matrix into symmetric traceless and antisymmetric Lorentz tensors and trace. 
using \eqref{spin_exp} and the
completeness relation of polarization vector $\varepsilon^\mu_\lambda(p)$ \cite{Jaffe1997,Jaffe:1988up}.

\chapter{Identities in Dirac theory}
\label{app:Gordan}
\section{Off-forward product of Dirac spinors }

For a spin-$1/2$ particle with mass $m_h$,  Poincare covariance, parity, and hermicity in principle allows for off-forward tensor structures with the Lorentz indices carried by $\{\gamma^\mu,\gamma^\mu\gamma^5,\bar{p}^\mu, q^\mu, g^{\mu\nu}, i\sigma^{\mu\nu},\epsilon^{\mu\nu\rho\lambda}\}$ alongside with the Dirac scalar matrices $\{1,\slashed{\bar p}, \slashed{q}, [\slashed{\bar p},\slashed{q}]\}$.  The linear combinations of those can be used to establish the Lorentz structure of baryonic matrix elements and form factors. Those Dirac scalar matrices can be further simplified by
 
\begin{equation}
    \bar{u}_{s'}(p')\slashed{\bar{p}} u_s(p)= m_h \bar{u}_{s'}(p')u_s(p)
\end{equation}
\begin{equation}
    \bar{u}_{s'}(p')\slashed{q} u_s(p)= 0
\end{equation}

Due to the sandwiching between the on-shell baryon spinors $\bar{u}_s(p' )$ and $u_s(p)$, among those basis of Gamma matrices, not all tensor structures are linearly independent; their relations are captured by the following identities

\begin{equation}
    \bar{u}_{s'}(p')\gamma^\mu u_s(p)=\bar{u}_{s'}(p')\left(\frac{\bar p^\mu}{m_h} +\frac{i\sigma^{\mu\nu}q_\nu}{2m_h}\right)u_s(p)
\end{equation}

\begin{equation}
\begin{aligned}
    \bar{u}_{s'}(p')\gamma^\mu\gamma^5 u_s(p)=\quad\bar{u}_{s'}(p')\left(\frac{i\sigma^{\mu\nu}\gamma^5\bar{p}_\nu}{m_h} +\frac{q^\mu\gamma^5}{2m_h} \right)u_s(p)
\end{aligned}
\end{equation}

\begin{equation}
    \bar{u}_{s'}(p')\left(\gamma^\mu \bar{p}^\nu-\gamma^\nu \bar{p}^\mu\right)\bar{u}_s(p)=\frac{i}2\epsilon^{\mu\nu\rho\lambda}\bar{u}_{s'}(p')q_\rho \gamma_\lambda\gamma^5\bar{u}_s(p)
\end{equation}

\begin{equation}
    \bar{u}_{s'}(p')i\sigma^{\mu\nu} \bar{u}_s(p)=\bar{u}_{s'}(p')\left(\frac{\gamma^\mu q^\nu-\gamma^\nu q^\mu}{2m_h}-\frac{i}{m_h}\epsilon^{\mu\nu\rho\sigma}\bar{p}_\rho\gamma_\sigma\gamma^5\right) \bar{u}_s(p)
\end{equation}

In forward limit, we have

\begin{equation}
\begin{aligned}
    &\bar{u}_s(p)\gamma^\mu u_s(p)=2p^\mu\\
    &\bar{u}_s(p)\gamma^\mu\gamma^5 u_s(p)=2m_hS^\mu\\
    &\bar{u}_s(p)\sigma^{\mu\nu} u_s(p)=2\epsilon^{\mu\nu\rho\lambda}S_\rho p_\lambda
\end{aligned}
\end{equation}

\section{Dirac algebra and Levi-Civita symbol}

Here we present several useful identities in Dirac algebra related Levi-Civita symbol with convention $\epsilon^{0123}=1$. 

Schouten identity of Levi-Civita symbol simplifies the Lorentz decomposition of the form factors. Given abitrary Lorentz covariant 4-vector $\{a^\mu,b^\mu, c^\mu\}$, this identity reads,
\begin{equation}
\begin{aligned}
&a^\mu \epsilon^{\nu\sigma\rho\lambda} b_\rho c_\lambda - a^\nu \epsilon^{\mu\sigma\rho\lambda} b_\rho c_\lambda + a^\sigma \epsilon^{\mu\nu\rho\lambda} b_\rho c_\lambda=(a\cdot b) \epsilon^{\mu\nu\sigma\gamma} c_\gamma - (a \cdot c) \epsilon^{\mu\nu\sigma\gamma} b_\gamma
\end{aligned}
\end{equation}

The Dirac sigma matrix $\sigma^{\mu\nu}$ can also be related to $\sigma^{\mu\nu}\gamma^5$ through the Levi-Civita symbol. In 3+1D Minkowski space, the identity reads
\begin{equation}
\frac i2\epsilon^{\mu\nu\rho\lambda}\sigma_{\rho\lambda}=\sigma_{\mu\nu}\gamma^5
\end{equation}

In 3+1D Minkowski space, the product of three gamma matrices leads to this identity which is useful to simplify the off-forward structure of Dirac spinor~\cite{Leader:2013jra}.

\[
\sigma^{\mu\nu} \gamma^{\rho}
= i g^{\nu\rho} \gamma^{\mu}
- i g^{\mu\rho} \gamma^{\nu}
- \epsilon^{\mu\nu\rho\sigma} \gamma_{\sigma} \gamma_{5},
\]
\[
\gamma^{\rho} \sigma^{\mu\nu}
= i g^{\mu\rho} \gamma^{\nu}
- i g^{\nu\rho} \gamma^{\mu}
- \epsilon^{\mu\nu\rho\sigma} \gamma_{\sigma} \gamma_{5},
\]

With these identities, several off-forward spinor products can be further simplified by

\begin{equation}
\begin{aligned}
&\bar u_{s'}(p')\,\frac i{2m_h^2}\epsilon^{\mu\nu\alpha\beta}\bar{p}_\alpha q_\beta\gamma^5\,u_s(p)
=\\
&\bar u_{s'}(p')\left[
-i\frac{q^2}{4m_h^2}\,\sigma^{\mu\nu}-\frac{\gamma^\mu q^\nu-\gamma^\nu q^\mu}{2m_h}-\frac{i\left(\sigma^{\mu\alpha}q^\nu-\sigma^{\nu\alpha}q^\mu\right)q_\alpha}{2m_h^2}
\right]u_s(p)
\end{aligned}
\end{equation}

\section{Equation of motion}

For the effective quark theory such as RILM, the equation of motion leads to the following three identities 
\cite{Freese:2019bhb,Bhoonah:2017olu}

\begin{equation}
\,\bar{\psi}\,\gamma^{[\mu} i\overleftrightarrow{\partial}^{\nu]} \psi
= \frac{1}{4} \,\epsilon^{\mu\nu\rho\sigma} \,\partial_{\rho}
\left( \bar{\psi}\,\gamma_{\sigma}\,\gamma_{5}\,\psi \right)
\end{equation}

\begin{equation}
\bar{\psi}\, i\sigma^{\lambda\mu}\gamma_5\, i\overleftrightarrow{\partial}_\mu \psi
= 2M\,\bar{\psi}\gamma^\lambda\gamma_5\psi
+ i\partial^\lambda(\bar{\psi}\gamma_5\psi)
\end{equation}

\begin{equation}
\epsilon_{\lambda\mu\nu\alpha}\bar{\psi}\, i\sigma^{\lambda\mu}\gamma_5\, i\overleftrightarrow{\partial}^{\nu} \psi
=2\,
\partial_\alpha(\bar{\psi}\psi).
\end{equation}
where $M$ is the corresponding quark mass in the given theory. These identities can be simply extended to gauge theory by replacing derivatives to covariant derivatives $\partial\rightarrow D$.

\chapter{Quark bubble functions for meson and diquark channels}
\label{app:vac_pol}

In Ch.~\ref{ch:low_QCD}, we derive the mass spectrum and various chiral properties of light hadrons using gap-like equations in terms of RPA bubble diagrams. The bubble diagrams in each channel are defined and evaluated as follows

\section{Identical quark mass}
\begin{itemize}
    \item \textbf{scalar-scalar vacuum polarization}

\begin{equation}
\begin{aligned}
   \Pi_{SS}=& -2i\int\frac{d^4k}{(2\pi)^4}\mathrm{Tr}\left[S(k)S(k-P)\right]\mathcal{F}(k)\mathcal{F}(P-k)
\end{aligned}
\end{equation}

\item\textbf{vector-vector vacuum polarization}

\begin{equation}
\begin{aligned}
\Pi^{\mu\nu}_{VV}=& -2i\int\frac{d^4k}{(2\pi)^4}\mathrm{Tr}\left[S(k)\gamma^\mu S(k-P)\gamma^\nu\right]\mathcal{F}(k)\mathcal{F}(P-k)\\
=&-\Pi_{VV}(P^2)\left(g^{\mu\nu}-\frac{P^\mu P^\nu}{P^2}\right)
\end{aligned}
\end{equation}

\item\textbf{pseudoscalar-pseudoscalar vacuum polarization}
\begin{equation}
\begin{aligned}
   \Pi_{PP}=& -2i\int\frac{d^4k}{(2\pi)^4}\mathrm{Tr}\left[S(k)i\gamma^5S(k-P)i\gamma^5\right]\mathcal{F}(k)\mathcal{F}(P-k)
\end{aligned}
\end{equation}

\item\textbf{axial-pseudoscalar vacuum polarization}
\begin{equation}
\begin{aligned}
   \Pi^{\mu}_{PA}=& -2i\int\frac{d^4k}{(2\pi)^4}\mathrm{Tr}\left[S(k)i\gamma^5S(k-P)\gamma^\mu\gamma^5\right]\mathcal{F}(k)\mathcal{F}(P-k)\\
   =&i\Pi_{PA}(P^2)\frac{P^\mu}{\sqrt{P^2}}
\end{aligned}
\end{equation}
and
\begin{equation}
\begin{aligned}
   \Pi^{\mu}_{AP}=& -2i\int\frac{d^4k}{(2\pi)^4}\mathrm{Tr}\left[S(k)\gamma^\mu\gamma^5S(k-P)i\gamma^5\right]\mathcal{F}(k)\mathcal{F}(P-k)\\
   =&-i\Pi_{PA}(P^2)\frac{P^\mu}{\sqrt{P^2}}
\end{aligned}
\end{equation}

\item\textbf{axial-axial vacuum polarization}

\begin{equation}
\begin{aligned}
   \Pi^{\mu\nu}_{AA}=& -2i\int\frac{d^4k}{(2\pi)^4}\mathrm{Tr}\left[S(k)\gamma^\mu\gamma^5S(k-P)\gamma^\nu\gamma^5\right]\mathcal{F}(k)\mathcal{F}(P-k)\\
   =&-\Pi^{(t)}_{AA}(P^2)\left(g^{\mu\nu}-\frac{P^\mu P^\nu}{P^2}\right)-\Pi^{(l)}_{AA}(P^2)\frac{P^\mu P^\nu}{P^2}
\end{aligned}
\end{equation}
\end{itemize}

With the identity $2k\cdot(P-k)=P^2-k^2-(P-k)^2$, the quark bubble functions $\Pi_X$ can be decomposed into three master integrals $I_1$, $I_2$, and $I_3$ which are defined by
\begin{align}
    &I_1(P^2)=\int\frac{d^4k}{(2\pi)^4}\frac{-i}{[(k-P/2)^2-M^2][(k+P/2)^2-M^2]}\mathcal{F}(k-P/2)\mathcal{F}(k+P/2)\\
    &I_2(P^2)=\int\frac{d^4k}{(2\pi)^4}\frac{i}{k^2-M^2}\mathcal{F}(k)\mathcal{F}(P-k)\\
    &I_3(P^2)=\int\frac{d^4k}{(2\pi)^4}\frac{4i(k\cdot P)^2/(P^2)^2}{[(k-P/2)^2-M^2][(k+P/2)^2-M^2]}\mathcal{F}(k-P/2)\mathcal{F}(k+P/2)
\end{align}

The vacuum polarization functions read
\begin{equation}
    \Pi_{SS}=4(P^2-4M^2)I_1(P^2)+8I_2(P^2)
\end{equation}

\begin{equation}
    \Pi_{PP}
   =4P^2I_1(P^2)+8I_2(P^2)
\end{equation}

\begin{equation}
    \Pi_{VV}
   =\frac{8}{3}(P^2+2M^2)I_1(P^2)+\frac{16}{3}I_2(P^2)
\end{equation}

\begin{equation}
\Pi_{PA}=8M\sqrt{P^2}I_1(P^2)
\end{equation}

\begin{equation}
    \Pi^{(l)}_{AA} =-16M^2I_1(P^2)+8I_2(P^2)-4P^2I_3(P^2)
\end{equation}

\begin{equation}
\begin{aligned}
    \Pi^{(t)}_{AA}=\frac{8}{3}(P^2-4M^2)I_1(P^2)+\frac{8}{3}I_2(P^2)+\frac{4}{3}P^2I_3(P^2)
\end{aligned}
\end{equation}

The summation of the quark bubble diagrams gives the transfer matrices $D_X(P^2)$ for each meson channel.

\begin{eqnarray}
    &D_{\sigma,a_0}(P^2)&=\frac1{N_c}\frac{ig_{\sigma,a_0}}{1-g_{\sigma,a_0}\Pi_{SS}}\nonumber\\[7pt]
    &D_{\eta',\pi}(P^2)&=\frac1{N_c}\frac{i\left[g_{\eta',\pi}-g_{f_1,a_1}+g_{\eta',\pi}g_{f_1,a_1}\left(\Pi_{PP}-\Pi^{(l)}_{AA}\right)\right]}{(1-g_{\eta',\pi}\Pi_{PP})(1-g_{f_1,a_1}\Pi^{(l)}_{AA})+g_{\eta',\pi}g_{f_1,a_1}\Pi^2_{PA}}\nonumber\\[7pt]
    &D_{\omega,\rho}^{\mu\nu}(P^2)&=\frac1{N_c}\frac{-ig_{\omega,\rho}}{1-g_{\omega,\rho}\Pi_{VV}}\left(g^{\mu\nu}-\Pi_{VV}\frac{P^\mu P^\nu}{P^2}\right)\nonumber\\[7pt]
    &D_{f_1,a_1}^{\mu\nu}(P^2)&=\frac1{N_c}\frac{-ig_{f_1,a_1}}{1-g_{f_1,a_1}\Pi^{(t)}_{AA}}\left(g^{\mu\nu}-\frac{P^\mu P^\nu}{P^2}\right)
\end{eqnarray}

\section{Different quark mass}
\label{vac_pol_2}
Similar calculatioin can be carried out for the bubble functions with different quark masses in each channel. One can replace the first propagator by $S_q$ and the second by $S_{q'}$ where the quark propagator is 

\begin{equation}
    S_q(k)=\frac{i(\slashed{k}+M_q(k))}{k^2-M_q^2(k)}\simeq\frac{i(\slashed{k}+M_q(k))}{k^2-M_q^2(0)}
\end{equation}
with the momentum dependent constituent mass $M_q(k)$ determined by the gap equation in \eqref{gap}.

With the identity $2k\cdot(P-k)=P^2-k^2-(P-k)^2$, the quark bubble functions $\Pi_X$ can be decomposed into three master integrals $I_1$, $I_2$, and $I_3$ which are defined by

\begin{align}
    &I_1(P^2)=\int\frac{d^4k}{(2\pi)^4}\frac{-i}{[(k-P/2)^2-M_{q'}^2][(k+P/2)^2-M_{q}^2]}\mathcal{F}(k-P/2)\mathcal{F}(k+P/2)\\
    &I_{2q}(P^2)=\int\frac{d^4k}{(2\pi)^4}\frac{i}{k^2-M_q^2}\mathcal{F}(k)\mathcal{F}(P-k)\\
    &I_3(P^2)=\int\frac{d^4k}{(2\pi)^4}\frac{4i(k\cdot P)^2/(P^2)^2}{[(k-P/2)^2-M_{q'}^2][(k+P/2)^2-M_{q}^2]}\mathcal{F}(k-P/2)\mathcal{F}(k+P/2)
\end{align}

Now the bubble functions read
\begin{align}
&\Pi^{qq'}_{SS}
   =2[P^2-(M_q+M_{q'})^2]I_1(P^2)+2I_{2q}(P^2)+2I_{2q'}(P^2)\\[8pt] 
&\Pi^{qq'}_{PP}
   =2[P^2-(M_q-M_{q'})^2]I_1(P^2)+2I_{2q}(P^2)+2I_{2q'}(P^2)\\[8pt] 
&\Pi^{qq'}_{VV}
   =\frac{4}{3}[P^2+2M_qM_{q'}-(M_q-M_{q'})^2]I_1(P^2)+\frac{4}{3}I_{2q}(P^2)+\frac{4}{3}I_{2q'}(P^2)\\[8pt] 
&\Pi^{qq'}_{PA}=2\frac{M_q+M_{q'}}{\sqrt{P^2}}\left[P^2+(M_q-M_{q'})^2\right]I_1(P^2)+2(M_q-M_{q'})[I_{2q}(P^2)-I_{2q'}(P^2)] \\[8pt]
& \Pi^{(l)qq'}_{AA} =-2(M_q+M_{q'})^2I_1(P^2)+2I_{2q}(P^2)+2I_{2q'}(P^2)-2P^2I_3(P^2)\\[8pt]
&\Pi^{(t)qq'}_{AA}=\frac{4}{3}\left[P^2-\frac12(M_q+M_{q'})^2-2M_qM_{q'}\right]I_1(P^2)+\frac{2}{3}I_{2q}(P^2)+\frac{2}{3}I_{2q'}(P^2)\nonumber\\
&\quad\qquad+\frac{2}{3}P^2I_3(P^2)
\end{align}


\chapter{Fierz identities for RPA}
\label{App:Fierz}

\subsubsection{Flavor space}
A useful identity to transform the $SU(2)$ flavor basis between $s$-exchange and $t$-exchange reads
\begin{equation}
    \delta_{ij}\delta_{kl}-\tau^a_{ij}\tau^a_{kl}=-\left(\delta_{il}\delta_{kj}-\tau^a_{il}\tau^a_{kj}\right)
\end{equation}
and
\begin{equation}
    \delta_{ij}\delta_{kl}+\tau^a_{ij}\tau^a_{kl}=2\delta_{il}\delta_{kj}
\end{equation}

In diquark channels, the identity can be rearranged as
\begin{equation}
    \delta_{ij}\delta_{kl}-\tau^a_{ij}\tau^a_{kl}=2\tau^2_{ik}\tau^2_{lj}
\end{equation}
and
\begin{equation}
    2\delta_{ij}\delta_{lk}=\tau^2_{ik}\tau^2_{lj}+(\tau^2\tau^a)_{ik}(\tau^a\tau^2)_{lj}
\end{equation}
with the identities

\begin{equation}
    \tau^2_{ik}\tau^2_{lj} = \delta_{ij}\delta_{kl} - \delta_{il}\delta_{kj},
\end{equation}
and
\begin{equation}
(\tau^2 \tau^a)_{ik}(\tau^a\tau^2)_{lj} = \delta_{ij}\delta_{kl} + \delta_{il}\delta_{kj}    
\end{equation}

\subsubsection{Color space}
Similar identity can also be derived for color basis in $SU(N_c)$. 
\begin{equation}
    \delta_{\alpha\beta}\delta_{\gamma\delta}=\frac{1}{N_c}\delta_{\alpha\delta}\delta_{\gamma\beta}+\frac{1}{2}\lambda^\alpha_{\alpha\delta}\lambda^\alpha_{\gamma\beta}
\end{equation}

In diquark channels, the color identity can be decomposed into antisymmetric part and symmetric part. 
\begin{equation}
    \delta_{\alpha\beta}\delta_{\gamma\delta}=\frac{1}{2}\beta^a_{A\alpha\gamma}\beta^a_{A\delta\beta}+\frac{1}{2}\beta^a_{S\alpha\gamma}\beta^a_{S\delta\beta}
\end{equation}
with the identities $$(\beta^a_{A})_{\alpha\gamma}(\beta^a_{A})_{\delta\beta}=N_c\left(\delta_{\alpha\beta}\delta_{\gamma\delta}-\delta_{\alpha\delta}\delta_{\gamma\beta}\right)/2$$ and  $$(\beta^a_{S})_{\alpha\gamma}(\beta^a_{S})_{\delta\beta}=N_c\left(\delta_{\alpha\beta}\delta_{\gamma\delta}+\delta_{\alpha\delta}\delta_{\gamma\beta}\right)/2$$
where $\beta_A^a$ is the color-antisymmetric $\bar{\bf{3}}$ generators and $\beta^a_{S}$ is the color-symmetric $\bf{6}$ generators. 

\subsubsection{Dirac space}
The Fierz identity in Dirac space is 

\begin{align}
    &\bar{\psi}\psi_{j\beta}\bar{\psi}_{k\gamma}\psi_{l\delta}-\bar{\psi}_{i\alpha}i\gamma^5\psi_{j\beta}\bar{\psi}_{k\gamma}i\gamma^5\psi_{l\delta}\nonumber\\
    &=-\frac{1}{2}\left[\bar{\psi}_{i\alpha}\psi_{l\delta}\bar{\psi}_{k\gamma}\psi_{j\beta}-\bar{\psi}_{i\alpha}i\gamma^5\psi_{l\delta}\bar{\psi}_{k\gamma}i\gamma^5\psi_{j\beta}\right]-\frac{1}{4}\bar{\psi}_{i\alpha}\sigma_{\mu\nu}\psi_{l\delta}\bar{\psi}_{k\gamma}\sigma^{\mu\nu}\psi_{j\beta}\\[10pt]
&\bar{\psi}_{i\alpha}\psi_{j\beta}\bar{\psi}_{k\gamma}\psi_{l\delta}+\bar{\psi}_{i\alpha}i\gamma^5\psi_{j\beta}\bar{\psi}_{k\gamma}i\gamma^5\psi_{l\delta}\nonumber\\
&=-\frac{1}{2}\left[\bar{\psi}_{i\alpha}\gamma^\mu\psi_{l\delta}\bar{\psi}_{k\gamma}\gamma_\mu\psi_{j\beta}-\bar{\psi}_{i\alpha}\gamma^\mu\gamma^5\psi_{l\delta}\bar{\psi}_{k\gamma}\gamma_\mu\gamma^5\psi_{j\beta}\right]\\[10pt] &\bar{\psi}_{i\alpha}\gamma^\mu\psi_{l\delta}\bar{\psi}_{k\gamma}\gamma_\mu\psi_{j\beta}+\bar{\psi}_{i\alpha}\gamma^\mu\gamma^5\psi_{l\delta}\bar{\psi}_{k\gamma}\gamma_\mu\gamma^5\psi_{j\beta}\nonumber\\
&=\left[\bar{\psi}_{i\alpha}\gamma^\mu\psi_{l\delta}\bar{\psi}_{k\gamma}\gamma_\mu\psi_{j\beta}+\bar{\psi}_{i\alpha}\gamma^\mu\gamma^5\psi_{l\delta}\bar{\psi}_{k\gamma}\gamma_\mu\gamma^5\psi_{j\beta}\right]\\[10pt]
&\bar{\psi}_{i\alpha}\sigma_{\mu\nu}\psi_{j\beta}\bar{\psi}_{k\gamma}\sigma^{\mu\nu}\psi_{l\delta}\nonumber\\
&=-3\left[\bar{\psi}_{i\alpha}\psi_{l\delta}\bar{\psi}_{k\gamma}\psi_{j\beta}-\bar{\psi}_{i\alpha}i\gamma^5\psi_{l\delta}\bar{\psi}_{k\gamma}i\gamma^5\psi_{j\beta}\right]+\frac{1}{2}\bar{\psi}_{i\alpha}\sigma_{\mu\nu}\psi_{l\delta}\bar{\psi}_{k\gamma}\sigma^{\mu\nu}\psi_{j\beta}
\end{align}
where $i$, $j$, $k$, $l$ is the flavor indices and $\alpha$, $\beta$, $\gamma$, $\delta$ is the color indices.

Analogously, we can also obtain the effective Hatree-Fock interaction for color-antisymmetric $\bar{\bf{3}}$ and color-symmetric $\bf{6}$ diquarks by rewriting the 4-fermi operators with their charge conjugation.

\begin{equation}
    \bar{\psi}_{i\alpha}\psi_{j\beta}\bar{\psi}_{k\gamma}\psi_{l\delta}\rightarrow\bar{\psi}_{i\alpha}\psi_{j\beta} \psi^T_{l\delta}CC\bar{\psi}^T_{k\gamma} 
\end{equation}

\begin{equation}
    \bar{\psi}_{i\alpha}\gamma^5\psi_{j\beta}\bar{\psi}_{k\gamma}\gamma^5\psi_{l\delta}\rightarrow\bar{\psi}_{i\alpha}\gamma^5\psi_{j\beta} \psi^T_{l\delta}C\gamma^5C\bar{\psi}^T_{k\gamma} 
\end{equation}

\begin{equation}
    \bar{\psi}_{i\alpha}\gamma^\mu\psi_{j\beta}\bar{\psi}_{k\gamma}\gamma_\mu\psi_{l\delta}\rightarrow-\bar{\psi}_{i\alpha}\gamma^\mu\psi_{j\beta} \psi^T_{l\delta}C\gamma_\mu C\bar{\psi}^T_{k\gamma} 
\end{equation}

\begin{equation}
    \bar{\psi}_{i\alpha}\gamma^\mu\gamma^5\psi_{j\beta}\bar{\psi}_{k\gamma}\gamma_\mu\gamma^5\psi_{l\delta}\rightarrow\bar{\psi}_{i\alpha}\gamma^\mu\gamma^5\psi_{j\beta} \psi^T_{l\delta}C\gamma_\mu\gamma^5C\bar{\psi}^T_{k\gamma} 
\end{equation}

\begin{equation}
    \bar{\psi}_{i\alpha}\sigma^{\mu\nu}\psi_{j\beta}\bar{\psi}_{k\gamma}\sigma^{\mu\nu}\psi_{l\delta}\rightarrow-\bar{\psi}_{i\alpha}\sigma^{\mu\nu}\psi_{j\beta} \psi^T_{l\delta}C\sigma^{\mu\nu}C\bar{\psi}^T_{k\gamma} 
\end{equation}

\chapter{Solving BSF equation for baryons}
\label{app:Baryon}

The color and isospin structure as a quark-diquark boundstate has been projected properly in Sec.~\ref{sec:bary}. For the spin structure, in this Appendix we follow the calculation in \cite{Mineo:2002bg} and diagonalize the Faddeev kernel in the static approximation in \eqref{baryon_eq} for the nucleon and delta baryons. The
nucleon vertex function given in \eqref{vertex_N} and \eqref{vertex_Delta} will be contructed in details. The diagonalization will be done in the rest frame of the baryon, and the solution will be boosted to a general frame.

\section{Nucleon}

Together with the Dirac index for the quark, the kernel $G_N\Pi_{N}$ in \eqref{G_N} and \eqref{Pi_N} is a $20 \times 20$ matrix, which can be reduced to $10\times10$ by projection to positive energy states. In rest frame, the projection is equivalent to the positive parity states. We directly diagonalize this kernel in the rest frame of the baryon by unitary transformations. It separates into a $6\times6$ block corresponding
to spin $1/2$ states (nucleon), and a $4 \times 4$ block for the spin $3/2$ states. 

In order to remove the Dirac matrices from the kernel $G_{N}\Pi_N$, we first apply the unitary transformation:

\begin{equation}
    G_N\rightarrow UG_N U=-\frac3M\begin{pmatrix}
        1 & -\sqrt{3} & -\sqrt{3} & -\sqrt{3} & -\sqrt{3} \\
        -\sqrt{3} & -1 &  1 &  1 &  1 \\
        -\sqrt{3} &  1 & -1 &  1 &  1 \\
        -\sqrt{3} &  1 &  1 & -1 &  1\\
        -\sqrt{3} &  1 &  1 &  1 & -1 \\
    \end{pmatrix}
\end{equation}
For simplicity, we have set the effective quark-diquark coupling to 1 and restore them at the end. The quark-diquark bubble function $\Pi_N$ in the rest frame after the same unitary transformation reads

\begin{equation}
\resizebox{0.9\textwidth}{!}{$
    \Pi_N\rightarrow U^\dagger\Pi_N U^\dagger=\begin{pmatrix}
        \Pi^+_0 & 0 & 0 & 0 & 0 \\
        0 & \Pi^+_1 & \Pi^+_2 & \Pi^+_2 & \Pi^+_2\\
        0 & \Pi^+_2 & \Pi^+_3 & 0 & 0 \\
        0 & \Pi^+_2 & 0 & \Pi^+_3 & 0\\
        0 & \Pi^+_2 & 0 & 0 & \Pi^+_3 \\
    \end{pmatrix}\frac{1-\gamma^0}2+\begin{pmatrix}
        \Pi^-_0 & 0 & 0 & 0 & 0 \\
        0 & \Pi^-_1 & \Pi^-_2 & \Pi^-_2 & \Pi^-_2\\
        0 & \Pi^-_2 & \Pi^-_3 & 0 & 0 \\
        0 & \Pi^-_2 & 0 & \Pi^-_3 & 0\\
        0 & \Pi^-_2 & 0 & 0 & \Pi^-_3 \\
    \end{pmatrix}\frac{1+\gamma^0}2
$}
\end{equation}
where unitary matrix is defined as $U=\mathrm{diag}(i\gamma^5,-\gamma^0,\gamma^1,\gamma^2,\gamma^3)$. The scalar part of the functions reads 
\begin{align}
    &\Pi^\pm_0(P^2)=-i\lambda^2_{qq_S}\int\frac{d^4k}{(2\pi)^4}\frac{\mp k^0-M}{k^2-M^2}\frac{1}{(P-k)^2-m^2_{qq_S}}\mathcal{F}(k)\\
    &\Pi^\pm_1(P^2)=-i\lambda^2_{qq_A}\int\frac{d^4k}{(2\pi)^4}\frac{\pm k^0-M}{k^2-M^2}\frac{1}{(P-k)^2-m^2_{qq_A}}\left(1-\frac{(P^0-k^0)^2}{m^2_{qq_A}}\right)\mathcal{F}(k)\\
    &\Pi^\pm_2(P^2)=-i\lambda^2_{qq_A}\int\frac{d^4k}{(2\pi)^4}\frac{\pm k^i}{k^2-M^2}\frac{1}{(P-k)^2-m^2_{qq_A}}\frac{(P-k)^0k^i}{m^2_{qq_A}}\mathcal{F}(k)\\
    &\Pi^\pm_3(P^2)=-i\lambda^2_{qq_A}\int\frac{d^4k}{(2\pi)^4}\frac{\mp k^0-M}{k^2-M^2}\frac{1}{(P-k)^2-m^2_{qq_A}}\left(1+\frac{(k^i)^2}{m^2_{qq_A}}\right)\mathcal{F}(k)
\end{align}

For the simplicity, we can solve the Faddeev equation in \eqref{baryon_eq} in rest frame and boost the solution to the general frame. The nucleon vertex function in the rest frame is defined as
\begin{equation}
\begin{aligned}
\label{vertex_N2}
    &\Gamma_N(S)=\begin{pmatrix}
        -i\alpha_1u_N(S) \\[4pt]
        \alpha_2\gamma^5u_N(S) \\[4pt]
        \alpha_3\sum_{\lambda,S'}\left\langle1\lambda;\frac12S'\big|\frac12S\right\rangle\varepsilon^1_\lambda u_N(S') \\[4pt]
        \alpha_3\sum_{\lambda,S'}\left\langle1\lambda;\frac12S'\big|\frac12S\right\rangle\varepsilon^2_\lambda u_N(S') \\[4pt]
        \alpha_3\sum_{\lambda,S'}\left\langle1\lambda;\frac12S'\big|\frac12S\right\rangle\varepsilon^3_\lambda u_N(S')
        
    \end{pmatrix}\\
    \rightarrow& U\Gamma_N(S)=\begin{pmatrix}
        \alpha_1\gamma^5u_N(S) \\[4pt]
        \alpha_2\gamma^5u_N(S) \\[4pt]
        \alpha_3\sum_{\lambda,S'}\left\langle1\lambda;\frac12S'\big|\frac12S\right\rangle\gamma^1\varepsilon^1_\lambda u_N(S') \\[4pt]
        \alpha_3\sum_{\lambda,S'}\left\langle1\lambda;\frac12S'\big|\frac12S\right\rangle\gamma^2\varepsilon^2_\lambda u_N(S') \\[4pt]
        \alpha_3\sum_{\lambda,S'}\left\langle1\lambda;\frac12S'\big|\frac12S\right\rangle\gamma^3\varepsilon^3_\lambda u_N(S') 
    \end{pmatrix}
\end{aligned}
\end{equation}
where the nucleon spinor in rest frame is defined as
\begin{equation}
    u_N(S)=\sqrt{m_N}\left(\begin{array}{c}
        \chi_S  \\
        \chi_S
    \end{array}\right)
\end{equation}
The last three terms are known as Clebsch Gordan projection $1\otimes\frac12\rightarrow\frac12$ for the vector–spinor combination. With the basis convention $\varepsilon^\mu_{\pm 1} = \left(0, \mp\frac{1}{\sqrt{2}}, -\frac{i}{\sqrt{2}}, 0 \right)$ and $\varepsilon^\mu_{0} = (0,0,0,1)$, they can be expressed in Lorentz covariant form as

\begin{equation}
    \sum_{\lambda, S'} 
\left\langle 1\lambda;\, \frac{1}{2} S' \,\middle|\, \frac{1}{2}, S \right\rangle 
\, \varepsilon^\mu_{\lambda}\, u_N(P,S')
= -\frac{1}{\sqrt{3}}\, \left(\gamma^\mu+\frac{P^\mu}{m_N} \right)\gamma^5\, u_N\!\left(P,S\right)
\end{equation}
with Wigner-Eckart theorem defined as

\begin{equation}
    \left\langle 1\lambda;\frac{1}{2}s' \middle| \frac{1}{2}s \right\rangle
=
\frac{(-1)^{\lambda+1}}{\sqrt{3}}
\left(\sigma^{[1]}_{-\lambda}\right)_{s,s'}
\end{equation}
and the spherical basis for Pauli matrices $\sigma^{[1]}_{-1}=-\sigma^-$, $\sigma^{[1]}_{0}=\sigma_3$, and $\sigma^{[1]}_{1}=\sigma^+$.

With this in mind, the vertex function in this basis is reducible. We can reduce the spinor by projecting into the irreducible basis spin-$1/2$ and spin-$3/2$. In this basis, it is clear to see the nucleon is a spin-$1/2$ couple to the scalar diquark and axial vector diquark.

\begin{equation}
    \Gamma_N(S)\rightarrow R U\Gamma_N(S)=\begin{pmatrix}
        \alpha_1 \\
        \alpha_2 \\
        \alpha_3 \\
        0 \\
        0 \\
    \end{pmatrix}\gamma^5u_N(S)
\end{equation}
with the normalization condition  $\alpha_1^2+\alpha_2^2+\alpha_3^2=1$.
The corresponding unitary matrix is defined as
\begin{equation}
    R=\mathrm{diag}(\left(\begin{array}{cc}
        1 & 0 \\
        0 & 1
    \end{array}\right),\left(\begin{array}{ccc}
        1/\sqrt{3} & 1/\sqrt{3} & 1/\sqrt{3} \\
        0 & -1/\sqrt{2} & 1/\sqrt{2} \\
        -2/\sqrt{6} & 1/\sqrt{6} & 1/\sqrt{6}
    \end{array}\right))
\end{equation}

The matrix transform the Faddeev kernel into block form 

\begin{equation}
\label{N_kernel}
    G_N\rightarrow RUG_NU R^\dagger=-\frac3M\left(\begin{array}{ccc|cc}
      1 & -\sqrt{3} & -3 & 0 & 0 \\
        -\sqrt{3} & -1 & \sqrt{3} & 0 & 0 \\
        -3 & \sqrt{3} & 1 & 0 & 0 \\
        \hline 
        0 & 0 & 0 & -2 & 0\\
        0 & 0 & 0 & 0 & -2 \\
    \end{array}\right)
\end{equation}
and
\begin{equation}
    \Pi_N\rightarrow RU^\dagger\Pi_NU^\dagger R^\dagger\frac{1-\gamma^0}2=\left(\begin{array}{ccc|cc}
      \Pi^+_0 & 0 & 0 & 0 & 0 \\
        0 & \Pi^+_1 & \sqrt{3}\Pi^+_2 & 0 & 0 \\
        0 & \sqrt{3}\Pi^+_2 & \Pi^+_3 & 0 & 0 \\
    \hline
        0 & 0 & 0 & \Pi^+_3 & 0\\
        0 & 0 & 0 & 0 & \Pi^+_3 \\
    \end{array}\right)\frac{1-\gamma^0}2
\end{equation}

It is natural to understand that the coefficient in rest frame stands for three components in the nucleon state: i) scalar diquark state couple to a quark ii) time component of the quark-axial-diquark state couple to a quark iii) spatial component of the axial-diquark state couple to a quark. The homogeneous equation for the 3 coefficients $\alpha_1, \alpha_2, \alpha_3$ leads to the eigenvalue equation for the nucleon mass. 

\begin{equation}
    -\frac3M\left(\begin{array}{ccc}
      1 & -\sqrt{3} & -3  \\
        -\sqrt{3} & -1 & \sqrt{3}  \\
        -3 & \sqrt{3} & 1 
    \end{array}\right)\left(\begin{array}{ccc}
      \Pi^+_0 & 0 & 0 \\
        0 & \Pi^+_1 & \sqrt{3}\Pi^+_2 \\
        0 & \sqrt{3}\Pi^+_2 & \Pi^+_3 \\
    \end{array}\right)\begin{pmatrix}
        \alpha_1 \\
        \alpha_2 \\
        \alpha_3 \\
    \end{pmatrix}=\begin{pmatrix}
        \alpha_1 \\
        \alpha_2 \\
        \alpha_3 \\
    \end{pmatrix}
\end{equation}

The largest eigenvalue determines the nucleon mass. If we neglect the mixing from axial vector diquark, the bound state equation reduces to a simple form

\begin{equation}
    -\frac3M \Pi^+_0(P^2=m^2_N)=1
\end{equation}

The covariant form of the nucleon vertex function can be obtained by boosting. The result reads

\begin{equation}
\begin{aligned}
    \Gamma_N(P,S)=&\begin{pmatrix}
        -i\alpha_1u_N(P,S) \\[4pt]
        \alpha_2\frac{P^\mu}{m_N}\gamma^5u_N(P,S) +\alpha_3\sum_{\lambda,s'}\left\langle1\lambda;\frac12S'\big|\frac12S\right\rangle\varepsilon^\mu_\lambda(P) u_N(P,S')
    \end{pmatrix}\\
    =&\begin{pmatrix}
        -i\alpha_1u_N(S) \\[4pt]
        \left(\alpha_2-\frac1{\sqrt{3}}\alpha_3\right)\frac{P^\mu}{m_N}\gamma^5u_N(P,S) -\frac{1}{\sqrt{3}}\alpha_3\gamma^\mu\gamma^5 u_N(P,S)
    \end{pmatrix}
\end{aligned}
\end{equation}
with the normalization condition
\begin{equation}
    \frac{1}{2P^0}\Gamma_N^\dagger(P,S)\Gamma_N(P,S)=\alpha_1^2+\alpha_2^2+\alpha_3^2=1
\end{equation}

\section{Delta baryon}
For the $\Delta$ baryon, the $16 \times 16$ matrix of the Faddeev kernel $G_\Delta\Pi_{\Delta}$ (see \eqref{G_Delta} and \eqref{Pi_Delta}) can be reduced to $8 \times 8$ matrix with positive parity projection. The direct diagonalization in the rest frame of the baryon separates the kernel into a $4\times4$ block corresponding
to spin $1/2$ states (nucleon), and a $4 \times 4$ block for the spin $3/2$ states. In order to remove the Dirac matrices, we first apply the unitary transformation:
\begin{equation}
    G_\Delta\rightarrow UG_\Delta U=\frac6M
    \begin{pmatrix}
         -1 & 1 & 1 & 1 \\
         1 & -1 & 1 & 1 \\
         1 & 1 & -1 & 1\\
         1 & 1 & 1 & -1 \\
    \end{pmatrix}
\end{equation}

\begin{equation}
    \Pi_\Delta\rightarrow U^\dagger\Pi_\Delta U^\dagger=
    \begin{pmatrix}
         \Pi_1^{+} & \Pi_2^{+} & \Pi_2^{+} & \Pi_2^{+} \\
         \Pi_2^{+} & \Pi_3^{+} & 0 & 0 \\
         \Pi_2^{+} & 0 & \Pi_3^{+} & 0\\
         \Pi_2^{+} & 0 & 0 & \Pi_3^{+} \\
    \end{pmatrix}\frac{1-\gamma^0}2+
    \begin{pmatrix}
         \Pi_1^{-} & \Pi_2^{-} & \Pi_2^{-} & \Pi_2^{-} \\
         \Pi_2^{-} & \Pi_3^{-} & 0 & 0 \\
         \Pi_2^{-} & 0 & \Pi_3^{-} & 0\\
         \Pi_2^{-} & 0 & 0 & \Pi_3^{-} \\
    \end{pmatrix}\frac{1+\gamma^0}2
\end{equation}
where $U=\mathrm{diag}(-\gamma^0,\gamma^1,\gamma^2,\gamma^3)$

In the rest frame, the vertex function for spin $\frac32$ is given by an expression. 
\begin{equation}
\label{vertex_Delta2}
    \Gamma_\Delta(S)=\begin{pmatrix}
        0 \\[4pt]
        \sum_{\lambda,s'}\left\langle1\lambda;\frac12s'\big|\frac32S\right\rangle\varepsilon^1_\lambda u_{s'}\\[4pt]
        \sum_{\lambda,s'}\left\langle1\lambda;\frac12s'\big|\frac32S\right\rangle\varepsilon^2_\lambda u_{s'}\\[4pt]
        \sum_{\lambda,s'}\left\langle1\lambda;\frac12s'\big|\frac32S\right\rangle\varepsilon^3_\lambda u_{s'}
        
    \end{pmatrix}\rightarrow U\Gamma_\Delta(S)=\begin{pmatrix}
        0 \\[4pt]
        
        \sum_{\lambda,s'}\left\langle1\lambda;\frac12s'\big|\frac32S\right\rangle\gamma^1\varepsilon^1_\lambda u_{s'} \\[4pt]
        \sum_{\lambda,s'}\left\langle1\lambda;\frac12s'\big|\frac32S\right\rangle\gamma^2\varepsilon^2_\lambda u_{s'}\\[4pt]
        \sum_{\lambda,s'}\left\langle1\lambda;\frac12s'\big|\frac32S\right\rangle\gamma^3\varepsilon^3_\lambda u_{s'}
    \end{pmatrix}
\end{equation}
where the Dirac spinor in the rest frame is defined as
\begin{equation}
    u_{s}=\sqrt{m_\Delta}\left(\begin{array}{c}
        \chi_s  \\
        \chi_s
    \end{array}\right)
\end{equation}
with the polarization basis convention 
\begin{align}
    \varepsilon^\mu_\pm=\mp(0,1,\pm i,0)/\sqrt{2}  && \varepsilon^\mu_0=(0,0,0,1)
\end{align}

The last three terms are known as Clebsch Gordan projection $1\otimes\frac12\rightarrow\frac32$ for the vector–spinor combination. The expression is similar to the last three components in \eqref{vertex_N2}, but with the diquark and quark spins coupled to total spin $3/2$ instead of $1/2$, giving a standard Rarita-Schwinger spinor form. By boosting, the covariant expression for Rarita–Schwinger spinor form reads

\begin{equation}
\begin{aligned}
\label{eq:RS_spinor}
&u^\mu_\Delta(P,S)=\sum_{\lambda,s'}\left\langle1\lambda;\frac12s'\big|\frac32S\right\rangle\varepsilon^\mu_\lambda(P) u_{s'}(P)
\end{aligned}
\end{equation}
where the spinor satisfies the projection

\begin{equation}
    u^\mu_\Delta(P,S)=
    \left(g_{\mu\nu}
- \frac{1}{3}\,\gamma_\mu \gamma_\nu
- \frac{1}{3 m_\Delta}\left(\gamma_\mu P_\nu - \gamma_\nu P_\mu\right)
- \frac{2}{3 m_\Delta^2} P^\mu P^\nu\right)u^\mu_\Delta(P,S)
\end{equation}

The expression of the vertex function in  \eqref{vertex_Delta2} is reducible. By projecting into the irreducible basis corresponding to spin-$1/2$ and spin-$3/2$, it is clear to see that $\Delta$ is a spin-$3/2$ state.
\begin{equation}
    \Gamma_\Delta(S=+3/2)\rightarrow R U\Gamma_\Delta(S=+3/2)=\frac12\begin{pmatrix}
        0 \\
        0 \\
        1 \\
        -\sqrt{3} \\
    \end{pmatrix}\gamma^5u_\downarrow
\end{equation}
and
\begin{equation}
    \Gamma_\Delta(S=+1/2)\rightarrow R U\Gamma_\Delta(S=+1/2)=\frac1{2}\begin{pmatrix}
        0 \\
        0 \\
        -\sqrt{3} \\
        -1 \\
    \end{pmatrix}\gamma^5u_\uparrow
\end{equation}

The corresponding unitary matrix is defined as
\begin{equation}
    R=\mathrm{diag}(1,\left(\begin{array}{ccc}
        1/\sqrt{3} & 1/\sqrt{3} & 1/\sqrt{3} \\
        0 & 1/\sqrt{2} & -1/\sqrt{2} \\
        -2/\sqrt{6} & 1/\sqrt{6} & 1/\sqrt{6}
    \end{array}\right))
\end{equation}

The unitary transform convert the Faddeev kernel into block form:
\begin{equation}
    G_\Delta\rightarrow RUG_\Delta U R^\dagger=\frac6M\left(\begin{array}{cc|cc}
         -1 & \sqrt{3} & 0 & 0 \\
         \sqrt{3} & 1 & 0 & 0 \\
        \hline 
         0 & 0 & -2 & 0\\
         0 & 0 & 0 & -2 \\
    \end{array}\right)
\end{equation}
and
\begin{equation}
    \Pi_\Delta\rightarrow RU\Pi_\Delta U R^\dagger\frac{1-\gamma^0}2=\left(\begin{array}{cc|cc}
         \Pi^+_1 & \sqrt{3}\Pi^+_2 & 0 & 0 \\
         \sqrt{3}\Pi^+_2 & \Pi^+_3 & 0 & 0 \\
    \hline
         0 & 0 & \Pi^+_3 & 0\\
         0 & 0 & 0 & \Pi^+_3 \\
    \end{array}\right)\frac{1-\gamma^0}2
\end{equation}
The diagonalization of the isospin-$3/2$ kernel proceeds in the same way by applying the unitary transformations as above. The positive parity part of the kernel then separates into a $2 \times 2$ block for $J=1/2$ (corresponding to the coupling of the time component and the space components of the axial vector diquark with the quark, respectively) and a diagonal $2 \times 2$ block
for $J=3/2$. The latter one has the same form as the lower $2 \times 2$ block of the kernel \eqref{N_kernel}, but with the factor $-2$ replaced by $4$ (from the isospin structure)

Now the delta baryon mass is determined by 
\begin{equation}
    -\frac{12}M\Pi^+_3(P^2=m^2_\Delta)=1
\end{equation}

\chapter{Light-cone wave functions}
\label{app:LCWF}

In this Appendix, we present the explicit forms of the light-cone wave functions following the framework of \cite{Ji:2003yj} using fermion creation operator $q^\dagger_{s,\alpha}(i)$ creating single quark state of flavor $q$ with spin $s$, color $\alpha$, and momentum $k^\mu_i=\left(x_i P^+,\frac{k_{i\perp}^2+M^2}{2x_iP^+},k_{i\perp}\right)$. The construction is carried out such that angular momentum conservation is preserved upon projection onto the light front. In particular, the total orbital angular momentum $l_z=\sum l_{zi}$ and the total parton helicity  $s_z=\sum s_{zi}$, combine to yield the hadron helicity, $\lambda=l_z+s_z$. Our formulation and classification are consistent with the general structure summarized in \cite{Ji:2003yj}


\section{Conventions used on the LF}
\label{Appx:LFspinor}
Throughout this paper, our conventions of the light front frame follows Kogut-Soper convention based on the Weyl chiral basis of the gamma matrices
\begin{eqnarray}
&\gamma^0=\begin{pmatrix}
0 & \mathds{1} \\
\mathds{1} & 0 \\
\end{pmatrix} ~\
&\gamma^{i}=\begin{pmatrix}
0 & \sigma^i \\
-\sigma^i  & 0 \\
\end{pmatrix}
\end{eqnarray}
The light front components are normalized to be
\begin{equation}
\gamma^\pm=\frac{\gamma^0\pm \gamma^3}{\sqrt{2}}
\end{equation}
with light front projection defined 
\begin{equation}
\frac{1}{2}\gamma^-\gamma^+=\begin{pmatrix}
1 & 0 & 0 & 0 \\
0 & 0 & 0 & 0 \\
0 & 0 & 0 & 0 \\
0 & 0 & 0 & 1 \\
\end{pmatrix}
\end{equation}
The LF spinors for the quarks are 
\begin{align}
\frac{1\pm\gamma^5}2u_s(k)=\frac{1}{\sqrt{\sqrt{2}\,k^+}}
\left[
\sqrt{2}\,k^+ \frac{\mathbf{1}\pm\sigma_3}{2}
+ M\,\mathbf{1}
\pm k_L \sigma^+
\pm k_R \sigma^-
\right]\chi_s
\end{align}
where Pauli matrices $\sigma^\pm=(\sigma_x+\pm i\sigma_y)/2$, $k_{L,R}=k^x_\perp\mp ik^y_\perp$, and $M$  the constituent quark mass.

In this Appendix, we present the explicit forms of the light-cone wave functions following the framework of \cite{Ji:2003yj}. The construction is carried out such that angular momentum conservation is preserved upon projection onto the light front. In particular, the total orbital angular momentum $l_z=\sum l_{zi}$ and the total parton helicity  $s_z=\sum s_{zi}$, combine to yield the hadron helicity, $\lambda=l_z+s_z$. Our formulation and classification are consistent with the general structure summarized in \cite{Ji:2003yj}


\section{Pseudoscalar meson}
We consider the $\pi^+$ as an illustrative example. The leading Fock-space component of the pion wave function is given by

\begin{equation}
\begin{aligned}
|\pi^+\rangle^{l_z=0} 
&= \int d[12]\, \psi^{l=0}_{\pi}(x,k_\perp)\, \frac{\delta_{\alpha\beta}}{\sqrt{N_c}}
\left[
u^\dagger_{\uparrow \alpha}(1)\, \bar d^\dagger_{\downarrow \beta}(2)
- 
u^\dagger_{\downarrow \alpha}(1)\, \bar d^\dagger_{\uparrow \beta}(2)
\right] |0\rangle
\\[6pt]
|\pi^+\rangle^{l_z=1} 
&= \int d[12]\, \psi^{l=1}_{\pi}(x,k_\perp)\, \frac{\delta_{\alpha\beta}}{\sqrt{N_c}}
\left[
k_{L}\, u^\dagger_{\uparrow \alpha}(1)\, \bar d^\dagger_{\uparrow \beta}(2)
+k_{R}\, u^\dagger_{\downarrow \alpha}(1)\, \bar d^\dagger_{\downarrow \beta}(2)
\right] |0\rangle
\end{aligned}
\end{equation}

where $k_{L,R}=k^x_\perp\mp ik^y_\perp$ and  $d[12]$ is the 2-body phase space integral defined in \eqref{eq:n-body} with fermion creation operator $q^\dagger_{s,\alpha}(i)$ creating single quark state of flavor $q$ with spin $s$, color $\alpha$, and momentum $k^\mu_i=\left(x_i P^+,\frac{k_{i\perp}^2+M^2}{2x_iP^+},k_{i\perp}\right)$.

The corresponding spatial wave functions in momentum space read

\begin{equation}
\begin{aligned}
    \psi^{l=0}_{\pi}(x,k_\perp)=&\frac{iM}{\sqrt{x\bar{x}}}\phi_\pi(x,k_\perp)\\
    \psi^{l=1}_{\pi}(x,k_\perp)=&-\frac{i}{\sqrt{x\bar{x}}}\phi_\pi(x,k_\perp)
\end{aligned}
\end{equation}

\section{Scalar meson}
For scalar mesons, we consider the $a_0^+$  as an illustrative example. The leading Fock-space component of the scalar meson wave function is given by 

\begin{equation}
\begin{aligned}
|a_0^+\rangle^{l_z=0} 
&= \int d[12]\, \psi^{l=0}_{a_0}(x,k_\perp)\, \frac{\delta_{\alpha\beta}}{\sqrt{N_c}}
\left[
u^\dagger_{\uparrow \alpha}(1)\, \bar d^\dagger_{\downarrow \beta}(2)
+ 
u^\dagger_{\downarrow \alpha}(1)\, \bar d^\dagger_{\uparrow \beta}(2)
\right] |0\rangle
\\[6pt]
|a_0^+\rangle^{l_z=1} 
&= \int d[12]\, \psi^{l=1}_{a_0}(x,k_\perp)\, \frac{\delta_{\alpha\beta}}{\sqrt{N_c}}
\left[
k_{L}\, u^\dagger_{\uparrow \alpha}(1)\, \bar d^\dagger_{\uparrow \beta}(2)
-k_{R}\, u^\dagger_{\downarrow \alpha}(1)\, \bar d^\dagger_{\downarrow \beta}(2)
\right] |0\rangle
\end{aligned}
\end{equation}

The corresponding wave functions in momentum space read

\begin{equation}
\begin{aligned}
    \psi^{l=0}_{a_0}(x,k_\perp)=&-\frac{M(x-\bar{x})}{\sqrt{x\bar{x}}}\phi_{a_0}(x,k_\perp)\\
    \psi^{l=1}_{a_0}(x,k_\perp)=&\frac{1}{\sqrt{x\bar{x}}}\phi_{a_0}(x,k_\perp)
\end{aligned}
\end{equation}
\section{Vector meson}

Strictly speaking, the $\rho$ meson is not an eigenstate of the QCD Hamiltonian but appears as a resonance. Nevertheless, it is still often treated as a quark–anti-quark bound state for practical purposes. As a vector meson, it has three helicity states $\lambda=0,\pm1$, corresponding to longitudinal and transverse polarizations. 

Here we first consider the longitudinal polarized $\rho^+$ for example. The leading Fock-space component of the rho meson wave function in the longitudinal polarization state is given by 

\begin{equation}
\begin{aligned}
|\rho^+,0\rangle^{l_z=0}
&= \int d[12]\, \psi^{l=0}_{\rho}(x,k_\perp)\, \frac{\delta_{\alpha\beta}}{\sqrt{N_c}}
\left[
u^\dagger_{\uparrow \alpha}(1)\, \bar d^\dagger_{\downarrow \beta}(2)
+
u^\dagger_{\downarrow \alpha}(1)\, \bar d^\dagger_{\uparrow \beta}(2)
\right] |0\rangle
\\[6pt]
|\rho^+,0\rangle^{l_z=1}
&= \int d[12]\, \psi^{l=1}_{\rho}(x,k_\perp)\, \frac{\delta_{\alpha\beta}}{\sqrt{N_c}}
\left[
k_{L}\, u^\dagger_{\uparrow \alpha}(1)\, \bar d^\dagger_{\uparrow \beta}(2)
-
k_{R}\, u^\dagger_{\downarrow \alpha}(1)\, \bar d^\dagger_{\downarrow \beta}(2)
\right] |0\rangle \, .
\end{aligned}
\end{equation}
where the corresponding spatial wave functions in momentum space read
\begin{equation}
\begin{aligned}
\psi^{l=0}
_\rho(x,k_\perp)=&-\sqrt{x\bar{x}}m_\rho\left(1+\frac{k_\perp^2+M^2}{m_\rho^2x\bar{x}}\right)\phi_\rho(x,k_\perp)\\
\psi^{l=1}_\rho(x,k_\perp)=&-\frac{m_\rho}{\sqrt{2}P^+}\frac{x-\bar{x}}{\sqrt{x\bar{x}}}\phi_\rho(x,k_\perp)
\end{aligned}
\end{equation}
The light cone wave functions for transverse polarization state read
\begin{equation}
    \begin{aligned}
|\rho^+,+1\rangle^{l_z=0}
&= \int d[12]\, \psi^{l=0}_{\rho}(x,k_\perp)\, \frac{\delta_{\alpha\beta}}{\sqrt{N_c}}
\left[
u^\dagger_{\uparrow i}(1)\, \bar d^\dagger_{\uparrow i}(2)
\right] |0\rangle
\\[6pt]
|\rho^+,+1\rangle^{l_z=1}
&= \int d[12]\,
\Bigg\{
\, \psi^{l=1}_{\rho,A}(1,2) \,k_{R} \, \frac{\delta_{\alpha\beta}}{\sqrt{N_c}}
\left[
u^\dagger_{\uparrow i}(1)\, \bar d^\dagger_{\downarrow i}(2)
+
u^\dagger_{\downarrow i}(1)\, \bar d^\dagger_{\uparrow i}(2)
\right]
\nonumber\\
&\hspace{2.8cm}
+
\psi^{l=1}_{\rho,S}(x,k_\perp) \,k_{R} \, \frac{\delta_{\alpha\beta}}{\sqrt{N_c}}
\left[
u^\dagger_{\uparrow i}(1)\, \bar d^\dagger_{\downarrow i}(2)
-
u^\dagger_{\downarrow i}(1)\, \bar d^\dagger_{\uparrow i}(2)
\right]
\Bigg\} |0\rangle
\\[6pt]
|\rho^+,+1\rangle^{l_z=2}
&= \int d[12]\, \psi^{l=2}_{\rho}(x,k_\perp)(k_{R})^2\, \frac{\delta_{\alpha\beta}}{\sqrt{N_c}}
\left[
u^\dagger_{\downarrow i}(1)\, \bar d^\dagger_{\downarrow i}(2)
\right] |0\rangle \, .
\end{aligned}
\end{equation}

where the corresponding wave functions in momentum space read
\begin{equation}
\psi^{l=0}_\rho(x,k_\perp)=-\frac{\sqrt{2}M}{\sqrt{x\bar{x}}}\phi_\rho(x,k_\perp)
\end{equation}

\begin{equation}
\psi^{l=1}_{\rho,A}(x,k_\perp)=-\frac{x-\bar{x}}{\sqrt{2x\bar{x}}}\phi_\rho(x,k_\perp)
\end{equation}

\begin{equation}
\psi^{l=1}_{\rho,S}(x,k_\perp)=-\frac{1}{\sqrt{2x\bar{x}}}\phi_\rho(x,k_\perp)
\end{equation}

\begin{equation}
\psi^{l=2}_{\rho}(x,k_\perp)=-\frac{2}{\sqrt{x\bar{x}}P^+}\phi_\rho(x,k_\perp)
\end{equation}

\section{Axial vector meson}
Finally in axial meson channel, we take $a_1^+$ state as an illustrative example. The leading Fock space component of the $a_1$ meson wave function in the longitudinal polarization state is given by

\begin{equation}
\begin{aligned}
|a_1^+,0\rangle^{l_z=0}
&= \int d[12]\, \psi^{l=0}_{a_1}(x,k_\perp)\, \frac{\delta_{\alpha\beta}}{\sqrt{N_c}}
\left[
u^\dagger_{\uparrow \alpha}(1)\, \bar d^\dagger_{\downarrow \beta}(2)
-
u^\dagger_{\downarrow \alpha}(1)\, \bar d^\dagger_{\uparrow \beta}(2)
\right] |0\rangle
\\[6pt]
|a_1^+,0\rangle^{l_z=1}
&= \int d[12]\, \psi^{l=1}_{a_1}(x,k_\perp)\, \frac{\delta_{\alpha\beta}}{\sqrt{N_c}}
\left[
k_{L}\, u^\dagger_{\uparrow \alpha}(1)\, \bar d^\dagger_{\uparrow \beta}(2)
+
k_{R}\, u^\dagger_{\downarrow \alpha}(1)\, \bar d^\dagger_{\downarrow \beta}(2)
\right] |0\rangle \, .
\end{aligned}
\end{equation}
where the corresponding wave functions in momentum space read
\begin{equation}
\begin{aligned}
\psi^{0,1}
_{a_1}(x,k_\perp)=&-\sqrt{x\bar{x}}m_{a_1}\left(1+\frac{k_\perp^2-M^2}{m_{a_1}^2x\bar{x}}\right)\phi_{a_1}(x,k_\perp)\\
\psi^{1,1}_{a_1}(x,k_\perp)=&-\frac{2M}{\sqrt{x\bar{x}}m_{a_1}}\phi_{a_1}(x,k_\perp)
\end{aligned}
\end{equation}
The transverse polarization state
\begin{equation}
    \begin{aligned}
|a_1^+,+1\rangle^{l_z=0}
&= \int d[12]\, \psi^{l=0}_{a_1}(x,k_\perp)\, \frac{\delta_{\alpha\beta}}{\sqrt{N_c}}
\left[
u^\dagger_{\uparrow i}(1)\, \bar d^\dagger_{\uparrow i}(2)
\right] |0\rangle
\\[6pt]
|a_1^+,+1\rangle^{l_z=1}
&= \int d[12]\,
\Bigg\{
\, \psi^{l=1}_{a_1,A}(1,2) \,k_{R} \, \frac{\delta_{\alpha\beta}}{\sqrt{N_c}}
\left[
u^\dagger_{\uparrow i}(1)\, \bar d^\dagger_{\downarrow i}(2)
+
u^\dagger_{\downarrow i}(1)\, \bar d^\dagger_{\uparrow i}(2)
\right]
\nonumber\\
&\hspace{2.8cm}
+
\psi^{l=1}_{a_1,S}(x,k_\perp) \,k_{R} \, \frac{\delta_{\alpha\beta}}{\sqrt{N_c}}
\left[
u^\dagger_{\uparrow i}(1)\, \bar d^\dagger_{\downarrow i}(2)
-
u^\dagger_{\downarrow i}(1)\, \bar d^\dagger_{\uparrow i}(2)
\right]
\Bigg\} |0\rangle
\\[6pt]
|a_1^+,+1\rangle^{l_z=2}
&= \int d[12]\, \psi^{l=2}_{a_1}(x,k_\perp)(k_{R})^2\, \frac{\delta_{\alpha\beta}}{\sqrt{N_c}}
\left[
u^\dagger_{\downarrow i}(1)\, \bar d^\dagger_{\downarrow i}(2)
\right] |0\rangle \, .
\end{aligned}
\end{equation}

where the corresponding wave functions in momentum space read
\begin{equation}
\begin{aligned}
\psi^{l=0}_{a_1}(x,k_\perp)=&-\frac{\sqrt{2}M}{\sqrt{x\bar{x}}}\phi_{a_1}(x,k_\perp)
\\
\psi^{l=1}_{a_1,A}(x,k_\perp)=&-\frac{x-\bar{x}}{\sqrt{2x\bar{x}}}\phi_{a_1}(x,k_\perp)\\
\psi^{l=1}_{a_1,S}(x,k_\perp)=&-\frac{1}{\sqrt{2x\bar{x}}}\phi_{a_1}(x,k_\perp)\\
\psi^{l=2}_{\rho}(x,k_\perp)=&-\frac{2}{\sqrt{x\bar{x}}P^+}\phi_{a_1}(x,k_\perp)
\end{aligned}
\end{equation}

\section{Proton}

We consider the proton $p^+$ as an illustrative example. Only the state with positive helicity will be discussed as the negative helicity state can be
obtained simply from the parity transformation. The leading Fock-space component of the proton wave function is given by

\begin{equation}
\begin{aligned}
|p\uparrow\rangle^{l_z=0}
&= \int d[123]\,
\left(
\psi_N^{(1)}(1,2,3)
+ i \epsilon_{\alpha\beta} k^\alpha_{1\perp} k_{2\perp}^\beta\,\psi^{(2)}_N(1,2,3)
\right)
\nonumber \\
&\quad \times
\frac{\epsilon^{\alpha\beta\gamma}}{\sqrt{2N_c}}\,
u^\dagger_{\alpha\uparrow}(1)
\left(
u^\dagger_{\beta\downarrow}(2)\,d^\dagger_{\gamma\uparrow}(3)
-
d^\dagger_{\beta\downarrow}(2)\,u^\dagger_{\gamma\uparrow}(3)
\right)
|0\rangle ,
\end{aligned}
\end{equation}

\begin{equation}
\begin{aligned}
|p\uparrow\rangle^{l_z=1}
&= \int d[123]\,
\left(
k_{1R}\,\psi^{(3)}_N(1,2,3)
+
k_{2R}\,\psi^{(4)}_N(1,2,3)
\right)
\nonumber \\
&\quad \times
\frac{\epsilon^{\alpha\beta\gamma}}{\sqrt{2N_c}}\,
\left(
u^\dagger_{\alpha\uparrow}(1)\,u^\dagger_{\beta\downarrow}(2)\,d^\dagger_{\gamma\downarrow}(3)
-
d^\dagger_{\alpha\uparrow}(1)\,u^\dagger_{\beta\downarrow}(2)\,u^\dagger_{\gamma\downarrow}(3)
\right)
|0\rangle .
\end{aligned}
\end{equation}

\begin{equation}
\begin{aligned}
|p\uparrow\rangle^{l_z=-1}
&= \int d[123]\,
k_{2L}\,\psi^{(5)}_N(1,2,3)
\nonumber \\
&\quad \times
\frac{\epsilon^{\alpha\beta\gamma}}{\sqrt{2N_c}}\,
u^\dagger_{\alpha\uparrow}(1)
\left(
u^\dagger_{\beta\uparrow}(2)\,d^\dagger_{\gamma\uparrow}(3)
-
d^\dagger_{\beta\uparrow}(2)\,u^\dagger_{\gamma\uparrow}(3)
\right)
|0\rangle ,
\\[1em]
|p\uparrow\rangle^{l_z=2}
&= \int d[123]\,
\left(k_{1R}\,k_{3R}\,\psi^{(6)}_N(1,2,3)+k_{2R}\,k_{3R}\,\psi^{(6)'}_N(1,2,3)\right)
\nonumber \\
&\quad \times
\frac{\epsilon^{\alpha\beta\gamma}}{\sqrt{2N_c}}\,
u^\dagger_{\alpha\downarrow}(1)
\left(
u^\dagger_{\beta\downarrow}(2)\,d^\dagger_{\gamma\downarrow}(3)-d^\dagger_{\beta\downarrow}(2)\,u^\dagger_{\gamma\downarrow}(3)
\right)
|0\rangle .
\end{aligned}
\end{equation}
where $k_{L,R}=k^x_\perp\mp ik^y_\perp$ and  $d[123]$ is the 3-body phase space integral defined in \eqref{eq:n-body}. With the useful identity $i\epsilon_{\alpha\beta} k^\alpha_{1\perp} k_{2\perp}^\beta k_{1R}=k_{1\perp}^2k_{2R}-k_{1\perp}\cdot k_{2\perp}k_{1R}$ and momentum conservation, we can rewrite the wave functions in \eqref{eq:LCWF_N} and \eqref{eq:LCWF_del} into these independent basis.

\begin{equation}
\begin{aligned}
    &\psi^{(1)}_N(1,2,3)=-
\frac{
2 \, M(M + m_N x_1)(x_2 + x_3)
}{\sqrt{x_1 x_2 x_3}}\alpha_1\, \phi_{N_s}(2,3,1)\\
&+
\frac{
2 x_1 k_{2\perp}^2  + 2k_{1\perp}\cdot k_{2\perp} (x_1 + x_3)
}
{\sqrt{x_1 x_2 x_3}}\alpha_1\, \phi_{N_s}(3,1,2)\\
&+\frac{4\left(M - m_N x_3\right)\left(k_{1\perp}\cdot k_{2\perp} - M^2 + m_N^2 x_1 x_2\right)}
{m_N \sqrt{x_1 x_2 x_3}}\left(\alpha_2-\frac{\alpha_3}{\sqrt{3}}\right)\phi_{N_a}(1,2,3)
\\
&+
\frac{2(M - m_N x_1)
\left(
k_{1\perp}\cdot k_{2\perp}
+ k^2_{2\perp}
+ M^2
- m_N^2 x_2 x_3
\right)}{m_N \sqrt{x_1 x_2 x_3}}
\left(\alpha_2-\frac{\alpha_3}{\sqrt{3}}\right)\phi_{N_a}(2,3,1)\\
&+
\frac{2M k^2_{2\perp}
}{m_N \sqrt{x_1 x_2 x_3}}\left(\alpha_2-\frac{\alpha_3}{\sqrt{3}}\right)\, \phi_{N_a}(3,1,2)\\
&-\frac{8\left(x_1k^2_{2\perp} + (x_1 + x_3)k_{1\perp}\cdot k_{2\perp} 
-M^2 x_3 - Mm_N x_1 x_2\right)
}
{\sqrt{x_1 x_2 x_3}}
\frac{\alpha_3}{\sqrt{3}}
\, \phi_{N_a}(1,2,3)\\
&-
\frac{4 \left(
x_1 k_{2\perp}^2
+ (x_1 + x_3) k_{1\perp}\cdot k_{2\perp} 
+M^2 x_1 + Mm_N x_2 x_3
\right)
}
{\sqrt{x_1 x_2 x_3}}\frac{\alpha_3}{\sqrt{3}}\, \phi_{N_a}(2,3,1)\\
&+
\frac{
4 M(M+x_2m_N)(x_1+x_3)}{\sqrt{x_1 x_2 x_3}}
\, \frac{\alpha_3}{\sqrt{3}}\phi_{N_a}(3,1,2)
\end{aligned}
\end{equation}

\begin{equation}
\begin{aligned}
&\psi^{(2)}_N(1,2,3)=
\frac{2(x_1 + x_3)}{\sqrt{x_1 x_2 x_3}}\alpha_1 \, \phi_{N_s}(3,1,2)+\frac{4(M - m_N x_3)}{m_N \sqrt{x_1 x_2 x_3}}\left(\alpha_2-\frac{\alpha_3}{\sqrt{3}}\right)\phi_{N_a}(1,2,3)\\
&+\frac{2(M - m_N x_1)}{m_N \sqrt{x_1 x_2 x_3}} \left(\alpha_2-\frac{\alpha_3}{\sqrt{3}}\right)\, \phi_{N_a}(2,3,1)-\frac{8 \, (x_1 + x_3)}{\sqrt{x_1 x_2 x_3}}\frac{\alpha_3}{\sqrt{3}}\phi_{N_a}(1,2,3)\\
&-
\frac{4(x_1 + x_3)}{\sqrt{x_1 x_2 x_3}}\frac{\alpha_3}{\sqrt{3}}\phi_{N_a}(2,3,1)
\end{aligned}
\end{equation}

\begin{equation}
    \begin{aligned}
\psi^{(3)}_N(1,2,3)=&-
\frac{
2(x_2M + m_N x_1x_2)
}{\sqrt{x_1 x_2 x_3}}
\alpha_1\,\phi_{N_s}(2,3,1)
\\
&-\frac{4\left(k_{2\perp}^2+M^2 
- m_N^2 x_1 x_2
\right)}{m_N \sqrt{x_1 x_2 x_3}}\left(\alpha_2-\frac{\alpha_3}{\sqrt{3}}\right)\phi_{N_a}(1,2,3)
\\
&+\frac{2M\left(M -m_N x_1\right)}
{m_N \sqrt{x_1 x_2 x_3}}\left(\alpha_2-\frac{\alpha_3}{\sqrt{3}}\right)\phi_{N_a}(2,3,1)\\
&+\frac{2k_{2\perp}^2}{m_N \sqrt{x_1 x_2 x_3}}
\left(\alpha_2-\frac{\alpha_3}{\sqrt{3}}\right)\phi_{N_a}(3,1,2)
\\
&
+\frac{8
m_N x_1 x_2}{\sqrt{x_1 x_2 x_3}}\frac{\alpha_3}{\sqrt{3}}\phi_{N_a}(1,2,3)-\frac{4 M x_1}{\sqrt{x_1 x_2 x_3}}\frac{\alpha_3}{\sqrt{3}}\phi_{N_a}(2,3,1)
\\[6pt]
&
+\frac{4\left(
 x_1M+ m_N x_1 x_2
\right)}{\sqrt{x_1 x_2 x_3}}\frac{\alpha_3}{\sqrt{3}}\phi_{N_a}(3,1,2)
\\[6pt]
\end{aligned}
\end{equation}

\begin{equation}
    \begin{aligned}
\psi^{(4)}_N(1,2,3)=&-
\frac{
2(x_2+x_3)\left(
M+ m_N x_1
\right)
}{\sqrt{x_1 x_2 x_3}}
\alpha_1\,\phi_{N_s}(2,3,1)-
\frac{
2M\left(
x_1 +x_3
\right)
}{\sqrt{x_1 x_2 x_3}}
\alpha_1\,\phi_{N_s}(3,1,2)\\
&
+\frac{4\left(
k_{1\perp}^2 + 2\, k_{1\perp} \cdot k_{2\perp}-M^2 + m_N^2 x_1 x_2
\right)}{m_N \sqrt{x_1 x_2 x_3}}\left(\alpha_2-\frac{\alpha_3}{\sqrt{3}}\right)\phi_{N_a}(1,2,3)
\\
&
-\frac{2\left(k_{1\perp}^2 + 2\, k_{1\perp} \cdot k_{2\perp}+M^2 - m_N^2 x_1 x_3
\right)}{m_N \sqrt{x_1 x_2 x_3}}\left(\alpha_2-\frac{\alpha_3}{\sqrt{3}}\right)\phi_{N_a}(3,1,2)
\\
&
+\frac{4(Mx_1 + m_N (x_1x_2 + x_1x_3))}{\sqrt{x_1 x_2 x_3}}\frac{\alpha_3}{\sqrt{3}}[2\phi_{N_a}(1,2,3)+\phi_{N_a}(3,1,2)]
\\[6pt]
\end{aligned}
\end{equation}

\begin{equation}
\begin{aligned}
    \psi^{(5)}_N(1,2,3)=&-\frac{2x_3(M+x_1m_N)}{\sqrt{x1 x2 x3}}\alpha_1\phi_{N_s}(2,3,1)\\
    &-\frac{2M (M-x_2 m_N)}{m_N \sqrt{x_1 x_2 x_3}}\left(\alpha_2-\frac{\alpha_3}{\sqrt{3}}\right)\phi_{N_a}(1,3,2)\\
    &+\frac{4M}{\sqrt{x_1 x_2 x_3}}\frac{\alpha_3}{\sqrt{3}}\phi_{N_a}(1,3,2)
 \end{aligned}
\end{equation}

\begin{equation}
\begin{aligned}
    \psi^{(6)}_N(1,2,3)=&-\frac{2x_2}{\sqrt{x_1x_2x_3}}\alpha_1\phi_{N_s}(2,3,1)- \frac{2M}{m_N \sqrt{x_1 x_2 x_3}}\left(\alpha_2-\frac{\alpha_3}{\sqrt{3}}\right)\phi_{N_a}(1,2,3)
\end{aligned}
\end{equation}

\begin{equation}
\begin{aligned}
    \psi^{(6)'}_N(1,2,3)=&-\frac{2M}{m_N \sqrt{x_1 x_2 x_3}}\left(\alpha_2-\frac{\alpha_3}{\sqrt{3}}\right)\phi_{N_a}(1,2,3)
\end{aligned}
\end{equation}

\section{Delta baryon}
For simplicity, we consider the delta-baryon $\Delta^{++}$ as an illustrative example. The leading Fock-space component of the proton wave function is given by

\begin{equation}
\begin{aligned}
|\Delta^{++}, \lambda = 3/2\rangle^{l_z=0}
&= \int d[123]\,
\left(
\psi^{(1)}_\Delta(1,2,3)
+ i \epsilon^{\alpha\beta} k_{1\alpha} k_{2\beta}\,\psi^{(2)}_\Delta(1,2,3)
\right)
\nonumber \\
&\quad \times
\frac{\epsilon^{\alpha\beta\gamma}}{\sqrt{2N_c}}\,
u^\dagger_{\alpha\uparrow}(1)\,
u^\dagger_{\beta\uparrow}(2)\,
u^\dagger_{\gamma\uparrow}(3)
|0\rangle ,
\\[1em]
|\Delta^{++}, \lambda= 3/2\rangle^{l_z=1}
&= \int d[123]\,
k_{1R}\,\psi^{(3)}_\Delta(1,2,3)
\nonumber \\
&\quad \times
\frac{\epsilon^{\alpha\beta\gamma}}{\sqrt{2N_c}}\,
u^\dagger_{\alpha\uparrow}(1)\,
u^\dagger_{\beta\uparrow}(2)\,
u^\dagger_{\gamma\downarrow}(3)
|0\rangle ,
\\[1em]
|\Delta^{++}, \lambda= 3/2\rangle^{l_z=2}
&= \int d[123]\,
\left(
k_{1R} k_{2R}\,\psi^{(4)}_\Delta(1,2,3)
+ k_{2R} k_{3R}\,\psi^{(5)}_\Delta(1,2,3)
\right)
\nonumber \\
&\quad \times
\frac{\epsilon^{\alpha\beta\gamma}}{\sqrt{2N_c}}\,
u^\dagger_{\alpha\uparrow}(1)\,
u^\dagger_{\beta\downarrow}(2)\,
u^\dagger_{\gamma\downarrow}(3)
|0\rangle ,
\\[1em]
|\Delta^{++}, \lambda= 3/2\rangle^{l_z=3}
&= \int d[123]\,
k_{1R}^2 k_{2R}\,
\psi^{(6)}_\Delta(1,2,3)
\nonumber \\
&\quad \times
\frac{\epsilon^{\alpha\beta\gamma}}{\sqrt{2N_c}}\,
u^\dagger_{\alpha\downarrow}(1)\,
u^\dagger_{\beta\downarrow}(2)\,
u^\dagger_{\gamma\downarrow}(3)
|0\rangle .
\end{aligned}
\end{equation}
and

\begin{equation}
\begin{aligned}
|\Delta^{++}, \lambda = 1/2\rangle^{l_z=0}
&= \int d[123]\,
\left(
\psi^{(1)}_\Delta(1,2,3)
+ i \epsilon^{\alpha\beta} k_{1\alpha} k_{2\beta}\,\psi^{(2)}_\Delta(1,2,3)
\right)
\nonumber \\
&\quad \times
\frac{\epsilon^{\alpha\beta\gamma}}{\sqrt{2N_c}}\,
u^\dagger_{\alpha\uparrow}(1)\,
u^\dagger_{\beta\uparrow}(2)\,
u^\dagger_{\gamma\downarrow}(3)
|0\rangle ,
\\[1em]
|\Delta^{++}, \lambda = 1/2\rangle^{l_z=1}
&= \int d[123]\,
k_{2R}\,\psi^{(3)}_\Delta(1,2,3)
\nonumber \\
&\quad \times
\frac{\epsilon^{\alpha\beta\gamma}}{\sqrt{2N_c}}\,
u^\dagger_{\alpha\uparrow}(1)\,
u^\dagger_{\beta\downarrow}(2)\,
u^\dagger_{\gamma\downarrow}(3)
|0\rangle ,
\\[1em]
|\Delta^{++}, \lambda = 1/2\rangle^{l_z=-1}
&= \int d[123]\,
k_{2L}\,\psi^{(4)}_\Delta(1,2,3)
\nonumber \\
&\quad \times
\frac{\epsilon^{\alpha\beta\gamma}}{\sqrt{2N_c}}\,
u^\dagger_{\alpha\uparrow}(1)\,
u^\dagger_{\beta\uparrow}(2)\,
u^\dagger_{\gamma\uparrow}(3)
|0\rangle ,
\\[1em]
|\Delta^{++}, \lambda = 1/2\rangle^{l_z=2}
&= \int d[123]\,
k_{1R} k_{2R}\,\psi^{(5)}_\Delta(1,2,3)
\nonumber \\
&\quad \times
\frac{\epsilon^{\alpha\beta\gamma}}{\sqrt{2N_c}}\,
u^\dagger_{\alpha\downarrow}(1)\,
u^\dagger_{\beta\downarrow}(2)\,
u^\dagger_{\gamma\downarrow}(3)
|0\rangle .
\end{aligned}
\end{equation}

For helicity $\lambda=+3/2$, the wave functions in momentum space reads

\begin{equation}
    \psi^{(1)}_\Delta(1,2,3)=-\frac{\sqrt{2} M\, \bar{x}_3\, (M+ m_N x_3)}{\sqrt{x_1 x_2 x_3}}\phi_\Delta(1,2,3)
\end{equation}

\begin{equation}
    \psi^{(2)}_\Delta(1,2,3)=0
\end{equation}

\begin{equation}
\begin{aligned}
    \psi^{(3)}_\Delta(1,2,3)=&-\frac{2\sqrt{2}\, M\, (x_1 + x_2)}{\sqrt{x_1 x_2 x_3}}\phi_\Delta(1,2,3)-\frac{2\sqrt{2}\, x_2\, (M+ m_N x_1)}{\sqrt{x_1 x_2 x_3}}\phi_\Delta(2,3,1)\\
    &-\frac{2\sqrt{2}\, x_1 \, (M+ m_N x_2)}{\sqrt{x_1 x_2 x_3}}\phi_\Delta(3,1,2)
\end{aligned}
\end{equation}

\begin{equation}
\begin{aligned}
    \psi^{(4)}_\Delta(1,2,3)=0
\end{aligned}
\end{equation}

\begin{equation}
\begin{aligned}
    \psi^{(5)}_\Delta(1,2,3)=&-\frac{\sqrt{2}\, x_1}{\sqrt{x_1 x_2 x_3}}\phi_\Delta(1,2,3)\\
    &-\frac{\sqrt{2}\, x_1}{\sqrt{x_1 x_2 x_3}}\phi_\Delta(3,1,2)
\end{aligned}
\end{equation}

\begin{equation}
\begin{aligned}
    \psi^{(6)}_\Delta(1,2,3)=0
\end{aligned}
\end{equation}

For helicity $\lambda=+1/2$, the wave functions in momentum space reads

\begin{equation}
\begin{aligned}
    &\psi^{(1)}(1,2,3)=-\sqrt{\frac{2}{3}}\frac{M\left(k_{3\perp}^2 + m_{\Delta}\bar{x}_3(M+m_{\Delta}x_3)\right)}{m_{\Delta}\sqrt{x_1x_2x_3}}\phi_\Delta(1,2,3)\\
    &-\sqrt{\frac{2}{3}}\frac{k_{1\perp}\cdot k_{3\perp}\left(M+m_{\Delta}(x_1-x_2)\right) + \left(M+m_{\Delta}x_1\right)\left(k_{3\perp}^2+M^2+m_{\Delta}^2x_2x_3\right)}{m_{\Delta}\sqrt{x_1x_2x_3}}\phi_\Delta(2,3,1)\\
    &-\sqrt{\frac{2}{3}}\frac{M^3+m_{\Delta}\left(k_{3\perp}^2x_1+M^2x_2\right)-k_{1\perp}\cdot k_{3\perp}\left(M-m_{\Delta}(x_1-x_2)\right)}{m_{\Delta}\sqrt{x_1x_2x_3}}\phi_\Delta(3,1,2)\\
    &-\sqrt{\frac{2}{3}}\frac{m_{\Delta}x_1x_3\left(M+m_{\Delta}x_2\right)}{\sqrt{x_1x_2x_3}}\phi_\Delta(3,1,2)
\end{aligned}
\end{equation}

\begin{equation}
\begin{aligned}
    \psi^{(2)}(1,2,3)=
    &\sqrt{\frac{2}{3}}\frac{M+m_{\Delta}(x_1-x_2)}{m_{\Delta}\sqrt{x_1x_2x_3}}\phi_\Delta(2,3,1)-\sqrt{\frac{2}{3}}\frac{M-m_{\Delta}(x_1-x_2)}{m_{\Delta}\sqrt{x_1x_2x_3}}\phi_\Delta(3,1,2)
\end{aligned}
\end{equation}

\begin{equation}
\begin{aligned}
\psi^{(3)}_\Delta(1,2,3)=&\sqrt{\frac{8}{3}}\frac{
k_{3\perp}^2 + M m_{\Delta} x_1 + m^2_{\Delta} x_1x_3}{m_{\Delta}\, \sqrt{x_1 x_2 x_3}}\phi_\Delta(1,2,3)\\
&+\sqrt{\frac{8}{3}}\frac{
k_{3\perp}^2+M^2 + m_{\Delta}^2 x_1 x_3}{m_{\Delta}\, \sqrt{x_1 x_2 x_3}}\phi_\Delta(3,1,2)\\
&+\sqrt{\frac{8}{3}}\frac{M\, (M + m_\Delta x_1)}{m_{\Delta}\, \sqrt{x_1 x_2 x_3}}\phi_\Delta(2,3,1)
\end{aligned}
\end{equation}

\begin{equation}
\begin{aligned}
    \psi^{(4)}_\Delta(1,2,3)=\sqrt{\frac{2}{3}}\frac{M^2 +Mm_{\Delta}(x_2-\bar{x}_2)}{m_{\Delta}\, \sqrt{x_1 x_2 x_3}}\phi_\Delta(3,1,2)
\end{aligned}
\end{equation}

\begin{equation}
\begin{aligned}
    \psi^{(5)}_\Delta(1,2,3)=&-\sqrt{\frac{2}{3}}\frac{M}{m_{\Delta}\,\sqrt{x_1 x_2 x_3}}\phi_\Delta(3,1,2)-\sqrt{\frac{2}{3}}\frac{M}{m_{\Delta}\,\sqrt{x_1 x_2 x_3}}\phi_\Delta(2,3,1)
\end{aligned}
\end{equation}

\chapter{Perturbative QCD remarks}
\label{App:pQCD}

\section{Cusp anomalous dimension}

Here we summarize the cusp anomalous dimension up to next-to-next-to-next-to-leading 
order (N$^3$LO) as~\cite{Korchemsky:1987wg,Moch:2004pa,Henn:2019swt,vonManteuffel:2020vjv} 
\begin{equation}
    \Gamma_{\mathrm{cusp}}(\alpha_s)=\sum_{n=1}^\infty \left(\frac{\alpha_s}{4\pi}\right)^n\Gamma_n\,,
\end{equation}
with the $n$-th order coefficients given by
\begin{equation}
    \Gamma_1=4C_F\,,
\end{equation}
\begin{equation}
\begin{aligned}
    \Gamma_2 =&8 C_F \left[ C_A \left( \frac{67}{18} - \frac{\pi^2}{6} \right) - \frac{5}{9} N_f \right]\\
    =&\frac{1072}{9} - \frac{16 \pi^2}{3} - \frac{160}{27} N_f\,,
\end{aligned}
\end{equation}
\begin{equation}
\begin{aligned}
    &\Gamma_3 = 352 \zeta(3) + \frac{176 \pi^4}{15} - \frac{2144 \pi^2}{9} + 1960 \\
    &+ N_f \left( -\frac{832 \zeta(3)}{9} + \frac{320 \pi^2}{27} - \frac{5104}{27} \right) - \frac{64}{81} N_f^2 \,,
\end{aligned}
\end{equation}

\begin{equation}
\begin{aligned}
    \Gamma_4 =& 
    \Bigg(-1536 \zeta(3)^2 - 704 \pi^2 \zeta(3) + 28032 \zeta(3) - \frac{34496 \zeta(5)}{3} 
    + \frac{337112}{9} \\
    &- \frac{178240 \pi^2}{27}+ \frac{3608 \pi^4}{5} - \frac{32528 \pi^6}{945}\Bigg)+ N_f \Bigg( 
    \frac{1664 \pi^2 \zeta(3)}{9} - \frac{616640 \zeta(3)}{81} \\
    &+ \frac{25472 \zeta(5)}{9} - \frac{1377380}{243} + \frac{51680 \pi^2}{81} - \frac{2464 \pi^4}{135} 
\Bigg)\\
&+ N_f^2 \Bigg( 
    \frac{16640 \zeta(3)}{81} + \frac{71500}{729} - \frac{1216 \pi^2}{243} - \frac{416 \pi^4}{405} 
\Bigg) + N_f^3 \Bigg( 
    \frac{256 \zeta(3)}{81} - \frac{128}{243} 
\Bigg)\,.\\
\end{aligned}
\end{equation}

The perturbative part of the CS kernel at minimal logarithmic scale $K^{(\rm pert)}_{\rm CS}(b_\perp=b_*, \mu=\mu_b)$ has been calculated up to four-loop order~\cite{Li:2016ctv,Vladimirov:2016dll,Moult:2022xzt,Duhr:2022yyp},
\begin{equation}
    K^{(\rm pert)}_{\rm CS}(b_*, \mu_b)=\sum_{n=1}^\infty\left(\frac{\alpha_s(\mu_b)}{4\pi}\right)^nk_n\,.
\end{equation}
 In this work, we use the perturbative CS kernel up to N$^3$LO,
\begin{align}
k_1 =& 0 \,,\\
k_2 =&  8 C_F \left[\left( \frac{7}{2} \zeta(3) - \frac{101}{27} \right) C_A + \frac{14}{27} N_f \right]\nonumber\\
=&-2(-56 \zeta(3) + \frac{1616}{27} - \frac{224}{81} N_f) \,,\\
k_3 =& -2\Bigg( \frac{176 \pi^2 \zeta(3)}{3} - \frac{24656 \zeta(3)}{9} + 1152 \zeta(5) \nonumber\\
&+ \frac{594058}{243} - \frac{6392 \pi^2}{81} - \frac{154 \pi^4}{45} \Bigg)  \nonumber\\
&-2 N_f \left( \frac{7856 \zeta(3)}{81} - \frac{166316}{729} + \frac{824 \pi^2}{243} + \frac{4 \pi^4}{405} \right) \nonumber\\
& -2 N_f^2 \left( \frac{64 \zeta(3)}{27} + \frac{3712}{2187} \right)\,.
\end{align}

\section{Small $b_T$ OPE for TMD}
The convolution is the operator product expansion (OPE) which factorize the small-$b$ behavior of the TMDs into the collinear PDFs $f^{q/h}$ convoluted with perturbative Wilson coefficients $C_{i/j}$ at scale $\mu_b$. The Sudakov factor accounts for the remaining scale dependence, evolving the TMD distribution back to $\mu$ from $\mu_b$. The perturbative Wilson coefficient for quarks up to $\mathcal{O}(\alpha_s)$ is given by \cite{Scimemi:2017etj}
\begin{equation}
\begin{aligned}
&C_{q/q}\left(x, b_T, \mu_b \right)=\delta(1-x)+\frac{\alpha_s}{4\pi}C_F \left[ -\frac{\pi^2}{6} \delta(1 - x) + 2(1 - x) \right]
\end{aligned}
\end{equation}


\end{document}